\documentclass[11pt]{cernrep} 
\usepackage{graphicx} 
\usepackage{here}
\usepackage{cite} 
\usepackage{amssymb}
\usepackage{amsmath}
\usepackage{epsf}
\usepackage{a4}
\usepackage{floatflt}  
\usepackage{fancybox}  
\usepackage{color}     
\usepackage{multirow}
\usepackage{myext}
\usepackage{psfig}
\usepackage{epsfig}
\usepackage{dcolumn}
\usepackage{psfrag}
\usepackage{rotating}
\usepackage{amsbsy}

\begin{document}

\bibliographystyle{unsrt}
\pagenumbering{arabic}
\pagestyle{plain}
\setcounter{page}{1}

\begin{center}
{\bf \Large \sc THE QCD/SM WORKING GROUP:}

\vspace{0.3cm}

{\bf \Large Summary Report}

\vspace{0.7cm}

Conveners:

\vspace{0.2cm}

{\sc 
W.~Giele$^{1}$, 
E.W.N.~Glover$^{2}$, 
I.~Hinchliffe$^{3}$, 
J.~Huston$^{4}$, 
E.~Laenen$^{5}$, 
E.~Pilon$^{6}$, 
and 
A.~Vogt$^{5}$.
}

\vspace{0.5cm}

Contributing authors:

\vspace{0.2cm}

{\sc
S.~Alekhin$^{7}$, 
C.~Bal\'azs$^{8}$, 
R.~Ball$^{9}$, 
T.~Binoth$^{9}$, 
E.~Boos$^{10}$,
M.~Botje$^{5}$, 
M.~Cacciari$^{11,12}$, 
S.~Catani$^{13}$, 
V.~Del~Duca$^{14}$, 
M.~Dobbs$^{15}$, 
S.D.~Ellis$^{16}$,
R.~Field$^{17}$,
D.~de~Florian$^{18}$, 
S.~Forte$^{19}$, 
E.~Gardi$^{13}$, 
T.~Gehrmann$^{13}$, 
A.~Gehrmann-De~Ridder$^{2}$, 
W.~Giele$^{1}$, 
E.W.N.~Glover$^{2}$, 
M.~Grazzini$^{20}$, 
J.-Ph.~Guillet$^{6}$, 
G.~Heinrich$^{2}$, 
J.~Huston$^{4}$, 
I.~Hinchliffe$^{3}$, 
V.~Ilyin$^{10}$, 
J.~Kanzaki$^{21}$,
K.~Kato$^{22}$, 
B.~Kersevan$^{23,24}$,
N.~Kidonakis$^{25}$, 
A.~Kulesza$^{26}$, 
Y.~Kurihara$^{27}$,
E.~Laenen$^{5}$, 
K.~Lassila-Perini$^{29}$,
L.~L\"onnblad$^{36}$,
L.~Magnea$^{28}$, 
M.~Mangano$^{13}$,
K.~Mazumudar$^{30}$,
S.~Moch$^{5}$, 
S.~Mrenna$^{1}$, 
P.~Nadolsky$^{25}$, 
P.~Nason$^{11}$, 
F.~Olness$^{25}$, 
F.~Paige$^{25}$, 
I~Puljak$^{31}$,
J.~Pumplin$^{4}$, 
E.~Richter-Was$^{32,33}$,
G.~Salam$^{34}$, 
R.~Scalise$^{25}$,  
M.~Seymour$^{35}$, 
T.~Sj\"ostrand$^{36}$, 
G.~Sterman$^{37}$, 
M.~T\"onnesmann$^{38}$, 
E.~Tournefier$^{39,40}$, 
W.~Vogelsang$^{37}$, 
A.~Vogt$^{5}$, 
R.~Vogt$^{3,41}$, 
B.~Webber$^{42}$, 
C.-P.~Yuan$^{4}$
and 
D.~Zeppenfeld$^{43}$.}

\vspace{0.7cm}

{\small
$^{1}$ FERMILAB, Batavia, IL-60650, U.S.A.\\
$^{2}$ Physics Dept., University of Durham, Durham, DH1 3LE, U.K.\\
$^{3}$ LBNL, Berkeley, CA 94720, U.S.A.\\
$^{4}$ Dept. of Physics and Astronomy, Michigan State University,
       East Lansing, MI 48824, U.S.A.\\ 
$^{5}$ NIKHEF, Theory Group, 1098 SJ, Amsterdam, The Netherlands.\\
$^{6}$ LAPTH, F-74941 Annecy-Le-Vieux, France.\\
$^{7}$ Inst. for High Energy Physics, Protvino, Moscow region 142281,
       Russia.\\
$^{8}$ Department of Physics and Astronomy, University of Hawaii, 
       Honolulu, HI 96822, U.S.A.\\
$^{9}$ Dept. of Physics and Astronomy, 
       University of Edinburgh, Edinburgh EH9 3JZ, Scotland.\\
$^{10}$  INP, Moscow State University, 119899 Moscow Russia.\\
$^{11}$ INFN, Sez. di Milano,I-20133 Milano, Italy.\\ 
$^{12}$ Dipartimento di Fisica, Universita di Parma, Italy.\\
$^{13}$ TH Division, CERN, CH-1211 Geneva 23, Switzerland.\\ 
$^{14}$ INFN, Sez. di Torino, I-10125 Torino, Italy\\  
$^{15}$ Dept. of Physics and Astronomy, University of Victoria, 
        Victoria, British  Columbia V8W 3P6, Canada.\\  
$^{16}$ Dept. of Physics, University of Washington, Seattle WA 98195, U.S.A.\\ 
$^{17}$ Department of Physics, University of Florida,
         Gainesville, FL 32611-8440, U.S.A. \\	   
$^{18}$ Departemento di Fisica, Universidad de Buenos Aires, Argentina.\\
$^{19}$ Dipartimento di Fisica "Edoardo Amaldi", Universita' di Roma 3, 
        INFN, 00146 Roma, Italy.\\
$^{20}$ INFN, Sez. di Firenze, I-50019 Sesto-Fiorentino, Firenze, Italy.\\
$^{21}$ IPNS, KEK Oho 1-1, Tsukuba Ibaraki 305-0081, Japan.\\ 
$^{22}$ Dept. of Physics Kogakuin, Univ. Nishi-Shinjuku, Tokyo,
        Japan.\\
$^{23}$ Faculty of Mathematics and Physics, University of Ljubljana, 
         SI-1000, Ljubljana, Slovenia.\\
$^{24}$ Jozeph Stefan Institute, SI-1000, Ljubljana, Slovenia.\\
$^{25}$ Dept. of Physics, Fondren Science Bldg., Southern Methodist University, 
        Dallas, TX 75275-0175, U.S.A.\\	
$^{26}$ Physics Dept., BNL, Upton, NY-19973, U.S.A.\\
$^{27}$ IPN, KEK, Tsukuba, Ibaraki 305-0081, Japan.\\
$^{28}$ Dipartimento di Fisica Teorica, University of Torino,  
        I-10125 Torino Italy.\\
$^{29}$ HIP, Helsinki, Finland.\\
$^{30}$ Experimental High Eneregy Physics Group, Tata Institute of Fundamental 
         Research, Mumbai 400 005, India.\\
$^{31}$ FESB, University of Split, 21 000 Split Croatia.\\
$^{32}$ Inst. of Computer Science, Jagellonian University, 30-055 Krak\'ov,    
        Poland.\\
$^{33}$ Henryk Niewodnicza$\tilde{\mbox{n}}$ski Institute of Nuclear Physics,
         High Energy Department, 30-055 Krak\'ov, Poland.\\
$^{34}$ LPTHE, Universit\'e Paris VI-VII, F-75252 Paris, France.\\
$^{35}$ Particle Physics Dept., Rutherford  Appelton Lab, 
        Chilton Didcot OX11 0QX, U.K.\\
$^{36}$ Department of Theoretical Physics, Lund University, 
        S-223 62 Lund, Sweden.\\
$^{37}$ SUNY, Stony Brook, NY 11794-3800, U.S.A.\\
$^{38}$ Max Planck Institut f\"r Physik (Werner-Heisenberg-Institut),
        80805 M\"unchen, Germany. \\
$^{39}$ ISN, F-338026 Grenoble Cedex, France.\\
$^{40}$ LAPP, F-74941 Annecy-le-Vieux Cedex, France.\\
$^{41}$ Dept. of Physics, Univ. of California, Davis, CA 95616, U.S.A.\\ 
$^{42}$ Cavendish Laboratory, Madingley Road Cambridge, Cambridge, CB3 0HE, U.K.\\
$^{43}$ Dept. of Physics, University of Winsconsin, Madison, WI 53706, U.S.A.
}

\vspace{0.3cm}

{\it Report of the Working Group on Quantum ChromoDynamics and the Standard 
Model for the Wokshop \\
\vspace{0.1cm}
``Physics at TeV Colliders", Les Houches, France, 21 May -- 1 June 2001.}
\end{center}

\vspace{0.3cm}

\begin{center}
{\bf \Large CONTENTS}
\end{center}

\noindent{\bf FOREWORD} \hfill 3\\
\indent

\noindent{\bf 1. PARTONS DISTRIBUTION FUNCTIONS} \hfill 4 \\
\indent Section coordinators: A.~Vogt and W.~Giele.\\
\indent Contributing authors: S.~Alekhin, M.~Botje, W.~Giele, J.~Pumplin, 
F.~Olness, G.~Salam, \\
\indent and A.~Vogt.

\vspace{0.3cm}

\noindent{\bf 2. HIGHER ORDERS} \hfill 23\\
\indent Section coordinator: E.W.N.~Glover.\\
\indent Contributing authors: T.~Binoth, V.~Del~Duca, T.~Gehrmann, 
A.~Gehrmann-De~Ridder, \\
\indent E.W.N.~Glover, J.-Ph.~Guillet and G.~Heinrich.

\vspace{0.3cm}

\noindent{\bf 3. QCD RESUMMATION} \hfill 46\\
\indent Section coordinator: E.~Laenen.\\
\indent Contributing authors: C.~Bal\'azs, R.~Ball, M.~Cacciari, S.~Catani, 
D.~de Florian, S.~Forte, \\
\indent E.~Gardi, M.~Grazzini, N.~Kidonakis, E.~Laenen, 
S.~Moch, P.~Nadolsky, P.Nason,\\
\indent A.~Kulesza, L.~Magnea, F.~Olness, R.~Scalise, 
G.~Sterman, W.~Vogelsang, R.~Vogt, \\
\indent and C.-P.~Yuan.

\vspace{0.3cm}

\noindent{\bf 4. PHOTONS, HADRONS AND JETS} \hfill 79\\
\indent Section coordinators: J.~Huston and E.~Pilon.\\
\indent Contributing authors: T.~Binoth, J.P.~Guillet, S.~Ellis, J.~Huston, 
K.~Lassila-Perini, \\
\indent M.~T\"onnesmann and E.~Tournefier.

\vspace{0.3cm}
 
\noindent{\bf 5. MONTE CARLO} \hfill 95\\
\indent Section coordinator: I.~Hinchliffe and J.~Huston.\\
\indent Contributing authors: C.~Bal\'azs, E.~Boos, M.~Dobbs, W.~Giele, 
I.~Hinchliffe, R.~Field, \\
\indent J.\ Huston, V.~Ilyin, J.~Kanzaki, B.~Kersevan, K.~Kato, Y.~Kurihara, 
L.~L\"onnblad,  \\
\indent M.~Mangano, K.~Mazumudar, S.~Mrenna, F.~Paige, I~Puljak, 
E.~Richter-Was, \\
\indent M.~Seymour, T.~Sj\"ostrand, M.~T\"onnesmann, B.~Webber, 
and D.~Zeppenfeld.

\vspace{0.3cm}
 
\noindent{\bf References} \hfill 139

\noindent
{\bf FOREWORD}

\vspace{0.3cm}

\noindent
Quantum Chromo-Dynamics (QCD), and more generally the physics of the Standard
Model (SM), enter in many ways in high energy processes at TeV Colliders, and
especially in hadron colliders (the Tevatron at Fermilab and the forthcoming
LHC at CERN), 

First of all, at hadron colliders, QCD controls the parton luminosity, which
rules the production rates of any particle or system with large invariant mass
and/or large transverse momentum.  Accurate predictions for any signal of
possible `New Physics' sought at hadron colliders, as well as the
corresponding backgrounds, require an improvement in the control of
uncertainties on the determination of PDF and of the propagation of these
uncertainties in the predictions. Furthermore, to fully exploit these new
types of PDF with uncertainties, uniform tools (computer interfaces,
standardization of the PDF evolution codes used by the various groups fitting 
PDF's) need to be proposed and developed.

The dynamics of colour also affects, both in normalization and shape, various
observables of the signals of any possible `New Physics' sought at the TeV
scale, such as, e.g. the production rate, or the distributions in transverse
momentum of the Higgs boson. Last, but not least, QCD governs many
backgrounds to the searches for this `New Physics'. Large and important QCD
corrections may come from extra hard parton emission (and the corresponding
virtual corrections), involving multi-leg and/or multi-loop amplitudes.  This
requires complex higher order calculations, and new methods have to be
designed to compute the required multi-legs and/or multi-loop corrections in a
tractable form. In the case of semi-inclusive observables, logarithmically
enhanced contributions coming from multiple soft and collinear gluon emission
require sophisticated QCD resummation techniques. Resummation is a catch-all
name for efforts to extend the predictive power of QCD by summing the large
logarithmic corrections to all orders in perturbation theory. In practice, the
resummation formalism depends on the observable at issue, through the type of
logarithm to be resummed, and the resummation methods. 

In parallel with this perturbative QCD-oriented working programme, the
implementation of both QCD/SM and New physics in Monte Carlo event generators 
is confronted with a number of issues which deserve uniformization or
improvements. The important issues are: 1) the problem of interfacing partonic
event generators to showering Monte-Carlos; 2) an implementation using this
interface to calculate backgrounds which are poorly simulated by the showering
Monte-Carlos alone; 3) a comparison of the {\tt  HERWIG} and {\tt PYTHIA}
parton shower models with the predictions of soft gluon resummation; 4)
studies of the underlying events at hadron colliders to check how well they
are modeled by the Monte-Carlo generators.

In this perspective, our Working Group devoted its activity to improvements of
the various QCD/SM ingredients relevant both for searches of `New Physics' and
estimates of the backgrounds to the latter at TeV colliders. This report
summarizes our work. Section 1 reports on the effort towards precision Parton
Distribution Functions (PDF's). Section 2 presents the issues worked out along
the two current frontiers of Higher Order QCD calculations at colliders, namely
the description of multiparton final states at Next-to-Leading Order (NLO), and
the extension of  calculations for precison observables beyond this order.
Section 3  `resummarizes\footnote{E.~Laenen \copyright, all rights reserved.}'
a large variety of questions concerning the relevance of resummation for
observables at TeV colliders. In parallel with these `general purpose tackling
angles', more specific studies, dedicated to jet physics and improved cone
algorithms, and to the QCD backgrounds to $H \to \gamma \gamma$, both
irreducible (isolated prompt photon pairs) and reducible (photon pion), are
presented in section 4. Finally, section 5 summarizes the activities of the
Intergroup on Monte Carlo issues, which are of practical interest for all three
Working Groups of the Workshop: HIGGS \cite{Cavalli:2002vs},  BSM
\cite{Azuelos:2002qw} and the present one.

%
%
\section{PARTON DISTRIBUTION FUNCTIONS\protect\footnote{Section
coordinators: A.~Vogt, W.~Giele}$^{,}$~\protect\footnote{Contributing authors:
S.~Alekhin, M.~Botje, W.~Giele, J.~Pumplin, F.~Olness, G.~Salam, A.~Vogt}}
\label{sec:pdf,qcdsm}

The experimental uncertainties in current and future hadronic colliders are 
decreasing to a level where more careful consideration has to be given to the 
uncertainties in the theoretical predictions. One important source of these 
uncertainties has its origin in Parton Distribution Functions (PDFs). The PDF 
uncertainties in turn are a reflection of those in the experimental data used 
as an input to the PDF fits and in the uncertainties of the theoretical 
calculations used to fit those data. As a consequence, sophisticated 
statistical methods and approximations have to be developed to handle the 
propagation of uncertainties. 
We will give a summary of the current status of several methods being pursued. 
To fully exploit these new types of PDF fits a uniform computer interface has 
been defined and developed. This code provides easy access to all the present 
and future fits. The code is available from the website {\bf\tt pdf.fnal.gov}. 
Such an interface is most efficient if the evolution codes of the various 
groups fiiting the PDFs are standardized to a sufficient degree. For this 
purpose detailed and accurate reference tables for the PDF evolution are 
provided at the end of this section. 

\subsection{Methods for estimating parton distribution function uncertainties
\protect\footnote{Contributing authors:
S.~Alekhin, M.~Botje, W.~Giele, J.~Pumplin, F.~Olness}}

\subsubsection{A mathematical framework}

At first sight PDF fitting is rather straightforward. However, a more
detailed look reveals many difficult issues. As the PDF uncertainties
will affect all areas of phenomenology at hadron colliders, 
a clear mathematical framework of a PDF fit is essential \cite{Giele:2001mr}. 
From this
formulation, all explicit methods can be derived. Also, the mathematical
description will make explicit all assumptions needed before one can 
make a fit. These assumptions are crucial and do not find their origin
in experimental results,  but rather in theoretical prejudice. Such 
assumptions are unavoidable as we fit a system with an infinite 
number of degrees of freedom (the PDF functional) 
to a finite set of data points.

We want to calculate the probability density function $P^{\cal O}_{pdf}$
which reflects the uncertainty in predicting the observable $\cal O$
due to the PDF uncertainties. The function $P^{\cal O}_{pdf}(x_e)$ gives 
the probability density to measure a value $x_e$ for observable $\cal O$.

To calculate the PDF probability density function for observable $\cal O$
we have to integrate over the functional space of all possible PDFs $V({\cal F})$.
The integration is weighted by three probability density functions:
the prior probability density function, $P_{prior}$, 
the experimental response function of the observable, $P_{exp}^{\cal O}$
and the probability density function of the fitted experiments, 
$P_{exp}^{input}$. The resulting formula is given by
\begin{equation}\label{eqn:pdf;qcdsm;Ifunctional}
P_{pdf}^{\cal O}(x_e) = \int_{V({\cal F})} d\,{\cal F}\ P_{prior}({\cal F})\times 
P_{exp}^{input}({\cal F})\times P_{exp}^{\cal O}\left(x_e|x_t({\cal F})\right)\ .
\end{equation}

The prior probability density function contains theoretical constraints
on the PDF functionals such as sum rules and other potential
fit constraints (e.g. an explicit $\alpha_S(M_Z)$ value).
The most crucial property of the prior function is that it  
defines the functional integral by imposing smoothness constraints
to make the number of degrees of freedom become finite. The simplest example
is an explicit finite parametrization of the PDF functionals
\begin{equation}
P_{pdf}^{\cal O}(x_e) = \int_{V(\{\lambda\})}\!\!\!\!\! 
d\,\lambda_1\,d\,\lambda_2\ldots\,d\,\lambda_n\ 
P_{prior}(\{\lambda\})\times P_{exp}^{input}(\{\lambda\})\times 
P_{exp}^{\cal O}\left(x_e|x_t(\{\lambda\})\right)\ ,
\end{equation}
where the PDF parameters are given by the list $\{\lambda\}$. Note that
through the functional parametrization choice we have restricted
the integration to a specific subset of potential PDF functionals \mbox{$\cal F$}.
Such a choice is founded on physics assumptions with no {\it a priori}
justification. The resulting phenomenology depends on this assumption.

The experimental response function forms the interface with the experimental
measurement. It gives the probability density to
measure a value $x_e$ given a prediction $x_t$.
The experimental response functions contain all information about
the experiment needed to analyze the measurement.
The prediction $x_t$ is an 
approximation using a perturbative estimate of the observable
amended with relevant nonperturbative corrections.
A simple example would be
\[ P_{exp}^{\cal O}(x_e|x_t ) =\frac{1}{\delta\sqrt{2\pi}}
\exp{\left(-\frac{1}{2}(x_e-x_t)^2/\delta^2\right)}\ ,\]
where we have a Gaussian response function 
with a one sigma uncertainty $\delta$.
Note that the form of the response function depends on the actual
experiment under consideration. It is sometimes convenient to
get a result that is independent of an experiment and its particular
detector: To obtain the
theoretical prediction for the probability density function of the
observable, one can simply replace
the experimental response function by a delta function
(i.e. assume a perfect detector)
\[P_{exp}^{theory}(x_e|x_t) = \delta(x_e-x_t)\ .\]

Finally the probability function for the fitted experiments is simply
a product of all the experimental response functions
\begin{equation}
P_{exp}^{input}({\cal F}) = 
\prod_i^{N_{exp}} P_{exp}^{(i)}(x_m^{(i)}|x_t^{(i)}({\cal F})) \ ,
\end{equation} 
where $x_m^{(i)}$ denotes the set of measurements provided by
experiment $(i)$.
This function measures the probability density that
the theory prediction based on PDF \mbox{$\cal F$}\ 
describes the combined experimental measurements.

Often the experimental uncertainties can be approximated by a Gaussian
form of the experimental reponse function, i.e. a $\chi^2$ 
description of the uncertainties:
\begin{equation}
P_{pdf}^{\cal O}(x_e)\propto \int_{V(\{\lambda\})}\!\!\!\!\! 
d\,\lambda_1\,d\,\lambda_2\ldots\,d\,\lambda_n\ 
e^{-\frac{1}{2}\sum_i \chi_i^2(x_m^{(i)}-x_t^{(i)}(\{\lambda\}))}\times 
e^{-\frac{1}{2}\chi_{\cal O}^2(x_e^{(i)}-x_t(\{\lambda\}))}\ ,
\end{equation}
where we have chosen a specific parametrization for the PDF functionals.
This approximation leads to a more traditional approach for determining
PDFs with uncertainties. These methods are outlined in sections
\ref{sec:pdf;qcdsm;covariance} and \ref{sec:pdf;qcdsm;offset}.
To go beyond the Gaussian approximation more elaborate methods are needed.
Sections \ref{sec:pdf;qcdsm;ransam} and \ref{sec:pdf;qcdsm;lagrange} 
describe techniques to accomplish this.

\subsubsection{Random sampling method}\label{sec:pdf;qcdsm;ransam}

This method attempts to calculate Eq. (\ref{eqn:pdf;qcdsm;Ifunctional}) without
any approximations by using random sampling, i.e. a Monte Carlo evaluation of 
the integral \cite{Giele:2001mr}. By generating a sufficiently 
large sample of PDFs Eq. (\ref{eqn:pdf;qcdsm;Ifunctional})
can be approximated by
\begin{equation}\label{eqn:pdf;qcdsm;Irandomsampling}
P_{pdf}^{\cal O}(x_e)\approx\frac{1}{N}
\sum_{i=1}^N P_{prior}({\cal F}_i)\times P_{exp}^{input}({\cal F}_i)\times  
P_{exp}^{\cal O}\left(x_e|x_t({\cal F}_i)\right)\ .
\end{equation}

The simple implementation of Eq. (\ref{eqn:pdf;qcdsm;Irandomsampling})
would lead to a highly inefficient
random sampling and an unreasonably large number of PDFs would be
required. By using an optimization procedure, this problem can be
solved. The optimization procedure on the functional level of
Eq. (\ref{eqn:pdf;qcdsm;Ifunctional}) is simply redefining the PDF \mbox{$\cal F$}\ to ${\cal F}^u$
such that the Jacobian of the transformation equals
\begin{equation}
\label{eqn:pdf;qcdsm;transform}
\frac{d\,{\cal F}^u}{d\,{\cal F}} = P_{prior}({\cal F})\times P_{exp}^{input}({\cal F})\ ,
\end{equation}
so that
\begin{equation}
P_{pdf}^{\cal O}(x_e) = 
\int_{V({\cal F}^u)}d\,{\cal F}^u\ P_{exp}^{\cal O}\left(x_e|x_t({\cal F}^u)\right)\ .
\end{equation}
Applying this to the random sampling evaluation gives
\begin{equation}\label{eqn:pdf;qcdsm;density}
P_{pdf}^{\cal O}(x_e)\approx\frac{1}{N}\sum_{i=1}^N 
P_{exp}^{\cal O}\left(x_e|x_t({\cal F}_i^u)\right)\ .
\end{equation}
In the random sampling approximation the redefined PDFs ${\cal F}_i^u$ 
are easily identified as the unweighted PDFs with respect to
the combined probability density 
$P_{prior}({\cal F})\times P_{exp}^{input}({\cal F})$. That is, the density of
${\cal F}_i^u$ is given by this combined probability density. As such,
each of the unweighted PDFs is equally likely.  
This is reflected in Eq. (\ref{eqn:pdf;qcdsm;density}), 
as the probability density function
of the observable is the average of the response function over
the unweighted PDFs.
Finally, we have to generate the set $\{{\cal F}_i^u\}$. 
One method is to use the Gaussian approximation for 
Eq. (\ref{eqn:pdf;qcdsm;transform}) simplifying the 
generation of the set $\{{\cal F}_i^u\}$ \cite{Giele:1998gw}. 

Another more general approach is to apply a Metropolis Algorithm on the
combined probability density function 
$P_{prior}({\cal F})\times P_{exp}^{input}({\cal F})$.
This approach will handle any probability function. Furthermore, in a
Metropolis Algorithm approach convergence does not depend on the number
of parameters used in the PDF parametrization. Also, complicated 
non-linear parameter subspaces which have constant probability will
be modelled correctly. These properties make it possible to use
large number of parameters and explore the issue of parametrization
dependence of the PDFs.

Once the set $\{{\cal F}_i^u\}$ is generated we can predict the probability
density function for any observable by averaging over the experimental
response function using Eq. (\ref{eqn:pdf;qcdsm;density}).

\subsubsection{Lagrange multiplier method}\label{sec:pdf;qcdsm;lagrange}

The values of the fit parameters that minimize $\chi^{2}$ provide by 
definition the best fit to the global set of data.  The dependence of 
$\chi^{2}$ on those parameters in the neighborhood 
of the minimum can be characterized in quadratic approximation by the 
matrix of second derivatives, which is known as the Hessian matrix.
The inverse of the Hessian matrix is the error matrix, and it 
forms the basis for traditional estimates of the uncertainties.

The traditional assumption of quadratic behavior of $\chi^{2}$ in all 
directions in parameter space is not necessarily a very good one in 
the case of PDF fitting, because there are ``flat directions'' in which 
$\chi^{2}$ changes very slowly, so large changes in certain combinations 
of parameters are not ruled out.  This difficulty is always present in 
the case of PDF fitting, because as more data become available to pin 
down the PDFs better, more flexible parametrizations of them are 
introduced to allow ever-finer details to be determined.

To some extent, the flat directions can be allowed for by an 
{\it iterative method} \cite{Pumplin:2000vx,Pumplin:2001ct}, whereby the 
eigenvector directions of the error matrix and their eigenvalues 
are found using a convergent series of successive approximations.
This iterative method is implemented as an extension to the standard 
minimization program {\tt minuit}.
The result is a collection of PDFs (currently 40 in number) that 
probe the allowed range of possibilities along each of the 
eigenvector directions.  The PDF uncertainty on any physical 
quantity---or on some feature of the PDFs themselves---can easily 
be computed after that quantity is evaluated for each of the 
eigenvector sets.  This method has been applied, for example, 
to find the uncertainty range for predictions of $W$ and $Z$ cross 
sections and their correlations \cite{Pumplin:2001ct}.

To completely overcome the need to assume quadratic behavior for 
$\chi^{2}$ as a function of the fitting parameters, one can use a 
{\it Lagrange Multiplier method} \cite{Pumplin:2000vx,Stump:2001gu} to directly 
examine how $\chi^{2}$ varies as a function of any particular variable 
that is of interest.  
The method is a classical mathematical technique that is more fully 
called Lagrange's ``method of undetermined multipliers.''  To explain 
it by example, suppose we want to find the effect of PDF uncertainties 
on the prediction for the Higgs boson production cross section.  
Instead of finding the PDF parameters that minimize $\chi^{2}$ (which 
measures the quality of fit to the global data), one minimizes 
$\chi^{2} + \lambda \sigma_H$, where $\sigma_H$ is the predicted 
Higgs cross section---which is of course also a function of the PDF
parameters.  The minimization is carried out for a variety of values 
of the Lagrange Multiplier constant $\lambda$.
Each minimization provides one point on the curve of $\chi^{2}$ versus 
predicted $\sigma_H$.  Once that curve has been mapped out, the 
uncertainty range is defined by the region for which the increase in 
$\chi^{2}$ above its minimum value is acceptable.

The essential feature of the Lagrange Multiplier method is that it 
finds the largest possible range for the predictions of a given 
physical quantity, such as $\sigma_H$, that are consistent with 
any given assumed maximum increase $\Delta\chi^{2}$ above the 
best-fit value of $\chi^{2}$.  This method has been applied, 
for example, to study the possible range of variation of the 
rapidity distribution for $W$ production, by extremizing various 
moments of that distribution \cite{Pumplin:2000vx,Stump:2001gu}.

\subsubsection{Covariance matrix method}\label{sec:pdf;qcdsm;covariance}

The covariance matrix method  
is based on the Bayesian approach to the treatment of the systematic errors, 
when the latter are considered as a random fluctuation like the 
statistical errors. 
Having no place for the detailed discussion of the advantages of this 
approach we refer to the 
introduction into this scope given in Ref.~\cite{D'Agostini:1995fv};
the only point we would like to underline here is 
that application of the Bayesian approach is especially justified 
in the analysis of the data sets with the numerous
sources of independent systematic errors, which is the case for the 
extraction of PDFs from existing experimental data.

Let the data sample have one common
additive systematic error\footnote{We consider the case of one source 
of systematic errors, generalization
on the many sources case is straightforward.}.
In this case following the Bayesian approach the measured values 
$y_i$ are given by   
\begin{equation}
y_i=f_i+\mu_i \sigma_i+\lambda s_i,
\label{eqn:pdf;qcdsm;CADDSET}
\end{equation}
where $f_i(\theta^0)$ is the theoretical model describing the data,
$\theta^0$  -- the fitted parameter of this model,
$\sigma_i$ -- statistical errors, 
$s_i$  -- systematic errors for each point, $\mu_i$ and $\lambda$
-- the independent random variables, and the index $i$ runs 
through the data points from 1 to $N$.
The only assumption we make
is that the average and the dispersion of these variables are zero and unity 
respectively. 
It is natural to assume that the $\mu_i$ are Gaussian distributed when the
data points are obtained from large statistical samples. Such an
assumption is often not justified for the distribution of $\lambda$

Within the covariance matrix approach the estimate of the 
fitted parameter $\hat\theta$ is obtained 
from a minimization of the functional 
\begin{equation}
\chi^2(\theta)=\sum_{i,j=1}^{N} (f_i(\theta)-y_i) E_{ij} (f_j(\theta)-y_j),
\label{eqn:pdf;qcdsm;CORCHI}
\end{equation}
where 
\begin{displaymath}
E_{ij}=\frac{1}{\sigma_i \sigma_j}
\Bigl(\delta_{ij} -
\frac{\rho_i\rho_j}{1+\rho^2}\Bigr)\ ,
\end{displaymath}
is the inverse of the covariance matrix  
\begin{equation}
C_{ij} =  s_i s_j+\delta_{ij}\sigma_i\sigma_j,
\label{eqn:pdf;qcdsm;COVA}
\end{equation}
$\rho$ is modulus of the vector $\vec\rho$ with $N$ components equal to 
$\rho_i={s_i}/{\sigma_i}$ 
and $\delta_{ij}$ is the Kronecker symbol. 
We underline that with the data model given by Eq. (\ref{eqn:pdf;qcdsm;CADDSET})
one does not need to assume a specific form for the distribution of $y_i$
in order to obtain the central value of this estimate.
In a linear approximation for $f_i(\theta)$ one also does not need such 
assumptions to estimate the dispersion.
In this approximation the dispersion reads \cite{Alekhin:2000es}
\begin{equation}
D_{\rm C}(\hat\theta)=\frac{1}{\phi^2}
\left[1+\frac{\rho^2 z^2}{1+\rho^2(1-z^2)}\right],
\label{eqn:pdf;qcdsm;DISPCA}
\end{equation}
where $\phi$ is modulus of the vector $\vec\phi$ 
with $N$ components equal to 
$\phi^i={f_i'(\theta_0)}/{\sigma_i}$, the symbol $'$ denotes the derivative
with respect to $\theta$, and 
$z$ is the cosine of the angle between $\vec\rho$ and $\vec\phi$.

To obtain the distribution of $\hat\theta$ one needs to know 
the complete set of its moments, which, in turn, requires similar information 
for the moments of $y_i$. At the same time 
from considerations similar to the proof of 
the Central Limit theorem of statistics (see. Ref.~\cite{James:1971})
one can conclude that the distribution of $\hat\theta$ is Gaussian, 
if all independent systematic errors are comparable in value and 
their number is large enough. More educated guesses of the form of 
the distribution can be performed with the help of the general approach 
described in subsection \ref{sec:pdf;qcdsm;ransam}.

The dispersion of fitted parameters obtained by the 
covariance method is 
different from the one obtained by the offset method described 
in subsection \ref{sec:pdf;qcdsm;offset}. Indeed, the dispersion of the 
parameter estimate obtained by the offset method
applied to the analysis of the data set given by 
Eq. (\ref{eqn:pdf;qcdsm;CADDSET})
is equal \cite{Alekhin:2000es}
\begin{equation}
D_{\rm O}(\hat\theta)=\frac{1}{\phi^2}\left(1+\rho^2 z^2\right).
\label{eqn:pdf;qcdsm;DISPSA}
\end{equation}
One can see that $D_{\rm O}$ is generally larger than $D_{\rm C}$.
The difference is especially visible for the case $N\gg 1$, when 
$\rho \gg 1$, if the systematic errors are not negligible as compared 
to the statistical ones. In this case and if $z\ne 1$
\begin{equation}
D_{\rm C}(\hat\theta)\approx\frac{1}{\phi^2(1-z^2)}\ ,
\label{eqn:pdf;qcdsm;dispmr}
\end{equation}
while
\begin{equation}
D_{\rm O}(\hat\theta)\approx\frac{\rho^2 z^2}{\phi^2}.
\label{eqn:pdf;qcdsm;disp0r}
\end{equation}
One can see that the standard deviation of the offset method
estimator rises linearly with
the increase of the systematics, while the 
covariance matrix dispersion saturates and the difference between them 
may be very large.

Some peculiarities arise in the case when the systematic errors 
are multiplicative, i.e. when the $s_i$ in
Eq. (\ref{eqn:pdf;qcdsm;COVA})
are given by $\eta_i y_i$, where $\eta_i$ are constants. 
As it was noted in Ref.~\cite{D'Agostini:1994uj} in this case 
the covariance matrix estimator may be biased.
The manifestation of this bias is that 
the fitted curve lays lower the data points on average, 
which is reflected by a distortion of the fitted parameters.
In order to minimize such bias one has to calculate
the covariance matrix of Eq. (\ref{eqn:pdf;qcdsm;COVA})  using the 
relation $s_i=\eta_i f_i(\theta)$. In this approach
the covariance matrix depends on the fitted parameters and hence 
has to be iteratively re-calculated during the fit.
This certainly makes calculation more time-consuming and 
difficult, but in this case the bias of the estimator is non-negligible 
as compared to the value of its standard deviation if the systematic
error on the fitted parameter is an order of magnitude
larger than the statistical error \cite{Alekhin:2000es}.
\subsubsection{Offset Method}\label{sec:pdf;qcdsm;offset}
With the offset method~\cite{Botje:2001fx}, already mentioned above,
the systematic errors are incorporated in the model prediction
\begin{equation}\label{eqn:pdf;qcdsm;modelpred}
 t_i(\theta,\lambda) = f_i(\theta) + \sum_k \lambda_k\; s_{ik}\ ,
\end{equation}
where we allow for several sources of systematic error $(k)$.  The
$\chi^2$, to be minimized in a fit, is defined as
\begin{equation}\label{eqn:pdf;qcdsm;xpsdef}
  \chi^2(\theta,\lambda)
   = \sum_i \left( \frac{y_i-t_i(\theta,\lambda)}{\sigma_i} \right)^2 
  + \sum_k \lambda_k^2.   
\end{equation}
It can be shown~\cite{Stump:2001gu} that leaving both $\theta$ and
$\lambda$ free in the fit is mathematically equivalent to the
covariance method described in the previous section.  However, there
is also the choice to fix the systematic parameters to their central
values $\lambda_k = 0$ which results in minimizing
\begin{equation}\label{eqn:pdf;qcdsm;offsetchi}
 \chi^2(\theta) = \sum_i \left( \frac{y_i-f_i(\theta)}{\sigma_i} \right)^2\ ,
\end{equation}
where only statistical errors are taken into account to get the best
value $\hat{\theta}$ of the parameters.  Because systematic errors are
ignored in the $\chi^2$ such a fit forces the theory prediction to be
as close as possible to the data.

The systematic errors on $\theta$ are estimated from fits where each
systematic parameter $\lambda_k$ is offset by its assumed error
($\pm1$) after which the resulting deviations $\Delta \theta$ are
added in quadrature.  To first order this lengthy procedure can be
replaced by a calculation of two Hessian matrices $M$ and $C$, defined
by
\begin{equation}\label{eqn:pdf;qcdsm;mcijdef}
  M_{ij} = \frac{1}{2} \frac{\partial^2 \chi^2}
  {\partial \theta_i \partial \theta_j} \qquad
  C_{ij} = \frac{1}{2} \frac{\partial^2 \chi^2}
  {\partial \theta_i \partial \lambda_j}\ .
\end{equation}
The statistical covariance matrix of the fitted parameters is then
given by
\begin{equation}\label{eqn:pdf;qcdsm;cstatdef}
  V_{\rm stat}^{\theta} = M^{-1}\ ,
\end{equation}
while a systematic covariance matrix can be defined
by~\cite{Pascaud:1995qs}
\begin{equation}\label{eqn:pdf;qcdsm;csysdef}
  V_{\rm syst}^{\theta} = M^{-1} C C^T M^{-1}\ ,
\end{equation}
where $C^T$ is the transpose of $C$.

Having obtained the best values and the covariance matrix of the
parameters, the covariance of any two functions $F(\theta)$ and
$G(\theta)$ can be calculated with the standard formula for linear
error propagation
\begin{equation}\label{eqn:pdf;qcdsm;fgmat}  
  V_{FG} = \sum_{ij} \frac{\partial F(\theta)}{\partial \theta_i}\;
  V_{ij}^{\theta}\;
  \frac{\partial G(\theta)}{\partial \theta_j}\ ,
\end{equation}
where $V^{\theta}$ is the statistical, systematic or, if the total
error is to be calculated, the sum of both covariance matrices.

Comparing Eqs. (\ref{eqn:pdf;qcdsm;xpsdef})  
and (\ref{eqn:pdf;qcdsm;offsetchi}) it is
clear that the parameter values obtained by the covariance and offset
methods will, in general, be different. This difference is accounted
for by the difference in the error estimates, those of the offset
method being larger in most cases. In statistical language this means
that the parameter estimation of the offset method is not {\em
efficient}. The method has a further disadvantage that the goodness of
fit cannot be directly judged from the $\chi^2$ which is calculated
from statistical errors only.

For a global QCD analysis of deep inelastic scattering data which uses
the offset method to propagate the systematic errors, we refer
to~\cite{Botje:1999dj} (see {\tt http://www.nikhef.nl/user/h24/qcdnum}
for the corresponding PDF set with full error information).

\subsection{The LHAPDF interface
\protect\footnote{Contributing authors:
S.~Alekhin, W.~Giele, J.~Pumplin}}

\subsubsection{Introduction}

The Les Houches Accord PDF (LHAPDF) interface package is designed to work with
PDF sets. A PDF set can consist of many individual member PDFs. While
the interpretation of the member PDFs depends on the particular set,
the LHAPDF interface is designed to accommodate PDFs with uncertainties
as well as ``central fit'' PDFs.
For PDFs with uncertainties the PDF set represents one ``fit'' to the
data. For instance, a random sampling PDF set would use Eq.
(\ref{eqn:pdf;qcdsm;density}).
In other words for each PDF in the set the observable is calculated. 
The set of resulting predictions of the observable build up the
probability density.
The individual member PDFs of the set are needed to calculate
the PDF uncertainty on the observable. All PDF sets are defined through
external files. This means that a new set can be added by simply downloading
its file while the LHAPDF interface code does not change. The evolution
code is not part of LHAPDF. The current default choice included and 
interfaced is QCDNUM \cite{QCDNUM}. 
Each group that contributes PDF sets can provide 
their own evolution code; or they can employ QCDNUM, which is available in
the package.

\subsubsection{The philosophy}

The Les Houches Accord Parton Distribution Function interface
was conceived at the Les Houches 2001 workshop in the PDF working group to
enable the usage of Parton Distribution Functions with uncertainties
in a uniform manner. 
When PDFs with uncertainties are considered, a ``fit'' to
the data no longer is described by a single PDF. Instead in its
most flexible implementation, a fit is represented by a PDF set
consisting of many individual PDF members. Calculating the observable
for all the PDF members enables one to reconstruct the uncertainty
on the observable. The LHAPDF interface was made with this in mind 
and manipulates PDF
sets. 

The LHAPDF interface can be viewed as a successor to PDFLIB and improvements
were added. To list some of the features:
\begin{itemize}
\item
  The handling of PDF sets to enable PDF fits that include uncertainties.
\item
  The default evolution codes are benchmarked and compared for accuracy.
  Apart from accuracy another important feature of the evolution code
  is speed. Currently the default for evolution program is QCDNUM.
\item
  All PDF sets are defined through external files in parametrized form.
  This means the files are compact compared to storing the PDFs in a 
  grid format. Also
  new PDF sets can be defined by constructing the PDF defining files.
  The actual LHAPDF code does not have to be changed.
\item
  The LHAPDF code is modular and default choices like the QCDNUM
  evolution code can be easily altered.
\end{itemize}

Note that the current ``best fit'' PDFs can be viewed as PDF sets with
one member PDF and can be easily defined through the PDF set external
file definition. Alternatively one can group these ``fits'' is single
sets (e.g. MRST98 set) as they often represent best fits given different
model assumptions and as such reflect theoretical modelling uncertainties.

The first version of the code is available in Fortran from 
{\bf\tt http://pdf.fnal.gov}.

\subsubsection{Interfacing with LHAPDF}
The interface of LHAPDF with an external code is easy. We will
describe the basic steps sufficient for most applications. The 
web site contains more detailed information about additional
function calls. The function calls described here will be not
be altered in any way in future versions. 
Including the LHAPDF evolution code into a program involves 
three steps:
\begin{enumerate}
\item First one has to setup the LHAPDF interface code:
\begin{center}{\bf\tt call InitPDFset({\it name})}\end{center}
  It is called only once at the beginning of the code.
  The string variable {\it name} is the file name of the 
  external PDF file that defines the PDF set.
  For the standard evolution code QCDNUM it will 
  either calculate or read from file the LO/NLO splitting 
  function weights. The calculation of the weights might
  take some time depending on the chosen grid size. However,
  after the first run a grid file is created. Subsequent
  runs will use this file to read in the weights so that the lengthy 
  calculation of these weights is avoided. The file depends
  on the grid parameters and flavor thresholds. This means different
  PDF sets can have different grid files. The name of the grid file
  is specified in the PDF setup file. 
\item To use a individual PDF member it has to be initialized:
  \begin{center}{\bf\tt call InitPDF({\it mem)}}\end{center}
  The integer {\it mem} specifies the member PDF number. This
  routine only needs to be called when changing to a new PDF
  member. The time needed to setup depends on the evolution code used
  For QCDNUM the grid size is the determining factor. Note that 
  {\it mem}=0 selects the ``best fit'' PDF.
\item Once the PDF member is initialized one can call the evolution
  codes which will use the selected member. 

  The function call
  \begin{center}{\bf\tt function alphasPDF({\it Q})}\end{center}
  returns the value of $\alpha_S(Q)$ at double precision scale $Q$. 
  Note that its value can change between different PDF members.

  The subroutine call
  \begin{center}{\bf\tt call evolvePDF({\it x,Q,f})}\end{center}
  returns the PDF momentum densities $f$ (i.e. $x\times$ PDF number density)
  at double precision momentum fraction $x$ and double precision scale $Q$.
  The double precision array {\it f(-6:6)} will contain the momentum PDFs
  using the labelling convention of table \ref{tab:pdf;qcdsm;labeling}.
  \begin{table}[t]\hspace{2cm}\label{tab:pdf;qcdsm;labeling}\begin{center}
    \begin{tabular}{|l|r|r|r|r|r|r|r|r|r|r|r|r|r|}\hline
        & -6 & -5 & -4 & -3 & -2 & -1 & 0 &
      1 & 2 & 3 & 4 & 5 & 6 \\ \hline
      {\it P} & $\bar t$ & $\bar b$ & $\bar c$ & $\bar s$ 
      & $\bar u$ & $\bar d$ &
      $g$ & $d$ & $u$ & $s$ & $c$ & $b$ & $t$ \\
      {\it $\overline{\mbox{P}}$} & $t$ & $b$ & $c$ & $s$ & $u$ & $d$ & $g$ &
      $\bar d$ & $\bar u$ & $\bar s$ & $\bar c$ & $\bar b$ & $\bar t$ \\  
      \hline
    \end{tabular}
    \caption{The flavor enumeration convention used in the LHAPDF interface.
      (Note: CTEQ code use a different labelling scheme internally,
      with 1 $\leftrightarrow$ 2 and -1 $\leftrightarrow$ -2,
      but will adopt the above standard in the Les Houches interface.)}
  \end{center}\end{table}
  As long as the member PDF is not changed (by the 
  {\bf\tt call InitPDF} of step 2)
  the evolution calls will always use the same PDF member.
\end{enumerate}
A few additional calls can be useful:
\begin{itemize}
\item 
  To get the number of PDF members in the set: 
  \begin{center}{\bf\tt call numberPDF({\it Nmem})}.\end{center}
  The integer {\it Nmem} will contain the number of PDF members (excluding
  the special ``best fit'' member, i.e. the member numbers run
  from 0 to {\it Nmem}). 
\item
  Optionally the different PDF members can have weights which are obtained
  by:
  \begin{center}{\bf\tt call weightPDF({\it wgt})}.\end{center}
  The double precision variable {\it wgt} is for unweighted PDFs
  set to $1/Nmem$ such that the sum of all PDF member weights is unity.
  For weighted sets the use of the weights has to be defined by the
  method description.
\item 
  To get the evolution order of the PDFs:
  \begin{center}{\bf\tt call GetOrderPDF({\it order})}.\end{center}
  The integer variable {\it order} is 0 for Leading Order, 1 for
  Next-to-Leading Order, etc.
\item 
  To get the evolution order of $\alpha_S$:
  \begin{center}{\bf\tt call GetOrderAs({\it order})}.\end{center}
  The integer variable {\it order} is 0 for Leading Order, 1 for
  Next-to-Leading Order, etc.
\item 
  It is possible that during the PDF fitting the renormalization
  scale was chosen different from the factorization scale. The
  ratio of the renormalization scale over the factorization
  scale used in the ``fit'' can be obtained by
  \begin{center}{\bf\tt call GetRenFac(muf)}.\end{center}
  The double precision variable {\it muf} contains the ratio.
  Usually {\it muf} is equal to unity.
\item
  To get a description of the PDF set:
  \begin{center}{\bf\tt call GetDesc()}.\end{center}
  This call will print the PDF description to the standard output stream
\item
  The quark masses can be obtained by:
  \begin{center}{\bf\tt call GetQmass(nf,mass)}.\end{center}
  The mass {\it mass} is returned for quark flavor {\it nf}.
  The quark masses are used in the $\alpha_S$ evolution.
\item
  The flavor thresholds in the PDF evolution can be obtained by:
  \begin{center}{\bf\tt call GetThreshold(nf,Q)}.\end{center}
  The flavor threshold {\it Q} is returned for flavor {\it nf}.
  If {\it Q=-1d0} flavor is not in the evolution (e.g. the top quark
  is usually not included in the evolution).
  If {\it Q=0d0} flavor is parametrized at the parametrization scale.
  For positive non-zero values of {\it Q} the value is set to the
  flavor threshold at which the PDF starts to evolve.
\item
  The call returns the number of flavors used in the PDF:
  \begin{center}{\bf\tt call GetNf(nfmax)}.\end{center}
  Usually the returned value for {\it nfmax} 
  is equal to five as the top quark is usually not considered in the PDFs.
\end{itemize}

\subsubsection{An example}
A very simple example is given below. It accesses all member PDFs in the
set {\tt mypdf.LHpdf} and print out the $\alpha_S(M_Z)$ value and the
gluon PDF at several $(x,Q)$ points.

\begin{verbatim}
      program example
      implicit real*8(a-h,o-z)
      character*32 name
      real*8 f(-6:6)
*
      name='mypdf.LHpdf'
      call InitPDFset(name)
*
      QMZ=91.71d0
      write(*,*)
      call numberPDF(N)
      do i=1,N
         write(*,*) '---------------------------------------------'
         call InitPDF(i)
         write(*,*) 'PDF set ',i
         write(*,*)
         a=alphasPDF(QMZ)
         write(*,*) 'alphaS(MZ) = ',a
         write(*,*)
         write(*,*) 'x*Gluon'
         write(*,*) '   x     Q=10 GeV     Q=100 GeV    Q=1000 GeV'
         do x=0.01d0,0.095d0,0.01d0
            Q=10d0
            call evolvePDF(x,Q,f)
            g1=f(0)
            Q=100d0
            call evolvePDF(x,Q,f)
            g2=f(0)
            Q=1000d0
            call evolvePDF(x,Q,f)
            g3=f(0)
            write(*,*) x,g1,g2,g3
         enddo
      enddo
*
      end
\end{verbatim}


\subsection{Reference results for the evolution of parton distributions%
\protect\footnote{Contributing authors: G. Salam, A. Vogt}} 

In this section we provide a new set of benchmark tables for the evolution of 
unpolarized parton distributions of hadrons in perturbative QCD. Unlike the 
only comparable previous study \cite{Blumlein:1996rp}, we include results for 
unequal factorization and renormalization scales, $\mu_{\rm f} \!\neq\! 
\mu_{\rm r}$, and for the evolution with a variable number of partonic flavours 
$N_{\rm f}$. Besides the standard LO and NLO approximations, we also present 
the evolution including the (still approximate) NNLO splitting functions and 
the corresponding non-trivial second-order matching conditions at the 
heavy-quark thresholds. Our reference results are computed using two entirely 
independent and conceptually different evolution programs which, however, agree 
to better than 1 part in $10^5$ for momentum fractions $10^{-8} < x < 0.9$.

\subsubsection{Evolution equations and their solutions}
\label{gsav-s1}

At N$^{\rm m}$LO the scale dependence (`evolution') of the parton distributions 
$ f_p (x,\mu_{\rm f}^2) \equiv p(x,\mu_{\rm f}^2) $, where $ p = q_i, 
\,\bar{q}_i\, , g $ with $ i = 1, \,\ldots , N_{\rm f} $, is governed by the 
$ 2N_{\rm f}\! +\! 1$ coupled integro-differential equations
\begin{equation}
\label{gsav-eq1}
 \frac {d\, f_p(x,\mu_{\rm f}^2)} {d \ln \mu_{\rm f}^2} \: = \: 
 \sum_{l=0}^m \, a_{\rm s}^{l+1}(\mu_{\rm r}^2)\:
 {\cal P}_{pp'}^{(l)}\bigg(x,\frac{\mu_{\rm f}^2}{\mu_{\rm r}^2}\bigg) 
 \otimes f_{p'}(x,\mu_{\rm f}^2) \:\: .
\end{equation}
Here $\otimes$ denotes the Mellin convolution in the fractional-momentum 
variable $x$,
%
%
and summation over $p'$ is understood. 
The scale dependence of the strong coupling $a_{\rm s} \equiv \alpha_{\rm s}/
(4\pi)$ is given by 
\begin{equation}
\label{gsav-eq2}
  \frac{d\, a_{\rm s}}{d \ln \mu_r^2} \: = \: \beta^{}_{\rm N^mLO}(a_{\rm s})
  \: = \: - \sum_{l=0}^m \, a_{\rm s}^{l+2} \,\beta_l \:\: .
\end{equation}
The general splitting functions ${\cal P}^{(l)}$ in Eq.~(\ref{gsav-eq1}) can be
reduced to the simpler expressions $P^{(l)}(x)$ at $\mu_{\rm r} = \mu_{\rm f}$. 
Up to NNLO ($\equiv$ N$^{\rm 2\,}$LO) the corresponding relations read
\begin{eqnarray}
\label{gsav-eq3}
  {\cal P}^{(0)} \bigg(x, \frac{\mu_{\rm f}^2}{\mu_{\rm r}^2} \bigg) 
  &\! =\! & P^{(0)}(x) \nonumber \\
  {\cal P}^{(1)} \bigg(x, \frac{\mu_{\rm f}^2}{\mu_{\rm r}^2} \bigg)
  &\! =\! & P^{(1)}(x) 
  - \beta_0 P^{(0)}(x)\, \ln \frac{\mu_{\rm f}^2}{\mu_{\rm r}^2} \\
  {\cal P}^{(2)} \bigg(x, \frac{\mu_{\rm f}^2}{\mu_{\rm r}^2} \bigg)
  &\! =\! & P^{(2)}(x) - \bigg\{ \beta_1 P^{(0)}(x) + 2\beta_0 P^{(1)}(x) 
    \bigg\} \ln \frac{\mu_{\rm f}^2}{\mu_{\rm r}^2} 
    + \beta_0^2 P^{(0)}(x)\, \ln^2 \frac{\mu_{\rm f}^2}{\mu_{\rm r}^2} \:\: .
  \nonumber
\end{eqnarray}
The generalization to higher orders is straightforward but not required for the
present calculations.

The LO and NLO coefficients $\beta_0$ and $\beta_1$ of the $\beta$-function in
Eq.~(\ref{gsav-eq2}) and the corresponding splitting functions $P^{(0)}(x)$ and 
$P^{(1)}(x)$ in Eq.~(\ref{gsav-eq3}) have been known for a long time, 
see Ref.~\cite{Furmanski:1982cw} and references therein. 
In the $\overline{\rm MS}$ scheme adopted here, the coefficient $\beta_2$ has 
been calculated in Refs.~\cite{Tarasov:1980au,Larin:1993tp}. The NNLO 
quantities $P^{(2)}(x)$ have not been computed so far, but approximate 
expressions have been constructed \cite{vanNeerven:1999ca,vanNeerven:2000uj,%
vanNeerven:2000wp} from the available partial results \cite{Larin:1994vu,%
Larin:1997wd,Retey:2000nq,Catani:1994sq,Blumlein:1996jp,Fadin:1998py,%
Ciafaloni:1998gs,Gracey:1994nn,Bennett:1998ch}.

An obvious and widespread approach to Eq.~(\ref{gsav-eq1}) is the direct 
numerical solution by discretization in both $x$ and $\mu_{\rm f}^2$. This 
method is also used by one of us (G.S.).
The parton distributions are represented on a grid with $n$ points
uniformly spaced in $\ln 1/x$.  Their values at arbitrary $x$ are 
defined to be equal to a $p^\mathrm{th}$ order interpolation of 
neighbouring grid points.  The splitting functions can then be 
represented as sparse matrices acting on the vector of grid points. At
initialization the program takes a set of subroutines for the splitting
functions and calculates the corresponding matrices with a Gaussian 
adaptive integrator.  At each value of $\mu_{\rm f}^2$ the derivatives 
of the parton distributions are calculated through matrix 
multiplication and the evolution is carried out with a Runge-Kutta 
method. The algorithm has partial overlap with those of 
Refs.~\cite{QCDNUM,Santorelli:1998yt,Ratcliffe:2000kp,Pascaud:2001bi} 
and is described in more detail in Appendix~F of 
Ref.~\cite{Dasgupta:2001eq}.

For the reference tables presented below, the program has been run
with $7^{th}$ order interpolation in $x$ and multiple $x$-grids: one 
for $10^{-8}<x<1$ with $750$ points, another for $0.135<x<1$ with $240$
points and a third for $0.6<x<1$ with $180$ points. A grid uniform in
$\ln \mu_{\rm f}^2$ has been used with $220$ points in the range 
$ 2\,\mbox{ GeV}^2 < \mu_{\rm f}^2 < 10^6 \mbox{ GeV}^2 $. Halving the 
density of points in both $x$ and $\mu_{\rm f}^2$ leaves the results 
unchanged at the level of better than $1$ part in $10^{5}$ in the range 
$10^{-8}<x<0.9$ (except close to sign-changes of parton distributions).

An important alternative to this direct numerical treatment of 
Eq.~(\ref{gsav-eq1}) is the Mellin-$N$ moment solution in terms of a power 
expansion. This method is employed by the second author (A.V.). Here
Eq.~(\ref{gsav-eq1}) is transformed to $N$-space (reducing the convolution to a 
simple product) and $\mu_{\rm f}^2$ is replaced by $a_{\rm s}$ as the 
independent variable, assuming that $\mu_{\rm f}^{}/\mu_{\rm r}^{}$ is a fixed 
number. Expanding the resulting r.h.s.\ into a power series in $a_{\rm s}$, 
one arrives at
\begin{equation}
\label{gsav-eq4}
  \frac {d\, f_p(N,a_{\rm s})} {d\, a_{\rm s}} \: = \: - \sum_{l=0}^\infty 
  \, a_{\rm s}^{l-1}\, R_{pp'}^{(l)}(N) \, f_{p'}(N,a_{\rm s}) 
\end{equation}
with
\begin{equation}
\label{gsav-eq5}
  R_{pp'}^{(0)}(N)  \: = \: \frac{1}{\beta_0}\, P_{pp'}^{(0)}(N)
  \:\: , \quad\quad
  R_{pp'}^{\, (k\geq 1)} \: = \: \frac{1}{\beta_0}\, {\cal P}_{pp'}^{(k)}(N)
  - \sum_{l=1}^{k} \,\frac{\beta_l}{\beta_0}\, R_{pp'}^{(k-l)}(N) \:\: .
\end{equation}
At N$^{\rm m}$LO only the coefficients $\beta_{\, l\leq m}$ and 
${\cal P}^{(l\leq m)}$ are retained in Eq.~(\ref{gsav-eq5}). The solution of 
Eq.~(\ref{gsav-eq4}) can be expressed as an expansion around the LO result
\begin{equation}
\label{gsav-eq6}
  f(N,\mu_{\rm f}^2) \: = \: 
  \bigg[ 1 + \sum_{k=1}^\infty \, a^k_{\rm s}\, U_k(N) \bigg] 
  \left( \frac{a_{\rm s}}{a^{}_0} \right)^{-R_0(N)}
  \bigg[ 1 + \sum_{k=1}^\infty \, a^k_0 \, U_k(N) \bigg]^{-1} 
  f(N,\mu_{\rm f,0}^2) \:\: ,
\end{equation}
where $\mu_{\rm f,0}^2$ is the initial scale for the evolution, and $a^{}_0 
\equiv a_{\rm s}(\mu_{\rm r}^2(\mu_{\rm f,0}^2))$.
It is understood in Eq.~(\ref{gsav-eq6}) that the matrix structure is 
simplified by switching to the appropriate flavour singlet and non-singlet
combinations. For the explicit construction of the remaining $2\!\times\! 2$
matrices $U_k$ the reader is referred to Section 5 of 
Ref.~\cite{Blumlein:1998em}. Finally the solutions $f_p(N,\mu_{\rm f}^2)$ are 
transformed back to $x$-space by
\begin{equation}
\label{gsav-eq7}
  f_p (x,\mu_{\rm f}^2) \: = \: \frac{1}{\pi}\, \int_0^\infty \! dz \:
  {\rm Im} \, \Big[ \, e^{i\phi} x^{- c - z \exp(i\phi)} 
  f_p (N\! =\! c\! +\! z \exp(i\phi),\, \mu_{\rm f}^2) \, \Big] \:\: .
\end{equation}

The Mellin inversions (\ref{gsav-eq7}) can be performed with a sufficient 
accuracy using a fixed chain of Gauss quadratures. Hence the quantities 
$R_0(N)$ and  $U_k(N)$ in Eq.~(\ref{gsav-eq6}) have to the computed only once 
for the corresponding support points at the initialization of the program,
rendering the $N$-space evolution competitive in speed with fast $x$-space
codes.
Except where the parton distributions become very small, an accuracy of 1 part
in $10^5$ or better is achieved by including contributions up to $k = 15$ in 
Eq.~(\ref{gsav-eq6}) and using at most 18 eight-point Gauss quadratures with,
e.g., $z_{\rm max} = 70$, $c = 2$, $\phi = 3\pi/4$ in Eq.~(\ref{gsav-eq7}).

The two methods for solving Eq.~(\ref{gsav-eq1}) discussed above are completely
equivalent, i.e., they do not differ even by terms beyond N$^{\rm m}$LO, 
provided that the coupling $a_{\rm s}$ evolves exactly according to 
Eq.~(\ref{gsav-eq2}). This condition is fulfilled in the present calculations.
Thus the results of our two programs can be compared directly, yielding a
powerful check of the correctness and accuracy of the codes. Note that the 
only previously published high-precision comparison of NLO evolution programs 
\cite{Blumlein:1996rp} used the truncated expansion of $a_{\rm s}^2
(\mu_{\rm r})$ in terms of inverse powers of $\ln \mu_{\rm r}^2 / \Lambda^2$ 
which does not exactly conform to Eq.~(\ref{gsav-eq2}). Consequently such 
direct comparisons were not possible in Ref.~\cite{Blumlein:1996rp}.

Following Ref.~\cite{Furmanski:1982cw}, the $N$-space solution (\ref{gsav-eq6})
has usually been subjected to a further expansion in the coupling constants, 
retaining only the terms up to order $m$ in the product of the $U$-matrices. 
Only the terms thus kept in Eq.~(\ref{gsav-eq6}) are free from contributions by 
the higher-order coefficients $\beta_{k>m}$ and $P^{(k>m)}$, and only these 
terms combine to factorization-scheme independent quantities when the parton 
distributions are convoluted with the N$^{\rm m}$LO partonic cross sections.
Reference results for the truncated solution will be presented elsewhere 
\cite{GSAV}.

\subsubsection{Initial conditions and heavy-quark treatment}
\label{gsav-s2}

The following initial conditions for the reference results have been set up
at the Les Houches meeting: The evolution is started at 
\begin{equation}
\label{gsav-eq8}
  \mu_{\rm f,0}^2 \: = \: 2 \mbox{ GeV}^2 \:\: .
\end{equation}
Roughly along the lines of the CTEQ5M parametrization \cite{Lai:1999wy}, the 
input distributions are chosen as
\begin{eqnarray}
\label{gsav-eq9}
  xu_v(x,\mu_{\rm f,0}^2)       &\! =\! & 5.107200\: x^{0.8}\: (1-x)^3  
    \nonumber \\
  xd_v(x,\mu_{\rm f,0}^2)       &\! =\! & 3.064320\: x^{0.8}\: (1-x)^4  
    \nonumber \\
  xg\,(x,\mu_{\rm f,0}^2)       &\! =\! & 1.700000\, x^{-0.1} (1-x)^5 
    \\
  x\bar{d}\,(x,\mu_{\rm f,0}^2) &\! =\! & .1939875\, x^{-0.1} (1-x)^6
    \nonumber\\
  x\bar{u}\,(x,\mu_{\rm f,0}^2) &\! =\! & (1-x)\: x\bar{d}\,(x,\mu_{\rm f,0}^2)
    \nonumber\\
  xs\,(x,\mu_{\rm f,0}^2)       &\! =\! & x\bar{s}\,(x,\mu_{\rm f,0}^2) 
    \: = \: 0.2\, x(\bar{u}+\bar{d}\,)(x,\mu_{\rm f,0}^2) 
    \nonumber  
\end{eqnarray}
where, as usual, $q_{i,v} \equiv q_i - \bar{q}_i$. The running couplings are 
specified via
\begin{equation}
\label{gsav-eq10}
  \alpha_{\rm s}(\mu_{\rm r}^2\! =\! 2\mbox{ GeV}^2) \: = \: 0.35 \:\: .
\end{equation}
For simplicity these initial conditions are employed regardless of the order of 
the evolution and the ratio of the renormalization and factorization scales. 
At LO this ratio is fixed to unity, beyond LO we use
\begin{equation}
\label{gsav-eq11}
  \mu_{\rm r}^2 \: = \: k_{\rm r} \mu_{\rm f}^2 \:\: , \quad\quad 
  k_{\rm r} \: = \:  0.5\, , \:\: 1\, , \:\: 2 \:\: .
\end{equation}
 
For the evolution with a fixed number $N_{\rm f} > 3$ of quark flavours the 
quark distributions not specified in Eq.~(\ref{gsav-eq9}) are assumed to vanish 
at $\mu_{\rm f,0}^2$, and Eq.~(\ref{gsav-eq10}) is understood to refer to the 
chosen value of~$N_{\rm f}$. For the evolution with a variable $ N_{\rm f} = 3 
\ldots 6 $, Eqs.~(\ref{gsav-eq9}) and (\ref{gsav-eq10}) always refer to three 
flavours. $N_{\rm f}$ is then increased by one unit at the heavy-quark pole 
masses taken as
\begin{equation}
\label{gsav-eq12}
  m_{\rm c} \: = \: \mu_{\rm f,0}      \: , \quad 
  m_{\rm b} \: = \: 4.5 \mbox{ GeV}^2  \: , \quad
  m_{\rm t} \: = \: 175 \mbox{ GeV}^2  \:\: ,
\end{equation}
i.e., the evolution is performed as discussed in Section \ref{gsav-s1} between 
these thresholds, and the respective matching conditions are invoked at 
$\mu_{\rm f}^2 = m_{\rm h}^2$, $h = c,\, b,\, t$. For the parton distributions 
these conditions have been derived in Ref.~\cite{Buza:1998wv}. Up to 
N$^{\rm m=2\,}$LO they read 
\begin{equation}
\label{gsav-eq13}
  l_i^{\,(N_{\rm f}+1)}(x,m_h^2) \: = \:  l_i^{\,(N_{\rm f})}(x,m_h^2) + 
  \delta_{m2} \: a_{\rm s}^2\: A^{\rm NS,(2)}_{qq,h}(x) \otimes 
  l_i^{\, (N_{\rm f})}(x,m_h^2) 
\end{equation}
where $l = q,\, \bar{q}$ and $i = 1,\ldots N_{\rm f}$, and
\begin{eqnarray}
\label{gsav-eq14}
  g^{\, (N_{\rm f}+1)}(x,m_h^2) \:\:\: &\! = \!\! & 
    g^{\, (N_{\rm f})}(x,m_h^2) + \nonumber \\[0.5mm] & & 
    \delta_{m2} \, a_{\rm s}^2\, \Big[  
    A_{\rm gq,h}^{S,(2)}(x) \otimes \Sigma^{\, (N_{\rm f})}(x,m_h^2) +
    A_{\rm gg,h}^{S,(2)}(x) \otimes g^{\, (N_{\rm f})}(x,m_h^2) \Big ] 
  \\[0.5mm]
  (h+\bar{h})^{\, (N_{\rm f}+1)}(x,m_h^2) \! &\! =\!\! & 
    \delta_{m2} \, a_{\rm s}^2\: \Big [ 
    \tilde{A}_{\rm hq}^{S,(2)}(x) \otimes \Sigma^{\, (N_{\rm f})}(x,m_h^2) +
    \tilde{A}_{\rm hg}^{S,(2)}(x) \otimes g^{\, (N_{\rm f})}(x,m_h^2) \Big  ] 
  \quad \nonumber
\end{eqnarray}
with $\Sigma^{\,(N_{\rm f})} \equiv \sum_{i=1}^{N_{\rm f}} (q_i + \bar{q}_i)$
and $h = \bar{h}$.  The coefficients $A^{(2)}$ can be found in Appendix B of 
ref.~\cite{Buza:1998wv} -- due to our choice of $\mu_{\rm f}^2 = m_{\rm h}^2$
for the thresholds only the scale-independent parts of the expressions are 
needed here -- from where the notation for these coefficients has been taken 
over. 
The corresponding N$^{\rm m\,}$LO relation for the coupling constant 
\cite{Larin:1995va,Chetyrkin:1997sg} is given by
\begin{equation}
\label{gsav-eq15}
  a_{\rm s}^{\, (N_{\rm f}+1)}(k_{\rm r} m_h^2) \: = \: 
  a_{\rm s}^{\, (N_{\rm f})}(k_{\rm r} m_h^2) + \sum_{n=1}^{m} \, 
  \Big( a_{\rm s}^{\, (N_{\rm f})}(k_{\rm r} m_h^2) \Big)^{n+1} 
  \sum_{l=0}^{n} \, c^{}_{n,l} \, \ln k_{\rm r} \:\: .
\end{equation}
The pole-mass coefficients $c_{n,l}$ in Eq.~(\ref{gsav-eq15}) can be inferred 
from Eq.~(9) of Ref.~\cite{Chetyrkin:1997sg}, where $4\, a_{\rm s}^{\, 
(N_{\rm f}-1)}$ is expressed in terms of $4\,a_{\rm s}^{\, (N_{\rm f})}$.
Note that we use $a_{\rm s}^{\, (N_{\rm f}+1)}(k_{\rm r} m_h^2)$ on the r.h.s.\
of Eq.~(\ref{gsav-eq14}).

\subsubsection{The benchmark results}
\label{gsav-s3}

We have compared the results of our two evolution programs, under the
conditions specified in Section \ref{gsav-s2}, at 500 $x$-$\mu_{\rm f}^2$
points covering the range $10^{-8} \leq x \leq 0.9$ and $2 \mbox{ GeV}^2 \leq
\mu_{\rm f}^2 \leq 10^6 \mbox{ GeV}^2 $. A representative subset of our results 
at $\mu_{\rm f}^2 = 10^4 \mbox{ GeV}^4$, a scale relevant to high-$E_T$ jets
at {\sc Tevatron} and close to $m_{\rm W}^2$, $m_{\rm Z}^2$ and, possibly,
$m_{\rm Higgs}^2$, is presented in Tables \ref{gsav-t1}--\ref{gsav-t5}.
These results are given in terms of the valence distributions, defined below 
Eq.~(\ref{gsav-eq9}), $L_{\pm} \equiv \bar{d} \pm \bar{u}$, and the quark-%
antiquark sums $q_+ \! \equiv q \! -\!\bar{q}$ for $q = s,\, c$ and, for the 
variable-$N_{\rm f}$ case, $b$.

For compactness an abbreviated notation is employed throughout the tables, 
i.e., all numbers $a\cdot 10^b$ are written as $a^b$. In the vast majority of 
the $x$-$\mu_{\rm f}^2$ points our results are found to agree to all five 
figures displayed, except for the tiny NLO and NNLO sea-quark distributions at 
$x = 0.9$, in the tables. In fact, the differences for $x < 0.9$ are not larger 
than $\pm 1$ in the sixth digit, i.e, the offsets are smaller than 1 part in 
$10^5$. Entries where these residual offsets lead to a different fifth digit 
after rounding are indicated by the subscript '$*$'. The number with the 
smaller modulus is then given in the tables, e.g., $1.1111_*^1$ should be read 
as $1.11115 \cdot 10^1$ with an uncertainty of $\pm 1$ in the last figure. 

\begin{figure}[hbt]
\vspace{2mm}
\begin{center}
\mbox{\includegraphics[height=12.0cm]{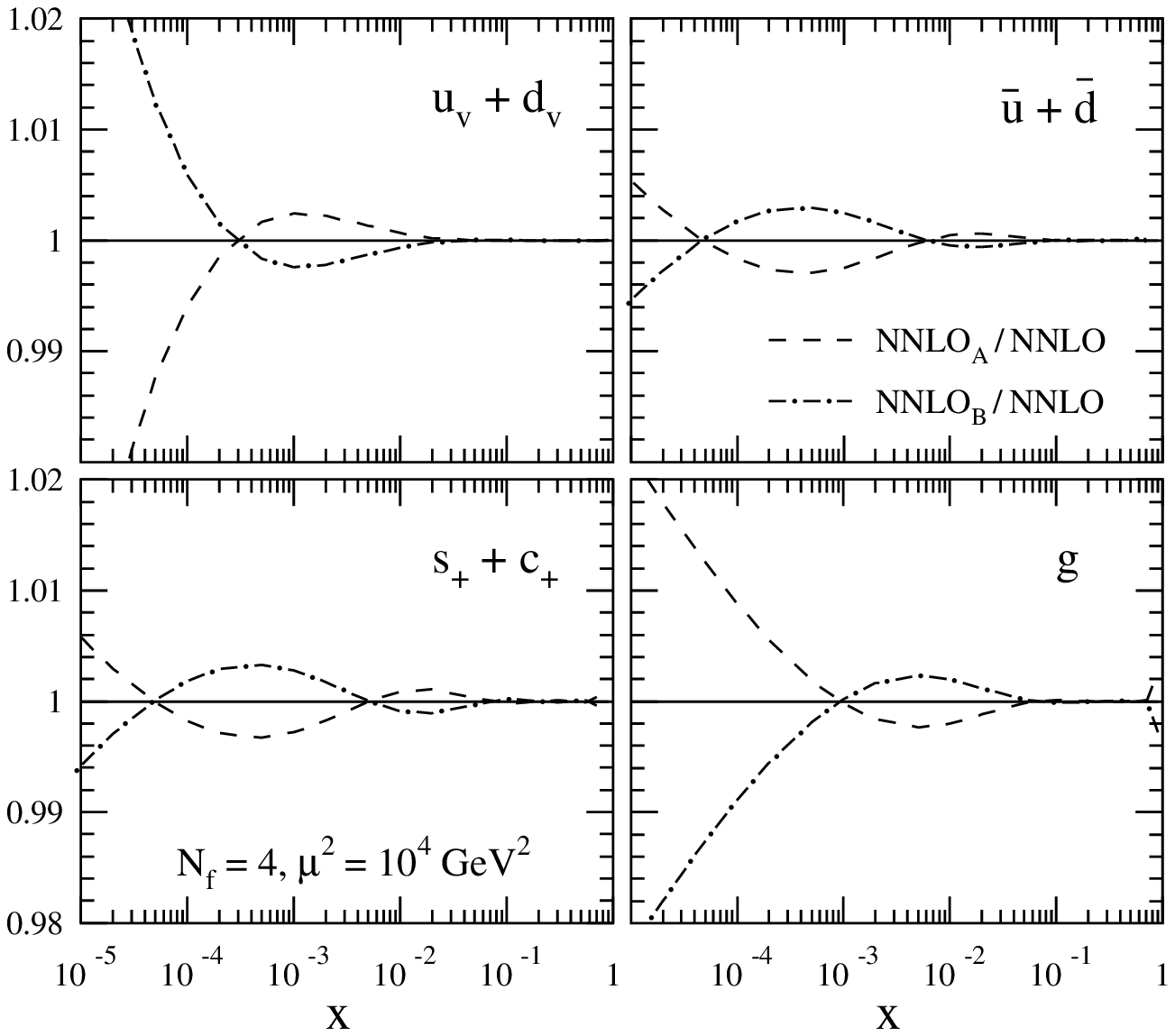}}
\end{center}
\vspace{-7mm}
\caption{The relative effects of the present uncertainties of the NNLO 
 splitting functions on the evolution of the input (\ref{gsav-eq8}) -- 
 (\ref{gsav-eq10}) for $\mu_{\rm r}=\mu_{\rm f}$, estimated by using the 
 extreme approximations `A' and `B' of Ref.~\cite{vanNeerven:2000wp} instead of 
 their average.}
\label{gsav-f1}
\end{figure}

As mentioned in Section \ref{gsav-s1}, the three-loop (NNLO) splitting 
functions $P^{(2)}(x)$ in Eq.~(\ref{gsav-eq3}) are not yet exactly known. For
the NNLO reference results in Tables \ref{gsav-t4} and \ref{gsav-t5} we have
resorted to the average of the two extreme approximations constructed in 
Ref.~\cite{vanNeerven:2000wp}. The remaining uncertainties of these results 
can be estimated by instead employing the extreme approximations itselves.
Their relative effects are illustrated, also at $\mu_{\rm f}^2 = 10^4 
\mbox{ GeV}^4$, in Fig.~\ref{gsav-f1}. The uncertainties of the NNLO evolution
to this scale turn out to be virtually negligible at $x \geq 0.05$ and to 
amount to less than $\pm 1\%$ down to $x \simeq 10^{-4}$.

\begin{table}[htp]
\caption{Reference results for the $N_{\rm f}=4$ (FFN) and the 
 variable-$N_{\rm f}$ (VFN) leading-order evolution for the initial conditions
 (\ref{gsav-eq8}) -- (\ref{gsav-eq10}), shown together with the input parton
 distributions (\ref{gsav-eq9}). The respective values for $\alpha_{\rm s}
 (\mu_{\rm r}^2\! =\! \mu_{\rm f}^2\! =\! 10^4 \mbox{ GeV}^2)$ read $0.117374$ 
 (FFN) and $0.122306$ (VFN). For the notation see the first two paragraphs of 
 Section \ref{gsav-s3}.} 
\label{gsav-t1}
\begin{center}
\vspace{5mm}
\begin{tabular}{||c||r|r|r|r|r|r|r|r||}
\hline \hline
\multicolumn{9}{||c||}{} \\[-3mm]
\multicolumn{1}{||c||}{$x$} &
\multicolumn{1}{c|} {$xu_v$} &
\multicolumn{1}{c|} {$xd_v$} &
\multicolumn{1}{c|} {$xL_-$} &
\multicolumn{1}{c|} {$xL_+$} &
\multicolumn{1}{c|} {$xs_+$} &
\multicolumn{1}{c|} {$xc_+$} &
\multicolumn{1}{c|} {$xb_+$} &
\multicolumn{1}{c||}{$xg$} \\[0.5mm] \hline \hline
\multicolumn{9}{||c||}{} \\[-3mm]
\multicolumn{9}{||c||}{Input, $\,\mu_{\rm f}^2 = 2 \mbox{ GeV}^2$} \\
\multicolumn{9}{||c||}{} \\[-0.3cm] \hline \hline
 & & & & & & & \\[-0.3cm]
$\! 10^{-7}\!$ &
$\! 1.2829^{-5}\!$ &$\! 7.6972^{-6}\!$ &$\! 9.7224^{-8}\!$ &$\! 3.8890^{+0}\!$ &
$\! 7.7779^{-1}\!$ &$\! 0.0   ^{+0}\!$ &$\! 0.0   ^{+0}\!$ &$\! 8.5202^{+0}\!$\\
$\! 10^{-6}\!$ &
$\! 8.0943^{-5}\!$ &$\! 4.8566^{-5}\!$ &$\! 7.7227^{-7}\!$ &$\! 3.0891^{+0}\!$ &
$\! 6.1782^{-1}\!$ &$\! 0.0   ^{+0}\!$ &$\! 0.0   ^{+0}\!$ &$\! 6.7678^{+0}\!$\\
$\! 10^{-5}\!$ &
$\! 5.1070^{-4}\!$ &$\! 3.0642^{-4}\!$ &$\! 6.1341^{-6}\!$ &$\! 2.4536^{+0}\!$ &
$\! 4.9072^{-1}\!$ &$\! 0.0   ^{+0}\!$ &$\! 0.0   ^{+0}\!$ &$\! 5.3756^{+0}\!$\\
$\! 10^{-4}\!$ &
$\! 3.2215^{-3}\!$ &$\! 1.9327^{-3}\!$ &$\! 4.8698^{-5}\!$ &$\! 1.9478^{+0}\!$ &
$\! 3.8957^{-1}\!$ &$\! 0.0   ^{+0}\!$ &$\! 0.0   ^{+0}\!$ &$\! 4.2681^{+0}\!$\\
$\! 10^{-3}\!$ &
$\! 2.0271^{-2}\!$ &$\! 1.2151^{-2}\!$ &$\! 3.8474^{-4}\!$ &$\! 1.5382^{+0}\!$ &
$\! 3.0764^{-1}\!$ &$\! 0.0   ^{+0}\!$ &$\! 0.0   ^{+0}\!$ &$\! 3.3750^{+0}\!$\\
$\! 10^{-2}\!$ &
$\! 1.2448^{-1}\!$ &$\! 7.3939^{-2}\!$ &$\! 2.8946^{-3}\!$ &$\! 1.1520^{+0}\!$ &
$\! 2.3041^{-1}\!$ &$\! 0.0   ^{+0}\!$ &$\! 0.0   ^{+0}\!$ &$\! 2.5623^{+0}\!$\\
$\! 0.1    \!$ &
$\! 5.9008^{-1}\!$ &$\! 3.1864^{-1}\!$ &$\! 1.2979^{-2}\!$ &$\! 4.9319^{-1}\!$ &
$\! 9.8638^{-2}\!$ &$\! 0.0   ^{+0}\!$ &$\! 0.0   ^{+0}\!$ &$\! 1.2638^{+0}\!$\\
$\! 0.3    \!$ &
$\! 6.6861^{-1}\!$ &$\! 2.8082^{-1}\!$ &$\! 7.7227^{-3}\!$ &$\! 8.7524^{-2}\!$ &
$\! 1.7505^{-2}\!$ &$\! 0.0   ^{+0}\!$ &$\! 0.0   ^{+0}\!$ &$\! 3.2228^{-1}\!$\\
$\! 0.5    \!$ &
$\! 3.6666^{-1}\!$ &$\! 1.1000^{-1}\!$ &$\! 1.6243^{-3}\!$ &$\! 9.7458^{-3}\!$ &
$\! 1.9492^{-3}\!$ &$\! 0.0   ^{+0}\!$ &$\! 0.0   ^{+0}\!$ &$\! 5.6938^{-2}\!$\\
$\! 0.7    \!$ &
$\! 1.0366^{-1}\!$ &$\! 1.8659^{-2}\!$ &$\! 1.0259^{-4}\!$ &$\! 3.8103^{-4}\!$ &
$\! 7.6207^{-5}\!$ &$\! 0.0   ^{+0}\!$ &$\! 0.0   ^{+0}\!$ &$\! 4.2810^{-3}\!$\\
$\! 0.9    \!$ &
$\! 4.6944^{-3}\!$ &$\! 2.8166^{-4}\!$ &$\! 1.7644^{-7}\!$ &$\! 4.3129^{-7}\!$ &
$\! 8.6259^{-8}\!$ &$\! 0.0   ^{+0}\!$ &$\! 0.0   ^{+0}\!$ &$\! 1.7180^{-5}\!$\\
\hline \hline
\multicolumn{9}{||c||}{} \\[-3mm]
\multicolumn{9}{||c||}{LO, $\, N_{\rm f} = 4\,$, 
                       $\,\mu_{\rm f}^2 = 10^4 \mbox{ GeV}^2$} \\
\multicolumn{9}{||c||}{} \\[-0.3cm] \hline \hline
 & & & & & & & \\[-0.3cm]
$\! 10^{-7}\!$ &
$\! 5.7722^{-5}\!$ &$\! 3.4343^{-5}\!$ &$\! 7.6527^{-7}\!$ &$\! 9.9465^{+1}\!$ &
$\! 4.8642^{+1}\!$ &$\! 4.7914^{+1}\!$ &$\! 0.0   ^{+0}\!$ &$\! 1.3162^{+3}\!$\\
$\! 10^{-6}\!$ &
$\! 3.3373^{-4}\!$ &$\! 1.9800^{-4}\!$ &$\! 5.0137^{-6}\!$ &$\! 5.0259^{+1}\!$ &
$\! 2.4263^{+1}\!$ &$\! 2.3685^{+1}\!$ &$\! 0.0   ^{+0}\!$ &$\! 6.0008^{+2}\!$\\
$\! 10^{-5}\!$ &
$\! 1.8724^{-3}\!$ &$\! 1.1065^{-3}\!$ &$\! 3.1696^{-5}\!$ &$\! 2.4378^{+1}\!$ &
$\! 1.1501^{+1}\!$ &$\! 1.1042^{+1}\!$ &$\! 0.0   ^{+0}\!$ &$\! 2.5419^{+2}\!$\\
$\! 10^{-4}\!$ &
$\! 1.0057^{-2}\!$ &$\! 5.9076^{-3}\!$ &$\! 1.9071^{-4}\!$ &$\! 1.1323^{+1}\!$ &
$\! 5.1164^{+0}\!$ &$\! 4.7530^{+0}\!$ &$\! 0.0   ^{+0}\!$ &$\! 9.7371^{+1}\!$\\
$\! 10^{-3}\!$ &
$\! 5.0392^{-2}\!$ &$\! 2.9296^{-2}\!$ &$\! 1.0618^{-3}\!$ &$\! 5.0324^{+0}\!$ &
$\! 2.0918^{+0}\!$ &$\! 1.8089^{+0}\!$ &$\! 0.0   ^{+0}\!$ &$\! 3.2078^{+1}\!$\\
$\! 10^{-2}\!$ &
$\! 2.1955^{-1}\!$ &$\! 1.2433^{-1}\!$ &$\! 4.9731^{-3}\!$ &$\! 2.0433^{+0}\!$ &
$\! 7.2814^{-1}\!$ &$\! 5.3247^{-1}\!$ &$\! 0.0   ^{+0}\!$ &$\! 8.0546^{+0}\!$\\
$\! 0.1    \!$ &
$\! 5.7267^{-1}\!$ &$\! 2.8413^{-1}\!$ &$\! 1.0470^{-2}\!$ &$\! 4.0832^{-1}\!$ &
$\! 1.1698^{-1}\!$ &$\! 5.8864^{-2}\!$ &$\! 0.0   ^{+0}\!$ &$\! 8.8766^{-1}\!$\\
$\! 0.3    \!$ &
$\! 3.7925^{-1}\!$ &$\! 1.4186^{-1}\!$ &$\! 3.3029^{-3}\!$ &$\! 4.0165^{-2}\!$ &
$\! 1.0516^{-2}\!$ &$\! 4.1379_*^{-3}\!$&$\!0.0   ^{+0}\!$ &$\! 8.2676^{-2}\!$\\
$\! 0.5    \!$ &
$\! 1.3476^{-1}\!$ &$\! 3.5364^{-2}\!$ &$\! 4.2815^{-4}\!$ &$\! 2.8624^{-3}\!$ &
$\! 7.3138^{-4}\!$ &$\! 2.6481^{-4}\!$ &$\! 0.0   ^{+0}\!$ &$\! 7.9240^{-3}\!$\\
$\! 0.7    \!$ &
$\! 2.3123^{-2}\!$ &$\! 3.5943^{-3}\!$ &$\! 1.5868^{-5}\!$ &$\! 6.8961^{-5}\!$ &
$\! 1.7725^{-5}\!$ &$\! 6.5549^{-6}\!$ &$\! 0.0   ^{+0}\!$ &$\! 3.7311^{-4}\!$\\
$\! 0.9    \!$ &
$\! 4.3443^{-4}\!$ &$\! 2.2287^{-5}\!$ &$\! 1.1042^{-8}\!$ &$\! 3.6293^{-8}\!$ &
$\! 1.0192^{-8}\!$ &$\! 4.8893_*^{-9}\!$&$\!0.0   ^{+0}\!$ &$\! 1.0918^{-6}\!$\\
\hline \hline
\multicolumn{9}{||c||}{} \\[-3mm]
\multicolumn{9}{||c||}{LO, $\, N_{\rm f} = 3\ldots 5\,$, 
                        $\,\mu_{\rm f}^2 = 10^4 \mbox{ GeV}^2$} \\
\multicolumn{9}{||c||}{} \\[-0.3cm] \hline \hline
 & & & & & & & \\[-0.3cm]
$\! 10^{-7}\!$ &
$\! 5.8771^{-5}\!$ &$\! 3.4963^{-5}\!$ &$\! 7.8233^{-7}\!$ &$\! 1.0181^{+2}\!$ &
$\! 4.9815^{+1}\!$ &$\! 4.9088^{+1}\!$ &$\! 4.6070^{+1}\!$ &$\! 1.3272^{+3}\!$\\
$\! 10^{-6}\!$ &
$\! 3.3933^{-4}\!$ &$\! 2.0129^{-4}\!$ &$\! 5.1142^{-6}\!$ &$\! 5.1182^{+1}\!$ &
$\! 2.4725^{+1}\!$ &$\! 2.4148^{+1}\!$ &$\! 2.2239^{+1}\!$ &$\! 6.0117^{+2}\!$\\
$\! 10^{-5}\!$ &
$\! 1.9006^{-3}\!$ &$\! 1.1229^{-3}\!$ &$\! 3.2249^{-5}\!$ &$\! 2.4693^{+1}\!$ &
$\! 1.1659^{+1}\!$ &$\! 1.1201^{+1}\!$ &$\! 1.0037^{+1}\!$ &$\! 2.5282^{+2}\!$\\
$\! 10^{-4}\!$ &
$\! 1.0186^{-2}\!$ &$\! 5.9819^{-3}\!$ &$\! 1.9345^{-4}\!$ &$\! 1.1406^{+1}\!$ &
$\! 5.1583^{+0}\!$ &$\! 4.7953^{+0}\!$ &$\! 4.1222^{+0}\!$ &$\! 9.6048^{+1}\!$\\
$\! 10^{-3}\!$ &
$\! 5.0893^{-2}\!$ &$\! 2.9576^{-2}\!$ &$\! 1.0730^{-3}\!$ &$\! 5.0424^{+0}\!$ &
$\! 2.0973^{+0}\!$ &$\! 1.8147^{+0}\!$ &$\! 1.4582^{+0}\!$ &$\! 3.1333^{+1}\!$\\
$\! 10^{-2}\!$ &
$\! 2.2080^{-1}\!$ &$\! 1.2497^{-1}\!$ &$\! 4.9985^{-3}\!$ &$\! 2.0381^{+0}\!$ &
$\! 7.2625^{-1}\!$ &$\! 5.3107^{-1}\!$ &$\! 3.8106^{-1}\!$ &$\! 7.7728^{+0}\!$\\
$\! 0.1    \!$ &
$\! 5.7166^{-1}\!$ &$\! 2.8334^{-1}\!$ &$\! 1.0428^{-2}\!$ &$\! 4.0496^{-1}\!$ &
$\! 1.1596^{-1}\!$ &$\! 5.8288^{-2}\!$ &$\! 3.5056^{-2}\!$ &$\! 8.4358^{-1}\!$\\
$\! 0.3    \!$ &
$\! 3.7597^{-1}\!$ &$\! 1.4044^{-1}\!$ &$\! 3.2629^{-3}\!$ &$\! 3.9592^{-2}\!$ &
$\! 1.0363^{-2}\!$ &$\! 4.0740^{-3}\!$ &$\! 2.2039^{-3}\!$ &$\! 7.8026^{-2}\!$\\
$\! 0.5    \!$ &
$\! 1.3284^{-1}\!$ &$\! 3.4802^{-2}\!$ &$\! 4.2031^{-4}\!$ &$\! 2.8066^{-3}\!$ &
$\! 7.1707^{-4}\!$ &$\! 2.5958^{-4}\!$ &$\! 1.3522^{-4}\!$ &$\! 7.4719^{-3}\!$\\
$\! 0.7    \!$ &
$\! 2.2643^{-2}\!$ &$\! 3.5134^{-3}\!$ &$\! 1.5468^{-5}\!$ &$\! 6.7201^{-5}\!$ &
$\! 1.7278^{-5}\!$ &$\! 6.3958^{-6}\!$ &$\! 3.3996^{-6}\!$ &$\! 3.5241^{-4}\!$\\
$\! 0.9    \!$ &
$\! 4.2047^{-4}\!$ &$\! 2.1529^{-5}\!$ &$\! 1.0635^{-8}\!$ &$\! 3.4998^{-8}\!$ &
$\! 9.8394^{-9}\!$ &$\! 4.7330^{-9}\!$ &$\! 2.8903^{-9}\!$ &$\! 1.0307^{-6}\!$\\
\hline \hline
\end{tabular}
\end{center}
\end{table}

\begin{table}[htp]
\caption{Reference results for the $N_{\rm f}=4$ next-to-leading-order 
 evolution for the initial conditions (\ref{gsav-eq8}) -- (\ref{gsav-eq10}). 
 The corresponding value of the strong coupling is $\alpha_{\rm s}
 (\mu_{\rm r}^2 \! =\! 10^4 \mbox{ GeV}^2) = 0.110902$. As in the leading-order 
 case, the valence distributions $s_v$ and $c_v$ vanish for the input 
 (\ref{gsav-eq9}). The notation is explained in the first two paragraphs of 
 Section \ref{gsav-s3}.}
\label{gsav-t2}
\vspace{4mm}
\begin{center}
\begin{tabular}{||c||r|r|r|r|r|r|r||}
\hline \hline
\multicolumn{8}{||c||}{} \\[-3mm]
\multicolumn{8}{||c||}{NLO, $\, N_{\rm f} = 4$, 
                       $\, \mu_{\rm f}^2 = 10^4 \mbox{ GeV}^2$} \\
\multicolumn{8}{||c||}{} \\[-0.35cm] \hline \hline
 & & & & & & & \\[-0.3cm]
\multicolumn{1}{||c||}{$x$} &
\multicolumn{1}{c|} {$xu_v$} &
\multicolumn{1}{c|} {$xd_v$} &
\multicolumn{1}{c|} {$xL_-$} &
\multicolumn{1}{c|} {$xL_+$} &
\multicolumn{1}{c|} {$xs_+$} &
\multicolumn{1}{c|} {$xc_+$} &
\multicolumn{1}{c||}{$xg$} \\ \hline \hline
\multicolumn{8}{||c||}{} \\[-3mm]
\multicolumn{8}{||c||}{$\mu_{\rm r}^2 = \mu_{\rm f}^2$} \\
\multicolumn{8}{||c||}{} \\[-0.3cm] \hline \hline
 & & & & & & & \\[-0.3cm]
 $10^{-7}$ & $1.0616^{-4}$ & $6.2328^{-5}$ & $4.2440^{-6}$ & $1.3598^{+2}$
           & $6.6913_*^{+1}$&$6.6195^{+1}$ & $1.1483_*^{+3}$ \\
 $10^{-6}$ & $5.4177^{-4}$ & $3.1719^{-4}$ & $1.9241^{-5}$ & $6.8396^{+1}$
           & $3.3342^{+1}$ & $3.2771^{+1}$ & $5.3911^{+2}$ \\
 $10^{-5}$ & $2.6870^{-3}$ & $1.5677^{-3}$ & $8.3575^{-5}$ & $3.2728^{+1}$
           & $1.5685^{+1}$ & $1.5231^{+1}$ & $2.3528^{+2}$ \\
 $10^{-4}$ & $1.2841^{-2}$ & $7.4558^{-3}$ & $3.4911^{-4}$ & $1.4746^{+1}$ 
           & $6.8355^{+0}$ & $6.4769^{+0}$ & $9.2872_*^{+1}$ \\
 $10^{-3}$ & $5.7926^{-2}$ & $3.3337^{-2}$ & $1.4162^{-3}$ &$6.1648_*^{+0}$ 
           & $2.6659^{+0}$ & $2.3878^{+0}$ & $3.1502^{+1}$ \\
 $10^{-2}$ & $2.3026^{-1}$ & $1.2928^{-1}$ & $5.3251^{-3}$ & $2.2527^{+0}$  
           & $8.4220_*^{-1}$&$6.5246^{-1}$ & $8.1066^{+0}$ \\
 $0.1$     & $5.5452^{-1}$ & $2.7336^{-1}$ & $1.0011^{-2}$ & $3.9336^{-1}$
           & $1.1489^{-1}$ & $6.0351^{-2}$ & $8.9867^{-1}$ \\
 $0.3$     & $3.5393^{-1}$ & $1.3158^{-1}$ & $3.0362^{-3}$ & $3.5848^{-2}$
           & $9.2030^{-3}$ & $3.3890^{-3}$ & $8.3451^{-2}$ \\
 $0.5$     & $1.2271^{-1}$ & $3.1967^{-2}$ & $3.8265^{-4}$ & $2.4126^{-3}$
           & $5.8424^{-4}$ & $1.6955^{-4}$ & $8.0473^{-3}$ \\
 $0.7$     & $2.0429^{-2}$ & $3.1473_*^{-3}$&$1.3701^{-5}$ & $5.3622^{-5}$
           & $1.2393^{-5}$ & $2.7807^{-6}$ & $3.8721^{-4}$ \\
 $0.9$     & $3.6096^{-4}$ & $1.8317^{-5}$ & $8.923 ^{-9}$ & $2.092 ^{-8}$
           & $4.039 ^{-9}$ &$-2.405^{-10}$ & $1.2127^{-6}$ \\
\hline \hline
\multicolumn{8}{||c||}{} \\[-3mm]
\multicolumn{8}{||c||}{$\mu_{\rm r}^2 = 2\, \mu_{\rm f}^2$} \\
\multicolumn{8}{||c||}{} \\[-0.3cm] \hline \hline
 & & & & & & & \\[-0.3cm]
 $10^{-7}$ & $9.2960^{-5}$ & $5.4699^{-5}$ & $3.3861^{-6}$ & $1.2214^{+2}$
           & $5.9987^{+1}$ & $5.9265^{+1}$ & $1.0911^{+3}$ \\
 $10^{-6}$ & $4.8463^{-4}$ & $2.8440^{-4}$ & $1.5820^{-5}$ & $6.1831^{+1}$
           & $3.0056^{+1}$ & $2.9483^{+1}$ & $5.1456^{+2}$ \\
 $10^{-5}$ & $2.4578^{-3}$ & $1.4374^{-3}$ & $7.1265^{-5}$ & $2.9845^{+1}$
           & $1.4240^{+1}$ & $1.3785^{+1}$ & $2.2580^{+2}$ \\
 $10^{-4}$ & $1.2018^{-2}$ & $6.9946^{-3}$ & $3.1111^{-4}$ & $1.3618^{+1}$
           & $6.2690^{+0}$ & $5.9088^{+0}$ & $8.9753^{+1}$ \\
 $10^{-3}$ & $5.5483^{-2}$ & $3.2009^{-2}$ & $1.3254^{-3}$ & $5.8076^{+0}$
           & $2.4848^{+0}$ & $2.2050^{+0}$ & $3.0729^{+1}$ \\
 $10^{-2}$ & $2.2595^{-1}$ & $1.2720^{-1}$ & $5.2141^{-3}$ & $2.1896^{+0}$
           & $8.0746^{-1}$ & $6.1564^{-1}$ & $8.0188^{+0}$ \\
 $0.1$     & $5.6007^{-1}$ & $2.7697^{-1}$ & $1.0180^{-2}$ & $3.9945^{-1}$
           & $1.1570^{-1}$ & $5.9661^{-2}$ & $9.1201^{-1}$ \\
 $0.3$     & $3.6474^{-1}$ & $1.3612^{-1}$ & $3.1588^{-3}$ & $3.7501^{-2}$
           & $9.6302^{-3}$ & $3.5499^{-3}$ & $8.6368^{-2}$ \\
 $0.5$     & $1.2843^{-1}$ & $3.3610^{-2}$ & $4.0510^{-4}$ & $2.5822^{-3}$
           & $6.3044^{-4}$ & $1.8999^{-4}$ & $8.4178^{-3}$ \\
 $0.7$     & $2.1779^{-2}$ & $3.3725^{-3}$ & $1.4798^{-5}$ & $5.9125^{-5}$
           & $1.3961^{-5}$ & $3.5593^{-6}$ & $4.0836^{-4}$ \\
 $0.9$     & $3.9817^{-4}$ & $2.0321^{-5}$ & $9.987 ^{-9}$ & $2.555 ^{-8}$
           & $5.586 ^{-9}$ & $7.930^{-10}$ & 1.3008$^{-6}$ \\
\hline \hline
\multicolumn{8}{||c||}{} \\[-3mm]
\multicolumn{8}{||c||}{$\mu_{\rm r}^2 = 1/2\, \mu_{\rm f}^2$} \\
\multicolumn{8}{||c||}{} \\[-0.3cm] \hline \hline
 & & & & & & & \\[-0.3cm]
 $10^{-7}$ & $1.2438^{-4}$ &$ 7.2817^{-5}$ & $5.5568^{-6}$ & $1.4556^{+2}$
           & $7.1706^{+1}$ &$ 7.0990^{+1}$ & $1.1468^{+3}$ \\
 $10^{-6}$ & $6.1759^{-4}$ &$ 3.6051^{-4}$ & $2.4322^{-5}$ & $7.3406^{+1}$
           & $3.5851^{+1}$ &$ 3.5282_*^{+1}$&$5.4041^{+2}$ \\
 $10^{-5}$ & $2.9770^{-3}$ &$ 1.7316^{-3}$ & $1.0121_*^{-5}$&$3.5158_*^{+1}$
           & $1.6903^{+1}$ &$ 1.6452^{+1}$ & $2.3663^{+2}$ \\
 $10^{-4}$ & $1.3820^{-2}$ &$ 7.9998^{-3}$ & $4.0093^{-4}$ & $1.5795^{+1}$
           & $7.3626^{+0}$ &$ 7.0057^{+0}$ & $9.3640^{+1}$ \\
 $10^{-3}$ & $6.0585^{-2}$ &$ 3.4766^{-2}$ & $1.5300^{-3}$ & $6.5284^{+0}$
           & $2.8504^{+0}$ &$ 2.5740^{+0}$ & $3.1795^{+1}$ \\
 $10^{-2}$ & $2.3422_*^{-1}$&$1.3114^{-1}$ & $5.4411^{-3}$ & $2.3221^{+0}$
           & $8.8022^{-1}$ &$ 6.9260^{-1}$ & $8.1613^{+0}$ \\
 0.1       & $5.4824^{-1}$ &$ 2.6954^{-1}$ & $9.8435^{-2}$ & $3.8787^{-1}$
           & $1.1419^{-1}$ &$ 6.0997^{-2}$ & $8.9361^{-1}$ \\
 0.3       & $3.4425^{-1}$ &$ 1.2760^{-1}$ & $2.9317^{-3}$ & $3.4294^{-2}$
           & $8.7599^{-3}$ &$ 3.1681^{-3}$ & $8.2031^{-2}$ \\
 0.5       & $1.1794^{-1}$ &$ 3.0618^{-2}$ & $3.6454^{-4}$ & $2.2530^{-3}$
           & $5.3541^{-4}$ &$ 1.4134^{-4}$ & $7.8595^{-3}$ \\
 0.7       & $1.9356^{-2}$ &$ 2.9698^{-3}$ & $1.2847^{-5}$ & $4.8328^{-5}$
           & $1.0666^{-5}$ &$ 1.6668^{-6}$ & $3.7624_*^{-4}$ \\
 0.9       & $3.3264^{-4}$ &$ 1.6800^{-5}$ & $8.124 ^{-9}$ & $1.573 ^{-8}$
           & $2.024 ^{-9}$ &$-1.870 ^{-9}$ & $1.1647^{-6}$ \\
\hline \hline
\end{tabular}
\end{center}
\end{table}

\begin{table}[htp]
\caption{As Table \ref{gsav-t2}, but for the variable-$N_{\rm f}$ evolution 
 using Eqs.~(\ref{gsav-eq12}) -- (\ref{gsav-eq15}).
 The corresponding values for the strong coupling $\alpha_{\rm s}(\mu_{\rm r}^2 
 \! =\! 10^4 \mbox{ GeV}^2)$ are given by $0.116461$, $0.116032$ and $0.115663$
 for $\mu_{\rm r}^2 / \mu_{\rm f}^2 = 0.5$, $1$ and $2$, respectively.} 
\label{gsav-t3}
\vspace{4mm}
\begin{center}
\begin{tabular}{||c||r|r|r|r|r|r|r|r||}
\hline \hline
\multicolumn{9}{||c||}{} \\[-3mm]
\multicolumn{9}{||c||}{NLO, $\, N_{\rm f} = 3\,\ldots 5\,$,
                       $\,\mu_{\rm f}^2 = 10^4 \mbox{ GeV}^2$} \\
\multicolumn{9}{||c||}{} \\[-0.3cm] \hline \hline
 & & & & & & & \\[-0.3cm]
\multicolumn{1}{||c||}{$x$} &
\multicolumn{1}{c|} {$xu_v$} &
\multicolumn{1}{c|} {$xd_v$} &
\multicolumn{1}{c|} {$xL_-$} &
\multicolumn{1}{c|} {$xL_+$} &
\multicolumn{1}{c|} {$xs_+$} &
\multicolumn{1}{c|} {$xc_+$} &
\multicolumn{1}{c|} {$xb_+$} &
\multicolumn{1}{c||}{$xg$} \\[0.5mm] \hline \hline
\multicolumn{9}{||c||}{} \\[-3mm]
\multicolumn{9}{||c||}{$\mu_{\rm r}^2 = \mu_{\rm f}^2$} \\
\multicolumn{9}{||c||}{} \\[-0.3cm] \hline \hline
 & & & & & & & \\[-0.3cm]
$\! 10^{-7}\!$ &
$\! 1.0927^{-4}\!$ &$\! 6.4125^{-5}\!$ &$\! 4.3925^{-6}\!$ &$\! 1.3787^{+2}\!$ &
$\! 6.7857^{+1}\!$ &$\! 6.7139^{+1}\!$ &$\! 6.0071^{+1}\!$ &$\! 1.1167^{+3}\!$\\
$\! 10^{-6}\!$ &
$\! 5.5533^{-4}\!$ &$\! 3.2498^{-4}\!$ &$\! 1.9829^{-5}\!$ &$\! 6.9157^{+1}\!$ &
$\! 3.3723^{+1}\!$ &$\! 3.3153^{+1}\!$ &$\! 2.8860^{+1}\!$&$\!5.2289_*^{+2}\!$\\
$\! 10^{-5}\!$ &
$\! 2.7419^{-3}\!$ &$\! 1.5989^{-3}\!$ &$\! 8.5701^{-5}\!$ &$\! 3.2996^{+1}\!$ &
$\! 1.5819^{+1}\!$ &$\! 1.5367^{+1}\!$ &$\! 1.2892^{+1}\!$ &$\! 2.2753^{+2}\!$\\
$\! 10^{-4}\!$ &
$\! 1.3039^{-2}\!$ &$\! 7.5664^{-3}\!$ &$\! 3.5582^{-4}\!$ &$\! 1.4822^{+1}\!$ &
$\! 6.8739^{+0}\!$ &$\! 6.5156^{+0}\!$ &$\! 5.1969^{+0}\!$ &$\! 8.9513^{+1}\!$\\
$\! 10^{-3}\!$ &
$\! 5.8507^{-2}\!$ &$\! 3.3652^{-2}\!$ &$\! 1.4329^{-3}\!$ &$\! 6.1772^{+0}\!$ &
$\! 2.6726^{+0}\!$ &$\! 2.3949^{+0}\!$ &$\! 1.7801^{+0}\!$ &$\! 3.0245^{+1}\!$\\
$\! 10^{-2}\!$ &
$\! 2.3128^{-1}\!$ &$\! 1.2978^{-1}\!$ &$\! 5.3472^{-3}\!$ &$\! 2.2500^{+0}\!$ &
$\! 8.4161^{-1}\!$ &$\! 6.5235^{-1}\!$ &$\! 4.3894^{-1}\!$ &$\! 7.7491^{+0}\!$\\
$\! 0.1    \!$ &
$\! 5.5324^{-1}\!$ &$\! 2.7252^{-1}\!$ &$\! 9.9709^{-3}\!$ &$\! 3.9099^{-1}\!$ &
$\! 1.1425^{-1}\!$ &$\! 6.0071^{-2}\!$ &$\! 3.5441^{-2}\!$ &$\! 8.5586^{-1}\!$\\
$\! 0.3    \!$ &
$\! 3.5129^{-1}\!$ &$\! 1.3046^{-1}\!$ &$\! 3.0061^{-3}\!$ &$\! 3.5463^{-2}\!$ &
$\! 9.1084^{-3}\!$ &$\! 3.3595^{-3}\!$ &$\! 1.9039^{-3}\!$ &$\! 7.9625^{-2}\!$\\
$\! 0.5    \!$ &
$\! 1.2130^{-1}\!$ &$\! 3.1564^{-2}\!$ &$\! 3.7719^{-4}\!$ &$\! 2.3775^{-3}\!$ &
$\! 5.7606^{-4}\!$ &$\! 1.6761^{-4}\!$ &$\! 1.0021^{-4}\!$ &$\! 7.7265^{-3}\!$\\
$\! 0.7    \!$ &
$\! 2.0101^{-2}\!$&$\! 3.0932^{-3}\!$ &$\! 1.3440^{-5}\!$ &$\! 5.2605^{-5}\!$ &
$\! 1.2166^{-5}\!$ &$\! 2.7408^{-6}\!$ &$\! 2.0095^{-6}\!$ &$\! 3.7574^{-4}\!$\\
$\! 0.9    \!$ &
$\!3.5232_*^{-4}\!$&$\! 1.7855^{-5}\!$ &$\! 8.680 ^{-9}\!$ &$\! 2.028 ^{-8}\!$ &
$\! 3.896 ^{-9}\!$ &$\!-2.666^{-10}\!$ &$\! 5.819^{-10}\!$ &$\! 1.1954^{-6}\!$\\
\hline \hline
\multicolumn{9}{||c||}{} \\[-3mm]
\multicolumn{9}{||c||}{$\mu_{\rm r}^2 = 2\, \mu_{\rm f}^2$} \\
\multicolumn{9}{||c||}{} \\[-0.3cm] \hline \hline
 & & & & & & & \\[-0.3cm]
$\! 10^{-7}\!$ &
$\! 9.5154^{-5}\!$ &$\! 5.5970^{-5}\!$ &$\! 3.4869^{-6}\!$ &$\! 1.2301^{+2}\!$ &
$\! 6.0424^{+1}\!$ &$\! 5.9703^{+1}\!$ &$\! 5.3916^{+1}\!$ &$\! 1.0568^{+3}\!$\\
$\! 10^{-6}\!$ &
$\! 4.9433^{-4}\!$ &$\! 2.8998^{-4}\!$ &$\! 1.6229^{-5}\!$ &$\! 6.2149^{+1}\!$ &
$\! 3.0215^{+1}\!$ &$\! 2.9643^{+1}\!$ &$\! 2.6100^{+1}\!$ &$\! 4.9744^{+2}\!$\\
$\! 10^{-5}\!$ &
$\! 2.4974^{-3}\!$ &$\! 1.4600^{-3}\!$ &$\! 7.2776^{-5}\!$ &$\! 2.9936^{+1}\!$ &
$\! 1.4286^{+1}\!$ &$\! 1.3831^{+1}\!$ &$\! 1.1768^{+1}\!$ &$\! 2.1783^{+2}\!$\\
$\! 10^{-4}\!$ &
$\! 1.2161^{-2}\!$ &$\! 7.0751^{-3}\!$ &$\! 3.1597^{-4}\!$ &$\! 1.3631^{+1}\!$ &
$\! 6.2759^{+0}\!$ &$\! 5.9160^{+0}\!$ &$\! 4.7997^{+0}\!$ &$\! 8.6372^{+1}\!$\\
$\! 10^{-3}\!$ &
$\! 5.5907^{-2}\!$ &$\! 3.2239^{-2}\!$ &$\!1.3377_*^{-3}\!$&$\! 5.8020^{+0}\!$ &
$\! 2.4824^{+0}\!$ &$\! 2.2029^{+0}\!$ &$\! 1.6701^{+0}\!$ &$\! 2.9488^{+1}\!$\\
$\! 10^{-2}\!$ &
$\! 2.2670^{-1}\!$ &$\! 1.2756^{-1}\!$ &$\! 5.2304^{-3}\!$ &$\! 2.1840^{+0}\!$ &
$\! 8.0522^{-1}\!$ &$\! 6.1376^{-1}\!$ &$\! 4.2146^{-1}\!$ &$\! 7.6713^{+0}\!$\\
$\! 0.1    \!$ &
$\! 5.5915^{-1}\!$ &$\! 2.7636^{-1}\!$ &$\! 1.0151^{-2}\!$ &$\! 3.9744^{-1}\!$ &
$\! 1.1509^{-1}\!$ &$\! 5.9318^{-2}\!$ &$\! 3.5492^{-2}\!$ &$\! 8.7075^{-1}\!$\\
$\! 0.3    \!$ &
$\! 3.6281^{-1}\!$ &$\! 1.3530^{-1}\!$ &$\! 3.1367^{-3}\!$ &$\! 3.7201^{-2}\!$ &
$\! 9.5529^{-3}\!$ &$\! 3.5209^{-3}\!$ &$\! 1.9869^{-3}\!$ &$\! 8.2697^{-2}\!$\\
$\! 0.5    \!$ &
$\! 1.2739^{-1}\!$ &$\! 3.3311^{-2}\!$ &$\! 4.0102^{-4}\!$ &$\! 2.5552^{-3}\!$ &
$\! 6.2392^{-4}\!$ &$\! 1.8815^{-4}\!$ &$\! 1.0842^{-4}\!$ &$\! 8.1131^{-3}\!$\\
$\! 0.7    \!$ &
$\! 2.1534^{-2}\!$ &$\! 3.3317^{-3}\!$ &$\! 1.4600^{-5}\!$ &$\! 5.8333^{-5}\!$ &
$\! 1.3778^{-5}\!$ &$\! 3.5193^{-6}\!$ &$\! 2.2960^{-6}\!$ &$\! 3.9788^{-4}\!$\\
$\! 0.9    \!$ &
$\! 3.9150^{-4}\!$ &$\! 1.9963^{-5}\!$ &$\! 9.797 ^{-9}\!$ &$\! 2.502 ^{-8}\!$ &
$\! 5.459 ^{-9}\!$ &$\! 7.574^{-10}\!$ &$\! 1.002 ^{-9}\!$ &$\! 1.2895^{-6}\!$\\
\hline \hline
\multicolumn{9}{||c||}{} \\[-3mm]
\multicolumn{9}{||c||}{$\mu_{\rm r}^2 = 1/2\, \mu_{\rm f}^2$} \\
\multicolumn{9}{||c||}{} \\[-0.3cm] \hline \hline
 & & & & & & & \\[-0.3cm]
$\! 10^{-7}\!$ &
$\! 1.2937^{-4}\!$ &$\! 7.5695^{-5}\!$ &$\! 5.8161^{-6}\!$ &$\! 1.4923^{+2}\!$ &
$\! 7.3543^{+1}\!$ &$\! 7.2829^{+1}\!$ &$\! 6.4225^{+1}\!$ &$\! 1.1239^{+3}\!$\\
$\! 10^{-6}\!$ &
$\! 6.3890^{-4}\!$ &$\! 3.7272^{-4}\!$ &$\! 2.5317^{-5}\!$ &$\! 7.4973^{+1}\!$ &
$\! 3.6635^{+1}\!$ &$\! 3.6068^{+1}\!$ &$\! 3.0846^{+1}\!$ &$\! 5.2763^{+2}\!$\\
$\! 10^{-5}\!$ &
$\! 3.0615^{-3}\!$ &$\! 1.7796^{-3}\!$ &$\! 1.0470^{-4}\!$ &$\! 3.5762^{+1}\!$ &
$\! 1.7205^{+1}\!$ &$\! 1.6755^{+1}\!$ &$\! 1.3756^{+1}\!$ &$\! 2.3006^{+2}\!$\\
$\! 10^{-4}\!$ &
$\! 1.4120^{-2}\!$ &$\! 8.1674^{-3}\!$ &$\! 4.1155^{-4}\!$ &$\! 1.5994^{+1}\!$ &
$\! 7.4628^{+0}\!$ &$\! 7.1064^{+0}\!$ &$\!5.5232_*^{+0}\!$&$\! 9.0599^{+1}\!$\\
$\! 10^{-3}\!$ &
$\! 6.1458^{-2}\!$ &$\! 3.5239^{-2}\!$ &$\! 1.5557^{-3}\!$ &$\! 6.5774^{+0}\!$ &
$\! 2.8758^{+0}\!$ &$\! 2.5999^{+0}\!$ &$\! 1.8773^{+0}\!$ &$\! 3.0589^{+1}\!$\\
$\! 10^{-2}\!$ &
$\! 2.3574^{-1}\!$ &$\! 1.3187^{-1}\!$ &$\! 5.4739^{-3}\!$ &$\! 2.3254^{+0}\!$ &
$\! 8.8301^{-1}\!$ &$\! 6.9613^{-1}\!$ &$\! 4.5587^{-1}\!$ &$\! 7.7996^{+0}\!$\\
$\! 0.1    \!$ &
$\! 5.4630^{-1}\!$ &$\! 2.6827^{-1}\!$ &$\! 9.7828^{-3}\!$ &$\! 3.8485^{-1}\!$ &
$\! 1.1348^{-1}\!$ &$\! 6.0825^{-2}\!$ &$\! 3.5524^{-2}\!$ &$\! 8.4801^{-1}\!$\\
$\! 0.3    \!$ &
$\! 3.4035^{-1}\!$ &$\! 1.2596^{-1}\!$ &$\! 2.8877^{-3}\!$ &$\! 3.3752^{-2}\!$ &
$\! 8.6313^{-3}\!$ &$\! 3.1345^{-3}\!$ &$\! 1.8308^{-3}\!$ &$\! 7.7817^{-2}\!$\\
$\! 0.5    \!$ &
$\! 1.1589^{-1}\!$ &$\! 3.0034^{-2}\!$ &$\! 3.5665^{-4}\!$ &$\! 2.2032^{-3}\!$ &
$\! 5.2398^{-4}\!$ &$\! 1.3891^{-4}\!$ &$\! 9.2307^{-5}\!$ &$\! 7.4943^{-3}\!$\\
$\! 0.7    \!$ &
$\! 1.8886^{-2}\!$ &$\! 2.8923^{-3}\!$ &$\! 1.2475^{-5}\!$ &$\! 4.6898^{-5}\!$ &
$\! 1.0350^{-5}\!$ &$\! 1.6179^{-6}\!$ &$\! 1.7069^{-6}\!$ &$\! 3.6221^{-4}\!$\\
$\! 0.9    \!$ &
$\! 3.2053^{-4}\!$ &$\! 1.6155^{-5}\!$ &$\! 7.787 ^{-9}\!$ &$\! 1.489 ^{-8}\!$ &
$\! 1.848 ^{-9}\!$ &$\!-1.884 ^{-9}\!$ &$\! 6.129^{-11}\!$ &$\! 1.1353^{-6}\!$\\
\hline \hline
\end{tabular}
\end{center}
\end{table}

\begin{table}[htp]
\caption{Reference results for the $N_{\rm f}\!=\!4$ next-next-to-leading-order
 evolution for the initial conditions (\ref{gsav-eq8}) -- (\ref{gsav-eq10}).
 The corresponding value of the strong coupling is $\alpha_{\rm s}
 (\mu_{\rm r}^2 \! =\! 10^4 \mbox{ GeV}^2) = 0.110141$. The valence 
 distributions $s_v$ and $c_v$ are equal for the input (\ref{gsav-eq9}). The 
 notation is explained in the first two paragraphs of Section \ref{gsav-s3}.}
\label{gsav-t4}
\vspace{4mm}
\begin{center}
\begin{tabular}{||c||r|r|r|r|r|r|r|r||}
\hline \hline
\multicolumn{9}{||c||}{} \\[-3mm]
\multicolumn{9}{||c||}{NNLO, $\, N_{\rm f} = 4\,$, 
                       $\,\mu_{\rm f}^2 = 10^4 \mbox{ GeV}^2$} \\
\multicolumn{9}{||c||}{} \\[-0.3cm] \hline \hline
 & & & & & & & \\[-0.3cm]
\multicolumn{1}{||c||}{$x$} &
\multicolumn{1}{c|} {$xu_v$} &
\multicolumn{1}{c|} {$xd_v$} &
\multicolumn{1}{c|} {$xL_-$} &
\multicolumn{1}{c|} {$xL_+$} &
\multicolumn{1}{c|} {$xs_v$} &
\multicolumn{1}{c|} {$xs_+$} &
\multicolumn{1}{c|} {$xc_+$} &
\multicolumn{1}{c||}{$xg$} \\[0.5mm] \hline \hline
\multicolumn{9}{||c||}{} \\[-3mm]
\multicolumn{9}{||c||}{$\mu_{\rm r}^2 = \mu_{\rm f}^2$} \\
\multicolumn{9}{||c||}{} \\[-0.3cm] \hline \hline
 & & & & & & & \\[-0.3cm]
$\! 10^{-7}\!$ &
$\! 1.4425^{-4}\!$ &$\! 8.9516_*^{-5}\!$&$\!6.0836^{-6}\!$ &$\! 1.3492^{+2}\!$ &
$\! 1.3116^{-5}\!$ &$\! 6.6385^{+1}\!$ &$\! 6.5669^{+1}\!$ &$\! 1.0105^{+3}\!$\\
$\! 10^{-6}\!$ &
$\! 6.7325^{-4}\!$ &$\! 4.1192^{-4}\!$ &$\! 2.6060^{-5}\!$ &$\! 6.9636^{+1}\!$ &
$\! 4.8617^{-5}\!$ &$\! 3.3965^{+1}\!$ &$\! 3.3396^{+1}\!$ &$\! 4.9543^{+2}\!$\\
$\! 10^{-5}\!$ &
$\! 3.0782^{-3}\!$ &$\! 1.8510^{-3}\!$ &$\! 1.0652^{-4}\!$ &$\! 3.3826^{+1}\!$ &
$\! 1.5280^{-4}\!$ &$\! 1.6236^{+1}\!$ &$\! 1.5784^{+1}\!$ &$\! 2.2406^{+2}\!$\\
$\! 10^{-4}\!$ &
$\! 1.3746^{-2}\!$ &$\! 8.0979^{-3}\!$ &$\! 4.1608^{-4}\!$ &$\! 1.5278^{+1}\!$ &
$\! 3.3973^{-4}\!$ &$\! 7.1041^{+0}\!$ &$\! 6.7468^{+0}\!$&$\!9.0854_*^{+1}\!$\\
$\! 10^{-3}\!$ &
$\! 5.9222^{-2}\!$ &$\! 3.4137^{-2}\!$ &$\! 1.5718^{-3}\!$ &$\! 6.3249^{+0}\!$ &
$\! 2.6800^{-4}\!$ &$\! 2.7483^{+0}\!$ &$\! 2.4713^{+0}\!$ &$\! 3.1353^{+1}\!$\\
$\! 10^{-2}\!$ &
$\! 2.3074^{-1}\!$ &$\! 1.2920^{-1}\!$ &$\! 5.5461^{-3}\!$ &$\! 2.2736^{+0}\!$ &
$\!-5.2053^{-4}\!$ &$\! 8.5420^{-1}\!$ &$\! 6.6542^{-1}\!$ &$\! 8.1341^{+0}\!$\\
$\! 0.1    \!$ &
$\! 5.5178^{-1}\!$ &$\! 2.7165^{-1}\!$ &$\! 1.0024^{-2}\!$ &$\! 3.9020^{-1}\!$ &
$\!-3.0680^{-4}\!$ &$\! 1.1386^{-1}\!$ &$\! 5.9780^{-2}\!$ &$\! 9.0567^{-1}\!$\\
$\! 0.3    \!$ &
$\! 3.5071^{-1}\!$ &$\! 1.3025^{-1}\!$ &$\! 3.0098^{-3}\!$ &$\! 3.5358^{-2}\!$ &
$\!-3.1905^{-5}\!$ &$\! 9.0479^{-3}\!$ &$\! 3.3060^{-3}\!$ &$\! 8.4184^{-2}\!$\\
$\! 0.5    \!$ &
$\! 1.2117^{-1}\!$ &$\! 3.1528^{-2}\!$ &$\! 3.7743^{-4}\!$ &$\! 2.3866^{-3}\!$ &
$\!-2.7207^{-6}\!$ &$\! 5.7966^{-4}\!$ &$\! 1.7171^{-4}\!$&$\!8.1127_*^{-3}\!$\\
$\! 0.7    \!$ &
$\! 2.0078^{-2}\!$ &$\! 3.0886^{-3}\!$ &$\! 1.3442^{-5}\!$ &$\! 5.4226^{-5}\!$ &
$\!-1.0121^{-7}\!$ &$\! 1.2936^{-5}\!$ &$\! 3.5305^{-6}\!$ &$\! 3.8946^{-4}\!$\\
$\! 0.9    \!$ &
$\! 3.5111^{-4}\!$ &$\! 1.7783^{-5}\!$ &$\! 8.870 ^{-9}\!$ &$\! 2.630 ^{-8}\!$ &
$\!-1.450^{-10}\!$ &$\! 7.115 ^{-9}\!$ &$\! 2.973 ^{-9}\!$ &$\! 1.2147^{-6}\!$\\
\hline \hline
\multicolumn{9}{||c||}{} \\[-3mm]
\multicolumn{9}{||c||}{$\mu_{\rm r}^2 = 2\, \mu_{\rm f}^2$} \\
\multicolumn{9}{||c||}{} \\[-0.3cm] \hline \hline
 & & & & & & & \\[-0.3cm]
$\! 10^{-7}\!$ &
$\! 1.2819^{-4}\!$ &$\! 7.8560^{-5}\!$ &$\! 5.2370^{-6}\!$ &$\! 1.3229^{+2}\!$ &
$\! 8.8691^{-6}\!$ &$\! 6.5070^{+1}\!$ &$\! 6.4352^{+1}\!$ &$\! 1.0296^{+3}\!$\\
$\! 10^{-6}\!$ &
$\! 6.1508^{-4}\!$ &$\! 3.7250^{-4}\!$ &$\! 2.2897^{-5}\!$ &$\! 6.7704^{+1}\!$ &
$\! 3.3358^{-5}\!$ &$\! 3.2997^{+1}\!$ &$\! 3.2427^{+1}\!$ &$\! 4.9907^{+2}\!$\\
$\! 10^{-5}\!$ &
$\! 2.8891^{-3}\!$ &$\! 1.7248^{-3}\!$ &$\! 9.5768^{-5}\!$ &$\! 3.2745^{+1}\!$ &
$\! 1.0595^{-4}\!$ &$\! 1.5694^{+1}\!$ &$\! 1.5241^{+1}\!$ &$\! 2.2382^{+2}\!$\\
$\! 10^{-4}\!$ &
$\! 1.3223^{-2}\!$ &$\! 7.7642^{-3}\!$ &$\! 3.8437^{-4}\!$ &$\! 1.4803^{+1}\!$ &
$\! 2.3744^{-4}\!$ &$\! 6.8647^{+0}\!$ &$\! 6.5065^{+0}\!$ &$\! 9.0293^{+1}\!$\\
$\! 10^{-3}\!$ &
$\! 5.8091^{-2}\!$ &$\! 3.3494^{-2}\!$ &$\! 1.4978^{-3}\!$ &$\! 6.1716^{+0}\!$ &
$\! 1.8990^{-4}\!$ &$\! 2.6702^{+0}\!$ &$\! 2.3924^{+0}\!$ &$\! 3.1117^{+1}\!$\\
$\! 10^{-2}\!$ &
$\! 2.2927^{-1}\!$ &$\! 1.2858^{-1}\!$ &$\! 5.4465_*^{-3}\!$&$\!2.2481^{+0}\!$ &
$\!-3.6267^{-4}\!$ &$\! 8.4001^{-1}\!$ &$\! 6.5031^{-1}\!$ &$\! 8.0964^{+0}\!$\\
$\! 0.1    \!$ &
$\! 5.5429^{-1}\!$ &$\! 2.7326^{-1}\!$ &$\! 1.0073^{-2}\!$ &$\! 3.9298^{-1}\!$ &
$\!-2.1631^{-4}\!$ &$\! 1.1440^{-1}\!$ &$\! 5.9718^{-2}\!$ &$\! 9.0854^{-1}\!$\\
$\! 0.3    \!$ &
$\! 3.5501^{-1}\!$ &$\! 1.3205^{-1}\!$ &$\! 3.0556^{-3}\!$ &$\! 3.6008^{-2}\!$ &
$\!-2.2642^{-5}\!$ &$\! 9.2226^{-3}\!$ &$\! 3.3770^{-3}\!$&$\!8.5020_*^{-2}\!$\\
$\! 0.5    \!$ &
$\! 1.2340^{-1}\!$ &$\! 3.2166^{-2}\!$ &$\! 3.8590^{-4}\!$ &$\! 2.4458^{-3}\!$ &
$\!-1.9414^{-6}\!$ &$\! 5.9487^{-4}\!$ &$\! 1.7699^{-4}\!$&$\!8.2294_*^{-3}\!$\\
$\! 0.7    \!$ &
$\! 2.0597^{-2}\!$ &$\! 3.1751^{-3}\!$ &$\! 1.3854^{-5}\!$ &$\! 5.5708^{-5}\!$ &
$\!-7.2724^{-8}\!$ &$\! 1.3244^{-5}\!$ &$\! 3.5361^{-6}\!$ &$\! 3.9686^{-4}\!$\\
$\! 0.9    \!$ &
$\! 3.6527^{-4}\!$ &$\! 1.8544^{-5}\!$ &$\! 9.204 ^{-9}\!$ &$\! 2.616 ^{-8}\!$ &
$\!-1.057^{-10}\!$ &$\! 6.700 ^{-9}\!$ &$\! 2.364 ^{-9}\!$ &$\! 1.2497^{-6}\!$\\
\hline \hline
\multicolumn{9}{||c||}{} \\[-3mm]
\multicolumn{9}{||c||}{$\mu_{\rm r}^2 = 1/2\, \mu_{\rm f}^2$} \\
\multicolumn{9}{||c||}{} \\[-0.3cm] \hline \hline
 & & & & & & & \\[-0.3cm]
$\! 10^{-7}\!$ &
$\! 1.6566^{-4}\!$ &$\! 1.0501^{-4}\!$ &$\! 7.1003^{-6}\!$ &$\! 1.3181^{+2}\!$ &
$\! 2.0996^{-5}\!$ &$\! 6.4832^{+1}\!$ &$\! 6.4117^{+1}\!$ &$\! 9.4582^{+2}\!$\\
$\! 10^{-6}\!$ &
$\! 7.4521^{-4}\!$ &$\! 4.6428^{-4}\!$ &$\! 2.9744^{-5}\!$ &$\! 6.9457^{+1}\!$ &
$\! 7.6503^{-5}\!$ &$\! 3.3877^{+1}\!$&$\!3.3309_*^{+1}\!$&$\!4.7543_*^{+2}\!$\\
$\! 10^{-5}\!$ &
$\! 3.2879^{-3}\!$ &$\! 2.0041^{-3}\!$ &$\! 1.1872^{-4}\!$ &$\! 3.4193^{+1}\!$ &
$\! 2.3809^{-4}\!$ &$\! 1.6421^{+1}\!$ &$\! 1.5970^{+1}\!$ &$\! 2.1922^{+2}\!$\\
$\! 10^{-4}\!$ &
$\! 1.4223^{-2}\!$ &$\! 8.4396^{-3}\!$ &$\! 4.5146^{-4}\!$ &$\! 1.5524^{+1}\!$ &
$\! 5.2716^{-4}\!$ &$\! 7.2281^{+0}\!$ &$\! 6.8715^{+0}\!$ &$\! 9.0112^{+1}\!$\\
$\! 10^{-3}\!$ &
$\! 5.9889^{-2}\!$ &$\! 3.4553^{-2}\!$ &$\! 1.6545^{-3}\!$ &$\! 6.4073^{+0}\!$ &
$\! 4.1428^{-4}\!$ &$\! 2.7908^{+0}\!$ &$\! 2.5144^{+0}\!$ &$\! 3.1340^{+1}\!$\\
$\! 10^{-2}\!$ &
$\! 2.3100^{-1}\!$ &$\! 1.2915_*^{-1}\!$&$\!5.6663^{-3}\!$ &$\! 2.2851^{+0}\!$ &
$\!-8.0979^{-4}\!$ &$\! 8.6079^{-1}\!$ &$\! 6.7251^{-1}\!$ &$\! 8.1526^{+0}\!$\\
$\! 0.1    \!$ &
$\! 5.5039^{-1}\!$ &$\! 2.7076^{-1}\!$ &$\! 1.0032^{-2}\!$&$\!3.8851_*^{-1}\!$ &
$\!-4.7547^{-4}\!$ &$\! 1.1333^{-1}\!$ &$\! 5.9500^{-2}\!$ &$\! 9.0801^{-1}\!$\\
$\! 0.3    \!$ &
$\! 3.4890^{-1}\!$ &$\! 1.2949^{-1}\!$ &$\! 2.9942^{-3}\!$ &$\! 3.5090^{-2}\!$ &
$\!-4.9326^{-5}\!$ &$\! 8.9666^{-3}\!$ &$\! 3.2669^{-3}\!$ &$\! 8.4307^{-2}\!$\\
$\! 0.5    \!$ &
$\! 1.2026^{-1}\!$ &$\! 3.1269^{-2}\!$&$\!3.7428_*^{-4}\!$ &$\! 2.3728^{-3}\!$ &
$\!-4.1969^{-6}\!$ &$\! 5.7784^{-4}\!$ &$\! 1.7391^{-4}\!$ &$\! 8.1102^{-3}\!$\\
$\! 0.7    \!$ &
$\! 1.9867^{-2}\!$ &$\! 3.0534^{-3}\!$ &$\! 1.3285^{-5}\!$ &$\! 5.4607^{-5}\!$ &
$\!-1.5564^{-7}\!$&$\!1.3275_*^{-5}\!$ &$\! 3.9931^{-6}\!$&$\!3.8822_*^{-4}\!$\\
$\! 0.9    \!$ &
$\! 3.4524^{-4}\!$ &$\! 1.7466^{-5}\!$ &$\! 8.828 ^{-9}\!$ &$\! 2.930 ^{-8}\!$ &
$\!-2.216^{-10}\!$ &$\! 8.837 ^{-9}\!$ &$\! 4.777 ^{-9}\!$ &$\! 1.2041^{-6}\!$\\
\hline \hline
\end{tabular}
\end{center}
\end{table}

\begin{table}[htp]
\caption{As Table \ref{gsav-t4}, but for the variable-$N_{\rm f}$ evolution
 using Eqs.~(\ref{gsav-eq12}) -- (\ref{gsav-eq15}).
 The corresponding values for the strong coupling $\alpha_{\rm s}(\mu_{\rm r}^2
 \! =\! 10^4 \mbox{ GeV}^2)$ are given by $0.115818$, $0.115605$ and $0.115410$
 for $\mu_{\rm r}^2 / \mu_{\rm f}^2 = 0.5$, $1$ and $2$, respectively. For
 brevity the small, but non-vanishing valence distributions $s_v$, $c_v$ and 
 $b_v$ are not displayed.}
\label{gsav-t5}
\vspace{3mm}
\begin{center}
\begin{tabular}{||c||r|r|r|r|r|r|r|r||}
\hline \hline
\multicolumn{9}{||c||}{} \\[-3mm]
\multicolumn{9}{||c||}{NNLO, $\, N_{\rm f} = 3\,\ldots 5\,$,
                       $\,\mu_{\rm f}^2 = 10^4 \mbox{ GeV}^2$} \\
\multicolumn{9}{||c||}{} \\[-0.3cm] \hline \hline
 & & & & & & & \\[-0.3cm]
\multicolumn{1}{||c||}{$x$} &
\multicolumn{1}{c|} {$xu_v$} &
\multicolumn{1}{c|} {$xd_v$} &
\multicolumn{1}{c|} {$xL_-$} &
\multicolumn{1}{c|} {$xL_+$} &
\multicolumn{1}{c|} {$xs_+$} &
\multicolumn{1}{c|} {$xc_+$} &
\multicolumn{1}{c|} {$xb_+$} &
\multicolumn{1}{c||}{$xg$} \\[0.5mm] \hline \hline
\multicolumn{9}{||c||}{} \\[-3mm]
\multicolumn{9}{||c||}{$\mu_{\rm r}^2 = \mu_{\rm f}^2$} \\
\multicolumn{9}{||c||}{} \\[-0.3cm] \hline \hline
 & & & & & & & \\[-0.3cm]
$\! 10^{-7}\!$ &
$\! 1.5053^{-4}\!$ &$\! 9.3396^{-5}\!$ &$\! 6.3350^{-6}\!$ &$\! 1.4236^{+2}\!$ &
$\! 7.0107^{+1}\!$ &$\! 6.8526^{+1}\!$ &$\! 5.8645^{+1}\!$ &$\! 1.0101^{+3}\!$\\
$\! 10^{-6}\!$ &
$\! 6.9800^{-4}\!$ &$\! 4.2692^{-4}\!$ &$\! 2.7015^{-5}\!$ &$\! 7.2905^{+1}\!$ &
$\! 3.5601^{+1}\!$ &$\! 3.4447^{+1}\!$ &$\! 2.8749^{+1}\!$ &$\! 4.9242^{+2}\!$\\
$\! 10^{-5}\!$ &
$\! 3.1691^{-3}\!$ &$\! 1.9048^{-3}\!$ &$\! 1.0987^{-4}\!$ &$\! 3.5102^{+1}\!$ &
$\! 1.6875^{+1}\!$ &$\! 1.6059^{+1}\!$ &$\! 1.3011^{+1}\!$ &$\! 2.2122^{+2}\!$\\
$\! 10^{-4}\!$ &
$\! 1.4048^{-2}\!$ &$\! 8.2711^{-3}\!$ &$\! 4.2655^{-4}\!$ &$\! 1.5694^{+1}\!$ &
$\! 7.3126^{+0}\!$ &$\! 6.7623^{+0}\!$ &$\! 5.2796^{+0}\!$ &$\! 8.8987^{+1}\!$\\
$\! 10^{-3}\!$ &
$\! 6.0085^{-2}\!$ &$\! 3.4610^{-2}\!$ &$\! 1.5988^{-3}\!$ &$\! 6.4190^{+0}\!$ &
$\! 2.7963^{+0}\!$ &$\! 2.4502^{+0}\!$ &$\! 1.8138^{+0}\!$ &$\! 3.0407^{+1}\!$\\
$\! 10^{-2}\!$ &
$\! 2.3240^{-1}\!$ &$\! 1.3001^{-1}\!$ &$\! 5.5869^{-3}\!$ &$\! 2.2762^{+0}\!$ &
$\! 8.5667^{-1}\!$ &$\! 6.6665^{-1}\!$ &$\! 4.5032^{-1}\!$ &$\! 7.7873^{+0}\!$\\
$\! 0.1    \!$ &
$\! 5.4993^{-1}\!$ &$\! 2.7035^{-1}\!$ &$\! 9.9604^{-3}\!$ &$\! 3.8527^{-1}\!$ &
$\! 1.1231^{-1}\!$ &$\! 6.4471^{-2}\!$ &$\! 3.7283^{-2}\!$ &$\! 8.5270^{-1}\!$\\
$\! 0.3    \!$ &
$\! 3.4622^{-1}\!$ &$\! 1.2833^{-1}\!$ &$\! 2.9571^{-3}\!$ &$\! 3.4600^{-2}\!$ &
$\! 8.8409^{-3}\!$ &$\! 4.0131^{-3}\!$ &$\! 2.1046^{-3}\!$ &$\! 7.8896^{-2}\!$\\
$\! 0.5    \!$ &
$\! 1.1868^{-1}\!$ &$\! 3.0811^{-2}\!$ &$\! 3.6760^{-4}\!$ &$\! 2.3198^{-3}\!$ &
$\! 5.6309^{-4}\!$ &$\! 2.3748^{-4}\!$ &$\! 1.2003^{-4}\!$ &$\! 7.6399^{-3}\!$\\
$\! 0.7    \!$ &
$\! 1.9486^{-2}\!$ &$\! 2.9902^{-3}\!$ &$\! 1.2964^{-5}\!$ &$\! 5.2335^{-5}\!$ &
$\! 1.2504^{-5}\!$ &$\! 5.6019^{-6}\!$ &$\! 2.8883^{-6}\!$ &$\! 3.7079^{-4}\!$\\
$\! 0.9    \!$ &
$\! 3.3522^{-4}\!$ &$\! 1.6933^{-5}\!$ &$\! 8.4201^{-9}\!$ &$\! 2.5098^{-8}\!$ &
$\! 6.8378^{-9}\!$ &$\! 4.3190^{-9}\!$ &$\! 2.6761^{-9}\!$ &$\! 1.1731^{-6}\!$\\
\hline \hline
\multicolumn{9}{||c||}{} \\[-3mm]
\multicolumn{9}{||c||}{$\mu_{\rm r}^2 = 2\, \mu_{\rm f}^2$} \\
\multicolumn{9}{||c||}{} \\[-0.3cm] \hline \hline
 & & & & & & & \\[-0.3cm]
$\! 10^{-7}\!$ &
$\! 1.3305^{-4}\!$ &$\! 8.1537^{-5}\!$ &$\! 5.4285^{-6}\!$ &$\! 1.3770^{+2}\!$ &
$\! 6.7776^{+1}\!$ &$\! 6.6452^{+1}\!$ &$\! 5.7708^{+1}\!$ &$\! 1.0172^{+3}\!$\\
$\! 10^{-6}\!$ &
$\! 6.3465^{-4}\!$ &$\! 3.8427^{-4}\!$ &$\! 2.3638^{-5}\!$ &$\! 7.0043^{+1}\!$ &
$\! 3.4167^{+1}\!$ &$\! 3.3187^{+1}\!$ &$\! 2.8117^{+1}\!$ &$\! 4.9102^{+2}\!$\\
$\! 10^{-5}\!$ &
$\! 2.9622^{-3}\!$ &$\! 1.7679^{-3}\!$ &$\! 9.8415^{-5}\!$ &$\! 3.3642^{+1}\!$ &
$\! 1.6144^{+1}\!$ &$\! 1.5435^{+1}\!$ &$\! 1.2685^{+1}\!$ &$\! 2.1913^{+2}\!$\\
$\! 10^{-4}\!$ &
$\! 1.3469^{-2}\!$ &$\! 7.9047^{-3}\!$ &$\! 3.9275^{-4}\!$ &$\! 1.5088^{+1}\!$ &
$\! 7.0083^{+0}\!$ &$\! 6.5145^{+0}\!$ &$\! 5.1475^{+0}\!$ &$\! 8.7878^{+1}\!$\\
$\! 10^{-3}\!$ &
$\! 5.8793^{-2}\!$ &$\! 3.3879^{-2}\!$ &$\! 1.5195^{-3}\!$ &$\! 6.2333^{+0}\!$ &
$\! 2.7018^{+0}\!$ &$\! 2.3753^{+0}\!$ &$\! 1.7742^{+0}\!$ &$\! 3.0063^{+1}\!$\\
$\! 10^{-2}\!$ &
$\! 2.3060^{-1}\!$ &$\! 1.2923^{-1}\!$ &$\! 5.4792^{-3}\!$ &$\! 2.2478^{+0}\!$ &
$\! 8.4084^{-1}\!$ &$\! 6.5026^{-1}\!$ &$\! 4.4324^{-1}\!$ &$\! 7.7467^{+0}\!$\\
$\! 0.1    \!$ &
$\! 5.5280^{-1}\!$ &$\! 2.7222^{-1}\!$ &$\!1.0021_*^{-2}\!$&$\! 3.8898^{-1}\!$ &
$\! 1.1313^{-1}\!$ &$\! 6.2921^{-2}\!$ &$\! 3.7050^{-2}\!$ &$\! 8.5900^{-1}\!$\\
$\! 0.3    \!$ &
$\! 3.5141^{-1}\!$ &$\! 1.3051^{-1}\!$ &$\! 3.0134^{-3}\!$ &$\! 3.5398^{-2}\!$ &
$\! 9.0559^{-3}\!$ &$\! 3.8724^{-3}\!$ &$\! 2.0992^{-3}\!$ &$\! 8.0225^{-2}\!$\\
$\! 0.5    \!$ &
$\! 1.2140^{-1}\!$ &$\! 3.1590^{-2}\!$ &$\! 3.7799^{-4}\!$ &$\! 2.3919^{-3}\!$ &
$\! 5.8148^{-4}\!$ &$\! 2.2373^{-4}\!$ &$\! 1.1917^{-4}\!$ &$\! 7.8099^{-3}\!$\\
$\! 0.7    \!$ &
$\! 2.0120^{-2}\!$ &$\! 3.0955^{-3}\!$ &$\! 1.3467^{-5}\!$ &$\! 5.4182^{-5}\!$ &
$\! 1.2896^{-5}\!$ &$\! 5.0316^{-6}\!$ &$\! 2.8148^{-6}\!$ &$\! 3.8098^{-4}\!$\\
$\! 0.9    \!$ &
$\! 3.5230^{-4}\!$ &$\! 1.7849^{-5}\!$ &$\! 8.835 ^{-9}\!$ &$\! 2.523 ^{-8}\!$ &
$\! 6.499 ^{-9}\!$ &$\! 3.377 ^{-9}\!$ &$\! 2.404 ^{-9}\!$ &$\! 1.2196^{-6}\!$\\
\hline \hline
\multicolumn{9}{||c||}{} \\[-3mm]
\multicolumn{9}{||c||}{$\mu_{\rm r}^2 = 1/2\, \mu_{\rm f}^2$} \\
\multicolumn{9}{||c||}{} \\[-0.3cm] \hline \hline
 & & & & & & & \\[-0.3cm]
$\! 10^{-7}\!$ &
$\! 1.7462^{-4}\!$ &$\! 1.1066^{-4}\!$ &$\! 7.4685^{-6}\!$ &$\! 1.4238^{+2}\!$ &
$\! 7.0121^{+1}\!$ &$\! 6.8041^{+1}\!$ &$\! 5.7115^{+1}\!$ &$\! 9.6557^{+2}\!$\\
$\! 10^{-6}\!$ &
$\! 7.7945^{-4}\!$ &$\! 4.8542^{-4}\!$ &$\!3.1111_*^{-5}\!$&$\! 7.4239^{+1}\!$ &
$\! 3.6270^{+1}\!$ &$\! 3.4778^{+1}\!$ &$\! 2.8492^{+1}\!$ &$\! 4.8135^{+2}\!$\\
$\! 10^{-5}\!$ &
$\! 3.4099^{-3}\!$ &$\! 2.0775^{-3}\!$ &$\! 1.2343^{-4}\!$ &$\! 3.6106^{+1}\!$ &
$\! 1.7379^{+1}\!$ &$\! 1.6353^{+1}\!$ &$\! 1.3052^{+1}\!$ &$\! 2.1979^{+2}\!$\\
$\! 10^{-4}\!$ &
$\! 1.4619^{-2}\!$ &$\! 8.6691^{-3}\!$ &$\! 4.6591^{-4}\!$ &$\! 1.6161^{+1}\!$ &
$\! 7.5479^{+0}\!$ &$\! 6.8868^{+0}\!$ &$\! 5.3331^{+0}\!$ &$\! 8.9300^{+1}\!$\\
$\! 10^{-3}\!$ &
$\! 6.1014^{-2}\!$ &$\! 3.5170^{-2}\!$ &$\! 1.6915^{-3}\!$ &$\! 6.5570^{+0}\!$ &
$\! 2.8669^{+0}\!$ &$\! 2.4816^{+0}\!$ &$\! 1.8360^{+0}\!$&$\!3.0619_*^{+1}\!$\\
$\! 10^{-2}\!$ &
$\! 2.3320^{-1}\!$ &$\! 1.3024^{-1}\!$ &$\! 5.7229^{-3}\!$ &$\! 2.2924^{+0}\!$ &
$\! 8.6595^{-1}\!$ &$\! 6.7561^{-1}\!$ &$\! 4.5538^{-1}\!$ &$\! 7.8182^{+0}\!$\\
$\! 0.1    \!$ &
$\! 5.4799^{-1}\!$ &$\! 2.6905^{-1}\!$ &$\!9.9481_*^{-3}\!$&$\! 3.8193^{-1}\!$ &
$\! 1.1125^{-1}\!$ &$\! 6.7100^{-2}\!$ &$\! 3.7702^{-2}\!$ &$\! 8.4914^{-1}\!$\\
$\! 0.3    \!$ &
$\! 3.4291^{-1}\!$ &$\! 1.2693^{-1}\!$ &$\! 2.9238^{-3}\!$ &$\! 3.4069^{-2}\!$ &
$\! 8.6866^{-3}\!$ &$\! 4.3918^{-3}\!$ &$\! 2.1434^{-3}\!$&$\!7.8107_*^{-2}\!$\\
$\! 0.5    \!$ &
$\! 1.1694^{-1}\!$ &$\! 3.0310^{-2}\!$ &$\! 3.6112^{-4}\!$ &$\! 2.2827^{-3}\!$ &
$\! 5.5538^{-4}\!$ &$\! 2.7737^{-4}\!$ &$\! 1.2414^{-4}\!$ &$\! 7.5373^{-3}\!$\\
$\! 0.7    \!$ &
$\! 1.9076^{-2}\!$ &$\! 2.9217^{-3}\!$ &$\! 1.2646^{-5}\!$ &$\! 5.2035^{-5}\!$ &
$\! 1.2677^{-5}\!$ &$\! 7.2055^{-6}\!$ &$\! 3.0900^{-6}\!$&$\!3.6439_*^{-4}\!$\\
$\! 0.9    \!$ &
$\! 3.2404^{-4}\!$ &$\! 1.6333^{-5}\!$ &$\! 8.231 ^{-9}\!$ &$\! 2.752 ^{-8}\!$ &
$\! 8.380 ^{-9}\!$ &$\! 6.769 ^{-9}\!$ &$\! 3.204 ^{-9}\!$ & $\!1.1424^{-6}\!$\\
\hline \hline
\end{tabular}
\end{center}
\end{table}




\newcommand{\beq}{\begin{equation}}
\newcommand{\eeq}{\end{equation}}

\newcommand{\bea}{\begin{eqnarray}}
\newcommand{\eea}{\end{eqnarray}}
\newcommand{\nn}{\nonumber}

\def\eqn#1{Eq.~(\ref{#1})}
\def\eqns#1#2{Eqs.~(\ref{#1}) and~(\ref{#2})}
\def\eqnss#1#2{Eqs.~(\ref{#1})-(\ref{#2})}
\def\fig#1{Fig.~{\ref{#1}}}
\def\sec#1{Section~{\ref{#1}}}
\def\app#1{Appendix~\ref{#1}}
\def\tab#1{Table~\ref{#1}}


\def\ket#1{|{#1}\rangle}
\def\bra#1{\langle{#1}|}
\def\braket#1#2{\langle #1 |#2 \rangle}
\def\ord{{\cal O} }
\def\e{\epsilon}
\def\eps{\epsilon}
\def\cm{{\cal M}}
\def\cM{{\cal M}}
\def\cmb{\overline{{\cal M}}}
\def\cmqq{{\cal M}_{q\bar q}}
\def\cmbqq{\overline{{\cal M}}_{q\bar q}}
\def\cmqqg{{\cal M}_{q\bar qg}}
\def\cmbqqg{\overline{{\cal M}}_{q\bar qg}}
\def\cmqqgg{ {\cal M}_{q\bar qgg}}
\def\cmqqqq{ {\cal M}_{q\bar qq\bar q}}
\def\asopi{\left(\frac{\alpha_s}{2\pi}\right)}
\def\asmuopi{\left(\frac{\alpha_s(\mu^2)}{2\pi}\right)}
\newcommand\sss{\scriptscriptstyle}
\newcommand\als{\alpha_{\sss S}} 
\newcommand\gs{g_{\sss S}}
\def\tgs{\tilde\gs}
\def\CF{C_{\sss F}}
\def\CA{C_{\sss A}}
\def\NF{N_{\sss F}}
\def\TF{T_{\sss R}}
\def\ca#1{{\cal A}^{( #1)}}
\def\caqq#1{{\cal A}^{( #1)}_{q\bar q}}
\def\caqqg#1{{\cal A}^{( #1)}_{q\bar qg}}
\def\caqqgg#1{{\cal A}^{( #1)}_{q\bar qgg}}
\def\caqqqq#1{{\cal A}^{( #1)}_{q\bar qq\bar q}}
\def\sab{s_{12}}
\def\sac{s_{13}}
\def\sbc{s_{23}}
\def\yab{y_{12}}
\def\yac{y_{13}}
\def\ybc{y_{23}}
\def\sij{s_{ij}}
\def\sabc{s_{123}}
\def\Ione{{\mbox{\boldmath $I$}^{(1)}(\e)}}
\def\dIone{{\mbox{\boldmath $\Delta I$}^{(1)}(\e)}}
\def\Ionep{{\mbox{\boldmath $I$}^{(1)}(2\e)}}
\def\Htwo{{\mbox{\boldmath $H$}^{(2)}(\e)}}
\def\MSbar{$\overline{{\rm MS}}$}



\section{HIGHER ORDERS\protect\footnote{Section coordinator:
    E.W.N.~Glover}$^{,}$~\protect\footnote{Contributing authors: 
    T.~Binoth, V.~Del~Duca, T.~Gehrmann, A.~Gehrmann-De~Ridder, 
    E.W.N.~Glover, J.-Ph.~Guillet and G.~Heinrich}}

In this report, we summarize the issues discussed and worked out during
the 'Working Group on Higher Orders'. The two current frontiers of higher
order QCD calculations at colliders are the description of multiparton
final states at next-to-leading order and the extension of calculations
for precision observables beyond this order.  Considerable progress has
been made on both issues, with the highlights being the first calculation
of the Yukawa-model one-loop six-point amplitude and the two-loop
corrections to the $e^+e^- \to 3$~jets matrix element. Developments
towards the construction of NNLO parton level Monte Carlo programs are
described for the example of $\gamma^* \to q\bar q$. Finally, the first
applications of two-loop matrix elements to the improved description
of the high energy limit of QCD are reported.


\subsection{Introduction}
Present and future collider experiments will confront  us with large sets of
data on multi--particle final states. This is particularly true for the hadron
colliders Tevatron  and LHC, which will be the machines operating at the 
highest attainable energies in the near future.  Hence the comparison of jet
observables to theoretical  predictions will become increasingly important. 

Since the theoretical predictions based on leading order (LO) calculations  
are typically plagued by large scale uncertainties,  it becomes necessary to
calculate the next-to-leading order (NLO) corrections  in order to make
meaningful predictions which match the experimental precision.  Indeed, the
$N$-jet cross section  in hadronic collisions  is proportional to $\alpha_s^N$
at leading order, which means that  theoretical uncertainties  are actually
amplified for growing $N$. 

For processes with relatively few jets, such as dijet production,
next-to-next-to-leading order (NNLO) perturbative predictions will be needed to
reduce the theoretical uncertainties and enable useful physics to be extracted
from the copious high-precision data.

In this report, we address issues connected with theoretical progress in
calculating higher order corrections both at NLO (in the context of $2 \to 4$
scattering processes) and at NNLO (in the context of $2\to 2$ processes).   At
present, making numerical predictions for these types of processes lies
well beyond our capabilities.   However, there has been very rapid progress in
the last two years and it is very likely that the technical stumbling blocks
will be removed.   

Our report is structured as follows.   First,  we address the issue of NLO
corrections to multi-particle final states. The main technical problems
associated with dimensionally regulated pentagon integrals were solved some time
ago~\cite{Bern:1994kr} and the next-to-leading order matrix
elements for $2\rightarrow 3$ processes have become available in the recent 
years\,\cite{Bern:1993em,Bern:1994kr,Kunszt:1994tq,Signer:1995nk,
Bern:1997ka,Bern:1998sd,Campbell:1997tv,Glover:1997eh,DelDuca:1999pa,
Beenakker:2001rj,Reina:2001bc,DelDuca:2001fn}.
However, the step to
$2\rightarrow 4$ or even higher processes at NLO has not been made yet. The
reason lies in the fact that the computation of the  corresponding amplitudes
is highly nontrivial.   Although the calculation techniques  for amplitudes
with an arbitrary number of external  legs  are
available\,\cite{Binoth:1999sp}, it turns out that a brute force approach is
not viable. In order to  avoid intractably large expressions in the calculation
of  six-point (or higher) amplitudes, it is indispensable to understand better
recombination and cancellation mechanisms at intermediate steps of the
calculation.   This issue is addressed in Sec.~\ref{sec:ho;nlo}, in the
context of the Yukawa model, where all external legs are  massless scalars
attached to a massless fermion loop.

In Sec.~\ref{sec:ho;nnlo} we consider the NNLO corrections to $2\to 2$
scattering processes.   The rationale for going beyond the next-to-leading
order is reviewed in Sec.~\ref{sec:ho;nnlorationale} while the  various
building blocks necessary for such a calculation are discussed in
Secs.~\ref{sec:ho;nnlopdf}-- \ref{sec:ho;two-loop}.   While the individual
components are in relatively good shape - the infrared limits are well studied,
many two-loop matrix elements exist and the NNLO evolution of parton
distributions is almost under control -- a systematic procedure for combining
them to give numerical predictions is not established.   Therefore in
Sec.~\ref{sec:ho;exercise} we examine the infrared singular structure of the
various pieces for $\gamma^* \to 2$~jets which is one of the simplest
non-trivial processes at NNLO.

Sec.~\ref{sec:ho;highenergy} contains a summary of the present status of the
analytic structure of QCD amplitudes in the limit of forward and backward
scattering. In these high energy limits, the scattering process  is  dominated
by the exchange of a particle in the $t$- (or $u$-)channel respectively. This
may be the gluon or the quark.   In both cases, the amplitude reggeises and the
large logarithms can be resummed to next-to-leading-logarithmic accuracy by
simple forms involving the reggeised particle exchange together with modified
vertex functions (or impact factors).  We evaluate the gluon and quark Regge
trajectories to two-loop accuracy and show that they are strikingly similar:
the gluon Regge trajectory can be obtained from the quark trajectory by mapping
$\CF \to \CA$.
 
A brief summary and outlook is given in Sec.~\ref{sec:ho;conc}.

\subsection{NLO}
\label{sec:ho;nlo}

For NLO amplitudes with $N\ge 6$ external particles,
standard calculational methods are not adequate because of
the complexity of intermediate expressions.

In the last few years, methods either directly based on string perturbation 
theory\,\cite{Bern:1992aq,Bern:1991ux,Bern:1993wt}  or on a world line
formulation of field theory  amplitudes\,\cite{Strassler:1992zr} have been used
to derive a number of ``master formulae'' for one-loop $N$--point amplitudes.
Those are generating functionals which  yield, for any $N$,  a closed parameter
integral expression for the amplitude. The resulting integral representations
are related to standard Feynman parameter integrals in a well-understood 
way\,\cite{Bern:1992an}. Nevertheless,  due to their superior organisation they
often allow one to exploit at the integral level properties of an amplitude
which normally would be seen only at later stages in a Feynman graph
calculation\,\cite{Schubert:2001he,Frizzo:2000ez}.

Although the string inspired formalism allows for an elegant formulation of
amplitudes in terms of  a manifest Lorentz structure, one is in general not at
all dispensed from  doing cumbersome algebraic work.  The complexity of doing
tensor reduction in momentum space translates into the need to reduce Feynman
parameter integrals with nontrivial numerators to genuine $N$--point scalar
integrals. These can be expressed in terms of box, triangle and bubble scalar
integrals.   Substantial cancellations appear in all these steps and progress
in finding efficient calculational methods relies on a better understanding of
these mechanisms.

Here we will sketch the calculation of the six-point one-loop  amplitude in the
Yukawa model where all external legs are  massless scalars attached to a
massless fermion loop~\cite{Binoth:2001vm}.  The interest of this model is related to the fact that 
the appearance of tensor integrals can be completely avoided. This can
immediately be seen from the string inspired master formulas for one-loop
$N$--point functions derived in\,\cite{Schubert:2001he}. In this way one can
study the reduction mechanisms for scalar  6-point functions without the
additional complications arising from the tensor reduction.  Having understood
these mechanisms one can proceed towards the computation of gauge theory
amplitudes.

\subsubsection{Calculation of a hexagon amplitude in the Feynman diagrammatic
approach}
 
The amplitude $\Gamma_{\rm yuk}$ can be written as a sum over $6!$ permutations
of the external momentum vectors $p_1,\dots,p_6$,
\begin{equation}\label{Eq:start}
\Gamma^{\phi}_{\rm yuk}[p_1,p_2,p_3,p_4,p_5,p_6] = -\frac{g^6}{(4\pi)^{n/2}}
\frac{1}{6} 
 \sum\limits_{\pi\in S_6} {\cal A}(p_{\pi_1},p_{\pi_2},
                  p_{\pi_3},p_{\pi_4},p_{\pi_5},p_{\pi_6}).
\end{equation} 
Each permutation corresponds to a single Feynman diagram. The amplitude 
for the trivial permutation is given by,
\begin{eqnarray}\label{amp_graph}
{\cal A}(p_1,p_2,p_3,p_4,p_5,p_6) &=& \int \frac{d^nk}{i \pi^{n/2}}
\frac{{\rm tr}(q_1,q_2,q_3,q_4,q_5,q_6)}{q_1^2q_2^2q_3^2q_4^2q_5^2q_6^2},
\end{eqnarray}
where $q_j = k - r_j = k-p_1-\dots-p_j $.
Working out the trace gives a sum of products of terms 
$q_k\cdot q_j$ which can be written as 
inverse propagators which cancel directly. 
This means that each graph can simply be represented as
a linear combination of scalar integrals. 
For the trivial permutation we find
\begin{eqnarray}\label{amp_si}
2 \,{\cal A}(p_1,p_2,p_3,p_4,p_5,p_6) =
  4\, I_3^n(p_{12},p_{34},p_{56}) 
 +4\, I_3^n(p_{23},p_{45},p_{61})  \nonumber\\
 +{\rm tr}(p_1,p_2)\, I_4^n(p_1,p_2,p_{34},p_{56}) 
 +{\rm tr}(p_2,p_3)\, I_4^n(p_2,p_3,p_{45},p_{61}) \nonumber\\
 +{\rm tr}(p_3,p_4)\, I_4^n(p_3,p_4,p_{56},p_{12})  
 +{\rm tr}(p_4,p_5)\, I_4^n(p_4,p_5,p_{61},p_{23}) \nonumber\\
 +{\rm tr}(p_5,p_6)\, I_4^n(p_5,p_6,p_{12},p_{34}) 
 +{\rm tr}(p_6,p_1)\, I_4^n(p_6,p_1,p_{23},p_{45})  \nonumber\\
 +{\rm tr}(p_1,p_4)\, I_4^n(p_1,p_{23},p_4,p_{56}) 
 +{\rm tr}(p_2,p_5)\, I_4^n(p_2,p_{34},p_5,p_{61}) \nonumber\\
 +{\rm tr}(p_3,p_6)\, I_4^n(p_3,p_{45},p_6,p_{12})  \nonumber\\
 +{\rm tr}(p_1,p_2,p_3,p_4)\, I_5^n(p_{56},p_1,p_2,p_3,p_4)  
 +{\rm tr}(p_2,p_3,p_4,p_5)\, I_5^n(p_{61},p_2,p_3,p_4,p_5)  \nonumber\\
 +{\rm tr}(p_3,p_4,p_5,p_6)\, I_5^n(p_{12},p_3,p_4,p_5,p_6)  
 +{\rm tr}(p_4,p_5,p_6,p_1)\, I_5^n(p_{23},p_4,p_5,p_6,p_1)  \nonumber\\
 +{\rm tr}(p_5,p_6,p_1,p_2)\, I_5^n(p_{34},p_5,p_6,p_1,p_2)  
 +{\rm tr}(p_6,p_1,p_2,p_3)\, I_5^n(p_{45},p_6,p_1,p_2,p_3)  \nonumber\\
 +{\rm tr}(p_1,p_2,p_3,p_4,p_5,p_6)\, I_6^n(p_1,p_2,p_3,p_4,p_5,p_6) .
\end{eqnarray}
The arguments of the $N$--point scalar integrals are the momenta
of the external legs. We use the abbreviation 
$p_{ijk\dots}=p_i+p_j+p_k+\dots$.  
The spinor traces can be expressed by Mandelstam variables 
defined by the 9 cuts of the hexagon graph, but the form given above
is not only most compact but also most convenient to proceed.

We also note
that the amplitude is free of infrared poles. This can be seen
by power counting for the soft and collinear poles. 

To sketch the explicit calculation of the hexagon amplitude, 
we will draw special attention to the cancellation
mechanisms at work. First one has to
reduce hexagon and pentagon integrals to box 
integrals.
Then the explicit expressions for the box integrals are inserted. 
Finally the coefficients of  different terms are combined and simplified
by using linear relations for the reduction coefficients. We note
that in none of these steps the size of the expression 
will blow up, while this would surely be the case in a brute force approach.

The reduction formula for the hexagon integral reads,
\begin{eqnarray}
\label{hexa_red}
I_6^n(p_1,p_2,p_3,p_4,p_5,p_6) 
=\frac{1}{\det(\hat S)}
\bigl\{\hspace{5cm}\nonumber \\
\bigl[ {\rm tr}(123456){\rm tr}(3456)-2 s_{34}s_{45}s_{56} {\rm tr}(6123) \bigr]\; I_5^n(p_{12},p_3,p_4,p_5,p_6)\nonumber \\
+\bigl[{\rm tr}(123456){\rm tr}(4561)-2 s_{45}s_{56}s_{61} {\rm tr}(1234) \bigr]\; I_5^n(p_{23},p_4,p_5,p_6,p_1)\nonumber\\
+\bigl[{\rm tr}(123456){\rm tr}(5612)-2 s_{56}s_{61}s_{12} {\rm tr}(2345) \bigr] \;I_5^n(p_{34},p_5,p_6,p_1,p_2)\nonumber\\
+\bigl[{\rm tr}(123456){\rm tr}(6123)-2 s_{61}s_{12}s_{23} {\rm tr}(3456) \bigr]\; I_5^n(p_{45},p_6,p_1,p_2,p_3)\nonumber\\
+\bigl[{\rm tr}(123456){\rm tr}(1234)-2 s_{12}s_{23}s_{34} {\rm tr}(4561) \bigr]\;I_5^n(p_{56},p_1,p_2,p_3,p_4)\nonumber\\
+\bigl[{\rm tr}(123456){\rm tr}(2345)-2 s_{23}s_{34}s_{45} {\rm tr}(5612)
\bigr]\; I_5^n(p_{61},p_2,p_3,p_4,p_5) \bigr\},  
\end{eqnarray}
where the momenta inside the traces are represented
by their indices only. The coefficients in front of
the 5--point integrals are called $b_j$ ($j \in \{1,\dots ,6\}$) 
in the following.
They are defined by the linear equation,
\begin{eqnarray}\label{hexa_coeffs}
( \hat S \cdot b)_j = 1  &\Leftrightarrow& b_j = 
\sum\limits_{k=1}^6 \hat S^{-1}_{kj}\;\mbox{ where }\;\hat S_{kj}=(r_k-r_j)^2\\
\det(\hat S)&=& 4 s_{12}s_{23}s_{34}s_{45}s_{56}s_{61} 
- {\rm tr}(123456)^2.\nonumber
\end{eqnarray}
The traces allow for a compact notation for the coefficients $b_j$.
The  Gram matrix
$G_{kl} = 2 \,r_l \cdot r_k$ is related to 
$\hat S$ by $\hat S_{kl}=-G_{kl}+r_k^2+r_l^2$. 
For $N\ge 6$ and 4-dimensional external momenta one has det$(G)=0$, 
which leads to a non-linear constraint
between the Mandelstam variables. We note that this
constraint is represented {\em linearly} in terms of the
coefficients $b_j$. One has,
\begin{eqnarray}\label{hexa_cons}
\det(G) = 0   \hspace{1cm} \Leftrightarrow  
\hspace{1cm} \sum\limits_{j=1}^6 b_j = 0.
\end{eqnarray} 
By solving eq.~(\ref{hexa_coeffs}) with Cramer's rule
one sees that the constraint (\ref{hexa_cons}) relates
sums of determinants of 5$\times$5 matrices. In
terms of Mandelstam variables these are huge expressions 
just representing zero.
The guideline to keep the sizes of expressions under control
in calculations of multi--particle processes
is thus to use representations of amplitudes where the $b_j$
are kept manifestly and to use relations (\ref{hexa_coeffs}) and 
(\ref{hexa_cons}) to perform cancellations as far as possible.

Applying the reduction formula (\ref{hexa_red}) above to reduce the 
hexagon, we observe that the coefficients of the hexagon and pentagon 
integrals in the amplitude combine in a nice way to form a 
resulting coefficient for a given pentagon which is again 
proportional to $b_j$. For example, the resulting coefficient of 
$I_5^n(p_{12},p_3,p_4,p_5,p_6)$ in (\ref{amp_si}) is 
\begin{eqnarray}
{\rm tr}(3456) + {\rm tr}(123456) \, b_1 = -2 s_{34}s_{45}s_{56} \, b_4\;,
\end{eqnarray}
analogous for all cyclic permutations. 

Now we reduce the pentagons to boxes 
using the reduction formula given in\,\cite{Binoth:1999sp}. 
We obtain,
\begin{eqnarray}
{\cal A}(p_1,p_2,p_3,p_4,p_5,p_6) =
  \frac{2}{3} \, I_3^n(p_{12},p_{34},p_{56})  \hspace{6cm}\nonumber\\
 + s_{12}\, I_4^n(p_1,p_2,p_{34},p_{56}) 
+\frac{{\rm tr}(14)}{4}I_4^n(p_1,p_{23},p_{4},p_{56})\nonumber\\
+\frac{b_2}{2E_2}\Big\{
s_{23}s_{34}\,[\,{\rm tr}(1234)-2s_{12}\,(s_{234}-s_{23})]\,
I_4^n(p_2,p_{3},p_{4},p_{561})\nonumber\\
+s_{12}s_{23}\,[\,{\rm tr}(1234)-2s_{34}\,(s_{123}-s_{23})]\,
I_4^n(p_1,p_{2},p_{3},p_{456})\nonumber\\
+{\rm tr}(1234)\,E_2\,I_4^n(p_1,p_{23},p_{4},p_{56})\nonumber\\
+s_{34}\,[-s_{123}\,{\rm tr}(1234)-2s_{12}s_{23}\,(s_{123}-s_{56})]\,
I_4^n(p_{3},p_{4},p_{56},p_{12})\nonumber\\
+s_{12}\,[-s_{234}\,{\rm tr}(1234)-2s_{23}s_{34}\,(s_{234}-s_{56})]\,
I_4^n(p_1,p_2,p_{34},p_{56})\Big\}\nonumber\\
+\quad 5 \mbox{ cyclic permutations},
\label{amp_box}
\end{eqnarray}
where $E_1=s_{123}s_{345}-s_{12}s_{45}$. The $E_j$ 
for $j>1$ are defined by cyclic permutation.
Note that $E_{j}=E_{j+3}$.

The amplitude is now expressed in terms of four functions: The triangle 
with all three legs off-shell, box integrals with two off-shell legs at 
adjacent corners ($I_4^n(p_1,p_2,p_{34},p_{56})$ and 5 permutations), 
box integrals with two  off-shell legs at 
opposite corners ($I_4^n(p_1,p_{23},p_{4},p_{56})$ and 2 permutations), 
and box integrals with one off-shell leg 
($I_4^n(p_1,p_{2},p_{3},p_{456})$ and 5 permutations). 
 
We  now collect and combine the coefficients of particular terms in the 
cyclic sum. Again nontrivial cancellations
happen. In particular, using $\sum_{j=1}^6 b_j = 0$ and $\hat S\cdot b=1$ , 
the coefficients of the  box integrals with two  off-shell legs at 
opposite corners add up to zero and 
all dilogarithms related to box functions cancel! 
Hence the only terms which survive are the triangle graphs and
some logarithmic terms stemming from the finite parts of the 
box integrals, such that we finally obtain,
\begin{eqnarray}\label{amp_final}
{\cal A}(p_1,p_2,p_3,p_4,p_5,p_6) = G(p_1,p_2,p_3,p_4,p_5,p_6)
+ 5 \; \mbox{cyclic permutations},
\end{eqnarray}
with,
\begin{eqnarray} G(p_1,p_2,p_3,p_4,p_5,p_6) = 
 \frac{2}{3} \,I^n_3(p_{12},p_{34},p_{56})\hspace{6cm} \nonumber \\
+ \left\{  \frac{b_1}{E_1} 
\left[{\rm tr}(6123)-2s_{61}(s_{123}-s_{12}) \right] 
+\frac{b_2}{E_2} 
\left[ {\rm tr}(1234)-2s_{34}(s_{123}-s_{23}) \right] \right\} 
\hspace{1cm}\nonumber \\ \times
\log\left( \frac{s_{12}}{s_{123}} \right)
         \log\left( \frac{s_{23}}{s_{123}} \right) \nonumber \\
+\left\{ -b_1 + \frac{b_2}{2 E_2} 
\left[ {\rm tr}(1234) - 2 s_{34} (s_{123}-s_{23})  \right]
+ \frac{b_6}{2 E_6}
\left[ {\rm tr}(5612)- 2 s_{56} (s_{345}-s_{61})\right]\right\}
\hspace{1cm}\nonumber \\ \times
\left[ \log\left( \frac{s_{12}}{s_{234}} \right) 
       \log\left( \frac{s_{56}}{s_{234}} \right)
     + \log\left( \frac{s_{34}}{s_{234}} \right) 
       \log\left( \frac{s_{12}}{s_{56}} \right) \right].\nonumber\\
       \label{ampsi}
\end{eqnarray}
Note that $G(p_1,p_2,p_3,p_4,p_5,p_6)$ has no spurious singularities. 
We checked that the numerator of expression (\ref{ampsi}) vanishes in 
the limits where its denominator  vanishes. 

Finally, the full amplitude is given by the sum over
permutations of the function $G$,
\begin{eqnarray}\label{Eq:final}
\Gamma_{\rm yuk}^{\phi}[p_1,p_2,p_3,p_4,p_5,p_6] 
= - \frac{g^6}{(4\pi)^2}\sum\limits_{\pi\in S_6}^{} G(p_{\pi_1},p_{\pi_2},
                  p_{\pi_3},p_{\pi_4},p_{\pi_5},p_{\pi_6}).
\end{eqnarray}

\subsubsection{Summary}

The Yukawa model is a good testing ground to study 
nontrivial cancellations appearing in scalar integral reductions
without additional complications due to a nontrivial
tensor structure. Focusing on the massless case and $N=6$ 
we sketched the explicit calculation of the amplitude. 
It has been outlined how cancellations can be made manifest
at each step of the calculation by using linear relations
between reduction coefficients. 
With the present method there is no explosion of terms typical 
for multi--leg  calculations. The final answer is surprisingly compact
and contains -- apart from 3-point functions with 
3 off-shell legs -- only some products of logarithms.

As a next step more realistic examples have to be considered
including gauge bosons and a nontrivial infrared structure. 
It is justified to speculate that
the recombination of scalar integrals
will work similarly, such that
efficient algorithms to calculate 
six-point amplitudes at one loop are in reach.  
Work on this subject is in preparation.

\subsection{NNLO}
\label{sec:ho;nnlo}

\subsubsection{Motivation}
\label{sec:ho;nnlorationale}

There are many reasons why extending perturbative calculations to NNLO is
vital in reducing the theoretical uncertainty.  In the following we list five of
them.

\vspace{0.2cm}

\noindent 
{\it Renormalisation scale uncertainty}

\vspace{0.15cm}

\noindent 
In many cases, the uncertainty from the pdf's and from the choice of the
renormalisation scale $\mu_R$ give uncertainties that are as big or bigger than
the experimental errors.  Of course, the theoretical prediction should be
independent of $\mu_R$.   However, a scale dependence is introduced by
truncating the perturbative series. The change due to varying the scale is
formally higher order.
If an observable ${\cal O}bs$ is known to order $\alpha_s^N$ then,
$$
\frac{\partial}{\partial \ln(\mu_R^2)} 
\sum_{0}^N A_n(\mu_R) \alpha_s^n(\mu_R) = {\cal O} \left(\alpha_s^{N+1} \right).
$$
Often the uncertainty due to uncalculated higher
orders is estimated by varying the renormalisation scale upwards and downwards
by a factor of two around a typical hard scale in the process.  
However, the variation only produces copies of the lower order terms,
\\
e.g.  $$
{\cal O}bs = A_0 \alpha_s(\mu_R) + \left(A_1 + b_0 A_0
\ln\left({\mu_R^2\over \mu_0^2}\right)\right)
\alpha_s(\mu_R)^2.$$
$A_1$ will contain generally contain infrared logarithms and constants that are
not present in $A_0$ and  therefore {\em cannot be predicted} by varying
$\mu_R$. For example, $A_0$ may contain infrared logarithms $L$ up to $L^2$,
while $A_1$ would contain these logarithms up to $L^4$.  $\mu_R$ variation is
{\em only an estimate} of higher order terms A large variation probably means
that {\em predictable} higher order terms are large.

\begin{figure}[t]
\label{fig:nnlo}
\begin{center}
\includegraphics[height=7cm]{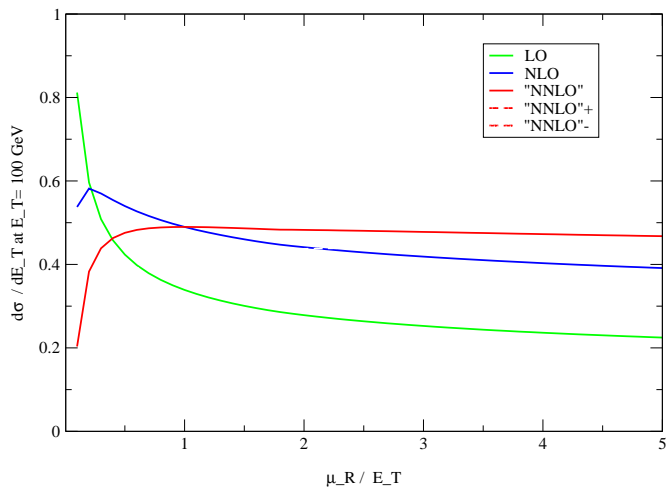}
\end{center}
\caption{Single jet inclusive distribution at $E_T = 100$~GeV and $0.1 < |\eta| < 0.7$ at
$\sqrt{s} = 1800$~GeV at LO (green), NLO (blue) and NNLO (red). The solid and dashed lines
how the NNLO prediction if $A_2=0$, $A_2= \pm A_1^2/A_0$ respectively. 
The same pdf's and $\alpha_s$ are used throughout.}
\end{figure}

To illustrate the improvement in scale uncertainty that may occur at NNLO, let
us consider the production of a central jet in $p\bar p$ collisions.   The
renormalisation scale dependence is entirely predictable,
\begin{eqnarray*}
\frac{d\sigma}{dE_T} &=& \alpha^2_s(\mu_R) {A_0}\\
&+&  \alpha^3_s(\mu_R) \left ({A_1} + 2 b_0 L {A_0}\right)\\
&+&  \alpha^4_s(\mu_R) \left ({A_2} + 3 b_0 L {A_1} + (3b_0^2L^2+2 b_1 L) { A_0} \right)
\end{eqnarray*}
with $ L = \log(\mu_R/E_T)$.  $A_0$ and $A_1$ are the known LO and NLO coefficients
while $A_2$ is not presently known.  Inspection of Fig.~\ref{fig:nnlo} shows
that the scale dependence is systematically reduced by increasing the number of
terms in the perturbative expansion.   At NLO, there is always a turning point
where the prediction is insenstitive to small changes in $\mu_R$.  If this
occurs at a scale far from the typically chosen values of $\mu_R$, the
$K$-factor (defined as $ K = 1 + \alpha_s(\mu_R) {A_1}/{A_0}$) will be
large.  At NNLO the scale dependence is clearly significantly reduced, although
a more quantitative statement requires knowledge of $A_2$. 

\vspace{0.2cm}

\noindent
{\it Factorisation scale dependence}

\vspace{0.15cm}

\noindent
Similar qualitative arguments can be applied to the factorisation scale inherent
in perturbative predictions for quantities with initial state hadrons. 
Including the NNLO contribution reduces the uncertainty due to the truncation of
the perturbative series.

\vspace{0.2cm}

\noindent
{\it Jet algorithms}

\vspace{0.15cm}

\noindent
There is also a mismatch between the number of hadrons and the number of
partons in the event.   At LO each parton has to model a jet and there is no
sensitivity to the size of the jet cone.   At NLO two partons can combine
to  make a jet giving sensitivity to the shape and size of the jet cone.
Perturbation theory is starting to reconstruct the parton shower within the
jet. This is further improved at NNLO where up to three partons can form a
single jet, or alternatively two of the jets may be formed by two partons. This
may lead to a better matching of the jet algorithm between theory and 
experiment.

\vspace{0.2cm}

\noindent
{\it Transverse momentum of the incoming partons}

\vspace{0.15cm}

\noindent
At LO, the incoming particles have no transverse momentum with respect to the
beam so that the final state is produced at rest in the transverse plane.  At
NLO, single hard radiation off one of the incoming particles gives the final
state a transverse momentum kick even if no additional jet is observed. In some
cases, this is insufficient to describe the data and one appeals to the
intrinsic transverse motion of the partons confined in the proton to explain
the data.    However, at NNLO, double radiation from one particle or  single
radiation off each incoming particle gives more complicated transverse momentum
to the final state and may provide a better, and more theoretically motivated, 
description of the data.

\vspace{0.2cm}

\noindent
{\it Power corrections}
\begin{figure}[t]
\label{fig:powerc}
\begin{center}
\includegraphics[height=6cm]{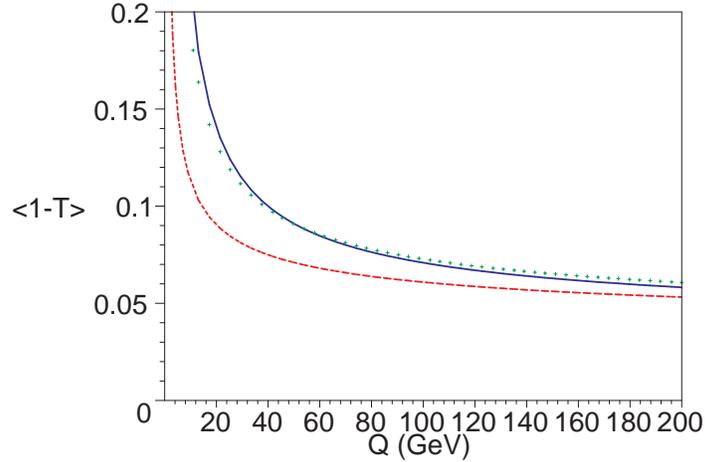}
\end{center}
\caption{The average value of $\langle 1-T\rangle$ given by Eq.~\ref{eq:omt}
showing the NLO prediction (dashed red), the NLO prediction with power correction of
$\lambda = 1$~GeV (solid blue) and an NNLO estimate with $a=3$ and a power correction
of $\lambda=0.5$~GeV (green dots).}
\end{figure}
Current comparisons of NLO predictions with experimental data generally reveal
the need for power corrections.   For example, in
electron-positron annihilation,  the experimentally measured average value of
1-Thrust lies well above the NLO predictions.   The difference can be accounted
for by a $1/Q$ power correction. While the form of the power correction can be
theoretically motivated, the magnitude is generally extracted from data and, to
some extent, can be attributed to uncalculated higher orders.   Including the
NNLO may therefore reduce the size of the phenomenological  power correction
needed to fit the data.

Before the calculation of the NNLO contribution it is not possible to make a more
quantitative statement. However to illustrate the qualitative point, 
let us take the simple example of an observable like
$\langle 1-T\rangle$ which can be modelled by the simplified series,
\begin{equation}
\label{eq:omt}
\langle 1-T\rangle = 0.33 \alpha_s(Q) + 1.0 \alpha_s(Q)^2 + a \alpha_s(Q)^3 +
\frac{\lambda}{Q},
\end{equation}
with $\alpha_s(Q) \sim 6\pi/23/\log(Q/\Lambda)$ and $\Lambda = 200$~MeV. Fig~\ref{fig:powerc} shows the NLO
perturbative prediction $a=0$,  $\lambda=0$ as well as the NLO prediction
combined with a power correction, $a=0$, $\lambda=1$~GeV which can be taken to
model the data.   If the NNLO coefficient turns out to be positive (which is by
no means guaranteed), then the size of the power correction needed to describe
the data will be reduced. For example, if we estimate the NNLO coefficient as
$a = 3$, which is large but perhaps not unreasonable, then the NLO prediction
plus power correction can almost exactly be reproduced with a power correction
of the same form, but $\lambda  = 0.6$~GeV.  We are effectively trading a
contribution of ${\cal O}(1/Q)$ for a contribution of $1/\log^3(Q/\Lambda)$.  
At present the data is insufficient to distinguish between these two functional
forms. 

\subsubsection{Parton densities at NNLO}
\label{sec:ho;nnlopdf}

Consistent NNLO predictions for processes involving hadrons in the  initial
state require not only the NNLO hard scattering  cross sections, but also
parton distribution functions which are accurate to this order. 

The evolution of parton distributions is governed at NNLO by the  three-loop
splitting functions, which are not fully known at present.  However, using the
available information on some of the lower Mellin 
moments~\cite{Larin:1994vu,Larin:1997wd,Retey:2000nq} and on the asymptotic
behaviour~\cite{Catani:1994sq},  
as well as some exactly known terms~\cite{Gracey:1994nn},
it is possible to  construct approximate expressions for these splitting 
functions~\cite{vanNeerven:2000wp}. These approximations (which are provided with an 
error band) can serve as a substitute until full results become 
available~\cite{Moch:2001fr}. 

The determination of NNLO parton distributions requires a global  fit to the
available data on a number of hard scattering  observables, with all
observables computed consistently at NNLO.  At present, the NNLO coefficient
functions are available only for  the inclusive  Drell-Yan
process~\cite{Hamberg:1991np,Harlander:2002wh} and for deep inelastic structure 
functions~\cite{Zijlstra:1992qd}. These two observables are by themselves  insufficient
to fully constrain all parton species. In the first  NNLO analysis of parton
distribution functions,  which was performed recently~\cite{Martin:2002dr}, these
processes were  therefore accompanied by several other observables only known
to NLO.  The resulting distributions~\cite{Martin:2002dr} illustrate some  important
changes in size and shape of the distribution functions  in going from NLO to
NNLO, visible in particular for the gluon distribution  function. 

It is clear that further progress on the determination of NNLO  parton
distributions requires a larger number of processes  (in particular jet
observables) to be treated consistently at NNLO.

\subsubsection{Infrared limits of one-loop and tree-level processes}
\label{sec:ho;nnloIR}

For simplicity let us consider a process with no initial state partons and with
$m$ partons in the final state at LO.
The NNLO contribution to the cross section can be decomposed as 
\begin{equation}
\sigma^{NNLO} = \int_{m+2} d\sigma^{RR} + \int_{m+1} d\sigma^{VR} 
+ \int_{m}
d\sigma^{VV}.
\end{equation}
Here $\int_n$ denotes the $n$-particle final state, while $d\sigma^{XY}$ denotes
the fully differential cross section with double radiation ($XY=RR$), single
radiation from one-loop graphs ($XY=VR$) and the double virtual contribution
($XY=VV$) that includes both the square of the one-loop graphs and the
interference of tree and two-loop diagrams. 
After renormalisation of the virtual matrix elements, each of the contributions
is UV finite.   However,  each of these terms is separately infrared
divergent and this manifests itself as poles in $\epsilon$.
Nevertheless, the Kinoshita-Lee-Nauenberg theorem states that
the infrared singularities must cancel for sufficiently inclusive physical
quantities.  The problem is to isolate the infrared poles and analytically
cancel them before taking the $\e \to 0$ limit.
Establishing a strategy for doing this requires a good understanding of  
$ d\sigma^{RR}$ and $d\sigma^{VR}$ in the infrared region where the additional
radiated particle(s) are unresolved.
 
\vspace{0.2cm}

\noindent
{\it Double unresolved limits of tree amplitudes}

\vspace{0.15cm}

\noindent
The infrared singular regions of tree amplitudes can be divided into several
categories, 
\begin{enumerate} 
\item three collinear particles, 
\item two pairs of collinear particles, 
\item two soft particles, 
\item one soft and two collinear particles, 
\item a soft quark-antiquark pair. 
\end{enumerate} 
In each of these limits, the tree-level $m+2$ particle amplitudes factorise and
yield an infrared singular factor multiplying the tree-level $m$-particle
amplitude.   The various limits have been well studied.   

The limit where a quark, gluon and photon simultaneously become collinear was
first studied in \cite{Gehrmann-DeRidder:1998gf} and then extended for generic
QCD processes in Ref.~\cite{Campbell:1998hg}  by directly taking the limit of
tree-level matrix elements. These limits have been subsequently rederived using
general gauge invariant methods and extended to include azimuthal
correlations between the particles~\cite{Catani:1998nv,Catani:1999ss,DelDuca:1999ha}. 

When two independent pairs of particles  are collinear, the singular limits can
be treated independently, and just lead to a trivial extension of the NLO
result - the product of two single collinear splitting functions.

The limit where the momenta of two of the gluons simultaneously become soft has
also been studied~\cite{Berends:1989zn,Catani}. At the amplitude level, the singular
behaviour factorises in terms of a process independent soft two-gluon
current~\cite{Catani:1999ss} which is the generalisation of the one-gluon
eikonal current.

The soft-collinear limit occurs when the momentum of one gluon becomes soft
simultaneously with two other partons becoming collinear.   Factorisation
formulae in this limit have been provided in both the azimuthally averaged
case~\cite{Campbell:1998hg} and  including the angular
correlations~\cite{Catani:1998nv,Catani:1999ss,DelDuca:1999ha}.

Finally, when the momentum of both quark and antiquark of a $q \bar q$ pair
become soft, the tree-level amplitude again factorises~\cite{Catani:1999ss}.

Taken together, these factorisation formulae describe all of the cases where
$m$-partons are resolved in a $m+2$ parton final state. Techniques for
isolating the divergences have not yet been established and there is an
on-going effort to develop a set of {\em local} subtraction counter-terms that
can be analytically integrated over the infrared regions.

\vspace{0.2cm}

\noindent
{\it Single unresolved limits of one-loop amplitudes}

\vspace{0.15cm}

\noindent
The soft- and collinear-limits of one-loop QCD amplitudes has also been
extensively studied. In the collinear limit, the one-loop $m+1$ particle
process factorises as a  one-loop $m$ particle amplitude multiplied by a tree
splitting function together with a tree $m$ particle amplitude multiplied by a
one-loop splitting  function.  The explicit forms of the splitting amplitudes
were first determined to ${\cal O}(\e^0)$~\cite{Bern:1994zx}.  However, because
the integral over the infrared phase space generates additional poles, the
splitting functions have been determined to all orders in
$\e$~\cite{Bern:1998sc,Kosower:1999xi,Bern:1999ry,Kosower:1999rx}. Similarly,
when a gluon becomes soft, there is a factorisation of the one-loop amplitude
in terms of the one-loop soft
current~\cite{Bern:1998sc,Bern:1999ry,Catani:2000pi}. Because of the similarity
of the factorisation properties  with the single unresolved particle limits of 
tree-amplitudes, it is relatively straightforward to isolate the infrared poles
through the construction of a set of {\em local} subtraction counter-terms. In
Sec.~\ref{sec:ho;exercise}, we illustrate how the infrared singularities from
the single unresolved limits of one-loop amplitudes combine with the
predictable infrared pole structure of the virtual contribution for the
explicit example of $\gamma^* \to q\bar q$.

\subsubsection{Two-loop matrix elements for scattering processes}
\label{sec:ho;two-loop}

In recent years, considerable progress has been made on the  calculation of
two-loop  virtual corrections to the multi-leg matrix elements relevant for 
jet physics, which describe either $1\to 3$ decay or $2\to 2$ scattering 
reactions.  Much of this progress is due to a number of important technical 
breakthroughs related to the evaluation of the large number of  different
integrals appearing in the two-loop four-point amplitudes.

It should be recalled that  perturbative corrections to many inclusive quantities
have been computed  to the two- and three-loop level already several years ago. 
From the technical  point of view, these inclusive calculations correspond to the
computation of  multi-loop two-point functions, for which many elaborate
calculational tools have been developed.  Using dimensional
regularization~\cite{Bollini:1972ui,Ashmore:1972uj,Cicuta:1972jf,'tHooft:1972fi}  with $d\neq 4$
dimensions as regulator for ultraviolet and infrared divergences,  the large number
of different integrals appearing in  multi-loop two-point functions can be reduced
to a small number of  so-called {\em master integrals} by using 
integration-by-parts
identities~\cite{'tHooft:1972fi,Tkachov:1981wb,Chetyrkin:1981qh}. These identities 
exploit the fact that the integral over the total derivative of any  of the loop
momenta vanishes in dimensional regularization. 

Integration-by-parts identities can also be obtained for  integrals appearing
in amplitudes with more than two external legs;  for these amplitudes, another
class of identities exists due to  Lorentz invariance of the amplitudes. These 
Lorentz invariance identities~\cite{Gehrmann:1999as} rely on the fact that  an
infinitesimal Lorentz transformation commutes with the loop integral, thus
relating different integrals. Using integration-by-parts  and Lorentz
invariance identities, all  two-loop Feynman amplitudes for $2\to 2$ scattering
or $1\to 3$ decay  processes  can be expressed as linear combinations of a
small number of  master integrals,
which have  to be computed by some different method.  Explicit
reduction formulae for on-shell two-loop four-point  integrals were derived
in~\cite{Smirnov:1999wz,Anastasiou:1999bn,Anastasiou:2000mf}. Computer algorithms for the  automatic reduction of all
two-loop four-point integrals  were described in~\cite{Gehrmann:1999as,Laporta:2001dd}. 

A related development was the proof of the equivalence of  integration-by-parts
identities for integrals with the same  total number of external and loop
momenta~\cite{Baikov:2000jg}. Consequently,  much of the tools developed for
the computation of three-loop propagator
integrals~\cite{Gorishnii:1989gt,Chetyrkin:1996ia} can be readily applied to
two-loop  vertex functions. As a first application, the two-loop  QCD
corrections to Higgs boson production in gluon-gluon fusion were 
computed~\cite{Harlander:2000mg} in the limit of large top quark mass. This
result allowed the complete NNLO
description~\cite{Catani:2001ic,Harlander:2001is,Harlander:2002wh} of inclusive
Higgs production at hadron colliders. 

The master integrals relevant to $2\to 2$ scattering or $1\to 3$ decay 
processes are massless, scalar two-loop  four-point functions with all legs
on-shell or a single leg off-shell.  Several techniques for the computation of
those functions  have been proposed in the literature, such as the application
of a Mellin-Barnes transformation to all propagators~\cite{Smirnov:1999gc,Tausk:1999vh} 
or the negative dimension approach~\cite{Anastasiou:1999cx,Suzuki:2001yf}.  Both techniques
rely on an explicit integration  over the loop momenta, with differences mainly
in the representation used for the propagators. These techniques were used
successfully to compute a number of master integrals. Employing the
Mellin-Barnes method, the on-shell planar double box
integral~\cite{Smirnov:1999gc,Anastasiou:2000kp}, the on-shell non-planar double box
integral~\cite{Tausk:1999vh}  and two double box
integrals with one leg off-shell~\cite{Smirnov:2000vy,Smirnov:2000ie}
were computed. Most recently, the same method was used to  derive the on-shell
planar double box integral~\cite{Smirnov:2001cm} with one internal mass scale.  The
negative dimension approach has been applied~\cite{Anastasiou:1999cx}  to compute the
class of two-loop box integrals which correspond to a one-loop bubble insertion
in one of the propagators of the one-loop box. 

A method for the analytic computation of master integrals avoiding the explicit
integration over the loop momenta is to derive differential equations in 
internal propagator masses or in external momenta for the master integral,  and
to solve these with appropriate boundary conditions.  This method has first
been suggested by Kotikov~\cite{Kotikov:1991pm} to relate  loop integrals with
internal masses to massless loop integrals.  It has been elaborated in detail
and generalized to differential  equations in external momenta
in~\cite{Remiddi:1997ny}; first  applications were presented
in~\cite{Caffo:1998du,Caffo:1999nk}. The
computation of master integrals from differential equations proceeds as 
follows. Carrying out the derivative with respect to an external invariant  on
the master integral of a given topology, one obtains a linear combination of a
number of  more complicated integrals, which can however be reduced to the 
master integral itself plus simpler integrals by applying the reduction 
methods discussed above. As a result, one obtains  an inhomogeneous linear
first order differential equation in each invariant for the master integral.

The inhomogeneous term in these differential equations contains only 
topologies simpler than the topology under consideration, which are considered
to be known  if working in a bottom-up approach.  The master integral is then
obtained by matching the general  solution of its differential equation to an
appropriate boundary condition. Quite in general, finding a boundary condition
is  a  simpler problem than evaluating the whole integral, since it depends on
a smaller number of kinematical variables. In some cases, the boundary
condition can even be determined from the differential equation itself. 

Using the differential equation technique, one of the on-shell 
planar double box integrals~\cite{Gehrmann:2000xj} as well as the full set of 
planar and non-planar off-shell double box integrals~\cite{Gehrmann:2000zt,Gehrmann:2001ck}
 were derived. 

A strong check on all these computations of master integrals is  given by the
completely numerical calucations of~\cite{Binoth:2000ps}, which are  based on an iterated
sector decomposition to isolate the infrared pole  structure. The methods
of~\cite{Binoth:2000ps} were applied to confirm {\em all}  of the above-mentioned
calculations. 

The two-loop four-point functions with all legs on-shell can be expressed  in
terms of Nielsen's polylogarithms~\cite{nielsen,lewin,Kolbig:1986qt,bit}.  In
contrast, the closed analytic  expressions for  two-loop four-point functions
with one leg off-shell contain  two new classes of functions: harmonic 
polylogarithms~\cite{Remiddi:1999ew,Gehrmann:2001pz} and two-dimensional
harmonic  polylogarithms (2dHPL's)~\cite{Gehrmann:2001jv}.  Accurate numerical
implementations for these functions~\cite{Gehrmann:2001pz,Gehrmann:2001jv} are
available. 

\vspace{0.2cm}

\noindent
{\it $2 \to 2$ Processes with all legs on-shell}

\vspace{0.15cm}

\noindent
With the explicit solutions of the integration-by-parts and  Lorentz-invariance
identities for on-shell two-loop four-point 
functions~\cite{Smirnov:1999wz,Anastasiou:1999bn,Anastasiou:2000mf} and the
corresponding master  
integrals~\cite{Gonsalves:1983nq,Kramer:1987sr,Smirnov:1999gc,Tausk:1999vh,Anastasiou:2000kp,Gehrmann:2000xj}, all
necessary ingredients  for the computation of two-loop corrections to  $2 \to 2$
processes with all legs on-shell are now available. In fact, only half a  year
elapsed between the completion of the full set of master 
integrals~\cite{Anastasiou:2000kp,Gehrmann:2000xj} and the calculation of the 
two-loop QED corrections to Bhabha-scattering~\cite{Bern:2000ie}.  Subsequently,
results were obtained for the two-loop  QCD corrections to all parton-parton
scattering processes~\cite{Anastasiou:2000kg,Anastasiou:2000ue,Anastasiou:2001sv,
Glover:2001af}. For gluon-gluon scattering, the two-loop helicity amplitudes have
also  been derived~\cite{Bern:2000dn,Bern:2002tk}. Moreover, two-loop corrections were
derived  to processes involving two partons and two real
photons~\cite{Bern:2001df,Anastasiou:2002zn}. Finally, light-by-light scattering in
two-loop QED and QCD was  considered in~\cite{Bern:2001dg}, these results were
extended to supersymmetric QED in~\cite{Binoth:2002xg}. 

It turns out that supersymmetry can provide strong checks on the  consistency
of the matrix elements. Calculations in this framework  do however require
modifications to the dimensional reguarization scheme, which were discussed in
detail in~\cite{Bern:2002zk}. 

At the same order in perturbation theory
as the two-loop matrix elements (which are 
obtained by contracting the two-loop and the tree level amplitudes),
one also finds contributions from the square of the one-loop amplitude. 
The evaluation of these contributions uses well-known one-loop techniques and 
is straightforward. For parton-parton scattering, these one-loop
self-interference contributions were computed 
in~\cite{Anastasiou:2000mv,Anastasiou:2001sv,Glover:2001rd}. 

The results for the two-loop QED matrix element for  Bhabha
scattering~\cite{Bern:2000ie} were used in~\cite{Glover:2001ev} to extract the  single
logarithmic contributions to the Bhabha scattering cross section,  thus
improving considerably on the accuracy of earlier~\cite{Arbuzov:1997qd} double
logarithmic results.

\vspace{0.2cm}

\noindent
{\it $2 \to 2$ Processes with one off-shell leg}

\vspace{0.15cm}

\noindent
With the reduction algorithm~\cite{Gehrmann:1999as} and full set of two-loop master 
integrals~\cite{Gehrmann:2000zt,Gehrmann:2001ck} for two-loop four-point functions  now available, it
is possible to compute the two-loop corrections  a number of $1\to 3$ decay and
$2\to 2$ scattering reactions with one  off-shell leg.

As a first result (and initiated during this workshop),
in~\cite{Garland:2001tf},  the two-loop QCD matrix element for  $e^+e^- \to
3$~jets and the  corresponding one-loop self-interference matrix element were
derived. The three jet production rate in electron-positron  collisions and
related event shape observables are in fact the  most precisely measured jet 
observables at present~\cite{Bethke:2000ai}. They will also play an important
role in future  QCD studies~\cite{Aguilar-Saavedra:2001rg} at a the proposed
high energy linear $e^+e^-$ collider. 

It is worthwhile to note that  besides its  phenomenological importance,  the
three-jet rate has also served as a theoretical testing ground for the
development of new techniques for higher order calculations in QCD: both the
subtraction~\cite{Ellis:1981wv} and the phase-space
slicing~\cite{Fabricius:1981sx} methods for the extraction of infrared
singularities from  NLO real radiation  processes were developed in the context
of the first three-jet calculations. The systematic formulation of 
phase-space  slicing~\cite{Giele:1992vf} as  well as the dipole
subtraction~\cite{Catani:1997vz} method were also first demonstrated for
three-jet observables, before being applied to other processes. It is very
likely that similar techniques at higher orders will first be developed in the
context of jet production in $e^+e^-$ annihilation, which in contrast to
hadron--hadron collisions or electron--proton scattering does not pose the
additional difficulty of the regularization of  initial state singularities.

Processes related to  $e^+e^- \to 3$~jets  by crossing symmetry  are
$(2+1)$-jet production in  deep inelastic $ep$ scattering and
vector-boson-plus-jet production at hadron  colliders.  Crossing of the 
$e^+e^- \to 3$~jets two-loop matrix element  to the kinematic regions relevant
for these scattering processes  requires the analytic continuation of 2dHPL's
outside their range of allowed arguments. This topic is currently under
investigation.

\subsubsection{Infrared structure for $\gamma^* \to 2$~jets at NNLO}
\label{sec:ho;exercise}

As an exercise in how to combine the various next-to-next-to-leading order
contributions, we consider the infrared singularity structure for  $\gamma^*
\to 2$~jets at NNLO.   This is the simplest process that we can imagine at
NNLO.   It involves the one- and two-loop two parton amplitudes, the one-loop
three parton amplitudes and the tree-level four parton amplitudes.    The aim
is to symbolically identify the origins of the infrared singularities of the
single and double radiation graphs.  Once this is achieved,  it should be
possible to construct {\em local} counter terms that can be subtracted
numerically from the three- and four-parton matrix elements to give infrared
finite contributions  and analytically integrated and combined with the
explicit singularity structure of the two-parton contribution.

\vspace{0.2cm}

\noindent
{\it The one- and two-loop two parton contribution}

\vspace{0.15cm}

\noindent
The renormalized $\gamma^* \to q\bar q$
amplitude can be written as
\begin{equation}
|\cmqq\rangle = \sqrt{4\pi\alpha}e_q  \left[
|\cmqq^{(0)}\rangle 
+ \asmuopi |\cmqq^{(1)}\rangle 
+ \asmuopi^2 |\cmqq^{(2)}\rangle 
+ {\cal O}(\alpha_s^3) \right] \;,
\end{equation}
where $\alpha$ denotes the electromagnetic coupling constant, 
$e_q$ the quark charge,
and the $|{\cal M}_{q\bar q}^{(i)}\rangle$ are the $i$-loop contributions to the 
renormalized amplitude. They are scalars in colour space. 

The squared amplitude, summed over spins, colours and quark flavours, 
is denoted by
\begin{equation}
\langle\cmqq|\cmqq\rangle = \sum |\cm (\gamma^* \to q\bar q)|^2 
= {\cal A}_{q\bar q}\; .
\end{equation}
The perturbative expansion of ${\cal A}_{q\bar q}$ at renormalization scale 
$\mu^2 = \sab$ reads,
\begin{equation}
{\cal A}_{q\bar q} = 4\pi\alpha\sum_q N e_q^2 \Bigg[\caqq{0}
+\asmuopi \caqq{2} +\asmuopi^2 \caqq{4} + {\cal O}(\alpha_s^3)\Bigg] \;,
\end{equation}
where $N$ is the number of colours
and where, 
\begin{eqnarray}
\label{eq:Aqq0}
\caqq{0} &=& \langle\cmqq^{(0)}|\cmqq^{(0)}\rangle 
= 4 (1-\e) \sab \;,\\
\label{eq:Aqq2}
\caqq{2} &=& 
\langle\cmqq^{(0)}|\cmqq^{(1)}\rangle +
\langle\cmqq^{(1)}|\cmqq^{(0)}\rangle \nonumber \\
\caqq{4} &=& 
\langle\cmqq^{(1)}|\cmqq^{(1)}\rangle +
\langle\cmqq^{(0)}|\cmqq^{(2)}\rangle +
\langle\cmqq^{(2)}|\cmqq^{(0)}\rangle \;,
\label{eq:Aqq4}
\end{eqnarray}
with~\cite{Kramer:1987sg,Matsuura:1988wt,Matsuura:1989sm},
\vspace{-0.5cm}
\begin{eqnarray}
\langle\cmqq^{(0)}|\cmqq^{(0)}\rangle 
&\equiv& \includegraphics[width=2.5cm]{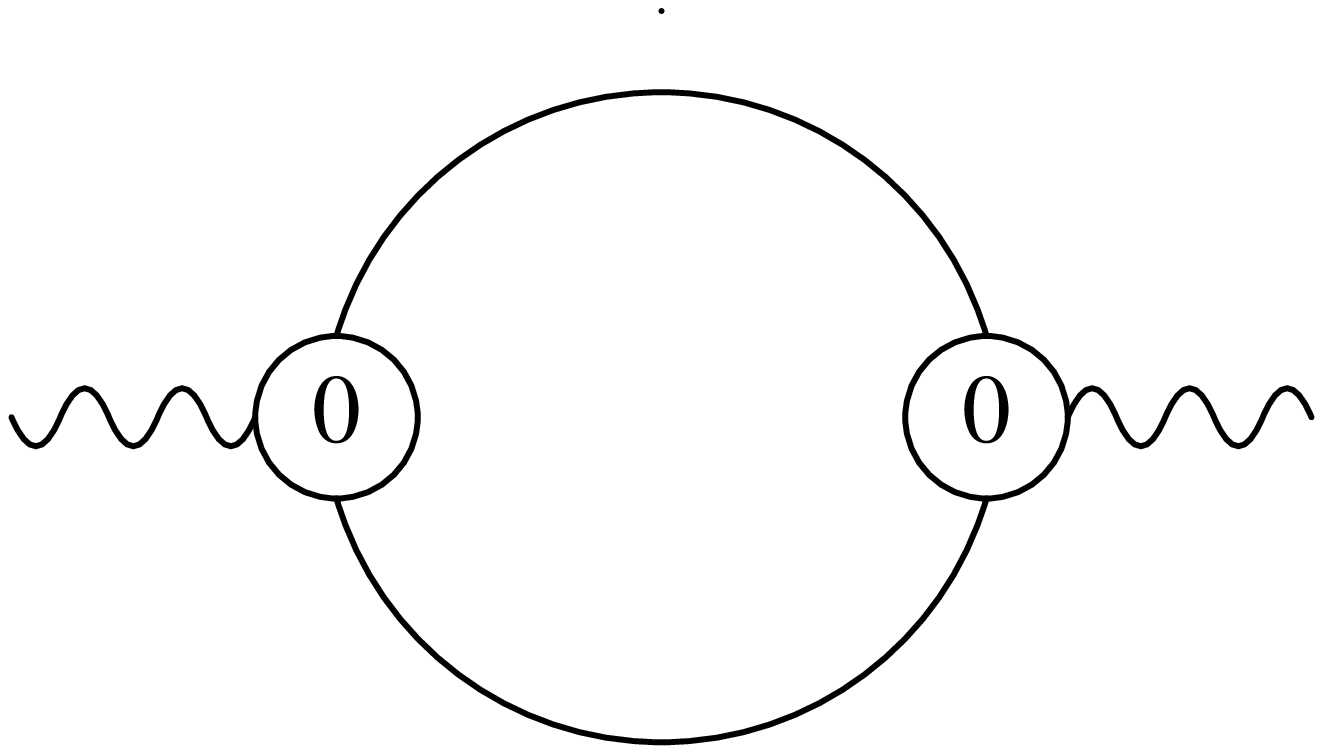} , \\
\langle\cmqq^{(0)}|\cmqq^{(1)}\rangle 
&\equiv& \includegraphics[width=2.5cm]{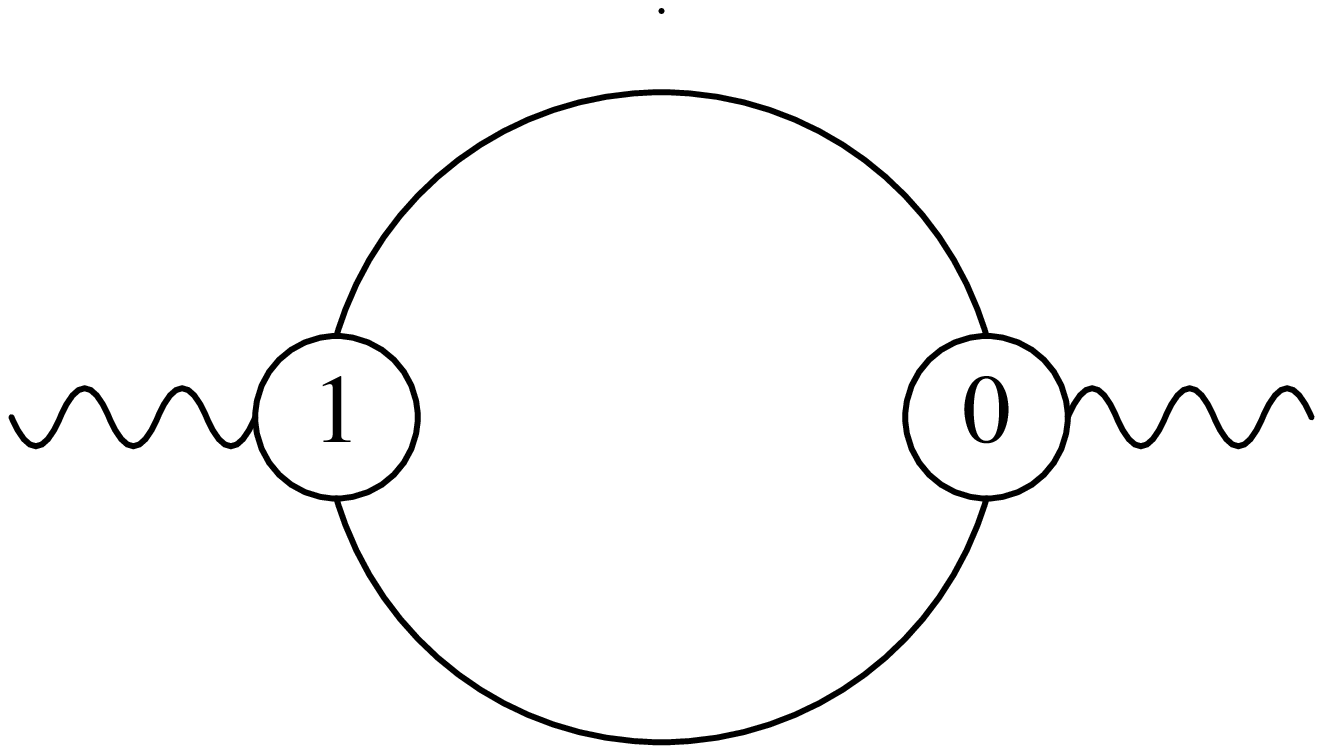} \nonumber \\
&=& \includegraphics[width=2.5cm]{m200.ps} \Bigg\{ \frac{1}{2}\, C_F\Bigg[
-\frac{2}{\e^2}-\frac{3}{\e} +\left( \frac{7\pi^2}{6}-8\right) 
\nonumber \\ &&
\hspace{3.5cm}
+\left(-16+\frac{7\pi^2}{4} +\frac{14}{3}\zeta_3\right)\, \e 
\nonumber \\ &&
\hspace{3.5cm}
+\left(-32 +\frac{14\pi^2}{3}+ 7\zeta_3-\frac{73\pi^4}{720}\right) \, \e^2
+ {\cal O}(\e^3)\Bigg] \Bigg\},\\
\langle\cmqq^{(1)}|\cmqq^{(1)}\rangle 
&\equiv& \includegraphics[width=2.5cm]{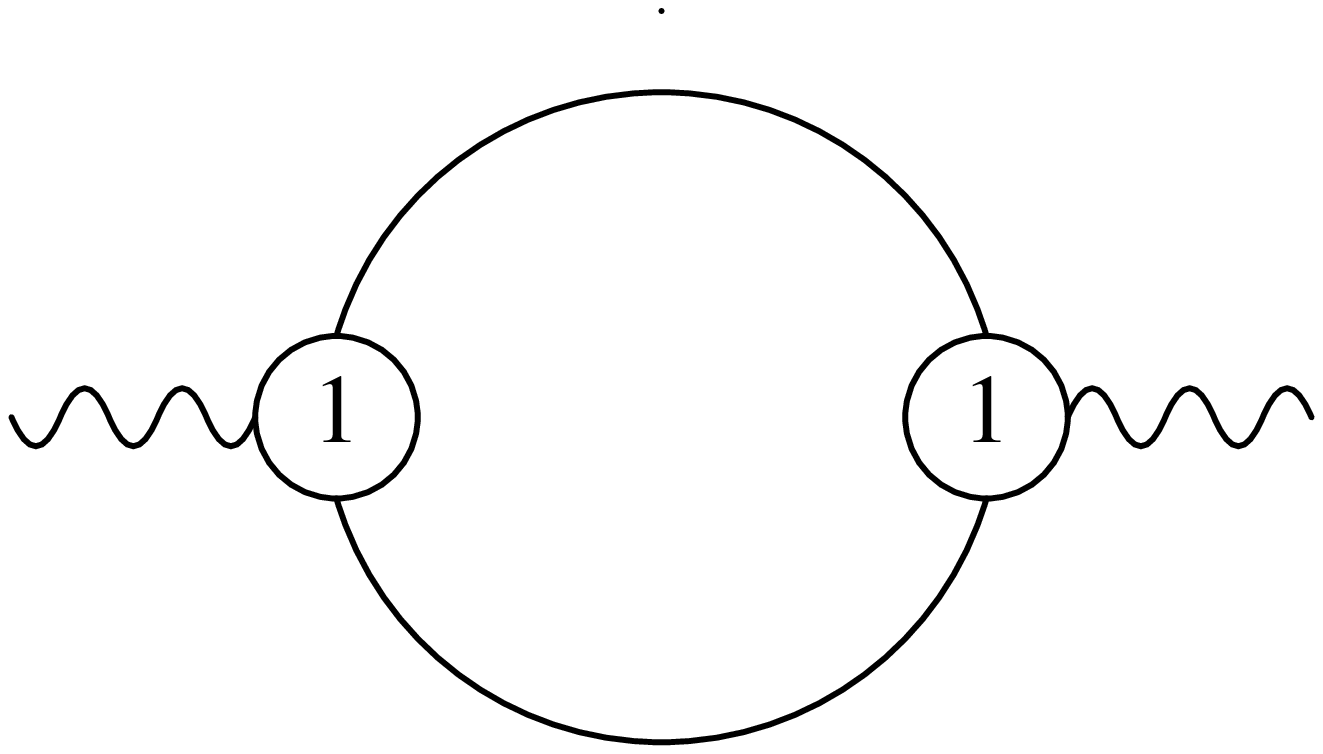} \nonumber \\
&=& \includegraphics[width=2.5cm]{m200.ps} \Bigg\{  C_F^2\Bigg[
\frac{1}{\e^4} + \frac{3}{\e^3} 
+ \left(\frac{41}{4}-\frac{\pi^2}{6}\right)\frac{1}{\e^2}
 + \left(28 -\frac{\pi^2}{2}-\frac{14}{3}\zeta_3\right)\frac{1}{\e} 
\nonumber \\ && 
\hspace{3.5cm}
+ \left(         72
          - \frac{41\pi^2}{24}
          - \frac{7\pi^4}{120}
          - 14 \zeta_3 \right)
+ {\cal O}(\e)
\Bigg] \Bigg\} ,\\
\langle\cmqq^{(0)}|\cmqq^{(2)}\rangle 
&\equiv& \includegraphics[width=2.5cm]{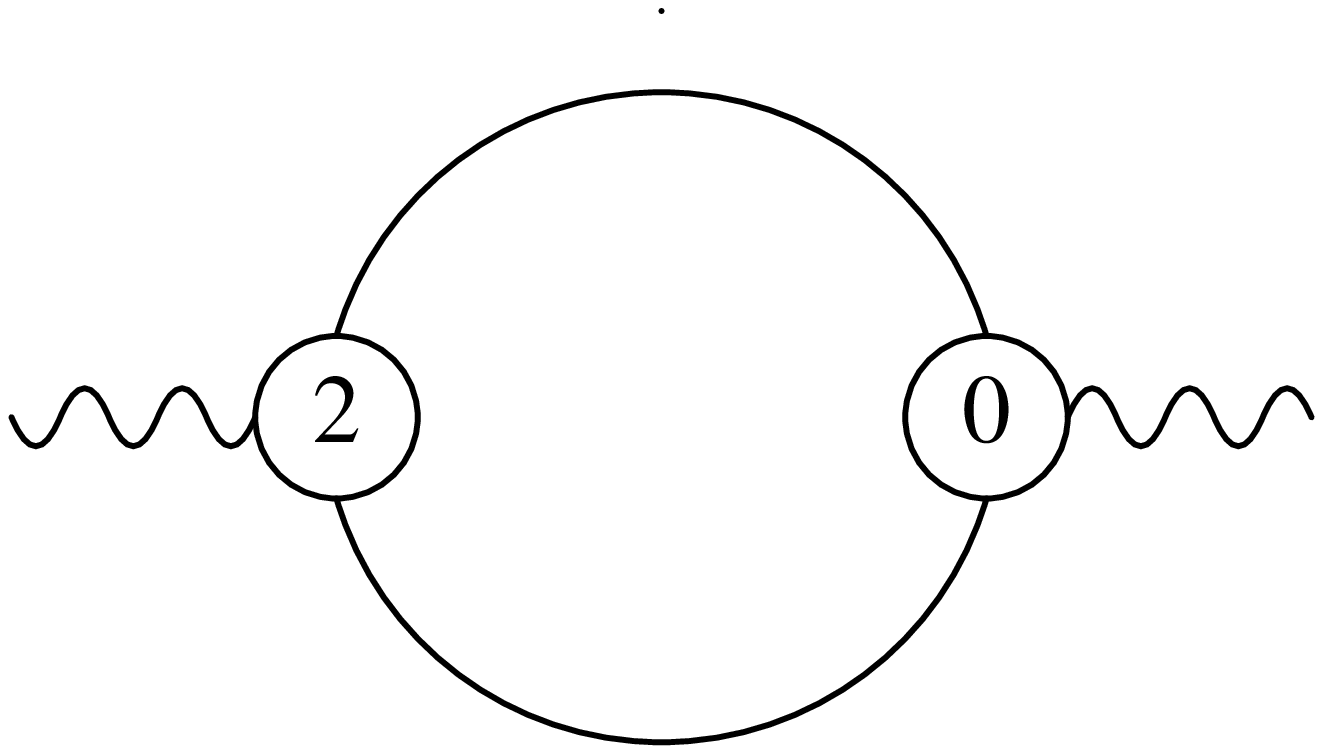} \nonumber \\
&=& \includegraphics[width=2.5cm]{m200.ps} \Bigg\{ 
\frac{1}{2\e^4} \, C_F^2  + \frac{1}{\e^3}\left[
\frac{3}{2}\, C_F^2 +\frac{11}{8} C_FC_A - \frac{1}{2} C_F T_R N_F
\right]\nonumber \\
&&\hspace{3cm}
     + \frac{1}{\e^2}\left[ 
\left( \frac{41}{8} - \frac{13\pi^2}{12}\right) C_F^2
+ \left( \frac{4}{9} +\frac{\pi^2}{24} \right) C_FC_A
-\frac{2}{9} C_F T_R N_F \right]\nonumber \\
&&\hspace{3cm}
     + \frac{1}{\e}\Bigg[ 
\left( \frac{221}{16} - 3\pi^2 -\frac{16}{3}\zeta_3\right) C_F^2
+ \left( \frac{65}{108} +\frac{\pi^2}{12} \right) C_F T_R N_F
\nonumber \\
&&\hspace{3.5cm}
+ \left( -\frac{961}{432} -\frac{11\pi^2}{48} +\frac{13}{4}\zeta_3\right)
                       C_FC_A\Bigg]\nonumber \\
&&\hspace{3cm}
+ \Bigg[
\left( \frac{1151}{32} - \frac{475\pi^2}{48} 
-\frac{29}{2}\zeta_3+\frac{59\pi^4}{144}\right) C_F^2
\nonumber \\
&&\hspace{3.5cm}
+ \left( -\frac{51157}{2592} +\frac{1061\pi^2}{432} 
   +\frac{313}{36}\zeta_3 - \frac{\pi^4}{45}\right)      C_FC_A
\nonumber \\
&&\hspace{3.5cm}
+ \left( \frac{4085}{648} -\frac{91\pi^2}{108} +\frac{1}{9} \zeta_3\right)
 C_F T_R N_F \Bigg]
+ {\cal O}(\e)
\Bigg\}.
\end{eqnarray}
\vspace{0.5cm}

\noindent We see that the NLO part $\caqq{2}$ is proportional to $\CF$ while the NNLO
contribution $\caqq{4}$ contains $\CF^2$, $\CF\CA$ and $\CF T_R \NF$ parts.

Catani~\cite{Catani:1998bh} has shown how to organize the 
infrared pole structure of the one- and two-loop contributions 
renormalized in the 
\MSbar\ scheme in terms of the tree and renormalized one-loop amplitudes,
$\ket{\cmqq^{(0)}}$ and $\ket{\cmqq^{(1)}}$ respectively by introducing two infrared
operators, $\Ione$ and $\Htwo$.  

The form of $\Ione$ can be simply understood by considering the total cross
section to ${\cal O}(\alpha_s)$ which is finite,
\vspace{-0.5cm}
\begin{equation}
\includegraphics[width=2.5cm]{m210.ps} 
+
\includegraphics[width=2.5cm]{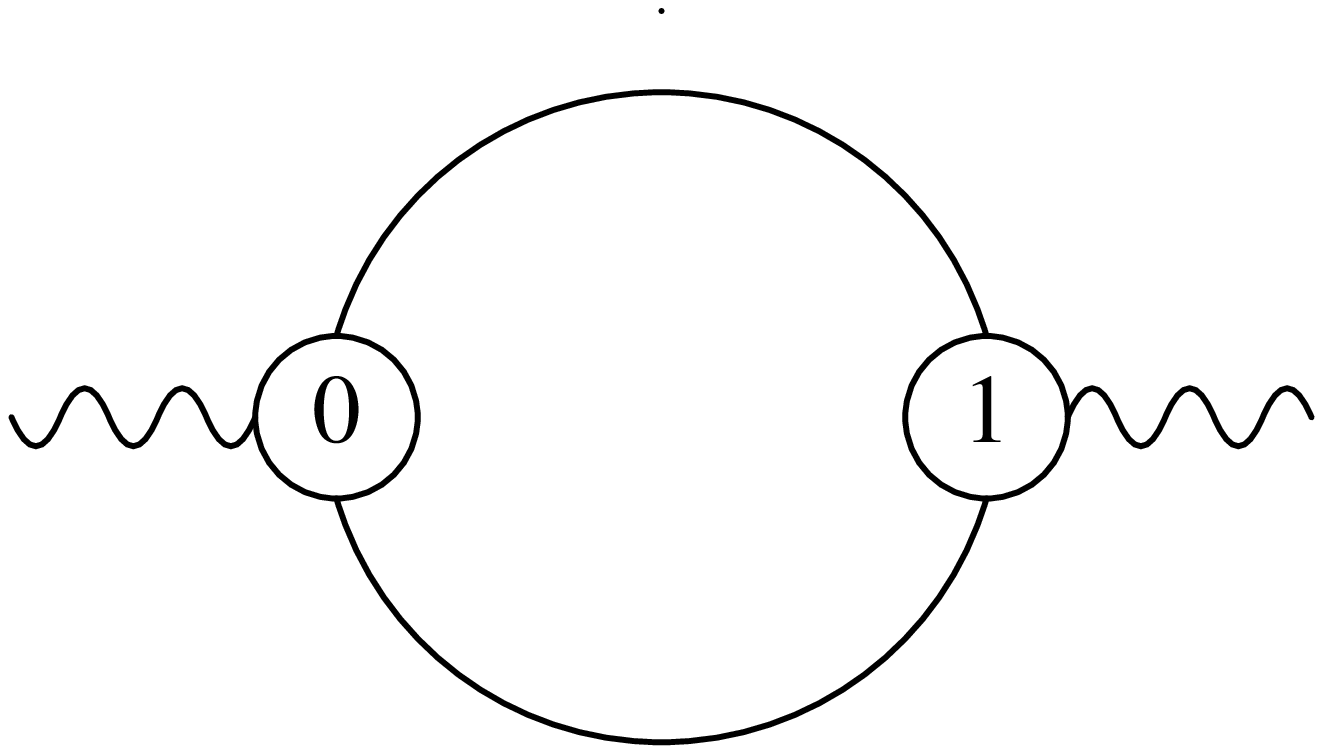} 
+
\includegraphics[width=2.5cm]{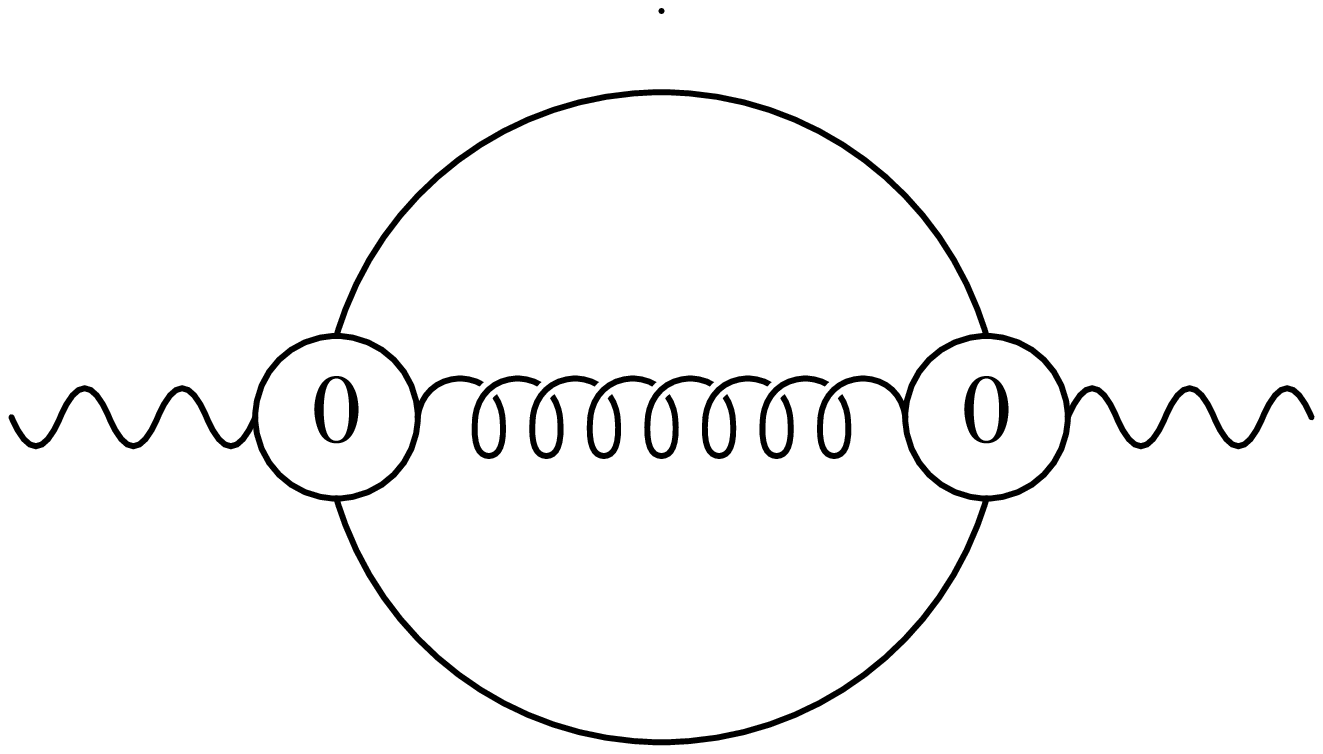} 
= {\rm finite}.
\label{eq:qqnlo}
\end{equation}
\vspace{0.5cm}

\noindent Taking soft and collinear limits of the real radiation graphs, 
and integrating over the unresolved phase space, we see that,
\vspace{-0.5cm}
\begin{eqnarray}
\includegraphics[width=2.5cm]{m300.ps} 
&\longrightarrow& 
\includegraphics[width=2.5cm]{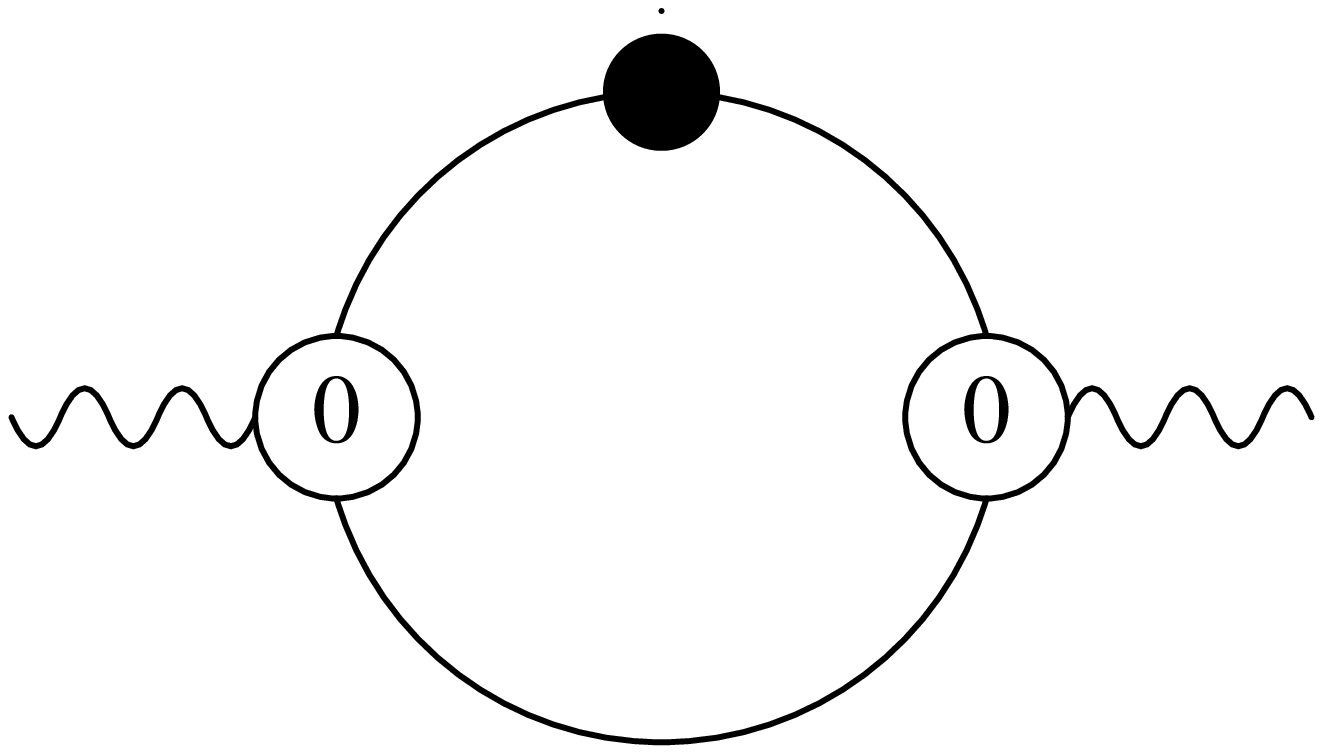}
+
\includegraphics[width=2.5cm]{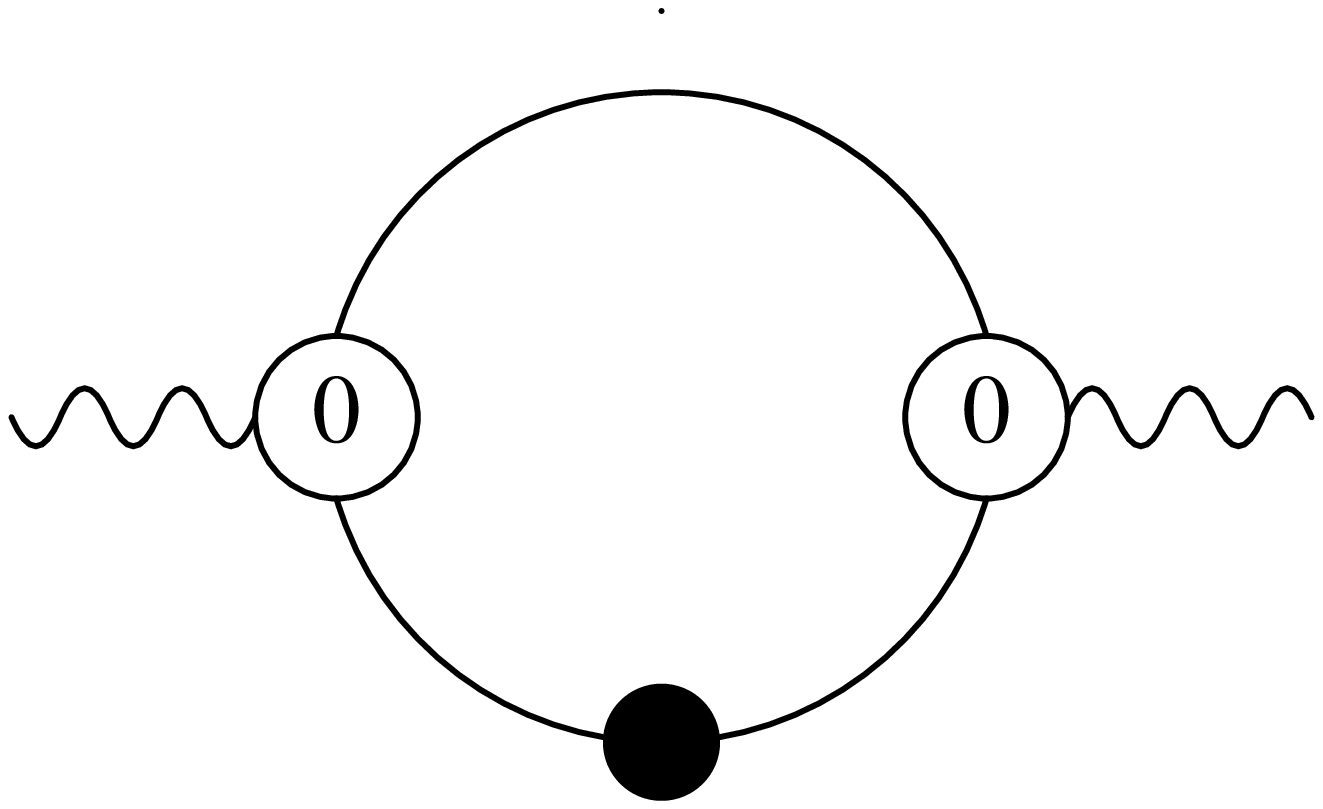}
+{\rm finite}
\nonumber \\
&=& 
-2 \Ione \includegraphics[width=2.5cm]{m200.ps} 
+{\rm finite}.
\label{eq:findi1}
\end{eqnarray}
\vspace{0.3cm}

\noindent
The symbol $\includegraphics[width=2cm]{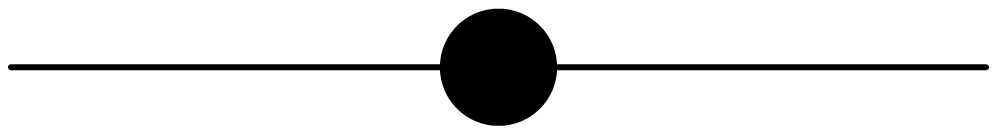}$
 represents the taking of the
infrared limit and integrating over the infrared phase space.   This operation
can be carried out in many ways. The most commonly used procedures are 
the subtraction formalism~\cite{Ellis:1981wv},
phase space slicing method~\cite{Fabricius:1981sx}, as well as 
systematic procedures~\cite{Giele:1992vf} and improved 
formulations of these methods, such as the dipole 
subtraction formalism~\cite{Catani:1997vz}.

In general, $\Ione$ contains colour correlations. However, 
for this particular process, there is only one colour structure present at
tree level which is simply $\delta_{ij}$. Adding higher loops does not
introduce additional colour structures into the amplitude, and $\Ione$ is therefore a
$1 \times 1$ matrix in colour space.  It is proportional to $\CF$ and
is given by
\begin{equation}
\Ione
=
- \frac{e^{\epsilon\gamma}}{2\Gamma(1-\epsilon)}  
\CF
\left(\frac{2}{\epsilon^2}+\frac{3}{\epsilon}\right)
\left(-\frac{\mu^2}{s_{12}+i0}\right)^{\epsilon}  ,\label{eq:I1}
\end{equation}
with $\mu^2 = \sab$.  The double pole is due to the soft gluon singularity while
the single pole is derived from the collinear quark-gluon splitting function.
Inserting Eq.~(\ref{eq:findi1}) into Eq.~(\ref{eq:qqnlo}) yields the infrared
singular behaviour of the one-loop amplitude,
\vspace{-0.5cm}
\begin{equation}
\includegraphics[width=2.5cm]{m210.ps} 
= 
\Ione\includegraphics[width=2.5cm]{m200.ps} 
+{\rm finite}.
\end{equation}
\vspace{0.5cm}
\noindent
According to Catani~\cite{Catani:1998bh}, the infrared behaviour of the square of the
one-loop
contribution is given by,
\vspace{-0.5cm}
\begin{equation}
\includegraphics[width=2.5cm]{m211.ps}=
2\Re \Biggl[
-\frac{1}{2}\left| \Ione \right | ^2 \includegraphics[width=2.5cm]{m200.ps}
+\Ione
\includegraphics[width=2.5cm]{m201.ps}  
\Biggr ] + {\rm
finite},
\label{eq:polesa}
\end{equation}
\vspace{0.5cm}
where $\Ione^\dag$ is obtained from Eq.~(\ref{eq:I1}) by reversing the sign of
$+i0$.  The two-loop contribution can be decomposed as,
\vspace{-0.5cm}
\begin{eqnarray}
\lefteqn{\includegraphics[width=2.5cm]{m220.ps} + 
\includegraphics[width=2.5cm]{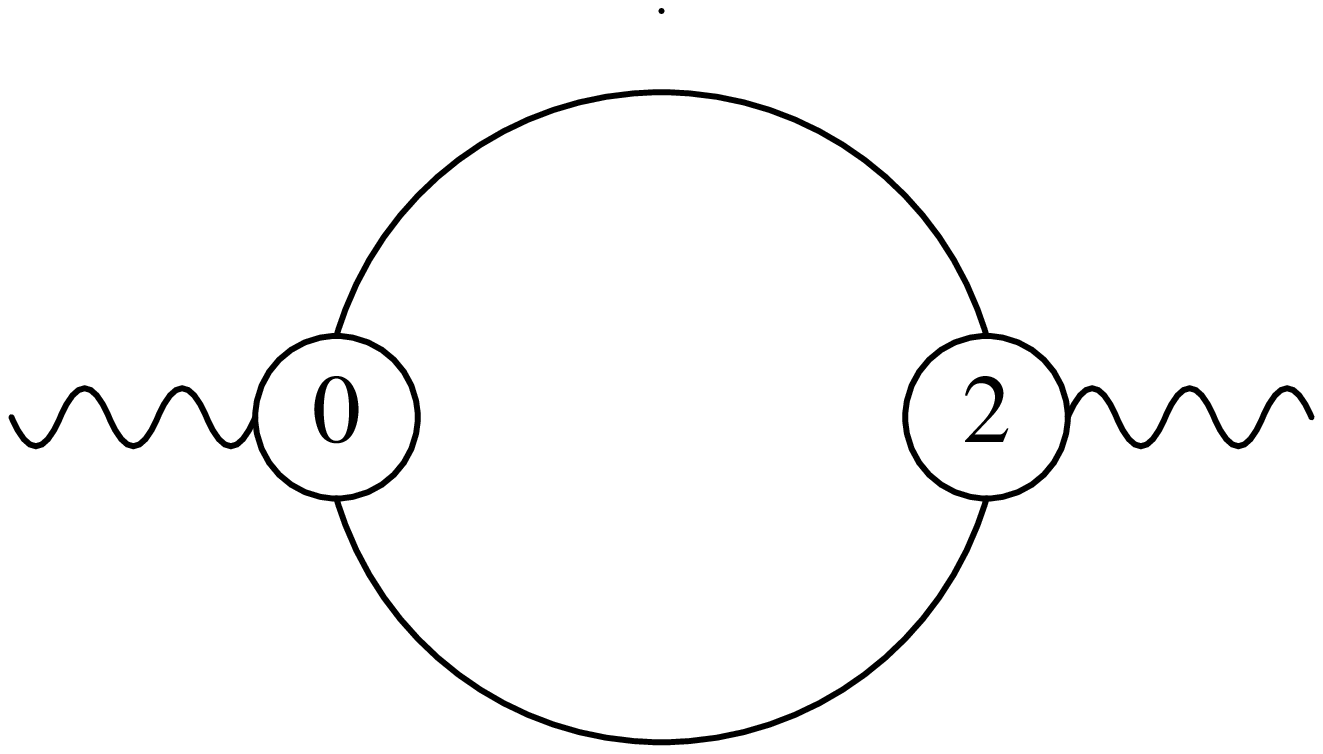}}\nonumber \\
 &=& 2\Re \Biggl [
-\frac{1}{2} \left(\Ione\right)^2 
\includegraphics[width=2.5cm]{m200.ps}  
+ \Ione \includegraphics[width=2.5cm]{m210.ps}
\nonumber \\
&&
-\frac{\beta_0}{\e} 
\Ione \includegraphics[width=2.5cm]{m200.ps}
+ e^{-\epsilon \gamma } \frac{ \Gamma(1-2\epsilon)}{\Gamma(1-\epsilon)} 
\left(\frac{\beta_0}{\epsilon} + K\right)\Ionep
\includegraphics[width=2.5cm]{m200.ps}\nonumber \\
&&
+\Htwo \includegraphics[width=2.5cm]{m200.ps} \Biggr ] + {\rm finite},
\label{eq:polesb}
\end{eqnarray}
with 
\begin{equation}
\beta_0 = \frac{11 \CA - 4 T_R \NF}{6},\qquad
K = \left( \frac{67}{18} - \frac{\pi^2}{6} \right) \CA - 
\frac{10}{9} T_R \NF.
\end{equation}
We note that the first line of the RHS of Eq.~(\ref{eq:polesb}) is
 proportional to
$\CF^2$ and produces fourth order poles in $\e$, 
while the second line contains terms like $\CF\CA$ and $\CF\TF\NF$ and generates 
third order poles in $\e$.
The last term of Eq.~(\ref{eq:polesb}) involving $\Htwo$
produces only a single pole in $\epsilon$ and is given by 
\begin{equation}
\label{eq:htwo}
\bra{\cm^{(0)}}\Htwo\ket{\cm^{(0)}} 
=\frac{e^{\epsilon \gamma}}{4\,\epsilon\,\Gamma(1-\epsilon)} H^{(2)} 
\braket{\cm^{(0)}}{\cm^{(0)}} \;,  
\end{equation}
where the constant $H^{(2)}$ is renormalization-scheme-dependent.
As with the single pole parts of $\Ione$,
the process-dependent
$H^{(2)}$ can be constructed by counting the number of
radiating partons present in the event.
In our case, there is only a 
 quark--antiquark pair present in the final
state, so that 
\begin{equation}
H^{(2)} =  2H^{(2)}_{q},
\end{equation}
where in the \MSbar\ scheme
\begin{equation}
H^{(2)}_q =       
         C_F^2  \left(
          - \frac{3}{8}
          - 6 \zeta_3
          + \frac{\pi^2}{2}
          \right)
       + C_F C_A  \left(
            \frac{245}{216}
          + \frac{13}{2} \zeta_3
          - \frac{23\pi^2}{48}
          \right) 
       + C_F T_R N_F  \left(
          - \frac{25}{54}
          + \frac{\pi^2}{12}
          \right)\;.
\end{equation}
The singularities present in Eqs.~(\ref{eq:polesa}) and (\ref{eq:polesb}) must
be cancelled against the infrared poles present in the one-loop three parton
and tree-level four parton contributions.

\vspace{0.2cm}

\noindent
{\it The one-loop three parton contributions}

\vspace{0.15cm}

\noindent
The renormalized $\gamma^* \to q\bar q g$
amplitude can be written as
\begin{equation}
|\cmqqg\rangle = \sqrt{4\pi\alpha}e_q \sqrt{\asmuopi} \left[
|\cmqqg^{(0)}\rangle 
+ \asmuopi |\cmqqg^{(1)}\rangle 
+ {\cal O}(\alpha_s^2) \right] \;,
\end{equation}
where the $|{\cal M}_{q\bar qg}^{(i)}\rangle$ are the $i$-loop contributions to the 
renormalized amplitude.   

The squared three-parton amplitude, summed over spins, 
colours and quark flavours, 
is denoted by
\begin{equation}
\langle\cmqqg|\cmqqg\rangle = \sum |\cm (\gamma^* \to q\bar qg)|^2 
= {\cal A}_{q\bar qg}\; .
\end{equation}
The perturbative expansion of ${\cal A}_{q\bar qg}$ at renormalization scale 
$\mu^2 = \sab$ reads,
\begin{equation}
{\cal A}_{q\bar qg} = 4\pi\alpha\sum_q N e_q^2 \Bigg[ 
+\asmuopi \caqqg{2} +\asmuopi^2 \caqqg{4} + {\cal O}(\alpha_s^3)\Bigg] \;,
\end{equation}
where,\vspace{-0.5cm}
\begin{eqnarray}
\label{eq:Aqqg2}
\caqqg{2} &=& \langle\cmqqg^{(0)}|\cmqqg^{(0)}\rangle 
=\includegraphics[width=2.5cm]{m300.ps},\\
&& \nonumber \\
\label{eq:Aqqg4}
\caqqg{4} &=& 
\langle\cmqqg^{(0)}|\cmqqg^{(1)}\rangle +
\langle\cmqqg^{(1)}|\cmqqg^{(0)}\rangle \nonumber \\
&=& 
\includegraphics[width=2.5cm]{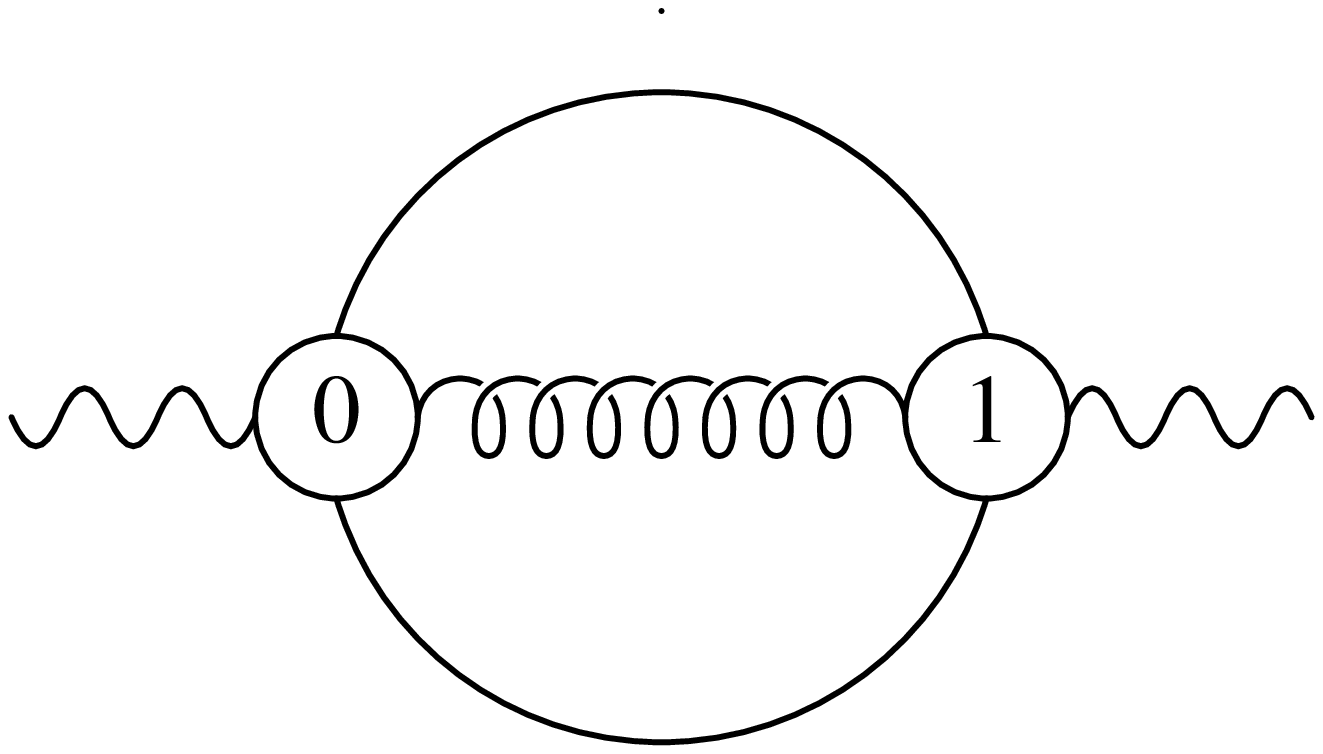}
+\includegraphics[width=2.5cm]{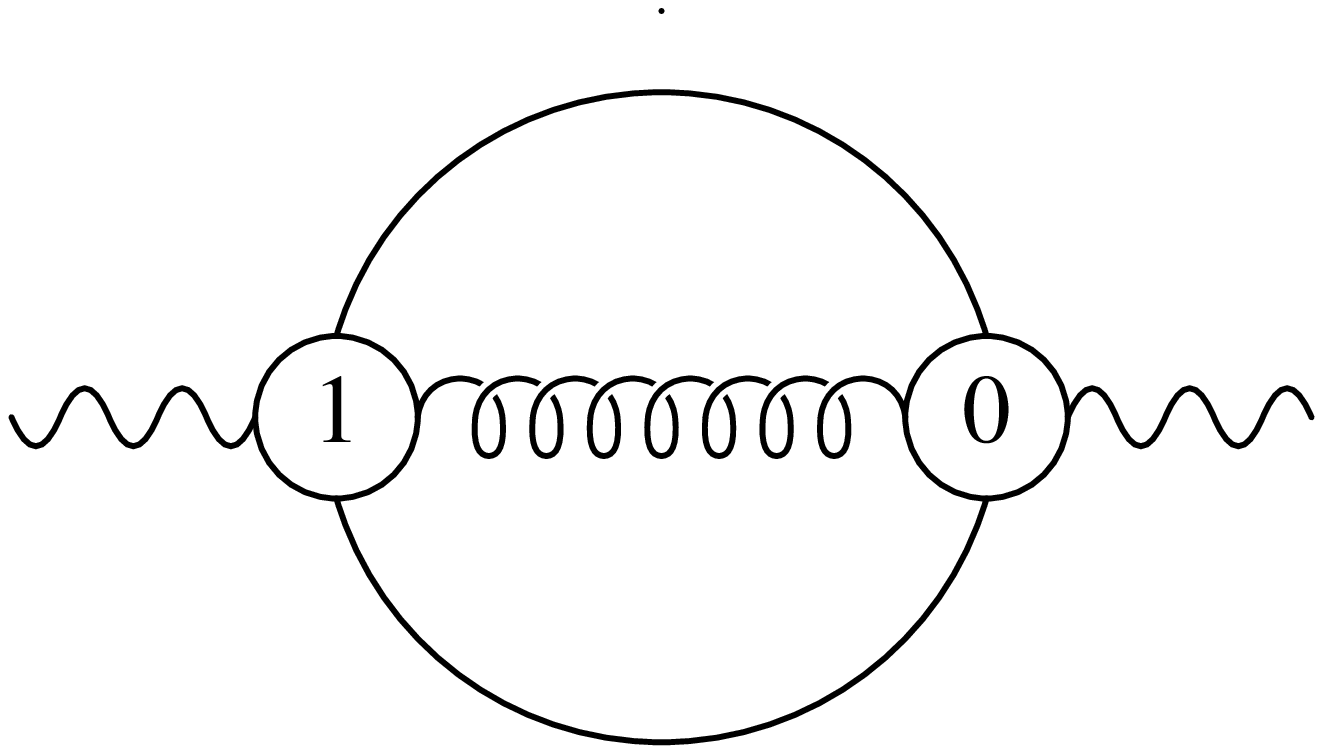}
-\frac{\beta_0}{\e} \includegraphics[width=2.5cm]{m300.ps}
\end{eqnarray}
\vspace{0.5cm}

\noindent 
$\caqqg{2}$ is proportional to $\CF$, while $\caqqg{4}$ contains $\CF^2$, 
$\CF\CA$ and $\CF T_R N_F$ terms~\cite{Ellis:1981wv,Fabricius:1981sx}.

The virtual unresolved contributions to two-jet final states at this order arise
from the one-loop corrections to $\gamma^*\to q \bar q g$ in the  limit where the
gluon becomes either collinear or soft. In both these  limits, one observes a
factorization of the {\it renormalized}  matrix element. In the collinear limit,
the one-loop $n+1$ parton  amplitude factorises as a one-loop $n$ parton amplitude
multiplied by a tree-level splitting function plus a tree-level $n$ parton
amplitude multiplied by a one-loop splitting 
function~\cite{Bern:1994zx,Bern:1998sc,Kosower:1999xi,Bern:1999ry,Kosower:1999rx}. 
In the soft limit, one observes the factorization  of the one-loop $n+1$ parton
amplitude into the one-loop $n$ parton amplitude  times the leading order soft
factor, plus the  $n$ parton tree level amplitude multiplied with the
soft gluon factor  at the one-loop
level~\cite{Bern:1998sc,Bern:1999ry,Catani:2000pi}. This one-loop correction to the
soft gluon factor contains only the colour factor $\CA$. In the infrared limit, we
therefore obtain the usual $\Ione$ factor  multiplying the one-loop two-parton
amplitude plus a new contribution $\dIone$ multiplying the tree-level contribution,
\vspace{-0.5cm} 
\begin{eqnarray} 
\includegraphics[width=2.5cm]{m310.ps} 
&\longrightarrow&  
\includegraphics[width=2.5cm]{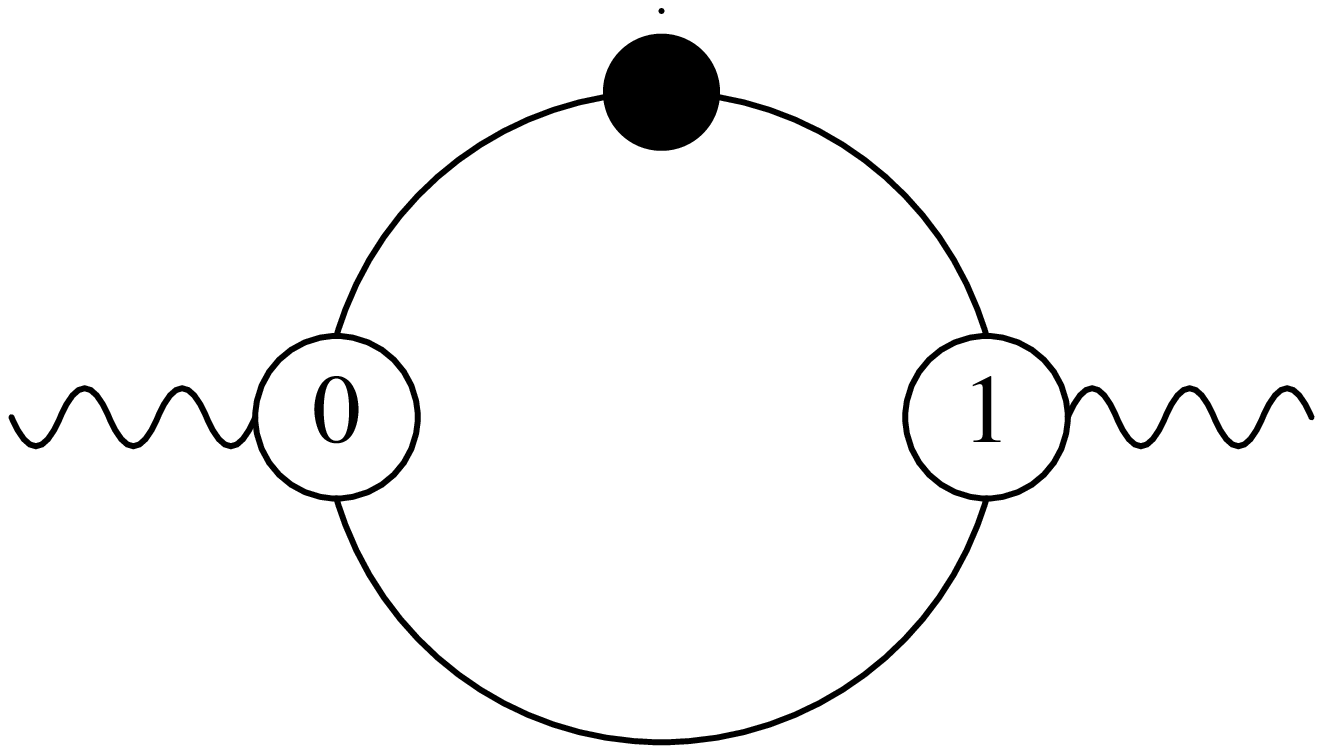} +
\includegraphics[width=2.5cm]{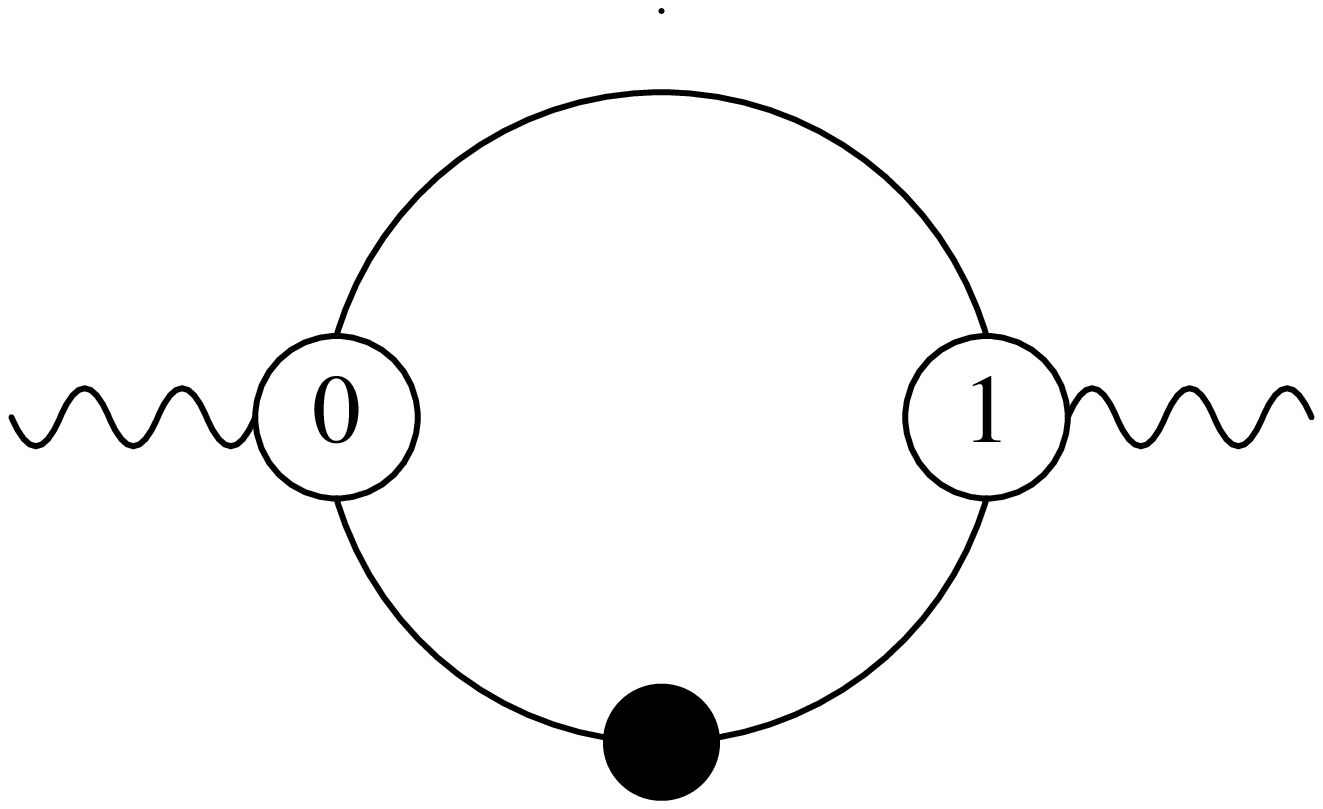} \nonumber \\ &+&
\includegraphics[width=2.5cm]{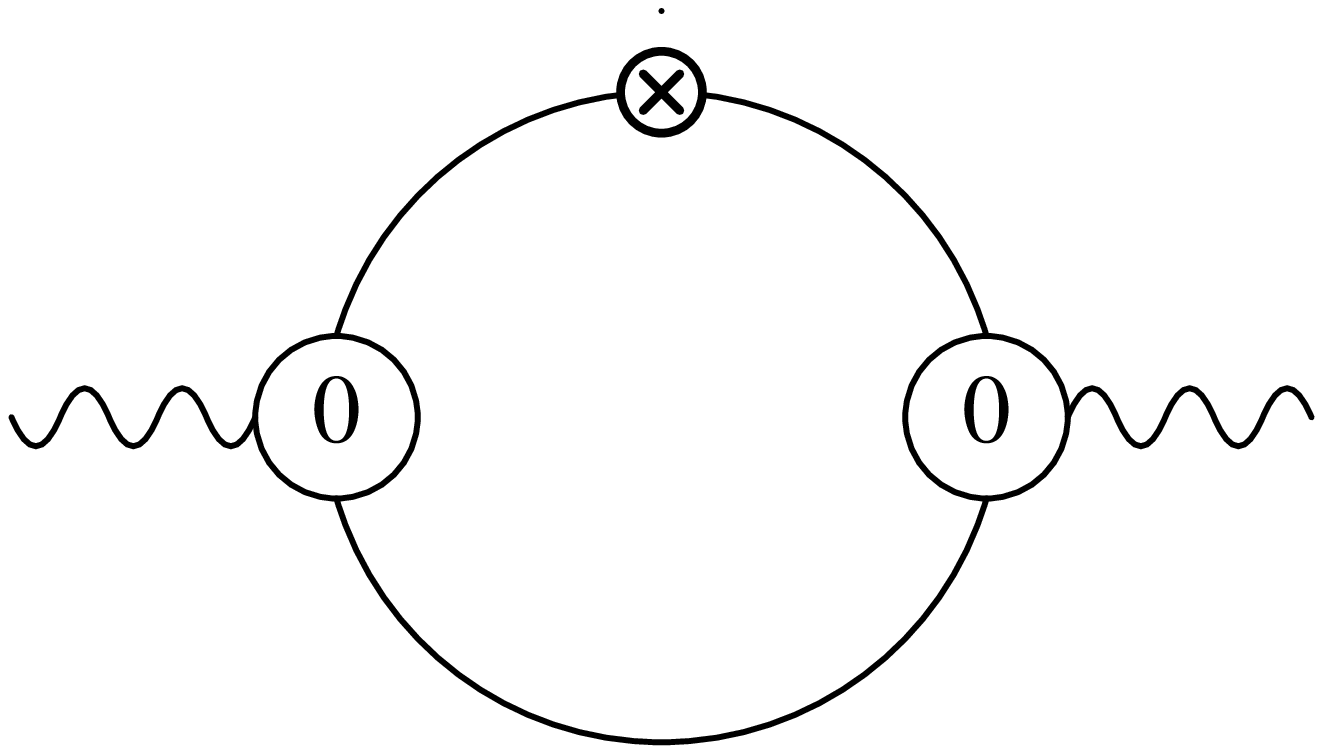} +
\includegraphics[width=2.5cm]{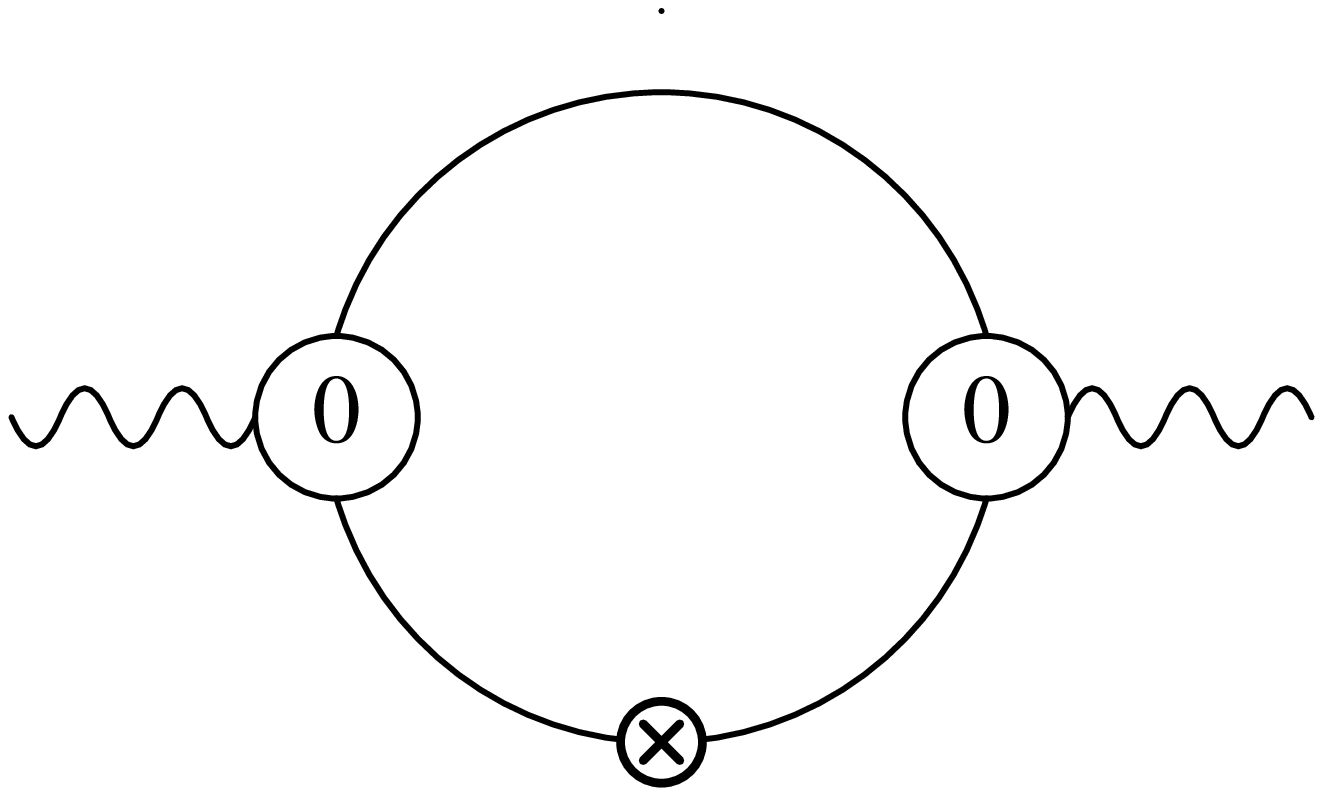} +{\rm finite}. 
\end{eqnarray}
\vspace{0.5cm}

\noindent
The first term is proportional to $\CF^2$, while 
$\dIone$ contains terms proportional to $\CF$ multiplied 
by double poles in $\e$
and terms proportional to $\CA$.  Altogether, the infrared structure from the 
one-loop three parton contribution is given by,
\vspace{-0.5cm}
\begin{eqnarray}
\label{eq:polesc}
\lefteqn{\includegraphics[width=2.5cm]{m310.ps}
+\includegraphics[width=2.5cm]{m301.ps}
-\frac{\beta_0}{\e} \includegraphics[width=2.5cm]{m300.ps}}\nonumber \\
&=& 
2\Re \Biggl[
-2 \Ione \includegraphics[width=2.5cm]{m210.ps} 
-2 \dIone \includegraphics[width=2.5cm]{m200.ps} \nonumber \\
&& +\frac{\beta_0}{\e} \Ione \includegraphics[width=2.5cm]{m200.ps}\Biggr ]
+{\rm finite}.
\end{eqnarray}

\noindent
When Eq.~(\ref{eq:polesc}) is 
taken together with Eqs.~(\ref{eq:polesa}) and (\ref{eq:polesb}) we see
that the terms of the type 
\begin{displaymath}
\Ione \includegraphics[width=2.5cm]{m210.ps}
\end{displaymath} 
and
\begin{displaymath}
\frac{\beta_0}{\e} \Ione \includegraphics[width=2.5cm]{m200.ps}
\end{displaymath} 
precisely cancel.

\vspace{0.2cm}

\noindent
{\it The tree-level four parton contributions}

\vspace{0.15cm}

\noindent
The renormalized $\gamma^* \to q\bar q +2$~parton
amplitudes can be written as
\begin{eqnarray}
|\cmqqgg\rangle &=& \sqrt{4\pi\alpha}e_q  \asmuopi  \left[
|\cmqqgg^{(0)}\rangle 
+ {\cal O}(\alpha_s^2) \right] , \nonumber \\
|\cmqqqq\rangle &=& \sqrt{4\pi\alpha}e_q  \asmuopi  \left[
|\cmqqqq^{(0)}\rangle 
+ {\cal O}(\alpha_s^2) \right] , 
\end{eqnarray}
where the $|{\cal M}_{q\bar qpp}^{(i)}\rangle$ are the $i$-loop contributions to the 
renormalized amplitude for $p=g,~q$.   

The squared four-parton amplitudes, summed over spins, 
colours and quark flavours, 
are denoted by $\cmqqgg$ and $\cmqqqq$, which have the perturbative expansions,
\begin{equation}
{\cal A}_{q\bar qpp} = 4\pi\alpha\sum_q N e_q^2 \Bigg[ 
\asmuopi^2 {\cal A}_{q\bar qpp}^{(4)} + {\cal O}(\alpha_s^3)\Bigg] \;,
\end{equation}
where,\vspace{-0.5cm}
\begin{eqnarray}
\label{eq:Aqqgg4}
\caqqgg{4} &=& \langle\cmqqgg^{(0)}|\cmqqgg^{(0)}\rangle 
=\includegraphics[width=2.5cm]{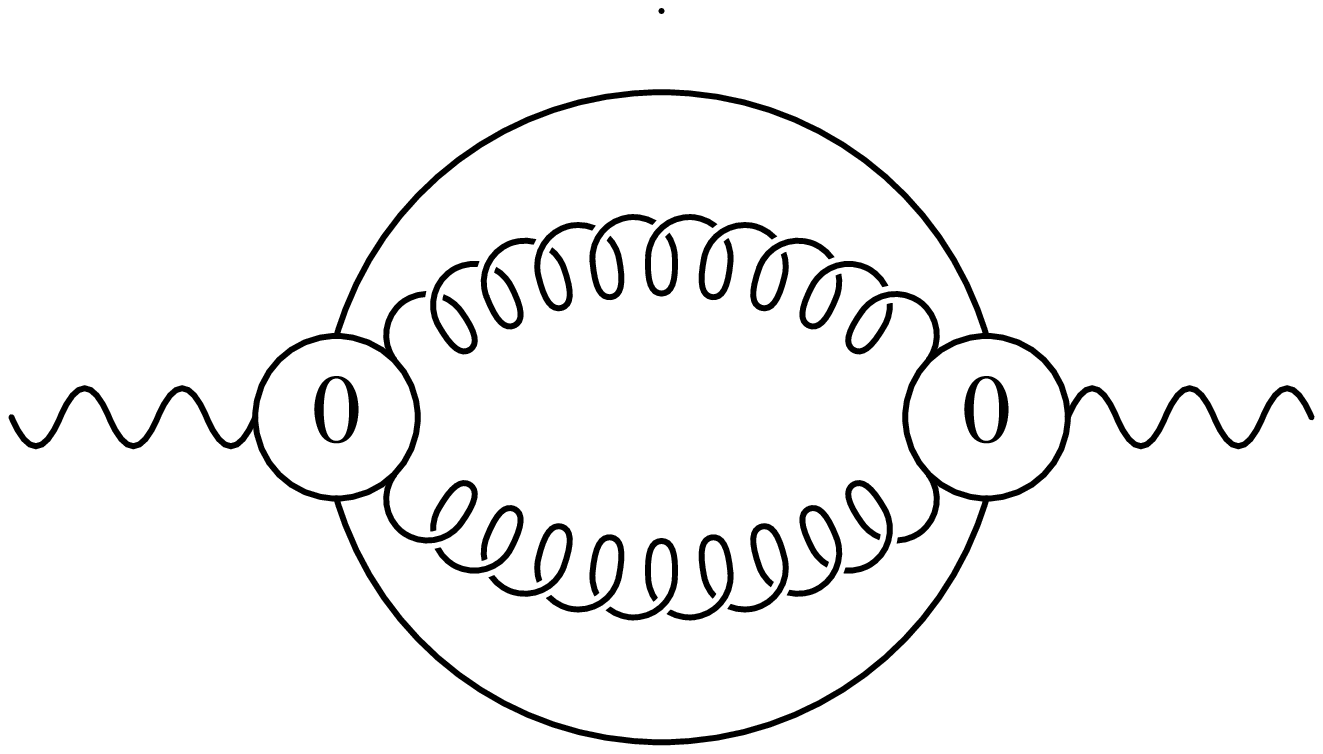},\\
&& \nonumber \\
\label{eq:Aqqqq4}
\caqqqq{4} &=& \langle\cmqqqq^{(0)}|\cmqqqq^{(0)}\rangle 
=\includegraphics[width=2.5cm]{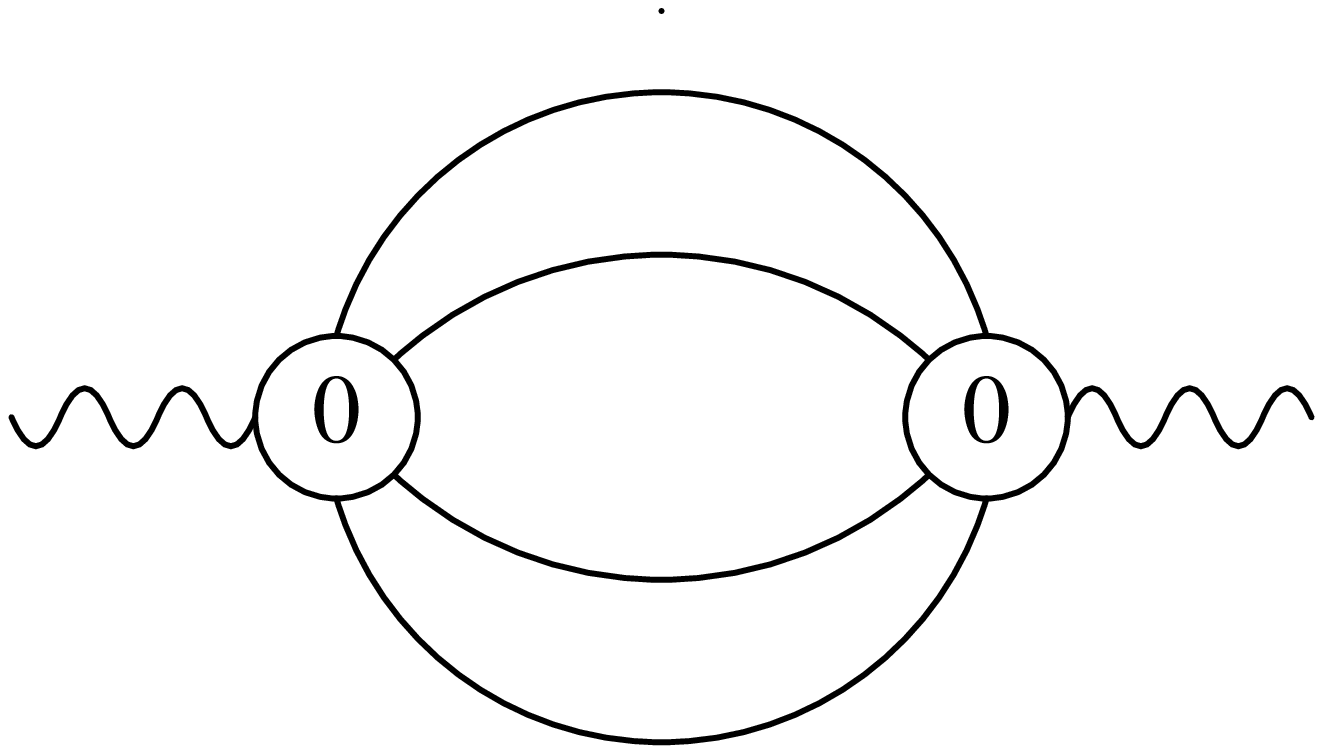}.
\end{eqnarray}
\vspace{0.5cm}

\noindent The infrared singularity structure of these matrix elements is well
documented in the double soft, triple collinear and independent double 
collinear
limits~\cite{Gehrmann-DeRidder:1998gf,Campbell:1998hg,Catani:1998nv,Catani:1999ss,DelDuca:1999ha}.  
At present, it is not clear how to efficiently isolate these singularities. Up
to now, the only calculation involving these elements  was carried out for the
case of  photon-plus-one-jet final  states in electron--positron
annihilation~\cite{Gehrmann-DeRidder:1998gf,Gehrmann-DeRidder:1997wx}, which
was performed in the  hybrid subtraction method, which is an extension of the 
phase space slicing procedure of~\cite{Giele:1992vf}.  These double unresolved
factors have also been used to compute the general form of 
logarithmically-enhanced contributions (up to next-to-next-to-leading
logarithmic accuracy) of the transverse momentum distributions of high-mass
systems in hadronic collisions~\cite{deFlorian:2000pr,deFlorian:2001zd}.

\vspace{0.2cm}

\noindent
{\it Summary}

\vspace{0.15cm}

\noindent
The infrared singular structure of the two-loop amplitudes is well
documented.   Concerning the singular structure of the 
one-loop three parton amplitudes, it
seems straightforward to evaluate and to construct counter terms.  There is a
precise cancellation with some of the singular terms in the one- and two-loop
two parton contributions.   However, more work is required to disentangle the
singular structure of the tree-level  four-parton contributions.

\subsection{The high energy limit of qcd at two loops}
\label{sec:ho;highenergy}

In this section we give a summary of the present status of
the analytic structure of QCD amplitudes in the limit of forward
and backward scattering. The gluon and quark Regge trajectories are
evaluated at two loop accuracy.

\subsubsection{Forward Scattering}

The scattering of two particles, with momenta $p_1 p_2\to p_3 p_4$
in their centre-of-mass frame, is parametrised by the invariants
$s=(p_1+p_2)^2$, $t=(p_2-p_3)^2$ and $u=(p_2-p_4)^2$. Since 
$t = - s (1-\cos\theta)/2$, where $\theta$ is
the scattering angle, the kinematic region where
the squared centre-of-mass energy is much greater than
the momentum transfer, $s\gg|t|$, defines the forward scattering.
It has been known for a long time that in the $s\gg|t|$ limit
the scattering amplitudes of a gauge
theory take a simple analytic form. In fact, in that limit a generic
scattering process is dominated by the exchange of the particle of
highest spin in the $t$ channel. That is a gluon in QCD
(which we take as representative of $SU(N)$
Yang-Mills theories), a photon in QED, and a graviton in quantum gravity.

In the scattering of two partons in QCD in the limit $s\gg|t|$,
the processes which yield the largest contribution are
quark-quark, quark-gluon and gluon-gluon scattering. The other
processes, like $q\ \bar{q}\to g\ g$, which do not feature gluon
exchange in the $t$ channel, are subleading in this limit.
The leading processes all have the same analytic form.
In fact, the tree amplitude for parton-parton scattering 
$i_a(p_2)\, j_b(p_1)\to i_{a'}(p_3)\,j_{b'}(p_4)$, with $i, j$ either a
quark or a gluon, may be written as~\cite{Kuraev:1976ge}
\begin{equation}
\cM^{(0) aa'bb'}_{ij\to ij} = 2  s
\left[\gs\, (T^c_r)_{aa'}\, C^{i(0)}(p_2,p_3) \right]
{1\over t} \left[\gs\, (T^c_r)_{bb'}\, C^{j(0)}(p_1,p_4) 
\right]\, ,\label{elas}
\end{equation}
where $a, a', b, b'$ label the parton colours, and $r$ represents either
the fundamental $(F)$ or the adjoint $(G)$ representations of $SU(N)$, with
$(T^c_G)_{ab} = i f^{acb}$ and ${\rm tr}(T^c_F T^d_F) = \delta^{cd}/2$.
The coefficient function $C^{i(0)}$ is process dependent; in helicity
space it just contributes a phase 
factor~\cite{DelDuca:1995zy,DelDuca:1999ha}: its square is 1.
By removing the colour factor and replacing the
strong with the electromagnetic coupling, \eqn{elas} holds also for
the forward scattering of two charged leptons in QED with exchange 
of a photon in the $t$ channel.

The amplitude (\ref{elas}) constitutes the leading term of the BFKL 
theory~\cite{Kuraev:1976ge,Kuraev:1977fs,Balitsky:1978ic}, which models
strong-interaction processes with two large and disparate scales,
by resumming the radiative corrections to parton-parton
scattering. This is achieved to leading logarithmic (LL) accuracy, in
$\ln(s/|t|)$, through the BFKL equation, {\em i.e.}
an integral equation obtained by computing the one-loop LL
corrections to the gluon exchange in the $t$ channel. These corrections are:
the emission of a gluon along the ladder~\cite{Lipatov:1976zz}, and 
the one-loop {\em gluon Regge trajectory} (see \eqn{exp1loop}).
To see how the latter comes about, we generalise
\eqn{elas} to include the virtual radiative corrections
and write the high-energy amplitude for parton-parton 
scattering, with exchange of a colour octet in the $t$
channel, as~\cite{Fadin:1993wh,Lipatov:1989bs}
\begin{eqnarray}
\lefteqn{\cM^{aa'bb'}_{ij\to ij} } \nonumber\\ 
&=& s
\left[\gs\, (T^c_{r_A})_{aa'}\, C^{i}(p_2,p_3) \right]
{1\over t} \left[\left({-s\over -t}\right)^{\alpha(t)} +
\left({s\over -t}\right)^{\alpha(t)}  \right]
\left[\gs\, (T^c_{r_A})_{bb'}\, C^{j}(p_1,p_4) \right]
\nonumber\\ &+& s
\left[\gs\, (T^c_{r_S})_{aa'}\, C^{i}(p_2,p_3) \right]
{1\over t} \left[\left({-s\over -t}\right)^{\alpha(t)} -
\left({s\over -t}\right)^{\alpha(t)}  \right] 
\left[\gs\, (T^c_{r_S})_{bb'}\, C^{j}(p_1,p_4) \right]\, ,\label{elasb}
\end{eqnarray}
with 
\begin{equation}
(T^c_{G_A})_{aa'} = i f^{aca'} \quad (T^c_{G_S})_{aa'} = d^{aca'}
\quad T^c_{F_A} = T^c_F \quad T^c_{F_S} = \sqrt{N_c^2-4\over N_c^2} T^c_F\, .
\end{equation}
The first (second) line of \eqn{elasb} corresponds to the exchange of a
colour octet in the $t$ channel, of negative (positive) signature under 
$s\leftrightarrow u$ exchange.
The function $\alpha(t)$ in \eqn{elasb} is the gluon Regge trajectory. 
It has the perturbative expansion
\begin{equation}
\alpha(t) = \tilde\gs^2(t) \alpha^{(1)} + 
\tgs^4(t) \alpha^{(2)} + \ord(\tgs^6)\,
,\label{alphb}
\end{equation}
%
while the coefficient function can be written as
\begin{equation}
C^{i} = C^{i (0)}(1 + \tgs^2(t) C^{i (1)} + \tgs^4(t) C^{i (2)}) + 
\ord(\tgs^6)\, .\label{fullv}
\end{equation}
In \eqns{alphb}{fullv}, we rescaled the coupling
\begin{equation}
\tgs^2(t) = \gs^2 \left({\mu^2\over -t}\right)^{\eps}
{1\over (4\pi)^{2-\epsilon}}\, {\Gamma(1+\epsilon)\,
\Gamma^2(1-\epsilon)\over \Gamma(1-2\epsilon)}\, ,\label{rescal}
\end{equation}
and used dimensional regularisation in $d=4-2\epsilon$ dimensions.
Note that in multiplying \eqn{elasb} by the tree amplitude, the second
line of \eqn{elasb} contributes only to quark-quark scattering,
since the colour factor $f^{ada'}$, occurring in any tree amplitude 
with an external gluon, acts as an $s$-channel projector.
Then we write the projection of the amplitude (\ref{elasb}) on the tree
amplitude as an expansion in $\tgs^2(t)$
\begin{equation}
\cM^{aa'bb'}_{ij\to ij} \cM^{(0) aa'bb'}_{ij\to ij}
= |\cM^{(0) aa'bb'}_{ij\to ij}|^2 \left( 1 +
\tgs^2(t)\ M^{(1) aa'bb'}_{ij\to ij} + \tgs^4(t) M^{(2) aa'bb'}_{ij\to ij}
+ \ord(\tgs^6) \right)\, ,\label{elasexpand}
\end{equation}
with $i, j = g, q$. 
\begin{figure}[t]
(a)
\begin{center}
~\includegraphics[width=2.5cm]{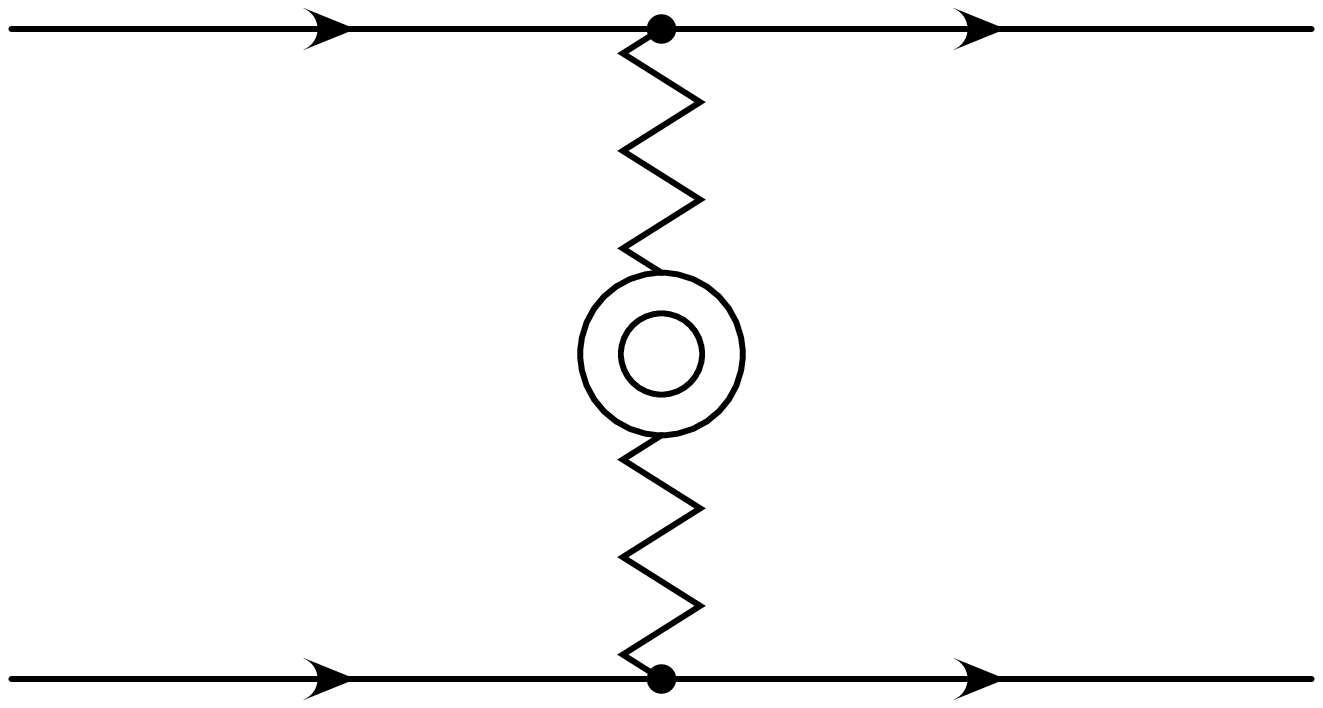}
\end{center}
(b)
\begin{center}
~\includegraphics[width=2.5cm]{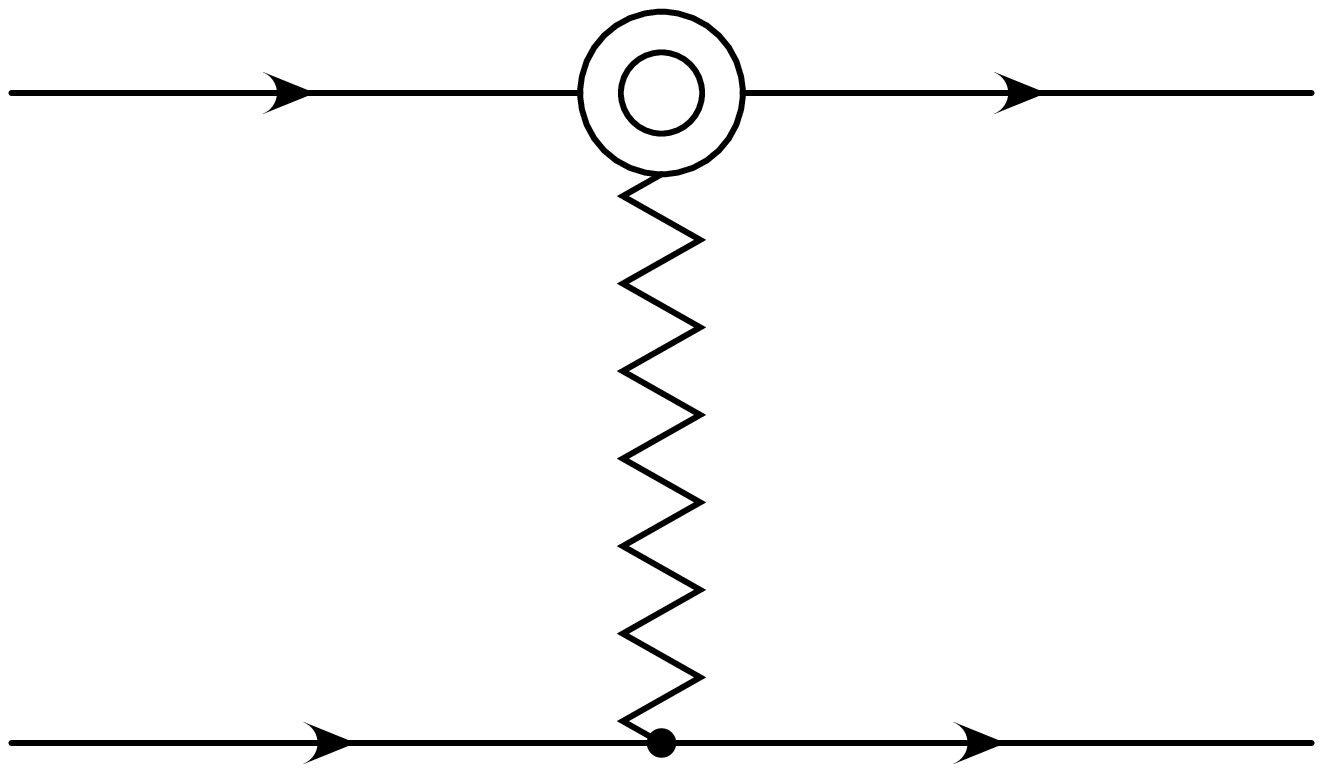}\hspace{2cm} 
~\includegraphics[width=2.5cm]{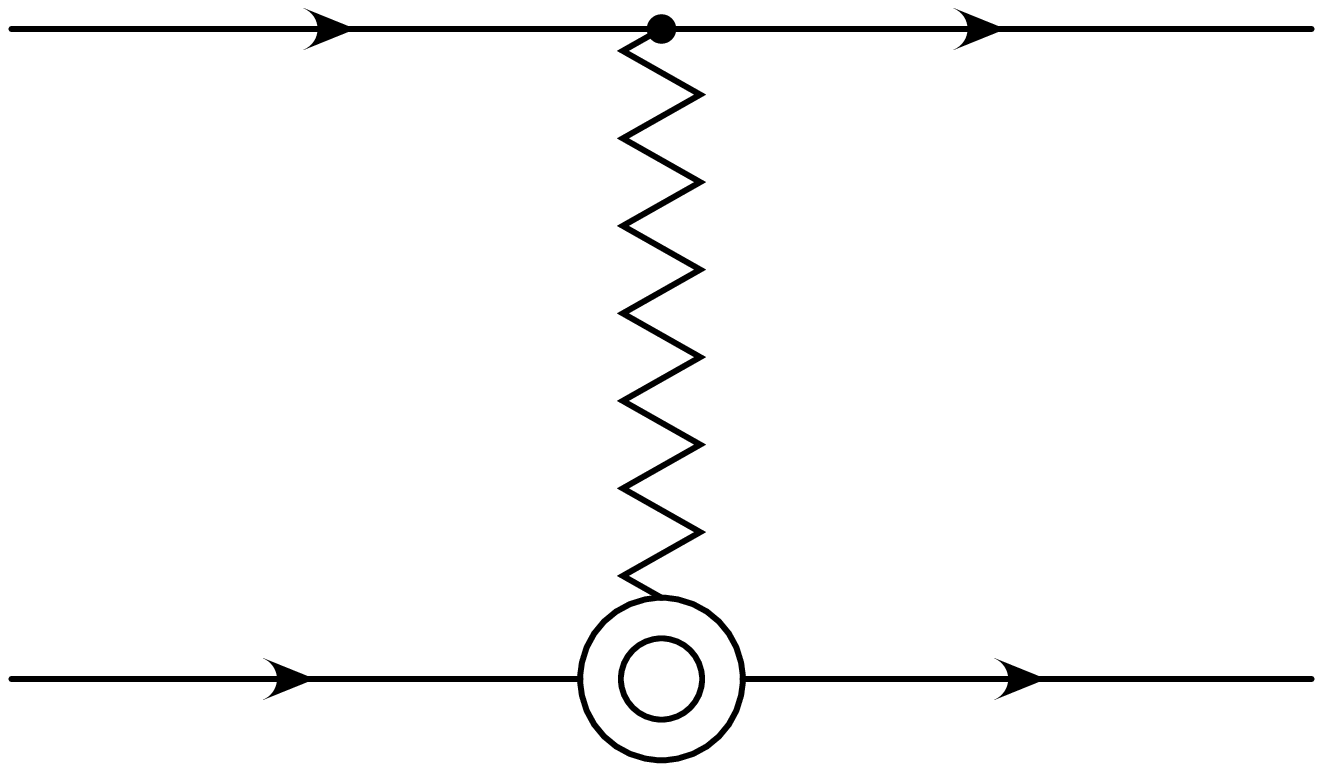} 
\end{center}
\caption{Schematic one-loop expansion of the factorised form for the 
high energy limit of the parton-parton scattering amplitude. 
The pairs of concentric circles represent the one-loop corrections to the 
coefficient function and Regge trajectory and the individual diagrams
represent  terms that contribute at (a) leading  and (b) next-to-leading
logarithmic order. }
\label{fig:oneloop}
\end{figure}
The one-loop coefficient of \eqn{elasexpand} is, 
\begin{equation}
M^{(1) aa'bb'}_{ij\to ij}
= \alpha^{(1)} \ln\left({s\over -t}\right) +\ C^{i(1)} + C^{j(1)}
- i{\pi\over 2} \left( 1 + K {N_c^2-4\over N_c^2} \right)
\alpha^{(1)}\, ,\label{exp1loop}
\end{equation}
where $K = 1$ for quark-quark scattering, and $K = 0$ in the other cases.
Schematically, this is illustrated in Fig.~\ref{fig:oneloop}.
In \eqn{exp1loop}
we used the usual prescription $\ln(-s) = \ln(s) - i\pi$, for $s > 0$.
The one-loop gluon Regge trajectory, $\alpha^{(1)} = 2\CA/\eps$, 
Fig.~\ref{fig:oneloop}(a), is independent of the type of parton 
undergoing the scattering process (it is {\em universal}).
Conversely, the one-loop coefficient 
functions, $C^{i (1)}$, Fig.~\ref{fig:oneloop}(b),
are process and IR-scheme dependent (see 
Ref.~\cite{DelDuca:2001gu} and references therein).
They can be used to construct the next-to-leading order (NLO)
impact factors, to be used in conjunction with the BFKL resummation
at next-to-leading-log (NLL) accuracy~\cite{Fadin:1998py}.
Note that \eqn{exp1loop} forms a system of three equations (given
by gluon-gluon, quark-quark and quark-gluon scattering)
and only two unknowns: the one-loop coefficients $C^{g(1)}$ and 
$C^{q(1)}$. Thus, we can use two equations to determine 
$C^{g(1)}$ and $C^{q(1)}$, and the third to show that high-energy 
factorisation holds to one-loop accuracy. Finally,
note that in \eqn{exp1loop} the contribution of the positive signature
gluon appears in the imaginary part, thus it can contribute only to
a next-to-next-to-leading order (NNLO) calculation.
\begin{figure}[t]
(a)
\begin{center}
~\includegraphics[width=2.5cm]{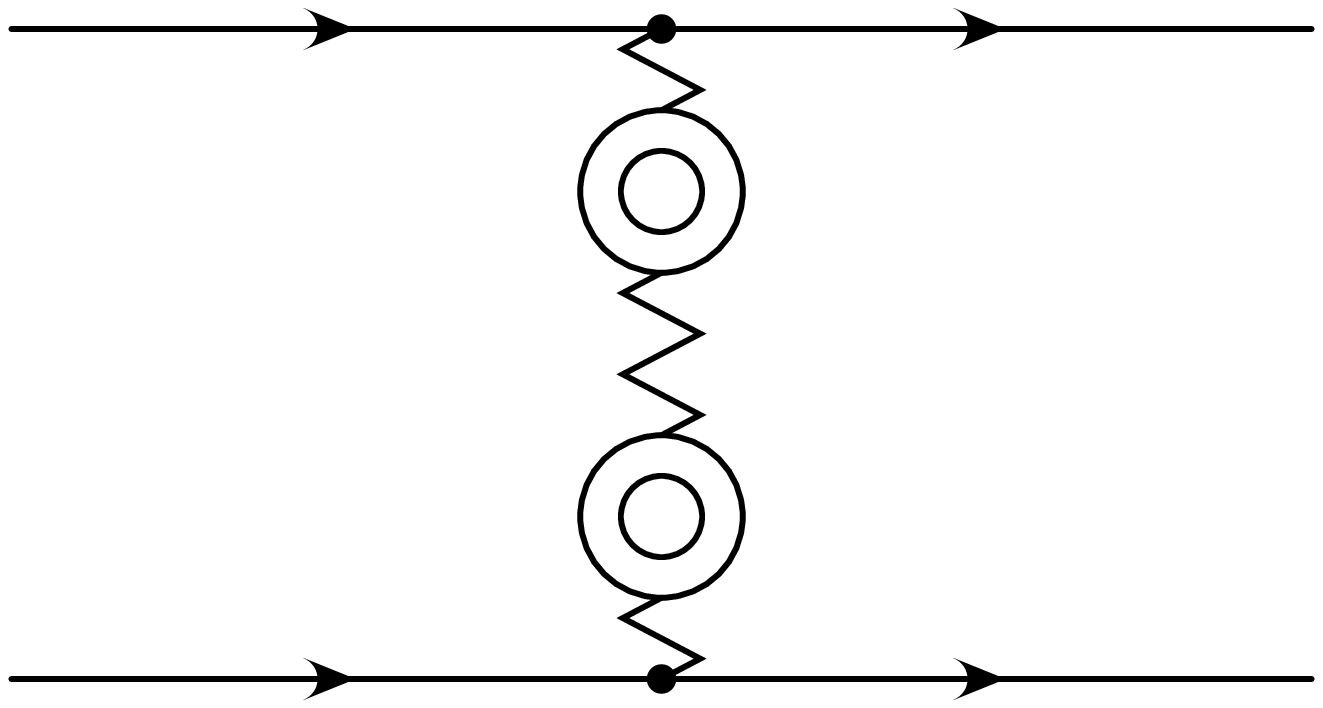}
\end{center}
(b)
\begin{center}
~\includegraphics[width=2.5cm]{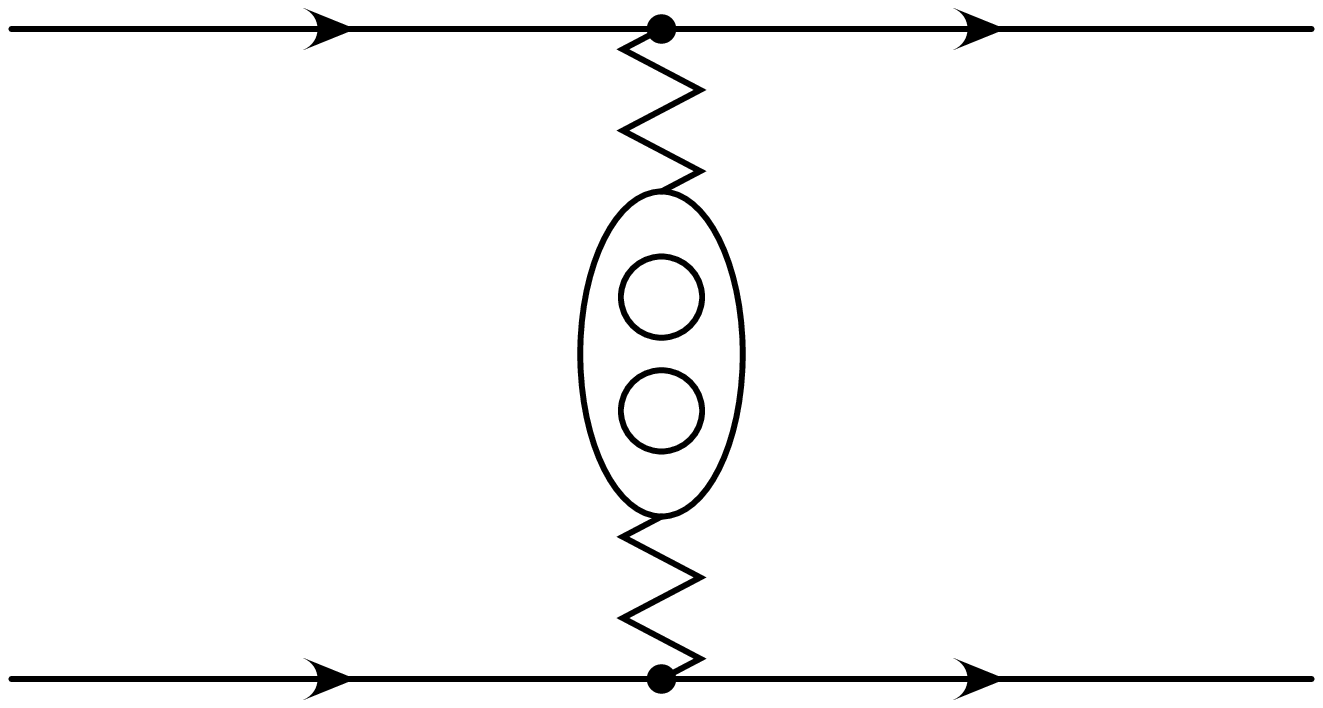}\hspace{1cm} 
~\includegraphics[width=2.5cm]{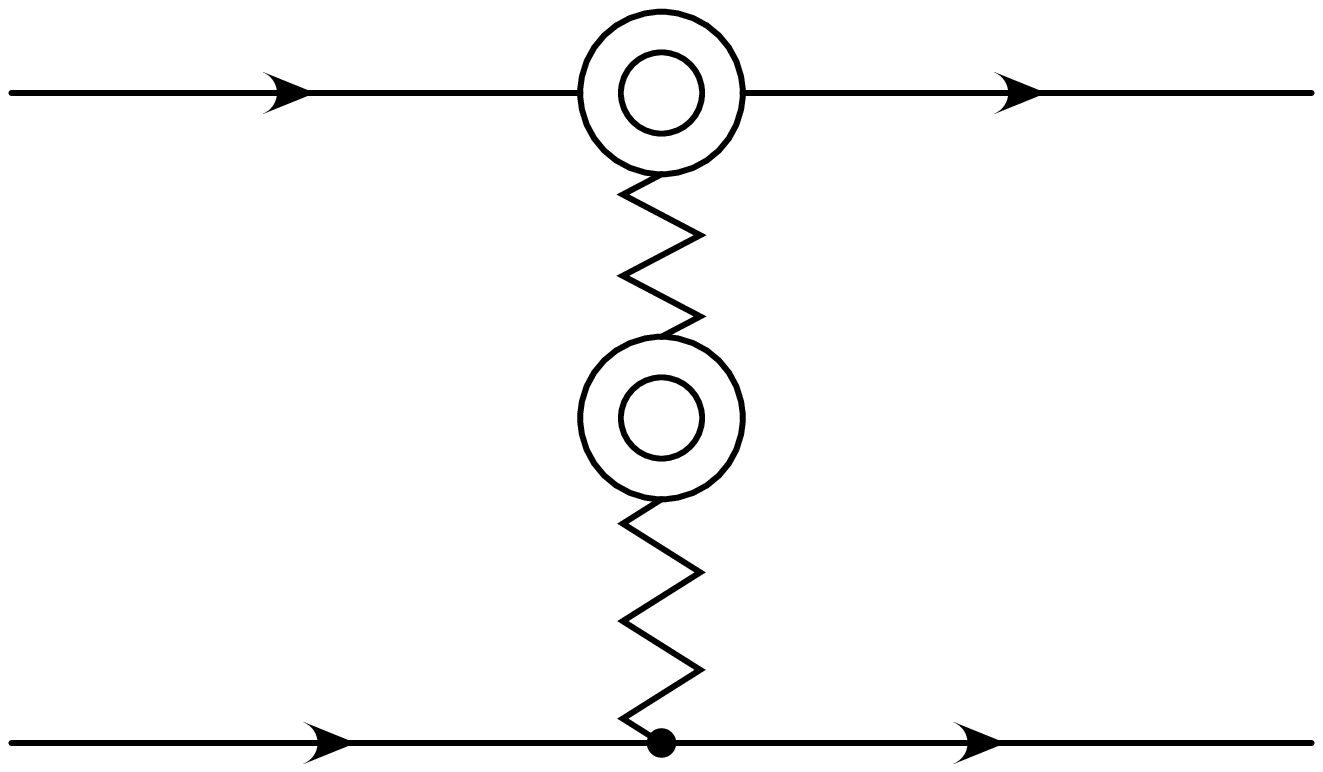}\hspace{1cm} 
~\includegraphics[width=2.5cm]{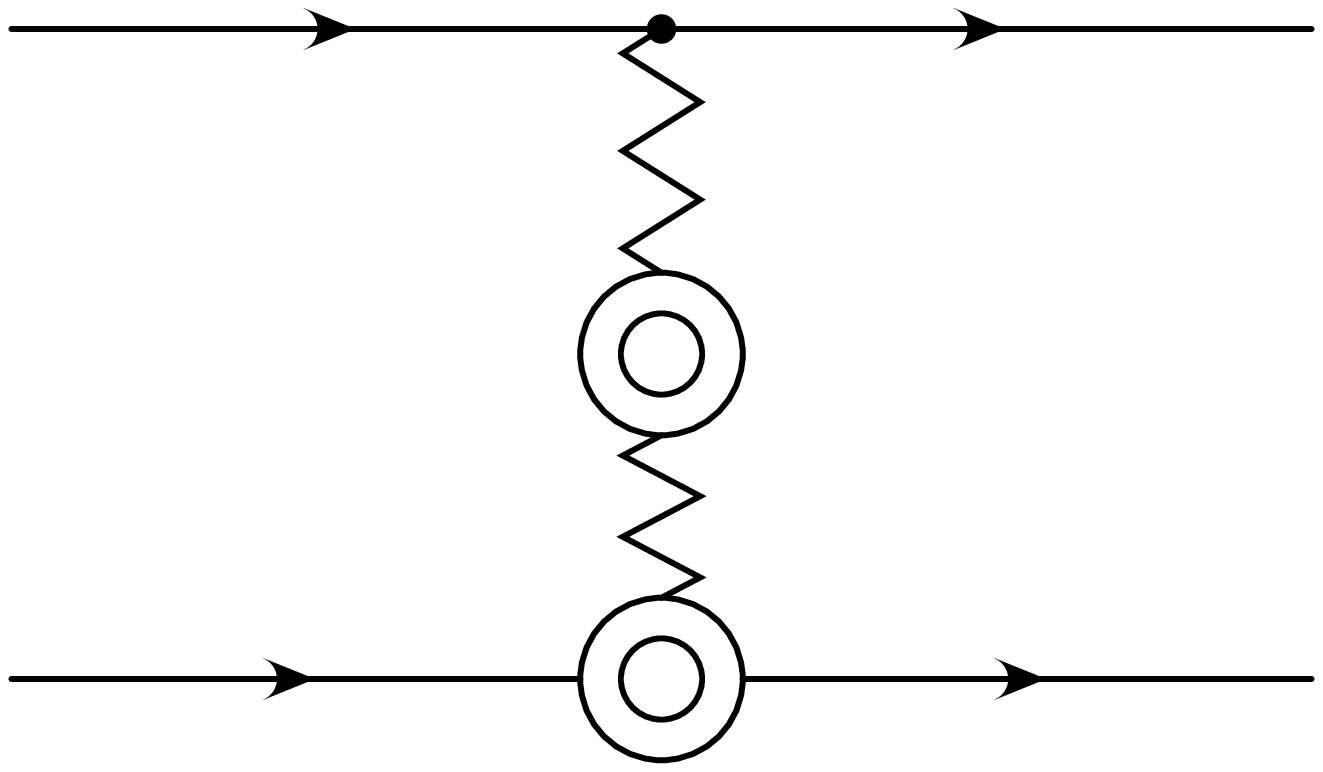} 
\end{center}
(c)
\begin{center}
~\includegraphics[width=2.5cm]{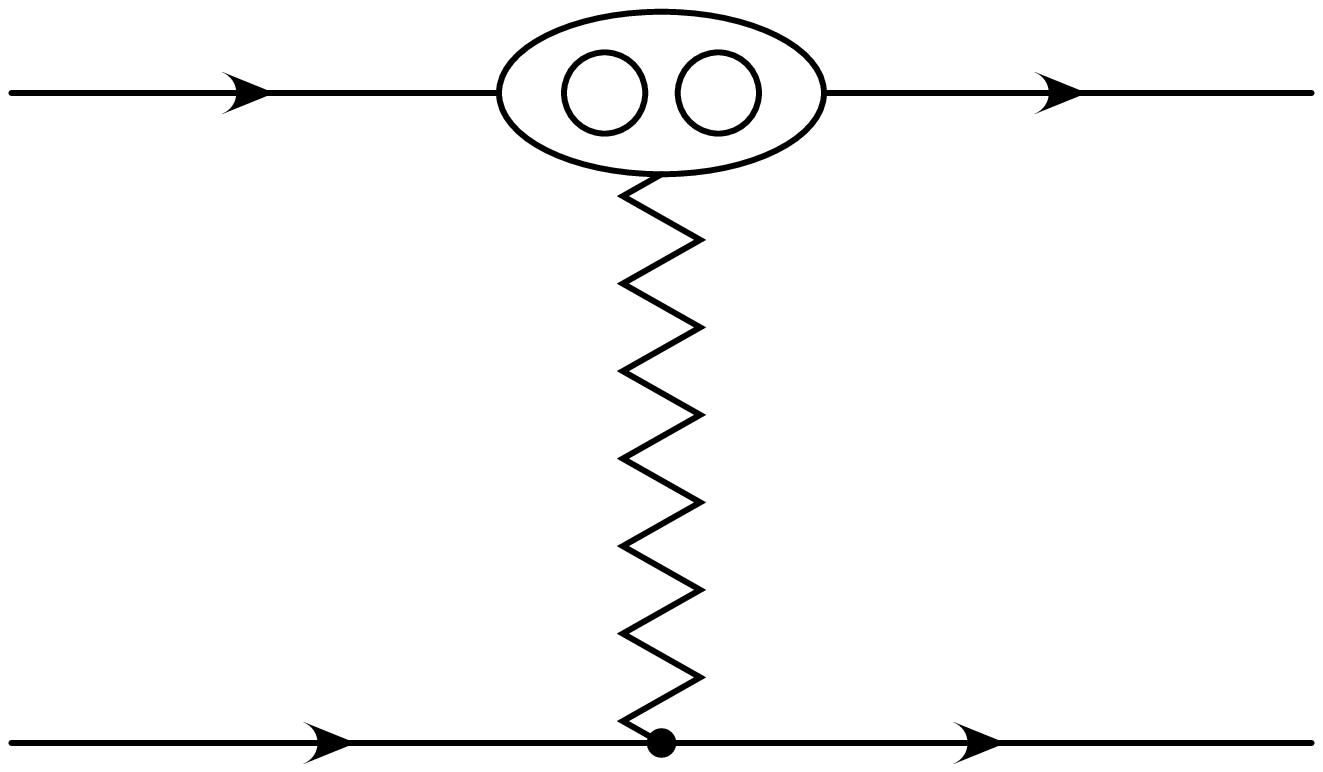}\hspace{1cm}  
~\includegraphics[width=2.5cm]{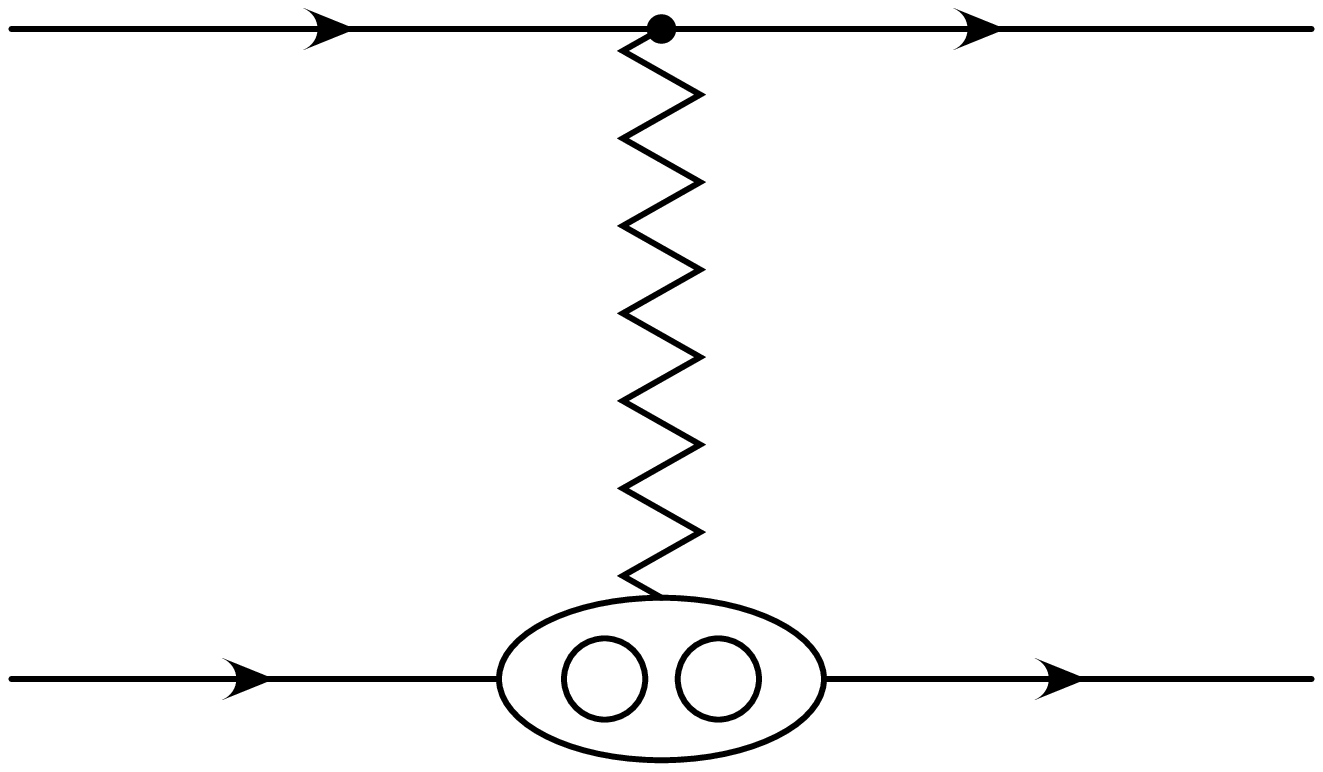}\hspace{1cm}  
~\includegraphics[width=2.5cm]{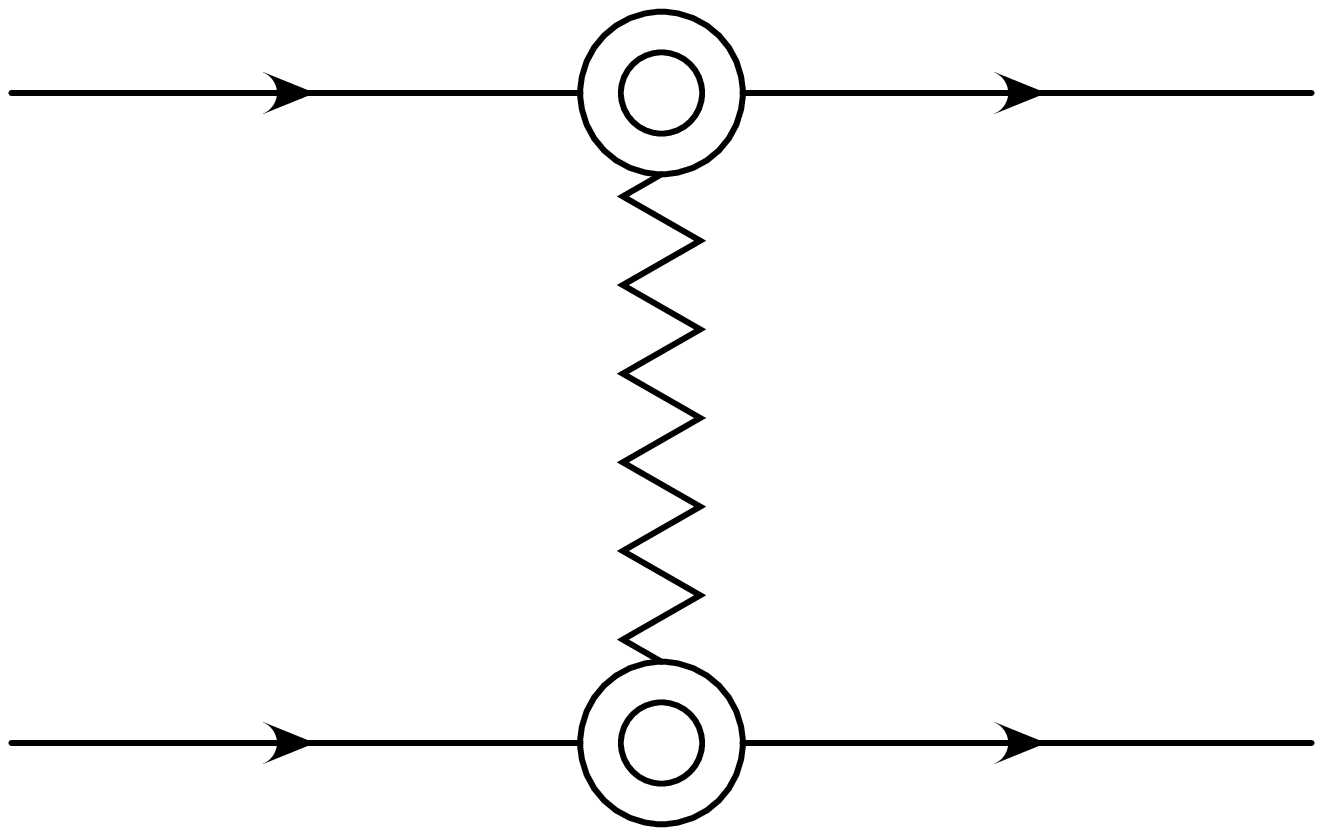} 
\end{center}
\caption{ Schematic two-loop expansion of the factorised form for the 
high energy limit of the parton-parton scattering amplitude. 
The combinations of ovals and circles represent the one-loop and 
two-loop corrections to the coefficient function and Regge trajectory and
the individual diagrams represent  terms that contribute at 
(a) leading, (b) next-to-leading and (c) next-to-next-to-leading
logarithmic order. }
\label{fig:twoloop}
\end{figure}
The two-loop coefficient of \eqn{elasexpand} is,
\begin{eqnarray}
M^{(2) aa'bb'}_{ij\to ij}
&=& {1\over 2} \left(\alpha^{(1)}\right)^2 \ln^2\left({s\over -t}\right) 
\nn\\  &+& \left[ \alpha^{(2)} + \left(C^{i(1)} + C^{j(1)}
\right) \alpha^{(1)} - 
i{\pi\over 2} \left( 1 + K {N_c^2-4\over N_c^2} \right)
\left(\alpha^{(1)}\right)^2
\right] \ln\left({s\over -t}\right) \nn\\ &+& 
\left[ C^{i(2)} + C^{j(2)} + 
C^{i(1)}\ C^{j(1)} - 
{\pi^2\over 4} \left( 1 + K {N_c^2-4\over N_c^2} \right)
\left(\alpha^{(1)}\right)^2 \right] \nn\\ 
&-& i{\pi\over 2} \left( 1 + K {N_c^2-4\over N_c^2} \right)
\left[ \alpha^{(2)} + \left(C^{i(1)} + 
C^{j(1)} \right) \alpha^{(1)} \right]
\, .\label{exp2loop}
\end{eqnarray}
Schematically, this is illustrated in Fig.~\ref{fig:twoloop}.
The first line of \eqn{exp2loop} is just the exponentiation of the one-loop
trajectory (Fig.~\ref{fig:twoloop}(a)). The second line of \eqn{exp2loop}
allows to determine $\alpha^{(2)}$, the two-loop gluon 
trajectory (the first diagram in Fig.~\ref{fig:twoloop}(b)), which
provides the virtual corrections to the NLL kernel of the BFKL equation. 
The two-loop gluon trajectory is universal, and has been computed in 
Ref.~\cite{Fadin:1995xg,Fadin:1996tb,Fadin:1996km,Blumlein:1998ib}.
Recently it was re-evaluated in a completely independent 
way~\cite{DelDuca:2001gu}, by taking the high energy limit of the two-loop
amplitudes for parton-parton 
scattering~\cite{Anastasiou:2001sv,Anastasiou:2000kg,
Anastasiou:2000ue,Glover:2001af}. 
The (unrenormalised) two-loop gluon Regge trajectory is
\begin{equation}
\alpha^{(2)} = \CA \left[ \beta_0 {2\over \eps^2} + K{2\over \eps} 
+ \CA \left({404\over 27} - 2\zeta_3\right) 
+ \NF \left(-{56\over 27}\right)\right]\, ,\label{eq:2loop}
\end{equation}
where
\begin{equation} 
\beta_0= {(11\CA-2\NF)\over 6}, \qquad\qquad K = \left({67\over 18} -
{\pi^2\over 6} \right) \CA - {5\over 9} \NF\, .\label{eq:beta}
\end{equation}
Note that at two loop accuracy, the gluon trajectory vanishes
in the Abelian limit, $\CA\to 0$. That is in agreement with our 
expectations: in the forward scattering of two charged leptons in QED, a log
of the type $\ln(s/t)$ occurs for the first time at the three loop
level. Thus we expect that only in $\alpha^{(3)}$ it occurs a term
which is independent of $\CA$.
However, note that in QED there is no analog of \eqn{elasb}, because
the photon does not reggeise~\cite{mandel}.

The third and fourth lines of \eqn{exp2loop} are respectively
the real (Fig.~\ref{fig:twoloop}(c)) and the imaginary parts of the constant 
term. Note that the gluon of
positive signature does not contribute to the BFKL resummation at LL and NLL
accuracy. In principle, the two-loop
coefficient functions $C^{i(2)}$ could be used to construct
the NNLO impact factors, if the BFKL resummation held to 
next-to-next-to-leading-log (NNLL) accuracy.
The real constant term of \eqn{exp2loop}
can be used to evaluate the two-loop coefficient functions $C^{g(2)}$
and $C^{q(2)}$. Like in \eqn{exp1loop}, we have three equations 
(given by the two-loop amplitudes for gluon-gluon, 
quark-quark and quark-gluon scattering~\cite{Anastasiou:2001sv,
Anastasiou:2000kg,Anastasiou:2000ue,Glover:2001af})
and only two unknowns. Thus we can
use two equations to evaluate $C^{g(2)}$ and $C^{q(2)}$, and the third
to check if high-energy factorisation holds at NNLO accuracy.
In fact, it was found that factorisation is 
violated~\cite{DelDuca:2001gu}. That, in turn, voids the evaluation 
of $C^{g(2)}$ and $C^{q(2)}$. The source of the violation is not clear 
at present. It might be due to the (yet unknown) exchange
of three or more reggeised gluons in \eqn{elasb}.

\subsubsection{Backward scattering}

In the scattering of two particles in their centre-of-mass frame,
$u = - s (1+\cos\theta)/2$, thus the
kinematic region where $s\gg |u|$ defines the backward scattering.
In this limit, we consider a scattering process with matter exchange, 
namely quark-gluon scattering, which proceeds via
the exchange of a quark in the crossed channel\footnote{By crossing symmetry,
this is equivalent to quark pair annihilation (or creation from) two
gluons at small angles.}. In this limit, the tree amplitude for quark-gluon 
scattering $q_a(p_2)\ g_b(p_1)\to q_{a'}(p_3)\ g_{b'}(p_4)$ can be written as
\begin{equation}
{\cM}^{(0)}_{qg\to gq} = -2i \left[\gs\, (T^b)_{a'i} C^{(0)}_{gq}(p_1,p_3) 
\right] \sqrt{s\over - u} \left[\gs\, (T^{b'})_{ia}\, C^{(0)}_{qg}(p_2,p_4) 
\right]\, ,\label{elasq}
\end{equation}
where $a, a'$ ($b, b'$) label the colour of the quarks (gluons). In
helicity space, the coefficient functions $C^{(0)}$ just contribute a
phase factor~\cite{Bogdan:2002sr}. 

Generalising \eqn{elasq} to include virtual radiative corrections,
it was shown that the quark reggeises as well.
In fact, in the limit $s\gg |u|$ the amplitude for quark-gluon
scattering with exchange in the $u$ channel of a colour triplet, with positive 
signature under $s\leftrightarrow t$ exchange, is~\cite{Fadin:1977jr}
\begin{equation}
{\cM}_{qg\to gq} = -i \left[\gs\, (T^b)_{a'i} C_{gq}(p_1,p_3) 
\right] \sqrt{s\over - u} \left[ \left({s\over - u}\right)^{\delta(u)}
+ \left({- s\over - u}\right)^{\delta(u)} \right]
\left[\gs\, (T^{b'})_{ia}\, C_{qg}(p_2,p_4) \right]\, .\label{allooppm}
\end{equation}
The function $\delta(u)$ is the {\em quark Regge trajectory}.
It has the perturbative expansion
\begin{equation}
\delta(u) = \tilde\gs^2(u) \delta^{(1)} + 
\tgs^4(u) \delta^{(2)} + \ord(\tgs^6)\, ,\label{delta}
\end{equation}
with $\tilde\gs^2$ given in \eqn{rescal}. 
The coefficient function $C$ has the same expansion as $C^i$ in
\eqn{fullv}, but for replacing $\tilde\gs^2(t)$ with $\tilde\gs^2(u)$,
and like in \eqn{elasexpand}, we can write the interference of \eqn{allooppm}
with the tree amplitude (\ref{elasq}) as an expansion in $\tilde\gs^2(u)$
\begin{equation}
\cM \cM^{(0) *} = |\cM^{(0)}|^2
\left( 1 + \tgs^2(u)\ M^{(1)} + \tgs^4(u) M^{(2)}
+ \ord(\tgs^6) \right)\, .\label{elasexpandpm}
\end{equation}
The one-loop coefficient of \eqn{elasexpandpm} is
\begin{equation}
M^{(1)}
= \delta^{(1)} \ln\left({s\over -u}\right) +\ 2 C^{(1)}
- i{\pi\over 2} \delta^{(1)}\, .\label{exp1loopplus}
\end{equation}
The one-loop quark trajectory is 
$\delta^{(1)} = 2\CF/\eps$~\cite{Fadin:1977jr}. Note that
$\delta^{(1)}$ does not vanish in the Abelian limit $\CF\to 1$.
In fact, in electron-photon scattering in QED,
the electron reggeises~\cite{gellmann,McCoy:1976fj}.
The coefficient function $C^{(1)}$
can be found in Ref.~\cite{Bogdan:2002sr,Fadin:2001dc}.
The two-loop coefficient of \eqn{elasexpandpm} is
\begin{eqnarray}
M^{(2)}
&=& {1\over 2} \left(\delta^{(1)}\right)^2 \ln^2\left({s\over -u}\right) 
+ \left[ \delta^{(2)} + 2 C^{(1)} \delta^{(1)} - 
i{\pi\over 2} \left(\delta^{(1)}\right)^2
\right] \ln\left({s\over -u}\right) \nn\\ &+& 
\left[ 2 C^{(2)} + \left( C^{(1)}\right)^2 -
{\pi^2\over 4} \left(\delta^{(1)}\right)^2  \right]
- i{\pi\over 2} \left[ \delta^{(2)} + 2 C^{(1)} \delta^{(1)} \right]
\, .\label{exp2loopplus}
\end{eqnarray}
The first term is just the exponentiation of the one-loop trajectory. 
Using the two-loop amplitude for quark-gluon 
scattering~\cite{Anastasiou:2001sv}, the second term of \eqn{exp2loopplus}
allows to determine $\delta^{(2)}$, the (unrenormalised)
two-loop quark Regge trajectory~\cite{Bogdan:2002sr}
\begin{equation}
\delta^{(2)} = \CF \left[ \beta_0 {2\over \eps^2} + K {2
\over \eps} + \CA \left({404\over 27} -2 \zeta_3\right)
+ \NF \left(-{56\over 27}\right) 
+ (\CF-\CA) \left(16\zeta_3\right) \right]\, ,\label{twolooptraj}
\end{equation}
with $\beta_0$ and $K$ as in \eqn{eq:beta}.
Note that \eqn{twolooptraj} has the remarkable feature that by
mapping $\CF\to\CA$, the two-loop gluon trajectory (\ref{eq:2loop})
is obtained. Since the forward and backward scattering are
seemingly unrelated, there is presently no understanding of why that
occurs.

The second line of \eqn{exp2loopplus} displays respectively the real
and the imaginary parts of the constant term.
If \eqn{allooppm} held at NNLO, by comparing the
two-loop amplitude for quark-gluon scattering~\cite{Anastasiou:2001sv} 
with the real part of the constant term of \eqn{exp2loopplus}
one could derive the two-loop coefficient function $C^{(2)}$.
However, \eqn{allooppm} is not expected to hold at NNLO, because
of possible unknown Regge cut contributions.

\subsubsection{Summary}

In conclusion, we have given a summary of the present status of the analytic
structure of QCD amplitudes in the limit of forward and backward scattering. We
have displayed the gluon and quark Regge trajectories at two loop accuracy.
They are strikingly similar: the gluon Regge trajectory can be obtained from
the quark  trajectory by mapping $\CF\to\CA$.

\subsection{CONCLUSION AND OUTLOOK}
\label{sec:ho;conc}

In this working group we addressed issues related to multiparticle states at
NLO and precision observables at NNLO. The motivation for this is very
straightforward: we need more accurate theoretical predictions in order to make
a sensible comparison with experiment, either to establish the existence of new
physics or to make more precise determinations of the parameters of the theory.
During the course of the workshop there
has been significant progress in this field.   In particular, work
initiated at the workshop have yielded the following identifiable results,
\begin{itemize}
\item 
The first Feynman diagrammatic calculation of a hexagon amplitude~\cite{Binoth:2001vm}. This was
performed   in the Yukawa model where all external legs are  massless scalars
attached to a massless fermion loop. This is a good model to study how the
enormous cancellations in the scalar integral reductions occur without the
complications of tensor structures in the numerator. The next step is to  apply
these methods to  realistic examples  including gauge bosons and a nontrivial
infrared structure.  However, it is reasonable to expect that the recombination
of scalar integrals will work similarly, such that efficient algorithms to
calculate  six-point amplitudes at one loop are in reach.  
\item 
The first evaluation of two-loop four point amplitudes with one off-shell
leg~\cite{Garland:2001tf} for the case $\gamma^* \to q\bar q g$.  This process 
is important in determining the strong coupling constant from hadronic data in
electron-positron annihilation.  By analytical continuation it should be
possible to extend these results to describe processes in hadron-hadron
collisions, $p\bar p \to V + {\rm jet}$,  and deep inelastic scattering,
$\gamma^* p \to (2+1)$~jets. 
\item 
The first determination of the two-loop quark Regge
trajectory~\cite{Bogdan:2002sr} together with confirmation of the two-loop
gluon Regge trajectory~\cite{DelDuca:2001gu}.   These trajectories control the
behaviour of the scattering amplitudes in
the high energy limit - or equivalently forward or backward scattering.
\end{itemize}

While the benefits of NLO calculations are generally well appreciated,  there
is some debate as to the motivation for NNLO calculations.   In 
Sec.~\ref{sec:ho;nnlorationale} we give reasons why NNLO
predictions should qualitatively improve the agreement between theory and
experiment, reduced renormalisation and factorisation scale dependence, 
reduced phenomenological power correction, better matching between the
parton-level and hadron-level jet algorithm and a better description of the
transverse momentum of the final state.

The basic ingredients for computing NNLO corrections are  by now either known
or conceptually understood.  However a method of how to combine the
individually infrared singular parts in a way that allows the construction of
general purpose NNLO parton level Monte Carlo programs has not been
established. This is clearly the next major task and requires considerable
effort.   During the workshop some preparatory steps were made in this 
direction for the relatively simple case of $\gamma^* \to q\bar q$ and the
general structure of the infrared singularities for this process is discussed
in Sec.~\ref{sec:ho;exercise}.  We expect that by the time of the next workshop,
there will be significant progress to report.

\vskip1cm
\noindent

\subsection*{Acknowledgements}

We would like to thank the organizers of ``Les Houches 2001" for creating a
stimulating and productive environment for the workshop. We would like to thank
our many and various collaborators for many stimulating and  thought provoking
discussions; T.B., J.-Ph.G. and G.H. thank C.~Schubert, V.D.D and E.W.N.G.
thank V.~Fadin and A.~Bogdan and    T.G. and E.W.N.G. thank L.~Garland,
A.~Koukoutsakis and E.~Remiddi. 




\newcommand\as{\alpha_{\mathrm{S}}}
\def\ltap{\raisebox{-.4ex}{\rlap{$\,\sim\,$}}
  \raisebox{.4ex}{$\,<\,$}}
\def\gtap{\raisebox{-.4ex}{\rlap{$\,\sim\,$}}
  \raisebox{.4ex}{$\,>\,$}}

\def\nn{\nonumber}
\def \be {\begin{equation}} 
\def \ee {\end{equation}} 
\def \ba {\begin{eqnarray}} 
\def \ea {\end{eqnarray}} 
\def \baa {\begin{eqnarray*}} 
\def \eaa {\end{eqnarray*}} 
\def \cusp {{\rm cusp}} 
\newcommand \ci [1] {\cite{#1}} 
\newcommand \bi [1] {\bibitem{#1}}
\def \lab #1 {\label{#1}} 
\newcommand\re[1]{(\ref{resum:#1})} 
\def \qqquad {\qquad\quad} 
\def  \qqqquad {\qquad\qquad}
\newcommand\lr[1]{{\left({#1}\right)}} 
\def \Tr {\mbox{Tr\,}} 
\def \tr {\mbox{tr\,}} 
\newcommand \vev [1] {\langle{#1}\rangle} 
\newcommand \VEV [1] {\left\langle{#1}\right\rangle} 
\newcommand \elket [1] {|{#1}\rangle} 
\newcommand \elbra [1] {\langle {#1}|} 
\def \e {\mbox{e}}
\def \CO {{\cal O}} 
\def \CP {{\cal P}} 
\def \CT {{\cal T}} 
\def \CF {{\cal F}} 
\def \W {\Sigma} 
\def \PT {{\rm PT}} 
\def \pint {\int\hspace{-1.19em}\not\hspace{0.6em}} 
\def \fracs #1#2 {\mbox{\small $\frac{#1}{#2}$}} 
\newcommand \partder [1] {{\partial \over\partial #1}} 
\def \bin #1#2 {{\left({#1}\atop{#2}\right)}}
\def\lapproxeq{{\ \lower 0.6ex \hbox{$\buildrel<\over\sim$}\ }}
\def\gapproxeq{{\ \lower 0.6ex \hbox{$\buildrel>\over\sim$}\ }}

\def \as {\relax\ifmmode\alpha_s\else{$\alpha_s${ }}\fi} 
\def \alpi {\frac \as \pi} 
\def \al #1 {\frac {\as({#1})}{\pi} } 
\def \ds #1 {\ooalign{$\hfil/\hfil$\crcr$#1$}} 
\def \MS {\overline{\rm MS}} 
\def \QCD {\mbox{{\tiny QCD}}} 
\def \GeV {\mbox{GeV}} 
\def \prt{perturbative } 
\def \nprt{nonperturbative }
\def \tomega {W} 
\def \d {{\rm d}} 
\def \bi {\bibitem} 
\def \CO {{\cal O}}
\def\xhat{\hat{x}} 
\def\zhat{\hat{z}} 
\def\qt{q_T}

\def\gappeq{\mathrel{\rlap {\raise.5ex\hbox{$>$}}
    {\lower.5ex\hbox{$\sim$}}}}
\def\lappeq{\mathrel{\rlap{\raise.5ex\hbox{$<$}}
    {\lower.5ex\hbox{$\sim$}}}}

\newcommand{\mycomm}[1]{\hfill\break
  $\phantom{a}$\kern-3.5em{\tt===$>$ \bf #1}\hfill\break}
\newcommand{\mycommA}[1]{\hfill\break
  $\phantom{a}$\kern-3.5em{\tt***$>$ \bf #1}\hfill\break}
\newcommand{\ksl}{\mbox{$k$\hspace{-0.5em}\raisebox{0.1ex}{$/$}}}
\newcommand{\psl}{\mbox{$p$\hspace{-0.4em}\raisebox{0.1ex}{$/$}}}
\newcommand{\pbsl}{\mbox{$\bar{p}$\hspace{-0.4em}\raisebox{0.1ex}{$/$}}}
\newcommand{\qsl}{\mbox{$q$\hspace{-0.45em}\raisebox{0.1ex}{$/$}}}

\def\beq{\begin{equation}} 
\def\eeq{\end{equation}} 
\def\MSbar {\hbox{$\overline{\hbox{\tiny MS}}\,$}} 
\def\eff{\hbox{\tiny eff}}
\def\res{\hbox{\tiny res}}
\def\FP{\hbox{\tiny FP}}
\def\PV{\hbox{\tiny PV}} 
\def\IR{\hbox{\tiny IR}} 
\def\UV{\hbox{\tiny  UV}} 
\def\ECH{\hbox{\tiny ECH}} 
\def\NP{\hbox{\tiny NP}}
\def\osg{\hbox{\tiny {\rm off-shell\,gluon}}} 
\def\QCD{\hbox{\tiny  QCD}} 
\def\CMW{\hbox{\tiny CMW}} 
\def\SDG{\hbox{\tiny SDG}}
\def\spins{\hbox{\tiny spins}} 
\def\SDG{\hbox{\tiny SDG}}
\def\pinch{\hbox{\tiny pinch}} 
\def\brem{\hbox{\tiny brem}}
\def\V{\hbox{\tiny V}} 
\def\BLM{\hbox{\tiny BLM}} 
\def\NLO{\hbox{\tiny NLO}} 
\def\DGE{\hbox{\tiny DGE}} 
\def\QED{\hbox{\tiny QED}}
\def\PT{\hbox{\tiny PT}} 
\def\PA{\hbox{\tiny PA}}
\def\1loop{\hbox{\tiny 1-loop}} 
\def\2loop{\hbox{\tiny 2-loop}}

\def\mysim{\kern -.1667em\lower0.8ex\hbox{$\tilde{\phantom{a}}$}}
\def\a{\bar{a}}

\def\Bar#1{\overline{#1}}                       
\newcommand{\opi}{\rm\scriptstyle 1PI}
\newcommand{\pim}{\rm\scriptstyle PIM}

\def\smallfrac#1#2{\hbox{${{#1}\over {#2}}$}} 
\def\as{\alpha_s}
\def\ah{\widehat\alpha_s} 
\def\bea{\begin{eqnarray}}
\def\eea{\end{eqnarray}} 
\def\blackbox{\vrule height7pt width5pt depth2pt} 
\def\matele#1#2#3{\langle {#1} \vert {#2} \vert {#3} \rangle } 
\def\VertL{\Vert_{\Lambda}}
\def\VertR{\Vert_{\Lambda_R}}
\def\Real{\Re e}\def\Imag{\Im m}
\def\SZP{\hbox{S0$'$}}
\def\DZP{\hbox{D0$'$}}
\def\DMP{\hbox{D-$'$}}
\def\MS{\hbox{$\overline{\rm MS}$}}
\def\ms{\hbox{$\overline{\scriptstyle\rm MS}$}}
\def\half{\hbox{${1\over 2}$}}
\def\third{\hbox{${1\over 3}$}}
\def\QMS{Q$_0$\MS} 
\def\QDIS{Q$_0$DIS}
\catcode`@=11 
\def\toinf#1{\mathrel{\mathop{\sim}\limits_{\scriptscriptstyle
      {#1\rightarrow\infty }}}}
\def\tozero#1{\mathrel{\mathop{\sim}\limits_{\scriptscriptstyle
      {#1\rightarrow0 }}}}
\def\slash#1{\mathord{\mathpalette\c@ncel#1}}
\def\c@ncel#1#2{\ooalign{$\hfil#1\mkern1mu/\hfil$\crcr$#1#2$}}
\def\lsim{\mathrel{\mathpalette\@versim<}}
\def\gsim{\mathrel{\mathpalette\@versim>}}
\def\@versim#1#2{\lower0.2ex\vbox{\baselineskip\z@skip\lineskip\z@skip
    \lineskiplimit\z@\ialign{$\m@th#1\hfil##$\crcr#2\crcr\sim\crcr}}}
\catcode`@=12 
\def\twiddles#1{\mathrel{\mathop{\sim}\limits_ {\scriptscriptstyle
      {#1\rightarrow \infty }}}}

\section{QCD RESUMMATION\protect\footnote{Section coordinator:
    E. Laenen}$^{,}$~\protect\footnote{Contributing authors: C.
    Bal\'azs, R. Ball, M. Cacciari, S. Catani, D. de Florian, S.
    Forte, E. Gardi, M. Grazzini, N. Kidonakis, E. Laenen, S. Moch, P.
    Nadolsky, P.Nason, A. Kulesza, L. Magnea, F. Olness, R. Scalise,
    G. Sterman, W. Vogelsang, R. Vogt, C.-P. Yuan}}\label{sec:resum,qcdsm}

\subsection{Introduction}

In our working group we investigated a variety of issues concerning
the relevance of resummation for observables at TeV colliders.
Resummation is a catch-all name for efforts to extend the predictive
power of QCD by summing large (logarithmic) corrections to all orders
in perturbation theory. In practice, the resummation formalism depends
on the observable at issue, through the type of logarithm to be
resummed, and the resummation methods.

A number of resummation formalisms (threshold resummation, $Q_T$ or
recoil resummation and any renormalization-group resummation) have now
matured to the point where one can employ them for precision physics.
It is known how to organize the associated logarithms to all orders
and to any accuracy, at least in principle.  Such resummation
formalisms therefore constitute a systematically improvable
calculational scheme, like standard perturbation theory.  It is also
known how to consistently match these resummations to finite order
perturbation theory. In our working group, the full
next-to-next-to-leading logarithmic threshold and $Q_T$ resummation
were performed for the inclusive Higgs production cross section and
its transverse momentum respectively. Further studies addressing the
value of resummation for precision physics were done for heavy quark
production cross sections and fragmentation functions. The
applicability of joint resummation, the combination of threshold and
recoil resummation, was examined in detail for electroweak
annihilation. The relation between small-$x$ resummation, the DGLAP
evolution equation, and precision analysis of the deep-inelastic
structure function at HERA was further explored. Detailed application
of these results to TeV colliders is still to come.

Resummed cross sections are inherently ambiguous because they require
a prescription to handle singularities due to very soft radiation.
These ambiguities take the form of power corrections, about which
there is still much to learn. Our working group has studied the
characteristics of various prescriptions, and the power corrections
that they imply.

Both within our working group, and in joint sessions with others,
numerous fruitful discussions took place based on short 
presentations by participants. These presentations and discussions 
addressed both the topics mentioned and reported on below, as well 
as the issues of resummation and Monte Carlo programs (by V. Ilyin)
and next-to-next-leading logarithmic threshold resummation for 
Drell-Yan and deep-inelastic scattering (by A. Vogt).

In general, the studies performed in the QCD resummation working
group, whose reports now follow, as well as the discussions held
strengthen the view that QCD resummation does, must, and will play an 
important part in the quantitative study of observables at TeV colliders.

\subsection{Higgs boson transverse momentum at the
  LHC\protect\footnote{Contributing authors: C. Bal\'azs, D. de
    Florian, A. Kulesza}}

\subsubsection{Introduction} 
 
The underlying mechanism of the electroweak symmetry breaking (EWSB)
is an uncovered sector of the Standard Model (SM), thus the physical
remnant of the spontaneous EWSB, the Higgs boson, is the primary
object of search at present and future colliders.
At the CERN Large Hadron Collider (LHC), a Standard Model (like) Higgs
boson can mainly be produced in 14 TeV center of mass energy
proton--proton collisions by the partonic subprocess $g g$ (via heavy
quark loop) $\to H X$ (see section A.1 of ref. \cite{Cavalli:2002vs}). 
The extraction
of the signal requires the accurate prediction of its production rate,
as well as the transverse momentum ($Q_T$) distribution of the Higgs
boson and its decay products and backgrounds, since the shape of these
distributions can dictate the analysis strategies for the search
\cite{Balazs:2000sz}.
To reliably predict these distributions, especially in the low to
medium $Q_T$ region where the bulk of the rate is, the effects of the
multiple soft--gluon emission have to be included. This, and the need
for the systematic inclusion of the higher order QCD corrections
require the extension of the standard hadronic factorization theorem
to the low $Q_T$ region. With a smooth matching to the usual
factorization formalism, it is possible to obtain a sound prediction
in the full $Q_T$ range.

\subsubsection{Low $Q_T$ Factorization}\label{sec:low-qt,resum,qcdsm} 
 
In this section the low transverse momentum factorization formalism is
summarized briefly.  We consider the case of the inclusive
hard-scattering process where the final-state system $F$ is produced
by the collision of the two hadrons $h_1$ and $h_2$.  The final state
$F$ is a generic system of non-strongly interacting particles, such as
{\em one} or {\em more} vector bosons $(\gamma^*, W, Z, \dots)$, Higgs
particles ($H$) and so forth.
 
When calculating fixed order QCD corrections to the cross section
the hadronic factorization theorem is invoked. While the transverse
momentum $Q_T$ of the produced system is of the order of its invariant
mass $Q$, this calculation is reliable. But the standard factorization
fails when the $Q_T \ll Q$, as a result of multiple soft and collinear
emission of gluons from the initial state. As an indication of this
problem, the ratio of the two very different scales appear in
logarithmic corrections of the form $\alpha_S^n/Q_T^2 \ln^{m}
Q^2/Q_T^2$ ($1<m<2n-1$), which spoil the convergence of fixed-order
calculation in the low $Q_T$ region.
%
These logarithmically-enhanced terms, not absorbed by the parton
distribution functions, have to be evaluated at higher perturbative
orders, and possibly resummed to all orders in the QCD coupling
constant $\as$.
%
 
 
To resolve the problem, the differential cross section is split into a
part which contains all the contribution from the logarithmic terms
(res.), and into a regular term (fin.):
\begin{eqnarray} 
\label{resum:eq1} 
       \frac{d\sigma_{F}}{dQ dQ_T^2 d\phi} =  
\left[ \frac{d\sigma_{F}}{dQ dQ_T^2 d\phi} \right]_{\rm res.} +  
\left[ \frac{d\sigma_{F}}{dQ dQ_T^2 d\phi} \right]_{\rm fin.} \, , 
\end{eqnarray} 
where $\phi$ denotes the remaining independent kinematical variables
of the final system.  Since the second term on the right hand side in
Eq.(\ref{resum:eq1}) does not contain potentially large logs, it can
be calculated using the usual factorization. The first term has to be
evaluated differently, keeping in mind that failure of the standard
factorization occurs because it neglects the transverse motion of the
incoming partons in the hard scattering.
 
In the Fourier conjugate transverse position (${\vec b}$) space the
resummed component of the cross section
\cite{Dokshitzer:1980hw}--\cite{Davies:1984hs}
can be written as
\begin{eqnarray} 
\label{resum:resgen} 
 \frac{Q^2\, d\sigma_{F}}{dQ^2 dQ_T^2d \phi} = \sum_{a,b} 
\int dx_1 \, dx_2\, db  \frac{b}{2} J_0(b Q_T) \; 
f_{a/h_1}\left(x_1,\frac{b_0^2}{b^2}\right)   
f_{b/h_2}\left(x_2,\frac{b_0^2}{b^2}\right) 
 s\, W_{ab}^{F}(x_1 x_2 s; Q, b, \phi) \,. 
\end{eqnarray} 
Here, the resummed partonic cross section $W_{ab}^{F}$ is
\begin{eqnarray} 
\label{resum:unw} 
W_{ab}^{F}(s; Q, b, \phi) &= \sum_c \displaystyle \int dz_1\,  
dz_2  
\; C_{ca}^{F}\left(\as(b_0^2/b^2), z_1\right)  
\; C_{{\bar c}b}^{F}\left(\as(b_0^2/b^2), z_2\right) 
\; \delta\left(Q^2 - z_1 z_2 s\right) \nonumber \\ 
&\cdot\, 
\displaystyle \frac{d\sigma_{c{\bar c}}^{(LO) \,F}(Q^2,\phi)}{d\phi} \; 
H_c^{F}\left(\as(Q^2), \phi\right) 
\;S_c(Q,b) \;\;, 
\end{eqnarray} 
where $\sigma_{c{\bar c}}^{(LO)}$ is the lowest order partonic
cross-section and $S_c(Q,b)$ the Sudakov form factor (with $c=q,g$)
\begin{equation} 
\label{resum:sudakov} 
S_c(Q,b)=\exp \left\{ -\int_{b_0^2/b^2}^{Q^2} \frac{dq^2}{q^2}  
\left[ A_c\left(\as(q^2)\right) \;\ln \left(\frac{Q^2}{q^2}\right)  
+ B_c\left(\as(q^2)\right) \right] \right\}\, . 
\end{equation} 
In the usual CSS approach \cite{Collins:1985kg}, the coefficient
function $H_c^{F}$ does not appear (i.e. $H_c^{F}\equiv 1$), with the
consequence that both the coefficient functions
$C_{ab}(\as(b_0^2/b^2),z)$ and the Sudakov form factor $S_c(Q,b)$
become process dependent, a certainly unpleasent feature.  As
discussed in Ref.~\cite{Catani:2000vq}, the inclusion of $H_c^F$ is
sufficient to make the Sudakov form factor $S_c(Q,b)$ and the
coefficient functions $C_{ab}(\as(b_0^2/b^2),z)$ process-independent.
In a similar way as it happens for parton densities, which have to be
defined by fixing a factorization scheme (e.g. the ${\overline {\rm
    MS}}$ scheme or the DIS scheme), there is an ambiguity in the
factors on the right-hand side of Eq.~(\ref{resum:unw}), which have to
be defined by choosing a `resummation scheme'. Note that the choice of
a `resummation scheme' amounts to defining $H_c^{F}$ (or $C_{ab}$) for
a {\em single} process. Having done that, the process-dependent factor
$H_c^{F}$ and the universal factors $S_c$ and $C_{ab}$ in
Eq.~(\ref{resum:unw}) are unambiguously determined for any other
process.
 
The resummation formula in Eq.~(\ref{resum:unw}) has a simple physical
origin.  When the final-state system $F$ is kinematically constrained
to have a small transverse momentum, the emission of accompanying
radiation is strongly inhibited, so that
mostly soft and collinear partons (i.e. partons with low transverse
momenta $Q_T$) can be radiated in the final state. The
process-dependent factor $H_c^{F}(\as(Q^2))$ embodies hard
contributions produced by virtual corrections at transverse-momentum
scales $Q_T \sim Q$. The form factor $S_c(Q,b)$ contains real and
virtual contributions due to soft (the function $A_c(\as)$ in
Eq.~(\ref{resum:sudakov})) and flavour-conserving collinear (the
function $B_c(\as)$ in Eq.~(\ref{resum:sudakov})) radiation at scales
$Q \gtap Q_T \gtap 1/b$. At very low momentum scales, $Q_T \ltap 1/b$,
only real and virtual contributions due to collinear radiation remain
(the coefficient functions $C_{ab}(\as(b_0^2/b^2),z)$).
 
The $A_c$ and $B_c$ functions, as well as the coefficients $C_{ab}$
are free of large logarithmic corrections and safely calculable
perturbatively as expansions in the strong coupling $\as$
\begin{equation} 
A_c(\alpha_S) = 
\sum_{n=1}^\infty  
\left( \frac{\alpha _S}\pi \right)^n A_c^{(n)}, 
 ~~~  
B_c(\alpha_S) = 
\sum_{n=1}^\infty  
\left( \frac{\alpha _S}\pi \right)^n B_c^{(n)}, 
 ~~~  
C_{ab}^F(\alpha_s,z) = 
\sum_{n=0}^\infty  
\left( \frac{\alpha _S}\pi \right)^n C^{F(n)}_{ab}(\alpha_S,z)\, , 
\end{equation} 
with a similar expansion for the coefficient function $H_c^F$.  More
detailed expression for the coefficients can be found in
\cite{Kodaira:1982nh,Catani:1988vd,Davies:1984hs,deFlorian:2000pr,deFlorian:2001zd
}.  The coefficients of the perturbative expansions $A_c^{(n)}$,
$B_c^{(n)}$ and $C_{ab}^{F(n)}(z)$ are the key of the resummation
procedure since their knowledge allows to perform the resummation to a
given {\it logarithmic} order: $A^{(1)}$ leads to the resummation of
leading logarithmic (LL) contributions, $\{ A^{(2)}, B^{(1)} \}$ give
the next-to-leading logarithmic (NLL) terms, $\{ A^{(3)}, B^{(2)},
C^{(1)}\}$ give the next-to-next-to-leading logarithmic (NNLL) terms,
and so forth\protect\footnote{A concensus has not been reached regarding 
the classification of the so-called LL, NLL, NNLL etc. terms, and their
corresponding $B$ contents. The above classification of subsect. 
\ref{sec:low-qt,resum,qcdsm}  is used in subsects.\ref{higgs-prod-nnll} and
\ref{softglue-higgs}.  Another classification is used in subsect. \ref{qt-css}.
The motivation for the latter is presented in
sect.\ref{sec:resumps,mc,qcdsm}.}\label{ftnote-classif}.

In particular, the coefficient $B^{(2)}$ has been recently computed
\cite{deFlorian:2000pr, deFlorian:2001zd} for both $q\bar{q}$ and $gg$
channels allowing to extend the analysis to NNLL accuracy.  Even
though there is no analytical result available for it, the coefficient
$A_{q,g}^{(3)}$ has been extracted numerically with a very good
precision in Ref.~\cite{Vogt:2000ci}.
After matching the resummed and fixed order cross sections, it is
expected that the normalization of the resummed cross section
reproduces the fixed order total rate (at which the $A$, $B$ and $C$
functions are evaluated), since when expanded and integrated over
$Q_T$ they deviates only in higher order terms.~\cite{Balazs:1997xd}

\subsubsection{Higgs $Q_T$ at the LHC using the CSS formalism}\label{qt-css}
 
The low $Q_T$ factorization formalism, described in the previous
section, is utilized to calculate the QCD corrections to the
production of Higgs
bosons at the LHC. 
In the low $Q_T$ region this calculation takes into account the
effects of the multiple--soft gluon emission including the Sudakov
exponent $S$, and the non--perturbative contributions ${\cal
  F}_{a/h}$. At the next-to-leading-logarithmic order (NLL) the
$A^{(1,2)}$, and $B^{(1,2)}$ coefficients are included in the Sudakov
exponent.  The normalization changing effect of the ${\cal
  O}(\alpha_S^3)$ virtual corrections are also taken into account by
including the coefficient ${C}_{gg}^{(1)}$, which ensures agreement
with the ${\cal O}(\alpha_S^3)$ total rate.  At the
leading-logarithmic order (LL) the coefficients $A^{(1)}$, and
$B^{(1)}$ and ${C}_{gg}^{(0)}$ are included. By matching to the ${\cal
  O}(\alpha_S^3)$ fixed order distributions a prediction is obtained
for the Higgs production cross section in the full $Q_T$ range which
is valid up to ${\cal O}(\alpha_S^3)$. The expressions for the
included $A$, $B$ and $C$ coefficients and further details of this
calculation are given in an earlier work.~\cite{Balazs:1997xd} The
analytic results are coded in the ResBos Monte Carlo event generator.

\begin{figure}[htbp] 
\begin{center} 
  \includegraphics[width=\textwidth]{LHC_mH125_GeV.epsi}
\end{center} 
\caption{ 
  Higgs boson transverse momentum distributions at the LHC calculated
  by ResBos (curves) and PYTHIA (histograms). The solid curve was
  calculated at NLL \protect\footnote{Notice that this is not the
    conventional definition of LL and NLL accuracy.} (including
  $A^{(1,2)}$, $B^{(1,2)}$, and $C^{(0,1)}$).  The dashed curve is LL
  (includes $A^{(1)}$, $B^{(1)}$, and $C^{(0)}$).  For PYTHIA, the
  original output with default input parameters rescaled by a factor
  of $K = 2$ (solid), and one calculated by the altered input
  parameter value $Q_{max}^2 = s$ (dashed) are shown.  The lower
  portion, with a logarithmic scale, also shows the high $Q_T$ region.
  In the last frame all are normalized to the solid curve.}
\label{resum:Fig:PYTHIA} 
\end{figure} 
 
Fig.~\ref{resum:Fig:PYTHIA} compares the Higgs boson transverse
momentum distributions at the LHC calculated by ResBos (curves) and by
PYTHIA (histograms from version 6.136). We use a Higgs mass of 125 GeV
and CTEQ5M parton distributions. The solid curve was calculated at NLL
(including $A^{(1,2)}$, $B^{(1,2)}$, and $C^{(0,1)}$).  The dashed
curve is LL (includes $A^{(1)}$, $B^{(1)}$, and $C^{(0)}$).  The shape
of these curves are quite similar reflecting the small uncertainty in
the shape of the resummed prediction. The normalization changes
considerably after including the sub-leading tower of logs, since both
the $C^{(1)}$ and the $B^{(2)}$ coefficients receive contributions
from the ${\cal O}(\alpha_S^3)$ virtual corrections, which are known
to be large.
 
In the low and intermediate $Q_T$ ($\lesssim 100$ GeV) region the
shape of the default PYTHIA histogram agrees better with the LL ResBos
curve, justifying the similar physics that goes into PYTHIA and the LL
calculation.  For high $Q_T$, the PYTHIA prediction falls under the
ResBos curve, since ResBos mostly uses the exact fixed order ${\cal
  O}(\alpha_S^3)$ matrix elements in that region, while PYTHIA still
relies on the multi--parton radiation ansatz. PYTHIA can be tuned to
agree with ResBos in the high $Q_T$ region, by changing the maximal
virtuality a parton can acquire in the course of the shower (dashed
histogram).  Matrix element corrections to this process in PYTHIA are
expected to fix this discrepancy in the near future.  Since showering
is attached to a process after the hard scattering takes place, and
the parton shower occurs with unit probability, it does not change the
total cross section for Higgs boson production given by the hard
scattering rate.  Thus, the total rate is given by PYTHIA at ${\cal
  O}(\alpha_S^2)$. Thus in Fig.~\ref{resum:Fig:PYTHIA} the dashed
PYTHIA histograms are plotted with their rate normalized to the NLL
ResBos curve.
 
\subsubsection{Higgs production to NNLL accuracy}\label{higgs-prod-nnll} 
 
As has been discussed in the introduction, the coefficients appearing
in Eq.~(\ref{resum:unw}) (with the exception of $A^{(n)}$) are
dependent on the resummation scheme.  Under a change from scheme $S$
to $S'$, the coefficients are modified as
\begin{eqnarray} 
H_{c\, S}^{(1)F} & \to & H_{c\, S'}^{(1)F} \;, \nonumber \\ 
C_{ab\, S}^{(1)}(z) & \to & C_{ab\, S'}^{(1)}(z)= 
C_{ab\, S}^{(1)}(z) +\frac{1}{2} \delta_{ab} \delta(1-z) \left( 
H_{a\, S}^{(1)F} - H_{a\, S'}^{(1)F} \right) \;, \nonumber \\ 
B_{c\, S}^{(2)} & \to & B_{c\, S'}^{(2)}= B_{c\, S}^{(2)} + \beta_0 \left( 
H_{c\, S}^{(1)F} - H_{c\, S'}^{(1)F} \right) 
\label{transf} 
\end{eqnarray} 
One possible scheme is the scheme-$H$\cite{Catani:2000vq}, where
$H_{g\, H}\equiv 1$ and the coefficients agree with the corresponding
to the CSS formulation for this process.  Another posibility consist
in the $\overline{MS}$ scheme, where $B^{(2)}$ is defined to be
proportional to the coefficient of the $\delta(1-z)$ term in the
two-loop $gg$ splitting function \cite{Catani:2000vq}.  Based on the
physical interpretation of the different pieces in the resummation
formulae in Eq.~(\ref{resum:unw}), it is possible to define the
`collinear resummation scheme', where only terms of collinear origin
(i.e. the ones originated from the $n-4$ component of the splitting
functions, see \cite{deFlorian:2000pr, deFlorian:2001zd} ) remain in
the coefficient function $C$.

\begin{figure}[htbp] 
\begin{center} 
  \includegraphics[width=0.8\textwidth]{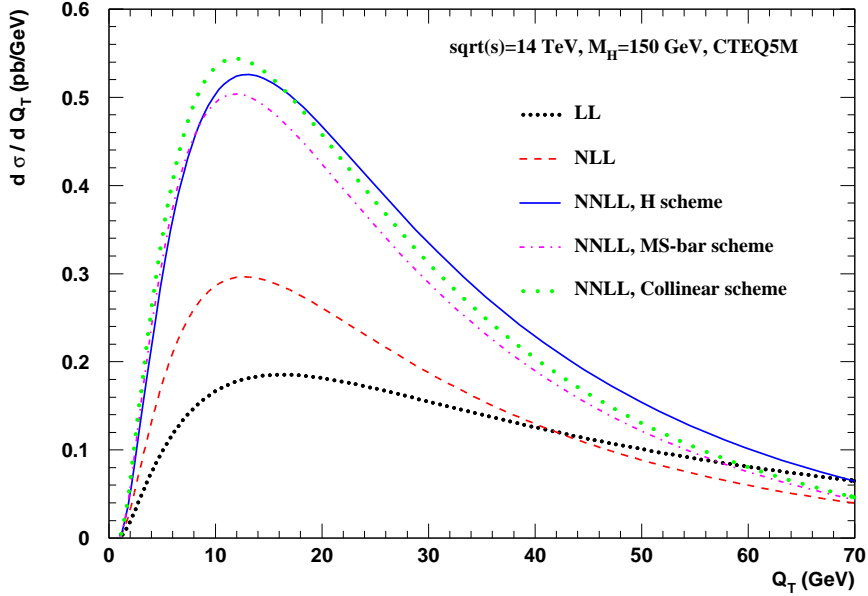}
\end{center} 
\caption{\label{resum:Fig:higgsnnll}{$Q_T$ distribution for Higgs
    production at LHC to LL, NLL and NNLL accuracy. The three NNLL
    lines correspond to the choice of different resummation schemes.
  }}
\end{figure}

Even though the cross section for Higgs production is a physical
observable and therefore independent on the chosen scheme, the
truncation of the perturbative expansion at the level of the resummed
coefficients introduces an explicit dependence on the scheme. We will
use this scheme dependence as a way to quantify the perturbative
stability of the resummed result, estimating the size of the
unaccounted higher order terms as well as of the non-perturbative
contributions.  It follows from~(\ref{transf}) that the scheme
dependence first enters at the NNLL level.  In Fig.
\ref{resum:Fig:higgsnnll} we show preliminary results for the $Q_T$
distribution for Higgs production at LHC ($M_H=150$ GeV) at LL, NLL
and NNLL, in the last case using the three schemes discussed above.

The results are obtained using the code developed
in~\cite{Kulesza:2001jc}, adapted for the case of inclusive Higgs
production in the $gg$ channel (in the limit of infinite top quark
mass) and to genuine NNLL accuracy, i.e. including up to $A^{(3)}$,
$B^{(2)}$, $C^{(1)}$ and $H^{(1)}$.

Matching to the perturbative result has been performed only at LO, but
this will not affect our conclusions since we are mainly interested in
the low $Q_T$ region where the resummed contribution completely
dominates the cross section.  The matching at ${\cal O}(\alpha_S^4)$
can be performed by using the NLO calculation of
\cite{deFlorian:1999zd}. Furthermore we have not included any
non-perturbative contribution and the parton distributions correspond
to the CTEQ5M set.
As can be observed, the NNLL corrections are rather large. The main
reason for that is the appearence of the $C^{(1)}$ and $H^{(1)}$
coefficients, which contain the large virtual (and soft) corrections
for Higgs production, which are known to considerably increase (up to
almost a factor of two) the inclusive cross section. Higher order
terms are expected to be smaller, at least from the observation of the
rather good perturbative stability concerning the dependence on the
resummation scheme at NNLL.\\
{\bf Acknowledgements:} C.B. and D.de F. thank the organizers of the
Les Houches conference for their financial support.  C.B. also thanks
I. Puljak for the PYTHIA curves. A.K. would like to thank W.~Vogelsang
for many useful discussions. D.de F. thanks M. Grazzini for
discussions.

\subsection{Soft-gluon effects in Higgs boson production at hadron
  colliders\protect\footnote{Contributing authors: S. Catani, D. de
    Florian, M. Grazzini and P. Nason}}\label{softglue-higgs}

The most important mechanism for SM Higgs boson production at hadron
colliders is gluon--gluon fusion through a heavy (top) quark loop.
NLO QCD corrections to this process are known to be large
\cite{Dawson:1991zj,Djouadi:1991tk,Spira:1995rr}: their effect
increases the LO inclusive cross section by about 80-100$\%$.  The NLO
corrections are weakly dependent on the mass $M_t$ of the top quark,
thus justifying the evaluation of higher-order terms within the
large-$M_t$ approximation.  Recently, the calculation of the NNLO
corrections in the large-$M_t$ limit has been completed
\cite{Harlander:2000mg,Catani:2001ic,Harlander:2001is,Catani:2001cr,Harlander:2002wh}.
Their effect is moderate at the LHC: in the case of a light Higgs, the
$K$-factor is about $2.3$--$2.4$, corresponding to an increase of
about 25$\%$ with respect to NLO.  The NNLO effect is more sizeable at
the Tevatron where $K\sim 3$, the increase being of about $50 \%$ with
respect to NLO.  The bulk of the NNLO contributions is due to soft and
collinear radiation
\cite{Catani:2001ic,Harlander:2001is,Catani:2001cr}, the hard
radiation \cite{Harlander:2002wh} giving only a small (typically
5-7$\%$) correction.  Multiple soft-gluon emission beyond NNLO can
thus be important, particularly at the Tevatron.  Here we investigate
the effects of resummation of soft (Sudakov) emission to all orders.
More details will be given elsewhere.

The cross section ${\hat \sigma}_{gg}$ for the partonic subprocess
$gg\to H+X$ at the centre--of--mass energy ${\hat s}=M_H^2/z$ can be
written as
\begin{equation}
\label{resum:spart}
{\hat \sigma}_{gg}({\hat s},M_H^2) = \sigma_0 \;z \;G_{gg}(z) \;,
\end{equation}
where $M_H$ is the Higgs mass, $\sigma_0$ is the Born-level cross
section and $G_{gg}$ is the perturbatively computable coefficient
function.  Soft-gluon resummation has to be performed (see
\cite{Catani:1996yz} and references therein) in the Mellin (or
$N$-moment) space ($N$ is the moment variable conjugate to $z$).  The
{\em all-order} resummation formula for the coefficient function
${G}_{gg}$ is \cite{Catani:2001ic,Kramer:1996iq}:
\begin{equation}
\label{resum:resformula} 
{G}_{gg, \, N}^{{\rm (res)}} =\as^2(\mu_R^2)\,
C_{gg}(\as(\mu^2_R),M_H^2/\mu^2_R;M_H^2/\mu_F^2) \, 
\Delta_{N}^{H}(\as(\mu^2_R),M_H^2/\mu^2_R;M_H^2/\mu_F^2)\; ,
\end{equation}
where $\mu_R$ and $\mu_F$ denote the renormalization and factorization
scales, respectively.  The function $C_{gg}(\as)$ contains all the
terms that are constant in the large-$N$ limit,
produced by hard virtual contributions and non-logarithmic soft
corrections.  It can be computed as a power series expansion in $\as$
as
\begin{equation}
\label{resum:coef}
C_{gg}(\as(\mu^2_R),M_H^2/\mu^2_R;M_H^2/\mu_F^2) =  
1 + \sum_{n=1}^{+\infty} \;  
\left( \frac{\as(\mu^2_R)}{\pi} \right)^n \; 
C_{gg}^{(n)}(M_H^2/\mu^2_R;M_H^2/\mu_F^2) \;\;,
\end{equation}
where the perturbative coefficients $C_{gg}^{(n)}$ are closely related
to those of the $\delta(1-z)$ contribution to $G_{gg}(z)$.  The
radiative factor $\Delta_{N}^{H}$ embodies the large logarithmic terms
due to soft-gluon radiation.  To implement resummation, the radiative
factor is expanded to a given logarithmic accuracy as
\begin{equation} 
\label{resum:deltannll} 
\Delta_N^{H} = 
\exp \Big\{ \frac{}{} \ln N \; g^{(1)}(\lambda) + 
g^{(2)}\!\left(\lambda,\frac{M_H^2}{\mu^2_R};\frac{M_H^2}{\mu_F^2}\right)  
+ \as(\mu^2_R) 
g^{(3)}\!\left(\lambda,\frac{M_H^2}{\mu^2_R};\frac{M_H^2}{\mu_F^2}\right)  
+ {\cal O}(\as^2(\as \ln N)^k) \Big\} , 
\end{equation} 
such that the functions $g^{(1)}, g^{(2)}$ and $g^{(3)}$ respectively
collect the leading logarithmic (LL), next-to-leading logarithmic
(NLL) and next-to-next-to-leading logarithmic (NNLL) terms with
respect to the expansion parameter $\lambda=\as(\mu^2_R) \ln N$.

NLL resummation \cite{Catani:2001ic} is controlled by three
perturbative coefficients, $A_g^{(1)}, A_g^{(2)}$ and $C^{(1)}_{gg}$.
The coefficients $A_g^{(1)}$ and $A_g^{(2)}$, which appear in the
functions $g^{(1)}$ and $g^{(2)}$, are well known (see
\cite{Catani:1996yz} and references therein).  The coefficient
$C^{(1)}_{gg}$ in Eq.~(\ref{resum:coef}) is extracted from the NLO
result.

At NNLL accuracy three new coefficients are needed
\cite{Catani:2001ic}: the coefficient $C_{gg}^{(2)}$ in
Eq.~(\ref{resum:coef}) and two coefficients, $D^{(2)}$ and
$A_g^{(3)}$, which appear in the NNLL function $g^{(3)}$.  The
functional form of $g^{(3)}$ was computed in Ref.~\cite{Vogt:2000ci}.
The coefficients $D^{(2)}$ and $C_{gg}^{(2)}$ are obtained
\cite{Catani:2001ic} from the NNLO result. The coefficient $A_g^{(3)}$
is not yet fully known: we use its exact $N_f^2$-dependence
\cite{Bennett:1998ch} and the approximate numerical estimate of
Ref.~\cite{vanNeerven:2000wp}.

Finally, to take into account the dominant collinear logarithmic
terms, the coefficient $C_{gg}^{(1)}$ in the resummation formula can
be modified as \cite{Catani:2001ic}
\begin{equation}
\label{resum:colterm}
C_{gg}^{(1)} \rightarrow C_{gg}^{(1)} + 2 A_g^{(1)} \; \frac{\ln N}{N} \;.
\end{equation}

\begin{figure}[htb]
\begin{center}
  \includegraphics[width=12cm,height=7.5cm,angle=0]{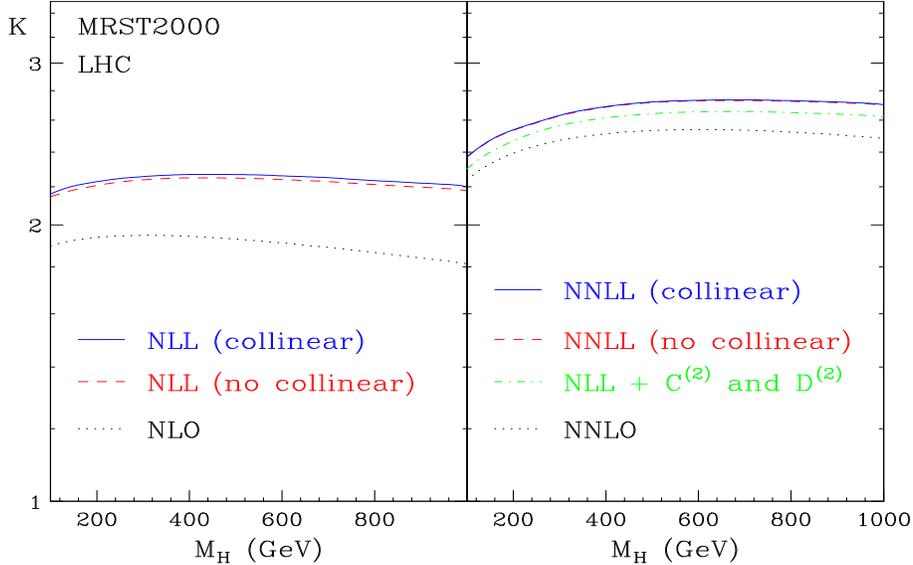}
\end{center}
\caption{\label{resum:fig:lhc}{Resummed K-factors at the LHC}.}
\end{figure}

\begin{figure}[htb]
\begin{center}
  \includegraphics[width=12cm,height=7.5cm,angle=0]{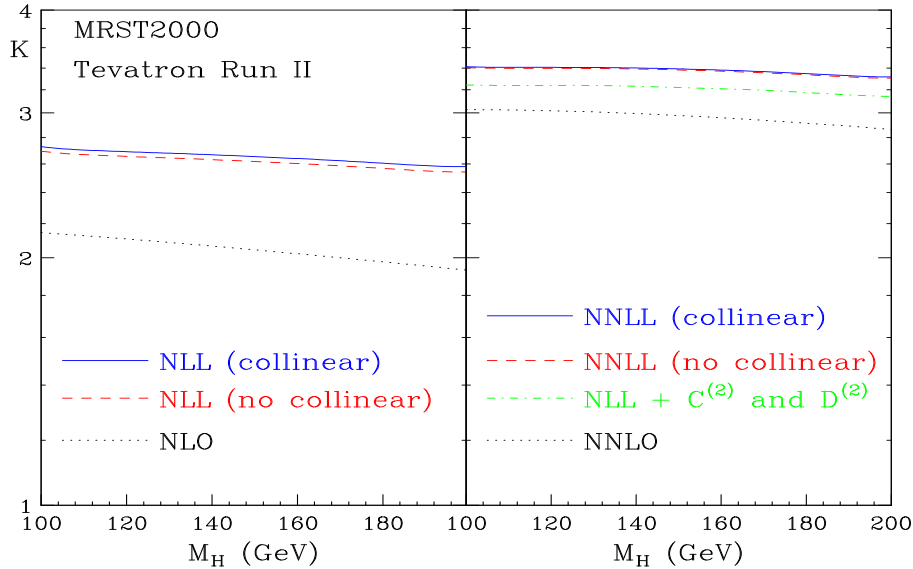}
\end{center}
\caption{\label{resum:fig:tev}{Resummed K-factors at the Tevatron Run II}.}
\end{figure}

In the following we present a preliminary study of the resummation
effect at the Tevatron and the LHC.  The hadron-level cross section is
obtained by convoluting the partonic cross section in
Eq.~(\ref{resum:spart}) with the parton distributions of the colliding
hadrons. We use the MRST2000 set \cite{Martin:2000gq}, which includes
approximated NNLO parton distributions, with $\as$ consistently
evaluated at one-loop, two-loop, three-loop order for the LO, NLO
(NLL), NNLO (NNLL) results, respectively.  All the results correspond
to the choice $\mu_F=\mu_R=M_H$.  The $K$-factors in
Figs.~\ref{resum:fig:lhc} and \ref{resum:fig:tev} are defined with
respect to the LO cross section. The resummed calculations are always
matched to the corresponding fixed-order calculations, i.e. we
consider the full fixed-order result and include higher-order
resummation effects.
In Fig.~\ref{resum:fig:lhc} we plot the K-factors at the LHC, as a
function of the Higgs mass.  On the left-hand side we show the NLL
result, matched to NLO, with and without the inclusion of the
collinear term in Eq.~(\ref{resum:colterm}).  The NLO result is
reported for comparison.  On the right-hand side we show the NNLL
results, again with and without the inclusion of the collinear term.
The matching is done to the NNLO result\footnote{More precisely, we
  include all the soft and virtual contributions and the hard terms in
  the form $(1-z)^n$ up to $n=1$. Higher powers of $(1-z)$ give very
  small effect \cite{Harlander:2002wh}.}  \cite{Harlander:2002wh}.  We
also plot the resummed NLL result obtained by including $C_{gg}^{(2)}$
and $D^{(2)}$ only.  We see that the inclusion of the collinear term
is numerically not very relevant.  Soft-gluon resummation at NLL
accuracy increases the NLO cross section by 13-20 $\%$, the effect
being more sizeable at high $M_H$.  Going from NNLO to NNLL the effect
is smaller, with an increase of $\sim 6$-9$\%$ in the full range of
$M_H$.  Note also that NLL resummation gives a good approximation of
the complete NNLO calculation.

In Fig.~\ref{resum:fig:tev} we report the analogous results at the
Tevatron.  Here the resummation effects are larger. Going from NLO to
NLL accuracy, the cross section increases by 25-30$\%$.  NNLL
resummation increases the NNLO cross section by $\sim 12$-$16\%$ when
$M_H$ varies in the range 100-200~GeV.  These results are not
unexpected \cite{Catani:2001cr}, since at the Tevatron the Higgs boson
is produced closer to threshold and the effect of multiple soft-gluon
emission is more important.

\subsection{Joint resummation in electroweak
  production\protect\footnote{Contributing authors: A. Kulesza, G.
    Sterman, W. Vogelsang}}

\subsubsection{Introduction}
The hadronic annihilation cross sections for electroweak boson
production ($\gamma^*$, W, Z, H) provide important test cases for
resummation techniques in QCD.  Soft-gluon emission leads to large
logarithmic contributions to perturbative cross sections, both near
threshold, when the incoming partons have just enough energy to
produce the observed boson, and near measured boson transverse
momentum $Q_T\ll Q$. All-order resummations of the leading and
next-to-leading terms in such contributions have been achieved
separately for the threshold and small-$Q_T$ cases. Very recently, a
formalism has been developed that encompasses both.  For electroweak
annihilation as well as QCD cross sections, the necessary analysis for
this combination, which we refer to as joint resummation, has been
carried out in \cite{Laenen:2000ij}. In this brief report, we develop
the application of the joint resummation formalism to the
phenomenology of electroweak annihilation.

The effects of resummation are closely bound to momentum conservation.
The singular corrections associated with soft gluon emission
exponentiate in the corresponding spaces of moments, impact parameter
for transverse momentum, and Mellin (or Laplace) moments in energy for
threshold resummation.  The transforms relax momentum and energy
conservation, while their inverses reimpose it.  In joint resummation,
both transverse momentum and energy conservation are respected. This
leads to two separate transforms in the calculation, and we will
discuss below how we invert these.  This is a non-trivial issue,
since, taking into account nonleading logarithms and the running of
the coupling, resummation leads in each case to a perturbative
expression in which the scale of the coupling reflects the value of
the transform variable. Because of the singularity of the perturbative
effective coupling at $\Lambda_{\rm QCD}$, the resulting expressions
are, strictly speaking, undefined.  A closer look, however, shows that
singular contributions appear only at nonleading powers of momentum
transfer. This is an example of how perturbative resummation can
suggest the way nonperturbative dynamics is expressed in infrared safe
hard scattering functions.  In effect, perturbation theory is
ambiguous, and the resolution of its ambiguities is, by definition,
nonperturbative
\cite{Contopanagos:1994yq,Webber:1994cp,Korchemsky:1995is}.  Each
scheme for dealing with these ambiguities constitutes a specification
of perturbation theory, and implies a parameterization of
nonperturbative effects.  We hope that a joint resummation affords a
more general approach to this problem.

\subsubsection{Joint resummation method}

In the framework of joint resummation~\cite{Laenen:2000ij}, the cross
section for electroweak annihilation is given as \begin{eqnarray}
\label{resum:crsec}
\frac{d\sigma_{AB}^{\rm res}}{dQ^2\,dQ_T^2} &=& \sum_a
\sigma_{a}^{(0)}\, \int_{C_N}\, \frac{dN}{2\pi i} \,\tau^{-N}\; \int
\frac{d^2b}{(2\pi )^2} \,
e^{i{\vec{Q}_T}\cdot {\vec{b}}}\, \nonumber \\
&\times& {\cal C}_{a/A}(Q,b,N,\mu,\mu_F )\; \exp\left[ \,E_{a\bar
    a}^{\rm PT} (N,b,Q,\mu)\,\right] \; {\cal
  C}_{\bar{a}/B}(Q,b,N,\mu,\mu_F) \; , \end{eqnarray} where $\tau =Q^2/S$, $Q$
denotes the invariant mass of produced boson, and $\sigma_{a}^{(0)}$
is the Born cross section.  The flavour-diagonal Sudakov exponent
$E_{a\bar a}^{\rm PT}$ is given to next-to-leading logarithmic (NLL)
accuracy by \begin{eqnarray}
\label{resum:jointsud}
\tilde{E}_{a\bar a}^{\rm PT} (N,b,Q,\mu,\mu_F) &=&
-\int_{Q^2/\chi^2}^{Q^2} \frac{d k_T^2}{k_T^2} \; \left[
  A_a(\as(k_T))\, \ln\left( \frac{Q^2}{k_T^2} \right) +
  B_a(\as(k_T))\right] \,.  \end{eqnarray} It has the classic form of the Sudakov
exponent in the recoil-resummed $Q_T$ distribution for electroweak
annihilation, with the same functions $A$ and $B$. The latter are
perturbative series in $\as$~\cite{Dokshitzer:1980hw,Collins:1981uk};
at NLL, one only needs the second (first) order expansion of $A$
($B$)~\cite{Kodaira:1982nh}.
The only difference between the standard Sudakov exponent for $Q_T$
resummation and Eq.~\ref{resum:jointsud} comes from the lower limit of
the integrand, i.e. the function $\chi$ that organizes the logarithms
of $N$ and $b$ in joint resummation: \begin{equation}
\label{resum:chinew}
\chi(\bar{N},\bar{b})=\bar{b} + \frac{\bar{N}}{1+\eta\,\bar{b}/
  \bar{N}}\; \qquad \qquad \bar{b}\equiv b Q {\rm e^{\gamma_E}}/2 \; .
\end{equation} This choice for $\chi(\bar{N},\bar{b})$ fulfills the requirement
of reproducing the LL and NLL terms in $N$ and $b$ when $N \rightarrow
\infty$ (at fixed $b$) and $b \rightarrow \infty$ (at fixed $N$).
Moreover, it also guarantees that in the limit $b\gg N$ the quantity
$E_{a\bar a}^{\rm PT}$ expanded up to ${\cal O}(\as)$ does not have
any subleading terms with singular behavior different from the one
present in the fixed-order result.

The functions ${\cal C}(Q,b,N,\mu,\mu_F )$ in Eq.~(\ref{resum:crsec})
\begin{equation}
\label{resum:cpdf}
{\mathcal C}_{a/H}(Q,b,N,\mu,\mu_F ) = \sum_{j,k} C_{a/j}\left(N,
  \alpha_s(\mu) \right)\, {\cal E}_{jk} \left(N,Q/\chi,\mu_F\right) \,
f_{k/H}(N ,\mu_F) \; , \end{equation} contain products of parton distribution
functions $f_{k/H}$ at scale $\mu_F$ with an evolution operator ${\cal
  E}_{jk}$ between scales $\mu_F$ and $Q/\chi$. Furthermore, as
familiar from both threshold and $Q_T$ resummation, they contain hard
coefficients $C_{a/j}(N,\as)$ that are again perturbative series in
$\as$ and have the first-order expansions \begin{eqnarray} C_{q/q}^{(1)}\left(
  N,\as \right) &=& 1+ \frac{\as}{4\pi} C_F \left(-8+\pi^2
  +\frac{2}{N(N+1)} \right) \; =\;
C_{\bar{q}/\bar{q}}^{(1)}\left( N,\as \right)\; ,\\
C_{q/g}^{(1)}\left( N,\as\right)
&=&\frac{\as}{2\pi}\frac{1}{(N+1)(N+2)}\; =\;
C_{\bar{q}/g}^{(1)}\left( N,\as\right)\; .  \end{eqnarray} To be consistent with
the NLL approximation for the $E_{a\bar a}^{\rm PT}$ part of the cross
section, the evolution matrix ${\cal E}$ is derived from the NLO
solutions of standard evolution equations~\cite{Furmanski:1982cw}.
Inclusion of the full evolution of parton densities (as opposed to
keeping only the $\sim \ln \bar{N}$ part of the anomalous dimension
that dominates near threshold, see Ref.~\cite{Kulesza:2002rh}) 
extends the joint formalism and leads to resummation of collinear logarithms not all of
which are associated with threshold corrections.

Owing to the presence of the Landau pole in the strong coupling, the
jointly resummed cross section~(\ref{resum:crsec}) requires
definitions for the inverse Mellin and Fourier transforms. In pure
threshold resummation, it was proposed to parameterize the inverse
Mellin contour as~\cite{Catani:1996yz} \begin{equation}
\label{resum:cont}
N = C + z {\rm e}^{\pm i \phi} \; , \end{equation} where $C$ lies to the right of
the singularities of the parton densities, but to the left of the
Landau pole. This is a natural definition of the contour since it
decouples the Landau pole from any finite-order expansion of the
resummed cross section.  For the jointly resummed cross section in
Eq.~(\ref{resum:crsec}), if the $b$ integration is carried out in the
standard way over the real axis, one will inevitably reach the Landau
pole. This is well-known also from pure $Q_T$ resummation and led to
the introduction of a (non-perturbative) upper value $b_{\rm max}$ for
the $b$ integral, along with a redefinition $b\to b_{\ast}=b/
\sqrt{1+b^2/b_{\rm max}^2}$. To avoid introducing a new parameter, we
treat the $b$ integral in a manner analogous to the $N$ integral
above~\cite{Laenen:2000de}: were the Landau pole not present we could,
instead of performing the $b$ integral along the real axis, use
Cauchy's theorem and divert it into complex $b$ space. Our
prescription will be to use the diverted contour also in the presence
of the Landau pole. Technically, this can be achieved by writing \begin{equation} 2
\pi\, \int_0^{\infty} \, db\,b\, \,J_0(bQ_T) \,f(b) = \pi\,
\int_0^\infty db\, b\, \left[\, h_1(bQ_T,v) + h_2(bQ_T,v)\,
\right]\,f(b) \, ,
\label{resum:J0split}
\end{equation} where the $h_i$ are related to Hankel functions and distinguish
positive and negative phases in Eq.\ (\ref{resum:J0split}). We then
can treat the $b$ integral as the sum of two contours, the one
associated with $h_1$ ($h_2$) corresponding to closing the contour in
the upper (lower) half $b$-plane~\cite{Laenen:2000de}.

\subsubsection{Transverse momentum distribution for $Z$ production}

\begin{figure}[h]
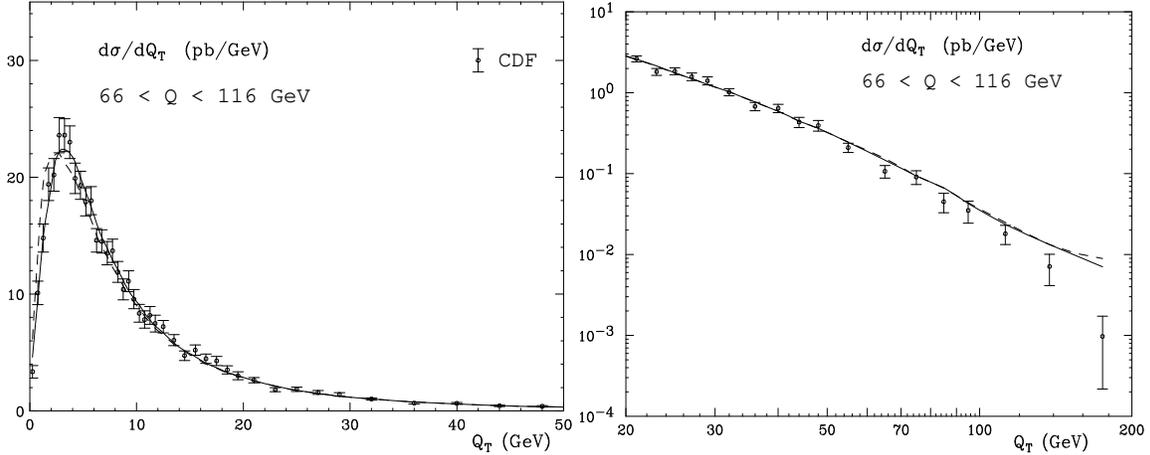

\begin{center}
  \includegraphics[width=6cm,height=7.5cm,angle=90]{cdf_lesh.epsi}
  \includegraphics[width=6cm,,height=7.5cm,angle=90]{cdf_lesh_largeqt.epsi}
\end{center}
\caption{CDF data \protect \cite{Affolder:1999jh}
  on Z production, compared to joint resummation predictions (matched
  to the ${\cal O} (\as)$ result according to
  Eq.~(\ref{resum:joint:match})) without \nprt smearing (dashed) and
  with Gaussian smearing using the nonperturbative parameter $g=0.8$
  GeV$^2$ (solid). The normalizations of the curves (factor of 1.069)
  have been adjusted in order to give an optimal description.  We use
  CTEQ5M parton distribution functions with renormalization and
  factorization scales $\mu=\mu_F=Q$.
\label{resum:fig:cdf}}
\end{figure}

A phenomenological application of the joint resummation formalism
requires matching the resummed distribution~(\ref{resum:crsec}) to the
fixed-order result $d \sigma^{\rm fixed} $, which we do as follows:
\begin{equation}
\frac{d \sigma}{d Q^2 d Q_T^2} = \frac{d \sigma^{\rm res}}{d Q^2 d Q_T^2}
-  \frac{d\sigma^{\rm exp(k)}}{d Q^2 d Q_T^2} +
\frac{d \sigma^{\rm fixed(k)}}{d Q^2 d Q_T^2} \,,
\label{resum:joint:match}
\end{equation}
with $d\sigma^{\rm exp(k)} $ denoting the $k$-th order expansion of
the resummed cross section. This way of matching in the conjugate
$(N,\,b)$ space evidently avoids any type of double counting. Given
the above prescription we calculate the $Q_T$ distribution for Z
production at the Tevatron and compare it with the latest CDF
data~\cite{Affolder:1999jh} in Fig.~\ref{resum:fig:cdf} (dashed
lines). We emphasize that the dashed lines are obtained without any
additional nonperturbative parameter.  Towards low $Q_T$, one expects
perturbation theory (as defined in our formalism) to fail without
nonperturbative input.  The expected form of such effects can be
derived from resummed perturbation theory
itself~\cite{Kulesza:2002rh,Korchemsky:1995is,Tafat:2001in} by examining the
full NLL jointly resummed exponent in the limit of small transverse
momentum of soft radiation.  The gross structure one finds is a simple
Gaussian function $-gb^2$ added to the Sudakov exponent
in~(\ref{resum:crsec}). A fit to the data gives a modest $g= 0.8$
GeV$^2$, similar to~\cite{Qiu:2000ga}.  We finally note that, unlike
standard $Q_T$ resummation, our resummed cross
section~(\ref{resum:joint:match}) stays positive even at very large
$Q_T$ so that an extra switching between a matched distribution and a
fixed order result is not required here.  We obtain a very good
agreement between data and the jointly resummed theoretical
distribution even out to large $Q_T$.

\subsection{Transverse momentum and threshold resummations in heavy
  quark production\protect\footnote{Contributing authors: N.
    Kidonakis, P.M. Nadolsky, F. Olness, R. Scalise, C.-P. Yuan}}

Theoretical analysis of QCD problems involving a single scale
parameter is highly developed; however, the problem becomes complex
when more than one distinct scale is involved. Unfortunately, most of
the interesting experimental measurements fall in the multi-scale
category.  In recent years, powerful theoretical tools have been
developed to describe such complex reactions.  The fundamental problem
with the analysis of multi-scale regimes in the context of
perturbation theory is that the truncated series in the strong
coupling $\alpha_s$ may converge poorly due to the presence of large
logarithms of dimensionless quantities.
The ``big bad logs'' can appear in a number of guises.  Logarithms of
the form ($\alpha_s^n/\qt^2) \ln^m(\qt^2/Q^2)$, $m \le 2n-1$ appear
when calculating differential distributions in transverse momentum.
Separately, we encounter logarithms of the form $\alpha_s^n
[\ln^m(1-z)/1-z]_+$, $m \le 2n-1$, with $z=Q^2/s$, in the production
of heavy quarks near threshold.  The solution to the problem is to
find all-order sums of large logarithmic terms from first principles
of the theory, such as gauge and renormalization group invariance.

The soft-gluon resummation formalism of Collins, Soper, Sterman (CSS)
\cite{Collins:1988ig,Collins:1985kg} sums large logs of the form
$\ln^m(\qt^2/Q^2)$; this formalism has enjoyed a number of theoretical
advances in recent years
\cite{Berger:2002mt}.  Logarithms of the second type, $\ln^m(1-z)$,
arise in the production of heavy quarks near threshold
\cite{Kidonakis:1997gm,Kidonakis:1996aq,Laenen:1998kp,Kidonakis:1999ze,Kidonakis:2000ui}.
More recently, new techniques have been developed to perform joint
resummation to handle both types of logs simultaneously
\cite{Laenen:2000ij}.

In this study, we address the impact of $q_T$ and threshold logarithms
on the heavy quark production. Our approach is to first examine the
$\qt$ and threshold resummation results separately, compare the
contributions in various kinematic regions, and finally consider the
joint resummation.  The dependence of results on the mass $M$ of the
heavy quarks gives an additional dimension to this problem.  Our work
on $q_T$ resummation is motivated, in part, by recently calculated
${\cal O}(\alpha_s)$ cross sections for heavy quark production in
semi-inclusive deep-inelastic scattering (SIDIS)
\cite{Kretzer:2001tc}, and by newly developed methods for resummation
of $q_T$ logarithms in massless SIDIS
\cite{Meng:1996yn,Nadolsky:2000ky,Nadolsky:1999kb}. These results can
be combined to obtain improved description of differential heavy-quark
distributions in a large range of collision energies.


Our calculations use the ACOT variable flavor number factorization
scheme \cite{Aivazis:1994pi,Collins:1998rz}, which itself is a
resummation of logarithms involving the heavy quark mass via the DGLAP
evolution equation. This scheme is successfully applied to obtain
accurate predictions for inclusive quantities (e.g. the charm
structure function $F_2^c(x,Q^2)$) both at asymptotically high
energies $s \gg M^2$ and near the threshold $s \approx M^2$.
 
For less inclusive observables, application of the ACOT scheme in a
fixed-order calculation is often not sufficient due to the presence of
logarithms different from $\ln(Q^2/M^2)$. Fixed-order calculations,
such as the calculation in \cite{Kretzer:2001tc}, have to combine
subprocesses with different numbers of final-state particles. This
combination leads to unphysical ``plus'' and ``delta-function''
distributions; it also leaves large logarithmic terms in the
coefficient functions.  The dynamical origin of this deficiency lies
in the intensive soft QCD radiation, which accompanies particle
reactions near boundaries between $n$- and $(n+1)$-particle
kinematics. In SIDIS, unphysical distributions appear in the current
fragmentation region, which in our notations corresponds to the limit
$q_T^2 \ll Q^2$. Note that this phenomenon occurs when the hard scale
of the reaction is much larger than the heavy-quark mass, {\it i.e.},
$Q^2 \gg M^2$.

The key result of the CSS resummation formalism is that all large
logarithms due to the soft radiation in the high-energy limit can be
resummed into a Sudakov exponent.  This result can be summarized in
the following master equation:\footnote{In such a short report, the
  equations presented can only be schematic; complete notations will
  be defined in our upcoming publication; see also
  \cite{Meng:1996yn,Nadolsky:2000ky,Nadolsky:1999kb}.}
\begin{equation}
\frac{d\sigma}{dQ^2\, dq_T^2}=
\frac{\sigma_0}{s} \ 
\int \frac{d^2 b}{(2\pi)^2} \ e^{i \vec q_T \cdot \vec b} \  
\left( C^{\mathrm{in}} \otimes f \right)
\left( C^{\mathrm{out}} \otimes d \right)
e^{-S} \ 
+ \sigma_{\mathrm{FO}} - \sigma_{\mathrm{ASY}} \, .
\label{resum:CSS}
\end{equation}
Here $b$ is the impact parameter, $f$ and $d$ are parton distribution
functions and fragmentation functions, respectively.  $C^{\mathrm{in}}$,
$C^{\mathrm{out}}$ contain perturbative corrections to contributions from the
incoming and outgoing hadronic jets, respectively.  The factor
$e^{-S}$ is the Sudakov exponential, which includes an all-order sum
of perturbative logarithms $\ln^m{q_T^2/Q^2}$ at $b\lesssim 1\mbox {
  GeV}^{-1}$ and nonperturbative contributions at $b\gtrsim 1\mbox{
  GeV}^{-1}$.  Finally, $\sigma_{\mathrm{FO}}$ is the fixed-order expression
for the considered cross section, while $\sigma_{\mathrm{ASY}}$ ({\it
  asymptotic piece}) is the perturbative expansion of the $b$-space
integral up to the same order of $\alpha_s$ as in $\sigma_{FO}$.  At
small $q_T$, where terms $\ln^m(q_T^2/Q^2)$ are large, $\sigma_{\mathrm{FO}}$
cancels well with $\sigma_{\mathrm{ASY}}$, so that the cross section
(\ref{resum:CSS}) is approximated well by the $b$-space integral.  At
$q_T \gtrsim Q$, where the logarithms are no longer dominant, the
$b$-space integral cancels with $\sigma_{\mathrm{ASY}}$, so that the cross
section (\ref{resum:CSS}) is equal to $\sigma_{\mathrm{FO}}$ up to higher order
corrections.

\begin{figure}[thbp]
  \includegraphics[width=0.8\textwidth]{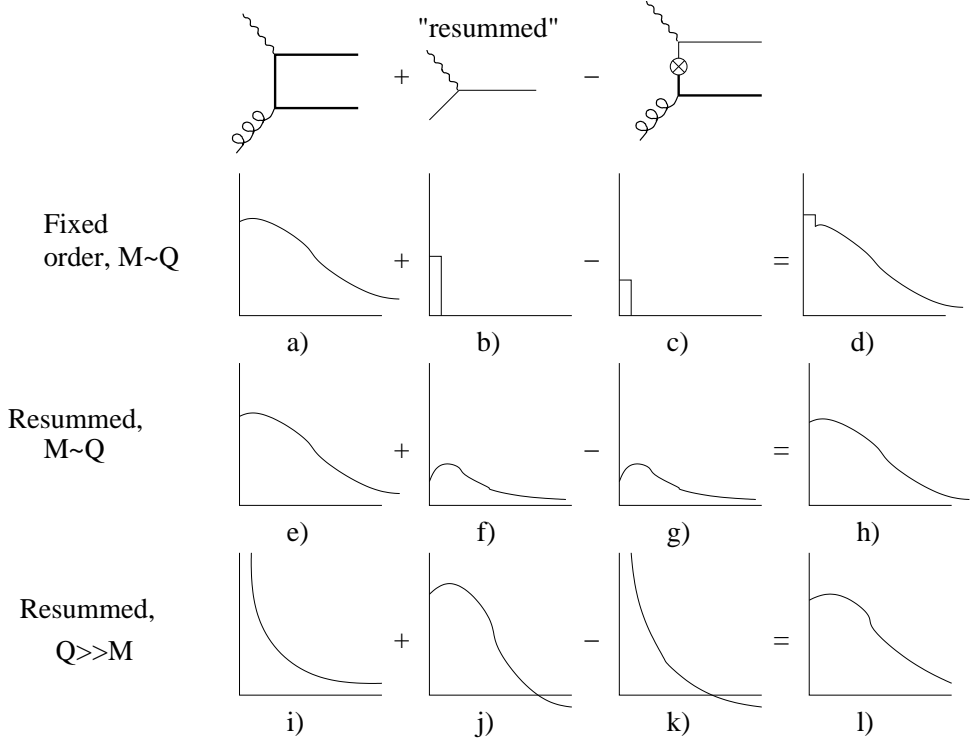}
\caption{ Balance of various terms in the ACOT scheme and resummed cross
  section. Graphs show $q_T$ on the $x$-axis and $d\sigma/dq_T^2$ on
  the $y$-axis.}
\label{resum:fig2}
\end{figure}
In extending the CSS resummation from the massless to the massive
case, we observe that the proof of the resummation formula in the
original papers \cite{Collins:1988ig,Collins:1985kg} does not require
quarks to be massless; nor does it rely on the usage of the
$\overline{MS}$ scheme for regularization of soft and collinear
singularities. Hence the reorganization of perturbative calculations
as in Eq.~(\ref{resum:CSS}) is also justified for massive quarks.
Note, however, that nonvanishing quark masses regulate collinear
singularities, so that at least some of the logarithms are not
dominant in the threshold region.
On the other hand, the discussion in Ref.~\cite{Collins:1998rz}
suggests that masses can be set to zero for those heavy quarks that
enter the hard scattering subprocess directly from the proton.

This approximation, which differs from the complete mass-dependent
cross section by higher-order terms, significantly simplifies the
analysis. We therefore drop heavy-quark masses in the hard scattering
subdiagrams with incoming heavy quark lines.  This approximation leads
to massless expressions for the perturbative Sudakov factor and
$C$-functions for quark-initiated subprocesses, and mass-dependent
$C$-functions for gluon-initiated subprocesses.

Figure~\ref{resum:fig2} qualitatively illustrates the balance of
various terms in the resummation formula in various regions of phase
space for the lowest-order contributions. First, consider a
fixed-order calculation in the simplified ACOT scheme near the
threshold. In this region, the cross section $d\sigma/dq_T^2$ is well
approximated by the ${\cal O}(\alpha_s)$ photon-gluon diagram
(Fig.~\ref{resum:fig2}a).  To this diagram, we add the lowest-order
$\gamma^* q$ term, which resums powers of $\ln(Q^2/M^2)$ and
contributes at $q_T = 0$ (Fig.~\ref{resum:fig2}b).  We also subtract
the overlap between the two diagrams (Fig.~\ref{resum:fig2}c), which
is approximately equal to, but not the same as, the $\gamma^* q$
contribution. The resulting distribution (Fig.~\ref{resum:fig2}d) is
close to the fixed-order result, but has discontinuities in the small
$q_T$ region.  These discontinuities are amplified when $Q$ increases.

In the resummed cross section, the fusion diagram still dominates near
the threshold, but now the resummed cross section and subtracted
asymptotic piece are smooth functions, which cancel well at all values
of $q_T$ (Fig.~\ref{resum:fig2}e-h). Hence, the distribution is
physical in the whole range of $q_T$. Finally, at $Q^2 \gg M^2$ the
small-$q_T$ region the $\gamma^* g$ fusion contribution is dominated
by the $1/q_T^2$ term (Fig.~\ref{resum:fig2}i).  Such singular terms
are summed through all orders in the $b$-space integral corresponding
to Fig.~\ref{resum:fig2}j and are canceled in the fusion contribution
by subtracting the asymptotic piece (Fig.~\ref{resum:fig2}k).

As mentioned above, threshold resummation organizes logs of the form
$\alpha_s^n [\ln^m(1-z)/1-z]_+$, which can yield a large contribution
when $z \rightarrow 1$
\cite{Kidonakis:1997gm,Kidonakis:1996aq,Laenen:1998kp,Kidonakis:1999ze,Kidonakis:2000ui}.
When combining the results of both the $\qt$ and threshold
resummation, an important consideration is whether there is an overlap
in resummed terms, which would result in double counting. While in a
general context this delicate problem makes the double resummation
non-trivial, this is not an issue for the low order case we are
examining here; for this reason, we can compute the contributions
separately and compare them without any complications.

Our analysis in this study will continue well past the time scale of
the present workshop, and detailed results will be described in a
separate publication. New experimental measurements will contain a
wealth of information of high precision -- if only we can extract and
analyze it. Theoretical tools must be refined to keep up with this
progress.  The resummation formalisms that we are studying here
provide an efficient way to go beyond low order perturbative
calculations and describe the heavy quark data accurately in a wide
range of kinematical variables. With these new methods at hand, we
will be well prepared for examination of new detailed information from
HERA, TEVATRON Run II, and the LHC.

\subsection{Soft gluon resummation for fragmentation
  functions\protect\footnote{Contributing author: M. Cacciari}}

When one is considering the issue of performing tests of perturbative
QCD (pQCD) predictions and of calculational techniques, production
processes involving heavy quarks (i.e. of mass $m \gg \Lambda_{QCD}$)
can be thought of as an ideal choice for at least two distinct
reasons.
\begin{itemize}
\item The large mass of the quark acts as a cutoff for collinear
  singularities, allowing perturbative calculations to give finite
  results without resorting to factorization into phenomenological
  functions, hence allowing, at least in principle, direct comparisons
  to experimental data.
\item When the quarks undergo a non-perturbative
  hadronization\footnote{This is the case for charm and bottom, while
    the top quarks decay weakly before hadronizing.} before being
  observed, say, as a heavy-light meson like $D$ or $B$, the extent to
  which this process modifies the perturbative predictions can be
  expected to be of the order of magnitude of (powers of) the ratio
  $\overline{\Lambda}/m$, where $\overline{\Lambda}$ is a hadronic
  scale of the order of a few hundred MeV . One can therefore hope to
  be less dependent on the inclusion of non-perturbative
  parametrizations (for instance parton distribution and fragmentation
  function) which, by introducing additional phenomenological degrees
  of freedom, can obscure the real predictive power of the underlying
  theoretical framework.
\end{itemize}

Such nice features do of course come at a price. Perturbative
calculations, while finite, will however contain in higher order
logarithms of the various physical scales entering the problem. For
instance, heavy quark hadroproduction at transverse momentum $p_T$ or
heavy quark fragmentation in $e^+e^-$ collisions at a center-of-mass
energy $Q$ will display $\alpha_s^2(\alpha_s^n\log^k(p_T/m))$ or
$\alpha_s^n\log^k(Q/m)$ (with $0\le k \le n$) terms respectively. When
$Q,\;p_T \gg m$ such logarithms, remnants of the screening of the
collinear singularities, grow large, eventually spoiling the
convergence of perturbation theory. Analogously, observing the heavy
quark close to a phase space boundary will constrain soft radiation
and result in Sudakov logarithms which can also grow very large.

Therefore, in order to really exploit the aforementioned possibly
superior qualities of heavy quark processes, one must first take care
of properly and accurately evaluating the perturbative result. In
particular, this means performing an all-order resummation of the
large logarithms previously mentioned, and matching to an exact fixed
order result for recovering the non-leading terms.

Resummation of ``collinear'' large logarithms to next-to-leading log
(NLL) accuracy was first performed in
Ref.~\cite{Mele:1990yq,Mele:1991cw} for $e^+e^-$ collisions. Sudakov
logarithms were also resummed to leading logarithmic order (LL).
Reference~\cite{Dokshitzer:1996ev} extended the resummation of Sudakov
logs to NLL level.

The papers previously mentioned only directly addressed heavy quark
production in $e^+e^-$ collisions.
Reference~\cite{Mele:1990yq,Mele:1991cw} did however put forward
convincing arguments about the existence of a process-independent {\sl
  heavy quark fragmentation function}. Making use of this
universality, NLL resummation of collinear logs was first performed
for hadron-hadron collisions in \cite{Cacciari:1994mq}, and then
extended to other processes as well
\cite{Cacciari:1996fs,Cacciari:1996ej}. These resummations where then
matched to complete next-to-leading (NLO) perturbative calculations
\cite{Cacciari:1998it,Cacciari:2001td} to provide reliable results in
the whole range of transverse momenta.

The universality has recently been fully exploited in
Ref.~\cite{Cacciari:2001cw}, where the heavy quark fragmentation
function has been extracted by making no reference to any explicit
production process.  The production cross section for a heavy quark of
mass $m$ at a scale $Q$ (or $p_T$) can therefore always be written as
a convolution of this process-independent function $D^{\rm
  ini}(\mu_F,m)$ and a process dependent coefficient function
$C(Q,\mu_F)$:\footnote{For simplicity we will not be showing parton
  indices and summation over them.}
\begin{equation}
\sigma(Q,m) = C(Q,\mu_F) \otimes D^{\rm ini}(\mu_F,m) + {\cal
O}((m/Q)^p) \; .
\end{equation}
The factorization scale $\mu_F$ separates the two functions, and
Altarelli-Parisi evolution can be used to resum the collinear
logarithms in $D^{\rm ini}(\mu_F,m)$ by evolving from an initial scale
$\mu_{0F} \simeq m$ up to the hard scale $\mu_F \simeq Q$, so that
$D^{\rm ini}(\mu_F,m) = E(\mu_F,\mu_{0F}) \otimes D^{\rm
  ini}(\mu_{0F},m)$.

Making use of this factorization it has been possible to resum the
Sudakov logarithms to NLL accuracy for $D^{\rm ini}(\mu_{0F},m)$ and
$C(Q,\mu_F)$ separately. The resummed result for $D^{\rm
  ini}(\mu_{0F},m)$ is, of course, also process independent (though
factorization-scheme dependent), and its Mellin moments can be written
(we use the ${\overline {\rm MS}}$ factorization-scheme) in the form
\cite{Cacciari:2001cw}:
\begin{eqnarray}
D_N^{{\rm ini}, \,S}(\alpha_s(\mu_0^2);\mu_0^2,\mu_{0F}^2,m^2) &\!\!\!=& \!\!\! 
\left\{ 1 + \frac{\alpha_s(\mu_0^2) C_F}{\pi}\left[
-\frac{\pi^2}{6} + 1 - \gamma_E^2 + \gamma_E + \left(\frac{3}{4} -
\gamma_E\right) \ln\frac{\mu_{0F}^2}{m^2}\right]\right\}\nonumber\\
\label{resum:sudHQini}
&\!\!\!\times&\!\!\! \exp \Bigl[ \ln N \;g^{(1)}_{\rm ini}(\lambda_0) + 
g^{(2)}_{\rm ini}(\lambda_0,m^2/\mu_0^2;m^2/\mu_{0F}^2) 
\Bigr]\, .
\label{resum:Sudres}
\end{eqnarray}
Explicit results for the $g_{\rm ini}^{(1)}$ and $g_{\rm ini}^{(2)}$
functions can be found in Ref.~\cite{Cacciari:2001cw}.

Contrary to that of the ``collinear'' $\log(p_T/m)$ terms, resummation
of the Sudakov logs is not of direct phenomenological importance for
studying inclusive $p_T$ distributions in heavy quark production at
the Tevatron or the LHC.  Indeed, unless one gets very close to phase
space limits, only low-$N$ moments of $D_N^{{\rm ini}}$, around $N
\sim 5$, are relevant\cite{Nason:1999ta}. Sudakov resummation, on the
other hand, starts producing visible effects around $N \sim 10$
\cite{Cacciari:2001cw}.

Having a reliable theoretical prediction for $D^{{\rm
    ini}}(\mu_{0F},m)$ (and, with the analogous NLL Sudakov
resummation of $C(Q,\mu_F)$ also performed in
Ref.~\cite{Cacciari:2001cw}, for the measurable single-inclusive
heavy-quark cross section in $e^+e^-$ collisions) is however useful
when extracting from experimental data the non-perturbative
contributions which must complement the perturbative calculation in
order to get a good description of the data. A generic ``hadron
level'' cross section for $B$ or $D$ meson production in $e^+e^-$ or
hadron collisions can be described by\footnote{The coefficient
  function $C(Q,\mu_F)$ and the heavy quark function $D^{\rm
    ini}(\mu_F,m)$ can of course be evaluated to different levels of
  theoretical accuracy, for instance including NLL Sudakov resummation
  and matching to complete NLO calculations, as in
  Ref.~\cite{Cacciari:2001cw}. Sudakov resummation for the coefficient
  function $C(Q,\mu_F)$ is at present only available for $e^+e^-$
  collisions. It is however less important for hadronic collisions,
  where phase space edges (and hence large-$N$ moments), are rarely
  experimentally probed (an exception to this might be $D$ meson
  photoproduction data at HERA).}
\begin{equation}
\sigma_{\rm had}(Q,m) = C(Q,\mu_F) \otimes D^{\rm ini}(\mu_F,m) 
\otimes D^{\rm
np}(\epsilon_1,...,\epsilon_n)\; .
\label{resum:hadlev}
\end{equation}
The function $D^{\rm np}(\epsilon_1,...,\epsilon_n)$ can represent
either a phenomenological parametrization or an attempt to establish
the form of such a contribution through the analysis of power
corrections.  It can be argued~\cite{Cacciari:1997wr,Cacciari:1997du}
that it will be, at least to some reasonable accuracy level, as
process independent as its perturbative counterpart $D^{\rm
  ini}(\mu_F,m)$. In any case, it will depend on a set of parameters
$(\epsilon_1,...,\epsilon_n)$ which, pending a full solution of
non-perturbative QCD, can only be determined by comparison with the
data.

\begin{figure}[thbp]
  \includegraphics[width=\textwidth]{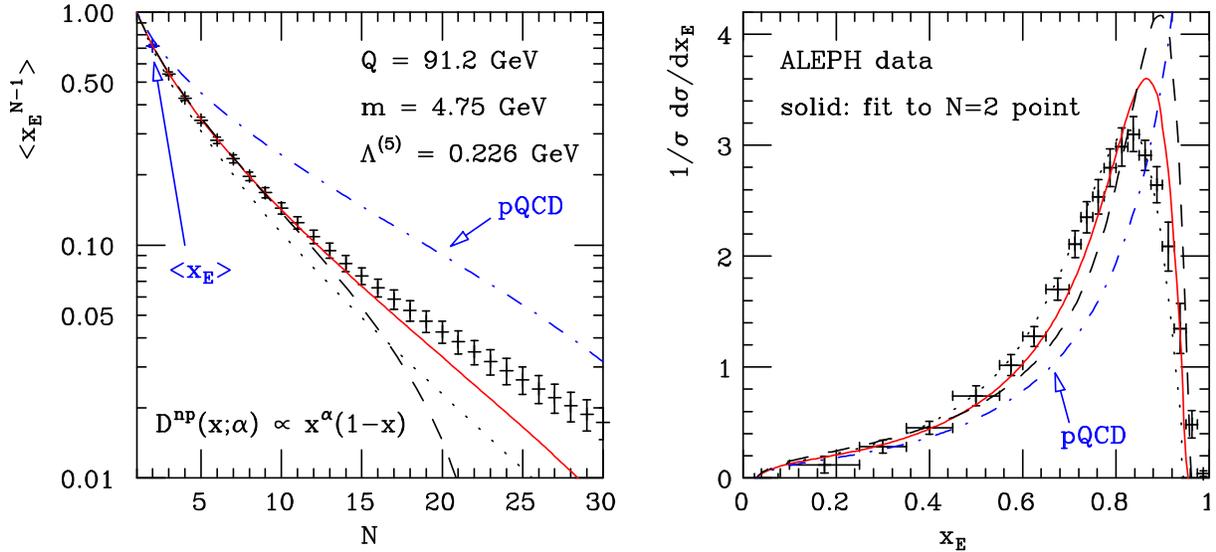}
\caption{ALEPH data for weakly-decaying $B$ mesons are fitted to a
  convolution of the perturbative prediction and a non-perturbative
  function (see text).  Solid line: $\alpha = 27.45$, fitted in $N$
  space, fixed by the $\langle x_E\rangle$ (i.e.  $N=2$) point only.
  Dashed line: $\alpha = 29.1$, fitted as before, but no Sudakov
  resummation in the pQCD part. Dotted line: $\alpha = 17.7$, fitted
  with resummation and using all points in $x_E$ space. The dot-dashed
  line is the purely perturbative contribution, as defined in
  Ref.~\cite{Cacciari:2001cw}.}
\label{resum:fig1}
\end{figure}

An example of such a determination is given in Fig.~\ref{resum:fig1}.
ALEPH $e^+e^-$ data~\cite{Heister:2001jg} for weakly-decaying $B$
meson production are fitted to the Kartvelishvili et
al.~\cite{Kartvelishvili:1978pi} non-perturbative, one-parameter
function\footnote{Despite its simplicity, this function is
  known~\cite{Heister:2001jg} to provide one of the best fits to the
  experimental data. Alternatively, one could use the form
\begin{equation}
D_N^{\rm np}(\overline{\Lambda})
 = \exp\left\{\sum_{k=1}^\infty \frac{(-1)^k}{k!\; k
}\left(\frac{\overline{\Lambda} (N-1)}{m}\right)^k\right\}\;,
\end{equation}
suggested by the structure of the Sudakov resummation for $D^{\rm
  ini}(\mu_F,m)$. Upon identyifing $\alpha$ with
$2m/\overline{\Lambda}$, the results of the two expressions are
virtually indistinguishable, even when including only a few terms of
the series.}  $D^{\rm np}(x;\alpha) = (\alpha+1)(\alpha+2) x^\alpha
(1-x)$.  The solid line is obtained by fitting Eq.~\ref{resum:hadlev}
to the $\langle x_E\rangle$ point only (the average energy fraction),
in $N$ space. The dashed line is obtained by performing the same kind
of fit but excluding the Sudakov resummation of
Eq.~\ref{resum:Sudres}. One can see this degrades the result when
comparing to $x_E$ data in the right panel.

While giving in this particular example a worse result, in $x_E$
space, than the one obtained by fitting the $x_E$ distribution itself
(dotted line in Fig.~\ref{resum:fig1}), the determination of the
non-perturbative parameter(s) from $N$ space data is however
theoretically more self-consistent. Non-perturbative effects increase
in size as $x_E$ gets closer to one. Hence when fitting in $x_E$ space
one is not only using many experimental points to fix a single (or a
few) parameter(s) while however not having the functional form on firm
theoretical grounds, but one is also adjusting the non-perturbative
contribution to fit points where its weight greatly differs. In $N$
space, on the other hand, the non-perturbative contribution increases
with $N$. Fitting only $\langle x_E\rangle$ means using the point
where it is smallest, and then one can still check if the shape of the
curve in $x_E$ is correctly predicted.

A theoretically sensible program for evaluating heavy meson cross
sections at the Tevatron or the LHC would therefore entail:
\begin{itemize}
\item[a)] Analysing the structure of the process-independent function
  $D^{\rm ini}(\mu_{0F},m)$, and trying to extract information about
  its power corrections, which build the non-perturbative function
  $D^{\rm np}$. Full results for this step are at present not yet
  available in the literature, and work is in progress.  Some of the
  characteristics of the non-perturbative contribution are however
  well established. It is for instance well known
  \cite{Jaffe:1994ie,Nason:1997pk} that the leading power correction
  is linear in $\overline{\Lambda}/m$, where $m$ is the heavy quark
  mass and $\overline{\Lambda}$ a hadronic scale of the order of a few
  hundred MeV.
\item[b)] Fitting the first few $N$ moments of $e^+e^-$ data, either
  with a theoretically motivated non-perturbative function (see
  previous item) or simply with a given functional form, and
  extracting the phenomenological parameters. Ideally, one would fit
  only as many moments as the number of free parameters.
\item[c)] Checking the resulting shape in $x_E$ space against $e^+e^-$
  data. Good agreement might not be necessary if one is only
  interested in predicting hadronic cross sections\footnote{We pointed
    out before that $N$ moments around $N\sim 5$ are more important to
    this aim, and we can see from Fig.~\ref{resum:fig1} that fitting
    in $x_E$ space actually leads to a {\sl poorer} description of
    low-$N$ moment data. The magnitude of the discrepancy is in this
    case of the order of 10\%, in agreement with the observations made
    in Ref.~\cite{Nason:1999ta}.}, but it provides a good guidance as
  to how sensible the chosen non-perturbative parametrization is.
\item[d)] Evaluating the hadronic cross sections at the Tevatron or
  the LHC using the fitted non-perturbative form, together with the
  proper\footnote{It is of course important that the perturbative
    terms included at this stage in $D^{\rm ini}(\mu_{0F},m)$ match
    the ones employed when fitting the non-perturbative contribution.}
  $D^{\rm ini}(\mu_{0F},m)$, and a specific coefficient function.  A
  matched approach like the one presented in
  Ref.~\cite{Cacciari:1998it} can also be used at this stage, to
  improve the result at transverse momentum values of the order of the
  heavy quark mass or below.
\end{itemize}

\subsection{Analytic resummation and power
  corrections\protect\footnote{Contributing author: L. Magnea}}

A characteristic feature of resummations is the fact that they
highlight the inherent limitations of perturbation theory. In QCD,
these limitations are particularly severe, reflecting the complex
nonperturbative structure of the theory: the perturbative expansion
for any IR safe observable must be divergent, and furthermore not
summable even \`a la Borel, as a consequence of the presence of the
Landau pole in the infrared evolution of the running coupling. In the
case of IR safe quantities, this ambiguity in the perturbative
prediction can be shown to be suppressed by powers of the hard scale,
and it is interpreted as a signal of the occurence of nonperturbative
corrections of the same parametric size.

Soft gluon resummations always lead to expressions in which the Landau
pole appears explicitly. On the other hand, it should be kept in mind
that resummations at a fixed logarithmic accuracy do not in general
yield correct estimates for the size of the relevant power correction.
This was first shown in Ref.~\cite{Beneke:1995pq} in the case of the
inclusive Drell--Yan cross section. In that case, threshold
resummation carried out at NLL level suggests a power correction
proportional to $N \Lambda/Q$, where $N$ is the Mellin variable
conjugate to the soft gluon energy fraction in the CM frame; it turns
out, however, that this power correction is cancelled by the inclusion
of nonlogarithmic terms, which can be exactly evaluated in the large
$n_F$ limit. This observation lead the authors of
Ref.~\cite{Catani:1996yz} to introduce the ``minimal prescription''
for the evaluation of the partonic cross section in momentum space;
with this prescription, the inverse Mellin transform of the partonic
cross section is evaluated picking a contour of integration located to
the left of the Landau pole: as a consequence, the presence of the
pole does not affect the result, and information about power
corrections must be supplied independently.

It should be emphasized, however, that a more precise relationship
between threshold resummations and power corrections can still be
established, by fully exploiting the factorization properties of soft
and collinear radiation. The refactorization of the partonic cross
section proposed in Ref.~\cite{Sterman:1987aj}, in fact, is valid up
to corrections that are suppressed by powers of $N$: thus if the
functions involved are evaluated maintaining this accuracy, one can
expect that they should encode correctly at least the information
concerning power corrections of the form $N^p (\Lambda/Q)^m$ for the
leading values of $p$ and $m$, while corrections suppressed by extra
powers of $N$ might still be missed. This fact has been verified in
Ref.~\cite{Sterman:1999gz}, where the cancellation of the leading
power correction for the Drell--Yan cross section, established in
Ref.~\cite{Beneke:1995pq} in the large $n_F$ limit, was reproduced. 
In the more general setting of joint resummation, the same result was
obtained in Ref.~\cite{Laenen:2000ij}, and yet a different approach to
reproduce it is discussed in the next section~\cite{Gardi:2001di}.

One may wonder to what extent phenomenological results are affected by
the inclusion of power suppressed effects, or, in other words, by the
choice of different methods to regulate the Landau singularity. The
minimal prescription can be seen as one possible regulator, operating
at the level of the inverse Mellin transform.
Recently~\cite{Magnea:2000ss,Magnea:2001ge} it was observed that
another gauge invariant regularization procedure is given simply by
dimensional regularization: using a dimensionally continued version of
the running coupling, one may show that the Landau singularity moves
away from the integration contour for sufficiently large values of $d
= 4 - 2 \epsilon$. Resummed expression are then analytic functions of
the coupling and of $\epsilon$, and the Landau singularity appears as
a cut. To see how this comes about, recall that in $d = 4 - 2
\epsilon$, the $\beta$ function acquires $\epsilon$ dependence, so
that
\begin{equation}
\beta(\epsilon, \alpha_s) \equiv \mu \frac{\partial \alpha_s}{\partial \mu} = 
- 2 \epsilon \alpha_s + \hat{\beta} (\alpha_s)~, 
\label{resum:beta}
\end{equation}
where $\hat{\beta} (\alpha_s) = - b_0 \alpha_s^2/(2 \pi) + {\cal
  O}(\alpha_s^3)$.  As a consequence, the running coupling also
becomes dimension dependent.  At one loop,
\begin{equation}
\overline{\alpha} \left(\frac{\mu^2}{\mu_0^2}, \alpha_s(\mu_0^2), 
\epsilon \right) = 
\alpha_s(\mu_0^2) \left[\left(\frac{\mu^2}{\mu_0^2}\right)^\epsilon - 
\frac{1}{\epsilon} \left(1 - \left(\frac{\mu^2}{\mu_0^2}\right)^\epsilon
\right) \frac{b_0}{4 \pi}\alpha_s(\mu_0^2) \right]^{-1}~.
\label{resum:loalpha}
\end{equation}
It is easy to see that the running coupling in
Eq.~(\ref{resum:loalpha}) has a qualitatively different behavior with
respect to its four dimensional counterpart. First of all, it vanishes
as $\mu^2 \to 0$ for $\epsilon < 0$, as appropriate for infrared
regularization. This is a consequence of the fact that the one loop
$\beta$ function, for $\epsilon < 0$, has two distinct fixed points:
the one at the origin in coupling space is now a Wilson--Fisher fixed
point, whereas the asymptotically free fixed point is located at
$\alpha_s = - 4 \pi \epsilon/b_0$. Furthermore, the location of the
Landau pole becomes $\epsilon$ dependent, and it is given by
\begin{equation}
\mu^2 = \Lambda^2 \equiv Q^2 \left(1 + \frac{4 \pi \epsilon}{b_0 
\alpha_s(Q^2)} \right)^{-1/\epsilon}~.
\label{resum:lapo}
\end{equation}
The pole is not on the real axis in the $\mu^2$ plane, {\it i.e.} not
on the integration contour of typical resummed formulas, provided
$\epsilon < - b_0 \alpha_s(Q^2)/(4 \pi)$. One may then perform the
scale integrals, and get analytic expressions in which the ambiguity
due to nonperturbative phenomena shows up as a cut originated by the
Landau pole in the integrand. This idea was applied in
Ref.~\cite{Magnea:2000ss} to the resummed quark form factor, which can
be evaluated in terms of simple analytic functions since it is a
function of a single scale.  Here we will briefly illustrate the
application of the formalism to the simplest resummed partonic cross
sections (DIS and Drell--Yan), at the leading-log
level~\cite{Magnea:2001ge}.

\subsubsection{An example: LL threshold resummation for DIS and Drell--Yan}
\label{resum:disdy}

Resummation of threshold ($x \to 1$) logarithms, both for DIS and
Drell-Yan, was performed at NNL level
in~\cite{Sterman:1987aj,Catani:1989ne}.  A formulation closer to the
present approach was later given in~\cite{Contopanagos:1997nh}.
Applying the latter formalism, one may express the Mellin transform of
the DIS structure function $F_2(x, Q^2/\mu^2,
\alpha_s(\mu^2),\epsilon)$, at the LL level, as a simple exponential
\begin{equation}
F_2 \left(N, \frac{Q^2}{\mu^2}, \alpha_s(\mu^2), \epsilon \right) = 
F_2 \left(1 \right) ~\exp \left[ \frac{C_F}{\pi} \int_0^1 d z 
\frac{z^{N - 1} - 1}{1 - z} 
\int_0^{(1 - z) Q^2} \frac{d \xi^2}{\xi^2}~
\bar{\alpha} \left(\frac{\xi^2}{\mu^2},\alpha_s(\mu^2),\epsilon
\right) \right]~.
\label{resum:dis1} \nonumber
\end{equation}
Integration of the running coupling around $\xi^2 = 0$ generates the
leading collinear divergences, which can be factorized by subtracting
the resummed parton distribution
\begin{equation}
\psi \left(N, \frac{Q^2}{\mu^2}, \alpha_s(\mu^2), 
\epsilon \right) = \exp \left[ \frac{C_F}{\pi} \int_0^1 d z 
\frac{z^{N - 1} - 1}{1 - z} \int_0^{Q^2} \frac{d \xi^2}{\xi^2}~
\bar{\alpha} \left(\frac{\xi^2}{\mu^2},\alpha_s(\mu^2),\epsilon
\right) \right]~. \label{resum:psims} \nonumber
\end{equation}
The IR and collinear finite resummed partonic DIS cross section is
then defined by taking the ratio of Eqs.~(\ref{resum:dis1}) and
(\ref{resum:psims}), as $\widehat{F}_2 = F_2/\psi$.

The integration over the renormalization scale $\xi$ is easily
performed by means of the change of variables $d \xi/\xi = d
\alpha/\beta (\epsilon, \alpha)$, obtaining the compact expression
\begin{equation}
\widehat{F}_2 \left(N, \frac{Q^2}{\mu^2}, \alpha_s(\mu^2), 
\epsilon \right) = \widehat{F}_2 \left(1\right) \exp \left[
- \frac{4 \pi C_F}{b_0} \int_0^1 \frac{z^{N - 1} - 1}{1 - z}
\log \left( \frac{\epsilon + a((1 - z) 
Q^2)}{\epsilon + a(Q^2)} \right) \right]~, 
\label{resum:dis2} \nonumber
\end{equation}
where $a(\mu^2) = b_0 \alpha_s (\mu^2)/(4 \pi)$.
Eq.~(\ref{resum:dis2}) is manifestly finite, though ambiguous due to
the cut, as $\epsilon \to 0$.

The expected power correction can now be evaluated by taking the limit
$\epsilon \to 0$ with $\alpha_s(Q^2)$ fixed. This limit depends on how
the cut is approached, and the size of the ambiguity is easily
evaluated.  One finds
\begin{equation}
\delta \widehat{F}_2 \left( N, \alpha_s (Q^2) \right) \propto N 
\frac{\Lambda^2}{Q^2}~ \left( 1 + {\cal O} \left(\frac{1}{N}\right) + 
{\cal O} \left(\frac{\Lambda^2}{Q^2} \right) \right)~,
\end{equation}
as expected in DIS.

The resummed expression for the Drell-Yan partonic cross section, at
the leading $\log N$ level, is very similar. One finds
\begin{equation}
\widehat{\sigma}_{DY} \left(N, \frac{Q^2}{\mu^2}, \alpha_s(\mu^2), 
\epsilon \right) = \frac{\sigma_{DY} \left(N, \frac{Q^2}{\mu^2}, 
\alpha_s(\mu^2), \epsilon \right)}{\psi^2 \left(N, \frac{Q^2}{\mu^2}, 
\alpha_s(\mu^2), \epsilon \right)}~,
\end{equation}
where $\sigma_{DY}$ differs from $F_2$ because of a factor of two in
the exponent (due to the presence of two radiating quarks in the
initial state for the DY process), and because phase space dictates
that the upper limit of the scale integration should be $(1 - z)^2
Q^2$ instead of $(1 - z) Q^2$. Thus one finds
\begin{equation}
\delta \widehat{\sigma}_{DY} \left( N, \alpha_s (Q^2) \right) 
\propto N \frac{\Lambda}{Q}~
\left( 1 + {\cal O} \left(\frac{1}{N}\right) + {\cal O} \left(
\frac{\Lambda}{Q} \right) \right)~.
\label{resum:dypc}
\end{equation}
Eq.~(\ref{resum:dypc}) is the result that must be expected from a LL
resummation, in agreement with~\cite{Contopanagos:1994yq}. On the
other hand, one may note that the cancellation of this leading power
ambiguity, in the context of the refactorization formalism of
Ref.~\cite{Sterman:1987aj}, is of a purely kinematical nature: the
non--logarithmic terms needed for the cancellation arise by matching
the kinematic variable used in computing $\sigma_{DY}$ (energy
fraction in the CM frame) with the one used in defining the parton
distribution $\psi$ (light--cone momentum fraction). The cancellation
then clearly survives dimensional continuation.

\subsubsection{Directions for further work}
\label{resum:wor}

It is interesting to notice that Eq.~(\ref{resum:dis2}) and its
generalizations to the Drell--Yan process and to subleading logarithms
can be seen as the first ingredients of a prescription to handle the
Landau pole in phenomenological applications, similar but not
identical to the minimal prescription. Performing the $z$ integration
for a fixed negative value of $\epsilon < - a(Q^2)$ one finds an
analytic function whose inverse Mellin transform can be computed
without having to worry about the Landau pole at all. To illustrate
this fact, note that if one introduces the well--known
substitution~\cite{Catani:1989ne}
\begin{equation}
\int_0^1 d z \frac{z^{N-1} - 1}{1 - z} f(z) \rightarrow 
- \int_0^{1 - 1/N} d z \frac{1}{1 - z} f(z)~,
\label{resum:ca32}
\end{equation}
valid at LL level, the integration in the exponent can be performed
analytically, even for finite $\epsilon$. At LL level and in the
$\overline{MS}$ scheme, for example, one finds that the exponent of
the resummed partonic Drell--Yan cross section reads
\begin{eqnarray}
E_{DY}^{(\overline{MS})} \left( N, \alpha_s (Q^2) \right) & = & 
- \frac{8 C_F}{b_0} \left[\log N \log \left( -\frac{a(Q^2)}{\epsilon} 
N^{2 \epsilon} \right) - \epsilon \log^2 N \right. \label{resum:edyms} \\ 
& - & \left. \frac{1}{2 \epsilon} \left({\rm Li}_2 \left(1 + 
\frac{\epsilon}{a(Q^2)} \right) - {\rm Li}_2 \left(\frac{a(Q^2) + 
\epsilon}{a(Q^2) N^{2 \epsilon}} \right) \right) \right]~. \nonumber
\end{eqnarray}
Eq.~(\ref{resum:edyms}) is essentially the dimensional continuation of
the function $g_1$ of Ref.~\cite{Catani:1989ne}, in the
$\overline{MS}$ scheme. A similar expression can be derived in the DIS
scheme, and slightly more cumbersome ones can be written for the NLL
function $g_2$ in either scheme. It would be interesting to develop
such a `dimensional prescritpion' in an actual phenomenological
application.

To conclude, we have observed that dimensional continuation can be
used to regulate in a gauge invariant way the Landau singularity,
which characterizes resummed expressions for QCD amplitudes and cross
sections.  Applying the formalism to the DIS and Drell--Yan cross
sections reproduces the known results for the expected power
corrections, and the method may be turned into a practical
prescription to evaluate inverse Mellin transforms bypassing the
Landau pole. Possible interesting generalizations include applications
to existing resummations for event shapes in $e^+ e^-$ annihilation
and for the production of coloured final states in hadronic
collisions. One may safely say that the relationship between
resummations and power corrections has not yet been fully explored.
New approaches, such as the one of Ref.~\cite{Gardi:2001di}, or a
sharpening of old tools, may yet yield a deeper insight both in the
theory and in the phenomenology of perturbative QCD.

\subsection{Threshold Resummation and Power Corrections 
  in the Drell-Yan Process by Dressed Gluon
  Exponentiation\protect\footnote{Contributing author: E. Gardi}}

The Drell-Yan (DY) process, where a lepton pair is produced in a
hadronic collision $h_a + h_b\longrightarrow l^+l^- + X$, is a
classical example where initial-state radiation determines the cross
section. At the perturbative level, \beq \frac{d\sigma}{d
  Q^2}=\left(\frac{4\pi\alpha_{\QED}^2}{9Q^2\, s}\right)
\sum_{i,j}\frac{dx_i}{x_i}\,\frac{dx_j}{x_j}\,f_a(x_i,Q^2)\,f_b(x_j,Q^2)
\, w_{ij}(z,Q^2).  \end{equation} Here $f_a(x_i,Q^2)$ are the twist-two parton
distribution functions, $w_{ij}(z,Q^2)$ is the partonic cross section
and $z=Q^2/(p+\bar{p})^2$ where $p$ and $\bar{p}$ are the momenta of
the incoming quark and antiquark.

We consider the threshold region where the invariant mass of the
produced lepton pair, $Q^2$, approaches the total center of mass
energy $s$, so at the partonic level $z\longrightarrow 1$.
Conseqently, the total energy of QCD radiation in the final state is
strongly constrained, the physical scale for gluon emission $(1-z)Q$
is low and that multiple emission has a significant r\^ole.
Perturbatively, this implies that fixed-order calculations are
insufficient, and that large Sudakov logs, $\ln (1-z)$, must be
resummed to all
orders~\cite{Sterman:1987aj,Collins:1988ig,Catani:1989ne,Contopanagos:1997nh,Catani:1991rr,Vogt:2000ci}.
It is clear that close enough to the threshold non-perturbative
corrections are enhanced as
well~\cite{Qiu:1991xx,Contopanagos:1994yq,Korchemsky:1995is,Beneke:1995pq,Akhoury:1997pb}:
they appear as powers of~$1/(Q(1-z))$ rather than as powers of~$1/Q$.
Contrary to structure functions in deep inelastic scattering, the DY
cross section does not have an operator product expansion. The
analysis of power correction must therefore rely on perturbative
tools, namely on renormalons. Infrared renormalons reflect the
sensitivity of Feynman diargams to the behaviour of the coupling in
the infrared. This sensitivity manifests itself in the large order
behaviour of the perturbative expansion.  Assuming that power
corrections which are associated with renormalons dominate, the form
of the power corrections can be deduce from the lowers order Feynman
diagrams with an off-shell
gluon~\cite{Beneke:1995qe,Ball:1995ni,Dokshitzer:1996qm}.  In the DY
case the lowest order Feynman diagram with an off-shell
gluon~\cite{Beneke:1995pq} (or the leading term in the flavour
expansion) reflects sensitivity of the form $1/(Q^2(1-z)^2)$. The
absence of a correction $1/(Q(1-z))$ is very intriguing, particularly
because such corrections do appear in other QCD observables, e.g. in
event-shape variables~\cite{Dokshitzer:1996qm} which share with the DY
case the structure of leading and next-to-leading Sudakov logs. (In
fact, the similarity is superficial~\cite{Gardi:2001di}. The
difference between the pattern of power corrections, as well as
sub-leading Sudakov logs, originates in the fact that in DY the
radiation is from initial-state partons and the constrain is on the
total energy, whereas in event shapes the radiation is from
final-state partons and the constrain is on the invariant mass of the
jet.).  The main problem with the renormalon method, however, is that
higher order Feynman diagrams, that are sub-leading in the flavour
expansion may generate stronger corrections~$\sim 1/(Q(1-z))$. The
diagrams which might be relevant are purely
non-Abelian~\cite{Korchemsky:1996iq}: the emission of a (dressed)
gluon off a virtual gluon, which is exchanged between the initial
partons.  Currently, this question remains open and we do not address
it here.

Since non-perturbative corrections, and particularly those associated
with the running coupling, cannot be unambiguously separated from the
resummed perturbation theory, whenever power corrections are
non-negligible, renormalon resummation must be employed. Perturbative
terms are alway parametrically larger than the ambiguous power
correction and they admit a different functional form. Therefore, the
resummation cannot be replaced by parametrization of power
corrections, but only supplemented by it. A most striking
demonstration of this fact is
provided~\cite{Gardi:1999dq,Gardi:2001ny,Gardi:2002bg} by the analysis
of event shape variables, where not only the magnitude of the power
correction but also the extracted value of $\alpha_s$ crucially depend
on the resummation.  In particular, in the calculation of differential
cross sections, such as the DY, in the threshold region, a fixed
logarithmic accuracy is insufficient. Power accuracy can be reached
only if the Sudakov exponent itself is evaluated to such accuracy.
This is the aim of DGE~\cite{Gardi:2001ny,Gardi:2002bg,Gardi:2001di}.

Let us now briefly describe the application of DGE in the case of
DY~\cite{Gardi:2001di}. The first stage is to evaluate the cross
section at the level of a single gluon emission. The gluon is assumed
to be off-shell. The gluon virtuality $k^2\equiv \lambda Q^2$ provides
the argument of the running coupling in the renormalon calculation.
Being interested in logarithmically enhanced terms, and since such
terms originate in the case under cosideration from the phase-space
region where {\em all} components of the gluon momentum are small, the
soft approximation can be used.  The partonic cross section is,
\begin{eqnarray}
\label{resum:DY_cs1}
\left.w(z,Q^2)\right\vert_{\osg}\simeq
\frac{C_F \alpha_s}{2\pi}
\,\int_{\beta_1}^{\beta_2}
\,d\beta\,\left[
\frac{2}{\left(1-z-\beta\right)\beta}
\right],
\end{eqnarray}
where $\beta=k_{+}/p_{+}=2k\bar{p}/2p\bar{p}$ is the longitudinal
momentum fraction of the gluon in the direction of the emitting quark
($p=(p_{+},0,0)$) in the gauge $A_{+}=0$, and the phase-space limits
(see~\cite{Gardi:2001di}) are $\beta_{1,2}=(1-z+\lambda\pm\Delta)/2$,
with $\Delta\equiv \sqrt{(1-z)^2-2\lambda(1+z)+\lambda^2}$.

The leading order result in Eq.~(\ref{resum:DY_cs1}) is promoted to a
resummed one by integrating over the running coupling.  The Borel
representation of the Single Dressed Gluon (SDG) partonic cross
section is \begin{equation} \left.w(z,Q^2)\right\vert_{\SDG}=\frac{C_F
  \alpha_s}{2\pi}\,\int_0^{\infty}d{u}\,\exp\left(-{{u} \ln
    Q^2/\bar{\Lambda}^2}\right)\,B_{\SDG}(z,u)\,
\frac{\sin\pi{u}}{\pi{u}}\,A_B(u), \end{equation} where $A_B(u)=1$ for the
1-loop running coupling, which is used below as an example, and
\begin{eqnarray}
\label{resum:B_SDG_DY}
B_{\SDG}(z,u) &=&
\,\frac{4}{1-z}\,\int_{1}^{(1-\sqrt{z})^2} \,d\lambda\,\lambda^{-u-1} \,\\
& &+\,\int_{0}^{(1-\sqrt{z})^2} {d\lambda} \,\lambda^{-u-1} \,
\,\left[\frac{4}{\sqrt{(1-z)^2-2\lambda(1+z)+\lambda^2}}-\frac{4}{1-z}
\right]. \nonumber
\end{eqnarray}
The modification of the lower integration limit over $\lambda$ for the
singular~$1/({1-z})$ term corresponds to factorization of gluons with
virtuality smaller than $Q^2$ into the parton distribution factors
$f_{a,b}(x_i,Q^2)$.

In Eq.~(\ref{resum:B_SDG_DY}) the phase-space limits are exact: the
upper integration limit was deduced from the condition $\beta_{2}\leq
\beta_1$, yielding~\hbox{$\Delta({\lambda_{\max}})=0$}, and therefore
$\lambda_{\max}=(1-\sqrt{z})^2$.  Since the integration over $\lambda$
is restricted to $\lambda \ll 1-z$, one can replace the integration
limit by $\lambda_{\max}\simeq (1-z)^2/4$ and approximate the
$\lambda$-dependent denominator in Eq.~(\ref{resum:B_SDG_DY}) by
$\sqrt{(1-z)^2-4\lambda}$. We thus find
\begin{eqnarray}
\label{resum:B_x_DY}
B_{\SDG}(z,u) &=&\frac{4}{1-z}\, \frac1u\,\left[
1-\,\frac{\sqrt{\pi}\,\Gamma(1-u)}{\Gamma(\frac12
-u)}\,\left(\frac{1-z}{2}\right)^{-2u}\right].
\end{eqnarray}
As was stressed in~\cite{Beneke:1995pq}, a further approximation where
also the term $4\lambda$ is neglected does not influence the leading
and next-to-leading Sudakov logs, however, since the integral in
Eq.~(\ref{resum:B_SDG_DY}) extends to $\lambda_{\max}\simeq
(1-z)^2/4$, it is not legitimate for power correction analysis.

At the second step of the calculation, the SDG result is
exponentiated: under the assumption of independent emission, the cross
section with any number of gluons can be written, in Mellin space,
$w(N,Q^2)\equiv \int_0^1 dz\, z^{N-1}\, w(z,Q^2) $, as
\begin{eqnarray}
\label{resum:DY_cs_exp}
\left.\ln w(N,Q^2)\right\vert_{\DGE}&=&\int_0^1 dz \left(z^{N-1}-1\right)\left. w(z,Q^2)\right\vert_{\SDG}\nonumber \\
&=&\frac{C_F \alpha_s}{2\pi}\,\int_0^{\infty}d{u}\,\exp\left(-{{u} \ln
Q^2/\bar{\Lambda}^2}\right)\,B_{N}(u)\,
 \frac{\sin\pi{u}}{\pi{u}}\,A_B(u),
\end{eqnarray}
where the Borel function is \begin{equation}
\label{resum:BN_DY} 
B_N(u)=\int_0^1 dz \left(z^{N-1}-1\right)B_{\SDG}(z,u)=2\left(e^{2u\ln
    N}-1\right)\Gamma(-u)^2-\frac4u\ln N.  \end{equation}

Eqs.~(\ref{resum:DY_cs_exp}) and (\ref{resum:BN_DY}) summarize our
final result for the DGE of the DY cross section. It contains both
perturbative and non-perturbative information. At the perturbative
level the exponent can be written in the standard way, in terms of
functions which sum all the contributions at a fixed logarithmic
accuracy, \begin{equation} \ln
\left.w(N,Q^2)\right\vert_{\DGE}=\frac{C_F}{2\beta_0}
\sum_{k=1}^{\infty}{{\bar{A}(Q^2)}^{k-2}}\,\,
f_{k}\!\left({\bar{A}}(Q^2)\ln N\right),
\label{resum:log_expansion}
\end{equation} where $\bar{A}(Q^2)\equiv \bar{\alpha}_s(Q^2)\beta_0/\pi$ and
1-loop running coupling was assumed.  The first two~functions
\begin{eqnarray}
\label{resum:f_k_12_DY}
f_{1}(\xi)&=& 2\,(1-2\xi)\,\ln(1 - 2\,\xi ) + 4\,\xi    \nonumber \\
f_{2}(\xi)&=& - 4\,\gamma \,\ln(1 - 2\,\xi )
\end{eqnarray}
are well
known~\cite{Sterman:1987aj,Collins:1988ig,Catani:1989ne,Contopanagos:1997nh,Catani:1991rr,Vogt:2000ci}.
We stress that in spite of the fact that the actual calculation is
done in the large $\beta_0$ limit, the DGE reuslt is {\em exact} to
next-to-leading logarithmic accuracy, provided that the 1-loop
coupling is replaced by the 2-loop coupling in the ``gluon
bremsstrahlung''
scheme~\cite{Catani:1991rr,Gardi:2001ny,Gardi:2002bg,Gardi:2001di}.
The large $\beta_0$ contribution to the higher order functions is
given by
\begin{eqnarray*}
\begin{array}{ll}
f_{3}(\xi)={1.33}/{(1-2\,\xi )} +6.58\,\xi \\
f_{4}(\xi)={2.12}/{(1-2\,\xi)^{2}}   \\
f_{5}(\xi)={4.00}/{(1 -2\,\xi)^{3}} - 19.48\,\xi  \\
f_{6}(\xi)={11.59}/{(1-2\,\xi)^{4}}  \\
f_{7}(\xi)={48.42}/{(1-2\,\xi)^{5}} +91.56\,\xi  \\
f_{8}(\xi)={238.80}/{(1-2\,\xi)^{6}}  \\
f_{9}(\xi)={1438.66}/{(1-2\,\xi)^{7}} - 527.14\,\xi  \\
f_{10}(\xi)={10078.0}/{(1-2\,\xi)^{8}}.
\end{array}
\end{eqnarray*}
We see that sub-leading logs are enhanced by factorially increasing
coefficients as well as an increasing singularity at $\xi=1/2$.
Therefore, truncation of this expansion has a significant impact on
the result. Such truncation would also induce renormalization scale
dependence. When power accuracy is required the sum, up to the minimal
term, or alternatively the Borel integral in
Eq.~(\ref{resum:DY_cs_exp}) must be computed.

Of course, due to the renormalon singularities of the Borel function
(\ref{resum:BN_DY}) the Borel integral (\ref{resum:DY_cs_exp}), as
written, is ill-defined. A prescription must be given.  Since the
full, non-perturbative result is unambiguous, power corrections can be
deduced from this ambiguity.  The first, crucial conclusion is that
the power corrections in the threshold region should exponentiate
together with the perturbative sum, so that the correction factorizes
in Mellin space: $w(N,Q^2) \longrightarrow w(N,Q^2) \, w_{\rm
  NP}(N,Q^2)$.  This is fully consistent with the formulation of the
resummation in terms of an evolution equation for a Wilson-line
operator~\cite{Korchemsky:1995is}, where the non-perturbative
correction appears the initial condition for the evolution.  The
singularities of the Borel integrand are located at {\em integer}
values of $u$.  Equation~(\ref{resum:BN_DY}) has double poles at all
integers, but the $\sin(\pi u)/\pi u$ factor of in
Eq.~(\ref{resum:DY_cs_exp}) leaves only simple poles. As observed in
Ref.~\cite{Beneke:1995pq}, this singularity structure implies that the
leading power correction at large $Q^2$ (and not too large $z$) is
$1/Q^2(1-z)^2$. Closer to $z=1$ sub-leading power corrections of the
form $1/(Q^2(1-z)^2)^n$, where $n$ is an integer, become important.
The non-perturbative correction factor in Mellin space
is~\cite{Gardi:2001di}, \begin{equation} w_{\rm
  NP}(N,Q^2)=\exp\left\{\sum_{n=1}^{\infty}\frac{C_F\,(-1)^n}{{\beta_0}\,n\,
    (n!)^2}\,\omega_n
  \,\left(\frac{{\bar{\Lambda}}^2N^2}{Q^2}\right)^n\right\}, \end{equation}
where $\omega_n$ are non-perturbative parameters.\\
{\bf Acknowledgements:} It a pleasure to thank Yuri Dokshitzer,
Gregory Korchemsky and Douglas Ross for very useful discussions.

\subsection{NNLO expansions of threshold-resummed heavy quark cross
  sections\protect\footnote{Contributing authors: N. Kidonakis, E.
    Laenen, S. Moch, R. Vogt}}

Long- and short-distance dynamics in QCD for inclusive hadronic
hard-scattering cross sections are factorized into universal,
non-perturbative parton distribution functions and fragmentation
functions, and perturbatively calculable hard scattering functions.
Remnants of long-distance dynamics in a hard scattering function can
become large in regions of phase space near partonic threshold and
dominate higher order corrections.  Such Sudakov corrections assume
the form of distributions that are singular at partonic threshold.
Threshold resummation organizes these double-logarithmic corrections
to all orders, thereby presumably extending the predictive power of
QCD to these phase space regions.  One use for resummed cross sections
is to provide, upon expansion in $\alpha_s$, estimates of finite
higher order corrections that are not yet known exactly. Here we
discuss next-to-next-to-leading order (NNLO) estimates for
double-differential heavy quark hadroproduction cross sections.  These
are based on expansions of their next-to-leading logarithmic resummed
versions
\cite{Kidonakis:1996aq,Kidonakis:1997gm,Laenen:1998qw,Kidonakis:2000ui,Kidonakis:2001nj}.

\subsubsection{Kinematics and threshold-singular functions}

The definition of the threshold depends on the observable to be
resummed. For double-differential cross sections various choices are
possible. In one-particle inclusive (1PI) kinematics, in our case
defined by the partonic kinematics ($ij =q\Bar{q},gg$),
\begin{equation}
\label{resum:eq:4}
 i(k_1) + j(k_2) \rightarrow {\rm{Q}}(p_1) +
X'[\Bar{\rm{Q}}](p_2')\,,
\end{equation}
and the corresponding invariants $s=2k_1\cdot k_2,\, t_1 = -2k_2\cdot
p_1,\, u_1 = -2k_1\cdot p_1,\; s_4 \equiv s+t_1+u_1$, the threshold
condition is
\begin{equation}
  \label{resum:eq:1}
s_4 =  0\,.
\end{equation}
The singular functions organized by threshold resummation are the
plus-distributions
\begin{equation}
  \label{resum:eq:2}
  \left[{\ln^{l}(s_4/m^2)\over s_4}\right]_+\,.
\end{equation}
Pair-invariant mass (PIM) kinematics is defined by
\begin{equation}
  \label{resum:eq:3}
   i(k_1) +j(k_2) \rightarrow {\rm{Q}}{\Bar{\rm{Q}}}(p') + X'(k)\,,
\end{equation}
with the variables $ p^{\prime \, 2} = M^2$ and $\cos\theta$, where $\theta$ is
the polar scattering angle in the partonic c.m. frame. The threshold
is set by
\begin{equation}
  \label{resum:eq:5}
   1-\frac{M^2}{s} \equiv 1-z = 0
\end{equation}
with the corresponding singular functions
\begin{equation}
  \label{resum:eq:6}
  \left[{\ln^{l}(1-z)\over 1-z}\right]_+\,.
\end{equation}
At threshold,
\begin{equation}
  \label{resum:eq:11}
t_1 = - \frac{M^2}{2} \left( 1 - \beta_M\, {\rm{cos}} \theta \right)\,,\qquad
u_1 = - \frac{M^2}{2} \left( 1 + \beta_M\, {\rm{cos}} \theta \right)\, ,
\end{equation}
with $\beta_M=\sqrt{1-4m^2/M^2}$.  We denote corrections as leading
logarithmic (LL) if $l=2i+1$ in Eqs.~(\ref{resum:eq:2}) and
(\ref{resum:eq:6}) at order $O(\alpha^{i+3}_s),\; i=0,1,\ldots$, as
next-to-leading logarithmic (NLL) if $l=2i$, etc.  Threshold
resummation is best performed in moment space, defined by the Laplace
transform with respect to $w_K$,
\begin{equation} {\tilde f}(N) = \int\limits_0^\infty dw_K
\,e^{-N w_K} f(w_K)\,,
\label{resum:laplacetfm}
\end{equation} 
where $w_{\opi} = s_4/m^2$ and $w_{\pim} = 1-z$.  In moment space the
singular functions become linear combinations of $\ln^k(\tilde{N})$
with $k\leq l+1$ and $\tilde{N}=N\exp(\gamma_E)$.

The NNLO double-differential partonic cross sections for which we
derived approximate results are
\begin{equation}
  \label{resum:eq:7}
 (\mathrm{\opi}):\; s^2 \frac{d^2 \sigma_{ij}(t_1,u_1,s_4)}
{dt_1\,du_1},\quad
 (\mathrm{\pim}):\;
  s \frac{d^2 \sigma_{ij}(s,M^2,\cos\theta)}{dM^2\,d\cos\theta}\,.
\end{equation}
We can only sketch the derivation here and refer to
\cite{Kidonakis:2001nj} for all details. The resummed
double-differential partonic cross section has the following
functional form in moment space (here in 1PI kinematics, for the PIM
result use Eq.~(\ref{resum:eq:11}) and multiply the right hand side by
$\beta_M/2$):
\begin{eqnarray}
\label{resum:eq:9}
&& \hspace{-5mm}
s^2 \frac{d^2 \tilde\sigma^{\rm res}_{ij}(t_1,u_1,N)}
{dt_1\,du_1}
\,=\,{\rm Tr}\Bigg\{
H_{ij}(m^2,m^2) \\ 
\nonumber & &\hspace{0mm} \times
{\rm \bar{P}} \exp\left[\int_m^{m/N} {d\mu'\over\mu'} 
(\Gamma^{ij}_S)^{\dagger}\left(\alpha_s(\mu^{\prime})\right)\right]
{\tilde S}_{ij}\!\left(1\right)
{\rm P} \exp\left[\int_m^{m/N} {d\mu'\over\mu'} 
\Gamma^{ij}_S\left(\alpha_s(\mu^{\prime})\right)\right] \Bigg\} \\
\nonumber & & \times
\exp\left(\tilde{E}_{i}(N_u,\mu,\mu_R)\right)\, 
\exp\left(\tilde{E}_{j}(N_t,\mu,\mu_R)\right)\,\, \exp\Bigg\{ 2\,
\int\limits_{\mu_R}^{m}{d\mu'\over\mu'}\,\, \Bigl(
\gamma_i\left(\alpha_s(\mu^{\prime})\right) +
\gamma_j\left(\alpha_s(\mu^{\prime})\right) \Bigr) \Bigg\}\, ,
\end{eqnarray}
where any $t_1$ andf $u_1$ dependence is suppressed.  The indicated
trace is in the space spanned by tensors that can couple the $SU(3)$
representations of the partons in (\ref{resum:eq:4}) to singlets:
$2$-dimensional for the $q\Bar{q}$ and $3$-dimensional for the $gg$
channel.  ($\rm \bar{P}$) $\rm P$ refers to (anti)-path-ordering in
$\mu'$.  The two-loop expansion of $\exp(\tilde{E}_{i})$ may be
written schematically as
\begin{eqnarray}
\label{resum:eq:8}
\exp(\tilde{E}_{i}(N_u,\mu,m))
\simeq 1 + \frac{\alpha_s}{\pi}\left(\sum_{k=0}^2 C^{i,(1)}_{k}\ln^k(N_u)  \right)
+ \left(\frac{\alpha_s}{\pi}\right)^2\left(
\sum_{k=0}^4 C^{i,(2)}_{k}\ln^k(N_u)
\right)
+\ldots
\end{eqnarray}
The coefficients $C^{i,(n)}_k$ are given in
Ref.~\cite{Kidonakis:2001nj}.  Similar expansions can be given for the
other $N$-dependent factors in Eq.~(\ref{resum:eq:9}). Momentum space
expressions to NNLL-NNLO are obtained by gathering together all terms
at ${\cal O}(\alpha_s^3)$ and ${\cal O}(\alpha_s^4)$, performing an
inverse Laplace transform, and matching the $N$-independent $H_{ij}$
and $\tilde{S}_{ij}(1)$ coefficients to known exact ${\cal
  O}(\alpha_s^3)$ results. The resulting, approximate, NNLO cross
sections in Eq.~(\ref{resum:eq:7}) have NNLL accuracy in the sense
stated below Eq.~(\ref{resum:eq:11}).  They are rather long and are
given explicitly in Ref.~\cite{Kidonakis:2001nj}.

\subsubsection{Numerical results}

We have so far performed numerical studies of these results for the
\textit{inclusive} partonic cross sections $\sigma_{ij}(s,m^2,\mu^2)$.
We must then attribute any differences in integrating either the PIM
or 1PI results to an ambiguity of our estimates. It is convenient to
express these inclusive partonic cross sections in terms of
dimensionless scaling functions $f^{(k,l)}_{ij}$ that depend only on
$\eta = s/4m^2 -1$, as follows:
\begin{eqnarray}
\label{resum:scalingfunctions}
\sigma_{ij}(s,m^2,\mu^2) &=& \frac{\alpha_s^2(\mu)}{m^2}
\sum\limits_{k=0}^{\infty} \,\, \left( 4 \pi \alpha_s(\mu) \right)^k
\sum\limits_{l=0}^k \,\, f^{(k,l)}_{ij}(\eta) \,\,
\ln^l\left(\frac{\mu^2}{m^2}\right) \, .
\end{eqnarray} 
From our results for the double-differential cross section we have
constructed LL, NLL, and NNLL approximations to $f^{(k,l)}_{ij}(\eta)$
for $k\leq 2,\, l\leq k$, and for both the $q{\Bar{q}}$ and $gg$
channel. For $k=1$ exact results are
known~\cite{Nason:1988xz,Beenakker:1989bq,Beenakker:1991ma}.  For
$k=2$ and $l=1,2$ we have derived exact results using renormalization
group methods. Our best NNLO estimate consists of all exactly known
scaling functions, together with the NNLL estimate of
$f^{(2,0)}_{ij}$.

We now show a few representative results. To exhibit the relevance of
threshold approximations we write the inclusive hadronic cross section
as a pointwise product in $\eta$ of the partonic cross section and the
parton flux $\Phi_{ij}$
\begin{eqnarray}
\label{resum:eq:10}
\sigma_{h_1h_2}(S,m^2) &=&  \sum\limits_{i,j = q,{\Bar{q}},g} \,\,
\int_{-\infty}^{\log_{10}(S/4m^2-1)} d\log_{10}\eta \, \frac{\eta}{1+\eta} 
\ln(10) \, \Phi_{ij}(\eta,\mu^2)\,\,
\sigma_{ij}(\eta,m^2,\mu^2)\, .
\end{eqnarray}
We use the two-loop expression of $\alpha_s$ and the CTEQ5M parton
distributions \cite{Lai:1999wy} at LO, NLO, and NNLO.

\begin{figure}[htp]
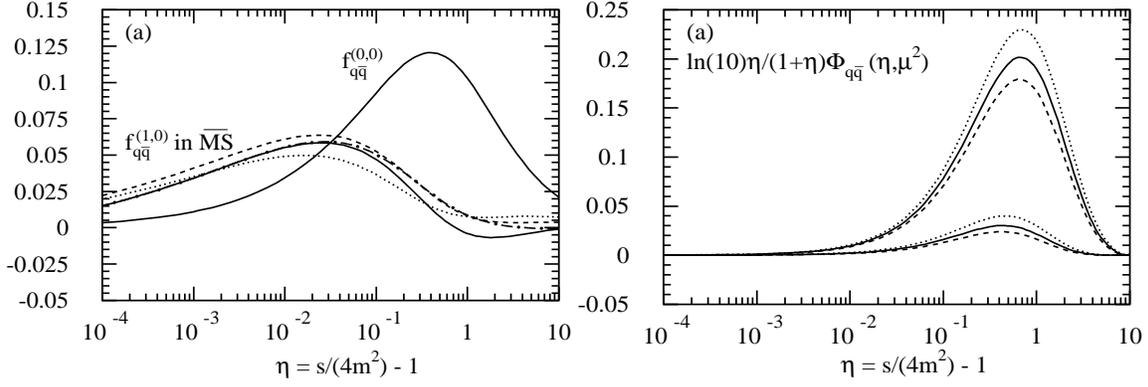

\begin{center}
  \includegraphics[width=5cm,height=7.5cm,angle=-90]{./fqq-1PI.epsi}
  \includegraphics[width=5cm,,height=7.5cm,angle=-90]{./phiqq.epsi}
\end{center}
\caption{(a) The $\eta$-dependence of the scaling functions 
  $f^{(k,0)}_{q{\Bar{q}}}(\eta),\;k=0,1$ in the
  ${\Bar{\rm{MS}}}$-scheme and 1PI kinematics.  We show the exact
  results for $f^{(k,0)}_{q{\Bar{q}}},\;k=0,1$ (solid lines), the LL
  approximation to $f^{(1,0)}_{q{\Bar{q}}}$ (dotted line), the NLL
  approximation to $f^{(1,0)}_{q{\Bar{q}}}$ (dashed line) and the NNLL
  approximation to $f^{(1,0)}_{q{\Bar{q}}}$ (dashed-dotted line).  (b)
  The $q\Bar{q}$ CTEQ5M parton flux for $t\Bar{t}$ production at the
  Tevatron (upper three curves, $\sqrt{S}=1.8$ TeV and $m=175$ GeV)
  and for $b\Bar{b}$ production at HERA-B (lower three curves
  $\sqrt{S}=41.6$ GeV and $m=4.75$ GeV).  The curves correspond to
  $\mu=m$ (solid curves), $\mu=m/2$ (dotted curves), and $\mu=2m$
  (dashed curves). }
\label{resum:fig:nnlo-part-flux}
\end{figure}

As an example, in Fig.~\ref{resum:fig:nnlo-part-flux} we show the
$q\Bar{q}$ NLO scaling function to various accuracies and the
corresponding partonic flux for $t\Bar{t}$ production at the Tevatron,
showing the values of $\eta$ where the integral in
Eq.~(\ref{resum:eq:10}) gets the most weight.
The partonic results show that the NNLL approximation at NLO
approximates the exact result very well. In
Ref.~\cite{Kidonakis:2001nj} we have studied the quality of the NNLL
approximation more extensively and find this conclusion to hold to
NNLO and for both channels. The other plot in
Fig.~\ref{resum:fig:nnlo-part-flux} shows that the flux selects
partonic processes that are reasonably close to threshold, making our
approximations phenomenologically relevant.

\begin{figure}[htp]
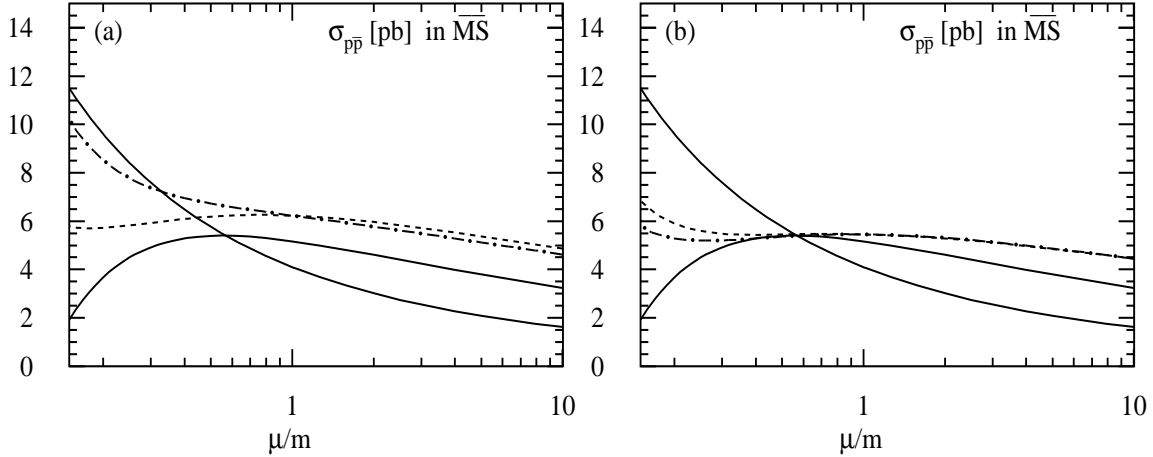

\begin{center}
  \includegraphics[width=6cm,height=7.5cm,angle=-90]{mu-1PI.epsi}
  \includegraphics[width=6cm,height=7.5cm,angle=-90]{mu-PIM.epsi}
\end{center}
\caption{The $\mu$-dependence of the top quark cross section 
  at the Tevatron with $\sqrt{S}=1.8$ TeV and $m=175$ GeV for the sum
  of the $q \bar{q}$ and $gg$ channels in the $\overline{\rm MS}$
  scheme. We show the Born (upper solid line at small $\mu/m$) and the
  exact NLO (lower solid line at small $\mu/m$) cross sections, the
  1PI approximate NNLL-NNLO cross section (dashed line) and the 1PI
  NNLO estimate with only $f^{(2,0)}_{q{\Bar{q}}}$ and
  $f^{(2,0)}_{gg}$ NNLL approximate (dashed-dotted line).  (b) The
  same as (a) in PIM kinematics.}
\label{resum:fig:nnlo-ttbar-scale}
\end{figure}

While QCD corrections to the top quark inclusive cross section at the
Tevatron are fairly modest, our NNLO threshold estimates help to
substantially reduce factorization scale dependence, as we show in
Fig.~\ref{resum:fig:nnlo-ttbar-scale}.  The latter is expected on
general grounds \cite{Oderda:1999im,Sterman:2000pu}.
We have checked that using presently available, almost NNLO densities
\cite{Martin:2000gq} lead to very similar results. Based on our
approximations we provide, by averaging the results from PIM and 1PI
kinematics, NNLL-NNLO estimates for the following inclusive top cross
sections
\begin{equation}
  \label{resum:eq:34}
  \sigma_{t\bar{t}}(1.8\,\mathrm{TeV})
 = 5.8
\pm 0.4 
\pm 0.1 \;\;\mathrm{pb}\,,
\end{equation}
\begin{equation}
\label{resum:eq:35}
  \sigma_{t\bar{t}}(2.0\,\mathrm{TeV})
 = 8.0  
\pm 0.6
\pm 0.1 \;\;\mathrm{pb}\,.
\end{equation}
For the HERA-B bottom cross section, we find
\begin{equation}
\label{resum:eq:36}
  \sigma_{b\bar{b}}(41.6\,\mathrm{GeV})
 = 30 
\pm  8 
\pm 10 \;\;\mathrm{nb}\,.
\end{equation}
The first set of errors indicates the kinematics ambiguity while the
second is an estimate of the scale uncertainty.  Note that the scale
uncertainty for top production is now considerably smaller than the
kinematics uncertainty.

\subsection{High energy resummation\protect\footnote{Contributing
    authors: R. Ball, S. Forte}}

\subsubsection{Perturbative QCD at HERA}

QCD has been tested at HERA~\cite{Forte:1999kh,Chekelian:2001pi} over
the last several years to an accuracy which is now comparable to that
of tests of the electroweak sector at LEP: perturbative QCD turns out
to provide an embarrassingly successful description of the HERA data,
even in kinematic regions where simple fixed--order perturbative
predictions should fail. This success is most strikingly demonstrated
by the comparison with the data of the scaling violations of structure
functions predicted by the QCD evolution
equations~\cite{Chekanov:2001qu,Adloff:2000qk}: the data agree with
the theory over five orders of magnitude in both $x$ and $Q^2$.

\begin{figure}
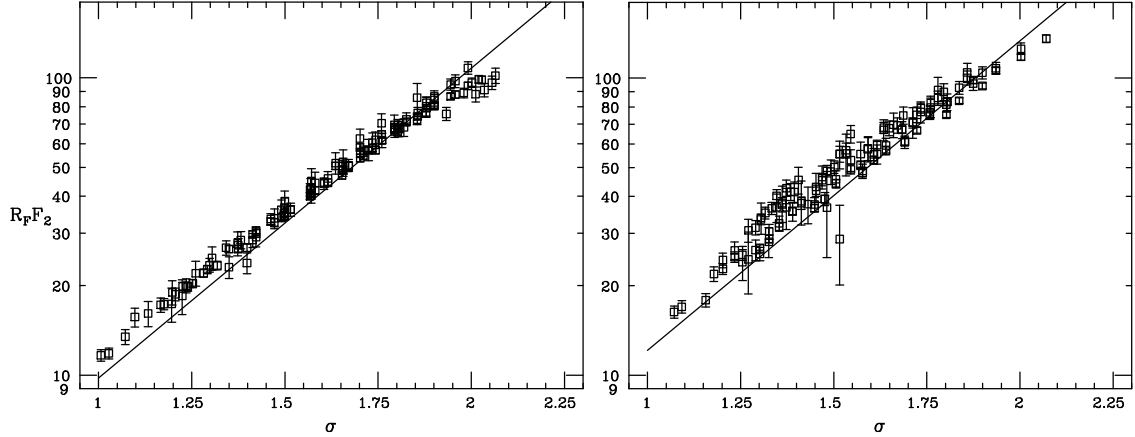

  \includegraphics[width=0.48\linewidth,clip]{scal4.ps}
  \includegraphics[width=0.45\linewidth,clip]{scal5.ps}
\caption{Double asymptotic scaling of the H1 data~\cite{Adloff:2000qk}. The
  scaling variable $\sigma\equiv\ln(x_0/ x)\ln (\as(Q_0^2)/\as(Q^2))$
  is defined with $x_0=0.1$, $Q_0=1~{\rm GeV}$; the rescaling factor
  $R_F$ is as in Ref.~\cite{Ball:1994du}. Only data with $\rho\ge 1$,
  $\sigma\ge 1$, $Q^2\ge4~{\rm GeV}^2$; $x\le0.03$ are plotted. Left:
  $Q^2\le m_b^2$; right: $Q^2> m_b^2$. The straight line is the
  asymptotic prediction.}
\label{resum:fig:bf1}
\end{figure}

The significance of this sort of result is somewhat obscured by the
need to fit the shape of parton distributions at a reference scale,
which might suggest that deviations from the predicted behaviour could
be accommodated by changing the shape of the parton distribution.
However, this is not true because of the predictive nature of the QCD
result: given the shape of partons at one scale, there is no freedom
left to fit the data at other scales. This predictivity is
particularly transparent in the small $x$ region, where the
fixed--order QCD result actually becomes asymptotically independent of
the parton distribution, apart from an overall normalization. Indeed,
the data for $\ln F_2$ plotted versus the variable
$\sigma\equiv\ln{x_0\over x}\ln {\as(Q^2_0)\over \as(Q^2)}$ are
predicted to lie on a straight line, with universal slope
$2\gamma=12/\sqrt{33-2 n_f}$ (double asymptotic
scaling~\cite{Ball:1994du,DeRujula:1974rf}).  The predicted scaling is
spectacularly borne out by the data, as shown in
Fig.~\ref{resum:fig:bf1}: in fact, the data are now so accurate that
one can see the change in slope when passing the $b$ threshold, and
indeed double scaling is only manifest if one separates data in the
regions where $\alpha_s$ runs with $N_f=4$ from those with
$N_f=5$.\footnote{The fact that the observed slope is somewhat smaller
  than the predicted one, especially at low $Q^2$, is due to NLO
  corrections~\cite{Forte:1995vs} as well as corrections due to the
  ``small'' eigenvalue of perturbative
  evolution~\cite{Mankiewicz:1997sd}.}  Equally good agreement with
fixed--order perturbation theory is seen when considering less
inclusive observables.

This agreement of the data with fixed--order perturbative QCD
computations is very surprising, in that the perturbative expansion
receives contributions of order $\as\ln{1\over x}$ so one would expect
higher--order corrections to be non--negligible whenever
$\as\ln{1\over x}\gsim 1$, i.e. in most of the HERA region. As is well
known, the resummation of leading $\ln{1\over x}$ (LLx) contributions
to gluon--gluon scattering, and thus to a wide class of hard
processes, including small $x$ scaling violations of structure
functions, is accomplished by means of the BFKL evolution
equation~\cite{Lipatov:1976zz,Fadin:1975cb,Kuraev:1976ge}. Matching
the BFKL approach to standard perturbative computation, however, is
nontrivial~\cite{Ball:1995vc,Ellis:1995gv}, while the BFKL equation
itself seems to be unstable towards the inclusion of higher order
corrections~\cite{Fadin:1998py}. Hence, the main problem in
understanding HERA physics, i.e. perturbative QCD at small $x$ is that
of establishing ``consistency of the BFKL approach with the more
standard DGLAP~\cite{Altarelli:1977zs,Gribov:1972ri} evolution
equations''~\cite{McLerran:2001sr}, which embody the leading $\ln Q^2$
(LLQ$^2$) resummation on which perturbative QCD is based.  This
problem is now
solved~\cite{Ball:1999sh,Altarelli:1999vw,Ciafaloni:1999yw}, and on
the basis of this solution it is possible to combine the available
information on perturbation theory at small $x$, and use it to explain
the unexpected success of fixed--order calculations.

\subsubsection{Duality}

Let us for definiteness consider the prototype problem of the
description of small $x$ scaling violations of parton distributions.
For simplicity, consider the case of a single parton distribution
$G(x,Q^2)$, which can be thought of as the dominant eigenvector of
perturbative evolution. Scaling violations are then described by the
Altarelli-Parisi equation satisfied by $G(x,Q^2)$, and thus summarized
by the Altarelli--Parisi splitting function
$P(x,\as)$~\cite{Altarelli:1977zs}.

The basic result which allows the study of scaling violations at small
$x$ is {\it duality} of perturbative
evolution~\cite{Ball:1997vf,Altarelli:1999vw,Altarelli:2000mh},
namely, the fact that, because the Altarelli-Parisi equation is an
integro--differential equation in the two variables $t\equiv\ln
Q^2/\Lambda^2$ and $\xi\equiv 1/x$, it can be equivalently cast in the
form of a differential equation in $t$ satisfied by the $x$--Mellin
transform
\begin{equation}
G(N,t)=\int^{\infty}_{0}\! d\xi\, e^{-N\xi}~G(\xi,t),
\label{resum:nmel}
\end{equation}
or a differential equation in $\xi$ satisfied by the $Q^2$--Mellin
transform
\begin{equation} 
G(\xi,M)=\int^{\infty}_{-\infty}\! dt\, e^{-Mt}~G(\xi,t)
\label{resum:mmel} 
\end{equation} 
of the parton distribution.  The pair of dual evolution equations are
\begin{eqnarray}
\frac {d}{dt}G(N,t)&=&\gamma(N,\as)~G(N,t)
\label{resum:tevol}\\
\frac {d}{d\xi}G(\xi,M)&=&\chi(M,\as)~G(\xi,M),
\label{resum:xevol}
\end{eqnarray}
where Eq.~(\ref{resum:tevol}) is the standard renormalization--group
equation, with anomalous dimension $\gamma(N,t)$, and
Eq.~(\ref{resum:xevol}) is essentially the BFKL equation. Duality is
the statement that the solutions of these two equations coincide to
all perturbative orders, up to power suppressed corrections, provided
their kernels are related by
\begin{equation}
\chi(\gamma(N,\as),\as)=N.\label{resum:dual}
\end{equation}
This means that the BFKL and Altarelli-Parisi equations describe the
same physics: it is the choice of the kernel to be used in the
evolution equation which determines which is the large scale which is
resummed.  We can then discuss the construction and resummation of the
kernel irrespective of the specific evolution equation where it is
used, with the understanding that the kernel can be equivalently
viewed as a $\gamma(N,\as)$ or a $\chi(M,\as)$, the two being related
by Eq.~(\ref{resum:dual}). Before doing this, we sketch how duality
can be proven order by order in perturbation theory.

\subsubsection{Fixed coupling}

Perturbative duality is most easy to prove when the coupling does not
run, since in this case the two scales $t$ and $\xi$ appear in the
Altarelli--Parisi equation in a completely symmetric way. It is
convenient to introduce the double--Mellin transform $G(N,M)$ of the
parton distribution. The solution to the Altarelli--Parisi equation in
$M,N$ space has the form (which can be e.g. obtained by performing an
$M$--mellin transform Eq.~(\ref{resum:mmel}) of the solution to the
renormalization--group Eq.~(\ref{resum:tevol}))
\begin{equation}
G(N,M)=\frac{G_0(N)}{M-\gamma(N,\as)},
\label{resum:tsoln}
\end{equation}
where $G_0(N)$ is a boundary condition at a reference scale $\mu^2$.

The inverse Mellin transform of Eq.~(\ref{resum:tsoln}) coincides with
the residue of the simple pole in the $M$ plane of $e^{t M} G(N,M)$,
and thus its scale dependence is entirely determined by the location
of the simple pole of $G(N,M)$~(\ref{resum:tsoln}) , namely, the
solution to the equation
\begin{equation}
M=\gamma(N,\as). 
\label{resum:pole}
\end{equation} 
The pole condition Eq.~(\ref{resum:pole}) can be equivalently viewed
as an implicit equation for $N$: $N=\chi(M,\as)$, where $\chi$ is
related to $\gamma$ by Eq.~(\ref{resum:dual}). Hence, the function
\begin{equation}
G(N,M)=\frac{F_0(N)}{N-\chi(M,\as)},
\label{resum:xisoln}
\end{equation}
corresponds to the same $G(t,x)$ as Eq.(\ref{resum:tsoln}), because
the location of the respective poles in the $M$ plane are the same,
while the residues are also the same, provided the boundary conditions
are matched by
\begin{equation}
 G_0(N)=-
{ F_0(\gamma(\as,N))\over\chi^\prime(\gamma(\as,N))}.
\label{resum:bcmatch}\end{equation}

Eq.~(\ref{resum:xisoln}) is immediately recognized as the $N$-Mellin
of the solution to the evolution equation~(\ref{resum:xevol}) with
boundary condition $F_0(M)$ (at some reference $x=x_0$), which is what
we set out to prove.  In general, the analytic continuation of the
function $\chi$ defined by Eq.~(\ref{resum:dual}) will be such that
Eq.~(\ref{resum:pole}) has more than one solution (i.e. $\gamma$ is
multivalued). In this case, poles further to the left in the $M$ plane
correspond to power--suppressed contributions, while poles to the
right correspond to contributions beyond perturbation theory (they do
not contribute when the inverse $M$--Mellin integral is computed along
the integration path which corresponds to the perturbative region).

It is easy to see that upon duality the leading--order
$\chi=\as\chi_0$ is mapped onto the leading singular
$\gamma=\gamma_s(\as/N)$, and conversely the leading--order
$\gamma=\as\gamma_0$ is mapped onto the leading singular
$\chi=\chi_s(\as/M)$. In general, the expansion of $\chi$ in powers of
$\as$ at fixed $M$ is mapped onto the expansion of $\gamma$ in powers
of $\as$ at fixed $\as/N$, and conversely.  So in particular at
LLQ$^2$ it is enough to consider $\gamma_0$ or $\chi_s$, and at LLx it
is enough to consider $\gamma_s$ or $\chi_0$. The running of the
coupling is a LLQ$^2$ but NLLx effect, so beyond LLx the discussion
given so far is insufficient.

\subsubsection{Running coupling}

The generalization of duality to the running coupling case is
nontrivial because the running of the coupling breaks the symmetry of
the two scales $\xi$ and $t$ in the Altarelli--Parisi equation.
Indeed, upon $M$--Mellin transform~(\ref{resum:mmel}) the usual
one--loop running coupling becomes the differential operator
\begin{equation}
\ah = \frac{\as}{1-\beta_0 \as \smallfrac{d}{dM}}+\cdots,
\label{resum:ahdef}
\end{equation}
where $d\as/dt= -\beta_0\as^2$.

Consider for simplicity the LLx $x$--evolution equation, i.e.
Eq.~(\ref{resum:xevol}) with $\chi=\as\chi_0(M)$, and include running
coupling effects by replacing $\as$ with the differential operator
Eq.~(\ref{resum:ahdef}). We can solve the equation perturbatively by
expanding the solution in powers of $\as$ at fixed $\as/N$: the
leading--order solution is given by Eq.~(\ref{resum:xisoln}), the
next--to--leading order is obtained by substituting this back into the
equation and retaining terms up to order $\beta_0\as$, and so
on~\cite{Ball:1999sh}. We can then determine the associate $G(N,t)$ by
inverting the $M$--Mellin, and try to see whether this $G(N,t)$ could
be obtained as the solution of a renormalization group (RG)
equation~(\ref{resum:tevol}).

The inverse Mellin is again given by the residue of the pole of $e^{t
  M} G(N,M)$ in the M--plane, where $G(N,M)$ is now the perturbative
solution. When trying to identify this with a solution to
Eq.~(\ref{resum:tevol}) there are two potential sources of trouble.
The first is that now the perturbative solution at order
$(\as\beta_0)^n$ has a $(2n+1)$--st order pole. Therefore, the
scale--dependence of the inverse Mellin is now a function of both
$\as$ and $t$, whereas the solution of a RG equation depends on $t$
only through the running of $\as$. Hence it is not obvious that a dual
anomalous dimension will exist at all. The second is that even if a
dual $\gamma$ does exist, it is not obvious that it will depend only
on $\chi$ and not also on the boundary condition $F_0(M)$
Eq.~(\ref{resum:xisoln}): in such case, the running of the coupling in
the $\xi$--evolution equation would entail a breaking of
factorization.

However, explicit calculation shows that it is possible to match the
anomalous dimension and the boundary condition order by order in
perturbation theory in such a way that both duality and factorization
are respected. Namely, the solution to the leading--twist running
coupling $x$--evolution Eq.~(\ref{resum:xevol}) with kernel $\ah
\chi_0$ and boundary condition $G_0(M)$ is the same as that of the
renormalization group Eq.~(\ref{resum:tevol}) with boundary conditions
and anomalous dimension given by
\begin{eqnarray}
&&\gamma(\as(t),\as(t)/N)=\gamma_s(\as(t)/N)+\as(t)\beta_0\Delta\gamma_{ss}(\as(t)/N)+\nonumber\\
&&\quad\qquad+
(\as(t)\beta_0)^2 \Delta\gamma_{sss}(\as(t)/N) +O(\as(t)\beta_0)^3\\
\label{resum:efgam}
&&G_0(\as,N)=G_0(N)+\as\beta_0 \Delta^{(1)}G_0(N)+
(\as\beta_0)^2\Delta^{(2)} G_0(N)
+O(\as\beta_0)^3,
\label{resum:efbc}
\end{eqnarray}
where the leading terms $\gamma_s$ and $G_0(N)$ are given by
Eqs.~(\ref{resum:dual}) and (\ref{resum:bcmatch}) respectively. The
subleading corrections are
\begin{eqnarray}
\Delta\gamma_{ss}&=&-
\frac{\chi_0''\chi_0}
{2\chi_0^{\prime \, 2}}\\
\Delta^{(1)}G_0(N)&=&{2{\chi_0'}^2
F_0-\chi_0\left(F_0'\chi_0''-\chi_0'F_0''\right)\over 2{\chi_0'}^3},
\label{resum:nlocorr}
\end{eqnarray}
where all derivatives are with respect to the arguments of $\chi_0(M)$
and $F_0(M)$, which are then evaluated as functions of $\gamma_s(\as/N
)$.  The sub--subleading correction to the anomalous dimension is
\begin{equation}
\Delta\gamma_{sss}=-\chi_0^2{15{\chi_0''}^3-16
\chi_0'\chi_0''\chi_0'''+3 {\chi_0'}^2\chi_0''''\over 24 {\chi_0'}^5},
\label{resum:nnlocorr}
\end{equation}
and we omit the very lengthy expression for $\Delta^{(2)}G_0(N)$.  The
fact that duality and factorization hold up to NNLLx is nontrivial,
and suggests that they should hold to all orders. An all--order proof
can be in fact constructed~\cite{Altarelli:2001ji}.

Once the corrections to duality Eq.~(\ref{resum:efgam}) are
determined, they can be formally re-interpreted as additional
contributions to $\chi$: namely, one can impose that the duality
Eq.~(\ref{resum:dual}) be respected, in which case the kernel to be
used in it is an ``effective'' $\chi$, obtained from the kernel of the
$x$--evolution Eq.~(\ref{resum:xevol}) by adding to it running
coupling corrections order by order in perturbation theory: $\chi_0$
will be free of such correction, $\chi_1$ will receive a correction
\begin{equation}
\Delta\chi_1=\beta_0{1\over
2}{\chi_0(M)\chi_0''(M)\over{\chi_0'}^2(M)},
\label{resum:efchi}
\end{equation}
and so forth. Applying duality to the known one--loop anomalous
dimensions $\gamma_0$ thus gives us the resummation of the all--order
singular contributions $\chi(\as/M)$ to this effective $\chi$, which
include the running coupling correction Eq.~(\ref{resum:efgam}) and
its higher--order generalizations.

\subsubsection{Resummation}
Because the first two orders of the expansion of $\chi$ in powers of
$\as$ at fixed $M$ and of the expansion of $\gamma$ in powers of $\as$
at fixed $N$ are known, it is possible to exploit duality of
perturbative evolution to combine this information into anomalous
dimension which accomplish the simultaneous resummation of leading and
next--to--leading logs of $x$ and $Q^2$.  In fact, it turns out that
both a small $M$ and a small $N$ resummation of anomalous dimensions
are necessary in order to obtain a stable perturbative expansion,
while unresummed anomalous dimensions lead to instabilities. Both
sources of instability are generic consequences of the structure of
the perturbative expansion, and could have been predicted before the
actual explicit computation~\cite{Fadin:1998py} of subleading
small-$x$ corrections.

\subsubsection{Small M}
The perturbative expansion of $\chi$ at fixed $M$ is very badly
behaved in the vicinity of $M\sim0$: at $M=0$, $\chi_0$ has a simple
pole, $\chi_1$ has a double pole and so on.  In practice, this spoils
the behaviour of $\chi$ in most of the physical region $0< M< 1$.
Because $1/M^k$ is the Mellin transform of ${\Lambda^2\over Q^2}
{1\over k!}\ln^{k-1}(Q^2/\Lambda^2)$, these singularities correspond
to logs of $Q^2$ which are left unresummed in a LLx or NLLx
approach~\cite{Salam:1998tj}.

The resummation of these contributions may be understood in terms of
momentum conservation, which implies that $\gamma(1,\as)=0$ (note our
definition of the $N$--Mellin transform~(\ref{resum:nmel}), and also
that $\gamma$ is to be identified with the large eigenvector of the
anomalous dimension matrix).  The duality Eq.~(\ref{resum:dual}) then
implies that a momentum--conserving $\chi$ must satisfy
$\chi(0,\as)=1$. This, together with the requirement that $\chi$
admits a perturbative expansion in powers of $\as$, implies that in
the vicinity of $M=0$, the generic behaviour of the kernel is
\begin{equation}
\chi_s\tozero M {\as\over\as+\kappa M}={\as\over\kappa M}-{\as^2\over
(\kappa M)^2}+{\as^3\over(\kappa M)^3}+\dots\quad,
\label{resum:momcons}
\end{equation} 
where $\kappa$ is a numerical constant which turns out to be
$\kappa=\pi/C_A$.  Hence we understand that there must be an
alternating--sign series of poles at $M=0$, which sums up to a regular
behaviour.  In fact, we can systematically resum singular
contributions to $\chi$ to all orders in $\as$ by including in $\chi$
the terms $\chi_s(\as/M)$ derived from the leading order
$\gamma_0(N)$, and similarly at next--to--leading order, and so on.
Because the usual anomalous dimension automatically respects momentum
conservation order by order in $\as$, in order to remove the small $M$
instability of the expansion of $\chi$ at fixed $M$, it is sufficient
to improve the expansion by promoting it to a ``double leading''
expansion which combines the expansions in powers of $\as$ at fixed
$M$ and at fixed $\as/M$~\cite{Altarelli:1999vw}.  For example, at
leading order $\chi=\as\chi_0(M)+\chi_s(\as/M)-{\rm d.c.}$, where the
subtraction refers to the double--counting of the $\as/M$ term which
is present both in $\as\chi_0$ and in $\chi_s(\as/M)$. This expansion
of $\chi$ is dual Eq.~(\ref{resum:dual}) to an analogous expansion of
$\gamma$, where at leading order
$\gamma=\as\gamma_0(M)+\gamma_s(\as/M)-{\rm d.c.}$, and so forth. Both
expansions are well behaved at small $M$, i.e. large $N$.  At this
level, it is already clear that the impact of the inclusion of
small-$x$ corrections is moderate: indeed, it turns out that the
double--leading kernel is quite close to the usual two--loop kernel,
except at the smallest values of $N$, i.e. in the neighbourhood of the
minimum of $\chi(M)$~\cite{Altarelli:1999vw}.

\subsubsection{Small N}

The improved double--leading expansion of the anomalous dimension
still requires resummation at small $N$. This is because, even though
the next--to--leading correction to the double--leading evolution
kernel is small for all fixed $M$, it is actually large if $N$ is
fixed and small.  This in turn follow from the fact that the leading
$\chi$ kernel has a minimum, so the small $N=\chi$ region corresponds
by duality Eq.~(\ref{resum:dual}) to the vicinity of the minimum where
the kernel is almost parallel to the $\gamma=M$ axis.

At small $N$, unlike at small $M$, there is no principle like momentum
conservation which may provide a fixed point of the expansion and thus
fix the all--order behaviour.  The only way out is thus to treat this
all--order behaviour as a free parameter. Namely, we introduce a
parameter $\lambda$ which is equal to the value of the all--order
kernel $\chi$ at its minimum, and then we expand about this all--order
minimum. In practice, this means that we reorganize the expansion of
$\chi$ according to~\cite{Ball:1999sh}
\begin{eqnarray}
\chi(M,\as)&=&\as \chi_0(M)+\as^2\chi_1(M)+\dots\nonumber\\
&=&\as \tilde \chi_0(M) +\as^2\tilde\chi_1(M)+\dots,
\label{resum:clamsub}
\end{eqnarray}
where
\begin{equation}
\as\tilde \chi_0(M,\as)\equiv \alpha_s
\chi_0(M)+\sum_{n=1}^\infty \alpha_s^{n+1} c_n,\qquad 
\tilde\chi_i(M)\equiv\chi_i(M)-c_i,
\label{resum:ctil}
\end{equation}
and the constants $c_i$ are chosen in such a way that
\begin{equation}
\lambda \equiv\as\tilde\chi_0(\half)=\as\chi_0(\half)+\Delta \lambda.
\label{resum:lamdef}
\end{equation}
is the all--order minimum of $\chi$.  Of course, in practice
phenomenological predictions will only be sensitive to the value of
$\lambda$ in the region where very small values of $N$ are probed,
i.e. at very small $x$.

\subsubsection{Phenomenology}

\begin{figure}
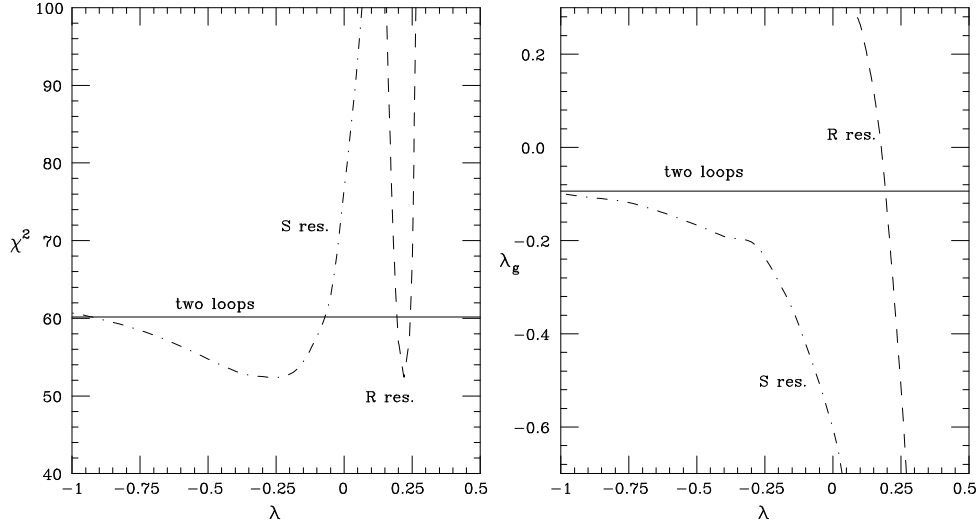

  \includegraphics[width=0.4\linewidth,clip]{chi2.ps}
  \includegraphics[width=0.4\linewidth,clip]{lamg.ps}
\caption{$\chi^2$ (left) and starting gluon slope $G(x, \hbox{4 GeV}^2)\sim
  x^{-\lambda_g}$ (right) for the fit~\cite{Altarelli:2000mh} to the
  95 H1 data~\cite{Adloff:2000qk} as a function of the resummation
  parameter $\lambda$ Eq.~(\ref{resum:lamdef}), for the two
  resummation prescriptions discussed in text. The fits are performed
  with $\as(M_z)=0.119$.}
\label{resum:fig:bf2}
\end{figure}

Using duality and the resummation discussed above, one can construct
resummed expressions for anomalous dimensions and coefficient
functions, and wind up with resummed expressions for physical
observables which may be directly compared to the data.  The need to
resum the small $N$ behaviour entails that phenomenological
predictions will necessarily depend on the parameter $\lambda$
Eq.~(\ref{resum:lamdef}). When the resummed double--leading expansion
is constructed, a further ambiguity arises in the treatment of
double--counting terms. This ambiguity is related to the nature of the
small $N$ singularities of the anomalous dimension, which control the
asymptotic small $x$ behaviour.  Specifically, according to the way
the double--counting is treated, the $N=0$ poles of the one-- and
two--loop result may survive in the resummed result (`S--resummation')
or not (`R--resummation').  Both alternatives are compatible with the
known low--order information on the evolution kernel, and can be taken
as two extreme resummation schemes which parametrize our ignorance of
higher order perturbative terms.  Since the resummed terms also have a
cut starting at $N=\lambda$, whether or not these low--$N$ poles are
present only makes a difference if $\lambda$ turns out to be small,
$\lambda\lsim0.3$.

The $\chi^2$ and starting gluon slope for a
fit~\cite{Altarelli:2000mh} to the recent H1 data~\cite{Adloff:2000qk}
for the deep--inelastic cross section are shown in
Fig.~\ref{resum:fig:bf2}, as a function of $\lambda$ and for the two
different resummation prescriptions. It is clear that if the
perturbative $N=0$ poles do not survive the resummation (R
resummation) then only a fine--tuned value of $\lambda\approx 0.2$ is
acceptable, whereas if they do survive (S resummation) essentially any
$\lambda\lsim 0$ gives a good fit.

Figure~\ref{resum:fig:bf2} demonstrates that it is possible to
accommodate the success of simple fixed--order approach within a fully
resummed scheme, and in fact the resummed calculation is in somewhat
better agreement with the data than the fixed order one. Even though
the effects of the resummation are necessarily small (otherwise the
success of the fixed order prediction could not be explained) they do
have a significant impact in the extraction of the parton
distribution: the gluon comes out to be significantly more
valence--like than in an unresummed fit.  Hence, the use of resummed
perturbation theory is crucial for the extraction of reliable parton
distributions at small $x$.

From a theoretical point of view, we see that current data already
pose very stringent constraints on the unknown high--orders of the
perturbative expansion: only a rather soft high--energy behaviour of
the deep-inelastic cross--section is compatible with the data. The
resummed parton distributions may in principle now be used to compute
hadronic cross-sections at the Tevatron and LHC: the resummation of
hadronic heavy quark production is discussed in some
detail in Ref.~\cite{Ball:2001pq}.\\
{\bf Acknowledgments:} A sizable part of this contribution is based on
work done in collaboration with G.~Altarelli. This work was supported
in part by EU TMR contract FMRX-CT98-0194 (DG 12 - MIHT).


\begin{thebibliography}{100}

\bibitem{Cavalli:2002vs}
D.~Cavalli et~al.
\newblock hep-ph/0203056.

\bibitem{Azuelos:2002qw}
G.~Azuelos et~al.
\newblock hep-ph/0204031.

\bibitem{Giele:2001mr}
W.~Giele, S.~Keller, and D.~Kosower.
\newblock hep-ph/0104052.

\bibitem{Giele:1998gw}
W.~Giele and S.~Keller.
\newblock {\em Phys. Rev.}, D58:094023, 1998.

\bibitem{Pumplin:2000vx}
J.~Pumplin, D.~Stump, and W.~Tung.
\newblock {\em Phys. Rev.}, D65:014011, 2002.

\bibitem{Pumplin:2001ct}
J.~Pumplin et~al.
\newblock {\em Phys. Rev.}, D65:014013, 2002.

\bibitem{Stump:2001gu}
D.~Stump et~al.
\newblock hep-ph/0101051.
\newblock {\em Phys. Rev.}, D65:014012, 2002.

\bibitem{D'Agostini:1995fv}
G.~D'Agostini.
\newblock hep-ph/9512295.

\bibitem{Alekhin:2000es}
S.~Alekhin.
\newblock hep-ex/0005042.

\bibitem{James:1971}
W.~Eadie, D.~Drijard, F.~James, M.~Roos, and B.~Sadoulet.
\newblock Statistical Methods in Experimental Physics, North Holland, 1971.

\bibitem{D'Agostini:1994uj}
G.~D'Agostini.
\newblock {\em Nucl. Instrum. Meth.}, A346:306--311, 1994.

\bibitem{Botje:2001fx}
M.~Botje.
\newblock hep-ph/0110123.

\bibitem{Pascaud:1995qs}
C.~Pascaud and F.~Zomer.
\newblock preprint LAL-95-05.

\bibitem{Botje:1999dj}
M.~Botje.
\newblock {\em Eur. Phys. J.}, C14:285--297, 2000.

\bibitem{QCDNUM}
M.~Botje.
\newblock Zeus Note 97--066.

\bibitem{Blumlein:1996rp}
J.~Bl{\"u}mlein et~al.
\newblock hep-ph/9609400.

\bibitem{Furmanski:1982cw}
W.~Furmanski and R.~Petronzio.
\newblock {\em Z. Phys.}, C11:293, 1982.

\bibitem{Tarasov:1980au}
O.V. Tarasov, A.A. Vladimirov, and A.Yu. Zharkov.
\newblock {\em Phys. Lett.}, B93:429, 1980.

\bibitem{Larin:1993tp}
S.A. Larin and J.A.M. Vermaseren.
\newblock {\em Phys. Lett.}, B303:334, 1993.

\bibitem{vanNeerven:1999ca}
W.~L. van Neerven and A.~Vogt.
\newblock {\em Nucl. Phys.}, B568:263, 2000.

\bibitem{vanNeerven:2000uj}
W.~L. van Neerven and A.~Vogt.
\newblock {\em Nucl. Phys.}, B588:345, 2000.

\bibitem{vanNeerven:2000wp}
W.~L. van Neerven and A.~Vogt.
\newblock {\em Phys. Lett.}, B490:111, 2000.

\bibitem{Larin:1994vu}
S.A. Larin, T.~van Ritbergen, and J.A.M. Vermaseren.
\newblock {\em Nucl. Phys.}, B427:41, 1994.

\bibitem{Larin:1997wd}
S.A. Larin, P.~Nogueira, T.~van Ritbergen, and J.A.M. Vermaseren.
\newblock {\em Nucl. Phys.}, B492:338, 1997.

\bibitem{Retey:2000nq}
A.~Retey and J.~A.~M. Vermaseren.
\newblock {\em Nucl. Phys.}, B604:281, 2001.

\bibitem{Catani:1994sq}
S.~Catani and F.~F.~Hautmann.
\newblock {\em Nucl. Phys.}, B427:475, 1994.

\bibitem{Blumlein:1996jp}
J.~Bl{\"u}mlein and A.~Vogt.
\newblock {\em Phys. Lett.}, B370:149, 1996.

\bibitem{Fadin:1998py}
V.S. Fadin and L.N. Lipatov.
\newblock {\em Phys. Lett.}, B429:127, 1998.

\bibitem{Ciafaloni:1998gs}
M.~Ciafaloni and G.~Camici.
\newblock {\em Phys. Lett.}, B430:349, 1998.

\bibitem{Gracey:1994nn}
J.A. Gracey.
\newblock {\em Phys. Lett.}, B322:141, 1994.

\bibitem{Bennett:1998ch}
J.F Bennett and J.A. Gracey.
\newblock {\em Nucl. Phys.}, B517:241, 1998.

\bibitem{Santorelli:1998yt}
Pietro Santorelli and Egidio Scrimieri.
\newblock {\em Phys. Lett.}, B459:599, 1999.

\bibitem{Ratcliffe:2000kp}
P.~G. Ratcliffe.
\newblock {\em Phys. Rev.}, D63:116004, 2001.

\bibitem{Pascaud:2001bi}
C.~Pascaud and F.~Zomer.
\newblock Preprint.
\newblock hep-ph/0104013.

\bibitem{Dasgupta:2001eq}
M.~Dasgupta and G.~P. Salam.
\newblock hep-ph/0110213.

\bibitem{Blumlein:1998em}
J.~Blumlein and A.~Vogt.
\newblock {\em Phys. Rev.}, D58:014020, 1998.

\bibitem{GSAV}
G.~P. Salam and A.~Vogt.
\newblock in preparation.

\bibitem{Lai:1999wy}
H.~L. Lai et~al.
\newblock {\em Eur. Phys. J.}, C12:375, 2000.

\bibitem{Buza:1998wv}
M.~Buza, Y.~Matiounine, J.~Smith, and W.~L. van Neerven.
\newblock {\em Eur. Phys. J.}, C1:301, 1998.

\bibitem{Larin:1995va}
S.~A. Larin, T.~van Ritbergen, and J.~A.~M. Vermaseren.
\newblock {\em Nucl. Phys.}, B438:278, 1995.

\bibitem{Chetyrkin:1997sg}
K.~G. Chetyrkin, B.~A. Kniehl, and M.~Steinhauser.
\newblock {\em Phys. Rev. Lett.}, 79:2184, 1997.

\bibitem{Bern:1994kr}
Z.~Bern, L.~J. Dixon, and D.~A. Kosower.
\newblock {\em Nucl. Phys.}, B412:751--816, 1994.

\bibitem{Bern:1993em}
Z.~Bern, L.~J. Dixon, and D.~A. Kosower.
\newblock {\em Phys. Lett.}, B302:299--308, 1993.

\bibitem{Kunszt:1994tq}
Z.~Kunszt, A.~Signer, and Z.~Trocsanyi.
\newblock {\em Phys. Lett.}, B336:529--536, 1994.

\bibitem{Signer:1995nk}
A.~Signer.
\newblock {\em Phys. Lett.}, B357:204--210, 1995.

\bibitem{Bern:1997ka}
Z.~Bern, L.~J. Dixon, D.~A. Kosower, and S.~Weinzierl.
\newblock {\em Nucl. Phys.}, B489:3--23, 1997.

\bibitem{Bern:1998sd}
Z.~Bern, L.~J. Dixon, and D.~A. Kosower.
\newblock {\em Nucl. Phys.}, B513:3--86, 1998.

\bibitem{Campbell:1997tv}
J.~M. Campbell, E.~W.~N. Glover, and D.~J. Miller.
\newblock {\em Phys. Lett.}, B409:503--508, 1997.

\bibitem{Glover:1997eh}
E.~W.~N. Glover and D.~J. Miller.
\newblock {\em Phys. Lett.}, B396:257--263, 1997.

\bibitem{DelDuca:1999pa}
V.~Del~Duca, W.~B. Kilgore, and F.~Maltoni.
\newblock {\em Nucl. Phys.}, B566:252--274, 2000.

\bibitem{Beenakker:2001rj}
W.~Beenakker et~al.
\newblock {\em Phys. Rev. Lett.}, 87:201805, 2001.

\bibitem{Reina:2001bc}
L.~Reina, S.~Dawson, and D.~Wackeroth.
\newblock {\em Phys. Rev.}, D65:053017, 2002.

\bibitem{DelDuca:2001fn}
V.~Del~Duca, W.B. Kilgore, C.~Oleari, C.~Schmidt, and D.~Zeppenfeld.
\newblock {\em Nucl. Phys.}, B616:367--399, 2001.

\bibitem{Binoth:1999sp}
T.~Binoth, J.~P. Guillet, and G.~Heinrich.
\newblock {\em Nucl. Phys.}, B572:361--386, 2000.

\bibitem{Bern:1992aq}
Z.~Bern and D.~A. Kosower.
\newblock {\em Nucl. Phys.}, B379:451--561, 1992.

\bibitem{Bern:1991ux}
Z.~Bern and D.~A. Kosower.
\newblock {\em Nucl. Phys.}, B362:389--448, 1991.

\bibitem{Bern:1993wt}
Z.~Bern, D.~C. Dunbar, and T.~Shimada.
\newblock {\em Phys. Lett.}, B312:277--284, 1993.

\bibitem{Strassler:1992zr}
M.~J. Strassler.
\newblock {\em Nucl. Phys.}, B385:145--184, 1992.

\bibitem{Bern:1992an}
Z.~Bern and D.~C. Dunbar.
\newblock {\em Nucl. Phys.}, B379:562--601, 1992.

\bibitem{Schubert:2001he}
C.~Schubert.
\newblock {\em Phys. Rept.}, 355:73--234, 2001.

\bibitem{Frizzo:2000ez}
A.~Frizzo, L.~Magnea, and R.~Russo.
\newblock {\em Nucl. Phys.}, B604:92--120, 2001.

\bibitem{Binoth:2001vm}
T.~Binoth, J.~P. Guillet, G.~Heinrich, and C.~Schubert.
\newblock {\em Nucl. Phys.}, B615:385--401, 2001.

\bibitem{Moch:2001fr}
S.~Moch, J.~A.~M. Vermaseren, and M.~Zhou.
\newblock {\em hep-ph/0108033}.

\bibitem{Hamberg:1991np}
R.~Hamberg, W.~L. van Neerven, and T.~Matsuura.
\newblock {\em Nucl. Phys.}, B359:343--405, 1991.

\bibitem{Harlander:2002wh}
R.V. Harlander and W.B. Kilgore.
\newblock {\em hep-ph/0201206}.

\bibitem{Zijlstra:1992qd}
E.~B. Zijlstra and W.~L. van Neerven.
\newblock {\em Nucl. Phys.}, B383:525--574, 1992.

\bibitem{Martin:2002dr}
A.~D. Martin, R.~G. Roberts, W.~J. Stirling, and R.~S. Thorne.
\newblock {\em hep-ph/0201127}.

\bibitem{Gehrmann-DeRidder:1998gf}
A.~Gehrmann-De~Ridder and E.~W.~N. Glover.
\newblock {\em Nucl. Phys.}, B517:269--323, 1998.

\bibitem{Campbell:1998hg}
J.~M. Campbell and E.~W.~N. Glover.
\newblock {\em Nucl. Phys.}, B527:264--288, 1998.

\bibitem{Catani:1998nv}
S.~Catani and M.~Grazzini.
\newblock {\em Phys. Lett.}, B446:143--152, 1999.

\bibitem{Catani:1999ss}
S.~Catani and M.~Grazzini.
\newblock {\em Nucl. Phys.}, B570:287--325, 2000.

\bibitem{DelDuca:1999ha}
V.~Del~Duca, A.~Frizzo, and F.~Maltoni.
\newblock {\em Nucl. Phys.}, B568:211--262, 2000.

\bibitem{Berends:1989zn}
F.~A. Berends and W.~T. Giele.
\newblock {\em Nucl. Phys.}, B313:595, 1989.

\bibitem{Catani}
S.~Catani.
\newblock 1992.
\newblock in Proceedings of the workshop on New technqiues for Calculating
  Higher Order QCD Corrections, report ETH-TH/93-01, Zurich (1992).

\bibitem{Bern:1994zx}
Z.~Bern, L.~J. Dixon, D.~C. Dunbar, and D.~A. Kosower.
\newblock {\em Nucl. Phys.}, B425:217--260, 1994.

\bibitem{Bern:1998sc}
Z.~Bern, V.~Del~Duca, and C.~R. Schmidt.
\newblock {\em Phys. Lett.}, B445:168--177, 1998.

\bibitem{Kosower:1999xi}
D.~A. Kosower.
\newblock {\em Nucl. Phys.}, B552:319--336, 1999.

\bibitem{Bern:1999ry}
Z.~Bern, V.~Del~Duca, W.~B. Kilgore, and C.~R. Schmidt.
\newblock {\em Phys. Rev.}, D60:116001, 1999.

\bibitem{Kosower:1999rx}
D.~A. Kosower and P.~Uwer.
\newblock {\em Nucl. Phys.}, B563:477--505, 1999.

\bibitem{Catani:2000pi}
S.~Catani and M.~Grazzini.
\newblock {\em Nucl. Phys.}, B591:435--454, 2000.

\bibitem{Bollini:1972ui}
C.~G. Bollini and J.~J. Giambiagi.
\newblock {\em Nuovo Cim.}, B12:20--25, 1972.

\bibitem{Ashmore:1972uj}
J.~F. Ashmore.
\newblock {\em Lett. Nuovo Cim.}, 4:289--290, 1972.

\bibitem{Cicuta:1972jf}
G.~M. Cicuta and E.~Montaldi.
\newblock {\em Nuovo Cim. Lett.}, 4:329--332, 1972.

\bibitem{'tHooft:1972fi}
G.~'t~Hooft and M.~J.~G. Veltman.
\newblock {\em Nucl. Phys.}, B44:189--213, 1972.

\bibitem{Tkachov:1981wb}
F.~V. Tkachov.
\newblock {\em Phys. Lett.}, B100:65--68, 1981.

\bibitem{Chetyrkin:1981qh}
K.~G. Chetyrkin and F.~V. Tkachov.
\newblock {\em Nucl. Phys.}, B192:159--204, 1981.

\bibitem{Gehrmann:1999as}
T.~Gehrmann and E.~Remiddi.
\newblock {\em Nucl. Phys.}, B580:485--518, 2000.

\bibitem{Smirnov:1999wz}
V.~A. Smirnov and O.~L. Veretin.
\newblock {\em Nucl. Phys.}, B566:469--485, 2000.

\bibitem{Anastasiou:1999bn}
C.~Anastasiou, E.~W.~N. Glover, and C.~Oleari.
\newblock {\em Nucl. Phys.}, B575:416--436, 2000.
\newblock Erratum: B585, 763, 2000.

\bibitem{Anastasiou:2000mf}
C.~Anastasiou, T.~Gehrmann, C.~Oleari, E.~Remiddi, and J.~B. Tausk.
\newblock {\em Nucl. Phys.}, B580:577--601, 2000.

\bibitem{Laporta:2001dd}
S.~Laporta.
\newblock {\em Int. J. Mod. Phys.}, A15:5087--5159, 2000.

\bibitem{Baikov:2000jg}
P.~A. Baikov and V.~A. Smirnov.
\newblock {\em Phys. Lett.}, B477:367--372, 2000.

\bibitem{Gorishnii:1989gt}
S.~G. Gorishnii, S.~A. Larin, L.~R. Surguladze, and F.~V. Tkachov.
\newblock {\em Comput. Phys. Commun.}, 55:381--408, 1989.

\bibitem{Chetyrkin:1996ia}
K.~G. Chetyrkin, J.~H. Kuhn, and A.~Kwiatkowski.
\newblock {\em Phys. Rept.}, 277:189--281, 1996.

\bibitem{Harlander:2000mg}
R.V. Harlander.
\newblock {\em Phys. Lett.}, B492:74--80, 2000.

\bibitem{Catani:2001ic}
S.~Catani, D.~de~Florian, and M.~Grazzini.
\newblock {\em JHEP}, 05:025, 2001.

\bibitem{Harlander:2001is}
R.V. Harlander and W.B. Kilgore.
\newblock {\em Phys. Rev.}, D64:013015, 2001.

\bibitem{Smirnov:1999gc}
V.~A. Smirnov.
\newblock {\em Phys. Lett.}, B460:397--404, 1999.

\bibitem{Tausk:1999vh}
J.~B. Tausk.
\newblock {\em Phys. Lett.}, B469:225--234, 1999.

\bibitem{Anastasiou:1999cx}
C.~Anastasiou, E.~W.~N. Glover, and C.~Oleari.
\newblock {\em Nucl. Phys.}, B565:445--467, 2000.

\bibitem{Suzuki:2001yf}
A.~T. Suzuki and A.~G.~M. Schmidt.
\newblock {\em J. Phys.}, A35:151--164, 2002.

\bibitem{Anastasiou:2000kp}
C.~Anastasiou, J.~B. Tausk, and M.~E. Tejeda-Yeomans.
\newblock {\em Nucl. Phys. Proc. Suppl.}, 89:262--267, 2000.

\bibitem{Smirnov:2000vy}
V.~A. Smirnov.
\newblock {\em Phys. Lett.}, B491:130--136, 2000.

\bibitem{Smirnov:2000ie}
V.~A. Smirnov.
\newblock {\em Phys. Lett.}, B500:330--337, 2001.

\bibitem{Smirnov:2001cm}
V.~A. Smirnov.
\newblock {\em Phys. Lett.}, B524:129--136, 2002.

\bibitem{Kotikov:1991pm}
A.~V. Kotikov.
\newblock {\em Phys. Lett.}, B267:123--127, 1991.

\bibitem{Remiddi:1997ny}
E.~Remiddi.
\newblock {\em Nuovo Cim.}, A110:1435--1452, 1997.

\bibitem{Caffo:1998du}
M.~Caffo, H.~Czyz, S.~Laporta, and E.~Remiddi.
\newblock {\em Nuovo Cim.}, A111:365--389, 1998.

\bibitem{Caffo:1999nk}
M.~Caffo, H.~Czyz, and E.~Remiddi.
\newblock {\em Nucl. Phys.}, B581:274--294, 2000.

\bibitem{Gehrmann:2000xj}
T.~Gehrmann and E.~Remiddi.
\newblock {\em Nucl. Phys. Proc. Suppl.}, 89:251--255, 2000.

\bibitem{Gehrmann:2000zt}
T.~Gehrmann and E.~Remiddi.
\newblock {\em Nucl. Phys.}, B601:248--286, 2001.

\bibitem{Gehrmann:2001ck}
T.~Gehrmann and E.~Remiddi.
\newblock {\em Nucl. Phys.}, B601:287--317, 2001.

\bibitem{Binoth:2000ps}
T.~Binoth and G.~Heinrich.
\newblock {\em Nucl. Phys.}, B585:741--759, 2000.

\bibitem{nielsen}
N.~Nielsen.
\newblock {\em Nova Acta Leopoldina (Halle).}, 90:123, 1909.

\bibitem{lewin}
L.~Lewin.
\newblock Polylogarithms and Associated Functions, North Holland (1981).

\bibitem{Kolbig:1986qt}
K.~S. Kolbig.
\newblock {\em SIAM J. Math. Anal.}, 17:1232--1258, 1986.

\bibitem{bit}
K.~S. Kolbig, J.A. Mignaco, and E.~Remiddi.
\newblock {\em BIT}, 10:38, 1970.

\bibitem{Remiddi:1999ew}
E.~Remiddi and J.~A.~M. Vermaseren.
\newblock {\em Int. J. Mod. Phys.}, A15:725--754, 2000.

\bibitem{Gehrmann:2001pz}
T.~Gehrmann and E.~Remiddi.
\newblock {\em Comput. Phys. Commun.}, 141:296--312, 2001.

\bibitem{Gehrmann:2001jv}
T.~Gehrmann and E.~Remiddi.
\newblock {\em Comput. Phys. Commun.}, 144:200--223, 2002.

\bibitem{Gonsalves:1983nq}
R.~J. Gonsalves.
\newblock {\em Phys. Rev.}, D28:1542, 1983.

\bibitem{Kramer:1987sr}
G.~Kramer and B.~Lampe.
\newblock {\em J. Math. Phys.}, 28:945, 1987.

\bibitem{Bern:2000ie}
Z.~Bern, L.~J. Dixon, and A.~Ghinculov.
\newblock {\em Phys. Rev.}, D63:053007, 2001.

\bibitem{Anastasiou:2000kg}
C.~Anastasiou, E.~W.~N. Glover, C.~Oleari, and M.~E. Tejeda-Yeomans.
\newblock {\em Nucl. Phys.}, B601:318--340, 2001.

\bibitem{Anastasiou:2000ue}
C.~Anastasiou, E.~W.~N. Glover, C.~Oleari, and M.~E. Tejeda-Yeomans.
\newblock {\em Nucl. Phys.}, B601:341--360, 2001.

\bibitem{Anastasiou:2001sv}
C.~Anastasiou, E.~W.~N. Glover, C.~Oleari, and M.~E. Tejeda-Yeomans.
\newblock {\em Nucl. Phys.}, B605:486--516, 2001.

\bibitem{Glover:2001af}
E.~W.~N. Glover, C.~Oleari, and M.~E. Tejeda-Yeomans.
\newblock {\em Nucl. Phys.}, B605:467--485, 2001.

\bibitem{Bern:2000dn}
Z.~Bern, L.~J. Dixon, and D.~A. Kosower.
\newblock {\em JHEP}, 01:027, 2000.

\bibitem{Bern:2002tk}
Z.~Bern, A.~De~Freitas, and L.J. Dixon.
\newblock {\em hep-ph/0201161}.

\bibitem{Bern:2001df}
Z.~Bern, A.~De~Freitas, and L.~J. Dixon.
\newblock {\em JHEP}, 09:037, 2001.

\bibitem{Anastasiou:2002zn}
C.~Anastasiou, E.~W.~N. Glover, and M.~E. Tejeda-Yeomans.
\newblock {\em hep-ph/0201274}.

\bibitem{Bern:2001dg}
Z.~Bern, A.~De~Freitas, L.~J. Dixon, A.~Ghinculov, and H.~L. Wong.
\newblock {\em JHEP}, 11:031, 2001.

\bibitem{Binoth:2002xg}
T.~Binoth, E.~W.~N. Glover, P.~Marquard, and J.~J. van~der Bij.
\newblock {\em hep-ph/0202266}.

\bibitem{Bern:2002zk}
Z.~Bern, A.~De~Freitas, L.~Dixon, and H.~L. Wong.
\newblock {\em hep-ph/0202271}.

\bibitem{Anastasiou:2000mv}
C.~Anastasiou, E.~W.~N. Glover, C.~Oleari, and M.~E. Tejeda-Yeomans.
\newblock {\em Phys. Lett.}, B506:59--67, 2001.

\bibitem{Glover:2001rd}
E.~W.~N. Glover and M.~E. Tejeda-Yeomans.
\newblock {\em JHEP}, 05:010, 2001.

\bibitem{Glover:2001ev}
E.~W.~N. Glover, J.~B. Tausk, and J.~J. Van~der Bij.
\newblock {\em Phys. Lett.}, B516:33--38, 2001.

\bibitem{Arbuzov:1997qd}
A.~B. Arbuzov, V.S. Fadin, E.A. Kuraev, L.N. Lipatov, N.P. Merenkov, and
  L.~Trentadue.
\newblock {\em Nucl. Phys.}, B485:457--502, 1997.

\bibitem{Garland:2001tf}
L.~W. Garland, T.~Gehrmann, E.~W.~N. Glover, A.~Koukoutsakis, and E.~Remiddi.
\newblock {\em Nucl. Phys.}, B627:107--188, 2002.

\bibitem{Bethke:2000ai}
S.~Bethke.
\newblock {\em J. Phys.}, G26:R27, 2000.

\bibitem{Aguilar-Saavedra:2001rg}
J.~A. Aguilar-Saavedra et~al.
\newblock {\em hep-ph/0106315}.

\bibitem{Ellis:1981wv}
R.~K. Ellis, D.~A. Ross, and A.~E. Terrano.
\newblock {\em Nucl. Phys.}, B178:421, 1981.

\bibitem{Fabricius:1981sx}
K.~Fabricius, I.~Schmitt, G.~Kramer, and G.~Schierholz.
\newblock {\em Zeit. Phys.}, C11:315, 1981.

\bibitem{Giele:1992vf}
W.~T. Giele and E.~W.~N. Glover.
\newblock {\em Phys. Rev.}, D46:1980--2010, 1992.

\bibitem{Catani:1997vz}
S.~Catani and M.~H. Seymour.
\newblock {\em Nucl. Phys.}, B485:291--419, 1997.
\newblock Erratum: B510, 503, 1997.

\bibitem{Kramer:1987sg}
G.~Kramer and B.~Lampe.
\newblock {\em Z. Phys.}, C34:497, 1987.
\newblock Erratum: C42, 504, 1989.

\bibitem{Matsuura:1988wt}
T.~Matsuura and W.~L. van Neerven.
\newblock {\em Z. Phys.}, C38:623, 1988.

\bibitem{Matsuura:1989sm}
T.~Matsuura, S.~C. van~der Marck, and W.~L. van Neerven.
\newblock {\em Nucl. Phys.}, B319:570, 1989.

\bibitem{Catani:1998bh}
S.~Catani.
\newblock {\em Phys. Lett.}, B427:161--171, 1998.

\bibitem{Gehrmann-DeRidder:1997wx}
A.~Gehrmann-De~Ridder, T.~Gehrmann, and E.~W.~N. Glover.
\newblock {\em Phys. Lett.}, B414:354--361, 1997.

\bibitem{deFlorian:2000pr}
D.~de~Florian and M.~Grazzini.
\newblock {\em Phys. Rev. Lett.}, 85:4678--4681, 2000.

\bibitem{deFlorian:2001zd}
D.~de~Florian and M.~Grazzini.
\newblock {\em Nucl. Phys.}, B616:247--285, 2001.

\bibitem{Kuraev:1976ge}
E.~A. Kuraev, L.~N. Lipatov, and V.~S. Fadin.
\newblock {\em Sov. Phys. JETP}, 44:443--450, 1976.

\bibitem{DelDuca:1995zy}
V.~Del~Duca.
\newblock {\em Phys. Rev.}, D52:1527--1534, 1995.

\bibitem{Kuraev:1977fs}
E.~A. Kuraev, L.~N. Lipatov, and V.~S. Fadin.
\newblock {\em Sov. Phys. JETP}, 45:199--204, 1977.

\bibitem{Balitsky:1978ic}
I.~I. Balitsky and L.~N. Lipatov.
\newblock {\em Sov. J. Nucl. Phys.}, 28:822--829, 1978.

\bibitem{Lipatov:1976zz}
L.~N. Lipatov.
\newblock {\em Sov. J. Nucl. Phys.}, 23:338--345, 1976.

\bibitem{Fadin:1993wh}
V.~S. Fadin and L.~N. Lipatov.
\newblock {\em Nucl. Phys.}, B406:259--292, 1993.

\bibitem{Lipatov:1989bs}
L.~N. Lipatov.
\newblock {\em Adv. Ser. Direct. High Energy Phys.}, 5, 1989.

\bibitem{DelDuca:2001gu}
V.~Del~Duca and E.~W.~N. Glover.
\newblock {\em JHEP}, 10:035, 2001.

\bibitem{Fadin:1995xg}
V.~S. Fadin, M.~I. Kotsky, and R.~Fiore.
\newblock {\em Phys. Lett.}, B359:181--188, 1995.

\bibitem{Fadin:1996tb}
V.~S. Fadin, R.~Fiore, and M.~I. Kotsky.
\newblock {\em Phys. Lett.}, B387:593--602, 1996.

\bibitem{Fadin:1996km}
V.~Fadin, R.~Fiore, and A.~Quartarolo.
\newblock {\em Phys. Rev.}, D53:2729--2741, 1996.

\bibitem{Blumlein:1998ib}
J.~Blumlein, V.~Ravindran, and W.~L. van Neerven.
\newblock {\em Phys. Rev.}, D58:091502, 1998.

\bibitem{mandel}
S.~Mandelstam.
\newblock {\em Phys. Rev.}, 137:B949, 1965.

\bibitem{Bogdan:2002sr}
A.~V. Bogdan, V.~Del~Duca, V.~S. Fadin, and E.~W.~N. Glover.
\newblock {\em hep-ph/0201240}.

\bibitem{Fadin:1977jr}
V.~S. Fadin and V.~E. Sherman.
\newblock {\em Zh. Eksp. Teor. Fiz.}, 72:1640--1658, 1977.

\bibitem{gellmann}
M.L. Gell-Mann, M.and~Goldberger, F.E. Low, E.~Marx, and F.~Zachariasen.
\newblock {\em Phys. Rev.}, 133:B145, 1964.

\bibitem{McCoy:1976fj}
B.~M. McCoy and T.~T. Wu.
\newblock {\em Phys. Rev.}, D13:484--507, 1976.

\bibitem{Fadin:2001dc}
V.~S. Fadin and R.~Fiore.
\newblock {\em Phys. Rev.}, D64:114012, 2001.

\bibitem{Balazs:2000sz}
C.~Balazs, J.~Huston, and I.~Puljak.
\newblock {\em Phys. Rev.}, D63:014021, 2001.

\bibitem{Dokshitzer:1980hw}
Y.L. Dokshitzer, D.~Diakonov, and S.~I. Troian.
\newblock {\em Phys. Rept.}, 58:269, 1980.

\bibitem{Davies:1984hs}
C.~T.~H. Davies and W.J. Stirling.
\newblock {\em Nucl. Phys.}, B244:337, 1984.

\bibitem{Collins:1985kg}
J.C. Collins, D.E. Soper, and G.~Sterman.
\newblock {\em Nucl. Phys.}, B250:199, 1985.

\bibitem{Catani:2000vq}
S.~Catani, D.~de~Florian, and M.~Grazzini.
\newblock {\em Nucl. Phys.}, B596:299, 2001.

\bibitem{Kodaira:1982nh}
J.~Kodaira and L.~Trentadue.
\newblock {\em Phys. Lett.}, 112B:66, 1982.

\bibitem{Catani:1988vd}
S.~Catani, E.~D'Emilio, and L.~Trentadue.
\newblock {\em Phys. Lett.}, B211:335, 1988.

\bibitem{Vogt:2000ci}
A.~Vogt.
\newblock {\em Phys. Lett.}, B497:228, 2001.

\bibitem{Balazs:1997xd}
C.~Balazs and C.~P. Yuan.
\newblock {\em Phys. Rev.}, D56:5558, 1997.

\bibitem{Kulesza:2001jc}
A.~Kulesza and W.J. Stirling.
\newblock {\em Eur. Phys. J.}, C20:349, 2001.

\bibitem{deFlorian:1999zd}
D.~de~Florian, M.~Grazzini, and Z.~Kunszt.
\newblock {\em Phys. Rev. Lett.}, 82:5209, 1999.

\bibitem{Dawson:1991zj}
S.~Dawson.
\newblock {\em Nucl. Phys.}, B359:283, 1991.

\bibitem{Djouadi:1991tk}
A.~Djouadi, M.~Spira, and P.~M. Zerwas.
\newblock {\em Phys. Lett.}, B264:440, 1991.

\bibitem{Spira:1995rr}
M.~Spira, A.~Djouadi, D.~Graudenz, and P.~M. Zerwas.
\newblock {\em Nucl. Phys.}, B453:17, 1995.

\bibitem{Catani:2001cr}
S.~Catani, D.~de~Florian, and M.~Grazzini.
\newblock {\em JHEP}, 01:015, 2002.

\bibitem{Catani:1996yz}
S.~Catani, M.L. Mangano, P.~Nason, and L.~Trentadue.
\newblock {\em Nucl. Phys.}, B478:273, 1996.

\bibitem{Kramer:1996iq}
M.~Kramer, E.~Laenen, and M.~Spira.
\newblock {\em Nucl. Phys.}, B511:523, 1998.

\bibitem{Martin:2000gq}
A.D. Martin, R.~G. Roberts, W.J. Stirling, and R.~S. Thorne.
\newblock {\em Eur. Phys. J.}, C18:117, 2000.

\bibitem{Laenen:2000ij}
E.~Laenen, G.~Sterman, and W.~Vogelsang.
\newblock {\em Phys. Rev.}, D63:114018, 2001.

\bibitem{Contopanagos:1994yq}
H.~Contopanagos and G.~Sterman.
\newblock {\em Nucl. Phys.}, B419:77, 1994.

\bibitem{Webber:1994cp}
B.~R. Webber.
\newblock {\em Phys. Lett.}, B339:148, 1994.

\bibitem{Korchemsky:1995is}
G.P. Korchemsky and G.~Sterman.
\newblock {\em Nucl. Phys.}, B437:415, 1995.

\bibitem{Collins:1981uk}
J.C. Collins and D.E. Soper.
\newblock {\em Nucl. Phys.}, B193:381, 1981.

\bibitem{Kulesza:2002rh}
A~Kulesza, G.~Sterman, and W.~Vogelsang.
\newblock hep-ph/0202251.

\bibitem{Laenen:2000de}
E.~Laenen, G.~Sterman, and W.~Vogelsang.
\newblock {\em Phys. Rev. Lett.}, 84:4296, 2000.

\bibitem{Affolder:1999jh}
T.~Affolder (CDF~Collab.).
\newblock {\em Phys. Rev. Lett.}, 84:845, 2000.

\bibitem{Tafat:2001in}
S.~Tafat.
\newblock {\em JHEP}, 05:004, 2001.

\bibitem{Qiu:2000ga}
J.-W. Qiu and X.-F. Zhang.
\newblock {\em Phys. Rev. Lett.}, 86:2724, 2001.

\bibitem{Collins:1988ig}
J.C. Collins, D.E. Soper, and G.~Sterman.
\newblock {\em Nucl. Phys.}, B308:833, 1988.

\bibitem{Berger:2002mt}
E.L. Berger et~al.
\newblock Summary of working group on QCD and strong interactions. To appear in
  the proceedings of APS / DPF / DPB Summer Study on the Future of Particle
  Physics (Snowmass 2001), Snowmass, Colorado, 30 Jun - 21 Jul 2001.

\bibitem{Kidonakis:1997gm}
N.~Kidonakis and G.~Sterman.
\newblock {\em Nucl. Phys.}, B505:321, 1997.

\bibitem{Kidonakis:1996aq}
N.~Kidonakis and G.~Sterman.
\newblock {\em Phys. Lett.}, B387:867, 1996.

\bibitem{Laenen:1998kp}
E.~Laenen and S.~Moch.
\newblock {\em Phys. Rev.}, D59:034027, 1999.

\bibitem{Kidonakis:1999ze}
N.~Kidonakis.
\newblock {\em Int. J. Mod. Phys.}, A15:1245, 2000.

\bibitem{Kidonakis:2000ui}
N.~Kidonakis.
\newblock {\em Phys. Rev.}, D64:014009, 2001.

\bibitem{Kretzer:2001tc}
S.~Kretzer, D.~Mason, and F.~Olness.
\newblock hep-ph/0112191.

\bibitem{Meng:1996yn}
R.~Meng, F.I. Olness, and D.E. Soper.
\newblock {\em Phys. Rev.}, D54:1919--1935, 1996.

\bibitem{Nadolsky:2000ky}
P.~M. Nadolsky, D.~R. Stump, and C.~P. Yuan.
\newblock {\em Phys. Rev.}, D64:114011, 2001.

\bibitem{Nadolsky:1999kb}
P.~Nadolsky, D.~R. Stump, and C.~P. Yuan.
\newblock {\em Phys. Rev.}, D61:014003, 2000.

\bibitem{Aivazis:1994pi}
M.A.G. Aivazis, J.C. Collins, F.I. Olness, and W.-K. Tung.
\newblock {\em Phys. Rev.}, D50:3102, 1994.

\bibitem{Collins:1998rz}
J.C. Collins.
\newblock {\em Phys. Rev.}, D58:094002, 1998.

\bibitem{Mele:1990yq}
B.~Mele and P.~Nason.
\newblock {\em Phys. Lett.}, B245:635, 1990.

\bibitem{Mele:1991cw}
B.~Mele and P.~Nason.
\newblock {\em Nucl. Phys.}, B361:626, 1991.

\bibitem{Dokshitzer:1996ev}
Yu.L. Dokshitzer, V.A. Khoze, and S.I. Troian.
\newblock {\em Phys. Rev.}, D53:89, 1996.

\bibitem{Cacciari:1994mq}
M.~Cacciari and M.~Greco.
\newblock {\em Nucl. Phys.}, B421:530, 1994.

\bibitem{Cacciari:1996fs}
M.~Cacciari and M.~Greco.
\newblock {\em Z. Phys.}, C69:459, 1996.

\bibitem{Cacciari:1996ej}
M.~Cacciari et~al.
\newblock {\em Nucl. Phys.}, B466:173, 1996.

\bibitem{Cacciari:1998it}
M.~Cacciari, M.~Greco, and P.~Nason.
\newblock {\em JHEP}, 05:007, 1998.

\bibitem{Cacciari:2001td}
M.~Cacciari, S.~Frixione, and P.~Nason.
\newblock {\em JHEP}, 03:006, 2001.

\bibitem{Cacciari:2001cw}
M.~Cacciari and S.~Catani.
\newblock {\em Nucl. Phys.}, B617:253, 2001.

\bibitem{Nason:1999ta}
P.~Nason et~al.
\newblock hep-ph/0003142.

\bibitem{Cacciari:1997wr}
M.~Cacciari, M.~Greco, S.~Rolli, and A.~Tanzini.
\newblock {\em Phys. Rev.}, D55:2736, 1997.

\bibitem{Cacciari:1997du}
M.~Cacciari and M.~Greco.
\newblock {\em Phys. Rev.}, D55:7134, 1997.

\bibitem{Heister:2001jg}
A.~Heister et~al.
\newblock {\em Phys. Lett.}, B512:30, 2001.

\bibitem{Kartvelishvili:1978pi}
V.G. Kartvelishvili, A.K. Likhoded, and V.A. Petrov.
\newblock {\em Phys. Lett.}, B78:615, 1978.

\bibitem{Jaffe:1994ie}
R.L. Jaffe and L.~Randall.
\newblock {\em Nucl. Phys.}, B412:79, 1994.

\bibitem{Nason:1997pk}
P.~Nason and B.R. Webber.
\newblock {\em Phys. Lett.}, B395:355, 1997.

\bibitem{Beneke:1995pq}
M.~Beneke and V.~M. Braun.
\newblock {\em Nucl. Phys.}, B454:253, 1995.

\bibitem{Sterman:1987aj}
G.~Sterman.
\newblock {\em Nucl. Phys.}, B281:310, 1987.

\bibitem{Sterman:1999gz}
G.~Sterman and W.~Vogelsang.
\newblock hep-ph/9910371.
\newblock 1999.

\bibitem{Gardi:2001di}
E.~Gardi.
\newblock {\em Nucl. Phys.}, B622:365, 2002.

\bibitem{Magnea:2000ss}
L.~Magnea.
\newblock {\em Nucl. Phys.}, B593:269, 2001.

\bibitem{Magnea:2001ge}
L.~Magnea.
\newblock hep-ph/0109168.

\bibitem{Catani:1989ne}
S.~Catani and L.~Trentadue.
\newblock {\em Nucl. Phys.}, B327:323, 1989.

\bibitem{Contopanagos:1997nh}
H.~Contopanagos, E.~Laenen, and G.~Sterman.
\newblock {\em Nucl. Phys.}, B484:303, 1997.

\bibitem{Catani:1991rr}
S.~Catani, B.R. Webber, and G.~Marchesini.
\newblock {\em Nucl. Phys.}, B349:635, 1991.

\bibitem{Qiu:1991xx}
J.-W. Qiu and G.~Sterman.
\newblock {\em Nucl. Phys.}, B353:105, 1991.

\bibitem{Akhoury:1997pb}
R.~Akhoury, M.~G. Sotiropoulos, and V.I. Zakharov.
\newblock {\em Phys. Rev.}, D56:377, 1997.

\bibitem{Beneke:1995qe}
M.~Beneke and V.~M. Braun.
\newblock {\em Phys. Lett.}, B348:513, 1995.

\bibitem{Ball:1995ni}
P.~Ball, M.~Beneke, and V.~M. Braun.
\newblock {\em Nucl. Phys.}, B452:563, 1995.

\bibitem{Dokshitzer:1996qm}
Y.L. Dokshitzer, G.~Marchesini, and B.~R. Webber.
\newblock {\em Nucl. Phys.}, B469:93, 1996.

\bibitem{Korchemsky:1996iq}
G.~P. Korchemsky.
\newblock in proceedings of 28th international conference on high-energy
  physics, warsaw, poland, july 1996.
\newblock 1996.

\bibitem{Gardi:1999dq}
E.~Gardi and G.~Grunberg.
\newblock {\em JHEP}, 11:016, 1999.

\bibitem{Gardi:2001ny}
E.~Gardi and J.~Rathsman.
\newblock {\em Nucl. Phys.}, B609:123, 2001.

\bibitem{Gardi:2002bg}
E.~Gardi and J.~Rathsman.
\newblock hep-ph/0201019.

\bibitem{Laenen:1998qw}
E.~Laenen, G.~Oderda, and G.~Sterman.
\newblock {\em Phys. Lett.}, B438:173, 1998.

\bibitem{Kidonakis:2001nj}
N.~Kidonakis, E.~Laenen, S.~Moch, and R.~Vogt.
\newblock {\em Phys. Rev.}, D64:114001, 2001.

\bibitem{Nason:1988xz}
P.~Nason, S.~Dawson, and R.K. Ellis.
\newblock {\em Nucl. Phys.}, B303:607, 1988.

\bibitem{Beenakker:1989bq}
W.~Beenakker, H.~Kuijf, W.L. van Neerven, and J.~Smith.
\newblock {\em Phys. Rev.}, D40:54, 1989.

\bibitem{Beenakker:1991ma}
W.~Beenakker, W.L. van Neerven, R.~Meng, G.A. Schuler, and J.~Smith.
\newblock {\em Nucl. Phys.}, B351:507, 1991.

\bibitem{Oderda:1999im}
G.~Oderda, N.~Kidonakis, and G.~Sterman.
\newblock hep-ph/9906338.

\bibitem{Sterman:2000pu}
G.~Sterman and W.~Vogelsang.
\newblock hep-ph/0002132.

\bibitem{Forte:1999kh}
S.~Forte.
\newblock hep-ph/9910397.

\bibitem{Chekelian:2001pi}
V.~Chekelian.
\newblock hep-ex/0107053.

\bibitem{Chekanov:2001qu}
S.~Chekanov et~al.
\newblock {\em Eur. Phys. J.}, C21:443, 2001.

\bibitem{Adloff:2000qk}
C.~Adloff et~al.
\newblock {\em Eur. Phys. J.}, C21:33, 2001.

\bibitem{Ball:1994du}
R.D. Ball and S.~Forte.
\newblock {\em Phys. Lett.}, B335:77, 1994.

\bibitem{DeRujula:1974rf}
A.~De~Rujula et~al.
\newblock {\em Phys. Rev.}, D10:1649, 1974.

\bibitem{Forte:1995vs}
S.~Forte and R.D. Ball.
\newblock {\em Acta Phys. Polon.}, B26:2097, 1995.

\bibitem{Mankiewicz:1997sd}
L.~Mankiewicz, A.~Saalfeld, and T.~Weigl.
\newblock {\em Phys. Lett.}, B393:175, 1997.

\bibitem{Fadin:1975cb}
V.~S. Fadin, E.~A. Kuraev, and L.~N. Lipatov.
\newblock {\em Phys. Lett.}, B60:50, 1975.

\bibitem{Ball:1995vc}
R.D. Ball and S.~Forte.
\newblock {\em Phys. Lett.}, B351:313, 1995.

\bibitem{Ellis:1995gv}
R.~K. Ellis, F.~Hautmann, and B.~R. Webber.
\newblock {\em Phys. Lett.}, B348:582, 1995.

\bibitem{Altarelli:1977zs}
G.~Altarelli and G.~Parisi.
\newblock {\em Nucl. Phys.}, B126:298, 1977.

\bibitem{Gribov:1972ri}
V.~N. Gribov and L.~N. Lipatov.
\newblock {\em Yad. Fiz.}, 15:781, 1972.

\bibitem{McLerran:2001sr}
L.D. McLerran.
\newblock hep-ph/0104285.

\bibitem{Ball:1999sh}
R.D. Ball and S.~Forte.
\newblock {\em Phys. Lett.}, B465:271, 1999.

\bibitem{Altarelli:1999vw}
G.~Altarelli, R.D. Ball, and S.~Forte.
\newblock {\em Nucl. Phys.}, B575:313, 2000.

\bibitem{Ciafaloni:1999yw}
M.~Ciafaloni, D.~Colferai, and G.~P. Salam.
\newblock {\em Phys. Rev.}, D60:114036, 1999.

\bibitem{Ball:1997vf}
R.D. Ball and S.~Forte.
\newblock {\em Phys. Lett.}, B405:317, 1997.

\bibitem{Altarelli:2000mh}
G.~Altarelli, R.D. Ball, and S.~Forte.
\newblock {\em Nucl. Phys.}, B599:383, 2001.

\bibitem{Altarelli:2001ji}
G.~Altarelli, R.D. Ball, and S.~Forte.
\newblock {\em Nucl. Phys.}, B621:359, 2002.

\bibitem{Salam:1998tj}
G.~P. Salam.
\newblock {\em JHEP}, 07:019, 1998.

\bibitem{Ball:2001pq}
R.~D. Ball and R.~K. Ellis.
\newblock {\em JHEP}, 05:053, 2001.

\bibitem{Blazey:2000qt}
G.C. Blazey et~al.
\newblock hep-ex/0005012.

\bibitem{Catani:2000zg}
S.~Catani et~al.
\newblock hep-ph/0005114.

\bibitem{Ellis:2002inprep}
S.D. Ellis, J.~Huston, and M.~T{\"o}nnesmann.
\newblock {\em in preparation}.

\bibitem{Huth:1990mi}
J.E. Huth et~al.
\newblock Toward a standardization of jet definitions.
\newblock Presented at Summer Study on High Energy Physics, Reaearch Directions
  for the Decade, Snowmass, CO, Jun 25 - Jul 13, 1990.

\bibitem{Ellis:1993tq}
S.D. Ellis and D.E. Soper.
\newblock {\em Phys. Rev.}, D48:3160--3166, 1993.

\bibitem{Catani:1992zp}
S.~Catani, Yu.L. Dokshitzer, and B.R. Webber.
\newblock {\em Phys. Lett.}, B285:291--299, 1992.

\bibitem{Catani:1993hr}
S.~Catani, Yu.L. Dokshitzer, M.H. Seymour, and B.R. Webber.
\newblock {\em Nucl. Phys.}, B406:187--224, 1993.

\bibitem{Ellis:1992qq}
S.D. Ellis, Z.~Kunszt, and D.E. Soper.
\newblock {\em Phys. Rev. Lett.}, 69:3615--3618, 1992.

\bibitem{Ellis:1993ik}
S.D. Ellis.
\newblock hep-ph/9306280.

\bibitem{Abbott:1997fc}
B.~Abbott, M.~Bhattacharjee, D.~Elvira, F.~Nang, and H.~Weerts.
\newblock FERMILAB-PUB-97-242-E.

\bibitem{Seymour:1998kj}
M.H. Seymour.
\newblock {\em Nucl. Phys.}, B513:269--300, 1998.

\bibitem{Marchesini:1992ch}
G.~Marchesini et~al.
\newblock {\em Comput. Phys. Commun.}, 67:465--508, 1992.

\bibitem{Corcella:2000bw}
G.~Corcella et~al.
\newblock {\em JHEP}, 01:010, 2001.

\bibitem{Corcella:1999qn}
G.~Corcella et~al.
\newblock hep-ph/9912396.

\bibitem{Abe:1992ui}
F.~Abe et~al.
\newblock {\em Phys. Rev.}, D45:1448--1458, 1992.

\bibitem{Abazov:2001ak}
V.M. Abazov et~al.
\newblock hep-ex/0106032.

\bibitem{Ellis:1992en}
S.D. Ellis, Z.~Kunszt, and D.E. Soper.
\newblock {\em Phys. Rev. Lett.}, 69:1496--1499, 1992.

\bibitem{Ellis:1990ek}
S.D. Ellis, Z.~Kunszt, and D.E. Soper.
\newblock {\em Phys. Rev. Lett.}, 64:2121, 1990.

\bibitem{Ellis:1989vm}
S.D. Ellis, Z.~Kunszt, and D.E. Soper.
\newblock {\em Phys. Rev.}, D40:2188, 1989.

\bibitem{Ellis:1989hv}
S.D. Ellis, Z.~Kunszt, and D.E. Soper.
\newblock {\em Phys. Rev. Lett.}, 62:726, 1989.

\bibitem{Sjostrand:1994yb}
Torbjorn Sjostrand.
\newblock {\em Comput. Phys. Commun.}, 82:74--90, 1994.

\bibitem{Binoth:1999qq}
T.~Binoth, J.~P. Guillet, E.~Pilon, and M.~Werlen.
\newblock {\em Eur. Phys. J.}, C16:311--330, 2000.

\bibitem{Armstrong:1994it}
ATLAS collaboration.
\newblock {ATLAS} {T}echnical {P}roposal.
\newblock CERN-LHCC-94-43.

\bibitem{ATLAS:1999tdr}
ATLAS collaboration.
\newblock {ATLAS} {D}etector and {P}hysics {P}erformance {TDR}.
\newblock CERN-LHCC-99-14.

\bibitem{CMS:1994tp}
CMS collaboration.
\newblock {CMS} {T}echnical {P}roposal.
\newblock CERN-LHCC-94-38.

\bibitem{CMS:1997tdr}
CMS collaboration.
\newblock {CMS} {ECAL} {TDR}.
\newblock CERN-LHCC-97-33.

\bibitem{Branson:2001pj}
J.~G. Branson et~al.
\newblock The {ATLAS} and {CMS} {C}ollaboration.
\newblock {\em hep-ph/0110021}.

\bibitem{Catani:2000jh}
S.~Catani et~al.
\newblock {P}roc. {CERN} {W}orkshop on {S}tandard {M}odel {P}hysics (and more)
  at the {LHC}, {G}eneva 1999,ed. by {G.~Altarelli} and {M.~Mangano}.
\newblock {\em hep-ph/0005025}.

\bibitem{Binoth:2001jd}
T.~Binoth.
\newblock {T}alk given at 36th {R}encontres de {M}oriond on {QCD} and
  {H}adronic {I}nteractions, {L}es {A}rcs, {F}rance, 17-24 {M}arch 2001.
\newblock {\em hep-ph/0105149}.

\bibitem{Gluck:1995uf}
M.~Gluck, E.~Reya, and A.~Vogt.
\newblock {\em Z. Phys.}, C67:433--448, 1995.

\bibitem{Binoth:2002wa}
T.~Binoth, J.~P. Guillet, E.~Pilon, and M.~Werlen.
\newblock {\em hep-ph/0203064}.

\bibitem{Kniehl:2000fe}
B.~A. Kniehl, G.~Kramer, and B.~Potter.
\newblock {\em Nucl. Phys.}, B582:514--536, 2000.

\bibitem{Pukhov:1999gg}
A.~Pukhov et~al.
\newblock hep-ph/9908288.

\bibitem{Ishikawa:1993qr}
T.~Ishikawa et~al.
\newblock Grace manual: Automatic generation of tree amplitudes in standard
  models: Version 1.0.
\newblock KEK-92-19.

\bibitem{Corcella:2001bw}
G.~Corcella et~al.
\newblock {\em JHEP}, 01:010, 2001.

\bibitem{Stelzer:1994ta}
T.~Stelzer and W.~F. Long.
\newblock {\em Comput. Phys. Commun.}, 81:357--371, 1994.

\bibitem{Baer:1999sp}
H.~Baer, F.~E. Paige, S.~D. Protopopescu, and X.~Tata.
\newblock hep-ph/0001086.

\bibitem{Berends:1991ax}
F.~A. Berends, H.~Kuijf, B.~Tausk, and W.~T. Giele.
\newblock {\em Nucl. Phys.}, B357:32--64, 1991.

\bibitem{Sjostrand:2000wi}
T.~Sjostrand et~al.
\newblock {\em Comput. Phys. Commun.}, 135:238--259, 2001.

\bibitem{Caravaglios:1999yr}
F.~Caravaglios, M.~L. Mangano, M.~Moretti, and R.~Pittau.
\newblock {\em Nucl. Phys.}, B539:215--232, 1999.

\bibitem{Mangano:2001xp}
M.~L. Mangano, M.~Moretti, and R.~Pittau.
\newblock hep-ph/0108069.

\bibitem{Belyaev:2000wn}
A.~S. Belyaev et~al.
\newblock hep-ph/0101232.

\bibitem{Sjostrand:2001yu}
T.~Sjostrand, L.~Lonnblad, and S.~Mrenna.
\newblock hep-ph/0108264.

\bibitem{Cranmer:2002kc}
K.~Cranmer et~al.
\newblock {\tt http://pheno.physics.wisc.edu/Software/MadCUP/}.

\bibitem{Mangano:2002ap}
M.L. Mangano et~al.
\newblock {\tt http://cern.ch/mlm/alpgen/alpgen.html}.

\bibitem{Ilyin:2002vi}
V.~Ilyin.
\newblock 2002.
\newblock Presented at the {\it Physics and Detectors for a 90 to 800 GeV
  Linear Collider Second Workshop of the Extended ECFA/DESY Study} in St. Malo,
  France.

\bibitem{Kersevan:2002dd}
B.~P. Kersevan and E.~Richter-Was.
\newblock hep-ph/0201302.

\bibitem{Dobbs:2001dq}
M.~Dobbs.
\newblock hep-ph/0111234.

\bibitem{Sjostrand:1989tj}
T.~Sjostrand et~al.
\newblock in {\it Z Physics at LEP 1}, vol.~3, edited by G. Altarelli, R.
  Kleiss and C. Verzegnassi, CERN 89-08 (Geneva, 1989), p.~327.

\bibitem{Dobbs:2001ck}
M.~Dobbs and J.~B. Hansen.
\newblock {\em Comput. Phys. Commun.}, 134:41--46, 2001.

\bibitem{Garren:2000st}
L.~Garren, I.~G. Knowles, T.~Sjostrand, and T.~Trippe.
\newblock {\em Eur. Phys. J.}, C15:205--207, 2000.

\bibitem{Plothow-Besch:1993qj}
H.~Plothow-Besch.
\newblock {\em Comput. Phys. Commun.}, 75:396--416, 1993.
\newblock refer to {\tt http://consult.cern.ch/writeup/pdflib/}.

\bibitem{Kersevan:2001ab}
B.~Kersevan and E.~Richter-Was.
\newblock 2001.
\newblock ATLAS Physics Communication, ATL-COM-PHYS-2001-013.

\bibitem{Kersevan:2001cd}
B.~Kersevan and E.~Richter-Was.
\newblock 2001.
\newblock ATLAS Physics Communication, ATL-COM-PHYS-2001-014.

\bibitem{Kersevan:2001ef}
B.~Kersevan and E.~Richter-Was.
\newblock 2001.
\newblock ATLAS Physics Communication, ATL-COM-PHYS-2001-025.

\bibitem{ATLAS:1999ab}
ATLAS Collaboration.
\newblock {\it ATLAS Detector and Physics Performance TDR}.
\newblock 1999.
\newblock CERN-LHCC/99-15.

\bibitem{Aachen1990}
G.~Jarlskog and D.~(editors) Rein.
\newblock Proceedings of the large hadron collider workshop, aachen.
\newblock 1990.
\newblock CERN 90-10/ECFA 90-133.

\bibitem{Kleiss1990}
B.~van Eijk and R.~Kleiss.
\newblock Proceedings of the large hadron collider workshop, aachen.
\newblock page 184, 1990.
\newblock CERN 90-10/ECFA 90-133.

\bibitem{vanEijk:1994ab}
B.~van Eijk et~al.
\newblock {\em Nuc. Phys. B}, 292:1, 1987.

\bibitem{DellaNegra1990}
M.~Della~Negra et~al.
\newblock Proceedings of the large hadron collider workshop, aachen.
\newblock page 509, 1990.
\newblock CERN 90-10/ECFA 90-133.

\bibitem{Campbell:2000bg}
J.~M. Campbell and R.~K. Ellis.
\newblock {\em Phys. Rev.}, D62:114012, 2000.

\bibitem{Gunion:1994ab}
J.~Gunion.
\newblock {\em Phys. Rev. Lett.}, 72:199, 1994.

\bibitem{Richterwas:1998ab}
E.~Richter-Was and M.~Sapinski.
\newblock 1998.
\newblock ATLAS Internal Note, ATL-PHYS-98-132.

\bibitem{Mangano:1993kq}
M.~L. Mangano, P.~Nason, and G.~Ridolfi.
\newblock {\em Nucl. Phys.}, B405:507--535, 1993.

\bibitem{Marchesini:1988cf}
G.~Marchesini and B.~R. Webber.
\newblock {\em Nucl. Phys.}, B310:461, 1988.

\bibitem{Knowles:1988vs}
I.~G. Knowles.
\newblock {\em Nucl. Phys.}, B310:571, 1988.

\bibitem{Mangano:2002ab}
M.~L. Mangano, M.~Moretti, and R.~Pittau.
\newblock http://mlm.home.cern.ch/mlm/wbb/wbb.html.

\bibitem{Mazzucato:2002ab}
F.~Mazzucato.
\newblock Studies on the standard model self-couplings.
\newblock ATLAS Internal Note, in preparation.

\bibitem{Hilgart:1993xu}
J.~Hilgart, R.~Kleiss, and F.~Le~Diberder.
\newblock 75:191--218, 1993.

\bibitem{Berends:1995xn}
F.~A. Berends, R.~Pittau, and R.~Kleiss.
\newblock {\em Comput. Phys. Commun.}, 85:437--452, 1995.

\bibitem{Berends:2000fj}
F.~A. Berends, C.~G. Papadopoulos, and R.~Pittau.
\newblock {\em Comput. Phys. Commun.}, 136:148--172, 2001.

\bibitem{Kleiss:1994qy}
R.~Kleiss and R.~Pittau.
\newblock {\em Comput. Phys. Commun.}, 83:141--146, 1994.

\bibitem{Lepage:1978sw}
G.~P. Lepage.
\newblock {\em J. Comput. Phys.}, 27:192, 1978.

\bibitem{Ohl:1999qm}
T.~Ohl.
\newblock hep-ph/9911437.

\bibitem{Jadach:1999vf}
S.~Jadach, B.~F.~L. Ward, and Z.~Was.
\newblock {\em Comput. Phys. Commun.}, 130:260--325, 2000.

\bibitem{Bengtsson:1985yx}
H.~U. Bengtsson and G.~Ingelman.
\newblock {\em Comput. Phys. Commun.}, 34:251, 1985.

\bibitem{Miu:1998ju}
G.~Miu and T.~Sjostrand.
\newblock {\em Phys. Lett.}, B449:313--320, 1999.

\bibitem{Corcella:1999gs}
G.~Corcella and M.~H. Seymour.
\newblock {\em Nucl. Phys.}, B565:227--244, 2000.

\bibitem{Chen:2001ci}
Y.~Chen, John~C. Collins, and N.~Tkachuk.
\newblock {\em JHEP}, 06:015, 2001.

\bibitem{Collins:2000gd}
J.~C. Collins and F.~Hautmann.
\newblock {\em JHEP}, 03:016, 2001.

\bibitem{Djouadi:2000gu}
A.~Djouadi et~al.
\newblock hep-ph/0002258.

\bibitem{Balazs:2000wv}
C.~Balazs and C.~P. Yuan.
\newblock {\em Phys. Lett.}, B478:192--198, 2000.

\bibitem{Bodwin:1985ft}
G.~T. Bodwin, S.~J. Brodsky, and G.~P. Lepage.
\newblock Factorization of the drell-yan cross-section.
\newblock Presented at 20th Rencontre de Moriond, Les Arcs, France, Mar 10-17,
  1985.

\bibitem{Bengtsson:1986gz}
M.~Bengtsson, T.~Sjostrand, and M.~van Zijl.
\newblock {\em Z. Phys.}, C32:67, 1986.

\bibitem{Gribov:1972rt}
V.~N. Gribov and L.~N. Lipatov.
\newblock {\em Yad. Fiz.}, 15:1218--1237, 1972.

\bibitem{Dokshitzer:1977sg}
Yu.~L. Dokshitzer.
\newblock {\em Sov. Phys. JETP}, 46:641--653, 1977.

\bibitem{Odorico:1980gg}
R.~Odorico.
\newblock {\em Nucl. Phys.}, B172:157, 1980.

\bibitem{Ellis:1996yz}
R.K. Ellis, W.J. Stirling, and B.R. Webber.
\newblock Qcd and collider physics; monogr. part. phys. nucl. phys. cosmol.
  (1996).

\bibitem{Sjostrand:1985xi}
T.~Sjostrand.
\newblock {\em Phys. Lett.}, B157:321, 1985.

\bibitem{Marchesini:1984bm}
G.~Marchesini and B.~R. Webber.
\newblock {\em Nucl. Phys.}, B238:1, 1984.

\bibitem{Dokshitzer:1978yd}
Y.~L. Dokshitzer, D.~Diakonov, and S.~I. Troian.
\newblock {\em Phys. Lett.}, B79:269--272, 1978.

\bibitem{Parisi:1979se}
G.~Parisi and R.~Petronzio.
\newblock {\em Nucl. Phys.}, B154:427, 1979.

\bibitem{Ellis:1981sj}
S.~D. Ellis, N.~Fleishon, and W.~J. Stirling.
\newblock {\em Phys. Rev.}, D24:1386, 1981.

\bibitem{Collins:1982zc}
J.~C. Collins and D.~E. Soper.
\newblock {\em Phys. Rev. Lett.}, 48:655, 1982.

\bibitem{Altarelli:1984pt}
G.~Altarelli, R.~K. Ellis, M.~Greco, and G.~Martinelli.
\newblock {\em Nucl. Phys.}, B246:12, 1984.

\bibitem{Altarelli:1985kp}
G.~Altarelli, R.~K. Ellis, and G.~Martinelli.
\newblock {\em Z. Phys.}, C27:617, 1985.

\bibitem{Arnold:1991yk}
P.~B. Arnold and R.~P. Kauffman.
\newblock {\em Nucl. Phys.}, B349:381--413, 1991.

\bibitem{Ellis:1998ii}
R.~K. Ellis and S.~Veseli.
\newblock {\em Nucl. Phys.}, B511:649--669, 1998.

\bibitem{Field:2001ab}
T.~Affolder et~al.
\newblock FERMILAB-PUB-01/211-E (to appear in Phys. Rev. D).

\bibitem{Field:2000ab}
R.~Field (for~the CDF~Collaboration).
\newblock The underlying event in large transverse momentum charged jets.
\newblock presented at DPF2000, Columbus, OH, August 11, 2000.

\bibitem{Tano:2001ab}
V.~Tano (for~the CDF~Collaboration.
\newblock The underlying event in jet and minimum bias events.
\newblock presented at ISMD2001, Datong, China, Sept. 2001.

\bibitem{Paige:1986vk}
F.~E. Paige and S.~D. Protopopescu.
\newblock Isajet 5.20: A monte carlo event generator for p p and anti-p p
  interactions.
\newblock Invited talk given at Workshop on Observable Standard Model Physics
  at the SSC: Monte Carlo Simulation ad Detector Capabilities, Los Angeles, CA,
  Jan 15-24, 1986.

\bibitem{Abe:1990td}
F.~Abe et~al.
\newblock {\em Phys. Rev.}, D41:2330, 1990.

\bibitem{Pumplin:1998ix}
J.~Pumplin.
\newblock {\em Phys. Rev.}, D57:5787--5792, 1998.

\bibitem{Butterworth:1996zw}
J.~M. Butterworth, J.~R. Forshaw, and M.~H. Seymour.
\newblock {\em Z. Phys.}, C72:637--646, 1996.

\bibitem{Butterworth:2001ab}
J.~M. Butterworth.
\newblock Talk presented by Jon Butterworth at Snowmass 2001.

\end{thebibliography}

\newcommand{\gaga}{{\gamma \gamma} }
\newcommand{\asgj}{{\alpha_s} }
\newcommand{\etag}{{\eta_\gamma} }
\newcommand{\phig}{{\phi_\gamma} }
\newcommand{\ptg}{{p_{T \,\gamma} } }
\newcommand{\ptq}{{p_{T \, p}} }
\newcommand{\ptgu}{{p_{T \, \gamma_1}} }
\newcommand{\ptgd}{{p_{T \, \gamma_2}} }
\newcommand{\mgg}{{M_{\gamma \gamma}} }
\newcommand{\qtgj}{{Q^t_{\gamma \gamma}} }
\newcommand{\Hgg}{{H\rightarrow\gamma \gamma} }
\newcommand{\bc}{\begin{center}}
\newcommand{\ec}{\end{center}}
\newcommand{\bmini}{\begin{minipage}{.48\linewidth}}
\newcommand{\emini}{\end{minipage}}

\section{PHOTONS, HADRONS AND JETS
         \protect\footnote{Section coordinators: J. Huston and E. Pilon}
        \protect\footnote{Contributing authors: T.~Binoth, S.D. Ellis, 
	J.-Ph.~Guillet, K.~Lassila-Perini, J. Huston, M. T\"onnesmann and 
	E.~Tournefier}}

Many signatures of both Standard Model physics and physics that lies beyond 
the Standard Model require the reconstruction of photons and jets in the final
state and the comparison of experimental data to theoretical predictions from
both Monte Carlos and from next-to-leading order (NLO) theory. There were two
contributions to the QCD/SM group dealing with these issues. The first
contribution discusses improvements to algorithms for the reconstruction of
experimental jets both at the Tevatron and at the LHC and comparisons to
perturbative QCD predictions at NLO. In the second contribution, predictions
for diphoton and photon-$\pi^o$ production at the LHC from the Monte Carlo
program {\tt PYTHIA} are compared to those from the NLO program {\tt DIPHOX}.
Distributions for kinematic variables of interest for searches for a Higgs
boson decaying into two photons are compared between the two programs.

\subsection{On building better cone jet algorithms%
        \protect\footnote{Contributing authors: S.D. Ellis, J. Huston,
                                                 M. T\"onnesmann}}

%
%


An important facet of preparations \cite{Blazey:2000qt,Catani:2000zg} for Run II
at the Tevatron, and for future data taking at the LHC, has been the study of
ways in which to improve jet algorithms. These algorithms are employed to map
final states, both in QCD perturbation theory and in the data, onto jets. The
motivating idea is that these jets are the surrogates for the underlying
energetic partons. In principle, we can connect the observed final states, in
all of their complexity, with the perturbative final states, which are easier to
interpret and to analyze theoretically. Of necessity these jet algorithms should
be robust under the impact of both higher order perturbative and
non-perturbative physics and the effects introduced by the detectors themselves.
The quantitative goal is a precision of order 1\,\% in the mapping between
theory and experiment. In this note we will provide a brief summary of recent
progress towards this goal. A more complete discussion of our results will be
provided elsewhere \cite{Ellis:2002inprep}. Here we will focus on cone jet
algorithms, which have formed the basis of jet studies at hadron colliders.

As a starting point we take the Snowmass Algorithm \cite{Huth:1990mi}, which was
defined by a collaboration of theorists and experimentalists and formed the
basis of the jet algorithms used by the CDF and D\O\ collaborations during Run I
at the Tevatron. Clearly jets are to be composed of either hadrons or partons
that are, in some sense, nearby each other. The cone jet defines nearness in an
intuitive geometric fashion: jets are composed of hadrons or partons whose
3-momenta lie within a cone defined by a circle in $(\eta, \phi)$. These are
essentially the usual angular variables, where
$\eta = \ln\,(\cot\,\theta/2)$ is the pseudorapidity and $\phi$ is the azimuthal
angle. This idea of being nearby in angle can be contrasted with an algorithm
based on being nearby in transverse momentum as illustrated by the so-called
$k_T$ Algorithm \cite{Ellis:1993tq,Catani:1992zp,Catani:1993hr} that has been
widely used at $e^+e^-$ and $ep$ colliders. We also expect the jets to be
aligned with the most energetic particles in the final state. This expectation
is realized in the Snowmass Algorithm by defining an acceptable jet in terms of
a ``stable'' cone such that the geometric center of the cone is identical to the
$E_T$ weighted centroid. Thus, if we think of a sum over final state partons or
hadrons defined by an index $k$ and in the direction $(\eta_k, \phi_k)$, a jet
($J$) of cone radius $R$ is defined by the following set of equations:

\begin{displaymath}
k\in J: (\phi_k - \phi_J)^2 + (\eta_k - \eta_J)^2\leq R^2,
\end{displaymath}
\begin{eqnarray}
E_{T,J} & = & \sum_{k\in J}E_{T,k},\nonumber\\
\phi_J  & = & \sum_{k\in J}\frac{E_{T,k}\,\phi_k}{E_{T,J}},\label{cone}\\
\eta_J  & = & \sum_{k\in J}\frac{E_{T,k}\,\eta_k}{E_{T,J}}.\nonumber
\end{eqnarray}

In these expressions $E_T$ is the transverse energy ($|\overrightarrow{p_T}|$
for a massless 4-vector). It is important to recognize that jet algorithms
involve two distinct steps. The first step is to identify the ``members'' of the
jet, i.e., the calorimeter towers or the partons that make up the stable cone
that becomes the jet. The second step involves constructing the kinematic
properties that will characterize the jet, i.e., determine into which bin the
jet will be placed. In the original Snowmass Algorithm the $E_T$ weighted
variables defined in Eq.~(\ref{cone}) are used both to identify and bin the jet.

In a theoretical calculation one integrates over the phase space corresponding
to parton configurations that satisfy the stability conditions. In the
experimental case one searches for sets of final state particles (and
calorimeter towers) in each event that satisfy the constraint. In practice
\cite{Blazey:2000qt,Catani:2000zg} the experimental implementation of the cone
algorithm has involved the use of various short cuts to minimize the search
time. In particular, Run I algorithms made use of seeds. Thus one looks for
stable cones only in the neighborhood of calorimeter cells, the seed cells,
where the deposited energy exceeds a predefined limit. Starting with such a seed
cell, one makes a list of the particles (towers) within a distance $R$ of the
seed and calculates the centroid for the particles in the list (calculated as in
Eq.~(\ref{cone})). If the calculated centroid is consistent with the initial
cone center, a stable cone has been identified. If not, the calculated centroid
is used as the center of a new cone with a new list of particles inside and the
calculation of the centroid is repeated. This process is iterated, with the cone
center migrating with each repetition, until a stable cone is identified or
until the cone centroid has migrated out of the fiducial volume of the detector.
When all of the stable cones in an event have been identified, there will
typically be some overlap between cones. This situation must be addressed by a
splitting/merging routine in the jet algorithm. This feature was not foreseen in
the original Snowmass Algorithm. Normally this involves the definition of a
parameter $f_\mathrm{merge}$, typically with values in the range
$0.5\leq f_\mathrm{merge}\leq 0.75$, such that, if the overlap transverse energy
fraction (the transverse energy in the overlap region divided by the smaller of
the total energies in the two overlapping cones) is greater than
$f_\mathrm{merge}$, the two cones are merged to make a single jet. If this
constraint is not met, the calorimeter towers/hadrons in the overlap region are
individually assigned to the cone whose center is closer. This situation yields
2 final jets.

The essential challenge in the use of jet algorithms is to understand the
differences between the experimentally applied algorithms and the theoretically
applied ones and hence understand the uncertainties. This is the primary concern
of this paper. It has been known for some time that the use of seeds in the
experimental algorithms means that certain configurations kept by the
theoretical algorithm are likely to be missed by the experimental one
\cite{Ellis:1992qq,Ellis:1993ik,Abbott:1997fc}. At higher orders in perturbation
theory the seed definition also introduces an undesirable (logarithmic)
dependence on the seed $E_T$ cut (the minimum $E_T$ required to be treated as a
seed cell) \cite{Seymour:1998kj}. Various alternative algorithms are described
in Ref.~\cite{Blazey:2000qt} for addressing this issue, including the Midpoint
Algorithm and the Seedless Algorithm. In the last year it has also been
recognized that other final state configurations are likely to be missed in the
data, compared to the theoretical result. In this paper we will explain these
new developments and present possible solutions. To see that there is a problem,
we apply representative jet algorithms to data sets that were generated with the
HERWIG Monte Carlo \cite{Marchesini:1992ch,Corcella:2000bw,Corcella:1999qn} and
then run through a CDF detector simulation. As a reference we include in our
analysis the JetClu Algorithm \cite{Abe:1992ui}, which is the algorithm used by
CDF in Run I. It employs both seeds and a property called ``ratcheting''. This
latter term labels the fact that the Run I CDF algorithm (unlike the
corresponding D\O\ algorithm) was defined so that calorimeter towers initially
found in a cone around a seed continue to be associated with that cone, even as
the center of the cone migrates due to the iteration of the cone algorithm. Thus
the final ``footprint'' of the cone is not necessarily a circle in
$(\eta, \phi)$ (even before the effects of splitting/merging). Since the cone is
``tied'' to the initial seed towers, this feature makes it unlikely that cones
will migrate very far before becoming stable. We describe results from JetClu
both with and without this ratcheting feature. The second cone algorithm studied
is the Midpoint Algorithm that, like the JetClu Algorithm, starts with seeds to
find stable cones (but without ratcheting). The Midpoint Algorithm then adds a
cone at the midpoint in $(\eta, \phi)$ between all identified pairs of stable
cones separated by less than $2R$ and iterates this cone to test for stability.
This step is meant to ensure that no stable ``mid-cones'' are missed, compared
to the theoretical result, due to the use of seeds. Following the recommendation
of the Run II Workshop, we actually use 4-vector kinematics for the Midpoint
Algorithm and place the cone at the midpoint in $(y, \phi)$, where $y$ is the
true rapidity. The third cone algorithm is the Seedless Algorithm that places an
initial trial cone at every point on a regular lattice in $(y, \phi)$, which is
approximately as fine-grained as the detector. It is not so much that this
algorithm lacks seeds, but rather that the algorithm puts seed cones
``everywhere''. The Seedless Algorithm can be streamlined by imposing the
constraint that a given trial cone is removed from the analysis if the center of
the cone migrates outside of its original lattice cell during the iteration
process. The streamlined version still samples every lattice cell for stable
cone locations, but is less computationally intensive. Our experience with the
streamlined version of this algorithm suggests that there can be problems
finding stable cones with centers located very close to cell boundaries. This
technical difficulty is easily addressed by enlarging the distance that a trial
cone must migrate before being discarded. For example, if this distance is
60\,\% of the lattice cell width instead of the default value of 50\,\%, the
problem essentially disappears with only a tiny impact on the required time for
analysis. In the JetClu Algorithm the value $f_\mathrm{merge} = 0.75$ was used
(as in the Run I analyses), while for the other two cone algorithms the value
$f_\mathrm{merge} = 0.5$ was used as suggested in Ref.~\cite{Blazey:2000qt}.
Finally, for completeness, we include in our analysis a sample $k_T$ Algorithm.

\begin{figure}[ptb]
\begin{center}
\includegraphics[width=\textwidth]{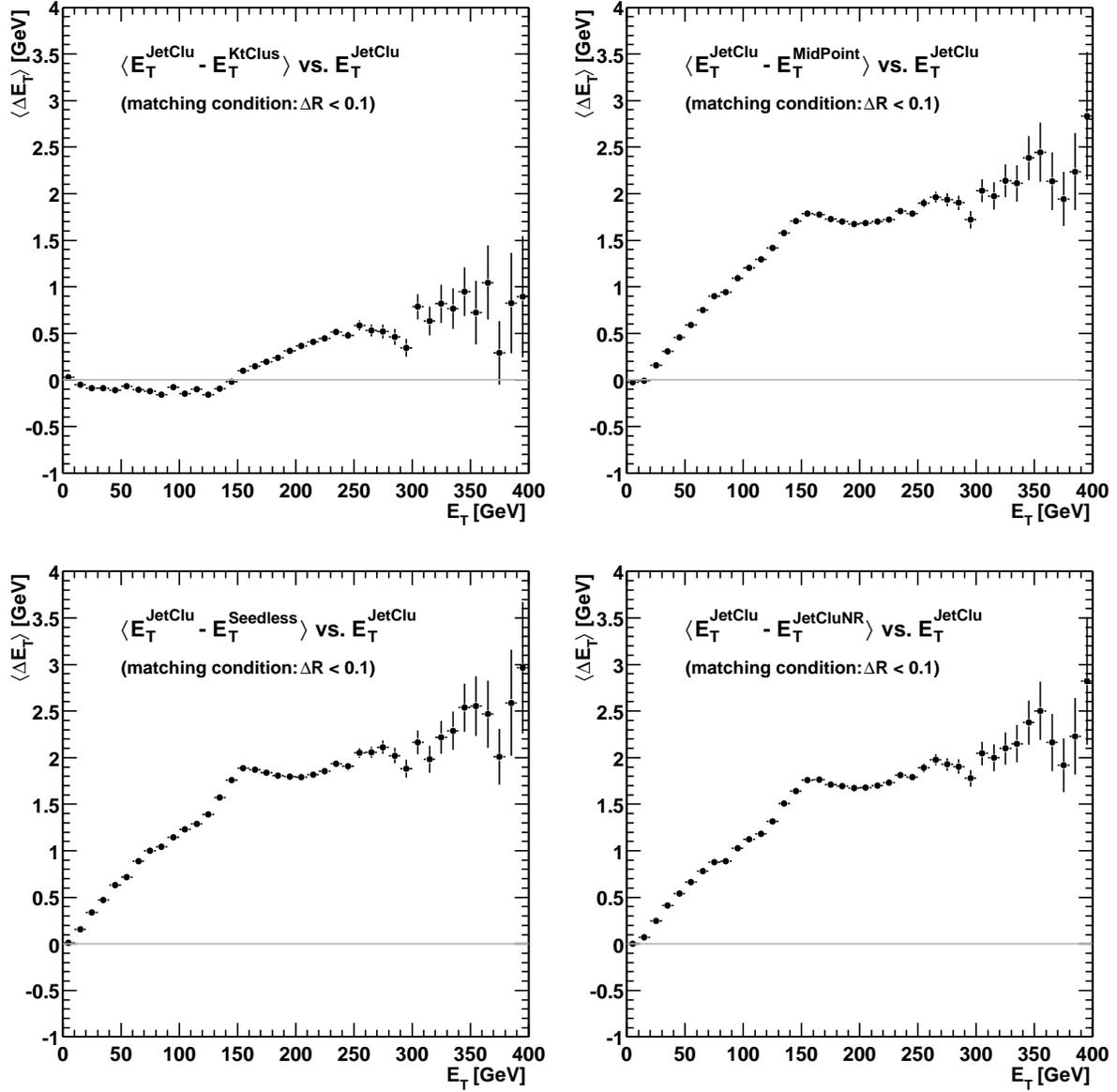}
\caption{Difference of $E_T$ for matched jets found with various jet algorithms
and compared to the JetClu CDF Run I algorithm. The events studied were
generated with HERWIG 6.1 and run through the CDF detector simulation.}
\label{deltaET}
\end{center}
\end{figure}

Starting with a sample of 250,000 events, which were generated with HERWIG 6.1
and run through a CDF detector simulation and which were required to have at
least 1 initial parton with $E_T > 200\,$GeV, we applied the various algorithms
to find jets with $R = 0.7$ in the central region ($|\eta| < 1$). We then
identified the corresponding jets from each algorithm by finding jet centers
differing by $\Delta R < 0.1$. The plots in Fig.~\ref{deltaET} indicate the
average difference in $E_T$ for these jets as a function of the jet $E_T$. (We
believe that some features of the indicated structure, in particular the
``knees'' near $E_T = 150\,$GeV, are artifacts of the event selection process.)

From these results we can draw several conclusions. First, the $k_T$ Algorithm
identifies jets with $E_T$ values similar to those found by JetClu, finding
slightly more energetic jets at small $E_T$ and somewhat less energetic jets at
large $E_T$. We will not discuss this algorithm further here except to note that
D\O\ has applied it in a study of Run I data \cite{Abazov:2001ak} and in that
analysis the $k_T$ Algorithm jets seem to exhibit slightly \emph{larger} $E_T$
than expected from NLO perturbation theory. The cone algorithms, including the
JetClu Algorithm without ratcheting, which is labeled JetCluNR, identify jets
with approximately 0.5\,\% to 1\,\% \emph{smaller} $E_T$ values than those
identified by the JetClu Algorithm (with ratcheting), with a corresponding
approximately 5\,\% smaller jet cross section at a given $E_T$ value. We believe
that this systematic shortfall can be understood as resulting from the smearing
effects of perturbative showering and non-perturbative hadronization.

To provide insight into the issues raised by Fig.~\ref{deltaET} we now discuss a
simple, but informative analytic picture. It will serve to illustrate the impact
of showering and hadronization on the operation of jet algorithms. We consider
the scalar function $F(\overrightarrow{r})$ defined as a function of the
2-dimensional variable $\overrightarrow{r} = (\eta, \phi)$ by the integral over
the transverse energy distribution of either the partons or the
hadrons/calorimeter towers in the final state with the indicated weight
function,

\begin{eqnarray}
F(\overrightarrow{r}) & = &
\frac{1}{2}\int d^2\rho\;
E_T(\overrightarrow{\rho})\cdot
\left(R^2 - (\overrightarrow{\rho} - \overrightarrow{r})^2\right)\cdot
\Theta\left(R^2 - (\overrightarrow{\rho} - \overrightarrow{r})^2\right)
\label{scalar}\\
& = &
\frac{1}{2}\sum_i E_{T,i}\cdot
\left(R^2 - (\overrightarrow{\rho_i} - \overrightarrow{r})^2\right)\cdot
\Theta\left(R^2 - (\overrightarrow{\rho_i} - \overrightarrow{r})^2\right).
\nonumber
\end{eqnarray}

The second expression arises from replacing the continuous energy distribution
with a discrete set, $i = 1\,...\,N$, of delta functions, representing the
contributions of either a configuration of partons or a set of calorimeter
towers (and hadrons). Each parton direction or the location of the center of
each calorimeter tower is defined in $\eta$, $\phi$ by
$\rho_i = (\eta_i, \phi_i)$, while the parton/calorimeter cell has a transverse
energy (or $E_T$) content given by $E_{T,i}$. This function is clearly related
to the energy in a cone of size $R$ containing the towers whose centers lie
within a circle of radius $R$ around the point $\overrightarrow{r}$. More
importantly it carries information about the locations of ``stable'' cones. The
points of equality between the $E_T$ weighted centroid and the geometric center
of the cone correspond precisely to the maxima of $F$. The gradient of this
function has the form (note that the delta function arising from the derivative
of the theta function cannot contribute as it is multiplied by a factor equal to
its argument)

\begin{equation}
\overrightarrow{\nabla}F(\overrightarrow{r}) = \sum_i E_{T,i}\cdot
(\overrightarrow{\rho_i} - \overrightarrow{r})\cdot
\Theta\left(R^2 - (\overrightarrow{\rho_i} - \overrightarrow{r})^2\right).
\label{gradient}
\end{equation}

This expression vanishes at points where the weighted centroid coincides with
the geometric center, i.e., at points of stability (and at minima of $F$, points
of extreme instability). The corresponding expression for the energy in the cone
centered at $\overrightarrow{r}$ is

\begin{equation}
E_C(\overrightarrow{r}) = \sum_i E_{T,i}\cdot
\Theta\left(R^2 - (\overrightarrow{\rho_i}-\overrightarrow{r})^2\right).
\label{energy}
\end{equation}

\begin{figure}[ptb]
\begin{center}
\includegraphics[width=\textwidth,bb=20 60 576 435,clip]{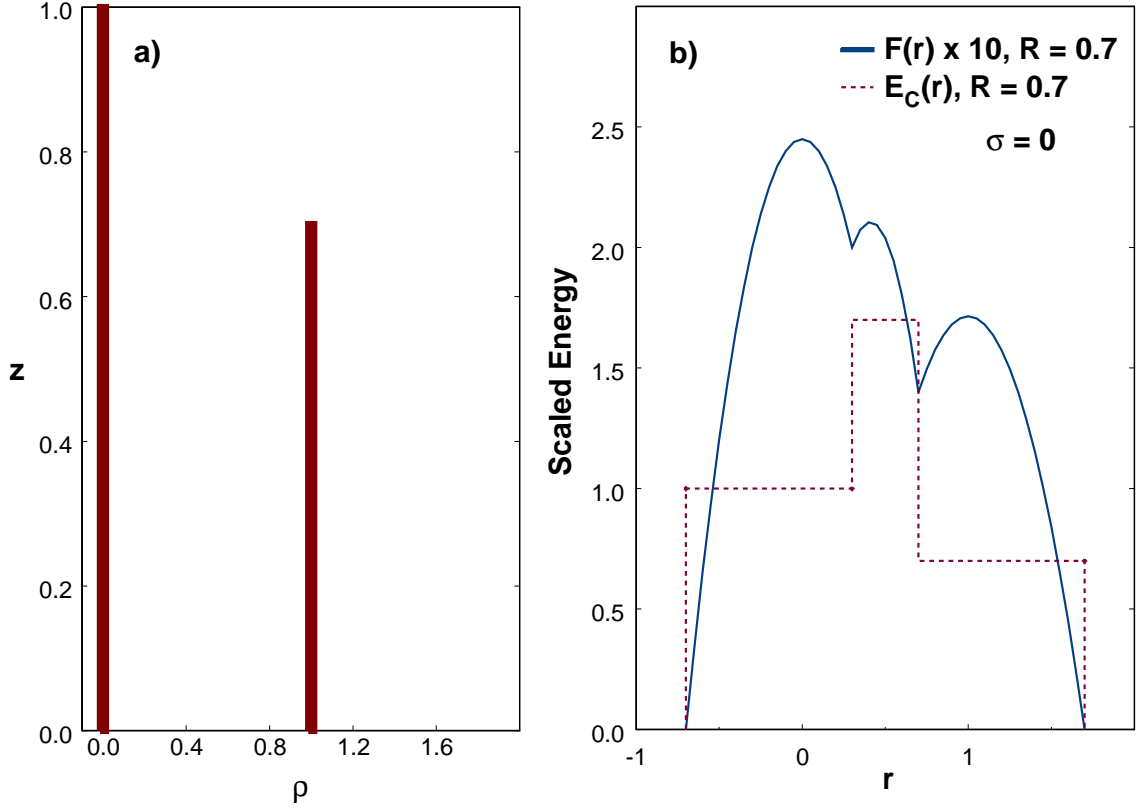}
\caption{2-Parton distribution: a) transverse energy distribution; b)
distributions $F(r)$ and $E_C(r)$ in the perturbative limit of no smearing.}
\label{edist2}
\end{center}
\end{figure}

To more easily develop our understanding of these equations consider a
simplified scenario (containing all of the interesting effects) involving 2
partons separated in just one angular dimension
$\overrightarrow{\rho}\rightarrow\rho$ ($\overrightarrow{r}\rightarrow r$) with
$\rho_2 - \rho_1 = d$. It is sufficient to specify the energies of the 2 partons
simply by their ratio, $z = E_2/E_1\leq 1$. Now we can study what sorts of 2
parton configurations yield stable cones in this 2-D phase space specified by
$0\leq z\leq 1$, $0\leq d\leq 2R$ (beyond $2R$ the 2 partons are surely in
different cones). As a specific example consider the case $\rho_1 = 0$,
$\rho_2 = d = 1.0$ and $z = 0.7$ with $R = 0.7$ (the typical experimental
value). The underlying energy distribution is illustrated in Fig.~\ref{edist2}a,
representing a delta function at $\rho = 0$ (with scaled weight 1) and another
at $\rho = 1.0$ (with scaled weight 0.7). This simple distribution leads to the
functions $F(r)$ and $E_C(r)$ indicated in Fig.~\ref{edist2}b. In going from the
true energy distribution to the distribution $E_C(r)$ the energy is effectively
smeared over a range given by $R$. In $F(r)$ the distribution is further shaped
by the quadratic factor $R^2 - (\rho_i - r)^2$. We see that $F(r)$ exhibits 3
local maxima corresponding to the expected stable cones around the two original
delta functions ($r_1 = 0$, $r_2 = 1$), plus a third stable cone in the middle
($r_3 = zd/(1 + z) = 0.41$ in the current case). This middle cone includes the
contributions from both partons as indicated by the magnitude of the middle peak
in the function $E_C(r)$. Note further that the middle cone is found at a
location where there is initially no energy in Fig.~\ref{edist2}a, and thus no
seeds. One naively expects that such a configuration is not identified as a
stable cone by the experimental implementations of the cone algorithm that use
seeds simply because they do not look for it. Note also that, since both partons
are entirely within the center cone, the overlap fractions are unity and the
usual merging/splitting routine will lead to a single jet containing all of the
initial energy ($1 + z$). This is precisely how this configuration was treated
in the NLO perturbative analysis of the Snowmass Algorithm
\cite{Ellis:1992en,Ellis:1990ek,Ellis:1989vm,Ellis:1989hv}, i.e., only the
leading jet, the middle cone, was kept).

\begin{figure}[ptb]
\begin{center}
\includegraphics[width=\textwidth,bb=20 55 576 430,clip]{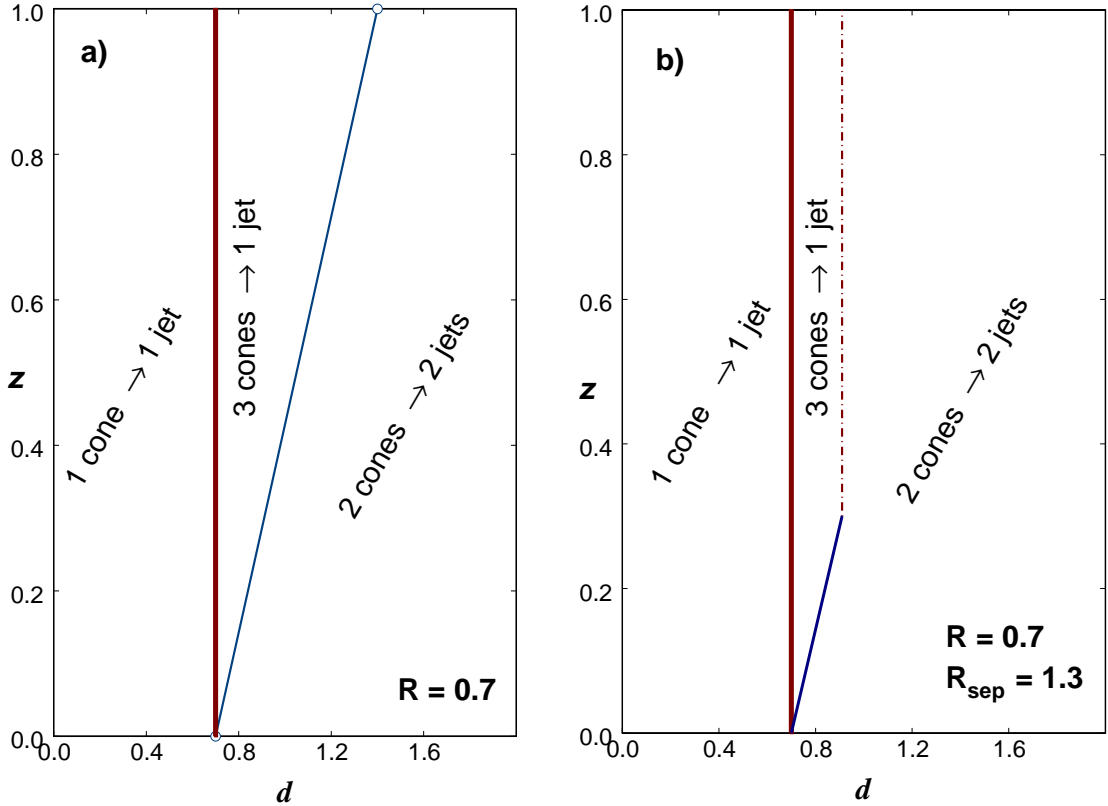}
\caption{Perturbation Theory Structure: a) $R_{sep} = 2$; b) $R_{sep} = 1.3$.}
\label{perthy4}
\end{center}
\end{figure}

Similar reasoning leads to Fig.~\ref{perthy4}a, which indicates the various 2
parton configurations found by the perturbative cone algorithm. For $d < R$ one
finds a single stable cone and a single jet containing both partons. For
$R < d < (1+z)R$ one finds 3 stable cones that merge to 1 jet, again with all of
the energy. For $d > (1+z)R$ we find 2 stable cones and 2 jets, each containing
one parton, of scaled energies $1$ and $z$. Thus, except in the far right region
of the graph, the 2 partons are always merged to form a single jet.

We expect that the impact of seeds in experimental algorithms can be (crudely)
simulated in the NLO calculations
\cite{Ellis:1992qq,Ellis:1993ik,Abbott:1997fc} by including a parameter
$R_{sep}$ such that stable cones containing 2 partons are not allowed for
partons separated by $d > R_{sep}\cdot R$. As a result cones are no longer
merged in this kinematic region. In the present language this situation is
illustrated in Fig.~\ref{perthy4}b corresponding to $R_{sep} = 1.3$,
$R\cdot R_{sep} = 0.91$. This specific value for $R_{sep}$ was chosen
\cite{Ellis:1992qq,Ellis:1993ik,Abbott:1997fc} to yield reasonable agreement
with the Run I data. The conversion of much the ``3 cones $\rightarrow$ 1 jet''
region to ``2 cones $\rightarrow$ 2 jets'' has the impact of lowering the
average $E_T$ of the leading jet and hence the jet cross section at a fixed
$E_{T,J}$. Parton configurations that naively produced jets with energy
characterized by $1 + z$ now correspond to jets of maximum energy 1. This is
just the expected impact of a jet algorithm with seeds. Note that with this
value of $R_{sep}$ the specific parton configuration in Fig.~\ref{edist2}a will
yield 2 jets (and not 1 merged jet) in the theoretical calculation. As mentioned
earlier this issue is to be addressed by the Midpoint and Seedless Algorithms in
Run II. However, as indicated in Fig.~\ref{deltaET}, neither of these two
algorithms reproduces the results of JetClu. Further, they both identify jets
that are similar to JetClu \emph{without} ratcheting. Thus we expect that there
is more to this story.

\begin{figure}[ptb]
\begin{center}
\includegraphics[width=\textwidth,bb=20 60 576 440,clip]{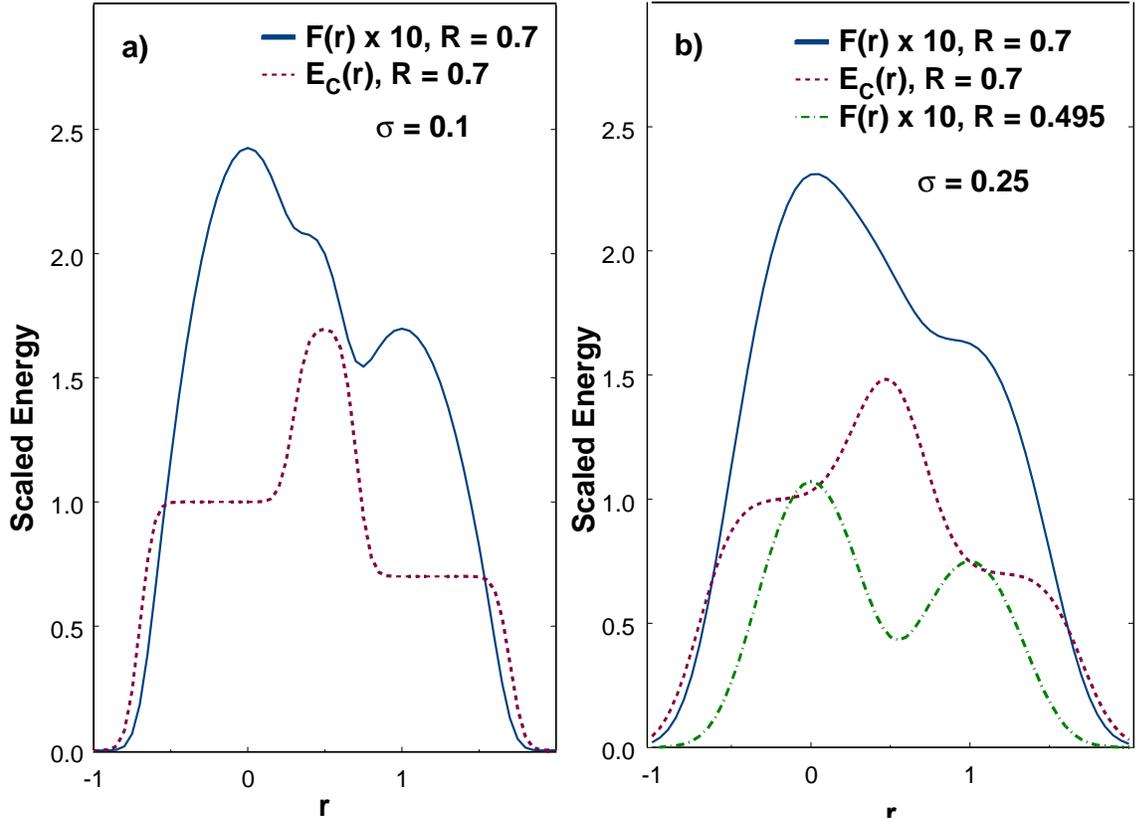}
\caption{The distributions $F(r)$ and $E_C(r)$ for smearing width a) 0.1; b)
0.25.}
\label{smearpt5}
\end{center}
\end{figure}

As suggested earlier, a major difference between the perturbative level, with a
small number of partons, and the experimental level of multiple hadrons is the
smearing that results from perturbative showering and nonperturbative
hadronization. For the present discussion the primary impact is that the
starting energy distribution will be smeared out in the variable $r$. We can
simulate this effect in our simple model using gaussian smearing, i.e., we
replace the delta functions in Eq.~(\ref{scalar}) with gaussians of width
$\sigma$. (Since this corresponds to smearing in an angular variable, we would
expect $\sigma$ to be a decreasing function of $E_T$, i.e., more energetic jets
are narrower. We also note that this naive picture does not include the expected
color coherence in the products of the showering/hadronization process.) The
first impact of this smearing is that some of the energy initially associated
with the partons now lies outside of the cones centered on the partons. This
effect, typically referred to as ``splashout'' in the literature, is
(exponentially) small in this model for $\sigma < R$. Here we will focus on less
well known but phenomenologically more relevant impacts of splashout. The
distributions corresponding to Fig.~\ref{edist2}b, but now with $\sigma = 0.10$
(instead of $\sigma = 0$), are exhibited in Fig.~\ref{smearpt5}a.

With the initial energy distribution smeared by $\sigma$, the distribution
$F(r)$ is now even more smeared and, in fact, we see that the middle stable cone
(the maximum in the middle of Fig.~\ref{edist2}b) has been washed out by the
increased smearing. Thus the cone algorithm applied to data (where such smearing
is present) may not find the middle cone that is present in perturbation theory,
not only due to the use of seeds but also due to this new variety of splashout
correction, which renders this cone unstable. Since, as a result of this
splashout correction, the middle cone is not stable, this problem is \emph{not}
addressed by either the Midpoint Algorithm or the Seedless Algorithm. Both
algorithms may look in the correct place, but they look for stable cones. This
point is presumably part of the explanation for why both of these algorithms
disagree with the JetClu results in Fig.~\ref{deltaET}.

\begin{figure}[ptb]
\begin{center}
\includegraphics[width=0.9\textwidth,bb=0 35 567 340,clip]{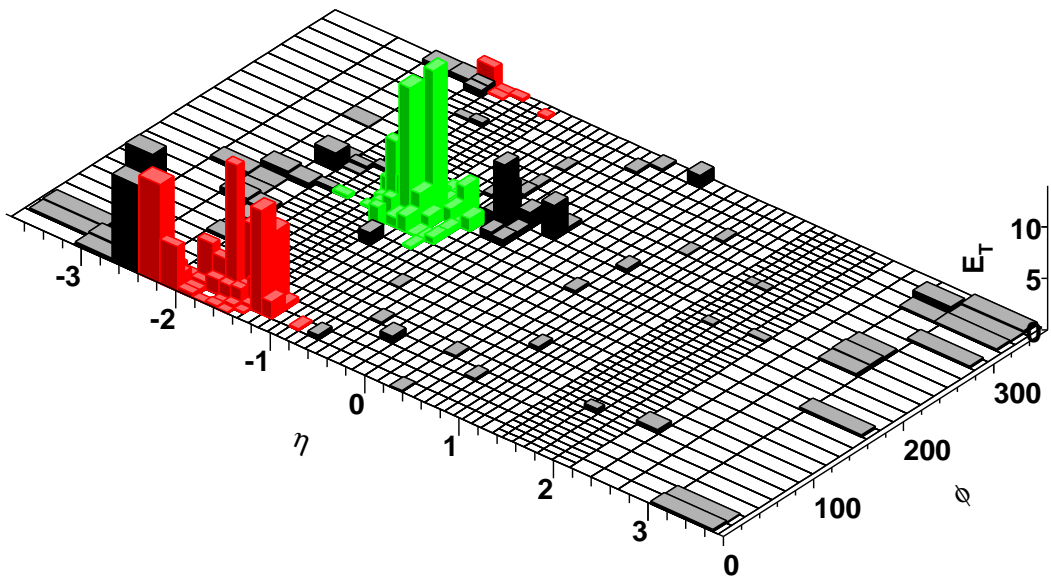}
\caption{Result of applying the Midpoint Algorithm to a specific Monte Carlo
event in the CDF detector.}
\label{lego}
\end{center}
\end{figure}

Our studies also suggest a further impact of the smearing of
showering/hadronization that was previously unappreciated. This new effect is
illustrated in Fig.~\ref{smearpt5}b, which shows $F(r)$, still for $z = 0.7$ and
$d = 1.0$, but now for $\sigma = 0.25$. With the increased smearing the second
stable cone, corresponding to the second parton, has now also been washed out,
i.e., the right hand local maximum has also disappeared. This situation is
exhibited in the case of ``data'' by the lego plot in Fig.~\ref{lego} indicating
the jets found by the Midpoint Algorithm in a specific Monte Carlo event. The
Midpoint Algorithm does not identify the energetic towers (shaded in black) to
the right of the energetic central jet as either part of that jet or as a
separate jet, i.e., these obviously relevant towers are not found to be in a
stable cone. The iteration of any cone containing these towers invariably
migrates to the nearby higher $E_T$ towers.

\begin{figure}[ptb]
\begin{center}
\includegraphics[width=0.6\textwidth]{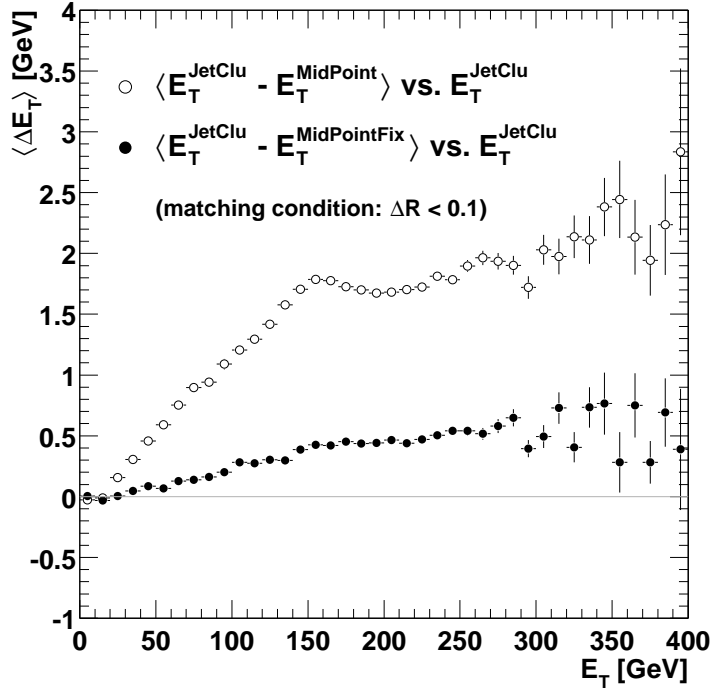}
\caption{The difference in the $E_T$ of identified central jets for the JetClu
and Midpoint Algorithms, both with and without the ``fix'' discussed in the
text. The events studied were generated with HERWIG 6.1 and run through the CDF
detector simulation.}
\label{deltaETfix}
\end{center}
\end{figure}

In summary, we have found that the impact of smearing and splashout is expected
to be much more important than simply the leaking of energy out of the cone.
Certain stable cone configurations, present at the perturbative level, can
disappear from the analysis of real data due to the effects of showering and
hadronization. This situation leads to corrections to the final jet yields that
are relevant to our goal of 1\,\% precision in the mapping between perturbation
theory and experiment. Compared to the perturbative analysis of the 2-parton
configuration, both the middle stable cone and the stable cone centered on the
lower energy parton can be washed out by smearing. Further, this situation is
not addressed by either the Midpoint Algorithm or the Seedless Algorithm. One
possibility for addressing the missing middle cone would be to eliminate the
stability requirement for the added midpoint cone in the Midpoint Algorithm.
However, if there is enough smearing to eliminate also the second (lower energy)
cone, even this scenario will not help, as we do not find two cones to put a
third cone between. There is, in fact, a rather simple ``fix'' that can be
applied to the Midpoint Algorithm to address this latter form of the splashout
correction. We can simply use 2 values for the cone radius $R$, one during the
search for the stable cones and the second during the calculation of the jet
properties. As a simple example, the 3rd curve in Fig.~\ref{smearpt5}b
corresponds to using $R/\sqrt{2} = 0.495$ during the stable cone discovery phase
and $R = 0.7$ in the jet construction phase. Thus the $R/\sqrt{2}$ value is used
only during iteration; the cone size is set to $R$ right after the stable cones
have been identified and the larger cone size is employed during the
splitting/merging phase. By comparing Figs.~\ref{smearpt5}b and \ref{edist2}b
we see that the two outer stable cones in the perturbative case are in
essentially the same locations as in the smeared case using the smaller cone
during discovery. The improved agreement between the JetClu results and those of
the Midpoint Algorithm with the last ``fix'' (using the smaller $R$ value during
discovering but still requiring cones to be stable) are indicated in
Fig.~\ref{deltaETfix}.

Clearly most, but not all, of the differences between the jets found by the
JetClu and Midpoint Algorithms are removed in the fixed version of the latter.
The small $R$ ``fix'' suggested for the Midpoint Algorithm can also be employed
for the Seedless Algorithm but, like the Midpoint Algorithm, it will still miss
the middle (now unstable) cone.

Before closing this brief summary of our results, we should say a few more words
about the Run I CDF algorithm that we used as a reference. In particular, while
ratcheting is difficult to simulate in perturbation theory, we can attempt to
clarify how it fits into the current discussion. As noted above, the JetClu
Algorithm is defined so that calorimeter towers initially found around a seed
stay with that cone, even as the center of the cone migrates due to the
iteration of the cone algorithm. For the simple scenario illustrated in
Fig.~\ref{edist2}a we assume that the locations of the partons are identified as
seeds, even when smearing is present. To include both ratcheting and the way it
influences the progress of the stable cone search, we must define 2 scalar
functions of the form of Eq.~(\ref{scalar}), one to simulate the search for a
stable cone starting at $\rho = 0$ and the second for the search starting at
$\rho = 1.0$. The former function is defined to include the energy within the
range $-R\leq\rho\leq +R$ independent of the value of $r$, while the second
function is defined to always include the energy in the range
$1.0 - R\leq\rho\leq 1.0 + R$. Analyzing the two functions defined in this way
suggests, as expected, that the search that begins at the higher energy seed
will always find a stable cone at the location of that seed, independent of the
amount of smearing. (If the smearing is small, there is also a stable cone at
the middle location but the search will terminate after finding the initial,
nearby stable cone.) The more surprising result arises from analyzing the second
function, which characterizes the search for a stable cone seeded by the lower
energy parton. In the presence of a small amount of smearing this function
indicates stable cones at both the location of the lower energy parton and at
the middle location. Thus the corresponding search finds a stable cone at the
position of the seed and again will terminate before finding the second stable
cone. When the smearing is large enough to wash out the stable cone at the
second seed, the effect of ratcheting is to ensure that the search still finds
a stable cone at the middle location suggested by the perturbative result,
$r_{3} = z\rho/(1 + z)$ (with a precision given by
$\sigma\cdot e^{-(R/\sigma)^2}$). This result suggests that the JetClu Algorithm
with ratcheting always identifies either stable cones at the location of the
seeds or finds a stable cone in the middle that can lead to merging (in the case
of large smearing). It is presumably just these last configurations that lead to
the remaining difference between the JetClu Algorithm results and those of the
``fixed'' Midpoint Algorithm illustrated in Fig.~\ref{deltaETfix}. We find that
the jets found by the JetClu Algorithm have the largest $E_T$ values of any of
the cone jet algorithms, although the JetClu Algorithm still does not address
the full range of splashout corrections.

In conclusion, we have found that the corrections due to the splashout effects
of showering and hadronization result in unexpected differences between cone jet
algorithms applied to perturbative final states and applied to (simulated) data.
With a better understanding of these effects, we have defined steps that serve
to improve the experimental cone algorithms and minimize these corrections.
Further studies are required to meet the goal of 1\,\% agreement between
theoretical and experimental applications of cone algorithms.

\subsection{Comparison of PYTHIA and DIPHOX for $\gaga$ and $\pi^{0} \gamma$ 
productions \protect\footnote{Contributing authors: T.~Binoth, J.-Ph.~Guillet, 
K.~Lassila-Perini and E.~Tournefier}}\label{intro}

Photon pair production plays a prominent role in the search  for a neutral Higgs
boson at the LHC. In this respect it is relevant to directly compare the
extensively used code PYTHIA~\cite{Sjostrand:1994yb} with a recent
next--to--leading order (NLO) code DIPHOX~\cite{Binoth:1999qq}. DIPHOX is a
computer code of partonic event generator type, describing the production of pairs
of particles in hadronic collisions at full NLO accuracy. PYTHIA is a computer
code of hadronic event generator type which fully describes an hadron-hadron
collision at leading--order (LO) accuracy. The comparisons performed for
the production of $\gamma \gamma$ and $\pi^0 \gamma$ are presented in subsects. 
\ref{sec:gamgam,gamjet,qcdsm} and \ref{sec:pigam,gamjet,qcdsm} respectivelly. 

\subsubsection{Comparison of PYTHIA and DIPHOX for $\gaga$ 
production}\label{sec:gamgam,gamjet,qcdsm}

Here, PYTHIA version 6.152~\cite{Sjostrand:1994yb} is compared to
DIPHOX for the $\gaga$ production and  K-factors are extracted. The scales used
for PYTHIA are the default ones, whereas for DIPHOX all the scales have been
chosen as $\sqrt{\ptgu + \ptgd}$. Note that a K-factor depends, among other
things, on the factorization and renormalization scale chosen and must be used
with care. In order to make easier the comparison, the direct contribution of
the $\gaga$ production (c.f. \cite{Binoth:1999qq}) has been split into two
parts.
\begin{itemize}
\item[a)] A first part, called``initial direct", groups together terms which
have no final state collinear singularities. It contains the Born term  $q +
\bar{q}\rightarrow \gamma + \gamma$, related higher order term $q +
\bar{q}\rightarrow \gamma + \gamma + g$ and the box term $g + g\rightarrow
\gamma + \gamma$.
\item[b)] A second part, called  ``Bremsstrahlung'', contains the left-over term of
the direct contribution, i.e. $q + g\rightarrow \gamma + \gamma + q$ and the LO
terms of the one fragmentation contribution: $q + g\rightarrow \gamma + q$, $q
+ \bar{q}\rightarrow \gamma + g$ where the $q/g$ fragments into a photon.
\end{itemize}

\vspace{0.2cm}

\noindent
{\it a) The initial direct contribution}

\vspace{0.2cm}

\indent
{\it a1) Comparison of PYTHIA without ISR/FSR and hadronisation with
DIPHOX at LO}\\
For this comparison, the CTEQ5L partonic densities and $\asgj$ at 1 loop are
used. To start with, a symmetric cut is applied on the transverse momentum of
the two photons $\ptg>20$~GeV. Then more realistic cuts (the $\Hgg$ selection
cuts) have been applied: $|\etag|<2.5$, $\ptgu>40$~GeV, $\ptgd>25$~GeV. As
shown in Table~\ref{LO} PYTHIA and DIPHOX agree at the few percent level. The
distributions of the variables $\etag$, $\ptg$ and $\mgg$  also agree well.
\begin{table}[h]
\begin{center}
  \begin{tabular}{|l|r|r||r|r|}
    \hline
    & \multicolumn{2}{|c|}{$\ptg>20$ GeV} & \multicolumn{2}{|c|}{$\Hgg$ selection cuts }  \\
    \hline
          & PYTHIA & DIPHOX & PYTHIA & DIPHOX\\
    Born  & 82.3 & 83.2 & 10.2 & 9.8\\
    \hline
    Box   & 82.3 & 83.2 & 5.7 & 5.6\\
    \hline
    Total & 164.6 & 166.4  & 15.9 & 15.5\\
    \hline
\end{tabular}
\end{center}
\caption{Comparison of DIPHOX and PYTHIA cross sections (in pb) 
for the Born and the 
Box terms with minimal cuts ($\ptg>20$ GeV) and after $\Hgg$ selection cuts.
\label{LO}}
\end{table}

\vspace{0.1cm}

\indent
{\it a2) NLO corrections to Born}\\
The K-factor K$_{\rm{NLO/LO}}$ is defined as the ratio of the NLO to the LO 
cross sections obtained with DIPHOX. The CTEQ5L pdf and $\asgj$ at  1 loop have
been used to compute the LO cross section whereas the CTEQ5M pdf and $\asgj$ at 2
loops have been used to compute the NLO cross section. The Table~\ref{DiNLO}
gives the value of the cross sections and of K$_{\rm{NLO/LO}}$ for different
stages of the $\Hgg$ selection.
\begin{table}[h]
\begin{center}
\begin{tabular}{|l|r|r|c|}
\hline
Cut                           & LO   & NLO  & K$_{\rm{NLO/LO}}$ \\
\hline
$\ptg>20$ GeV                 & 83.2 & 94.9 & 1.14 \\
\hline
+ $|\etag|<2.5$               & 32.5 & 41.0 & 1.26 \\ 
\hline
+ $\ptg>25$ GeV               & 18.8 & 23.6 & 1.26 \\
\hline
+ Max($\ptgu$,$\ptgd$)$>$40 GeV & 5.7 &10.4 & 1.83 \\
\hline
\end{tabular}
\end{center}
\caption{Comparison of LO and NLO cross section (in pb) for the Born term
at different stage of the $\Hgg$ selection.\label{DiNLO} }
\end{table}
The asymmetric cut results in an enhancement of the K-factor. The reason is the
following: at LO the two photons have the same $p_T$ whereas this is not the
case at NLO because the $p_T$  balance is distorted. Therefore, at LO, the cut
Max($\ptgu$,$\ptgd$)$>$40 GeV  is equivalent to a symmetric cut of 40 GeV on
both photons.  This cut acts more efficiently on LO than on NLO contributions.
After all selection cuts, and if one restricts  in addition to $\mgg>90$~GeV,
K$_{\rm{NLO/LO}}$ is  independent of $\mgg$  and is equal to 1.57.\\

\begin{figure}
  \begin{minipage}{.48\linewidth}
    \bc 
    \mbox{\hspace{-.8cm}\includegraphics[height=90mm]{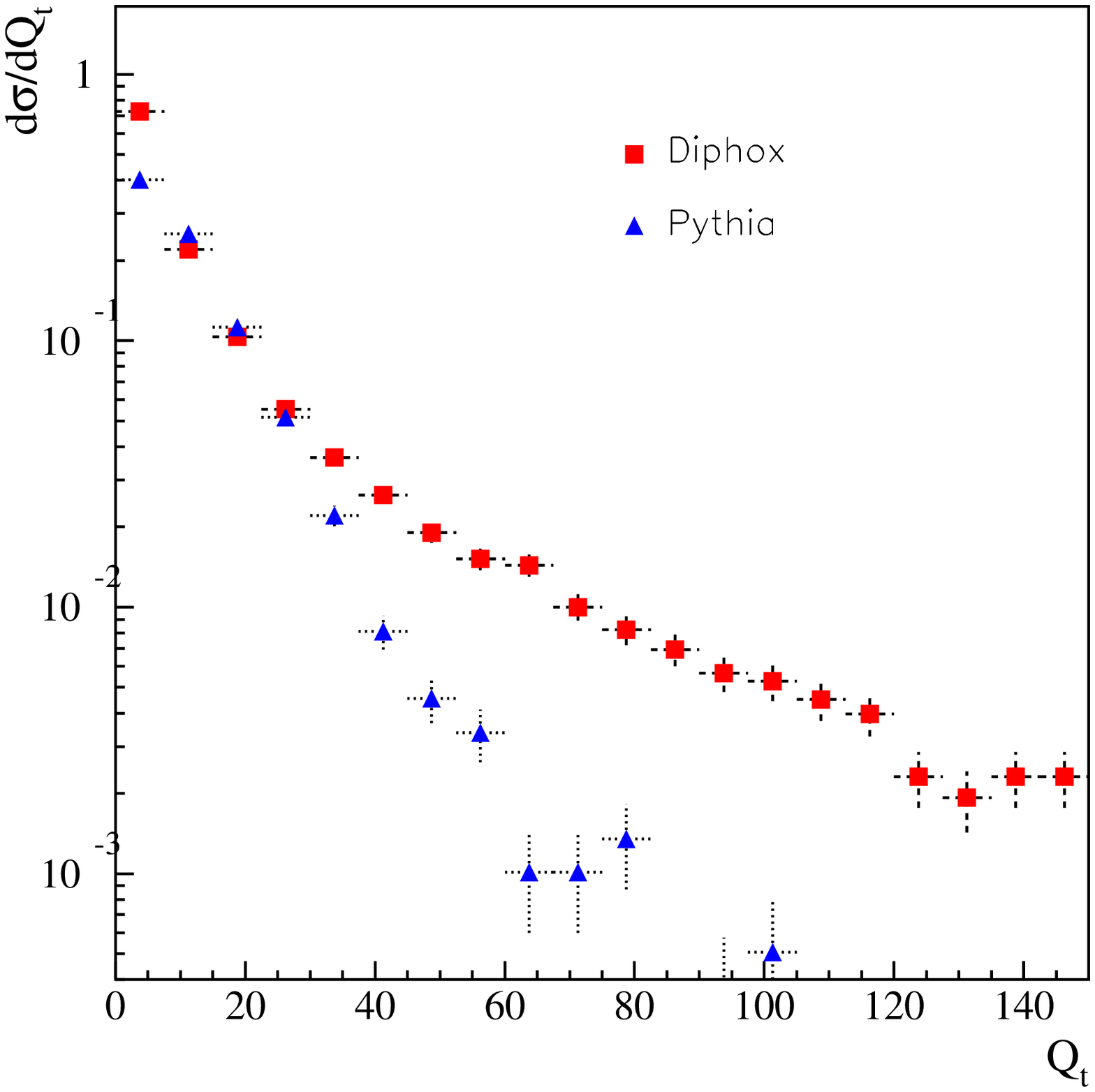}}
    \ec
    \caption{$Q_t$ distribution for PYTHIA with ISR and DIPHOX NLO after all selection cuts
      for the Born contribution.\label{Qtborn}}
    \end{minipage} \hspace{.3cm}
    \begin{minipage}{.48\linewidth}
    \bc \vspace{.7cm}
    \mbox{\includegraphics[height=81mm]{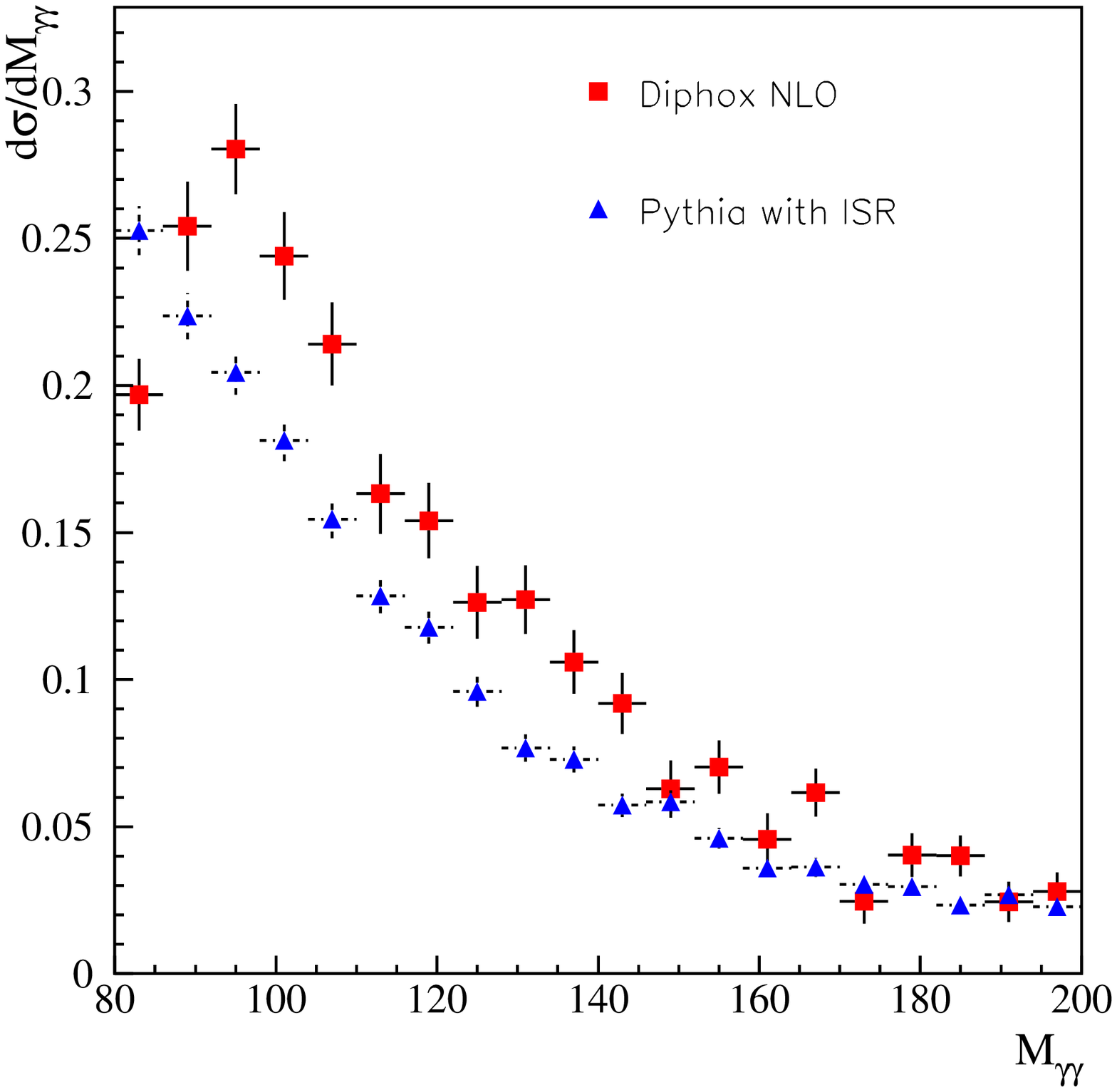}}
    \ec
    \caption{$\mgg$ distribution for PYTHIA with ISR and DIPHOX NLO after all selection cuts
      for the Born plus Box contribution.\label{mggborn+box}}
    \end{minipage} 
\end{figure}

ATLAS and CMS have based their prediction on PYTHIA with ISR. In order to
obtain the correction which has to be applied to their numbers
(K$_{\rm{NLO/ISR}}$) we  have compared PYTHIA with ISR to DIPHOX at NLO. The
cross sections are given in Table~\ref{PytDi}. PYTHIA  with ISR includes part
of the NLO corrections; therefore the effect of the  asymmetric cut on $\ptg$
is reduced and  K$_{\rm{NLO/ISR}}$ is smaller than K$_{\rm{NLO/LO}}$. Since
there is no hard radiation in PYTHIA, the cross section in the high $q_T$
region  is underestimated as is shown in Fig.~\ref{Qtborn}.
\begin{table}[h]
\begin{center}
\begin{tabular}{|r|l|l|c|}
\hline
Cut                           & PYTHIA with ISR   & DIPHOX at NLO  & K$_{\rm{NLO/ISR}}$ \\
\hline
$\ptg>20$ GeV                 & 68.8 & 94.9 & 1.38 \\
\hline
+ $|\etag|<2.5$               & 27.6 & 41.0 & 1.49 \\ 
\hline
+ $\ptg>25$ GeV               & 16.0 & 23.6 & 1.48 \\
\hline
+ Max($\ptgu$,$\ptgd$)$>$40 GeV & 6.9 &10.4 & 1.51 \\
\hline
\end{tabular}
\end{center}
\caption{Comparison of PYTHIA (CTEQ5L+ $\asgj$ at 1 loop) with ISR and DIPHOX 
NLO cross section (in pb) for the Born term
at different stage of the $\Hgg$ selection.\label{PytDi} }
\end{table}

%
\indent
{\it a3) The Box contribution}\\
Since there is no higher order calculation for the Box contribution, we have
studied the effect of switching on ISR and of including  the running of $\asgj$
at 2 loops. PYTHIA with ISR results in a cross section larger by 40$\%$ after
all selection cuts.  This increase is again an effect of the asymmetric cuts.
For $\mgg>80$~GeV the ratio is equal to 1.19. The inclusion of $\asgj$ at 2 loops
decreases the cross section by $23\%$. Therefore, the effect of including ISR
in PYTHIA almost cancels the effect of the  running of $\asgj$. Thus we keep the
LO calculation with $\asgj$ at one loop for the DIPHOX 'NLO' calculation in the
following.

\vspace{0.2cm}

\indent
{\it a4) K-factor for the direct $\gaga$ production}\\
In this section, we compare the cross sections obtained by PYTHIA with ISR
(which correspond to the experiment's Monte--Carlo) to the NLO prediction from
DIPHOX for the direct $\gaga$ production (i.e. Born plus Box). PYTHIA is used
with CTEQ5L and $\asgj$ at 1 loop. The DIPHOX 'NLO' contribution includes Born at
NLO with CTEQ5M and $\asgj$ at 2 loops (as in {\it a2)}) and Box at LO
with CTEQ5L and $\asgj$ at 1 loop (as in {\it a3)}). 
The cross sections  are given in Table~\ref{sum} and Fig.~\ref{mggborn+box}
shows the $\mgg$ distribution after all selection cuts for PYTHIA and DIPHOX.

\begin{table}[h]
\begin{center}
\begin{tabular}{|l|r|r|r||r|r|r|c|}
\hline
Cut      & \multicolumn{3}{|c|}{PYTHIA with ISR}   & \multicolumn{3}{|c|}{DIPHOX 'NLO'} & K$_{\rm{NLO/ISR}}$ \\
\hline
                       & Box & Born &Total& Box & Born &Total& \\
\hline
$\ptg>20$ GeV                 & 60.3 & 68.8 &129.1 & 83.2 & 94.9 & 178.1&1.38 \\
\hline
+ $|\etag|<2.5$               & 41.4 & 27.6 & 69.0 &56.3 & 41.0 & 97.3 &1.41 \\ 
\hline
+ $\ptg>25$ GeV               & 18.9 & 16.0 & 34.9 & 25.7 & 23.6 & 49.3& 1.41 \\
\hline
+ Max($\ptgu$,$\ptgd$)$>$40 GeV & 6.6 & 6.9 & 13.5 &4.7 & 10.4 & 15.1& 1.12 \\
\hline
+ $\mgg>80$GeV                & 5.6 & 6.4 & 12.0&4.7 & 9.5 & 14.2&1.18 \\
\hline
\end{tabular}
\end{center}
\caption{Comparison of PYTHIA (CTEQ5L+ $\asgj$ at 1 loop) with ISR and DIPHOX 
'NLO' cross section (in pb) at different stage of the $\Hgg$ selection for Born plus
Box contributions.
The K factor is the ratio of the total cross sections for PYTHIA with 
ISR and DIPHOX 'NLO'.
\label{sum} }
\end{table}

\begin{figure}
  \bmini
  \bc
  \mbox{\hspace{-.8cm}\includegraphics[height=90mm]{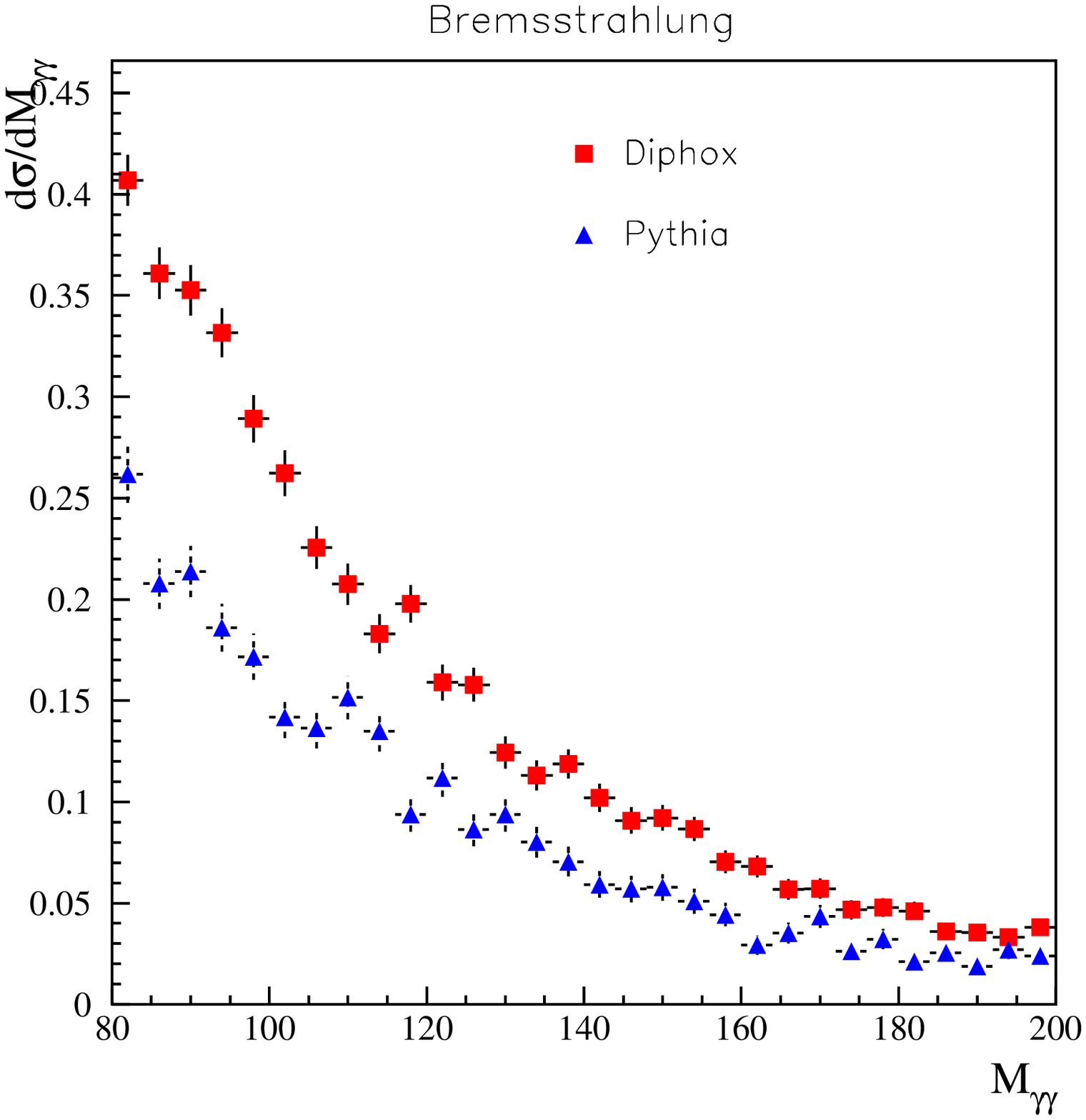}}
  \ec
  \caption{$\mgg$ distribution for PYTHIA and DIPHOX after all selection cuts
    for the Bremsstrahlung contribution.\label{mggbrem}}
  \emini \hspace{.3cm}
  \bmini
  \bc
  \mbox{\includegraphics[height=90mm]{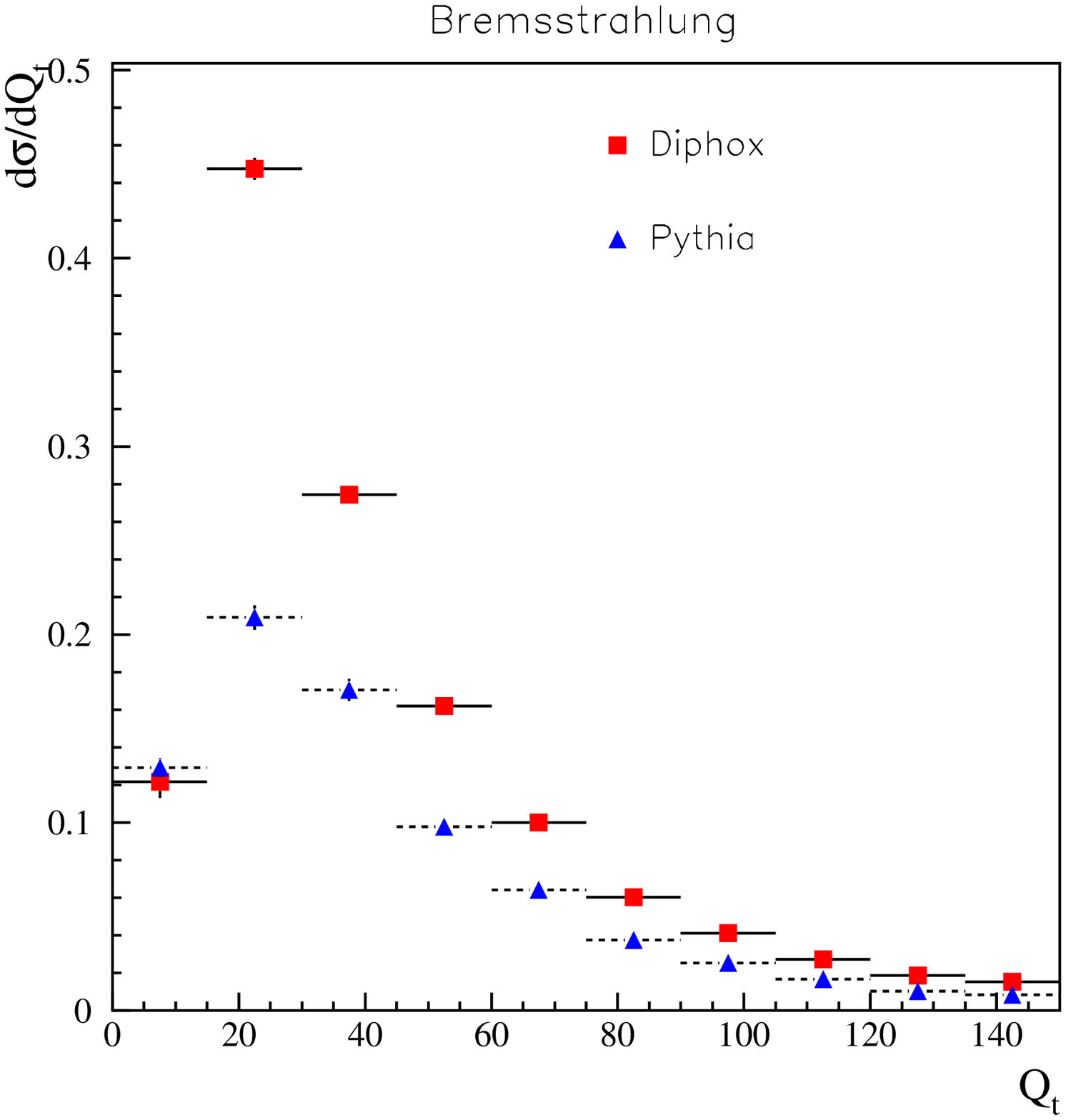}}
  \ec
  \caption{$Q_t$ distribution for PYTHIA and DIPHOX after all selection cuts
    for the Bremsstrahlung contribution.\label{qtbrem}}
  \emini
\end{figure}

\noindent 
{\it b) The Bremsstrahlung contribution}\\
In PYTHIA the Bremsstrahlung contribution is obtained with the processes
$qg\rightarrow q\gamma$ and $q\bar{q}\rightarrow g \gamma$ with QCD ISR and
FSR. The second photon is produced mainly by QED FSR. The cross sections are
summarized in Table~\ref{bremtab}. Note that the isolation cut reduces the
cross section by less than a factor 2. All the cross sections are obtained with
CTEQ5L and $\asgj$ at 1 loop. ISR/FSR in PYTHIA does not produce as high $\ptg$
as DIPHOX NLO, therefore the cross section is smaller in PYTHIA. Moreover the
asymmetric cut on $\ptg$ enhances this difference: since PYTHIA is a quasi
$2\rightarrow 2$ generator it does not produce high $p_T$ difference between
the photons.\\  An isolation cut is defined by imposing that $\Delta R > 0.4$
{\bf or} $\ptq<10$~GeV.  For DIPHOX, $\ptq$ is either the $p_T$ of the parton
$p$ produced with the photon pair or the transverse energy  of the residue of
the fragmentation flowing along the photon direction.  $\Delta R$ is  the
smallest distance between the parton $p$ produced with the photon pair and the
photons \footnote{In the case of the two to two kinematics, $\Delta R = 0$},
$\Delta R = \rm{Min} ((\eta_p-\eta_{\gamma_{1,2}})^2 + (\phi_p -
\phi_{\gamma_{1,2}})^2)$. For PYTHIA, $\ptq = || \sum_i \vec{p}_{T \, i} ||$
where the sum runs on all the partons produced in the shower going with the
photon and $\Delta R$ is  the smallest distance in rapidity-azimuthal angle
plane between the parton produced in the $2 \rightarrow 2$ hard scattering and
the photons. The isolation cut also slightly enhances the K-factor. It might be
due to the fact that  the quark tends to be more collinear to the radiated
photon in  PYTHIA. Figure~\ref{mggbrem} shows the $\mgg$ distribution after all
selection cuts and Figure~\ref{qtbrem} shows the $q_T$ distribution.
\begin{table}[h]
\begin{center}
\begin{tabular}{|l|r|r|c|}
\hline
Cut      &              PYTHIA (ISR/FSR)   & DIPHOX  & K$^{\rm{brem}}_{\rm{DIPHOX/PYTHIA}}$ \\
\hline
$|\etag|<2.5$,  $\ptg>25$ GeV & 22.2 & 27.5 &  1.24\\ 
\hline
+ Max($\ptgu$,$\ptgd$)$>$40 GeV & 14.9 & 24.7  & 1.66 \\
\hline
+ $\mgg>80$GeV                & 11.9 & 19.7  & 1.66 \\
\hline
$\Delta r > 0.4$ or $\ptq<10$~GeV & 7.4 & 12.8 & 1.72 \\
\hline
\end{tabular}
\end{center}
\caption{Comparison of PYTHIA (CTEQ5L+ $\asgj$ at 1 loop) with ISR/FSR and 
DIPHOX  cross section (in pb) at different stage of the $\Hgg$ selection
for the Bremsstrahlung contribution.
\label{bremtab} }
\end{table} 
%

\noindent
{\it Conclusion}\\
The contribution of the two photon background for the $H\rightarrow \gaga$
search  at the LHC has been estimated at NLO with DIPHOX and compared to PYTHIA
Monte--Carlo. The K-factor has to be used with care since it depends on the
selection cuts notably on the isolation criteria.  Note that the asymmetric cut
on $\ptg$ is responsible  for an enhancement of the K-factor.

%
%

%
%

\subsubsection{Comparison of PYTHIA and DIPHOX for $\pi^0 \, \gamma$ production}
\label{sec:pigam,gamjet,qcdsm}
Because of the huge jet rates at the LHC, any photonic observable is heavily
contaminated by neutral pions which appear as fake photons inside the
electro-magnetic calorimeter. Thus, concerning the search for a light neutral
Higgs boson in the mass window between 80 to 140 GeV, a detailed understanding
of not only $\gamma\gamma$ but also $\gamma\pi^0$ and $\pi^0\pi^0$ rates is
mandatory
\cite{Armstrong:1994it,ATLAS:1999tdr,CMS:1994tp,CMS:1997tdr,Branson:2001pj,Catani:2000jh}. 
The experimental studies for LHC depend heavily on Monte--Carlo event
generators like  PYTHIA which treat QCD observables on the leading order level
along with the modelling of hadronization and radiation effects. A comparison
with theoretical results which  include up-to-date knowledge is a way to test
the reliability of these programs. An adequate tool to do such a comparison is
the DIPHOX code which is a partonic event generator designed for the pair
production of hadrons and/or photons in hadronic collisions at full
next--to--leading order \cite{Binoth:1999qq,Binoth:2001jd}.

Here we present a comparison of $\pi^0\gamma$ observables  relevant for Higgs
search, namely the invariant mass, $M_{\gamma\pi}$  and the transverse momentum
distribution of the pair, $q_T$, with a special emphasis on isolation
criteria.  Both distributions were calculated with PYTHIA
6.152\cite{Sjostrand:1994yb} and DIPHOX. In the former, in order to allow a
comparison, the multiple interactions within the p-p collision were switched
off and the pile-up effects from the collisions within the same bunch crossing
were not taken into account. The comparisons were made with the initial and
final state radiation allowed.

\begin{figure}[th]
\begin{center}
\includegraphics[width=14cm]{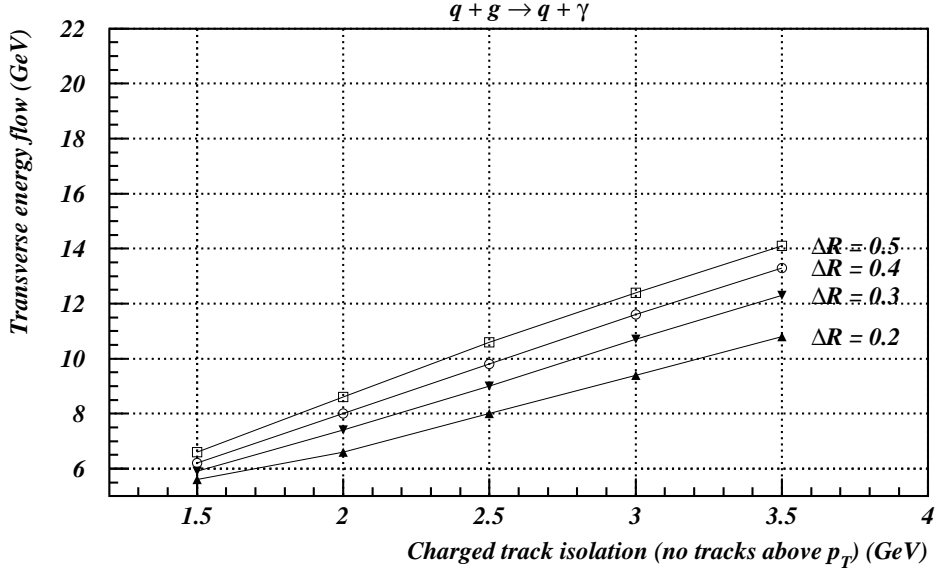}
\end{center}
\caption{\label{Fig:isol_crit}{\em Comparison between the 
cone isolation criterion vs. charged track isolation.}}
\end{figure}

Two isolation criteria were implemented: the transverse energy 
flow isolation and the charged track isolation. In the former,
a threshold is set to the sum of $E_T$ of all particles in the
isolation cone $\Delta R = \sqrt{(\Delta \eta)^2 + (\Delta \phi)^2}$.
In the latter, a threshold is set to the maximum $p_T$ of any
charged particle in the isolation cone. The correlation 
between the two isolation criteria is plotted in Fig.~\ref{Fig:isol_crit}.
As DIPHOX generates only partonic events with a subsequent  collinear
fragmentation of partons into pions, only  the transverse energy  flow
isolation can be implemented. From Fig.~\ref{Fig:isol_crit} it can be inferred
that the maximal hadronic transverse energy inside a cone   lies between 5 and
15 GeV in an experimentally realistic situation. As a reminder, this is the
transverse energy stemming from the partonic reaction only and not from
multiple interactions and/or pile-up.  In this window, the partonic reaction, 
$q g\rightarrow q\gamma$, is by far dominant in the cross section. This 
motivates us to focus on this case only. For the following plots we fixed the
isolation cone size to be $\Delta R=0.4$.

With PYTHIA the process $q + g \rightarrow q + \gamma$ was generated and a
requirement was made that the photon candidates (real photon or a hard $\pi^0$
from the quark jet) have $p_T >$ 25, 40 GeV, and that they are within the
pseudo-rapidity range of $|\eta| < 2.5$. A invariant mass cut of $80 \,
\mbox{GeV} < M_{\pi\gamma} <140\, \mbox{GeV}$ was imposed.  Furthermore, it was
required that one of the photon candidates is a  $\pi^0$, thus rejecting events
were the second photon comes from the jet fragmentation.  With DIPHOX the same
reaction, including next-to-leading corrections, was calculated. The
next-to-leading order parton distribution functions  GRV94\cite{Gluck:1995uf}
were used to compare with the PYTHIA default value.  Only direct photons  were
considered in this study, since in the case of severe isolation, the
contribution from fragmentation is suppressed below the 10 per cent level. For
the fragmentation scale we used the value $M_{\gamma\pi}/8$. This non-canonical
choice is dictated by the fact that in the case  of isolation cuts the typical
fragmentation scale should be  governed by values $\sim E_{T max}$
\cite{Binoth:2002wa}.  

\begin{figure}[th]
  \bmini
  \bc
  \mbox{\hspace{-.8cm}\includegraphics[height=90mm]{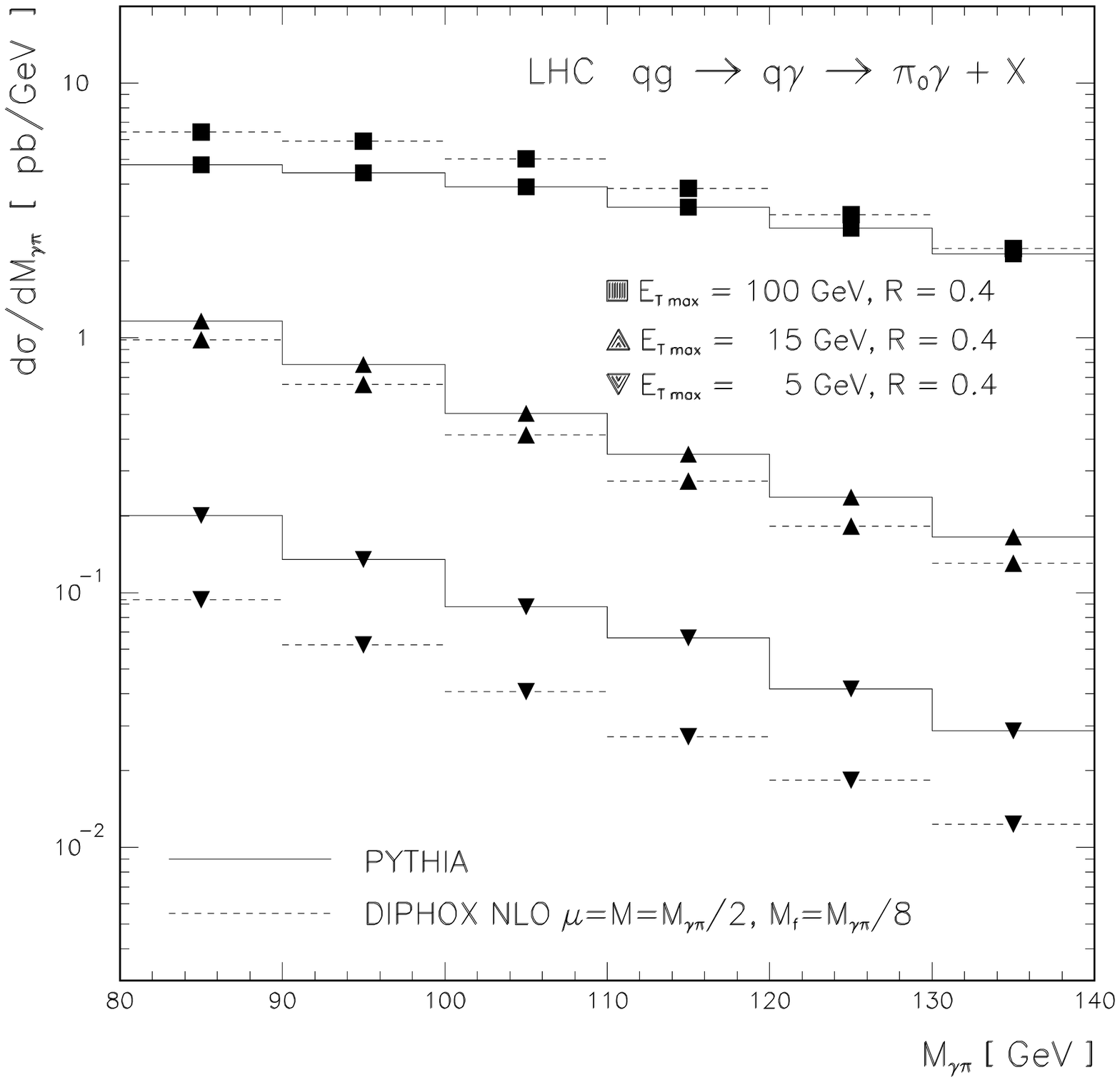}}
  \ec
  \caption{\label{Fig:mpg}{\em Comparison PYTHIA vs. DIPHOX:
   The invariant $\gamma\pi$-mass spectrum 
   for different cone isolation criteria and standard LHC cuts.}}
  \emini \hspace{.3cm}
  \bmini
  \bc
  \mbox{\includegraphics[height=90mm]{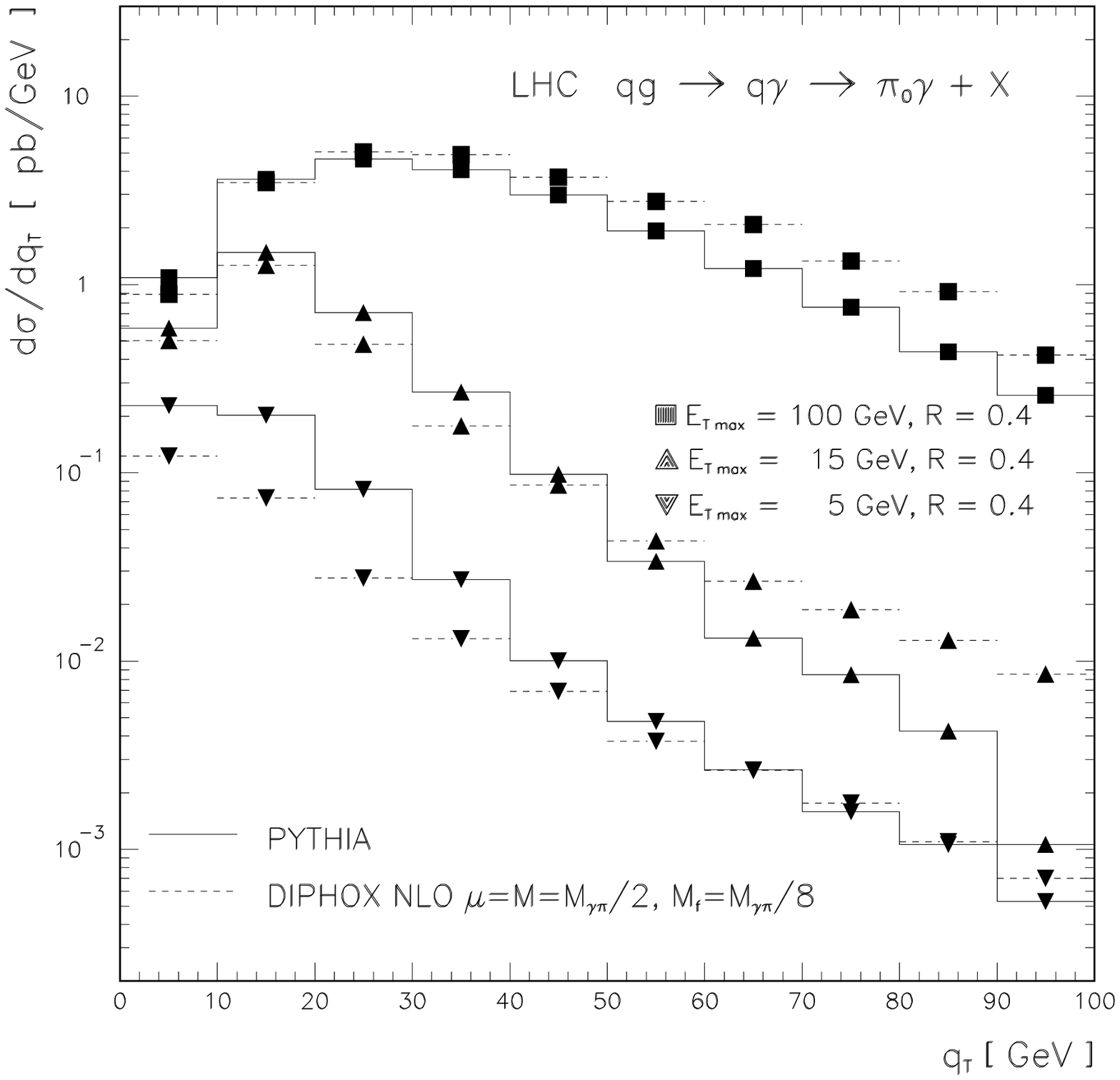}}
  \ec
\caption{\label{Fig:qt}{\em Comparison PYTHIA vs. DIPHOX:
The transverse momentum spectrum of $\gamma\pi$ pairs 
for different cone isolation criteria and standard LHC cuts.}}
  \emini
\end{figure}

In Fig.~\ref{Fig:mpg} the comparison between PYTHIA and DIPHOX is shown for the
invariant mass distribution of the pion photon pair. Good agreement is found
for the case of very loose isolation,  $E_{T max}=100$ GeV and the more
relevant value of  $E_{T max}=15$ GeV.     Note that the NLO prediction is
plagued by large scale uncertainties around $\pm 30$ to $40$ per cent, as the
hard experimental cuts spoil compensations of higher order terms with opposite
signs. In the case of very strict isolation, $E_{T max}=5$ GeV, PYTHIA produces
somewhat higher rates. The discrepancy has presumably three sources which have
to do with the small value of $E_{T max}$. Small $E_{T max}$ means that the
fragmentation variable $z$, the ratio of transverse pion momentum to parent
parton momentum, is pressed towards one, $z>z_{min}=p_{T min}/(p_{T min}+E_{T
max})\sim 0.8$.  In this regime, first of all, the available fragmentation
functions are not well constrained by  experimental data \cite{Kniehl:2000fe}
and there may be differences in the  fragmentation/hadronization models for
large $z$  implemented in PYTHIA and DIPHOX.  Second, large logarithms 
($\sim\log(1-z)^2$) are present  in the next-to-leading order calculation
which    may render the perturbative contributions unreliable.  Third, the
above mentioned fact that the fragmentation scale choice should be related to 
$E_{T max}$, indeed, indicates that the fragmentation scale should be chosen
smaller here, which  increases the DIPHOX prediction in the right way. With the
value $M_f$=$M_{\gamma\pi}/16$ the DIPHOX curve shifts upwards by about 40 per
cent leading to a better agreement. Altogether one can conclude that PYTHIA and
DIPHOX show a  reasonable agreement in the invariant mass distribution apart
from the regime of very hard isolation which deserves further investigation
from the theoretical side \cite{Binoth:2002wa}.  
In Fig.~\ref{Fig:qt}, the $q_T$ spectrum is plotted for the same  isolation
criteria. For $E_{T max}=15, 100$ GeV good agreement  is found for the used
scales for the first bins.  The shape for these contributions is not quite the
same, with the PYTHIA prediction being slightly steeper. This is to be
expected, as the NLO calculation  encoded in DIPHOX contains $2\rightarrow 3$
matrix elements, enhancing the tail of the distribution. As a reminder, the
tail in  PYTHIA is filled by a parton showering model which is not as reliable
as explicit higher order matrix elements. Still, the disagreement  in shape for
the relevant $q_T$ domain seems not too alarming in the plotted range.  Note
that in the case of hard isolation the PYTHIA prediction of the last 4 bins is
plagued by low  statistics. The apparent overall discrepancy in the case $E_{T
max}=5$ GeV can be understood by the same reasoning as given above for the 
$M_{\gamma\pi}$ spectrum. 

In conclusion, the comparison  between PYTHIA and the NLO DIPHOX code shows a
reasonable agreement. Whereas infrared insensitive spectra, such as the
invariant mass distribution,  differ only by an overall normalisation which can
be accounted for by an adequate scale choice, infrared sensitive observables
like the transverse momentum distribution of the pair   are typically  steeper
in PYTHIA due to missing NLO matrix elements. In the case of very hard
isolation an apparent difference  between the PYTHIA and DIPHOX predictions can
be traced back to a different modelling of the production of high $p_T$ pions,
a too high fragmentation scale choice and/or subtleties in the interplay
between higher order corrections and severe experimental cuts. This will be
extensively  discussed elsewhere \cite{Binoth:2002wa}.

%
\newcommand*{\eg}{{\it e.g.},\ }
\newcommand*{\ie}{{\it i.e.},\ }
\newcommand*{\etal}{{\it et al.\ }}
\newcommand*{\gev}{{\rm\,GeV}}
\newcommand*{\gevc}{{\rm\,GeV/c}}
\newcommand*{\mevc}{{\rm\,MeV/c}}
\newcommand*{\tev}{{\rm\,TeV}}
\newcommand*{\pt}{$p_T$}
\newcommand*{\bigpt}{$P_T$}
\newcommand*{\UE}{``underlying event"}
\newcommand*{\BBR}{``beam-beam remnants"}
\newcommand*{\MB}{Min-Bias}
\newcommand*{\pthard}{$p_T({\rm hard})$}
\newcommand*{\ptcut}{$p_T\!>\!0.5\,{\rm GeV/c}$}
\newcommand*{\etacut}{$|\eta|\!<\!1$}
\newcommand*{\hardcut}{$p_T\!({\rm hard})>3{\rm\,GeV/c}$}
\newcommand*{\hardzero}{$p_T\!({\rm hard})>0{\rm\,GeV/c}$}
\newcommand*{\ptchj}{$P_T\!({\rm chgjet}\#1)$}
\newcommand*{\aveN}{$\langle\! N_{\rm chg}\!\rangle$}
\newcommand*{\avePT}{$\langle P_T{\rm sum}\!\rangle$}
\newcommand*{\etaphi}{$\eta$-$\phi$}
%
\def\bea{\begin{eqnarray}}
\def\eea{\end{eqnarray}}
\def\bes{\begin{eqnarray*}}
\def\ees{\end{eqnarray*}}
\newcommand{\MEG}{MEG}
\def\q2{Q^2}              
\def\qt{Q_T}
\newcommand{\SHG}{SHG}

\newenvironment{myitemize}[2]
{ \begin{small}
  {\large\tt\bf #1} { \sc #2 } \vspace{.4cm} 
  \begin{itemize}
  \setlength{\itemsep}{2pt} 
  \setlength{\parsep}{0pt}
  \setlength{\topsep}{0pt}
  \setlength{\partopsep}{0pt}
}
{\end{itemize}\end{small}}
\newcommand{\myitem}[3]{ \item {\footnotesize\tt #1}~{\bf\tt #2}: {\it
    #3} }

\section{Monte Carlo\protect\footnote{Section coordinators:
I. Hinchliffe, J. Huston}$^{,}$~\protect\footnote{Contributing authors: 
C. Bal\'azs, E.\ Boos, M.\ Dobbs, W.\ Giele, 
I.\ Hinchliffe, Rick Field, J.\ Huston, V.\ Ilyin, 
J.\ Kanzaki, B. Kersevan, K.\ Kato, Y.\ Kurihara, 
L.\ L\"onnblad, K.\ Mazumudar, M.\ Mangano, S.\ Mrenna, 
F.\ Paige, I. Puljak, E.\ Richter-Was,
M.\ Seymour, T.\ Sj\"ostrand,  M. T\"onnesmann. B.\ Webber, D.\ Zeppenfeld}}
\label{sec:mc,qcdsm}

\subsection{Introduction}

The Monte-Carlo intergroup focused on four main issues: the problem of
interfacing partonic event generators to showering Monte-Carlos, an
implementation using this interface to calculate backgrounds which are poorly
simulated by the showering Monte-Carlos alone, a comparison of the {\tt 
HERWIG} and {\tt PYTHIA} parton  shower models with the predictions of a soft
gluon resummation program (ResBos),  and studies of the underlying events at
hadron colliders and how well they are modeled by the Monte-Carlo generators.

Section~\ref{interface} discusses a strategy whereby generic Fortran
common blocks are presented for use by High Energy
Physics event generators for the transfer of event configurations from
parton level generators to showering and hadronization event
generators.

Section~\ref{acer} discusses the {\bf AcerMC} Monte Carlo Event
Generator which is dedicated to the generation
of the Standard Model background processes at $pp$ LHC collisions. 
The program itself provides a library of the massive matrix elements 
and phase space modules for the generation of 
selected processes: $q \bar q \to W(\to \ell \nu) b \bar b$, $q \bar q \to W(\to \ell \nu) t \bar t$,
$gg, q \bar q \to Z/\gamma^*(\to \ell \ell) b \bar b$,
$gg, q \bar q \to Z/\gamma^*(\to \ell \ell, \nu \nu, b \bar b) t \bar t$, 
the QCD $gg, q \bar q \to t \bar t b \bar b$ and EW  $gg \to (Z/W/\gamma \to)
t \bar t b \bar b$. The hard process event, 
generated with these modules, can be completed by the initial and final 
state radiation, hadronization and decays, simulated with either the
{\tt PYTHIA} or {\tt HERWIG} Monte Carlo Event Generators. Interfaces to both
these generators are provided in the distribution version. The {\bf AcerMC} also uses several other external libraries: {\tt CERNLIB}, {\tt HELAS}, {\tt VEGAS}.

In Section~\ref{Huston}, predictions for the $p_T$ distribution for a Higgs particle were generated using two approaches: (1) a soft-gluon resummation technique, using the program {\tt ResBos} and (2) a parton shower technique using the Monte Carlo programs  {\tt HERWIG} and {\tt PYTHIA}. An understanding of the 
kinematics of the Higgs boson, or of any other Standard Model or non-
Standard Model particle, and the characteristics of any jets associated with its production, is of great interest for physics at the Tevatron, the LHC or any future hadron colliders. The transverse momentum distribution of the Higgs boson depends primarily on the details  of the soft gluon emission from the initial state partons. The effects of these soft gluon emissions can be described either by a
resummation calculation or by a Monte Carlo parton shower formalism. Comparisons between the two techniques for several different Higgs masses and for several center-of-mass energies show relatively good agreement between the {\tt ResBos} predictions and those of {\tt HERWIG} and recent versions of {\tt PYTHIA}.

In Section~\ref{field}, the behavior of the \UE\ in hard scattering proton-antiproton collisions at $1.8\tev$ is studied and 
compared with the QCD Monte-Carlo models. The ``hard scattering" component consists of the outgoing two ``jets" plus initial and final-state radiation. The \UE\ is everything 
except the two outgoing hard scattered ``jets" and consists of the \BBR\ plus possible contributions 
from the ``hard scattering" arising from initial and final-state radiation.  In addition multiple parton 
scattering might contribute to the \UE.  The data indicate that
neither {\tt ISAJET} or {\tt HERWIG} produce 
enough charged particles (with \ptcut) from the ``beam-beam remnant" component and that 
{\tt ISAJET} produces too many charged particles from initial-state radiation.  The ``tuning" of {\tt ISAJET} and 
{\tt PYTHIA} to fit the \UE\ is explored.

In addition to these activities, discussions were held with the
``standard model'' and ``beyond the standard model'' groups in order
to assess their needs. Many of these needs are expressed in terms of
the desire to have new processes added to the showering Monte-Carlo
event generators. The interface discussed in Section \ref{interface}
should make this easier for users to implement the processes themselves, obviating
the need to have the Monte-Carlo authors do it. Among the important
items mentioned are:

\begin{itemize}
\item The production of $WZ$ final states should include the
  contributions from virtual $W$ and $Z$ as these are vital for SUSY
  searches at the Tevatron
\item  Resummation of the $log(m_H/m_b)$  and $log(m_H/p_t)$  terms in
      the $gg\to H b \overline{b}$ process.
\item  Radion phenomenology
\item A better understanding of the $gg\to H$ process and in
  particular the effect of a jet veto
\item Correct $\tau$ polarization in Higgs decay.
\end{itemize}

\subsection{Generic User Process Interface for Event Generators}
\label{interface}

Modularization of High Energy Particle Physics event generation is
becoming increasingly useful as the complexity of Monte Carlo programs
grows. To accommodate this trend, several authors of popular Monte
Carlo and matrix element 
programs attending the {\it Physics at TeV Colliders Workshop}
in Les Houches, 2001 have agreed on a generic format for the transfer
of parton level event configurations from matrix element event generators 
(\MEG) to showering and hadronization event generators (\SHG).

\begin{center} \color{blue}
\begin{tabular}{rcl}
  CompHEP\cite{Pukhov:1999gg}     & &        \\
  Grace\cite{Ishikawa:1993qr}     & & {\tt HERWIG}\cite{Corcella:2001bw} \\
  MadGraph\cite{Stelzer:1994ta}   & \hspace{1cm} $\Rightarrow$
                                  \hspace{1cm} & \tt ISAJET\cite{Baer:1999sp} \\
  VecBos\cite{Berends:1991ax}     & & {\tt PYTHIA}\cite{Sjostrand:2000wi} \\
  WbbGen\cite{Caravaglios:1999yr} & & \ldots \\
  \ldots    & &        \\
\end{tabular}
\color{black} \end{center}

Events generated this way are customarily called user (or
user-defined) processes, to distinguish them from the internal
processes that come with the \SHG. Specific solutions are already in
use, including an interface of WbbGen with
{\tt HERWIG}~\cite{Mangano:2001xp} and an interface of CompHEP with
{\tt PYTHIA}~\cite{Belyaev:2000wn}---that experience is exploited here.


Since the specification of the user process interface in May 2001 at
Les Houches, the interface has been (or is being) implemented in a
number of \MEG\ and \SHG\ programs.  An implementation has been
included in {\tt PYTHIA}~6.2, described (with an example) in
Ref.~\cite{Sjostrand:2001yu}.  A {\tt HERWIG} implementation is in progress
and will appear in Version~6.5.  \MEG\ implementations exist
for the Madison Collection of User Processes (MADCUP)~\cite{Cranmer:2002kc},
ALPGEN~\cite{Caravaglios:1999yr,Mangano:2002ap},
and CompHEP~\cite{Ilyin:2002vi} (to be available publicly soon). An
implementation in MadGraph is in preparation.  Other \MEG\
implementations include Refs.~\cite{Kersevan:2002dd,Dobbs:2001dq}.

The user process interface discussed here is not intended as a
replacement for \mbox{ {\tt HEPEVT}}~\cite{Sjostrand:1989tj}, 
which is the standard Fortran
common block for interface between generators and analysis/detector
simulation. The user process common blocks address the communication
between two event generators only, a \MEG\ one and a \SHG\ one, and not the 
communication of event generators with the outside world.

In the course of a normal event generation run, this communication occurs 
at two stages: (1) at initialization, to establish the basic parameters 
of the run as a whole and (2) for each new event that is to be 
transferred from the \MEG\ to the \SHG. Each of these two stages here
corresponds to its own Fortran common block.\footnote{ An interface in 
C++ has been developed in Ref.~\cite{Dobbs:2001ck} and contains similar 
information content as that discussed here.} These common blocks are 
described in detail in the next two sections, followed by some examples.

One can also foresee that each stage will be associated with its own
subroutine, called from the \SHG, where information is put in the
respective common block, based on output from the \MEG. The details of
these subroutines are likely to be specific to a given \MEG\ and may
also be specific to a given \SHG.  The subroutine names {\tt UPINIT} and
{\tt UPEVNT} (each with no arguments) were chosen for the
{\tt PYTHIA}~6.2 implementation.  They are intended to
be generic (the usual {\tt PY} prefixes are omitted), and only dummy
versions are packaged with the program.  It is recommended that other
\SHG\ authors use the same dummy routine names (with zero arguments)
such that for simple cases which do not require intervention 
`by hand', \MEG\ authors will be able to interface several \SHG s with a
single set of routines. Example routines are presented in the {\tt PYTHIA}
documentation~\cite{Sjostrand:2001yu}.

In general, a user process run may consist of a number of subprocesses,
each denoted by a unique integer identifier. If the user
wishes to have the \SHG\ unweight events using acceptance-rejection and/or
mix together events from different processes, then the user process
author will need to supply a subroutine that is able to return an event of
the requested subprocess type to the \SHG\ on demand. 
The author may choose to organize the subroutine to
generate the event `on the fly', or to read the event from a file
stream (with a separate file stream for each subprocess).
The \SHG\ will also need information about the subprocess cross section
and/or maximum event weight to select which process is generated next
and for acceptance-rejection. This information will need to be
known from the onset (and could, for example, be determined in advance
from an initialization run). Alternatively, the user may already have 
a proper mixture of subprocesses from the \MEG\ and only wish the \SHG\
to process events in the order they are handed in. We therefore allow for 
several different event weight models.

If extra information is needed for a specific user implementation,
then a implementation-specific common block should be
created. The meaning of the user process common block
elements should not be overloaded, as this would defeat the generic
purpose.

The descriptions in this paper are intended for event generator
authors and may appear complex---most of the details will be
transparent to the casual user.

\subsubsection{`User Process' Run Information}

        The run common block contains information which pertains to a
        collection of events. \\
        In general this information is process dependent and it is
        impossible to include everything in a generic common
        block. Instead only the most general information is included
        here, and it is expected that users will have to intervene
        `by hand' for many cases (i.e.\ a user may need to specify
        which cutoffs are used to avoid singularities, which jet 
        clustering algorithm has been used to recombine partons in a
        next-to-leading-order calculation, the effective parton masses, 
        \ldots).

\begin{verbatim}
      integer     MAXPUP
      parameter ( MAXPUP=100 )
      integer     IDBMUP, PDFGUP, PDFSUP, IDWTUP, NPRUP, LPRUP
      double precision EBMUP, XSECUP, XERRUP, XMAXUP
      common /HEPRUP/ IDBMUP(2), EBMUP(2), PDFGUP(2), PDFSUP(2),
     +                IDWTUP, NPRUP, XSECUP(MAXPUP), XERRUP(MAXPUP), 
     +                XMAXUP(MAXPUP), LPRUP(MAXPUP)
\end{verbatim}

\vspace{.5cm}

\begin{myitemize}{HEPRUP}{ `User Process' Run Common Block }

   \myitem{parameter} { MAXPUP=100 }
        {maximum number of different processes to be interfaced at one
        time} 

\hspace{-.2cm} {\bf \underline{ Beam Information } } \\
        Beam particle 1 (2) is defined as traveling along the +Z (--Z)
        direction.
   \myitem{integer} { IDBMUP(2) }
        {ID of beam particle 1 and 2 according to the Particle Data
        Group convention~\cite{Garren:2000st}}
   \myitem{double} { EBMUP(2) }
        {energy in GeV of beam particles 1 and 2}
   \myitem{integer} { PDFGUP(2) }
        { the author group for beam 1 and 2, according to the Cernlib
        PDFlib~\cite{Plothow-Besch:1993qj} specification }
   \myitem{integer} { PDFSUP(2) }
        { the PDF set ID for beam 1 and 2, according to the Cernlib
        PDFlib specification } \\
    For $e^+e^-$ or when the \SHG\ defaults are
        to be used, PDFGUP=--1, PDFSUP=--1 should be specified. \\
    The PDFlib enumeration of PDFs is sometimes out of date, but it is the
        only unique integer labels for PDFs available. In the case
        where a PDF not included in PDFlib is being used, this
        information will have to be passed `by hand'.

\hspace{-.2cm} {\bf \underline{ Process Information } }

   \myitem{integer} { IDWTUP }
        {master switch dictating how the event weights
        (XWGTUP) are interpreted} \\
        The user is expected to pick the most appropriate event weight 
        model for a run,
        given the \MEG\ input at hand and the desired output. Normally 
        the \SHG\ should be able to handle all of the models.\\   
        A summary of the IDWTUP switches is presented in
        Table~\ref{t_IDWTUP}.
        \begin{itemize}
        \setlength{\itemsep}{-4pt} \setlength{\parsep}{0pt}
        \setlength{\topsep}{0pt} \setlength{\partopsep}{0pt} 

        \item[+1] Events are weighted on input and the \SHG\ is asked
        to produce events with weight +1 as output. XSECUP and XERRUP
        need not be provided, but are calculated by the \SHG. XWGTUP
        is a dimensional quantity, in pb, with a mean value converging
        to the cross section of the process.  The \SHG\ selects the
        next subprocess type to be generated, based on the relative
        size of the XMAXUP(i) values. The user-supplied interface routine
        must return an event of the requested type on demand from the
        \SHG, and the maximum weight XMAXUP (or a reasonable
        approximation to it) must be known from the onset.
        A given event is accepted with a probability
        XWGTUP/XMAXUP(i). In case of rejection, a new event type and a
        new event are selected.  If XMAXUP(i) is chosen too low, such
        that XWGTUP violates the XMAXUP(i), the \SHG\ will issue a
        warning and update XMAXUP(i) with the new maximum weight
        value.
        If events of some types are already available unweighted, then
        a correct mixing of these processes is ensured by putting
        XWGTUP = XMAXUP(i). In this option also the internal \SHG\
        processes are available, and can be mixed with the external
        ones.  All weights are positive definite. $k$-factors may be
        included on an event by event basis by the user process 
        interface by re-scaling the XWGTUP for each event.

        \item[--1] Same as above (IDWTUP=+1), but the event weights
        may be either positive or negative on input, and the \SHG\
        will produce events with weight +1 or --1 as 
        output.\footnote{Negative-weight events may occur e.g.\ in 
        next-to-leading-order calculations. They should cancel against
        positive-weight events in physical distributions. The details
        of this cancellation are rather subtle when considered in
        the context of showers and hadronization, however, and a
        proper treatment would require more information than discussed 
        here. The negative-weight options should therefore be used with 
        some caution, and the negative-weight events should be a 
        reasonably small fraction of the total event sample.}
        A given event would be accepted with a probability
        $|$XWGTUP$|$/$|$XMAXUP(i)$|$ and assigned weight {\tt
        sign}(1,XWGTUP), where the {\tt sign} function transfers the
        sign of XWGTUP onto 1. A
        physics process with alternating cross section sign must
        be split in two IDPRUP types,\footnote{The motivation for this
        requirement is best understood with a simple example: imagine
        two subprocesses with the same cross section. The first
        process includes events with both positive and negative event
        weights such that two events out of three have weight +1 and the
        third --1. All events from the second process have positive
        weight +1. In this scenario these two processes should be
        `mixed' with proportions 3:1 to account for the cancellations
        that occur for the first process. The proportions for the
        mixing are communicated to the \SHG\ by supplying the positive
        and negative contributions to the cross section separately.
        }
        based on the sign of
        the cross section, such that all events of a particular IDPRUP
        have the same event weight sign.  Also the XMAXUP(i) values must
        be available for these two IDPRUP types separately, so that
        $|$XMAXUP(i)$|$ gives the relative mixing of event types, with
        event acceptance based on $|$XWGTUP$|$/$|$XMAXUP(i)$|$.

        \item[+2] Events are weighted on input and the \SHG\ is asked
        to produce events with weight +1 as output. The \SHG\ selects
        the next subprocess type to be generated, based on the
        relative size of the XSECUP(i) values.  The user-supplied
        interface routine must return an event of the requested type
        on demand from the \SHG. The cross sections XSECUP(i) must be
        known from the onset.
        A given event is accepted with a probability
        XWGTUP/XMAXUP(i).  In case of rejection, a new event of
        the same type would be requested.  In this scenario only the
        ratio XWGTUP/XMAXUP(i) is of significance. If events of
        some types are already available unweighted, then a correct
        mixing of these processes is ensured by putting XWGTUP =
        XMAXUP(i). A $k$-factor can be applied
        to each process by re-scaling the respective XSECUP(i)
        value at the beginning of the run, but cannot be given
        individually for each event. In this option also the internal
        \SHG\ processes are available, and can be mixed with the user
        processes.

        \item[--2] Same as above (IDWTUP=+2), but the event weights
        may be either positive or negative on input, and the \SHG\
        will produce events with weight +1 or --1 as output.  A
        physics process with alternating cross section sign must
        therefore be split in two IDPRUP types, based on the sign of
        the cross section, such that all events of a particular IDPRUP
        have the same event weight sign.  Also the XSECUP(i) and
        XMAXUP(i) values must be available for these two IDPRUP types
        separately, so that $|$XSECUP(i)$|$ gives the relative mixing
        of event types, with event acceptance based on
        $|$XWGTUP$|$/$|$XMAXUP(i)$|$ and the total cross section of
        the two IDPRUP types combined given by XSECUP(i)+XSECUP(j).

        \item[+3] Events are unweighted on input such that all events
        come with unit weight XWGTUP=+1. The \SHG\ will only ask for
        the next event.  If any mixing or unweighting is desired, it
        will have to be performed by the user process interface.  The 
        \SHG\ will not reject any events (unless it encounters other 
        kinds of problems).  If a $k$-factor is desired, it is the
        responsibility of the user process interface. When events are  
        read sequentially from an already existing file, this would    
        imply one common $k$-factor for all processes. In this option it 
        is not possible to mix with internal \SHG\ processes.

        \item[--3] Same as above (IDWTUP=+3), but the event weights
        may be either +1 or --1 on input. A single process identifier
        (IDPRUP) may include events with both positive and negative
        event weights.

        \item[+4] Same as (IDWTUP=+3), but events are weighted
        on input and the average of the event weights (XWGTUP) is
        the cross section in pb. When histogramming results on analyzed
        events, these weights would have to be used.
        The \SHG\ will only ask for the next event
        and will not perform any mixing or unweighting. Neither XSECUP
        nor XMAXUP needs to be known or supplied. In this
        option it is not possible to mix with internal \SHG\
        processes.

        \item[--4] Same as (IDWTUP=+4), but event weights may be either
        positive or negative on input and the average of the event
        weights (XWGTUP) is the cross section. A single process identifier
        (IDPRUP) may include events with both positive and negative
        event weights.

        \end{itemize} 

\begin{table}
\begin{center}
\begin{tabular}{c c c c c} \hline
        & event selection& control of        & XWGTUP &        \\
 IDWTUP & according to   & mixing/unweighting& input & output \\ \hline
+1      & XMAXUP         & \SHG\        & + weighted     & +1 \\
--1     & XMAXUP         & \SHG\        & $\pm$ weighted & $\pm$1 \\
+2      & XSECUP         & \SHG\        & + weighted     & +1 \\
--2     & XSECUP         & \SHG\        & $\pm$ weighted & $\pm$1 \\
+3      & ---            & user interface & +1             & +1 \\
--3     & ---            & user interface & $\pm$1         & $\pm$1 \\
+4      & ---            & user interface & + weighted     & + weighted \\
--4     & ---            & user interface & $\pm$ weighted & $\pm$ weighted \\
\hline
\end{tabular}
\caption{\label{t_IDWTUP} Summary of the options available for the
master weight switch IDWTUP.}
\end{center}
\end{table}

   \myitem{integer} { NPRUP }
        { the number of different user subprocesses } \\
        i.e.\ LPRUP and other arrays will have NPRUP entries, LPRUP(1:NPRUP)

   \myitem{double} { XSECUP(J) }
        { the cross section for process J in pb } \\
        This entry is mandatory for IDWTUP=$\pm$2.

   \myitem{double} { XERRUP(J) }
        { the statistical error associated with the cross section 
        of process J in pb }\\
        It is not expected that this information will be used by the
        \SHG, except perhaps for printouts.

   \myitem{double} { XMAXUP(J) }{ the maximum XWGTUP for process J } \\
        For the case of weighted events (IDWTUP=$\pm$1,$\pm$2), this entry is
        mandatory---though it need not be specified to a high degree
        of accuracy. If too small a number is specified, the \SHG\
        will issue a warning and increase XMAXUP(J).
        This entry has no meaning for IDWTUP=$\pm$3,$\pm$4. 

   \myitem{integer} { LPRUP(J) } {a listing of all user process IDs
        that can appear in IDPRUP of HEPEUP for this run} \\
        When
        communicating between the user process and \SHG, the LPRUP
        code will be used. Example: if LPRUP(1)=1022, then the \SHG\
        will ask for an event of type 1022, not 1.


\end{myitemize}

\begin{center}{\rule[1mm]{5in}{1mm}}\end{center} 

\subsubsection{`User Process' Event Information}

\begin{verbatim}
      integer MAXNUP
      parameter ( MAXNUP=500 )
      integer NUP, IDPRUP, IDUP, ISTUP, MOTHUP, ICOLUP
      double precision XWGTUP, SCALUP, AQEDUP, AQCDUP, 
     +                 PUP, VTIMUP, SPINUP
      common /HEPEUP/ NUP, IDPRUP, XWGTUP, SCALUP, AQEDUP, AQCDUP,
     +                IDUP(MAXNUP), ISTUP(MAXNUP), MOTHUP(2,MAXNUP),
     +                ICOLUP(2,MAXNUP), PUP(5,MAXNUP), VTIMUP(MAXNUP),
     +                SPINUP(MAXNUP)
\end{verbatim}

\vspace{.5cm}

\begin{myitemize}{HEPEUP}{ `User Process' Event Common Block }

  \myitem{parameter} { MAXNUP=500 }
        { maximum number of particle entries }

  \myitem{integer} { NUP } 
        {number of particle entries in this event} \\
        An event with NUP=0 denotes the case where the user process is unable
        to provide an event of the type requested by the \SHG\
        (i.e.\ if the user process
        is providing events to the \SHG\ by reading them sequentially
        from a file and the end of the file is reached).

  \myitem{integer} { IDPRUP } 
        {ID of the process for this event} \\
        The process ID's are not intended to be generic. The entry is
        a hook which the event generators can use to
        translate into their own scheme, or use in print statements
        (e.g.\ so that cross section information can be shown per
        process). \\
        When IDWTUP$=\pm1,\pm2$ the next process to be generated 
        is selected by the \SHG, and so IDPRUP is set by the \SHG. For 
        IDWTUP$=\pm3,\pm4$ the process is selected by the \MEG, and
        IDPRUP is set by the \MEG.

  \myitem{double} { XWGTUP }
        { event weight } \\
        \underline{weighted events}:
        if the user process supplies
        weighted events and the \SHG\ is asked to produce 
        unweighted events, this number
        will be compared against XMAXUP in the run common block HEPRUP
        for acceptance-rejection. \\
        \underline{unweighted events}:
        if the user process supplies events which have already been
        unweighted, this number should be set to +1 (-1 for negative
        weight events in e.g. a NLO calculation).

        The precise definition of XWGTUP depends on the master weight
        switch IDWTUP in the run common block. More information is
        given there.

  \myitem{double} { SCALUP } 
        {scale of the event in GeV, as used for calculation of  PDFs} \\
        If the scale has not been defined, this should be denoted
        by setting the scale to --1.

  \myitem{double} { AQEDUP } 
        {the QED coupling $\alpha_\mathrm{QED}$ used for this event 
        (e.g.\ $\frac{1}{128}$) }

  \myitem{double} { AQCDUP } 
        {the QCD coupling $\alpha_\mathrm{QCD}$ used for this event }

        When $\alpha_\mathrm{QED}$ and/or $\alpha_\mathrm{QCD}$ is not
        relevant for the process, or in the case where the user
        process prefers to let the \SHG\ use its defaults,
        AQEDUP=--1 and/or AQCDUP=--1 should be specified.
      
\hspace{-.2cm} {\bf \underline{ ID, Status, and Parent-Child History } }
  \myitem{integer} { IDUP(I) } 
        {particle ID according to Particle Data Group 
        convention~\cite{Garren:2000st}} \\
        undefined (and possibly non-physical) ``particles'' should be
        assigned IDUP=0 (i.e.\ the $WZ$ particle in the example given
        in the MOTHUP description below)

  \myitem{integer} { ISTUP(I) } {status code} 
        \begin{itemize}
        \setlength{\itemsep}{-4pt} \setlength{\parsep}{0pt}
        \setlength{\topsep}{0pt} \setlength{\partopsep}{0pt} 
        \item[--1] Incoming particle
        \item[+1] Outgoing final state particle
        \item[--2] Intermediate space-like propagator defining an 
        $x$ and $Q^2$ which should be preserved
        \item[+2] Intermediate resonance, Mass should be preserved
        \item[+3] Intermediate resonance, for documentation only\footnote{
                Treatment of ISTUP(I)=+3 entries may be generator 
                dependent (in particular see
                Ref.~\cite{Sjostrand:2001yu} 
                for the special treatment in {\tt PYTHIA}).}
        \item[--9] Incoming beam particles at time $t=-\infty$
        \end{itemize} 
        The recoil from a parton shower (including
        photon emission) needs to be absorbed by particles in the
        event. Without special instructions, this can alter the mass
        of intermediate particles. The ISTUP flag +2 allows the user
        process to specify which intermediate states should have their
        masses preserved, i.e.\ for $e^+e^- \rightarrow Z^0 h^0
        \rightarrow q \bar{q} b \bar{b}$, the $Z^0$ and $h^0$ 
        would be flagged with ISTUP=+2. \\
        The primary application of the ISTUP=--2 status code is deep
        inelastic scattering (a negative number is chosen for this
        status code because the propagator in some sense can be
        thought of as incoming). See the example below. \\
        The status code ISTUP=--9 specifying incoming beams is not
        needed in most cases because the beam particle energy and identity
        is contained in the HEPRUP run information common block. The
        primary application of this status code will be non-collinear
        beams and/or runs for which the beam energy varies event by
        event (note that using the --9 status code to vary the machine
        energy may produce problems if the \SHG\
        is asked to combine separate processes).  The
        use of ISTUP=--9 entries is optional, and is only necessary
        when the information in HEPRUP is insufficient. If entries
        with ISTUP=--9 are specified, this information will over-ride
        any information in HEPRUP.

  \myitem{integer} { MOTHUP(2,I) } 
        {index of first and last mother} \\
        For decays, particles will normally have only one mother.\\ In
        this case either MOTHUP(2,I)=0 or MOTHUP(2,I)=MOTHUP(1,I).
        Daughters of a $2\rightarrow n$ process have 2 mothers.  This
        scheme does not limit the number of mothers, but in practice
        there will likely never be more than 2 mothers per particle.

        The history (intermediate particles) will be used by the
        \SHG s to decipher which
        combinations of particles should have their masses fixed and
        which particle decays should be ``dressed'' by the parton
        shower. Example: for $q\bar{q}'\rightarrow W^-Zg\rightarrow l^-
        \nu l^+ l^- g$, intermediate ``particles'' $WZ$, $W$, and $Z$
        could be specified with ISTUP=+2. Here the $WZ$ ``particle''
        would have its own entry in the common block with IDUP=0.  The
        showering generator would preserve the invariant masses of
        these ``particles'' when absorbing the recoil of the parton
        shower.

        In a case like $e^+e^- \rightarrow \mu^+\mu^-\gamma$
        proceeding via a
        $\gamma^*/Z^0$, where the matrix element contains an
        interference term between initial and final-state emission,
        this ambiguity in the parent-child history of the $\gamma$ has
        to be resolved explicitly by the user process.

  \hspace{-.2cm} {\bf \underline{ Color Flow } } \\
        A specific choice of color flow for a particular event is often
        unphysical, due to interference effects. However,
        \SHG s require a specific color state from which to begin
        the shower---it is the responsibility of the user process to
        provide a sensible choice for the color flow of a particular event.

  \myitem{integer} { ICOLUP(1,I) }
        {integer tag for the color flow line passing through the color of
        the particle}
  \myitem{integer} { ICOLUP(2,I) }
        {integer tag for the color flow line passing through the
        anti-color of the particle} \\
        The tags can be viewed as numbering the different color lines in
        the $N_C\rightarrow \infty$ limit.
        The color/anti-color of a particle are defined with respect to
        the physical time order of the process so as to allow a unique
        definition of color flow also through intermediate particles.\\
        This scheme is chosen because it has the fewest ambiguities, and
        when used with the history information, it supports Baryon
        number violation (an example is given below). \\
        To avoid confusion it is recommended that integer tags larger than 
        MAXNUP (i.e.\ 500) are used. The actual value of the tag has no meaning
        beyond distinguishing the lines in a given process.
      
  \hspace{-.2cm} {\bf \underline{ Momentum and Position } }
  \myitem{double} { PUP(5,I) }
        { lab frame momentum $(P_x, P_y, P_z, E, M)$ of particle in GeV } \\
        The mass is the `generated mass' for this particle, 
        $M^2=E^2-|\vec{p}|^2$ (i.e.\ not
        necessarily equal to the on-shell mass). The mass may 
        be negative, which denotes negative $M^2$ (i.e.\
        $M=2$ implies $M^2=4$ whereas $M=-2$ implies $M^2=-4$). \\
        Both $E$ and $M$ are needed for numerical reasons, the user
        should explicitly calculate and provide each one.

  \myitem{double} { VTIMUP(I) }
        { invariant lifetime $c\tau$ (distance from production to decay) in
        mm} \\
        Combined with the directional information from the momentum,
        this is enough to determine vertex locations.  Note that this
        gives the distance of travel for the particle from birth to
        death, in this particular event, and not its distance from the
        origin.

\hspace{-.2cm} {\bf \underline{ Spin / Helicity} }
  \myitem{double} { SPINUP(I) }
        { cosine of the angle between the spin-vector of particle I
          and the 3-momentum of the decaying particle, 
        specified in the lab frame } \\
        This scheme is neither general nor complete, but is chosen as
        the best compromise.
        The main foreseen application is $\tau$'s with a specific
        helicity. Typically a   
        relativistic $\tau^-$ ($\tau^+$) from a $W^-$ ($W^+$) has 
        helicity --1 (+1) (though this might be changed by the boost
        to the lab frame), so SPINUP(I)= --1 (+1). The use of a floating
        point number allows for the extension to the non-relativistic
        case. 
        Unknown or unpolarized particles should be given SPINUP(I)=9. The
        lab frame is the frame in which the four-vectors are
        specified.

\end{myitemize}

%
%
%



\subsection*{Example: hadronic $t \bar{t} $ production }
\parbox{1.4in}{\includegraphics[height=1in]{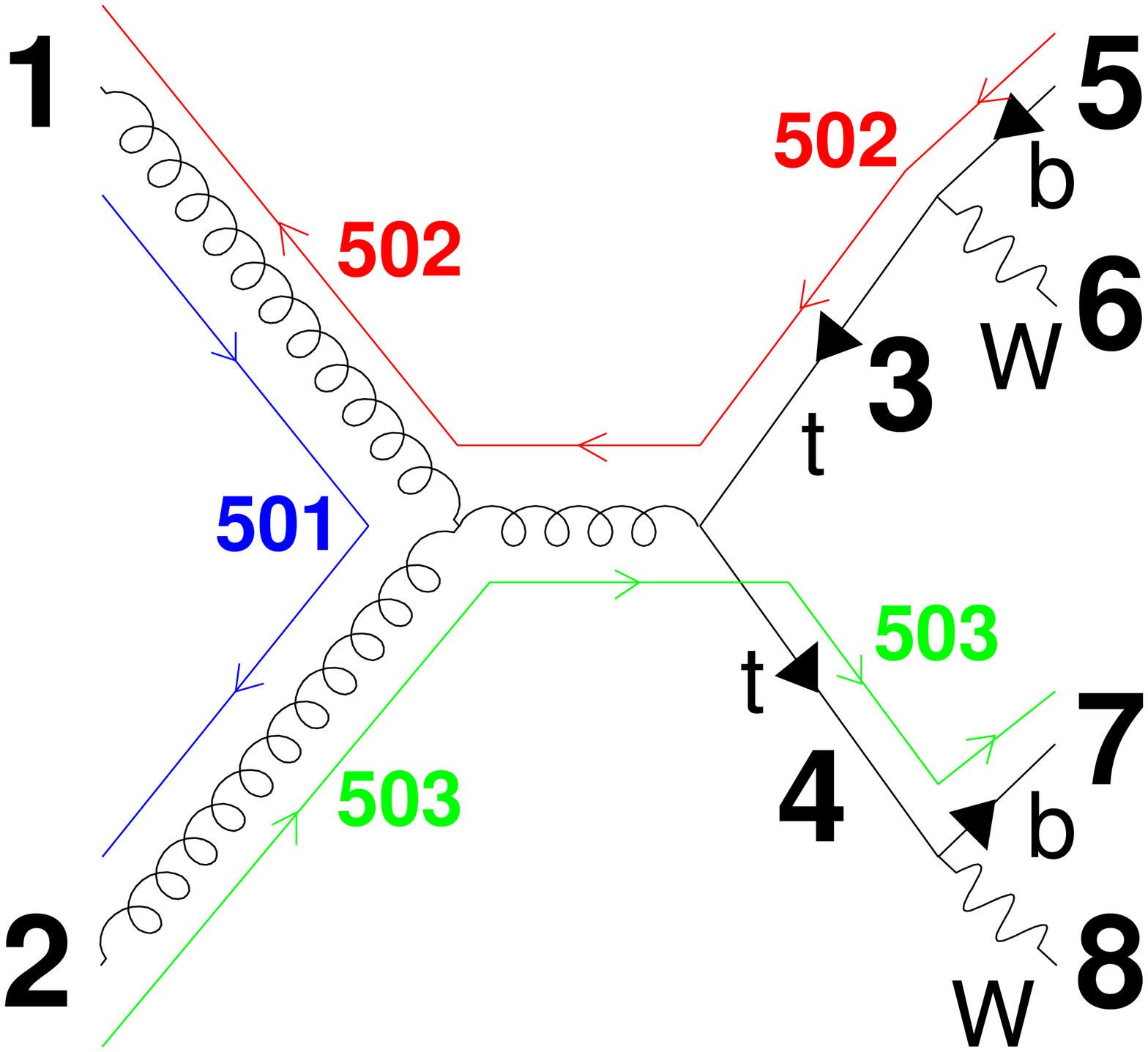}}
\parbox{4.5in}{{\tiny \begin{tabular}{ccccccc}
I & ISTUP(I) & IDUP(I)   & MOTHUP(1,I) & MOTHUP(2,I) 
                                        & ICOLUP(1,I) & ICOLUP(2,I) \\ \hline
1 &--1 &  21~($g$)               & 0   & 0         & 501 & 502     \\
2 &--1 &  21~($g$)               & 0   & 0         & 503 & 501     \\
3 & +2 & --6~($\bar{t}$)         &  1 & 2        &  0  & 502     \\
4 & +2 &  6~($t$)                &  1 & 2        & 503 &  0      \\
5 & +1 & --5~($\bar{b}$)         &  3 & 3        &  0  & 502     \\
6 & +1 & --24~($W^-$)            &  3 & 3        &  0  &  0      \\
7 & +1 &  5~($b$)                &  4 & 4        & 503 &  0      \\
8 & +1 &  24~($W^+$)             &  4 & 4        &  0  &  0      \\
\hline
\end{tabular}}}
%

The $t$ and $\bar{t}$ are given ISTUP=+2, which informs the \SHG\
to preserve their invariant masses when
showering and hadronizing the event. An intermediate s-channel gluon
has been drawn in the diagram, but since this graph cannot be usefully
distinguished from the one with a t-channel top exchange,
an entry has not been included for it in the event record.

The definition of a line as `color' or `anti-color' depends on the
orientation of the graph. This ambiguity is resolved  by defining 
color and anti-color according to the physical time order. 
A quark will always have its color tag ICOLUP(1,I) filled,
but never its anti-color tag ICOLUP(2,I). The reverse is true for an
anti-quark, and a gluon will always have information in both
ICOLUP(1,I) and ICOLUP(2,I) tags.

Note the difference in the treatment by the parton shower of the
above example, and an identical final state, where the intermediate
particles are not specified:
\begin{center}{\tiny 
\begin{tabular}{ccccccc}
I & ISTUP(I) & IDUP(I)   & MOTHUP(1,I) & MOTHUP(2,I) 
                                        & ICOLUP(1,I) & ICOLUP(2,I) \\ \hline
1 &--1        &  21~($g$)       &  0 & 0        & 501 & 502     \\
2 &--1        &  21~($g$)       &  0 & 0        & 503 & 501     \\
3 & +1        & --5~($\bar{b}$) &  1 & 2        &  0  & 502     \\
4 & +1        & --24~($W^-$)    &  1 & 2        &  0  &  0      \\
5 & +1        &  5~($b$)        &  1 & 2        & 503 &  0      \\
6 & +1        &  24~($W^+$)     &  1 & 2        &  0  &  0      \\
\hline
\end{tabular}}
\end{center}
In this case the parton shower will evolve the $b,\bar{b}$ without
concern for the invariant mass of any pair of particles. 
Thus the parton shower may alter the invariant mass of the $Wb$
system (which may be undesirable if the $Wb$ was generated from a top
decay).


\subsection*{Example: $gg \rightarrow gg$ }
\parbox{1.4in}{\includegraphics[height=1.2in]{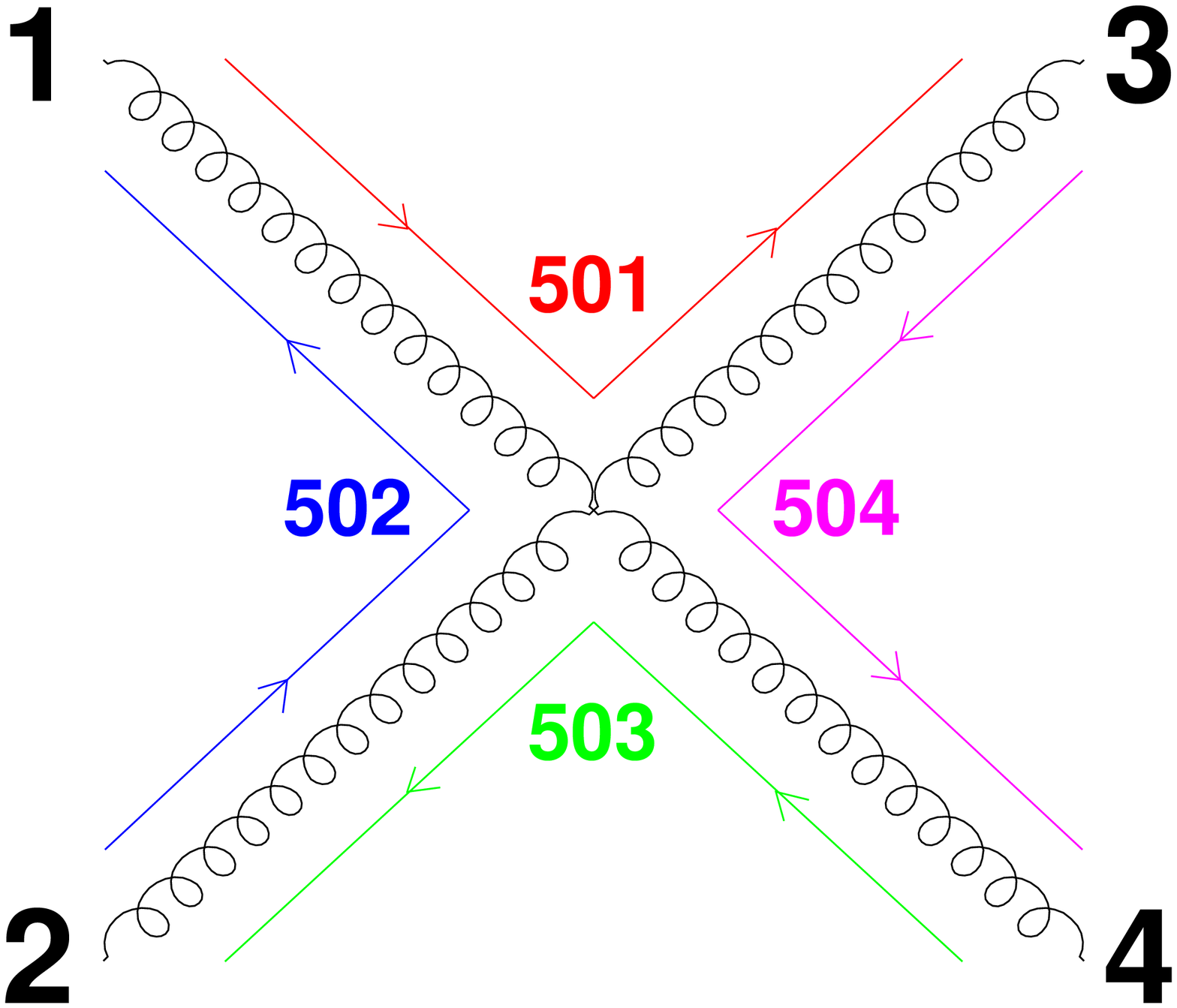}}
\parbox{4.5in}{\tiny \begin{tabular}{ccccccc}
I & ISTUP(I) & IDUP(I) & MOTHUP(1,I) & MOTHUP(2,I) 
                                & ICOLUP(1,I) & ICOLUP(2,I) \\ \hline
1 &--1        &  21~($g$)    &  0 & 0        &   501 & 502   \\
2 &--1        &  21~($g$)    &  0 & 0        &   502 & 503   \\
3 & +1        &  21~($g$)    &  1 & 2        &   501 & 504   \\
4 & +1        &  21~($g$)    &  1 & 2        &   504 & 503   \\
\hline
\end{tabular}}

\subsection*{Example: Baryon number violation in decays}
\parbox{1.4in}{\includegraphics[height=1.2in]{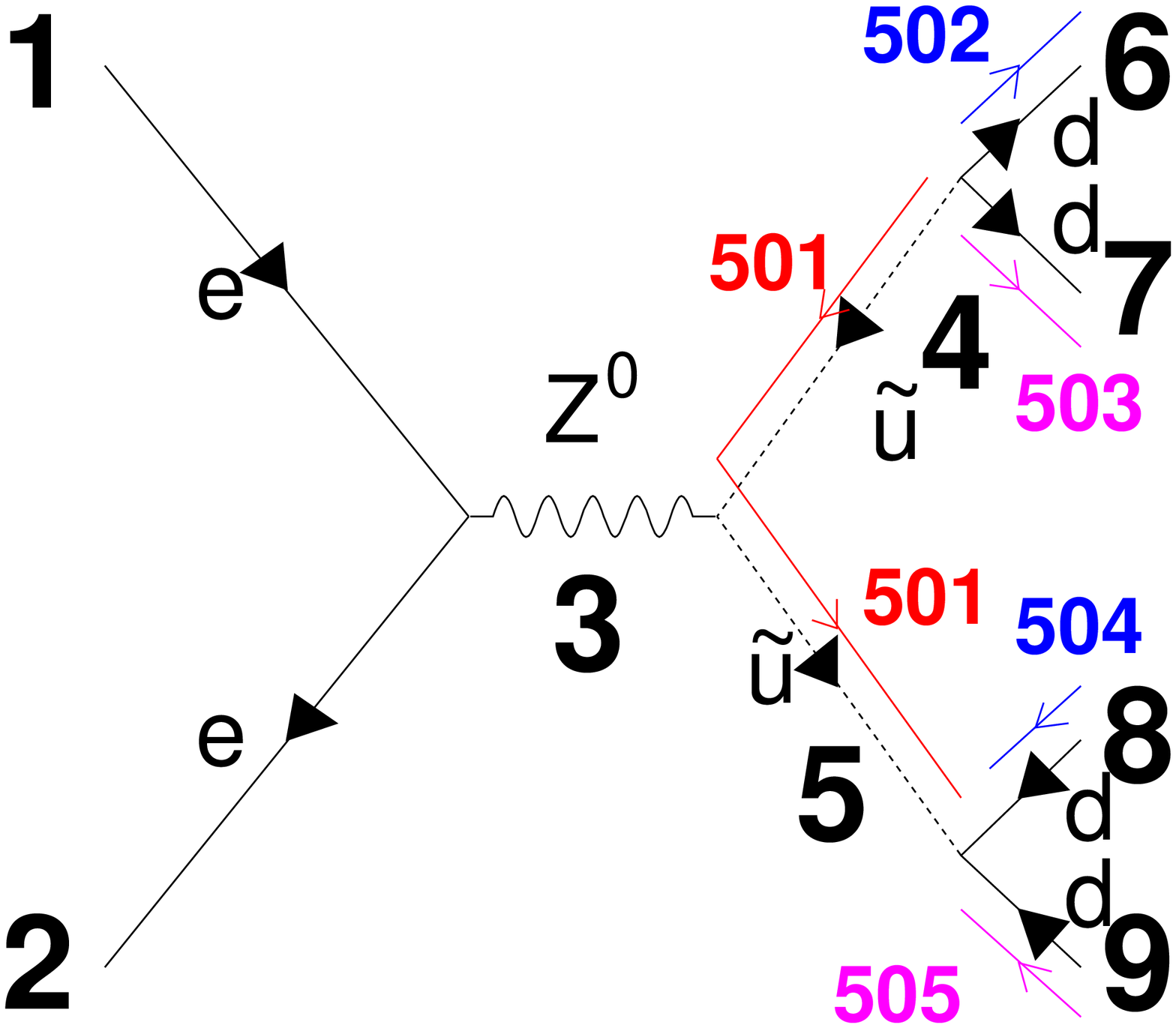}}
\parbox{4.5in}{\tiny \begin{tabular}{ccccccc}
I & ISTUP(I) & IDUP(I)   & MOTHUP(1,I) & MOTHUP(2,I) 
                                        & ICOLUP(1,I) & ICOLUP(2,I) \\ \hline
1 &--1  &  11~($e^-$)                            &  0 & 0    &  0   &   0   \\
2 &--1  & --11~($e^+$)                           &  0 & 0    &  0   &   0   \\
3 & +2  &  23~($Z^0$)                            &  1 & 2    &  0   &   0   \\
4 & +2  &--1000002~($\stackrel{\sim}{\bar{u}}$)  &  3 & 3    &  0   &  501  \\
5 & +2  &  1000002~($\stackrel{\sim}{u}$)        &  3 & 3    & 501  &   0   \\
6 & +1  &  1~($d$)                               &  4 & 4    & 502  &   0   \\
7 & +1  &  1~($d$)                               &  4 & 4    & 503  &   0   \\
8 & +1  & --1~($\bar{d}$)                        &  5 & 5    &  0   &  504  \\
9 & +1  & --1~($\bar{d}$)                       &  5 & 5    &  0   &  505  \\
\hline
\end{tabular}
}%

Three ``dangling'' color lines intersect at the vertex joining the
$\stackrel{\sim}{\bar{q}},q,q'$ 
(and $\stackrel{\sim}{q},\bar{q},\bar{q}'$), which
corresponds to a Baryon number source (sink) of +1 (-1), and will be
recognizable to the \SHG s.

\subsection*{Example: Baryon number violation in production}
\parbox{1.5in}{\includegraphics[height=1.2in]{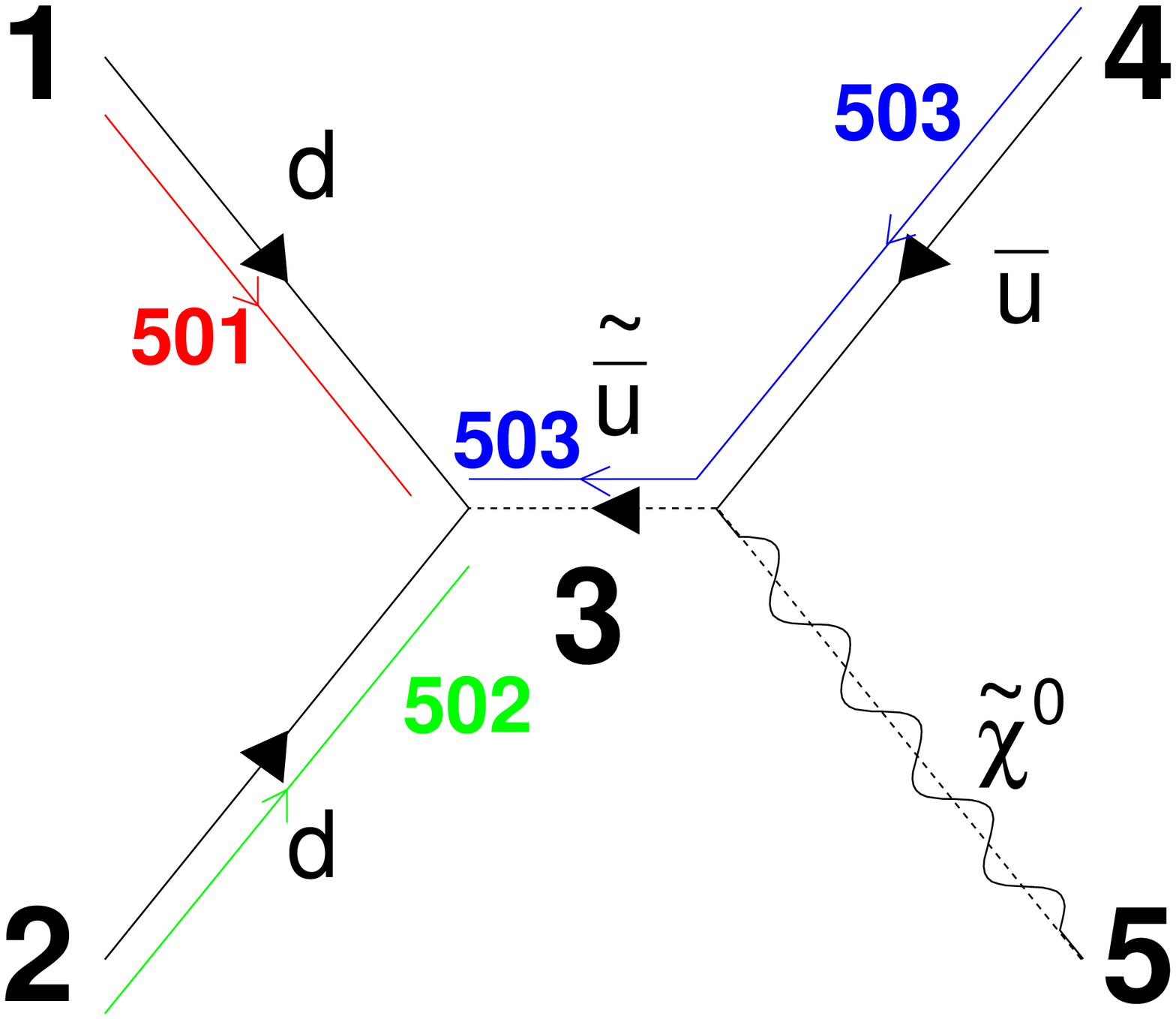}}
\parbox{4.5in}{\tiny \begin{tabular}{ccccccc}
I & ISTUP(I) & IDUP(I)               & MOTHUP(1,I) & MOTHUP(2,I) 
                                        & ICOLUP(1,I) & ICOLUP(2,I) \\ \hline
1 &--1 &  1~($d$)                               & 0 & 0   & 501  & 0        \\
2 &--1 &  1~($d$)                               & 0 & 0   & 502  & 0        \\
3 & +2 & --1000002~($\stackrel{\sim}{\bar{u}}$) & 1 & 2   &  0   & 503      \\
4 & +1 & --2~($\bar{u}$)                        & 3 & 3   &  0   & 503      \\
5 & +1 &  1000022~($\stackrel{\sim}{\chi}^0$)   & 3 & 3   &  0   & 0        \\
\hline
\end{tabular}}

\subsection*{Example: deep inelastic scattering }
\label{example_dis.eps}
\parbox{1.5in}{\includegraphics[height=1.2in]{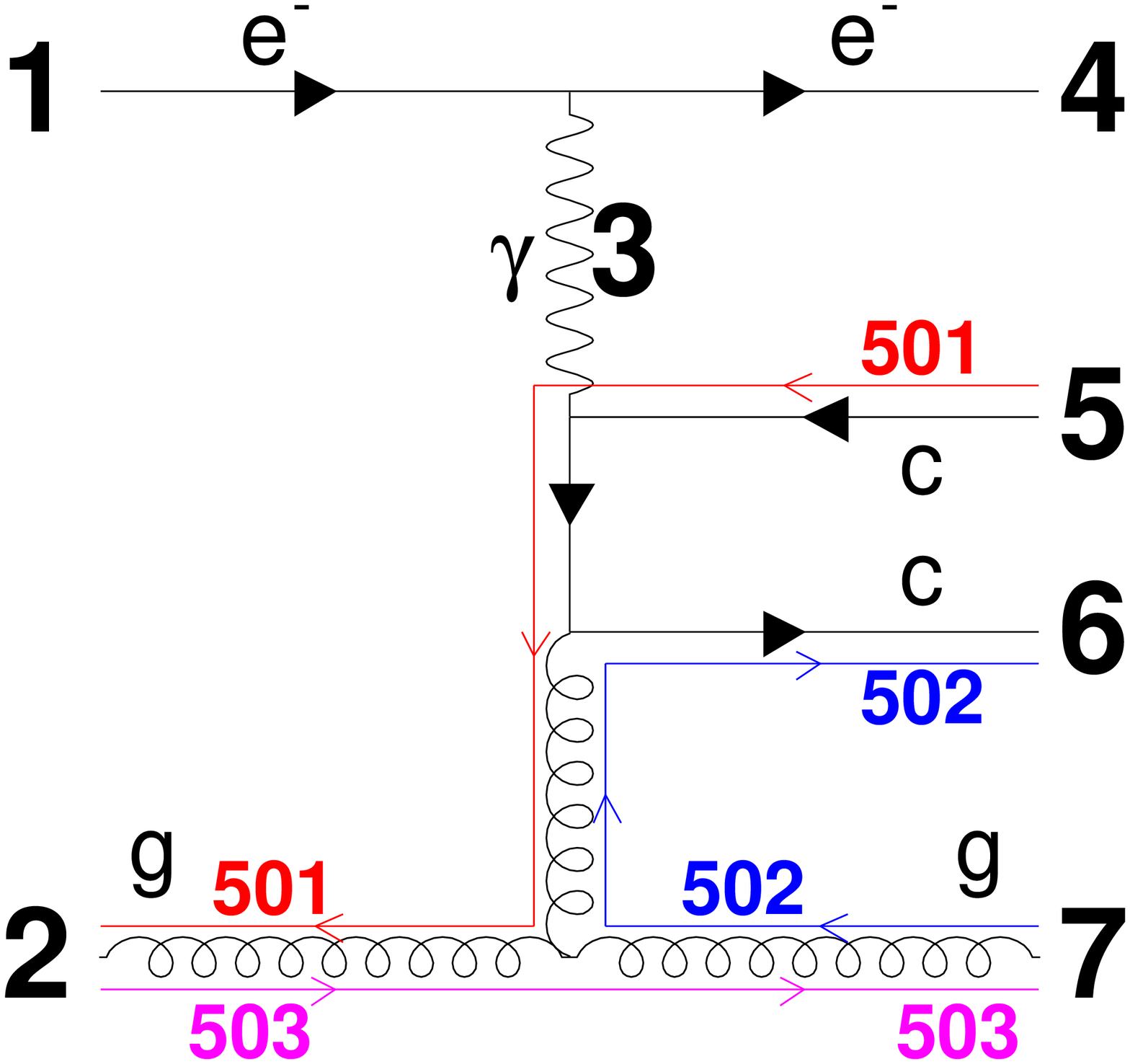}}
\parbox{4.5in}{\tiny \begin{tabular}{ccccccc}
I & ISTUP(I) & IDUP(I) & MOTHUP(1,I) & MOTHUP(2,I) 
                                & ICOLUP(1,I) & ICOLUP(2,I) \\ \hline
1 &--1        &  11~($e^-$)    &  0 &  0       &   0   &  0    \\
2 &--1        &  21~($g$)      &  0 &  0       &  503  & 501   \\
3 &--2        &  22~($\gamma$) &  1 &  0       &   0   &  0    \\
4 & +1        &  11~($e^-$)    &  1 &  0       &   0   &  0    \\
5 & +1        &  -4~($\bar{c}$)&  2 &  3       &   0   & 501   \\
6 & +1        &  4~($c$)       &  2 &  3       &  502  &  0    \\
7 & +1        &  21~($g$)      &  2 &  3       &  503  & 502   \\
\hline
\end{tabular}}

For DIS, the $x$ and $q^2$ of the $\gamma$ should not be altered by
the parton shower, so the $\gamma$ is given ISTUP=--2. We have not
specified the internal quark and gluon lines which will be dressed by
the parton shower, such that the partonic event configuration may be
drawn as follows, \\
\parbox{1.6in}{\includegraphics[height=1.5in]{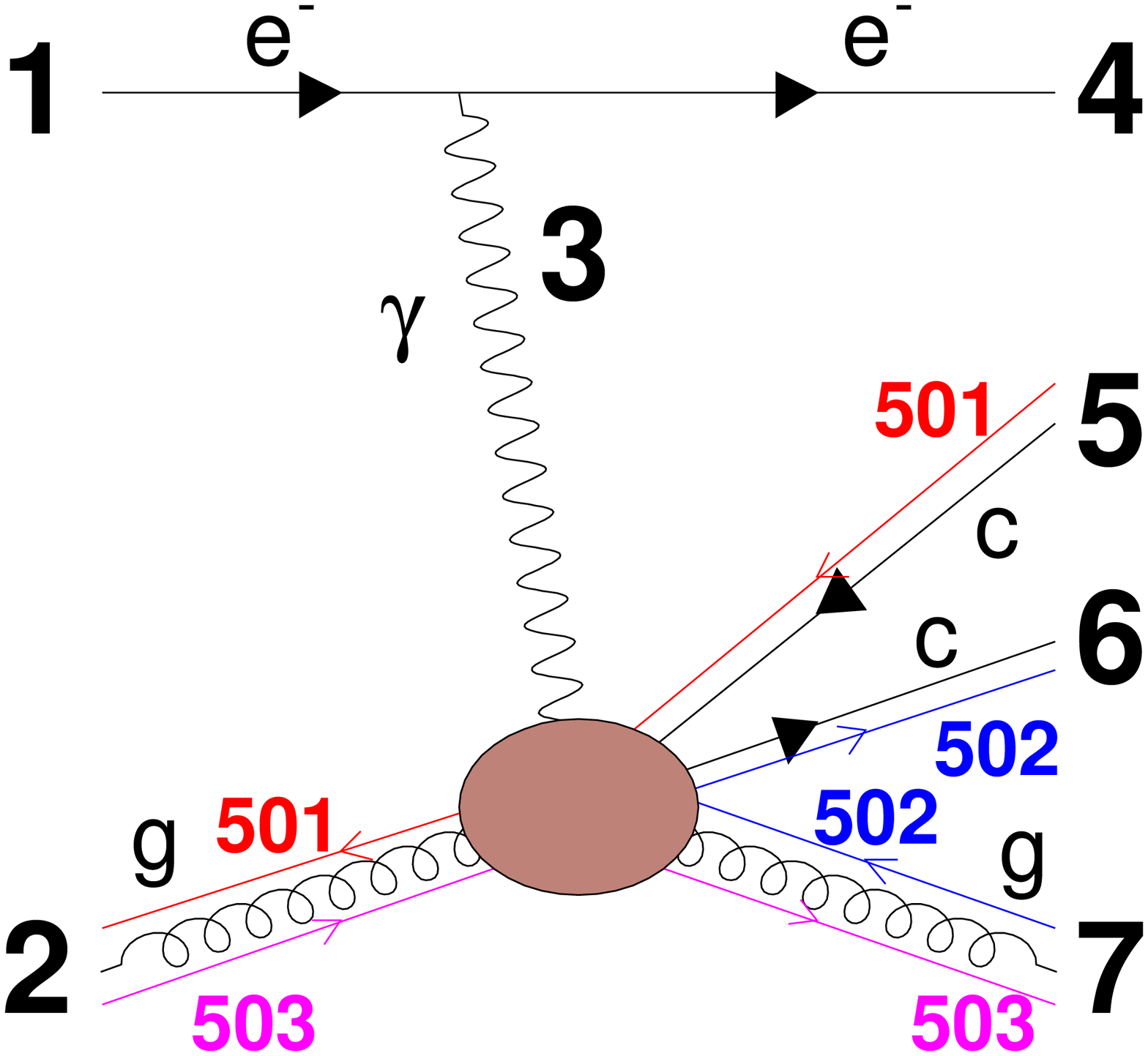}}
\parbox{4.5in}{If information about the quark and gluon propagators is desired (i.e\
for human readability), then those entries may be included with status
code ISTUP=+3.}

\subsection{The Monte Carlo Event Generator AcerMC }
\label{acer}

\unboldmath

Despite the existence of a large repertoire of  processes implemented in
universal generators such as {\tt PYTHIA} or {\tt HERWIG}, a number of  Standard Model
background processes crucial for studying expected physics potential of the LHC
experiments are still missing.  For some of these processes, the matrix element
expressions are rather lengthy and/or complex, and to achieve a reasonable generation
efficiency, it is necessary to tailor the phase-space selection procedure to the dynamics
of the process.  The practical solution could  therefore be to produce a choice of dedicated
matrix-element-based generators with standardized interfaces (such as the one proposed in
section \ref{interface}) to the more general Monte Carlo programs such as {\tt HERWIG} or {\tt PYTHIA},  which are then used to complete the event generation.

The {\bf AcerMC} Monte Carlo Event Generator \cite{Kersevan:2002dd} follows up on this idea.
It is dedicated to the simulation of  Standard Model background processes in 
LHC collisions. The program itself provides a library of the massive matrix elements 
and phase space modules for the generation of a few selected $2 \to 4$ processes.
The hard process event, generated with these modules, can be completed by the addition of 
initial and final state radiation, hadronization and decays, simulated with either 
{\tt PYTHIA~6.2} \cite{Sjostrand:2001yu} or {\tt HERWIG 6.3} \cite{Corcella:2001bw} 

Interfaces of {\tt AcerMC 1.0} to both  {\tt PYTHIA~6.2} and {\tt HERWIG 6.3} are prepared following the standard proposed in Section
\ref{interface},
and are provided in the distribution version. The {\tt AcerMC 1.0} also uses
several other external libraries: {\tt CERNLIB}, {\tt HELAS}, {\tt VEGAS}.
The matrix element codes have been derived with the help of the {\tt MADGRAPH/HELAS} package.
The  typical  efficiency achieved for the generation of unweighted events 
is up to {\bf 30\%}, rather impressively high given the complicated topology of the 
implemented processes.

The very first version of this library was  interfaced to {\tt PYTHIA 6.1 } within the
standard of the so called  {\it external processes} in {\tt PYTHIA}
\cite{Kersevan:2001ab, Kersevan:2001cd, Kersevan:2001ef}.
Since then, after upgrading to the {\tt AcerMC 1.0} standard, the efficiencies have
significantly improved due to an additional optimization step in the phase space
generation.  Also in the new version, the interface standard has been changed from {\tt PYTHIA 6.1 } to {\tt
PYTHIA~6.2 } conventions, an interface to the {\tt HERWIG~6.3} generator was introduced, and the
native {\tt AcerMC 1.0} calculations of the $\alpha_{QED}$ and $\alpha_{QCD}$ couplings
were coded to allow for consistent benchmarking between results obtained with {\tt
PYTHIA~6.2 } and {\tt HERWIG~6.3}.  We also added the $q \bar q \to W(\to \ell \nu) t \bar t$ and
$gg, q \bar q \to Z/\gamma^*(\to \ell \ell, \nu \nu, b \bar b) t \bar t$ and electro-weak 
$gg \to (Z/W/\gamma \to) b \bar b t \bar t$  processes, which have been
implemented for the first time in the {\tt AcerMC 1.0} library presented here.

It is not necessarily the case that the the lowest order
 matrix element calculations  
for a given topology represents the total expected background of a given type. This is
particularly true concerning the heavy flavour content of an event.  The heavy flavour in a
given event might  occur in the hard process of a much simpler topology, as the effect
of including higher order QCD corrections via the shower mechanism. This is  the case, for example, 
for the presence of b-quarks  in  inclusive Z-boson or W-boson production. 
$Wb \bar b$ or $Zb \bar b$ final states can be calculated through higher order radiative corrections to inclusive W and Z production (through parton showering), or with the use of explicit $Wb \bar b$ and $Zb \bar b$ matrix elements. 
The matrix-element-based calculation
itself is a very good reference point to compare with parton shower approaches using
different fragmentation/hadronization models.  It also helps to  study matching
procedures between calculations at a fixed $\alpha_{QCD}$ order and parton shower
approaches. For exclusively hard topologies matrix-element-based calculations
usually represent a much better approximation than the parton shower ones.

The physics processes implemented in the {\tt AcerMC 1.0} library represent a set of important
Standard Model background processes.  These processes are all  key backgrounds for
discoveries in  channels characterized by the presence of  heavy flavour jets and/or
multiple isolated leptons \cite{ATLAS:1999ab}.  The  Higgs boson searches,  $t \bar t
H$, $ZH, WH$ (with $H \to b \bar b$),  $gg \to H$ with ($H\to ZZ^* \to 4 \ell$), and  $b\bar
b h/H/A$ (with $h/H/A \to \mu \mu $) are the most obvious examples of such channels.  We will
briefly discuss the physics motivations for the  processes of interest and the 
implementations that are available:

\boldmath {\bf $Z/\gamma^*(\to \ell \ell)  b \bar b$ production} \unboldmath 
has, over the last several years,  has been recognized as one of the most substantial backgrounds for  several Standard Model (SM) and Minimal Supersymmetric
Standard Model (MSSM) Higgs boson decay modes, as well as for the observability of  SUSY particles.  There is a rather wide spectrum of {\it regions of interest} for this
background.  In all cases, the leptonic $Z/\gamma^*$ decay is relevant; events with
di-lepton invariant mass below, at, or  above  the Z-boson mass could be of interest.  This process enters an analysis
either by the accompanying b-quarks being tagged as b-jets, or by the presence of leptons
from the b-quark semi-leptonic decays in these events. 

A good understanding of those backgrounds and the availability of a credible Monte Carlo
generator which allows the study of the  expected acceptances for different final states
topologies, is crucial. Despite a large effort expended at the time of the
Aachen Workshop \cite{Aachen1990}, such well established Monte Carlo generators were missing
for several years\footnote{
The matrix element for the $gg \to Z b \bar b \to b \bar b \ell \ell$ production has been
published in \cite{Kleiss1990} and, at the  time of the Aachen Workshop, implemented into the
{\tt EUROJET} Monte Carlo \cite{vanEijk:1994ab}.  Since that generator did not allow for the
possibility for having a fully generated hadronic event, with initial and final state
radiation and hadronization, it was
interfaced to {\tt PYTHIA 5.6} \cite{Sjostrand:1994yb} Monte Carlo for the analyses presented in \cite{DellaNegra1990}.  This program, however, is no longer
supported.
The same matrix element has been directly implemented into {\tt PYTHIA 5.7}
\cite{Sjostrand:1994yb}. However, with this implementation the algorithm for the phase space
generation has never worked credibly and thus it was removed from  
{\tt PYTHIA 6.1} \cite{Sjostrand:2001yu}.}.
Recently, the massless matrix elements for the $gg, q \bar q \to Z b \bar b$ processes have been
implemented in the general purpose Monte Carlo program {\tt MCFM} \cite{Campbell:2000bg}.  In
that implementation radiative corrections to this process have also been addressed. 
The massive matrix elements, with an interface to {\tt PYTHIA 6.1}, became 
available in \cite{Kersevan:2001cd}. The {\bf AcerMC} library discussed here
includes an even more efficient implementation of the algorithm presented
in \cite{Kersevan:2001cd}.
The same process is also implemented in the very recent  
version of {\tt HERWIG (6.3)} \cite{Corcella:2001bw}.

\boldmath {\bf $Z/\gamma^*(\to \ell \ell, \nu \nu, b \bar b)  t \bar t$ production}
 \unboldmath at the LHC is an irreducible background to a Higgs search in  association with a 
top-quark pair \cite{Gunion:1994ab}. With the  $Z/\gamma^*(\to b \bar b)$ decay, it is also
an irreducible resonant background to a Higgs search in the $t \bar t H$  production channel
in which the Higgs boson decays to a b-quark pair \cite{Richterwas:1998ab}. 

\boldmath {\bf $W (\to \ell \nu) b \bar b$ production} \unboldmath at the LHC is
recognized as a substantial irreducible 
background for both   Standard Model (SM) and Minimal Supersymmetric
Standard Model (MSSM) Higgs boson searches,  in the associated production mode
$WH$, followed by the decay $H \to b \bar b$.
The massive matrix element for the $q \bar q \to W g^*(\to b \bar b)$ process
 has been calculated \cite{Mangano:1993kq} and interfaced with the {\tt HERWIG 5.6} Monte Carlo \cite{Marchesini:1988cf,Knowles:1988vs,Catani:1991rr} several years ago.
A more recent implementation of the $Wbb$ + multi-jet final states 
is available from \cite{Mangano:2002ab}. Recently, the massless matrix element
has been implemented in the general purpose Monte Carlo program
{\tt MCFM} \cite{Campbell:2000bg}, where the radiative corrections
to this process are also addressed. Yet another implementation of the
$q \bar q \to W ( \to \ell \nu)g^* (\to b \bar b)$ massive matrix elements,
with an interface to {\tt PYTHIA 6.1} became available in \cite{Kersevan:2001ab}.
The {\bf AcerMC} library discussed here  includes a more efficient implementation of
the algorithm presented in \cite{Kersevan:2001ab}. 

\boldmath {\bf $W (\to \ell \nu) t \bar t$ production } \unboldmath 
at the LHC, has,  to our knowledge,  not been implemented
in any publicly available code so far. It is of interest
\footnote{ We thank M. A. Mangano for bringing
this process to our attention and for providing benchmark numbers for verifying 
the total cross-section} because it contributes 
an overwhelming background \cite{Mazzucato:2002ab} to the measurement of the 
Standard Model Higgs  self-couplings at LHC in the most promising 
channel $pp \to HH \to WWWW$.

\boldmath {\bf $t \bar t b \bar b$ production  } \unboldmath
at  the LHC is a dominant irreducible background for both Standard Model (SM) and Minimal
Super-symmetric Standard Model (MSSM) Higgs boson searches in  associated production, $ t
\bar t H$, followed by the decay $H \to b \bar b$.  The potential for the observability of
this channel has been carefully studied and documented in \cite{ATLAS:1999ab} and
\cite{Richterwas:1998ab}.  The proposed analysis requires identifying four b-jets,
reconstruction of both top-quarks in the hadronic and leptonic modes and the visibility of a
peak in the invariant mass distribution of the remaining b-jets.  The irreducible $t \bar
t b \bar b$ background contributes about 60-70\% of the total background from  $t \bar
t$ events ($t \bar t b \bar b$, $t \bar t b j$, $t \bar t j j$).
In the {\bf AcerMC} library, we have implemented both QCD and EW processes leading to the
$t \bar t b \bar b$ final state, namely $gg, q \bar q \to t \bar t b \bar b$ and
$gg \to (Z/W/\gamma \to) t \bar t b \bar b$. The contribution from  EW processes, 
never studied thus far, is surprisingly important in the mass range of the b-quark system 
around 120~GeV, see \cite{Kersevan:2002dd}.
It would seem that the analyses documented in \cite{ATLAS:1999ab}
and \cite{Richterwas:1998ab} might need revisiting.

These complete the list of the native {\bf AcerMC 1.0} processes implemented so far (see
Table~\ref{AcerMC:Tab1}).  Having all these different production processes implemented in
a consistent framework, which can also be directly used  for generating standard {\tt
PYTHIA} or {\tt HERWIG} processes, represents a very convenient environment for several
phenomenological studies related to LHC physics.

\begin{Tabhere}
\vspace{-0.4cm}
\newcommand{\lstrut}{{$\strut\atop\strut$}}
  \caption {\em Matrix-element-based processes implemented  in the{\bf AcerMC} library.
}\label{AcerMC:Tab1}
\vspace{2mm} 
\begin{center}
\begin{tabular}{|c|c|c|} \hline \hline
Process id & Process specification & Efficiency for generation \\ 
           &                       &  of  unweighted events \\ 
\hline\hline
 1      &  $g + g  \to   t \bar t b \bar b $   &  20.2 \%  \\
\hline
 2      &  $q + \bar q  \to   t \bar t b \bar b $   &  26.3 \%  \\
\hline
 3      &  $q + \bar q  \to W(\to \ell \nu)  b \bar b $   &  33.0 \%  \\
\hline
 4      &  $q + \bar q  \to W(\to \ell \nu)  t \bar t $   &  21.0 \%  \\
\hline
 5      &  $g + \bar g  \to Z(\to \ell \ell)  b \bar b $   &  39.0 \%  \\
\hline
 6      &  $q + \bar q  \to Z(\to \ell \ell)  b \bar b $   &   31.7 \% \\
\hline
 7      &  $g +  g  \to Z(\to \ell \ell, \nu \nu, b \bar b)  t \bar t $   &   28.2 \%  \\
\hline
 8      &  $q + \bar q  \to Z(\to \ell \ell, \nu \nu, b \bar b)  t \bar t $   &   34.6 \%  \\
\hline
 9      &  $g +  g  \to (Z/W/\gamma \to) b \bar b  t \bar t $   &   11.2 \%  \\
\hline \hline
\end{tabular}
\end{center}
\end{Tabhere}

\boldmath
\subsubsection{Monte Carlo algorithm} 
\unboldmath

{\bf AcerMC 1.0} produces unweighted events with colour flow information using the
{\tt MADGRAPH/HELAS} \cite{Stelzer:1994ta} package, {\tt PDFLIB} \cite{Plothow-Besch:1993qj} and either the native
or {\tt PYTHIA 6.2}/{\tt HERWIG 6.3} coded running couplings $\alpha_s$ and $\alpha_{\rm
QED}$ (user's choice), for matrix element calculation and native (multi-channel based)
phase space generation procedures.  The generated events are then passed to either the {\tt
PYTHIA 6.2} or {\tt HERWIG 6.3} event generators, where the fragmentation and
hadronization procedures, as well as the initial and final state radiation, are added and
final unweighted events are produced.
\vspace{0.2cm}

\boldmath
\noindent{\bf The Matrix Element Calculation}
\unboldmath
\vspace{0.2cm}

The {\sf FORTRAN}-coded squared matrix elements of the processes were obtained by using the {\tt
MADGRAPH/HELAS} \cite{Stelzer:1994ta} package, taking properly into account the masses and
helicity contributions of the participating particles. The particle masses, charges and
coupling values that were passed to {\tt MADGRAPH} were taken from the interfaced libraries
({\tt PYTHIA/HERWIG}) to preserve the internal consistency of the event generation
procedure. In addition, the (constant) coupling values of $\alpha_s$ and $\alpha_{\rm
QED}$ were replaced with the appropriate running functions that were either taken from the
interfaced generators or provided by the {\bf AcerMC} code according to user settings.
In addition, a slightly modified {\tt MADGRAPH/HELAS} code was used for obtaining the colour
flow information of the implemented processes.
\vspace{0.2cm}

\boldmath
\noindent{\bf The Four Fermion Phase Space Generation}
\unboldmath
\vspace{0.2cm}

The four-fermion phase space corresponding to the processes discussed was modeled
using the importance sampling technique based on the procedures implemented in
the $e^+e^-$ event generators {\tt FERMISV} \cite{Hilgart:1993xu}, {\tt EXCALIBUR}
\cite{Berends:1995xn} and {\tt NEXTCALIBUR} \cite{Berends:2000fj}. For each implemented
process, a sequence of different kinematic diagrams ({\it channels}) modeling
the expected event topologies was constructed and the relative weights between the
contributions of each sampling channel was subsequently obtained by using a
multi-channel self-optimising approach \cite{Kleiss:1994qy}. Eventually, additional
smoothing of the phase space was obtained by using a modified {\tt VEGAS}
routine to improve the generation efficiency. 

The procedure of multi-channel importance sampling used in the event generation can
briefly be outlined as follows. An analytically integrable function
$g(\vec{\Phi})$, which aims to approximate the peaking behaviour of the differential
cross-section dependence on various kinematic quantities, is introduced into the
differential cross-section equation as:
\begin{equation}
d\sigma = s(\vec{\Phi})\, d\vec{\Phi} = \frac{s(\vec{\Phi})}{g(\vec{\Phi})} \cdot
g(\vec{\Phi})\, d\vec{\Phi} = w(\vec{\Phi})\,g(\vec{\Phi})\, d\vec{\Phi},
\end{equation}
where  $d\vec{\Phi}$ denotes the (four-)particle phase space and 
$s(\vec{\Phi})$ summarizes the matrix element, flux and structure functions,
all of which depend on the chosen phase space point.  The function $g(\vec{\Phi})$
is required to be unitary, i.e. a normalized probability density:
\begin{equation}
\int g(\vec{\Phi}) d\vec{\Phi} = 1.
\end{equation}
Since the peaking behaviour of $s(\vec{\Phi})$ can be very complex, due to the several possible
topologies introduced by a large number of contributing Feynman diagrams, 
the function $g(\vec{\Phi})$
is composed of a weighted sum of several channels $g_i(\vec{\Phi})$, each of which is adapted to a
certain event topology:
\begin{equation}
\int g(\vec{\Phi})  = \sum_i \alpha_i \cdot g_i(\vec{\Phi}). 
\end{equation}
The values of the relative weights $\alpha_i$ are determined from a multi-channel
self-optimization procedure  in order to minimize the variance of the 
weights, $w(\vec{\Phi})$ \cite{Kleiss:1994qy}. The phase space points are than sampled from
the function $g(\vec{\Phi})$, first by randomly choosing a channel $i$ according to the
relative frequencies $\alpha_i$, and then deriving the required four momenta from the
chosen $g_i(\vec{\Phi})$ using unitary algorithms. 

The modeling of the kinematic channels has heavily relied on the procedures developed in the {\tt
NEXTCALIBUR} program \cite{Berends:2000fj}; nevertheless, many additions and improvements
were made. The detailed description of the implementations of the four-momenta sampling in all
existing kinematic channels is omitted for the sake of brevity; an example of the
extended/added procedures used in {\bf AcerMC}, as given below, should serve as a representative
illustration,  For further details on the applied method and
unitary algorithms the reader is referred to the original papers
(e.g. \cite{Berends:2000fj,Hilgart:1993xu}).  

\noindent {\bf Example: Breit-Wigner Function with s-dependent Width}
\vspace{0.2cm}

In some topologies of  processes involving $W^\pm$ or $Z^0$ bosons, a bias of the matrix
element towards large values in the high $s^*_{W/Z}$ region is evident. This, in turn, means
that a more accurate description of the tail of the $s^*_W$ distribution is needed. Consequently, the
Breit-Wigner sampling function was replaced by:
\begin{equation}
BW_s(s^*_W)=\frac{s^*_W}{(s^*_W - M_W^2)^2 + M_W^2 \Gamma_W^2},
\label{e:bwnew}
\end{equation}
which is proportional to the (more accurate) Breit-Wigner function with an $s^*_W$ dependent
width (W in the above formula denotes either a $W^\pm$ or a $Z^0$ boson). 

After some calculation the whole unitary procedure can thus be listed as follows:
\begin{itemize}
\item Introduce a new variable  $\eta = (s^*_W-M_W^2)/(M_W\;\Gamma_W)$. The integral of
the above function thus gives:
\begin{equation}
F(\eta)=  \{ \frac{M_W^2}{M_W \Gamma_W} \cdot {\rm atan}(\eta) \}  + 
\{ \frac{1}{2} \{ \log(\eta^2+1) \}, =
F_1(\eta) + F_2(\eta)
\label{e:bwint}
\end{equation}
\item Calculate the kinematic limits $\eta_{\rm min}$ and $\eta_{\rm max}$.
\item Calculate the {\it normalisation} factors $\Delta_1=F_1(\eta_{\rm
max})-F_1(\eta_{\rm min})$, $\Delta_2=F_2(\eta_{\rm max})-F_2(\eta_{\rm min})$
and $\Delta_s = \Delta_1 + \Delta_2$; the term $\Delta_2$ can actually be
negative and thus does not represent a proper normalisation.

\item Obtain a (pseudo-)random number $\rho_1$. 
\item If $\rho_1 \leq \Delta_2/\Delta_s$ then:
\begin{itemize}
\item Obtain a (pseudo-)random number $\rho_2$;
\item Construct $\eta$ as:
\begin{eqnarray*}
X & = & \Delta_2 \cdot \rho_2 + F_2(\eta_{\rm min}),\\
\eta & = & \sqrt(e^{2X}-1),
\end{eqnarray*}
which is the inverse of the (normalized) cumulant $(F_2(\eta)-F_2(\eta_{\rm
min}))/\Delta_2$.
\item Note that the condition $\rho_1 \leq \Delta_2/\Delta_s$ can be fulfilled
only if $\Delta_2 \geq 0$, which means that $\eta_{\rm max}$ is positive and
greater than $\eta_{\rm min}$.
\end{itemize}
\item Conversely, if $\rho_1 > \Delta_2/\Delta_s$ then:
\begin{itemize}
\item Obtain a (pseudo-)random number $\rho_2$;
\item Construct $\eta$ as:
\begin{eqnarray*}
X & = & \Delta_1 \cdot \rho_2 + F_1(\eta_{\rm min}),\\
\eta & = & \tan(\frac{M_W \Gamma_W}{M_W^2}\cdot X)
\end{eqnarray*}
which is the inverse of the (normalized) cumulant $(F_1(\eta)-F_1(\eta_{\rm
min}))/\Delta_1$.
\item If the obtained $\eta$ is less than zero then calculate the normalized
probability densities:
\begin{eqnarray*}
P_1 & = & \frac{1}{\Delta_1} \cdot \{\frac{M_W^2}{M_W \Gamma_W} \cdot
\frac{1}{1+\eta^2} \} \\
P_s & = &  \frac{1}{\Delta_s} \cdot \{\frac{M_W^2}{M_W \Gamma_W} \cdot
\frac{1}{1+\eta^2} + \frac{\eta}{1+\eta^2} \}
\end{eqnarray*}
\item Obtain a (pseudo-)random number $\rho_3$;
\item If $\rho_3 > P_s/P_1$ map $\eta \to -\eta$.
\item If the new $\eta$ falls outside the kinematic limits $[\eta_{\rm
min},\eta_{\rm max}]$ the event is rejected.
\item Note also that the last mapping can only occur if the original $\eta$ was
negative, since $P_s < P_1$ only in the region $\eta < 0$. 
\item Calculate the value of $s^*_W$ using the inverse of $\eta$ definition:
\begin{equation}
s^*_W = (M_W\;\Gamma_W) \cdot \eta + M_W^2
\end{equation}
\end{itemize}
The weight corresponding to the sampled value $\eta$ is exactly:
\begin{equation}
\Delta_s \cdot \frac{(s^*_W - M_W^2)^2 + M_W^2 \Gamma_W^2}{s^*_W},
\end{equation}
which is the (normalized) inverse of Equation \ref{e:bwnew} as requested.
\end{itemize}
\begin{figure}
\vspace{-0.8cm}
\begin{center}
\mbox{
     \includegraphics[width=7cm]{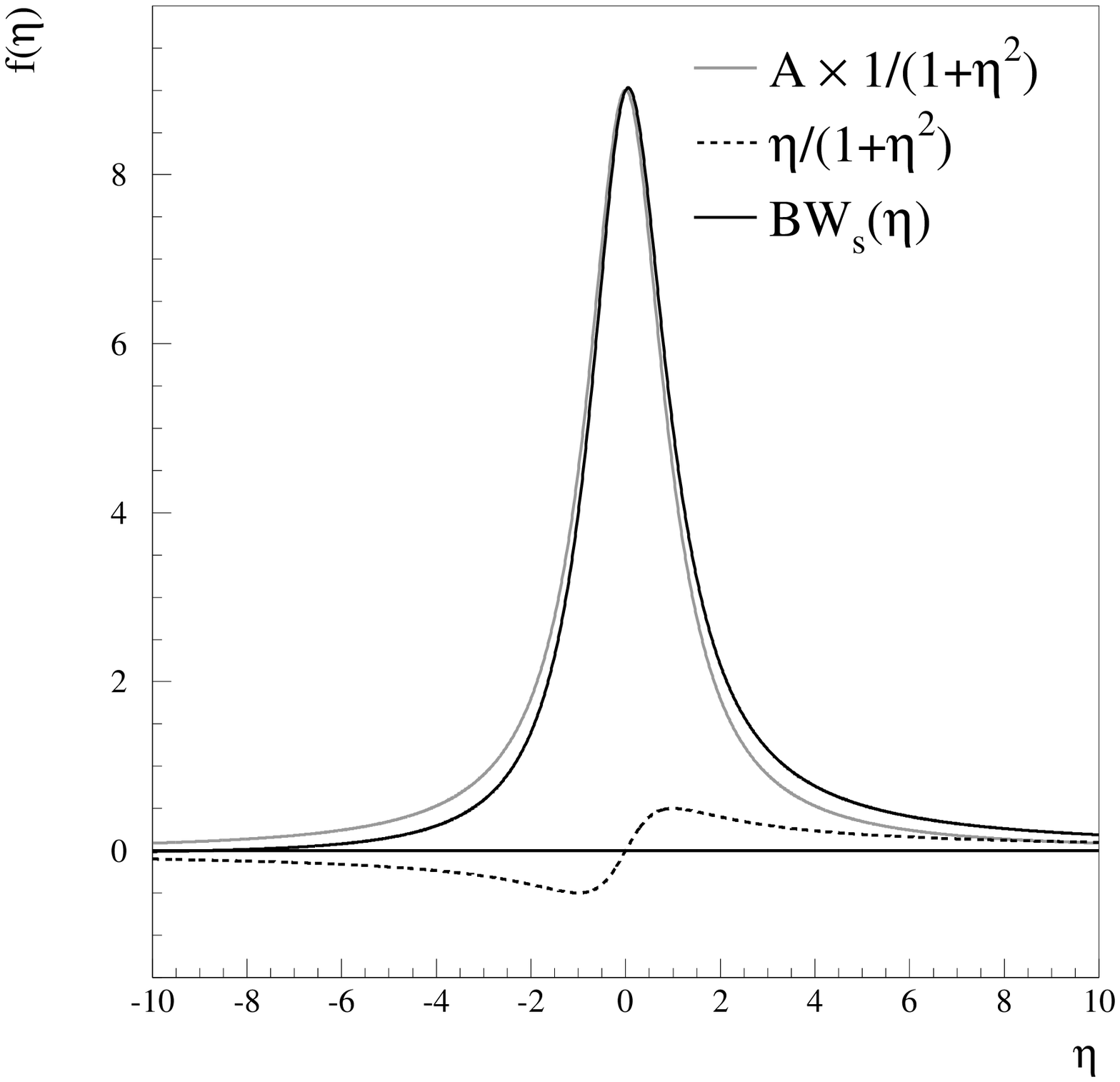} 
     \includegraphics[width=7cm]{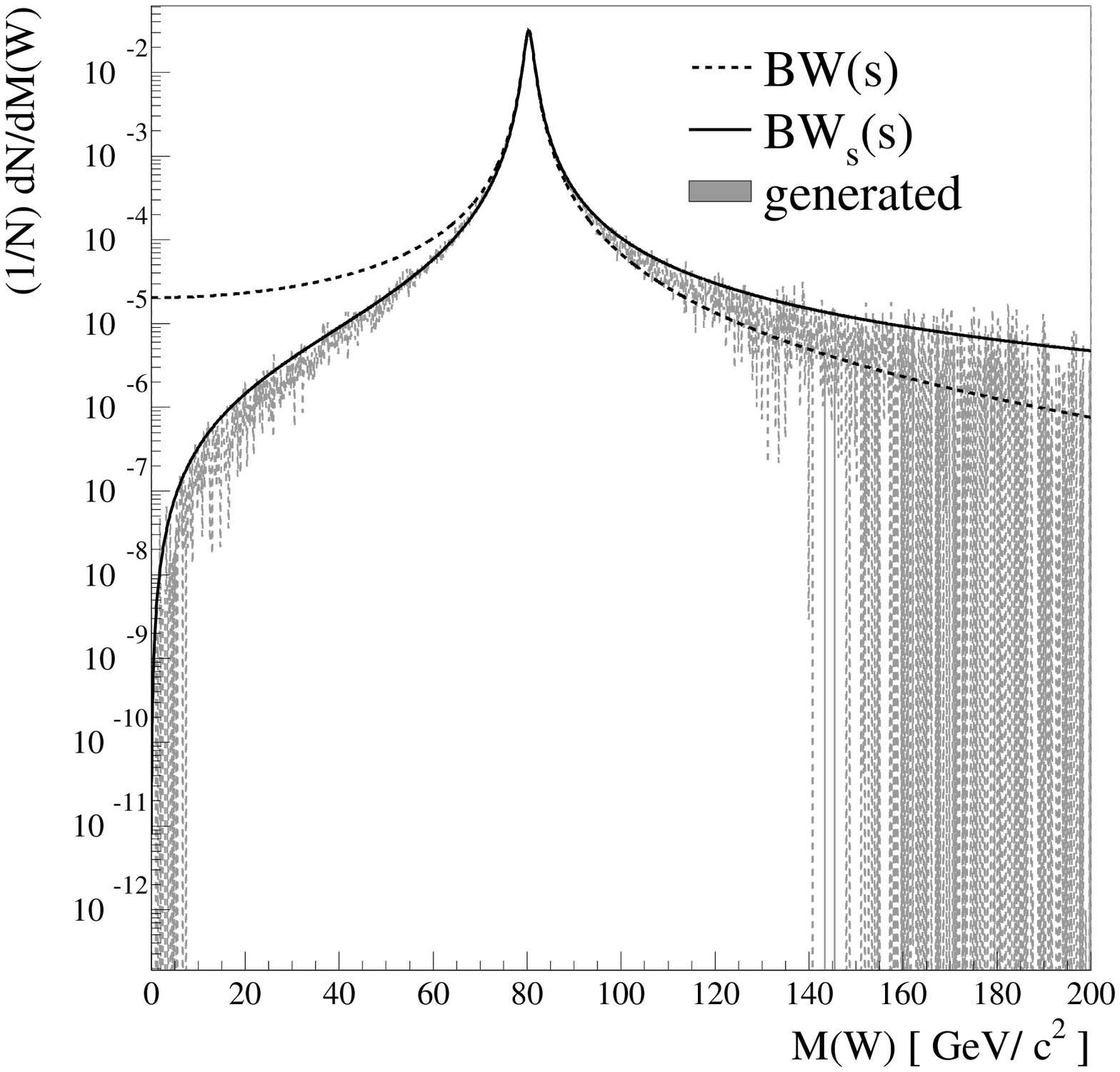} 
}
\end{center}
\vspace{-0.8cm}
\caption{\em {\bf Left} Comparisons of the two functional terms contributing to 
${\rm BW}_s(\eta)$ given by Equation \ref{e:bwnew}. Note that the scaling factor $A$ is
chosen in view of making the contributions more transparent; it is much too small compared
to the real case of $W^\pm/Z^0$ bosons. {\bf Right} Comparison of the (normalized)
distributions of differential cross-section for the process $q \bar q \to W b \bar b$
(dashed) and sampling functions (solid line) with respect to the variables obtained by
importance sampling, as described in the text.\label{f:bwcomp}
}
\vspace{0.2cm}
\end{figure}

Using the above re-sampling procedure, the whole approach remains completely unitary,
i.e. no events are rejected when either there are no limits set on the value of $\eta$ or they are
symmetric, $|\eta_{\rm min}|=\eta_{\rm max}$. In the contrary case, a small fraction of
sampling values is rejected.

As it turns out in subsequent generator level studies, this generation procedure
provides a much better agreement with the  differential distributions than the
{\it usual} (width independent) Breit-Wigner; an example obtained for the  $q \bar q \to W b
\bar b$ process is shown in Figure~\ref{f:bwcomp}. The evident consequence is
that the unweighting efficiency is substantially improved due to the reduction
of the event weights in the high $s^*_W$ region.

\noindent {\bf Modified {\tt VEGAS} Algorithm}
\vspace{0.2cm}

\begin{figure}
\vspace{-1.1cm}
\begin{center}
\mbox{
     \includegraphics[width=7cm]{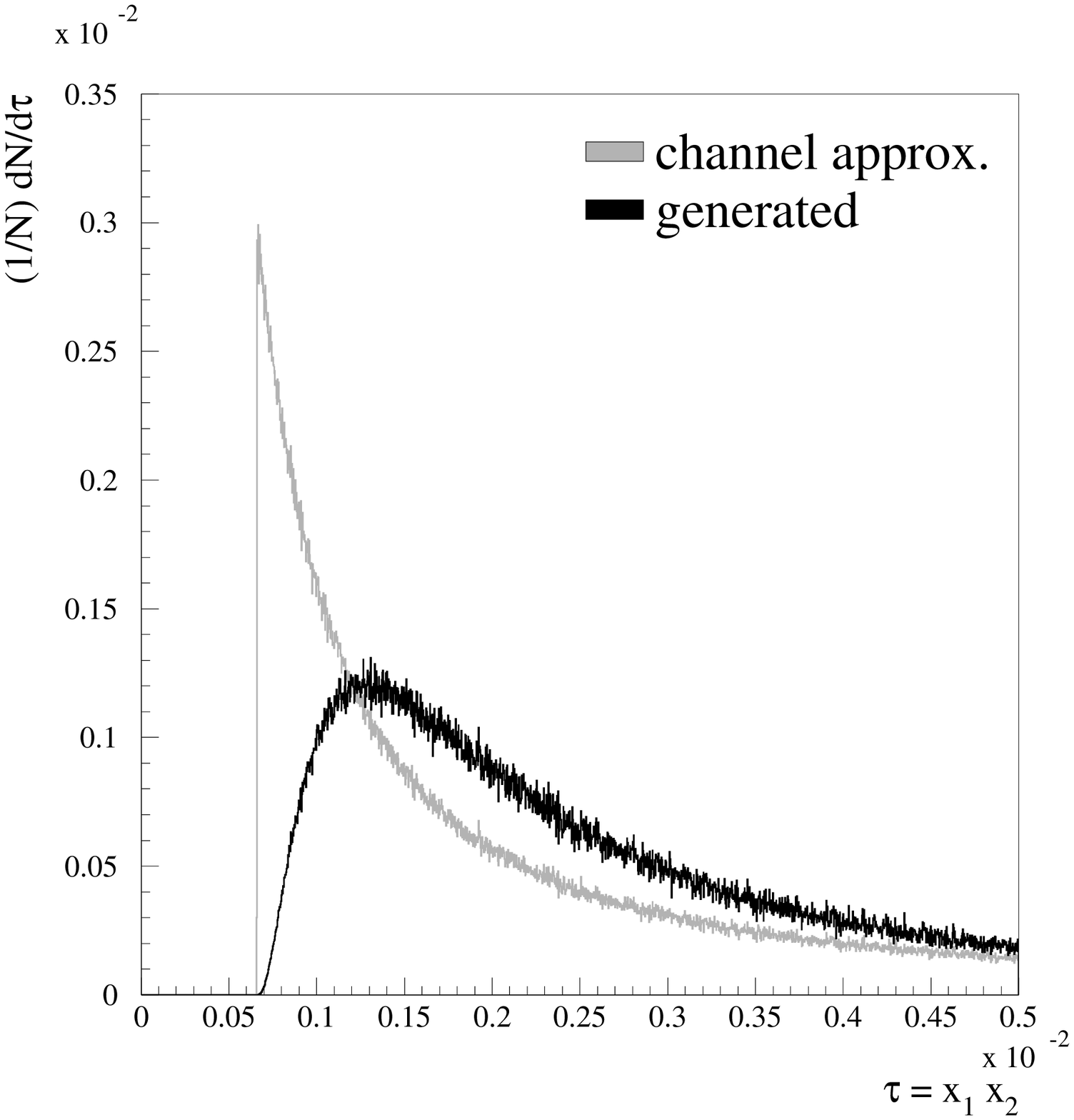}
     \includegraphics[width=7cm]{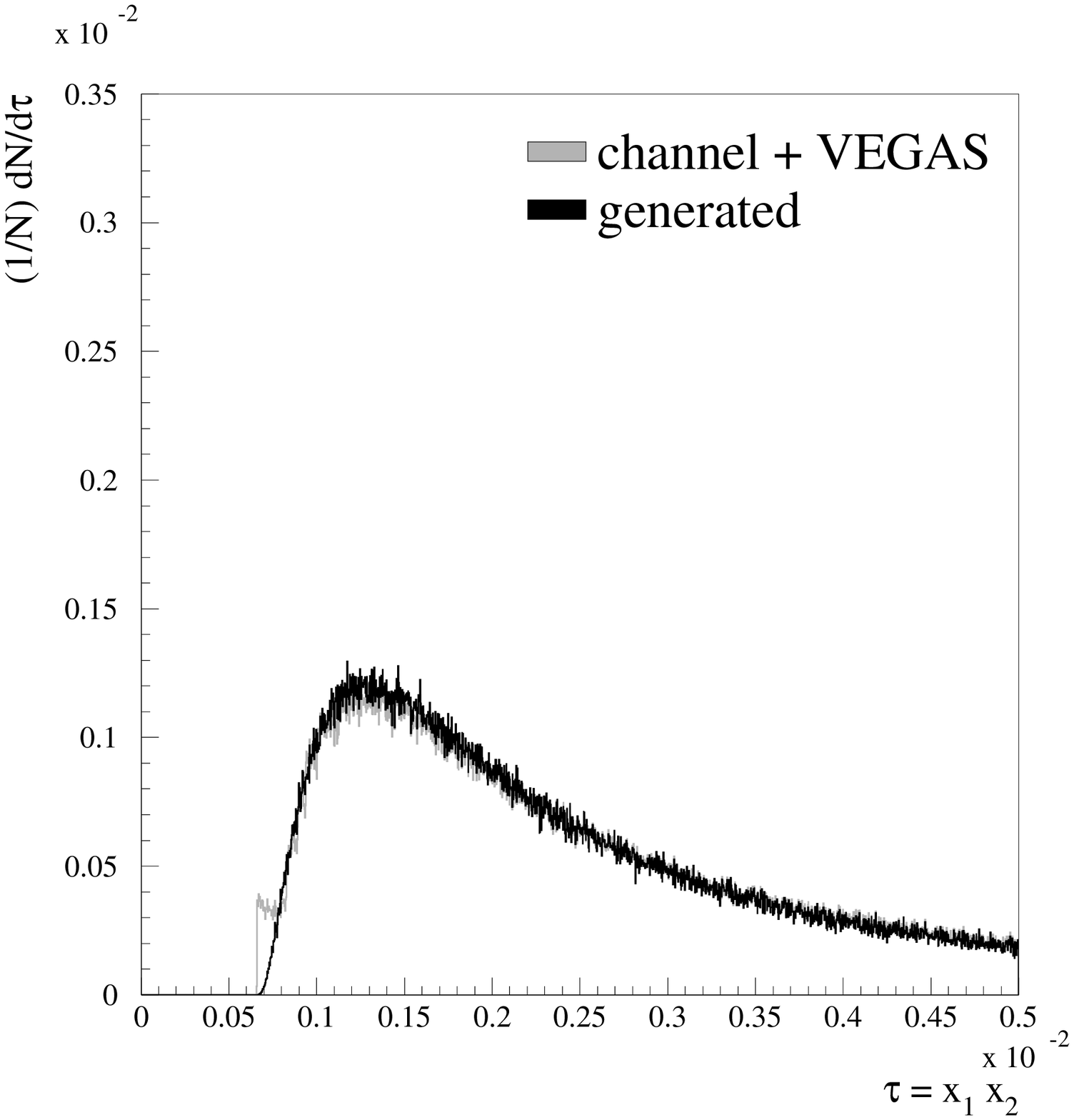}
}
\end{center}
\vspace{-0.4cm}
\caption{\em
Comparison between the sampling distribution for the $\tau = \hat{s}/s \in [\tau_{\rm
min},1]$ variable before and after the application of modified {\tt VEGAS} \cite{Lepage:1978sw}
smoothing procedure (see text).\label{f:tauveg}}
\end{figure}

Using the  multi-channel approach previously described, the total generation (unweighting)
efficiency amounts to about $3-10\%$ depending on the complexity of the chosen process.
 In order to further improve the efficiency, a set of modified {\tt VEGAS}
\cite{Lepage:1978sw} routines was used as a (pseudo-)random number generator for
sampling the peaking quantities in each kinematic channel. After training
all the sampling grids (of dimensions 4-7, depending on the kinematic channel), the
generation efficiency increased to values of above $20\%$. The motivation for this approach
is, that for  unitary algorithms, only a very finite set of simple sampling functions is
available, since the functions have to have simple analytic integrals for which an inverse
function also exists. Consequently, the non-trivial kinematic distributions can not be
adequately described by simple functions  in the whole sampling domain (e.g. the
$\tau$ distribution, c.f. Figure~\ref{f:tauveg}) and some additional smoothing might be
necessary.
\vspace{0.5cm}

\begin{figure}
\begin{center}
\mbox{
     \includegraphics[width=8cm]{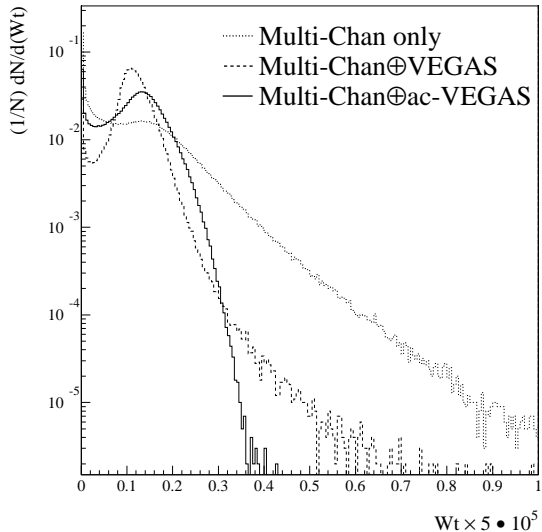}
}
\end{center}
\vspace{-0.5cm}
\caption{\em
The distribution of event weights using  the Multi-Channel approach only (dotted histogram)
 and after application of the {\tt VEGAS} (dashed histogram) and {\tt ac-VEGAS} (full histogram) 
algorithms in the $ g g \to (Z^0 \to) l \bar{l} b \bar{b}$ process.\label{f:weights} 
}
\end{figure}

 In addition, the random number distributions should, due to the applied
importance sampling, have a reasonably flat behaviour to be approached by an adaptive
algorithm such as {\tt VEGAS}\footnote{At this point  a disadvantage of using the
adaptive algorithms of the {\tt VEGAS} type should be stressed, namely that these are
burdened with the need of training them on usually very large samples of events before
committing them to event generation.}. The principal modification of {\tt VEGAS},
besides adapting it to function as a (pseudo-)random number generator instead of
the usual {\it integrator}, was based on the discussions \cite{Ohl:1999qm,Jadach:1999vf} that in the case
of event generation, i.e.the unweighting of events to  weight one, reducing the maximal
value of event weights is in principle of higher importance than achieving the
minimal weight variance. Thus, the learning algorithm was modified accordingly.
By observing the distributions of the event weights before and after the inclusion
of the modified {\tt ac-VEGAS} algorithm (Fig.~\ref{f:weights}), it is evident
that {\tt ac-VEGAS} quite efficiently clusters the weights at lower values.

\boldmath
\noindent{\bf Colour Flow Information}
\unboldmath
\vspace{0.2cm}

Before the generated events are passed to {\tt PYTHIA/HERWIG} for further treatment,
additional information on the colour flow/connection of the event has to be obtained.  To
provide an illustration, the method of the colour flow determination is described for the 
processes $ g g \to t \bar{t} b \bar{b}$.

\begin{figure}
\vspace{-0.6cm}
\begin{center}
\mbox{
      \includegraphics[width=8cm]{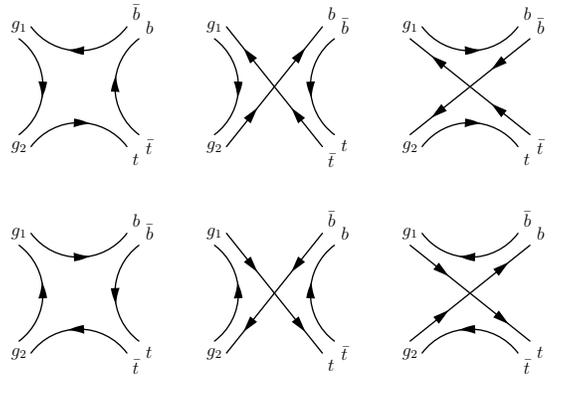}
]}
\end{center}
\vspace{-0.6cm}
\caption{\em 
A diagrammatic representation of the six colour flow configurations in the process
$g g \to t \bar{t} b \bar{b}$.\label{f:cols}
}
\end{figure}

For the process $ g g \to t \bar{t} b \bar{b}$ six colour flow configurations are
possible, as shown in Figure~\ref{f:cols}. With 36 Feynman diagrams contributing to the
process and at least half of them participating in two or more colour flow configurations,
calculations by hand would prove to be very tedious. Consequently, a slightly modified
colour matrix summation procedure from {\tt MADGRAPH} \cite{Stelzer:1994ta} was used to
determine the colour flow combinations of the diagrams and the corresponding colour
factors. The derived squared matrix elements for the separate colour flow combinations
$|{\mathcal{M}_{\rm flow}}|^2$ were used as sampling weights on an event-by-event basis to
decide on a colour flow configuration of the event before it is passsed on to {\tt PYTHIA/HERWIG}
for showering and fragmentation. The procedure was verified to give identical results
for the colour flow combinations and corresponding colour factors when applied to
the processes published in \cite{Bengtsson:1985yx}.

\boldmath
\subsubsection{How to use the package} 
\unboldmath

The {\bf AcerMC 1.0} package consists of the library of the matrix-element-based generators
for the selected processes, interfaces to the {\tt PYTHIA 6.2} and {\tt HERWIG 6.3}
generators, and two main programs: {\tt demo\_hw.f} and {\tt demo\_py.f}. The  makefiles provided
allow the user to build executables with either of these generators:
{\tt demo\_hw.exe} or {\tt demo\_py.exe}.

There are two steering input files: {\tt run.card} and {\tt acermc.card}, which have the
same form for both executables. 
The {\tt run.card} file provides switches for modifying the generated process, the number of events, 
the structure functions, and predefined option for hadronisation/fragmentation, random number, etc..
The  {\tt acermc.card} file provides switches for modifying more specialized settings for
the  {\bf AcerMC 1.0} library itself. Once the user decides on the setup for the generated process,
only the {\tt run.card} very likely needs to be modified for the job submission.

The same executables can be also used for generating standard  {\tt PYTHIA 6.2}/{\tt HERWIG 6.3}
processes. Examples of how to run such jobs are provided as well,
in respectively {\tt demo\_hw.f} and {\tt demo\_py.f}.
If the user requires that the  {\bf AcerMC} library is not used,  $ttH$ production will 
be generated with  {\tt demo\_py.exe}, and the  {\tt HERWIG 6.3} implementation of the $Zbb$ production
will be generated with {\tt demo\_hw.exe}. Only in this case will the {\tt run.card} file be
read, so the user should  implemente her/his steering there or create another xxx.card file,
and add the respective code to the {\tt demo\_xx.f}.

The {\bf AcerMC 1.0} matrix-element-based generators are very highly optimized, using
multi-channel optimization and additional improvement with the {\tt VEGAS} grid.  The
generation modules require three kinds of the input data to perform the generation of
unweighted events: {\bf (1)} A file containing the list of the values of relative channel
weights obtained by the multi-channel optimization.  {\bf (2)} A file containing the
pre-trained {\tt VEGAS} grid.  {\bf (3)} A file containing the maximum weight $wt_{\rm
max}$, $\epsilon$-cutoff maximum weight $wt_{\rm max}^\epsilon$ and the 100 events with
the highest weights. This means that in the case of changing the default running conditions,
such as the structure functions or centre-of-mass energy, in order to recover the initial efficiency for
event generation, the user should repeat the process of preparation of the internal data files
with the inputs for the phase-space generator modules.  Pre-trained data sets obtained
using $\sqrt{s}=14\;$ TeV, {\tt PYTHIA} default $\alpha_s(Q^2)$ and $\alpha_{\rm
QED}(Q^2)$ and the CTEQ5L (parametrized) structure function set are already provided for each
implemented process.

The number of required input files might at first glance seem large, considering
that many event generators do not require any input files for operation; the
difference is not so much in the complexity of the phase space generation as
in the fact that many event generators require a {\it warming run} instead.
That is,  before the generation of unweighted events is performed a certain number of
weighted events (typically of the order of $10^4$) is generated in order to
obtain the relative multi-channel weights (in case multi-channel phase space
generation is used) and/or the optimized {\tt VEGAS} grid and/or an estimate of
the maximal weight. Such an approach can have an advantage when event generation
is very fast and the phase space regions with the highest weights are well known
(as done for the $2 \to 2$ processes in {\tt PYTHIA}); on the other hand, when
the phase space topology of the process is more complex and the event generation
is comparatively slow, generating a relatively small number of e.g. $10^4$
weighted events {\it every time} a generator is started can become CPU wasteful
and/or inaccurate in terms of maximum weight estimation.

 Reasonably accurate
estimation of the latter is mostly crucial for correct event unweighting; event
generators using the {\it  warming-up} method for maximal weight search often find still 
higher weights during the normal run and reset the maximal weight
accordingly. In this case however, the statistically correct approach would be to
reject all events generated beforehand and start the event generation anew.
This is almost never implemented due to the CPU consumption and the possibility
of hitting a weak singularity. With a small pre-sampled set the
generator can, however, badly under-estimate the maximum weight and a large number
of events can be accepted with a too-high probability. The only hope
of obtaining correct results  in such cases is that the weight {\it plateau} will be hit
sufficiently early in the event generation process. Consequently, such an approach
can be very dangerous when generating  small numbers of events\footnote{{\it Small}
is a somewhat relative quantifier, since the size of an representative sample 
should depend on the phase space dimension, i.e. the number of particles in the
final state. For example, with 4 particles in the final state, $10^5$ events can still
be considered  relatively small statistics.}.

In contrast to the {\it warming-up} approach, we have decided that
using separate training runs with large numbers of weighted events to obtain the
optimized grids and maximum weight estimates is preferable. In case the user wants
to produce data sets for non-default settings, this can easily be done
by configuring the switches in the {\tt acermc.card}

\boldmath
\subsubsection{Outlook and conclusions} 
\unboldmath

We have presented here the {\bf AcerMC 1.0} Monte Carlo Event Generator,which is based on a library
of  matrix-element-based generators and an interface to the universal event generator
{\tt PYTHIA~6.2} and {\tt HERWIG 6.3}. The interface is based on the standard proposed
in section \ref{interface}. 

The presented library fulfils following goals:
\begin{itemize}
\item
It provides a possibility to generate a few Standard Model background processes which
were recognised as being very dangerous for searches for  {\it New Physics} at LHC,
and for which generation was either unavailable or not straightforward thus far.
\item
Although the hard process event is generated with a matrix-element-based generator, the provided
interface allows a complete event to be generated  with either {\tt PYTHIA~6.2} or {\tt HERWIG 6.3}
initial and final state radiation, multiple interaction, hadronization, fragmentation 
and decays.
\item
The interface can be also used for studying systematic differences between 
 {\tt PYTHIA~6.2} or {\tt HERWIG~6.3} predictions for the underlying QCD processes.
\end{itemize}

These complete the list of the native {\bf AcerMC} processes implemented so far is:
$q \bar q \to W(\to \ell \nu) b \bar b$,
 $gg, q \bar q \to Z/\gamma^*(\to \ell \ell) b \bar b$, 
QCD $gg, q \bar q \to t \bar t b \bar b$ and EW $gg \to (Z/W/\gamma^* \to) t \bar t b \bar b$,
 $q \bar q \to W(\to \ell \nu) t \bar t$ and
 $gg, q \bar q \to Z/\gamma^*(\to \ell \ell, \nu \nu, b \bar b) t \bar t$.
We plan to extend this crucial list of processes,
gradually in the near future.

\subsection{Comparisons of Higgs Boson Properties with Soft Gluon Emission:
  Analytic and Parton Showering Methods}\label{Huston}

In the near future, experiments at the Tevatron and the LHC will
search for evidence of both the Higgs boson and new phenomena 
that supersede the Standard Model.
Important among the tools that will be used in these searches are
event generators based on parton showering (PS-EG's).  The most
versatile and popular of these are the Monte Carlos {\tt HERWIG}
\cite{Corcella:2001bw,Marchesini:1988cf,Knowles:1988vs,Catani:1991rr}, {\tt ISAJET}  \cite{Baer:1999sp}, and {\tt PYTHIA} 
\cite{Sjostrand:2000wi,Sjostrand:2001yu}.
PS-EG's are useful because they accurately describe the emission of
multiple soft gluons (which is, in effect, an all orders problem in
QCD) and also allow a direct connection with non--perturbative models
of hadronization.  In the parton shower, energy--momentum is conserved
at every step, and realistic predictions can be made for arbitrary
physical quantities.
However, the prediction of the  total cross section is only accurate to leading
order accuracy, and, thus, can demonstrate a sizable dependence on
the choice of scale used for the parton distribution functions (PDF's)
or coupling constants (particularly $\alpha_s$).  Also, in general,
they do not accurately describe kinematic configurations where a hard
parton is emitted at a large angle with respect to other partons.  In
distinction to PS-EG's are certain analytic calculations which account
for multiple soft gluon emission and higher order corrections to the
hard scattering simultaneously using resummation.  
The resummation technique systematically includes towers of logarithms
which are formally of the same order using the renormalization group.
These calculations,
however, integrate over the kinematics of the soft gluons, and, thus,
are limited in their predictive power.  They can, for example,
describe the kinematics of a heavy gauge boson produced in hadron
collision, but cannot predict the number or distribution of jets that
accompany it.  

Much recent work has focused on 
correcting the parton shower predictions to reproduce the
hard emission limit, where the exact leading order
matrix element gives an accurate description,
with work proceeding on 
extending this correction to
next-to-leading order~\cite{Miu:1998ju,Corcella:1999gs,Chen:2001ci,Collins:2000gd}. 
In order to match this precision, it is also
important to verify that the PS-EG programs correctly reproduce
the expected logarithmic structure in their simulation of multiple,
soft gluon emission.
The best approximation we have of the
expected logarithmic structure is represented by resummation calculations.

Since 
Standard Model Higgs boson production is a primary
focus of the physics program
at the Tevatron and  the LHC, and since several PS-EG and analytic resummation
predictions exist, we have chosen $gg\to H$ production as
a benchmark for evaluating the consistency and accuracy of 
these two approaches.
In particular, the transverse momentum of the Higgs boson 
$Q_T^H$ depends primarily
on the details of the soft gluon emission from the initial state
partons. 
The $gg$ initial state is particularly interesting, since the large color
charge may emphasize any differences that might exist in parton
shower/resummation implementations.  
Furthermore, for the $gg$ induced process, the details of non--perturbative
physics (e.g. intrinsic $k_T$) are less important \cite{Balazs:2000sz,Catani:2000zg}.

To this end, we have compared Higgs production using {\tt HERWIG},
{\tt PYTHIA} (several recent versions) and ResBos \cite{Balazs:1997xd} for 
Higgs masses of 125 and 500 GeV at center-of-mass energies of 1.96 TeV 
(${{\rm p}\bar{\rm p}}$), 14 and 40 TeV (${\rm pp}$).  
The two different masses and three
different center of mass energies provide a wide variation of kinematics 
that test the showering/resummation processes.
This work extends our results published in \cite{Djouadi:2000gu,Catani:2000zg}, 
\cite{Balazs:2000wv} and \cite{Balazs:2000sz}.

\subsubsection{Parton Showers}

PS-EG's are based on the factorization theorem \cite{Bodwin:1985ft}, which,
roughly, states that physical observables in any sensible gauge theory
are the product of short--distance functions and long--distance
functions.  The short--distance functions are calculable in
perturbation theory.  The long--distance functions are fit at a scale,
but their evolution to any other scale is also calculable in
perturbation theory.

A standard application of the factorization theorem is to describe
heavy boson production at a hadron collider to a fixed order in
$\alpha_s$.  The production cross section is obtained by convoluting
the partonic subprocesses evaluated at the scale $Q$ with the PDF's
evaluated at $Q$.  The partons involved in the hard collision must be
sufficiently virtual to be resolved inside the proton, and a natural
choice for the scale $Q$ is the mass of the heavy boson \cite{Bengtsson:1986gz}.
Prior to the hard collision, however, the partons are not resolvable
in the proton (i.e. the proton is intact) and have virtualities at a
much lower scale $Q_0$ of the order of 1 GeV.  The connection between
the partons at the low scale $Q_0$ and those at the high scale $Q$ is
described by the DGLAP evolution equations \cite{Gribov:1972rt,Altarelli:1977zs,Dokshitzer:1977sg}.  
The DGLAP
equations include the most important kinematic configurations of the
splittings $a \to b c$, where $a,b$ and $c$ represent different types
of partons in the hadron ($q,g$, etc.).  Starting from a measurement
of the PDF's at a low scale $Q_0$, a solution of the DGLAP equations
yields the PDF's at the hard scale $Q$.  Equivalently, starting with a
parton $c$ involved in a hard collision, it is also possible to
determine probabilistically which splittings generated $c$.  In the
process of de--evolving parton $c$ back to the valence quarks in the
proton, a number of spectator partons (e.g. parton $b$ in the
branching $a\to bc$) are resolved.  These partons constitute a shower
of soft jets that accompany the heavy boson, and influence its
kinematics.

The shower described above occurs with unit probability and does not
change the total cross section for heavy boson production calculated
at the scale $Q$ \cite{Odorico:1980gg}.  The showering can be attached to the
hard--scattering process based on a probability distribution {\it
  after} the hard scattering has been selected.  Once kinematic cuts
are applied, the transverse momentum and rapidity of the heavy boson
populate regions never accessed by the differential partonic cross
section calculated at a fixed order.  This is consistent, since the
fixed--order calculation was inclusive and
was never intended to describe the detailed kinematics of the
heavy boson.  The parton shower, in effect, resolves the structure of
the inclusive state of partons denoted as $X$.  In practice, the fixed
order partonic cross section (without showering) can still be used to
describe properties of the decay leptons as long as the measurable is
not highly correlated with the heavy boson kinematics.

Here, we review parton showering schematically.  More details can be
found, for example, in Ref. \cite{Ellis:1996yz}.  First, for simplicity, consider the case
of final state or forward showering, where the parton virtuality $Q$
evolves forward to the low scale $Q_0$.  The basis for developing a
probabilistic picture of final state showering is the DGLAP equation
for the fragmentation functions: \bea Q {\partial \over \partial Q}
D_a(x,Q) = \int_{x}^{1-\epsilon}{dz \over z}{\alpha_{abc}(z,Q) \over
  \pi}
\hat{P}_{a\to bc}(z) D_b(x/z,Q) \nonumber \\
-D_a(x,Q) \int_{x}^{1-\epsilon}{dz}{\alpha_{abc}(z,Q) \over \pi}
\hat{P}_{a\to bc}(z), \eea where $\hat{P}_{a\to bc}$ is an
unregularized splitting function, $\alpha_{abc}$ is the coupling times
color factor, and $\epsilon$ is a cutoff.  The equation can be
rewritten as \bea {\partial \over \partial \ln Q^2} \left(
  D_a(x,Q)/\Delta(Q) \right) = \int_{x}^{1}{dz \over
  z}{\alpha_{abc}(z,Q) \over 2\pi} \hat{P}_{a\to bc}(z)
(D_b(x/z,Q)/\Delta(Q)) \nonumber \eea or, after integrating both sides
of the expression, \bea D_a(x,t') = D_a(x,t) \Delta(t') +
\int_{t'}^{t}\int_{x}^{1} dt'' {dz\over z} {\Delta(t') \over
  \Delta(t'')} {\alpha_{abc}(z,t'') \over 2\pi} \hat{P}_{a\to bc}(z)
D_b(x/z,t''), \eea where $t=\ln Q^2,$ with similar definitions for
$t'$ and $t''$.  
The function 
\bea 
   \Delta(t^\prime) = \exp\left(
  -\int_{t_0}^{t'}\int_{\epsilon}^{1-\epsilon} dt'' dz
  {\alpha_{abc}(z,t'') \over 2\pi} \hat{P}_{a\to bc}(z) \right) 
\eea
is called the Sudakov form factor, 
and gives the probability of evolving from the scale $t^\prime$ to $t_0$ 
with no resolvable branchings, 
where  $t_0$ is a cutoff scale for the showering.
The Sudakov $\Delta(t^\prime)$ contains all the information necessary to reconstruct a
shower, since it encodes the change in virtuality of a parton until a
resolvable showering occurs.
Showering is reduced to iterative solutions of the equation
$r=\Delta(t^\prime)/\Delta(t'')$, where $r$ is a random number uniformly
distributed in the interval $[0,1]$, until a solution for $t^\prime$ is
found which is below the cutoff $t_0$.  

For the case of initial state radiation, several modifications are
necessary.  The fragmentation function is replaced by a parton
distribution function, and the evolution proceeds backwards from a
large, negative scale $-|Q^2|$ to a small, negative cutoff scale
$-|Q_0^2|$.  There are two equivalent formulations of backwards
showering based on the probabilities 
\bea 
  \exp\left(-\int_{t'}^{t}\int_{\epsilon}^{1-\epsilon} dt'' dz
  {\alpha_{abc}(z,t'') \over 2\pi} \hat{P}_{a\to bc}(z) {x'
    f_a(x',t')\over x f_b(x,t') } \right), x'=x/z, 
\label{algo1}
\eea \cite{Sjostrand:1985xi},
and 
\bea 
{\Delta(t') \over f_b(x,t')} {f_a(x,t'') \over \Delta(t'') }
\label{algo2}
\eea 
\cite{Marchesini:1984bm}.
After choosing the change in virtuality, a particular backwards
branching is selected from the probability function based on their
relative weights (a summation over all possible branchings $a\to bc$
is implied these expressions), and the splitting variable is a solution
to the equation 
\bea 
  \int_\epsilon^{x/x'} {dz\over z}
  \hat{P}_{a\to bc}(z)f(x/z,t') = r \int_\epsilon^{1-\epsilon} {dz\over
  z} \hat{P}_{a\to bc}(z)f(x/z,t'), 
\eea 
where $r$ is a random number.
The details of how a full shower is reconstructed in {\tt PYTHIA},
 for example, can be found in
Ref.~\cite{Sjostrand:1994yb}.  The structure of the shower can be complex: the
transverse momentum of the heavy boson is built up from the whole
series of splittings and boosts, and is known only at the end of the
shower, after the final boost.

The PS-EG formulation outlined above is fairly independent of the hard
scattering process considered.  Only the initial choice of partons and
possibly the high scale differs.  Therefore, this formalism can be
applied universally to many different scattering problems.  In effect,
soft gluons are not sensitive to the specifics of the hard scattering,
only the color charge of the incoming partons.  This statement is true
to leading logarithm.

The parton showering of {\tt PYTHIA} obeys a strict ordering in
virtuality: the parton that initiates a hard scattering has a larger
(negative for initial state showers) virtuality than any other parton in the shower.  Parton
showers in {\tt HERWIG} proceed via a coherent branching process in which
a strict angular ordering is imposed on sequential gluon emissions:
the evolution variable is not virtuality, but a generalized virtuality
$\xi$.  
For an initial state shower, with parton splitting $c\to ba$, and
where $a$ has the largest virtuality, the variable $\xi=(p_b \cdot
p_c)/(E_bE_c)$.
At all values of $x$, the coherent branching algorithm
correctly sums the leading logarithmic contributions. At large $x\sim 1$, it
also sums the next-to-leading order contributions \cite{Catani:1991rr}, with
an appropriate definition of the splitting kernel.
(The exact definition of LL and NLL will be given later.)
Because of
the demonstrated importance of coherence effects, {\tt PYTHIA} includes
an additional veto on showers which are not also angular--ordered.
Note, however, that this does not make the two schemes equivalent --
some late emissions in a {\tt HERWIG} shower can have virtuality larger
than previous emissions.

\subsubsection{Analytic results}

At hadron colliders, the partonic cross sections for heavy boson
production can receive substantial corrections at higher orders in
$\alpha_s$.  This affects not only the total production rate, but also 
the kinematics of the heavy boson.  At leading order, 
the heavy boson has a $\delta(Q_T^2)$ distribution in $Q_T^2$.  At
next--to--leading order, the real emission of a single gluon generates
a contribution to $d\sigma/dQ_T^2$ that behaves as
$Q_T^{-2}\alpha_s(Q_T^2)$ and $Q_T^{-2}\alpha_s(Q_T^2)\ln(Q^2/Q_T^2)$
while the soft, and virtual corrections are
proportional to $-\delta(Q_T^2)$.  At higher orders, the most singular
terms follow the pattern of
$\alpha_s^n(Q_T^2)\sum_{m=0}^{2n-1}\ln^m(Q^2/Q_T^2)$ $=\alpha_s^n
L\equiv V^n$.  The logarithms arise from the incomplete cancellation
of the virtual and real QCD corrections. This cancellation becomes
complete for the integrated spectrum, where the real gluon can become
arbitrarily soft and/or collinear to other partons.  The pattern of
singular terms suggest that perturbation theory should be performed in
powers of $V^n$ instead of $\alpha_s^n$.  This reorganization of the
perturbative series is called resummation.

The first studies of soft gluon emission resummed the leading
logarithms \cite{Dokshitzer:1978yd,Parisi:1979se}, leading to a suppression of the cross
section at small $Q_T$.  The suppression underlines the importance of
including sub--leading logarithms \cite{Ellis:1981sj}.  The most rigorous
approach to the problem of multiple gluon emission is the
Collins--Soper--Sterman (CSS) formalism for transverse momentum
resummation \cite{Collins:1982zc}, which resums all of the important
logarithms.  This is achieved after a Fourier transformation with
respect to $Q_T$ in the transverse coordinate $b$, so that the series 
involving the delta function and terms $V^n$ simplifies to the form of an
exponential.  Hence, the soft gluon emission is resummed or
exponentiated in this $b$--space formalism.  The
Fourier transformation is the result of expressing the
transverse--momentum conserving delta functions
$\delta^{(2)}(\vec{Q}_T - \sum \vec{k}_{T_i})$ in their Fourier
representation.  Also, the exponentiation is accomplished through the
application of the renormalization group equation.  
Despite the successes of the $b$--space
formalism, there are several drawbacks. Most notable for the present
study is that it integrates out the soft gluon dynamics and does not
have a simple physical interpretation.

The CSS formalism was used by its authors to predict both the total
cross section to NLO and the kinematic distributions of the
heavy boson to resummed NLL order \cite{Collins:1985kg} at hadron colliders.  A
similar treatment was presented using the AEGM formalism \cite{Altarelli:1984pt},
that does not involve a Fourier transform, but is evaluated directly
in transverse momentum $Q_T$ space.  When expanded in $\alpha_s$, the 
two formalisms are equivalent to the NNNL order, and agree with
the NLO fixed order calculation of the total cross section \cite{Altarelli:1985kp}.
A more detailed numerical comparison of the two predictions can be
found in Ref. \cite{Arnold:1991yk}.  The AEGM formalism has been
reinvestigated, and an approximation to the $b$--space formalism has
been developed in $Q_T$--space which retains much of its predictive 
features \cite{Ellis:1998ii}.

In the $b$--space formalism, the differential cross section of the
heavy boson produced in association with soft gluons is: 
\bea
 {d\sigma(h_1 h_2 \to B^{(*)}X) \over dQ^2\,dQ^2_T\,dy} = {1\over (2
 \pi)^2} \int_{}^{} d^2 b \, e^{i {\vec b}\cdot{\vec Q_T}}
 \widetilde{W}(b,Q,x_1,x_2) + Y(Q_T,Q,x_1,x_2).  
\label{Eq:CSSFormula}
\eea 
where $Q$, $Q_T$ and $y$ describe the kinematics of the off--shell
heavy boson $B^{(*)}$. The function $Y$ is free of $\ln(Q^2/Q_T^2)$ and 
corrects for the soft gluon approximation in the high $Q_T$ region. 
The function $\widetilde W$ has the form: 
\bea 
 {\widetilde W}(b,Q,x_1,x_2) = e^{-S(b,Q)} 
 \left( C_{il} \otimes f_{l/h_1} \right)(x_1,b)
 \left( C_{jl} \otimes f_{l/h_2} \right)(x_2,b)
 H_{ij}(Q,y), 
\eea 
where 
\bea & &
 S(b,Q,C_1,C_2) = \int_{C_1^2/b^2}^{C_2^2Q^2} {d {\bar \mu}^2\over
 {\bar \mu}^2} \left[ \ln\left({C_2^2Q^2\over {\bar \mu}^2}\right)
 A\big(\alpha_s({\bar \mu})\big) + B\big(\alpha_s({\bar \mu})\big)
 \right], 
\eea 
and 
\bea 
 \left( C_{jl} \otimes f_{l/h_1} \right)
 (x_1,\mu) = \int_{x_1}^{1} {d \xi_1 \over \xi_1} \, C_{jl}\left( {x_1 \over
 \xi_1}, C_1, C_2, \mu=C_3/b\right) f_{l/h_1}\left(\xi_1, \mu=C_3/b\right).  
\eea 
In these expressions, $C_1$, $C_2$ and $C_3$ are renormalization scales, 
$H$ is a function that describes the hard scattering, and $A$, $B$ and 
$C$ are calculated perturbatively in powers of $\alpha_s$: 
\bes
 \{A,B\}=\sum_{n=1}^{\infty} \left({\alpha_s(\mu)\over\pi}\right)^n
 \{A^{(n)},B^{(n)}\} ~~~ {\rm and} ~~~
  C_{ij} =\sum_{n=0}^{\infty} \left({\alpha_s(\mu)\over\pi}\right)^n C^{(n)}.
\ees 
The functions $C^{(n)}$ are mostly responsible for the change in the total
production cross section at higher orders.  In fact, $\left( C\otimes f\right)$ 
is simply a redefinition of the parton distribution
function obtained by convoluting the standard ones with an
ultraviolet--safe function. These generalized parton distributions encode
both the longitudinal momentum fraction and the transverse recoil of the 
initial state parton.

We remove $C_1, C_2$ and $C_3$ from these expressions by
choosing their canonical values \cite{Collins:1982zc}, which also removes
large logarithms from the expansion of the resummed expression. 
At leading order in $C_{ij}$, the resummed part of the expression for the 
production of an on--shell heavy boson simplifies considerably to: 
\bea
  {d\sigma(h_1 h_2 \to B^{(*)}X) \over dQ^2_T} = \sigma_0 \int_{}^{} {d^2 b
  \over (2 \pi)^2} \, e^{i {\vec b}\cdot{\vec Q_T}} e^{-S(b,Q)}
  {f(x_1,b) f(x_2,b) \over f(x_1,Q) f(x_2,Q)}, 
\eea 
where
\bes 
  \sigma_0 = \kappa \int_{}^{} {dx_1 \over x_1} f(x_1,Q) f(x_2,Q), 
\ees 
and $\kappa$ contains physical constants.  
The expression contains two factors, the total cross section
at leading order $\sigma_0$, and a cumulative probability function in
$Q_T^2$ that describes the transverse momentum of the heavy boson
(the total integral over $Q_T^2$ transforms
$e^{i\vec{b}\cdot\vec{Q}_T}$ to $\delta^{(2)}(\vec{b})$).  Except for
the complication of the Fourier transform, the term $e^{-S/2}
f(x,b)/f(x,Q)$ is analogous to $\Delta(Q) f(x,Q')/\Delta(Q') f(x,Q)$
of the PS-EG of Eq.~(\ref{algo2}).

Equation (\ref{Eq:CSSFormula}), which is formulated in $b$--space, has a 
similar structure in $Q_T$--space.
This is surprising, since the $b$--space result depends critically on
the conservation of total transverse momentum.  To NNNL accuracy,
however, the $Q_T$ space expression agrees exactly with the $b$--space
expression, and has the form \cite{Ellis:1998ii}: 
\bea 
  {d\sigma(h_1 h_2 \to
  B^{(*)}X) \over dQ^2\,dQ^2_T\,dy} = {d\over
  dQ_T^2}\widetilde{W}(Q_T,Q,x_1,x_2) + Y(Q_T,Q,x_1,x_2).  
\eea 
Again ignoring $Y$, we can rewrite this expression as: 
\bea {d\sigma(h_1 h_2
  \to B^{(*)}X) \over dQ_T^2 } = \sigma_1 \left({d\over dQ_T^2}
  \left[e^{-S(Q_T,Q)} {\left(C\otimes f\right)(x_1,Q_T) \left(C\otimes
        f\right)(x_2,Q_T) \over \left(C\otimes f\right)(x_1,Q)
      \left(C\otimes f\right)(x_2,Q)} \right]\right), \eea \bes
\sigma_1 = \kappa \int_{}^{} {dx_1 \over x_1} \left(C\otimes
  f\right)(x_1,Q) \left(C\otimes f\right)(x_2,Q).  \ees The factor
$\sigma_1$ is the total cross section to a fixed order, while the rest
of the function yields the probability that the heavy boson has a
transverse momentum $Q_T$.

\subsubsection{Methodology for comparison}\label{sec:resumps,mc,qcdsm}

To make a comparison between the distributions from analytic and parton
showering calculations, we must quantify the differences in theoretical input
and identify what approximations have been made in each one.   >From the
discussion of the analytic resummation calculations, we see that the emission
of multiple soft gluons is described best by perturbation theory not in terms
of $\alpha_s$, but in powers of $\alpha_s$ times logarithms of large  numbers.
Typically, the logarithms are classified  according to the orders of $\alpha_s$
and the power of logarithms in  $(i)$ the Sudakov exponent, or in $(ii)$ the
perturbative expansion of  the resummed $\qt$ distribution.

In classification $(i)$\footnote{See also sect. \ref{sec:low-qt,resum,qcdsm}}, 
it is argued that terms with $A^{(n)}$ are leading
compared to  terms with $B^{(n)}$ because the former are enhanced by a large
logarithm.  Also, $X^{(n)}$ is leading compared to $X^{(n+1)}$ because the
latter is  suppressed by $\alpha_s$ for $X = A, B$ or $C$. The comparison of
$C$ to $A$ and  $B$ is somewhat {\it ad hoc}, because $C$ does not appear in
the Sudakov in all  resummation schemes. In this approach, the lowest order
resummed result (LL) has  only $A^{(1)}$ and $C^{(0)}$. Additionally, the next
order (NLL) has $A^{(2)}$, $B^{(1)}$ and $C^{(0)}$, because the first two
are suppressed by an $\alpha_s$  and the last by the inverse of a large
logarithm when compared to the leading  terms. This is an elegant and simple
classification scheme, because it  relies only on the structure of the Sudakov
exponent. On the other hand, it has an {\it ad hoc} mixing $\alpha_s$ and
$\log^{-1}$--suppressed terms, and  depends on the resummation scheme in
dealing with $C^{(n)}$.

In classification $(ii)$, the resummed cross section is expanded and reorganized in 
towers of logarithms as 
  \bea
  \frac{d\sigma}{dQ_T^2} \sim 
\begin{array}{llll}
\alpha_s^1 ( L_1^1 ) &   +      &          &   \\
\alpha_s^2 ( L_2^1   &+ L_2^3 ) & +        &   \\
\alpha_s^3 ( L_3^1   &+ L_3^3   & + L_3^5 )& + \cdots,
\end{array}
  \eea
where $L_i$ are various linear combinations of $\ln(Q/Q_T)$.
Remarkably, each column of logs receives contribution only from $A^{(n)}$, 
$B^{(n)}$ and $C^{(n-1)}$ ($n$ numbers the columns). Since $X^{(n)}$ 
and $X^{(n+1)}$ are clearly ordered by $\alpha_s$ ($X = A,B,C$),
the first column is called the leading tower of logs, i.e. LL, the next NLL, etc. 
This is a more--involved classification, but closer to the spirit of the 
resummation, and more precise. In the rest of this work, we use the naming 
convention $(ii)$ when comparing to the analytic resummation.
Thus, when we say that the coherent branching algorithm in {\tt HERWIG} correctly sums the 
leading logarithmic contributions, we are referring to the $A^{(1)}$ and $B^{(1)}$ terms.

\begin{figure}[!ht]
\centering
  \resizebox{10cm}{10cm}{ 
\includegraphics*{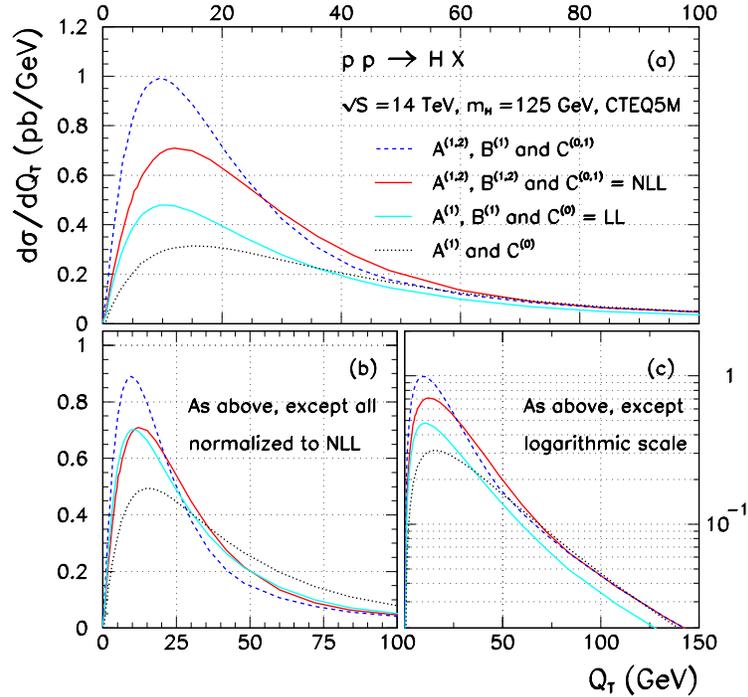}
    }
    \caption{\em (top) Absolute $Q_T$ distribution of the Higgs boson from an analytic resummation
calculation to different orders of accuracy in perturbation theory;
      (left, bottom) Same as (top), but normalized to the same total
cross section; (right, bottom) Same as (left,bottom), but on a logarithmic scale.  
All curves are for the production of a 125 GeV Higgs boson
at the LHC based on the CTEQ4M parton distribution functions.}
    \label{fig:abc}
\end{figure}

Figure~\ref{fig:abc} demonstrates the different predictions
for different choices of organizing the perturbative expansion.
The top frame shows the absolute distributions of $Q_T^H$ when the
perturbative
coefficients include: $A^{(1,2)},~B^{(1,2)},~C^{(0,1)}$ (solid red),
$A^{(1,2)},~B^{(1)}$ and $C^{(0,1)}$ (long-dashed blue), $A^{(1)},~B^{(1)}$
and
$C^{(0)}$ (solid light blue), and only $A^{(1)}$ and $C^{(0)}$ (short-dashed
black).
The change from LL to NLL results in an increase of the normalization of the
cross section as well as a slight shift of the peak towards higher
transverse
momentum, but the shapes are very similar. The lower left frame shows the
same
distributions, but all normalized to the same production rate. This
eliminates
the normalization dependence and focuses on the variation of the shape. The
lower
right frame, shown in logarithmic scale, amplifies the differences at high
$Q_T$'s.

\subsubsection{Numerical comparisons}

In the following plots, we show several numerical comparisons of ResBos,
{\tt HERWIG} and three versions of {\tt PYTHIA}.  Why three versions of {\tt PYTHIA}?
This reflects the evolution of the program.  
The version {\tt PYTHIA}-5.7 was used for almost all LHC Monte Carlo analyses for Higgs production.
The showering in {\tt PYTHIA}-5.7, as described earlier, is virtuality ordered, 
with an additional requirement of angular ordering.  No other kinematic restrictions
are imposed.  In later versions, beginning with {\tt PYTHIA}-6, each parton emission
was required to match the kinematic constraints of the NLO process, i.e. for
the first emission in the parton shower accompanying $gg\to H$,
the kinematic constraints $\hat{s}+\hat{t}+\hat{u}=M_H^2$ must be satisfied.
Since $\hat{s}$ and $\hat{t}$ are the two variables of the virtuality--ordered
shower, a cut on $\hat{u}$ restricts their range.  The plots labeled 
by {\tt PYTHIA}6.1 demonstrate the effect of the  $\hat{u}$ cut.
Finally, in {\tt PYTHIA}-6.2, an additional hard matrix element correction was
applied to the parton shower.  In this approach, the maximum virtuality of 
the shower is increased from its nominal value at $Q=m_H$ up to the largest
kinematically allowed value.  Furthermore, each parton emission is corrected
by the ratio of the exact matrix element squared at NLO to the approximate
matrix element squared given by the parton shower approximation.
Thus, showers generated using {\tt PYTHIA}-6.2 should have a closer agreement
to the ResBos predictions at high $Q_T^H$ (where ResBos relies on
the NLO prediction).  In contrast, showers induced in {\tt HERWIG} will still
have a cutoff set by $m_H$.  Note, however, that the {\tt PYTHIA} prediction
still has the LO normalization, and a full rescaling is necessary.  
This may be appropriate if the $K$--factor correction to the LO prediction
is the same as the one for the NLO emission.  

\begin{figure}
\centering
  \resizebox{15cm}{10cm}{
    \includegraphics*[0,50][650,600]{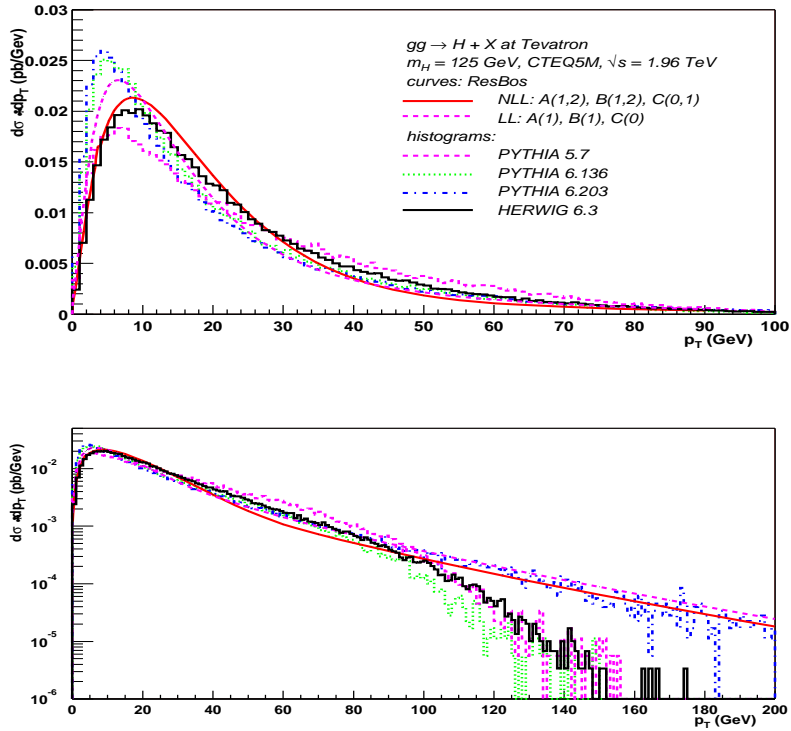} }
    \caption{\em Transverse momentum $Q_T$ distributions of a 125 GeV 
Higgs boson produced at the Tevatron for {\tt HERWIG}, different versions
of {\tt PYTHIA} and different perturbative orders of ResBos.  The
bottom plot is for an expanded $Q_T$ range.  All curves are
normalized to the same total cross section.}
    \label{fig:tev125}
\end{figure}

\begin{figure}
\resizebox{8cm}{10cm}{
\includegraphics*[0,50][600,650]{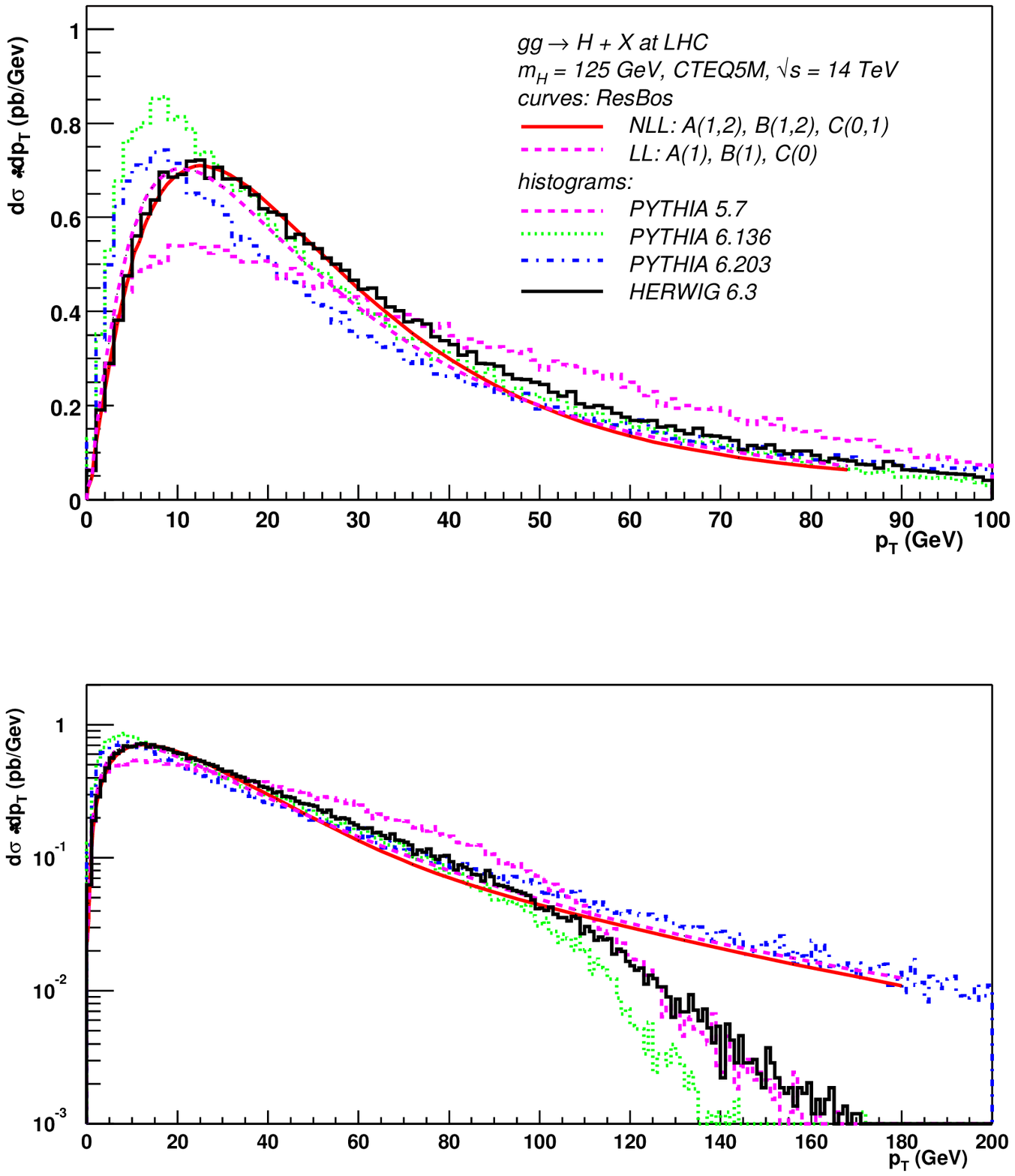}}
\resizebox{8cm}{10cm}{
\includegraphics[0,50][600,650]{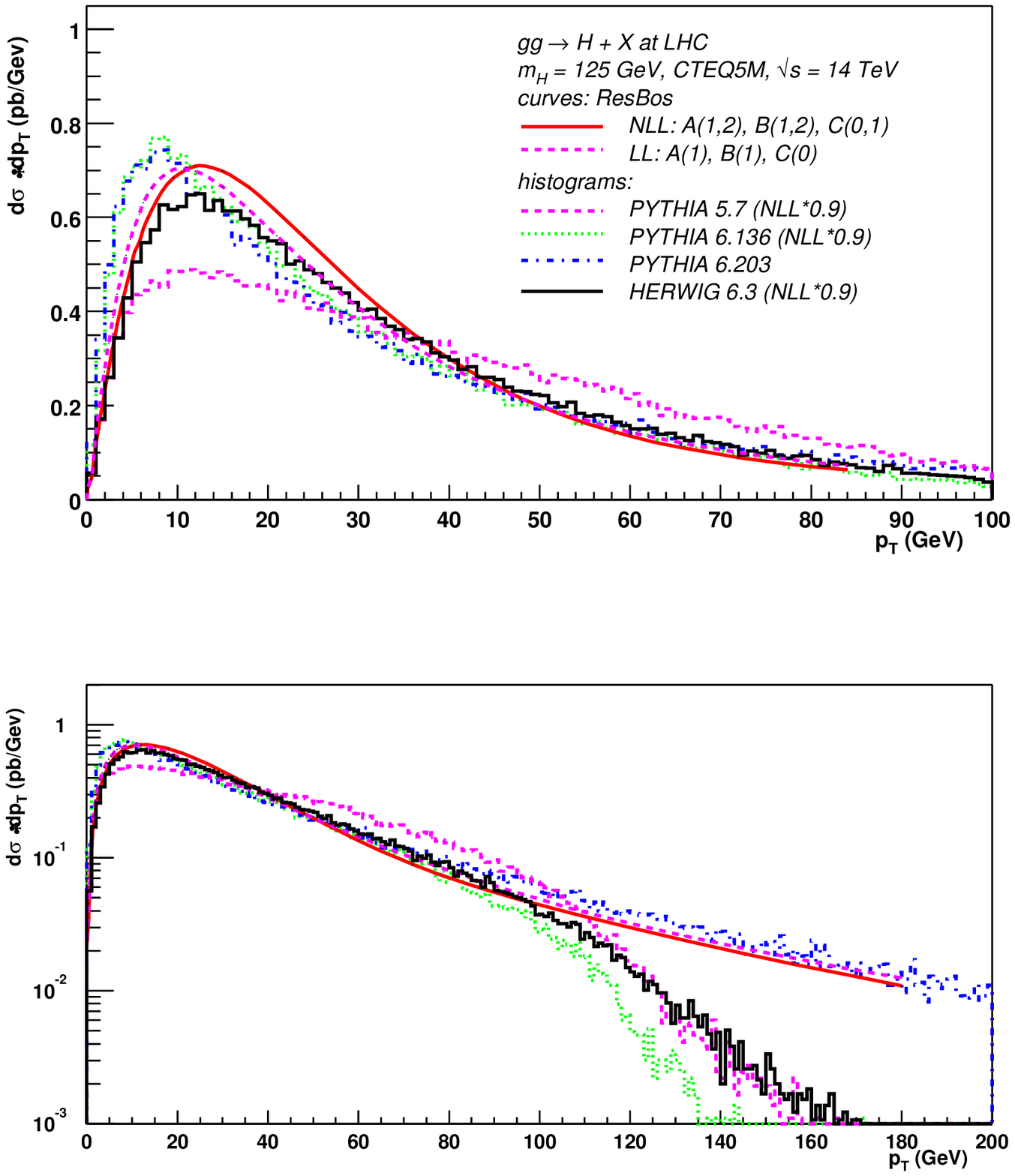}}
\caption{\em Comparison of light Higgs boson production at
the LHC. The right hand plots have  {\tt PYTHIA} 5.7, 6.1 and 
{\tt HERWIG} distributions normalized
to 90\% of the ResBos NLL cross section. }
    \label{fig:lhc125}
\end{figure}

Figures~\ref{fig:tev125} and \ref{fig:lhc125} show the predicted $Q_T$ 
distributions for
production of a Standard Model Higgs boson with mass $m_H=125$ GeV
at collider energies of 1.96, and 14 TeV. 
For all plots, the distributions have been normalized to the same cross
section (ResBos NLL); without this normalization, the
PS-EG predictions would be about a factor of 2 lower than the ResBos
curves.  
Two
ResBos curves are shown: LL (including $A^{(1)}$, $B^{(1)}$ and $C^{(0)}$) and
NLL (also including $A^{(2)}$, $B^{(2)}$ and $C^{(1)}$ ). The inclusion of the NLL terms leads to a
slightly harder $Q_T$ distribution, as discussed previously. The
ResBos curves appear close to the {\tt HERWIG} predictions, and somewhat less
close to the predictions of {\tt PYTHIA} (versions 6.1 and after).

In general, the PS-EG predictions are in fair agreement with
the analytic resummation ones for low $Q_T$, where multiple, soft
gluon emission is the most important.
The agreement of {\tt HERWIG} with the ResBos curves becomes even
better if the shape comparison are made by normalizing the
cross sections in the low $Q_T$ region alone, away from the effects of
the exact matrix element for Higgs plus jet. 
This is illustrated in Figures~\ref{fig:lhc125}.
The {\tt PYTHIA}-6+ predictions peak at a noticeably lower value
of $Q_T$ than either ResBos or {\tt HERWIG}.
A striking feature of the plots is the change
induced in {\tt PYTHIA}-6.1 relative to {\tt PYTHIA}-5.7,
indicating the importance of kinematic constraints.
Note also that the average Higgs transverse momentum increases with
increasing center-of-mass energy, due to the increasing phase space
available for gluon emission. 
At high $Q_T$, the exact matrix element for Higgs plus
jet, present both in the most recent version of {\tt PYTHIA} as well as
ResBos, correctly describes Higgs production at transverse momenta
on the order of the Higgs mass or larger, while a pure parton
showering description of the high $Q_T$ end is inadequate. 
The change observed in 6.2,
only visible at large values of $Q_T$, is the
result of the 
matrix element (Higgs + jet) corrections to the parton shower.

\begin{figure}
\resizebox{8cm}{10cm}{
\includegraphics[0,0][600,600]{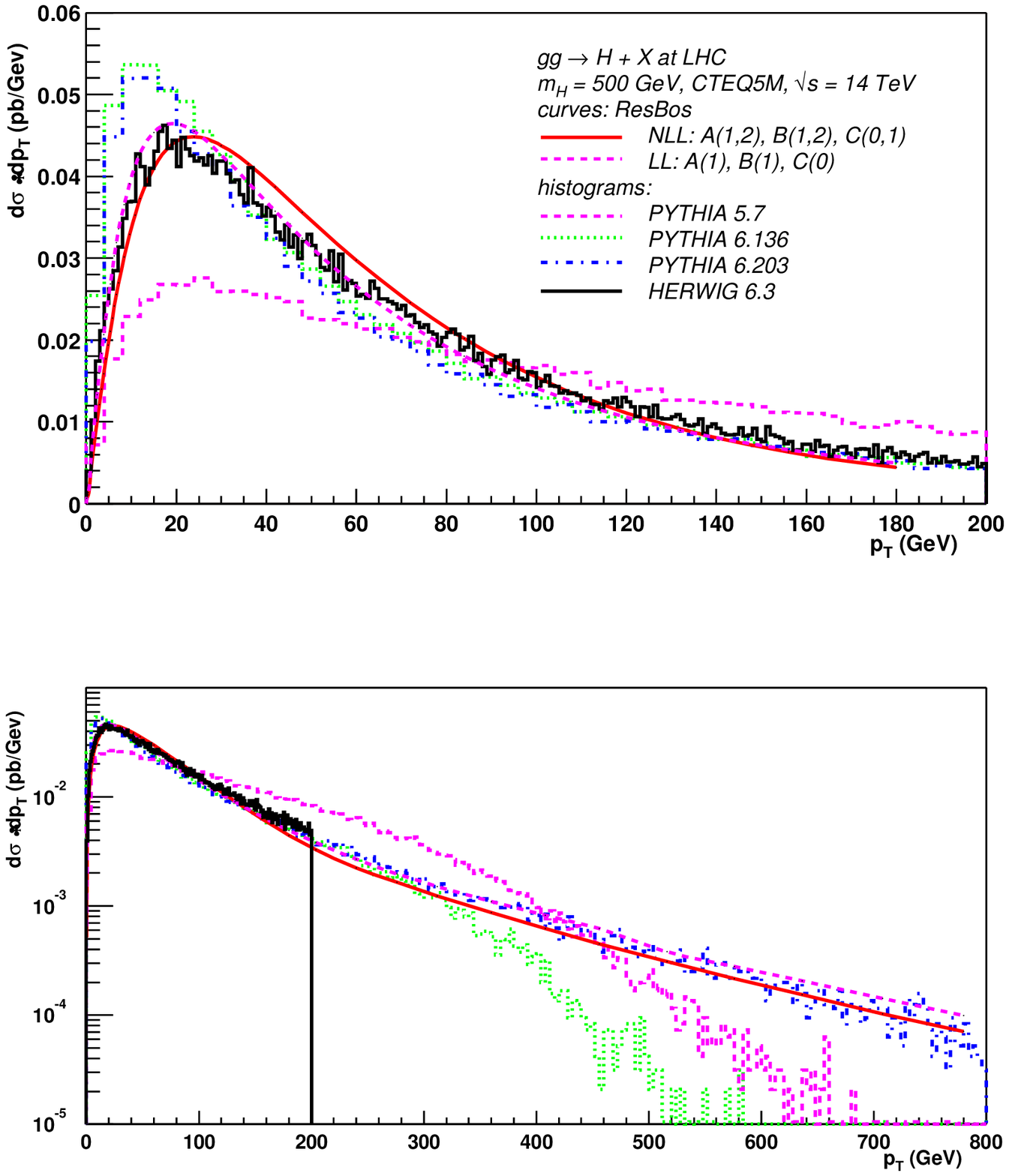}}
\resizebox{8cm}{10cm}{\includegraphics[0,0][600,600]{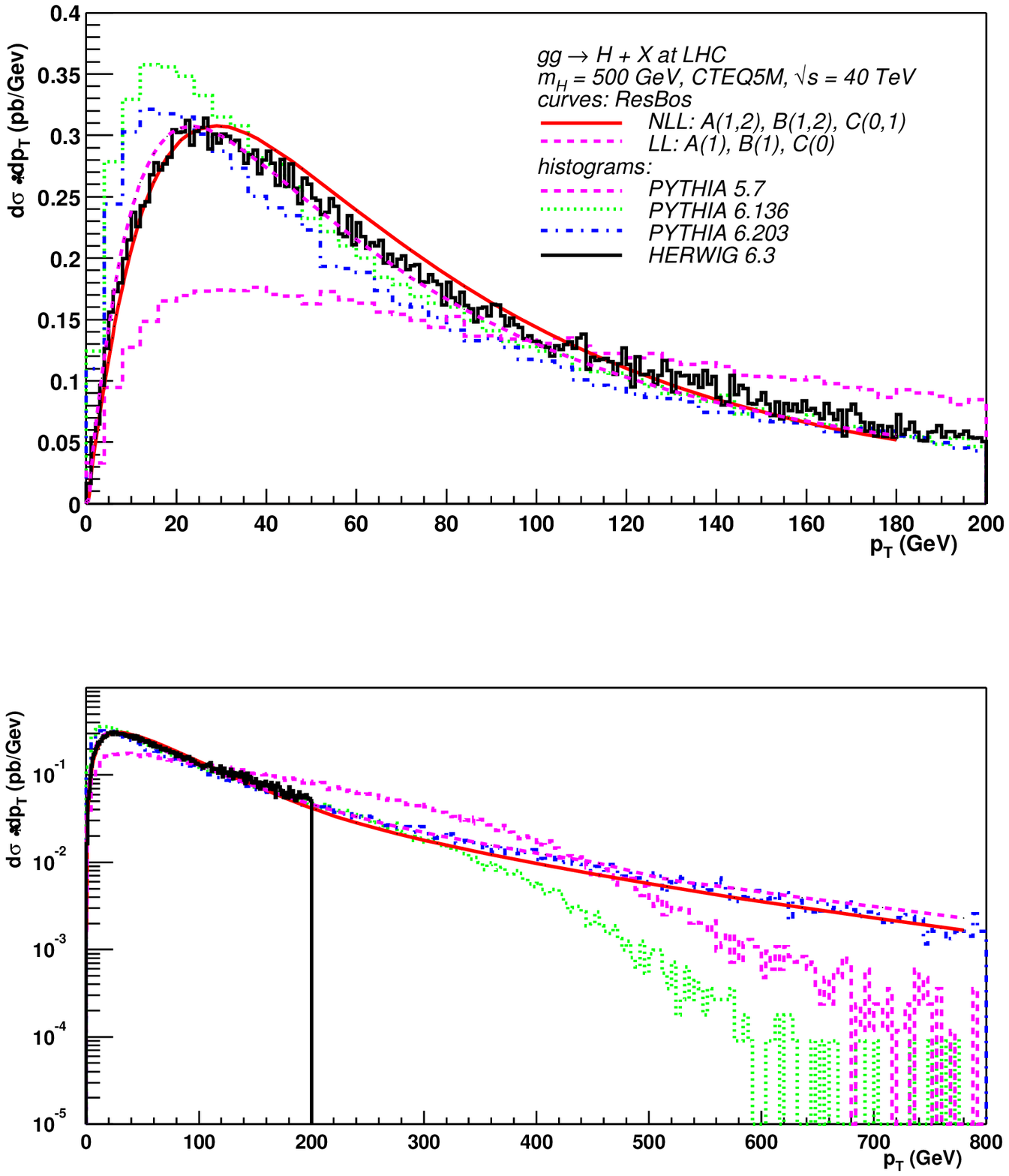}}
\caption{\em Comparison of Higgs boson production at
the LHC and 40 TeV $m_H=500$ GeV.}
    \label{fig:lhc500}
\end{figure}

The final comparison plot, 
Figure~\ref{fig:lhc500},  shows the $Q_T$ distributions for a 500 GeV
Higgs boson
generated through $gg$ fusion at 14 TeV (pp) and 40 TeV (pp).
Comparisons are made of the ResBos(LL and NLL), {\tt PYTHIA} and
{\tt HERWIG}  predictions. As in
Figures~\ref{fig:tev125} and \ref{fig:lhc125} the distributions have been scaled to have the same
total cross section. The average transverse momentum for a 500 GeV
Higgs is noticeably larger than that for a 125 GeV Higgs in all of the
predictions, as expected, since the hard scale for the
process is $m_H$.  
The agreement between the {\tt HERWIG} (version 6.1 and
later) predictions and the ResBos  curves is better than for the
predictions with a 125 GeV mass Higgs. The agreement
between {\tt HERWIG} and the ResBos curves remains very good.
For high $Q_T$, the {\tt PYTHIA} 6.2 prediction, with matrix element
corrections, agrees with the ResBos prediction.

\subsubsection{Properties of showers}

In the previous section, there were notable differences
in the predictions of {\tt PYTHIA} and {\tt HERWIG}, with 
{\tt HERWIG} giving a superior description of the Higgs
boson properties for low $Q_T^H$.  Therefore, we have
investigated several basic properties of the parton emissions
to determine the cause of this discrepancy.  It is already
known that the {\tt HERWIG} shower is of the coherent kind, whereas
{\tt PYTHIA} is virtuality ordered with approximate angular ordering
superimposed.  On the other hand, the same was true in
{\tt PYTHIA}5.7, but closer agreement with {\tt HERWIG} is
obtained after making the $\hat{u}$ cut.  Perhaps a more
careful analysis of the {\tt HERWIG} kinematics will result
in even better agreement.

\begin{figure}[!ht]
\resizebox{8cm}{8cm}{
\includegraphics*{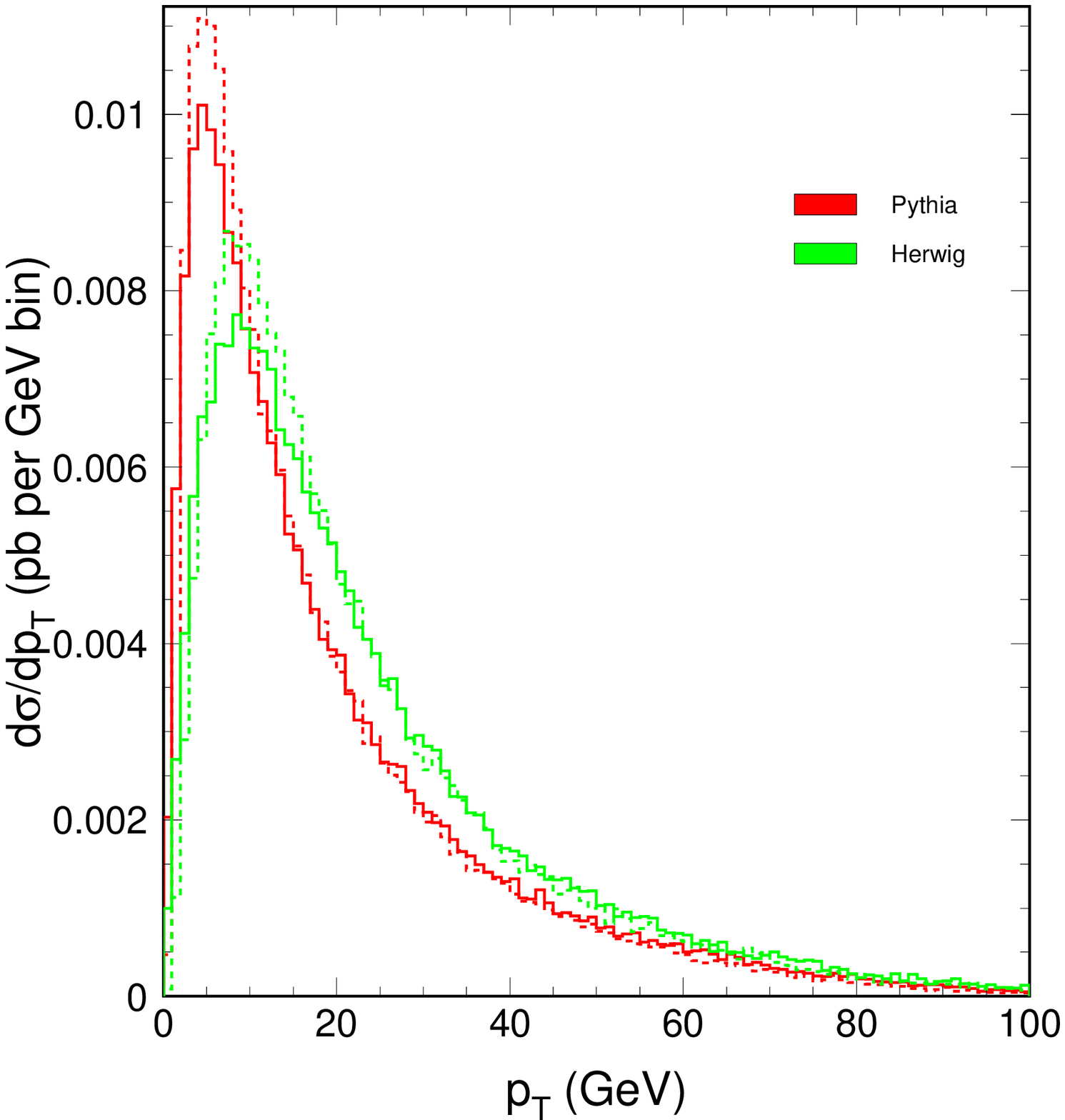}}
\resizebox{8cm}{8cm}{
\includegraphics*{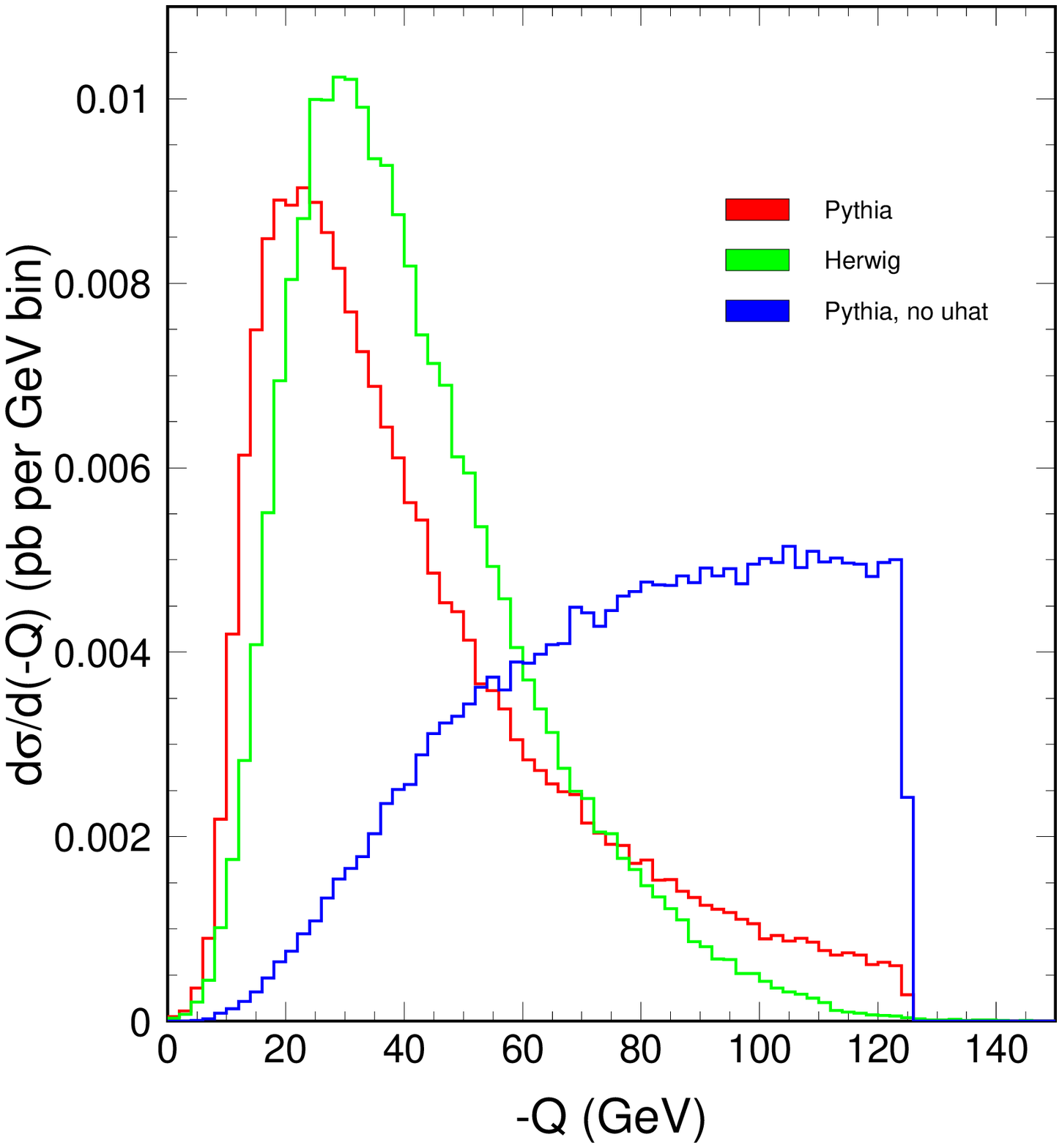}}
\caption{\em Comparison of showering properties for light Higgs
boson production at the Tevatron.  The leftmost plot shows
the transverse momentum of the Higgs boson and the first
parton emission for {\tt PYTHIA} (red) and {\tt HERWIG} (green).
The rightmost plot shows the largest (negative) virtuality
of parton in the shower.  The blue curve shows the effect of having no $\hat{u}$ cut.}
\label{fig:showering}
\end{figure}

Figure~\ref{fig:showering} shows
a comparison of showering properties for light Higgs
boson production at the Tevatron.  
For the purpose of these plots, the matrix element corrections
in {\tt PYTHIA} have been turned off.
The leftmost plot shows
the transverse momentum of the Higgs boson and the first
parton emission for {\tt PYTHIA} (red) and {\tt HERWIG} (green).
Note that the leading emission describes the Higgs boson $Q_T$ from
the full shower very well down to fairly low values $\sim 10$ GeV.
The main discrepancy between {\tt PYTHIA} and {\tt HERWIG} occurs in
the same kinematic region, exactly where the properties of
several emissions become important.  

The rightmost plot shows the largest (negative) virtuality
of a parton in the shower.  
The blue curve shows the effect of having no $\hat{u}$ cut, as
in the older version of {\tt PYTHIA}.  Clearly, {\tt PYTHIA} without
the $\hat{u}$ cut does
not have the $Q^{-2}$ behavior expected from the derivative
of the Sudakov.
While the agreement between {\tt PYTHIA}-6.2 and {\tt HERWIG} is markedly
improved, {\tt PYTHIA} still has a residual enhancement near the upper
scale $Q=m_H$ in this example.  This requires further investigation.

To summarize:-

\begin{itemize}
\item   The differences between ResBos LL and NLL are in the direction expected and are relatively
subtle.

\item   The two newer versions of {\tt PYTHIA}  and {\tt HERWIG} both approximately agree with  the predictions
of ResBos LL/NLL, with the {\tt HERWIG} shape agreement  being somewhat  better in the low
$Q_T$ region.

\item   The agreement of {\tt HERWIG} with ResBos becomes better as:
        $(a)$    the center of mass energy increases and
        $(b)$    the Higgs mass  increases
        the agreement seems to be better with LL.
\end{itemize}

\subsection{Studies of underlying events using CDF data}
\label{field}

\begin{figure}
\begin{center}
\includegraphics[scale=0.8]{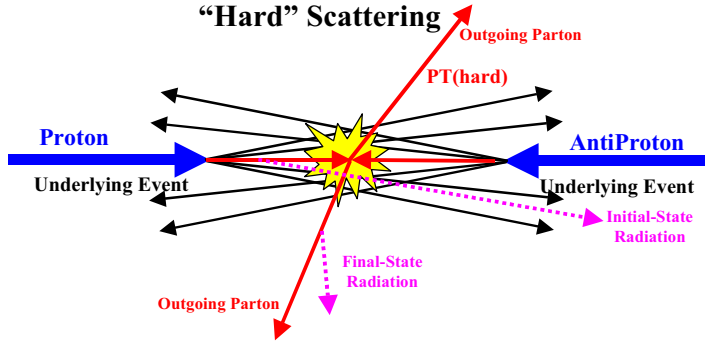}\end{center}
\caption{\em Illustration of the way the QCD Monte-Carlo models simulate a proton-antiproton collision in 
which a hard $2$-to-$2$ parton scattering with transverse momentum, \pthard, has occurred.  The 
resulting event contains particles that originate from the two outgoing partons 
({\it plus initial and final-state radiation}) and particles that come from the breakup of the 
proton and antiproton ({\it\BBR}).  The ``hard scattering" component consists of the outgoing two ``jets" 
plus initial and final-state radiation. The \UE\ is everything except the two outgoing hard scattered ``jets" 
and consists of the \BBR\ plus possible contributions from the "hard scattering" arising from 
initial and final-state radiation.
\label{rdf_fig1}}
\end{figure}

Fig.~\ref{rdf_fig1} illustrates the way QCD Monte-Carlo models simulate a proton-antiproton collision in which 
a "hard" $2$-to-$2$ parton scattering with transverse momentum, \pthard, has occurred.  The resulting event contains 
particles that originate from the two outgoing partons ({\it plus initial and final-state radiation}) and particles 
that come from the breakup of the proton and antiproton (\ie \BBR).  The ``hard scattering" component consists 
of the outgoing two ``jets" plus initial and final-state radiation. The \UE\ is everything 
except the two outgoing hard scattered ``jets" and consists of the \BBR\ plus possible contributions 
from the ``hard scattering" arising from initial and final-state radiation. 

The \BBR\ are what is left over after a parton is knocked out of each of the initial two beam 
hadrons.  It is the reason hadron-hadron collisions are more ``messy" than electron-positron annihilations and no one 
really knows how it should be modeled.  For the QCD Monte-Carlo models the \BBR\ are an 
important component of the \UE.  Also, it is possible that multiple parton scattering contributes to the 
\UE.  {\tt PYTHIA} \cite{Sjostrand:2000wi} models the \UE\ in proton-antiproton collision 
by including multiple parton interactions. In addition to the hard $2$-to-$2$ parton-parton scattering and the \BBR, 
sometimes there is a second ``semi-hard" $2$-to-$2$ parton-parton scattering that contributes particles to the \UE.

Of course, from a certain point of view there is no such thing as an \UE\ in a proton-antiproton 
collision.  There is only an ``event" and one cannot say where a given particle in the event originated.  On the other 
hand, hard scattering collider ``jet" events have a distinct topology.  On the average, the outgoing hadrons 
``remember" the underlying the $2$-to-$2$ hard scattering subprocess.  An average hard scattering event consists of a 
collection (or burst) of hadrons traveling roughly in the direction of the initial beam particles and two collections of 
hadrons (\ie ``jets") with large transverse momentum.  The two large transverse momentum ``jets" are roughly back 
to back in azimuthal angle.  One can use the topological structure of hadron-hadron collisions to study the 
\UE\ \cite{Field:2001ab,Field:2000ab,Tano:2001ab}. The ultimate goal is to understand the physics of the \UE, but since it is very 
complicated and involves both non-perturbative as well as perturbative QCD it seems unlikely that this will happen 
soon.  In the mean time, we would like to tune the QCD Monte-Carlo models to do a better job fitting the \UE.  
The \UE\ is an unavoidable background to most collider observables.  To find ``new" physics 
at a collider it is crucial to have Monte-Carlo models that simulate accurately ``ordinary" hard-scattering collider 
events.  This report  will compare collider observables that are sensitive to the \UE\ with the QCD 
Monte-Carlo model predictions of {\tt PYTHIA} 6.115 \cite{Sjostrand:2000wi}, {/tt HERWIG} 5.9 \cite{Marchesini:1988cf,Knowles:1988vs,Catani:1991rr}, and {\tt ISAJET} 7.32 \cite{Paige:1986vk}
and discuss the tuning of {\tt ISAJET} and {\tt PYTHIA}.

\begin{figure}
\begin{center}
\includegraphics[scale=0.8]{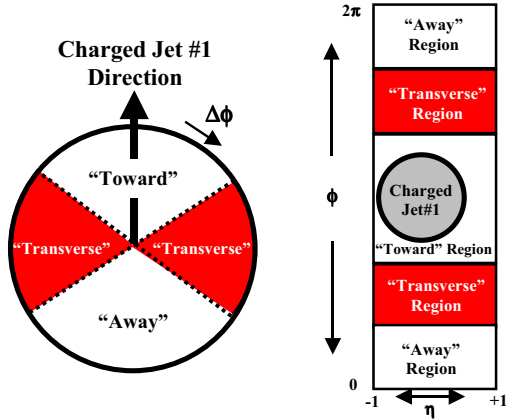}\end{center}
\caption{\em Illustrations of correlations in azimuthal angle $\Delta\phi$ relative to the direction of the 
leading charged jet in the event, chgjet\#1.  The angle $\Delta\phi=\phi-\phi_{\rm chgjet\#1}$  
is the relative azimuthal angle between charged particles and the direction of chgjet\#1.  The``toward" 
region is defined by $|\Delta\phi|<60^\circ$ and \etacut, while the ``away" 
region is $|\Delta\phi|>120^\circ$ and \etacut.   The ``transverse" region is defined by 
$60^\circ<|\Delta\phi|<120^\circ$  and \etacut.  Each region has an 
area in \etaphi\ space of $4\pi/3$.
On an event by event basis, we define ``transMAX (``transMIN") to 
be the maximum (minimum) of the two ``transverse" pieces, $60^\circ<\Delta\phi<120^\circ$ and \etacut, 
and $60^\circ<-\Delta\phi<120^\circ$ and \etacut. ``TransMAX" and ``transMIN" each have an area in \etaphi\ space 
of $2\pi/3$.  The sum of  ``TransMAX" and ``transMIN" is the total ``transverse" region with area $4\pi/3$.
\label{rdf_fig3}}
\end{figure}

\subsubsection{The ``Transverse" Region}

In a proton-antiproton collision large transverse momentum outgoing partons manifest themselves, in the laboratory, 
as a clusters of particles ({\it both charged and neutral}) traveling in roughly the same direction.  These clusters are 
referred to as ``jets".  In this analysis we examine only the charged particle component of ``jets".  Our philosophy in 
comparing the QCD Monte-Carlo models with data is to select a region where the data is very ``clean" so that ``what 
you see is what you get" ({\it almost}).  Hence, we consider only charged particles measured by the CDF central tracking 
chamber (CTC) in the region \ptcut\ and \etacut, where the track finding efficiency is high and uniform 
(estimated to be $92\%$ efficient) and we restrict ourselves to charged particle jets with transverse momentum less than 
$50\gevc$.  The data presented here are uncorrected.  Instead the theoretical Monte-Carlo models are corrected for 
the track finding efficiency by removing, on the average, $8\%$ of the charged particles.  The theory curves have an 
error ({\em statistical plus systematic}) of about $5\%$.   Thus, to within $10\%$ ``what you see is what you get".

\begin{figure}
\begin{center}
\includegraphics[scale=0.5]{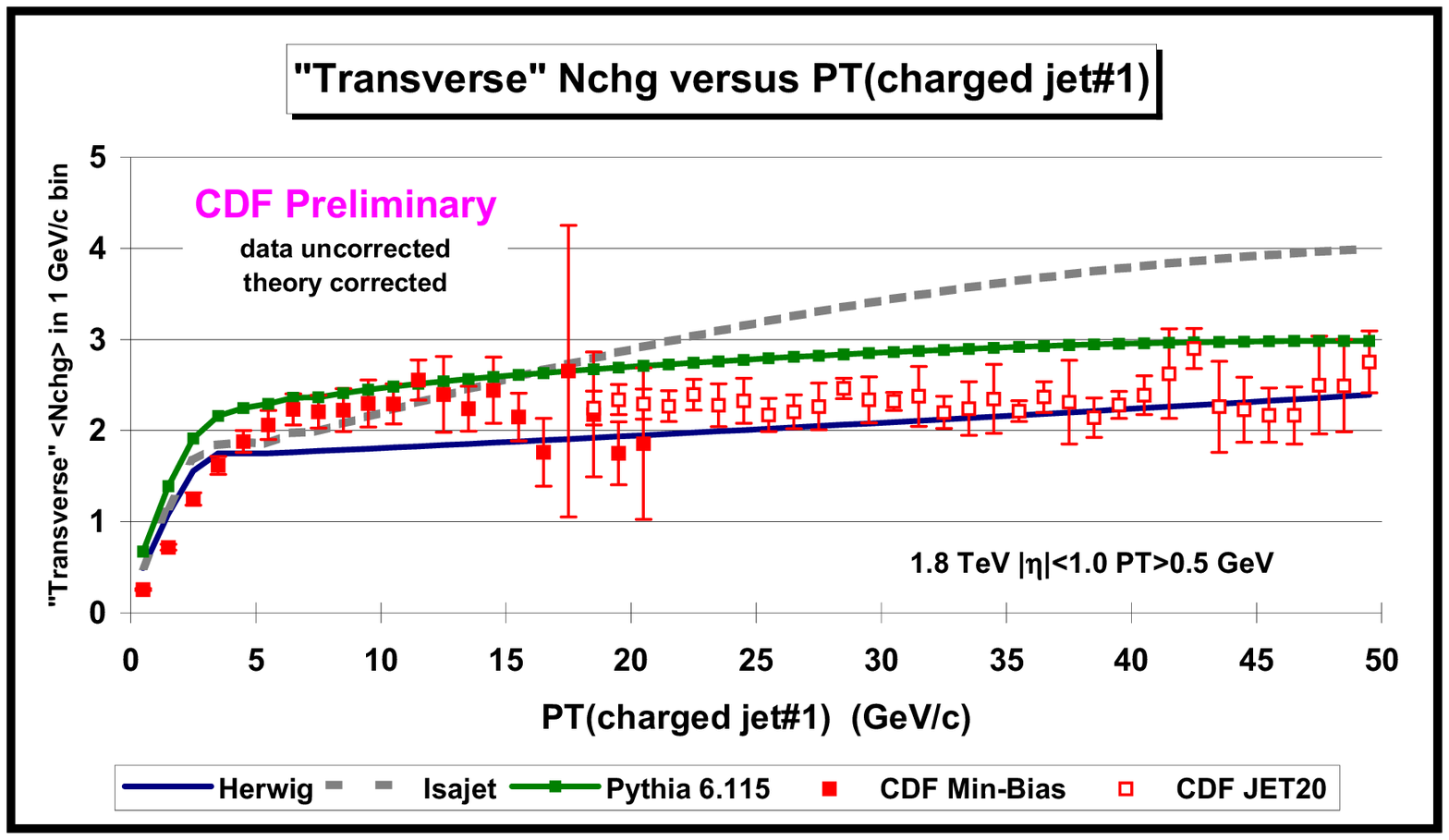}\end{center}
\caption{\em Data on the average number of charged particles (\ptcut, \etacut) in the ``transverse'' region defined in 
Fig.~\ref{rdf_fig3} as a function of transverse momentum of the leading charged jet compared with the QCD 
Monte-Carlo predictions of {\tt HERWIG 5.9}, {\tt ISAJET} 7.32, and {\tt PYTHIA} 6.115 with their default parameters and 
with \hardcut. Each point corresponds to the \aveN\  in a $1\gevc$ bin. The solid (open) points are the 
\MB\ (JET20) data. The theory curves are corrected for the track finding efficiency and have an 
error ({\it statistical plus systematic}) of around $5\%$.
\label{rdf_fig5}}
\end{figure}

\begin{figure}
\begin{center}
\includegraphics[scale=0.5]{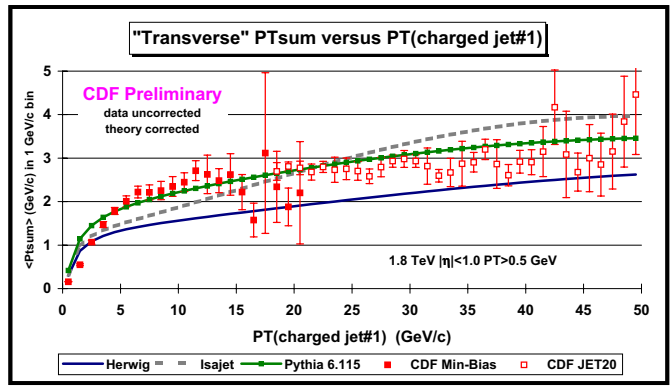}\end{center}
\caption{\em As for Fig~\ref{rdf_fig5} except that the average scalar
  $P_T$ is shown. \label{rdf_fig6}
}
\end{figure}

Charged particle ``jets" are defined as clusters of charged particles (\ptcut, \etacut) in ``circular regions" of 
\etaphi\ space with radius $R = 0.7$.   Every charged particle in the event is assigned to a ``jet", 
with the possibility that some jets might consist of just one charged particle.  The transverse momentum of a 
charged jet, $P_{T}\!({\rm chgjet})$, is the {\it scalar} \pt\ sum of the particles in the jet.  We use the 
direction of the leading charged particle jet to define correlations in azimuthal angle, $\Delta\phi$.  
The angle $\Delta\phi=\phi-\phi_{\rm chgjet\#1}$  is the relative azimuthal angle between a charged 
particle and the direction of chgjet\#1.  The``toward" region is defined by $|\Delta\phi|<60^\circ$ and \etacut, 
while the ``away" region is $|\Delta\phi|>120^\circ$ and \etacut.   The ``transverse" region is defined by 
$60^\circ<|\Delta\phi|<120^\circ$  and \etacut.  The three regions ``toward", ``transverse", and ``away" are 
shown in Fig.~\ref{rdf_fig3}.  Each region has an area in \etaphi\ space of $4\pi/3$.  As illustrated in 
Fig.~\ref{rdf_fig3}, the ``toward" region contains the leading charged particle jet, while the ``away" region, 
on the average, contains the ``away-side" jet.  The ``transverse" region is perpendicular to the plane of the 
hard $2$-to-$2$ scattering and is therefore very sensitive to the \UE. 

\begin{figure}
\begin{center}
\includegraphics[scale=0.5]{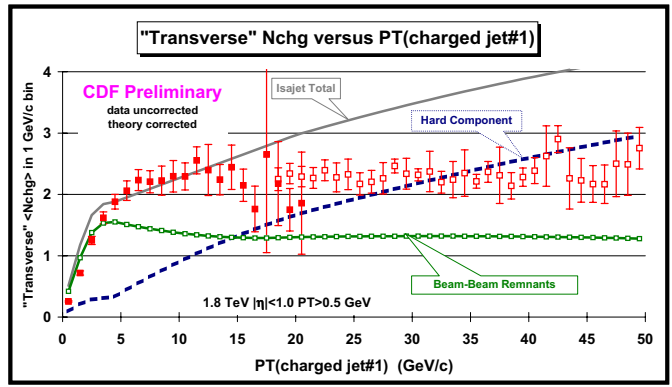}\end{center}
\caption{\em Data from Fig.~\ref{rdf_fig5} compared with the 
QCD Monte-Carlo predictions of {\tt ISAJET} 7.32 (default parameters and \hardcut). The predictions of {\tt ISAJET} are divided 
into two categories: charged particles that arise from the break-up of the beam and target ({\it beam-beam remnants}), 
and charged particles that result from the outgoing jets plus initial and final-state radiation 
({\it hard scattering component}). The theory curves are corrected for the track finding efficiency and have an 
error ({\it statistical plus systematic}) of around $5\%$.\label{rdf_fig7}
}
\end{figure}

Fig.~\ref{rdf_fig5} and Fig.~\ref{rdf_fig6} compare the ``transverse" \aveN\ and the ``transverse" \avePT, 
respectively, with the QCD 
Monte-Carlo predictions of {\tt HERWIG}, {\tt ISAJET}, and {\tt PYTHIA} 6.115 with their default parameters and \hardcut. The solid 
points are \MB\ data and the open points are the JET20 data. The JET20 data connect 
smoothly to the \MB\ data and allow us to study observables over the range $0.5 < P_T\!({\rm chgjet}\#1) < 50\gevc$.  
The average number of charged particles in the ``transverse" region doubles in going from \ptchj $=1.5\gevc$ to $2.5\gevc$ 
and then forms an approximately constant ``plateau" for \ptchj $>5\gevc$. If we 
suppose that the \UE\ is uniform in azimuthal angle $\phi$ and pseudo-rapidity $\eta$, the observed $2.3$ charged 
particles at \ptchj $=20\gevc$ translates to $3.8$ charged particles per unit pseudo-rapidity with \ptcut\
(multiply by $3$ to get $360^\circ$, divide by $2$ for the two units of pseudo-rapidity, 
multiply by $1.09$ to correct for the track 
finding efficiency).  We know that if we include all $p_T > 50$ MeV/c there are, on the average, about four charged 
particles per unit rapidity in a ``soft" proton-antiproton collision at 1.8 TeV \cite{Abe:1990td}.  The data in 
Fig.~\ref{rdf_fig5} imply that in the 
\UE\ of a hard scattering there are, on the average, about $3.8$ charged particles per unit rapidity with \ptcut!  
Assuming a charged particle \pt\ distribution of $e^{-2p_T}$ (see Fig.~\ref{rdf_fig25}) implies that there are 
roughly $10$ 
charged particles per unit pseudo-rapidity with $p_T >  0$ in the \UE\ (factor of e).  Since we examine 
only those charge particles with \ptcut, we cannot accurately extrapolate to low \pt, however, it is clear that 
the \UE\ has a charge particle density that is at least a factor of two larger than the four charged 
particles per unit rapidity seen in ``soft" proton-antiproton collisions at this energy.  

The \MB\ data were collected with a very ``loose" trigger.  The CDF \MB\ trigger requirement removes 
elastic scattering and most of the single and double diffraction events, but keeps essentially all
the ``hard-scattering" events.  
In comparing with the QCD Monte-Carlo models we do require that the models satisfy the CDF \MB\ trigger, 
however, for \ptchj $> 5\gevc$ essentially all the generated events satisfy the \MB\ trigger (\ie the 
\MB\ trigger is unbiased for large \pt\ ``jets").  If we had enough \MB\ events we would not need the JET20 data, 
but because of the fast fall-off of the cross section we run out of statistics at around \ptchj $=20\gevc$ (that 
is why the \MB\ data errors get large at around $20\gevc$).  The JET20 data were collected by requiring at least 
$20\gev$ of energy ({\it charged plus neutral}) in a cluster of calorimeter cells.  We do not use the calorimeter 
information, but instead look only at the charged particles measured in the CTC in the same way we do for the \MB\
data.  The JET20 data is, of course, biased for low \pt\ jets and we do not show the JET20 data below 
\ptchj\ around $20\gevc$.  At large \ptchj\ values the JET20 data becomes unbiased and, in fact, we 
know this occurs at around $20\gevc$ because it is here that it agrees with the ({\it unbiased}) \MB\ data.  

\begin{figure}
\begin{center}
\includegraphics[scale=0.5]{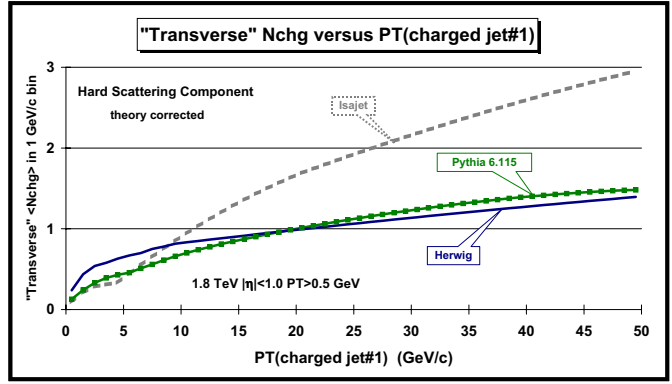}\end{center}
\caption{\em The ``hard scattering" component ({\it outgoing jets plus initial and final-state radiation}) of the number 
of charged particles (\ptcut, \etacut) in the ``transverse" region defined in Fig.~\ref{rdf_fig3} as a function of 
the transverse momentum of the leading charged jet from the QCD Monte-Carlo predictions of 
{\tt HERWIG 5.9}, {\tt ISAJET} 7.32, and {\tt PYTHIA} 6.115 with their default parameters and with \hardcut. The curves are corrected 
for the track finding efficiency and have an error ({\it statistical plus systematic}) of around $5\%$.\label{rdf_fig10}
}
\end{figure}

\begin{figure}
\begin{center}
\includegraphics[scale=0.6]{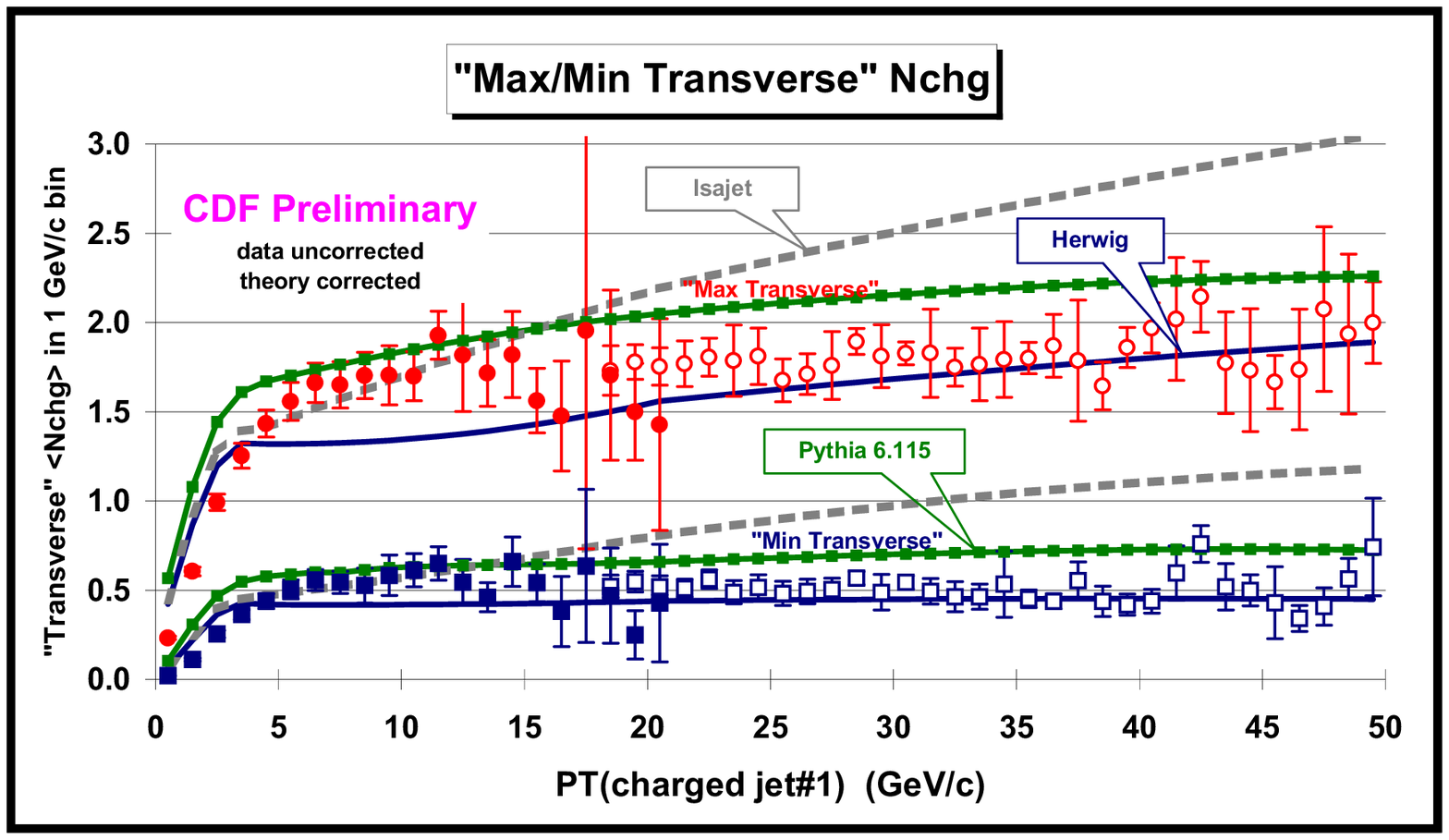}\end{center}
\caption{\em Data on the average number of ``transMAX" and ``transMIN" charged particles (\ptcut, \etacut) as a function 
of the transverse momentum of the leading charged jet compared with the QCD Monte-Carlo predictions of 
{\tt HERWIG 5.9}, {\tt ISAJET} 7.32, and {\tt PYTHIA} 6.115 with their default parameters and with \hardcut. The solid (open) points are 
the \MB\ (JET20) data. The theory curves are corrected for the track finding efficiency and have an error 
({\it statistical plus systematic}) of around $5\%$.\label{rdf_fig12}
}
\end{figure}

We expect the ``transverse" region to be composed predominately of particles that arise from the break-up of the 
beam and target and from initial and final-state radiation. This is clearly the case for the QCD Monte-Carlo models 
as can be seen in Figs.~7-9, where the predictions for the ``transverse" region are divided into two categories: charged 
particles that arise from the break-up of the beam and target ({\it beam-beam remnants}), and charged particles that 
result from the outgoing jets plus initial and final-state radiation ({\it hard scattering component}). For {\tt PYTHIA} the 
``beam-beam remnant" contribution includes contributions from multiple parton scattering.  It is interesting to see 
that in the QCD Monte-Carlo models it is the \BBR\ that are producing the approximately constant ``plateau".  
The contributions from initial-state and final-state radiation increase as \ptchj\ 
increases.  In fact, for {\tt ISAJET} it is the sharp rise in the initial-state radiation component that is causing the 
disagreement with the data for \ptchj $>20\gevc$.  The hard scattering component of {\tt HERWIG} and {\tt PYTHIA} 
does not rise nearly as fast as the hard scattering component of {\tt ISAJET}.  

There are two reasons why the hard scattering component of {\tt ISAJET} is different from {\tt HERWIG} and {\tt PYTHIA}.  The 
first is due to different fragmentation schemes.  {\tt ISAJET} uses independent fragmentation, which produces too many 
soft hadrons when partons begin to overlap.  The second difference arises from the way the QCD Monte-Carlos 
produce ``parton showers".  {\tt ISAJET} uses a leading-log picture in which the partons within the shower are ordered 
according to their invariant mass.  Kinematics requires that the invariant mass of daughter partons be less than the 
invariant mass of the parent.  {\tt HERWIG} and {\tt PYTHIA} modify the leading-log picture to include ``color coherence 
effects" which leads to ``angle ordering" within the parton shower.  Angle ordering produces less high \pt\ radiation 
within a parton shower which is what is seen in Fig.~\ref{rdf_fig10}.

Of course, the origin of an outgoing particle (``beam-beam remnant" or ``hard-scattering") is not an experimental 
observable.  Experimentally one cannot say where a given particle comes from.  However, we do know the origins of 
particles generated by the QCD Monte-Carlo models and Figs.~7-9 show the 
composition of the ``transverse" region as predicted by {\tt ISAJET}, {\tt HERWIG}, and {\tt PYTHIA}. 

\begin{figure}
\begin{center}
\includegraphics[scale=0.6]{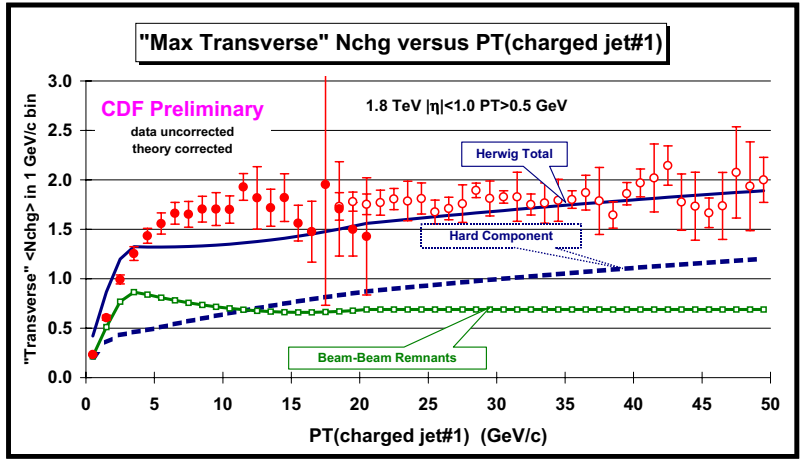}\end{center}
\caption{\em Data on the average number of ``transMAX" charged particles (\ptcut, \etacut) as a function of the 
transverse momentum of the leading charged jet compared with the QCD Monte-Carlo predictions of 
{\tt HERWIG 5.9} (default parameters and \hardcut). The predictions of {\tt HERWIG} are divided into two categories: 
charged particles that arise from the break-up of the beam and target ({\it beam-beam remnants}), and charged particles 
that result from the outgoing jets plus initial and final-state radiation ({\it hard scattering component}).  The theory 
curves are corrected for the track finding efficiency and have an error ({\it statistical plus systematic}) of 
around $5\%$.\label{rdf_fig14}
}
\end{figure}

\begin{figure}
\begin{center}
\includegraphics[scale=0.5]{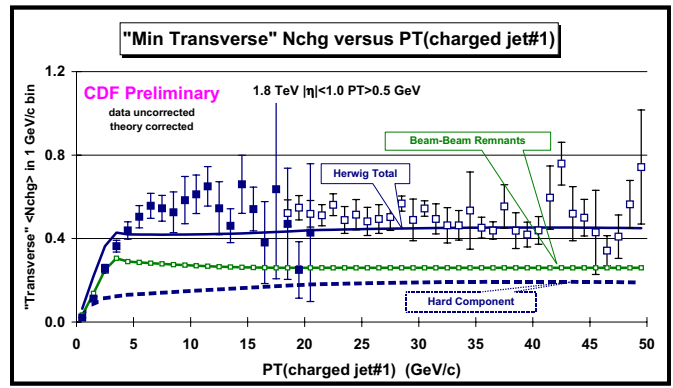}\end{center}
\caption{\em As for Fig~\ref{rdf_fig14} except that ``transMIN" is shown. \label{rdf_fig15}
}
\end{figure}

\subsubsection{Maximum and Minimum ``Transverse" Regions}

We now break up the ``transverse" region into two pieces.  As illustrated in Fig.~\ref{rdf_fig3}, on an event 
by event basis, we 
define ``transMAX" (``transMIN") to be the maximum (minimum) of the two ``transverse" pieces, 
$60^\circ<\Delta\phi<120^\circ$, \etacut, and $60^\circ<-\Delta\phi<120^\circ$, \etacut.  Each has an area 
in \etaphi\ space of $2\pi/3$ and what we previously referred 
to as the ``transverse" region is the sum of  ``transMAX" and ``transMIN".  One expects that ``transMAX"  will pick 
up more of the initial and final state radiation while ``transMIN" should be more sensitive to the ``beam-beam 
remnant" component of the \UE.  Furthermore, one expects that the ``beam-beam remnant" component 
will nearly cancel in the difference, ``transMAX" minus ``transMIN".  If this is true then the difference, ``transMAX" 
minus ``transMIN", would be more sensitive to the ``hard scattering" component (\ie initial and final-state radiation).  
I believe that this idea was first proposed by Bryan Webber and then implemented in a paper by Jon Pumplin \cite{Pumplin:1998ix} and then investigated in CDF by V. Tano~\cite{Tano:2001ab}. .
Fig.~\ref{rdf_fig12}  show the data on the \aveN\ for 
the``transMAX" and ``transMIN" region as a function of the \ptchj\ compared with QCD Monte-Carlo predictions of 
{\tt HERWIG}, {\tt ISAJET}, and {\tt PYTHIA} with their default parameters and \hardcut.
The data on  \avePT, show  similar behaviour. 
 Fig.~\ref{rdf_fig14} and  Fig.~\ref{rdf_fig15}, 
 show the data 
on \aveN\  for ``transMAX",  and ``transMIN", 
compared with QCD Monte-Carlo predictions of {\tt HERWIG}.   The predictions of {\tt HERWIG} are divided into two 
categories: charged particles that arise from the break-up of the beam and target ({\it beam-beam remnants}), 
and charged 
particles that result from the outgoing jets plus initial and final-state radiation ({\it hard scattering component}).  
It is 
clear from these plots that in the QCD Monte-Carlo models the ``transMAX" is more sensitive to the ``hard scattering 
component" of the \UE\ while ``transMIN" is more sensitive to the \BBR, especially 
at large values of \ptchj.   For example, for {\tt HERWIG} at \ptchj $=40\gevc$ the hard scattering 
component makes up $62\%$ of the ``transMAX" \aveN\ with $38\%$ coming from the \BBR.  On the 
other hand, the hard scattering component makes up only $42\%$ of the ``transMIN" \aveN\ with $58\%$ coming from 
the \BBR\ at \ptchj $=40\gevc$.  Taking difference between ``tansMAX" and ``transMIN" does 
not completely remove the ``beam-beam remnant" component, but reduces it to only about $32\%$ at \ptchj $=40\gevc$.

\begin{figure}
\begin{center}\includegraphics[scale=0.35]{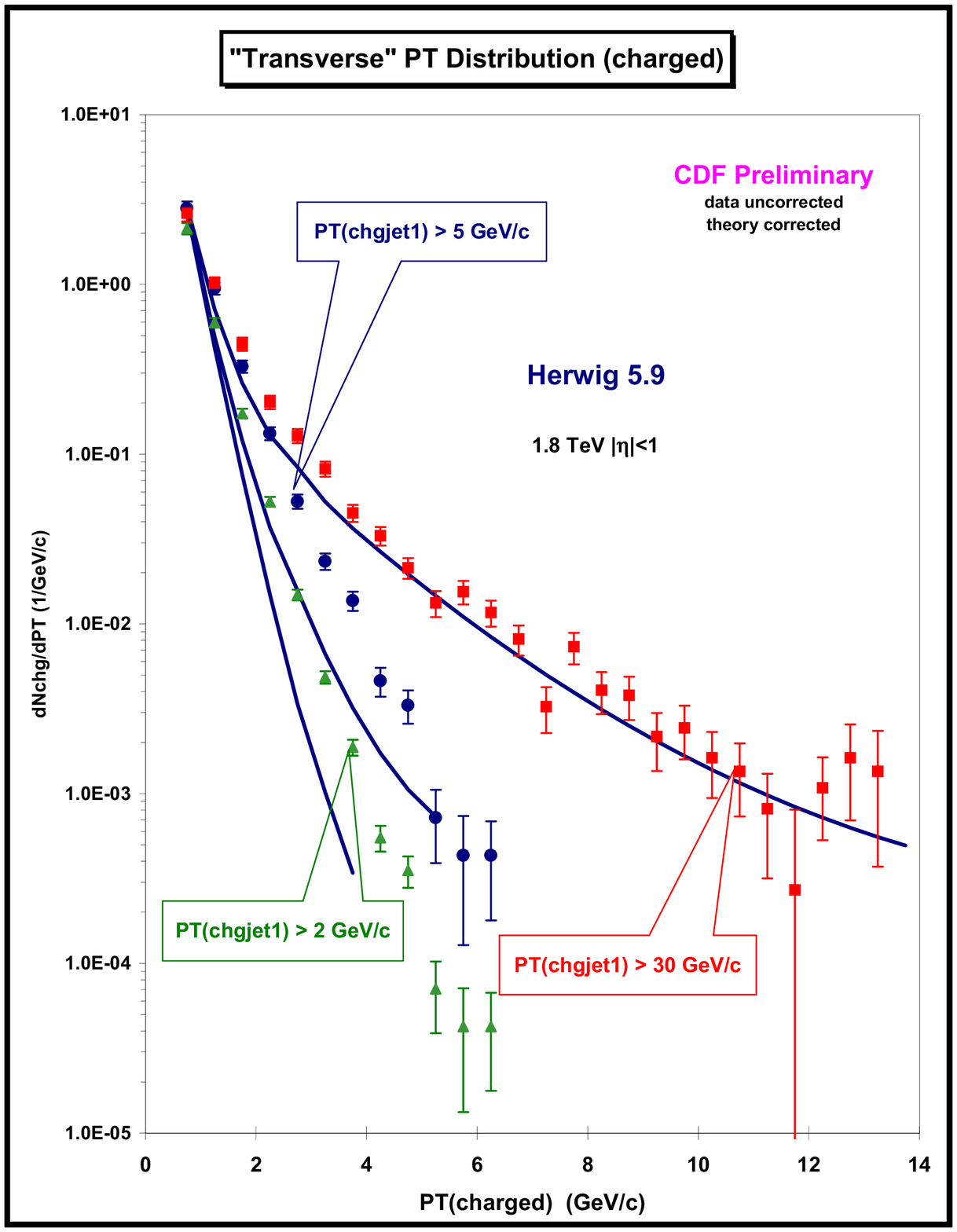}\includegraphics[scale=0.35]{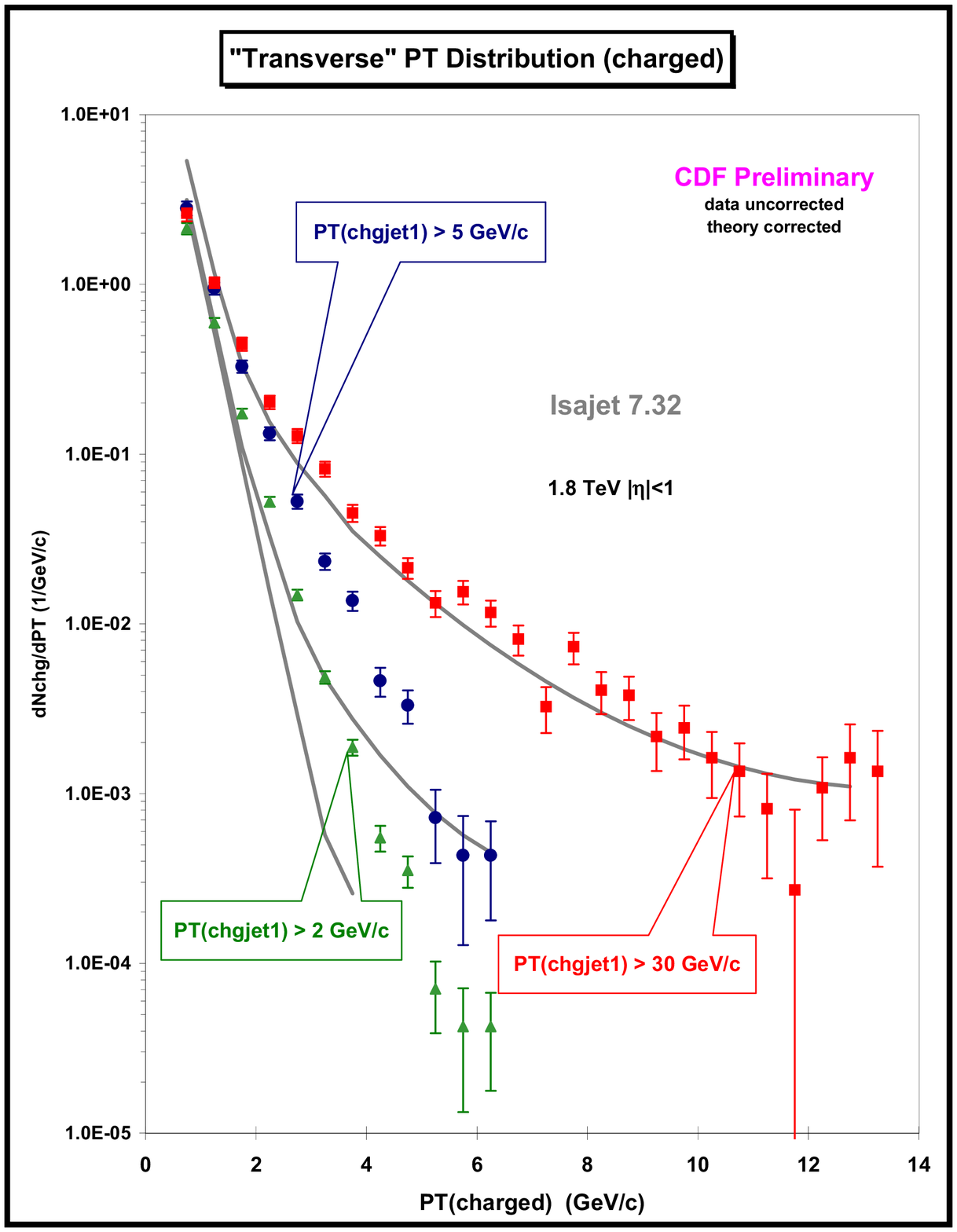}\end{center}
\caption{\em Data on the transverse momentum distribution of charged particles (\ptcut, \etacut) in the ``transverse'' region 
for \ptchj $>2\gevc$, $5\gevc$, and $30\gevc$, where chgjet\#1 is the leading charged particle jet. Each point 
corresponds to $dN_{chg}/dp_T$ and the integral of the distribution gives the average number of charged particles 
in the transverse region, $\langle N_{chg}({\rm transverse})\rangle$. The data are compared with the QCD Monte-Carlo 
model predictions of {\tt HERWIG 5.9} (default parameters and \hardcut).  The theory curves are corrected for the track 
finding efficiency and have an error ({\it statistical plus
  systematic}) of around $5\%$. Right plot is  compared with the QCD Monte-Carlo 
model predictions of {\tt ISAJET} 7.32
}
\label{rdf_fig17}
\end{figure}

\begin{figure}
\begin{center}\includegraphics[scale=0.35]{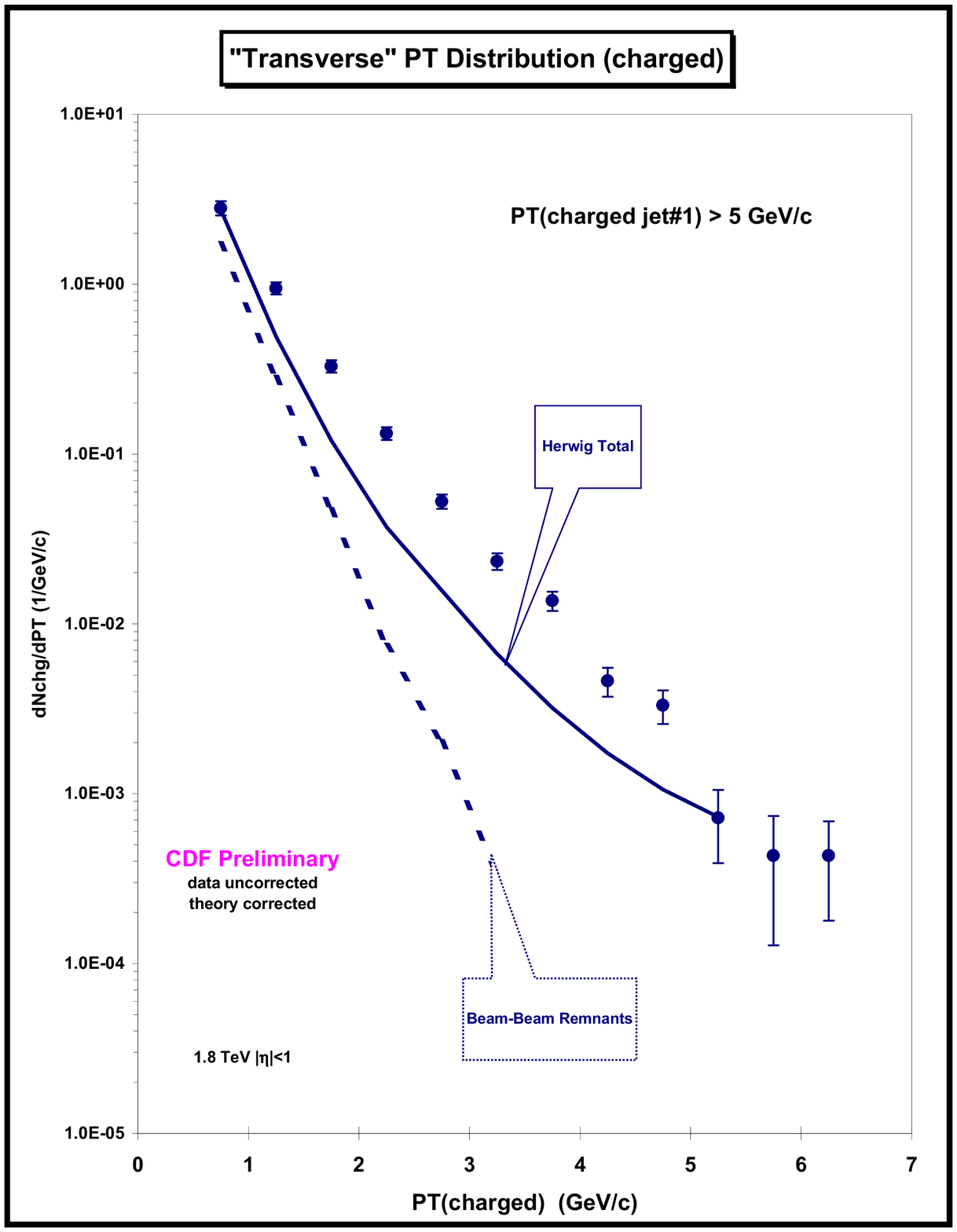}\includegraphics[scale=0.35]{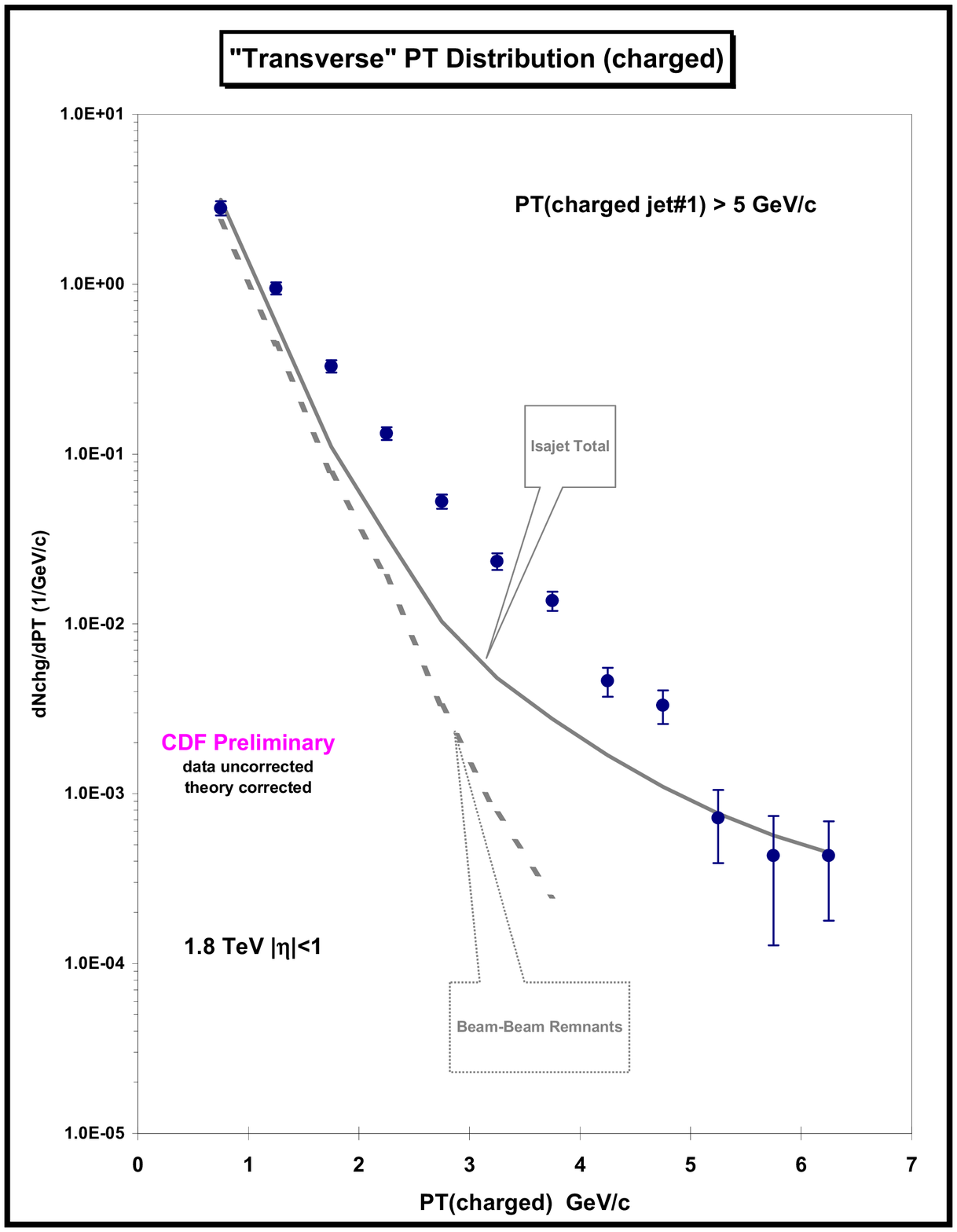}\end{center}
\caption{\em The data from Fig.~\ref{rdf_fig17}
for \ptchj $> 5\gevc$ compared with the QCD Monte-Carlo model
predictions of {\tt HERWIG 5.9} (left) and {\tt ISAJET} (right)
(default parameters and \hardcut). 
\label{rdf_fig19}
}
\end{figure}

\begin{figure}
\begin{center}
\includegraphics[scale=0.35]{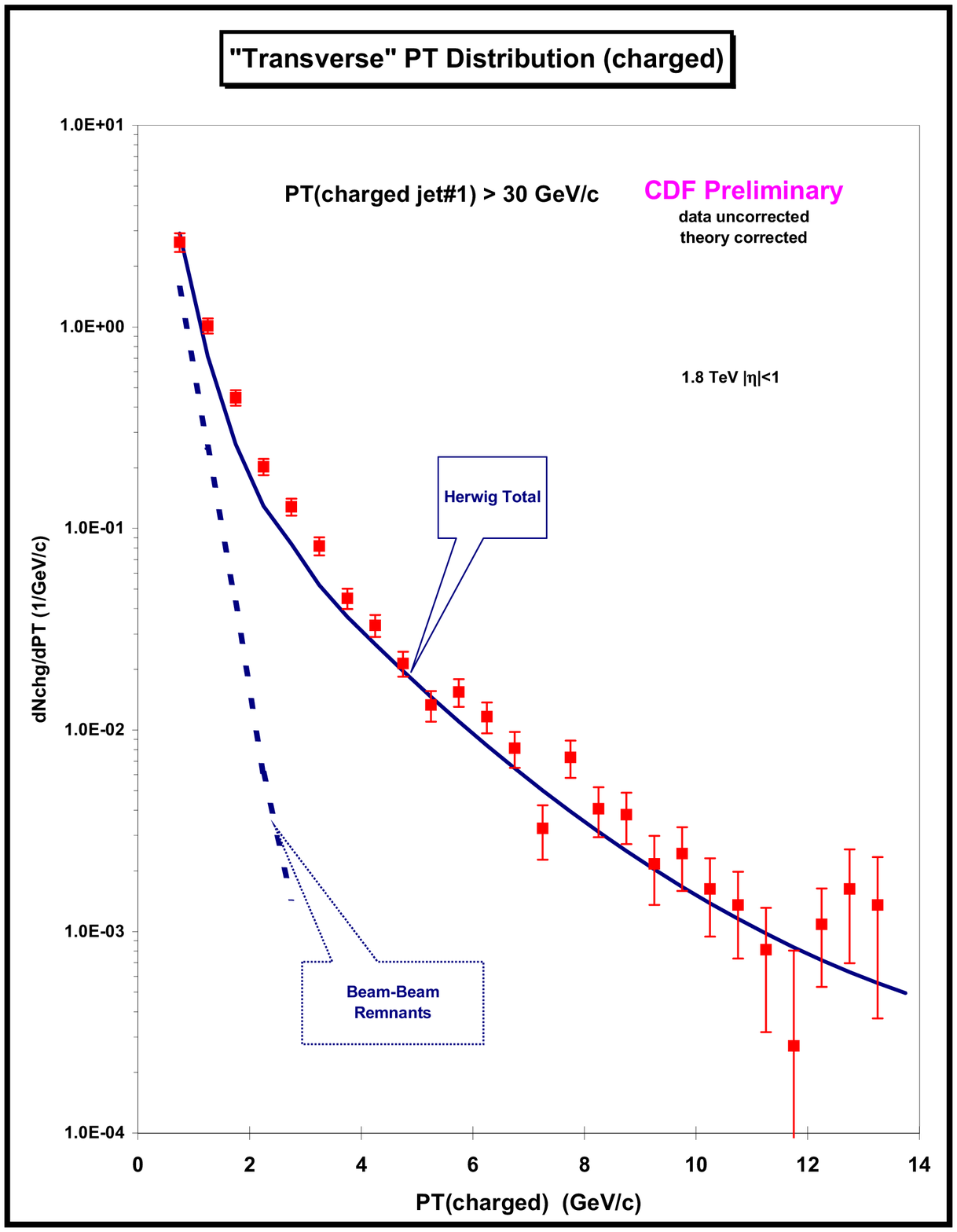}\end{center}
\caption{\em Data from Fig.~\ref{rdf_fig17}
for \ptchj $> 30\gevc$ compared with the QCD Monte-Carlo model predictions of {\tt HERWIG 5.9} 
(default parameters and \hardcut).  The theory curves are corrected for the track finding efficiency and have an 
error ({\it statistical plus systematic}) of around $5\%$. The solid curve is the total (``hard scattering" plus 
\BBR) and the dashed curve shows the contribution arising from the break-up of the beam 
particles (\BBR).\label{rdf_fig21}
}
\end{figure}

\subsubsection{The Transverse Momentum Distribution in the ``Transverse" Region}

Fig.~\ref{rdf_fig17} shows the data on the transverse momentum distribution of charged particles in the ``transverse" 
region for \ptchj $> 2\gevc$, $5\gevc$, and $30\gevc$. Each point corresponds to $dN_{\rm chg}/dp_T$ and the integral 
of the distribution gives the average number of charged particles in the ``transverse" region, \aveN, shown in 
Fig.~\ref{rdf_fig5}.  Fig.~\ref{rdf_fig5} shows only mean values, while Fig.~17 shows the 
distribution from which the mean is computed.  In Fig.~\ref{rdf_fig17} the data are compared with the QCD hard 
scattering Monte-Carlo models predictions {\tt HERWIG} and {\tt ISAJET} Since 
these distributions fall off sharply as \pt\ increases, it is essentially only the first few points at low \pt\ that 
determine the mean.  The approximately constant plateau seen in Fig.~\ref{rdf_fig5} is a result of the low \pt\ points in 
Fig.~\ref{rdf_fig17}  not changing much as \ptchj\ changes.  However, the high \pt\ points do 
increase considerably as \ptchj\ increases. This effect cannot be seen by simply examining the average number 
of  ``transverse" particles.  Fig.~\ref{rdf_fig17} shows the growth of 
the ``hard scattering component" at large \pt\ in the "transverse region" (\ie three or more hard scattering jets).

For the QCD Monte-Carlo models the \pt\ distribution in the ``transverse" region, at low values of \ptchj, is 
dominated by the ``beam-beam remnant" contribution with very little ``hard scattering" component.  This can be seen 
in Fig.~\ref{rdf_fig19}  which shows both the ``beam-beam remnant" component together 
with the total overall predictions of {\tt HERWIG} and {\tt ISAJET}, respectively, for \ptchj $>5\gevc$.  For the QCD 
Monte-Carlo models the \pt\ distribution in the ``transverse" region, at low values of \ptchj, measures directly 
the \pt\ distribution of the \BBR.  Both {\tt ISAJET} and {\tt HERWIG} have the wrong \pt\ dependence in the ``transverse" region 
due to a ``beam-beam remnant" component that falls off too rapidly as \pt\ increases.  It is, of course, understandable 
that the Monte-Carlo models might be slightly off on the parameterization of the \BBR.  This component cannot be 
calculated from perturbation theory and must be determined from data.

\begin{figure}
\begin{center}
\includegraphics[scale=0.375]{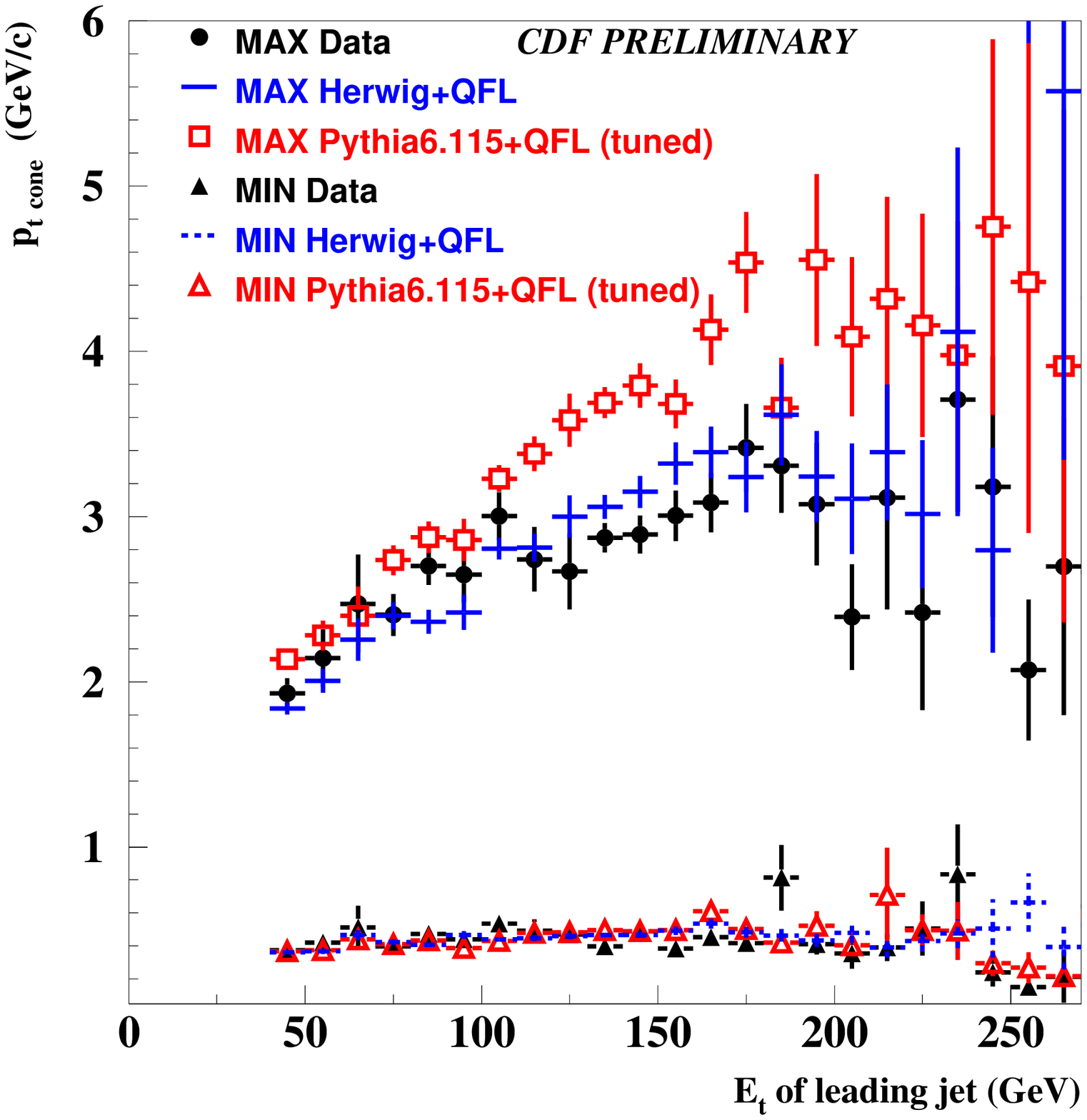}
\includegraphics[scale=0.375]{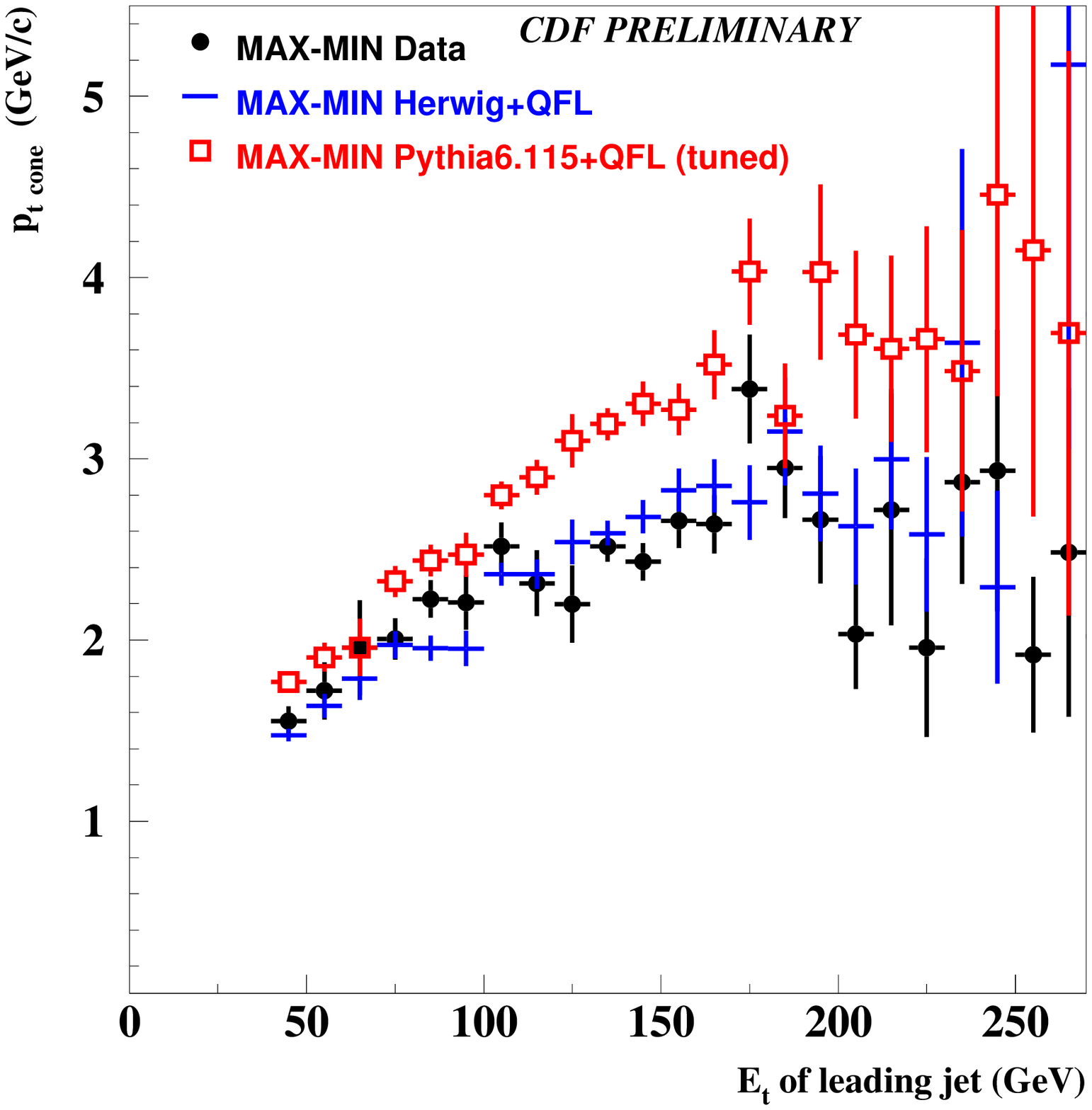}
\end{center}
\caption{\em On the left is shown data on the average {\it scalar} \pt\ sum of charged particles ($p_T > 0.4\gevc$, \etacut) within the 
maximum (MAX) and minimum (MIN) ``transverse cones" versus the transverse energy of 
the leading ({\it highest $E_T$}) ``calorimeter jet" compared with the QCD Monte-Carlo model predictions of {\tt HERWIG} and {\tt PYTHIA}.On the right is shown the difference (MAX-MIN) for each event.   
\label{rdf_fig24}
}
\end{figure}

Fig.~\ref{rdf_fig21} shows both the ``beam-beam remnant" component together with the overall prediction of {\tt HERWIG} for 
\ptchj $>30\gevc$.  Here the QCD Monte-Carlo models predict a large ``hard scattering" component 
corresponding to the production of more than two large \pt\ jets. {\tt HERWIG}, {\tt ISAJET}, and {\tt PYTHIA} all do well at 
describing the high \pt\ tail of this distribution.  However, Fig.~\ref{rdf_fig17} shows that {\tt ISAJET} produces too 
many charged particles at low \pt\ which comes from an overabundance of soft particles produced in the 
hard scattering. Fig.~\ref{rdf_fig17} shows that the large rise in the transverse charged multiplicity 
from the hard scattering component of {\tt ISAJET} seen in Fig.~\ref{rdf_fig7} comes from soft particles.  This is 
to be expected from a model that employs independent fragmentation such as {\tt ISAJET}.  Independent fragmentation 
does not differ much from color string or cluster fragmentation for the hard particles, but independent fragmentation 
produces too many soft particles.

\begin{figure}
\begin{center}
\includegraphics[scale=0.4]{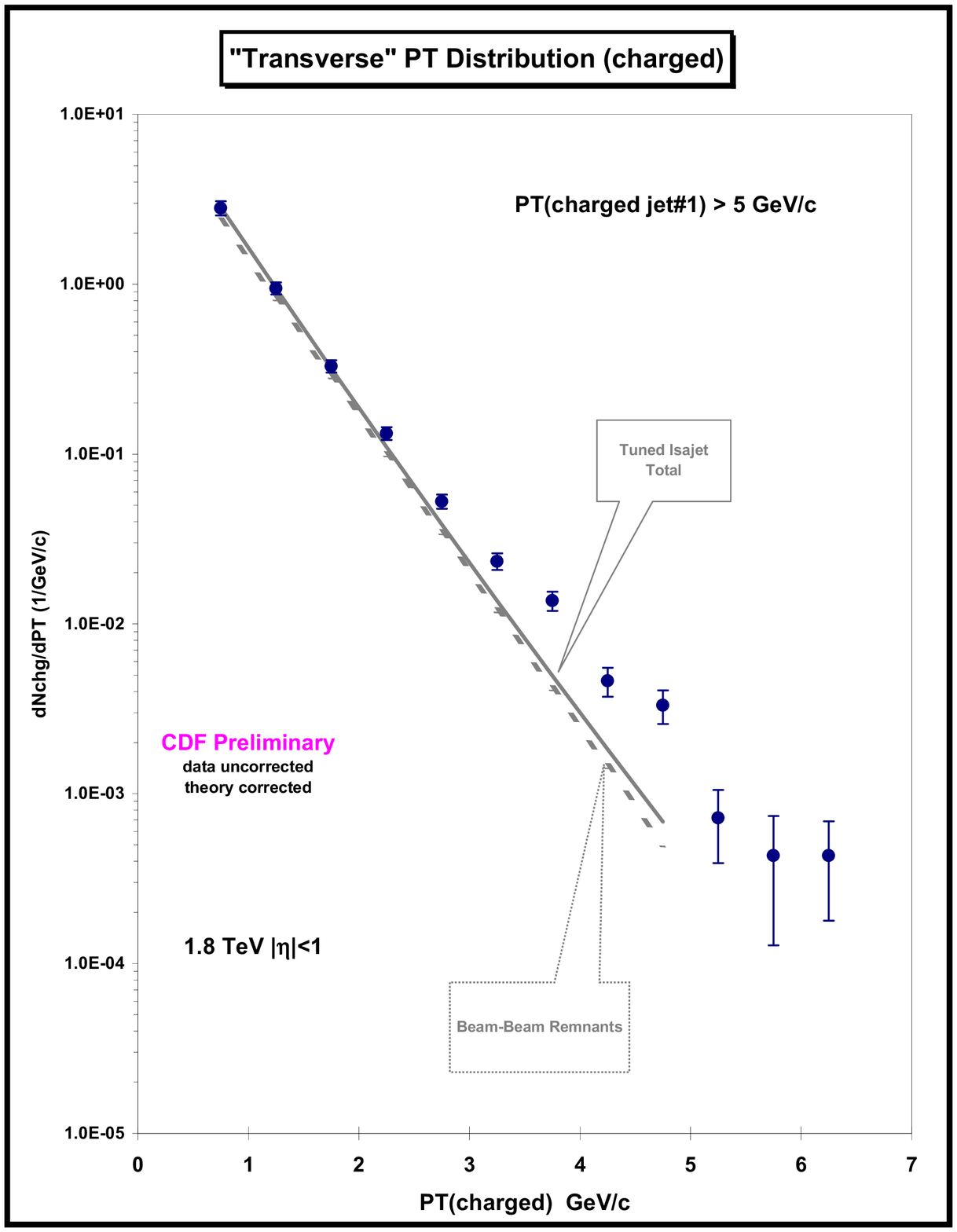}\includegraphics[scale=0.4]{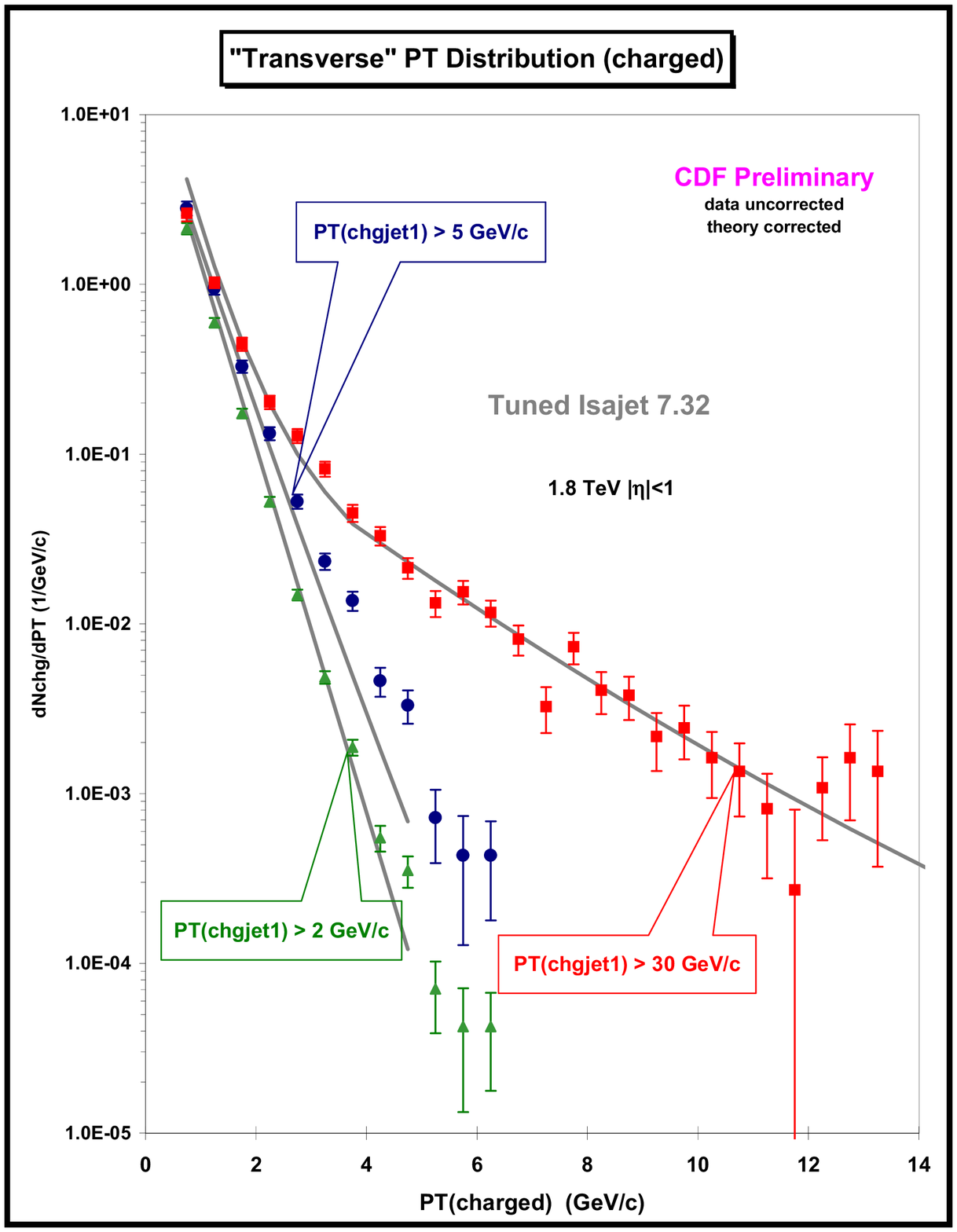}\end{center}
\caption{\em Data from Fig.~\ref{rdf_fig17}
for \ptchj $> 5\gevc$ compared with the QCD Monte-Carlo model predictions of a ``tuned" version of {\tt ISAJET} 7.32
(CUTJET$=12\gevc$, \hardcut).  For the ``tuned" version of {\tt ISAJET} the \pt\ distribution of the \BBR\ is generated 
according to $e^{-bp_T}$, where $b=2/(\gevc)$.  The theory curves are corrected for the track finding efficiency 
and have an error ({\it statistical plus systematic}) of around $5\%$. The solid curve is the total 
(``hard scattering" plus \BBR) and the dashed curve shows the contribution arising from the break-up of the 
beam particles (\BBR).\label{rdf_fig25}
}
\end{figure}

Note that the transverse momentum distribution of the ``beam-beam remnant" component for both {\tt HERWIG} and {\tt ISAJET}  
does not change in going from \ptchj $>5\gevc$ to \ptchj $>30\gevc$.  

\subsubsection{``Transverse Regions" Versus ``Transverse Cones"}

In a complementary CDF analysis, Valeria Tano \cite{Tano:2001ab} has studied the \UE\ in hard scattering processes 
by defining ``transverse cones" instead of ``transverse regions".   
The ``transverse cones" (with 
radius in \etaphi\ space of $R = 0.7$) are located at the same pseudo-rapidity as the leading jet but 
with azimuthal angle $\Delta\phi=+90^\circ$ and $\Delta\phi=-90^\circ$ relative to the leading ``jet".    
In the cone analysis the ``jet" is a ``calorimeter jet" ({\it charged plus neutrals}) defined using 
the standard CDF cluster algorithm.  Maximum (MAX) and minimum (MIN) 
``transverse" cones are determined, on an event-by-event basis, similar to the ``transMAX" and ``transMIN" regions 
described in Section III.  Each ``transverse cone" has an area in \etaphi\ space of $\pi R^2 = 0.49\pi$ 
(compared with $0.67\pi$).
Fig.~\ref{rdf_fig24} shows data at $1.8\tev$ on the average {\it scalar} \pt\ sum of charged particles 
($p_T > 0.4\gevc$, \etacut) within the 
MAX and MIN ``transverse cones" and the value of MAX-MIN for each event versus the transverse energy of the leading ({\it highest $E_T$}) ``calorimeter jet". (A similar analysis has been carried out at 630 GeV and a publication reporting the results from both energies is in preparation.) The data is 
compared with  QCD hard scattering Monte-Carlo models predictions from  {\tt HERWIG} and {\tt PYTHIA}.  The ``transverse cone" analysis 
covers the range $50 < E_T({\rm calorimeter\ jet}\#1) < 300\gev$, while the ``transverse region" analysis examines only 
charged particles and covers the range $0 <$\ptchj $<50\gevc$.  One cannot directly compare the two 
analysis, but if one scales the low $E_T({\rm jet}\#1)$ points in Fig.~\ref{rdf_fig24} by the ratio of 
areas $0.67\pi/0.49\pi=1.36$, one gets 
approximate agreement with the high \ptchj\ points.  Fig.~\ref{rdf_fig24} indicates that both {\tt HERWIG} and {\tt PYTHIA} correctly describe the MIN cone distributions but that {\tt PYTHIA} generates too much energy for the MAX cone. Both analyses together provide a  
good handle on the \UE\ in hard scattering processes.

\subsubsection{Tuning the Models to Fit the ``Underlying Event"}

\subsubsection{Tuning {\tt ISAJET}}
{\tt ISAJET} generates the \pt\ distribution of the \BBR\ according the power-law distribution 
$1/(1+P_T^2/b)^4$, where $b$ is chosen to give a mean \pt\ of primary particles (\ie before decay) of 
$450\mevc$. Fig.~\ref{rdf_fig19} indicates that this yields a \pt\ distribution that falls off too rapidly.  
Since one does not know a priori how to parameterize the \pt\ distribution of the \BBR, it is interesting to 
see if we can modify {\tt ISAJET} to do a better job at fitting the data.  Fig.~\ref{rdf_fig25}  
shows the \pt\ distribution of the ``beam-beam remnant" component together with the total overall predictions of 
a ``tuned" version {\tt ISAJET}.   For the ``tuned" version, {\tt ISAJET} is modified to generate the \pt\ distribution of 
the \BBR\ according to an exponential distribution of the form $e^{-bp_T}$, where $b=2/(\gevc)$. Also, for the 
tuned version of {\tt ISAJET} the parameter CUTJET is increased from its default value of $6\gevc$ to $12\gevc$ in 
order to reduce the amount of initial-state radiation.

\begin{figure}
\begin{center}
\includegraphics[scale=0.5]{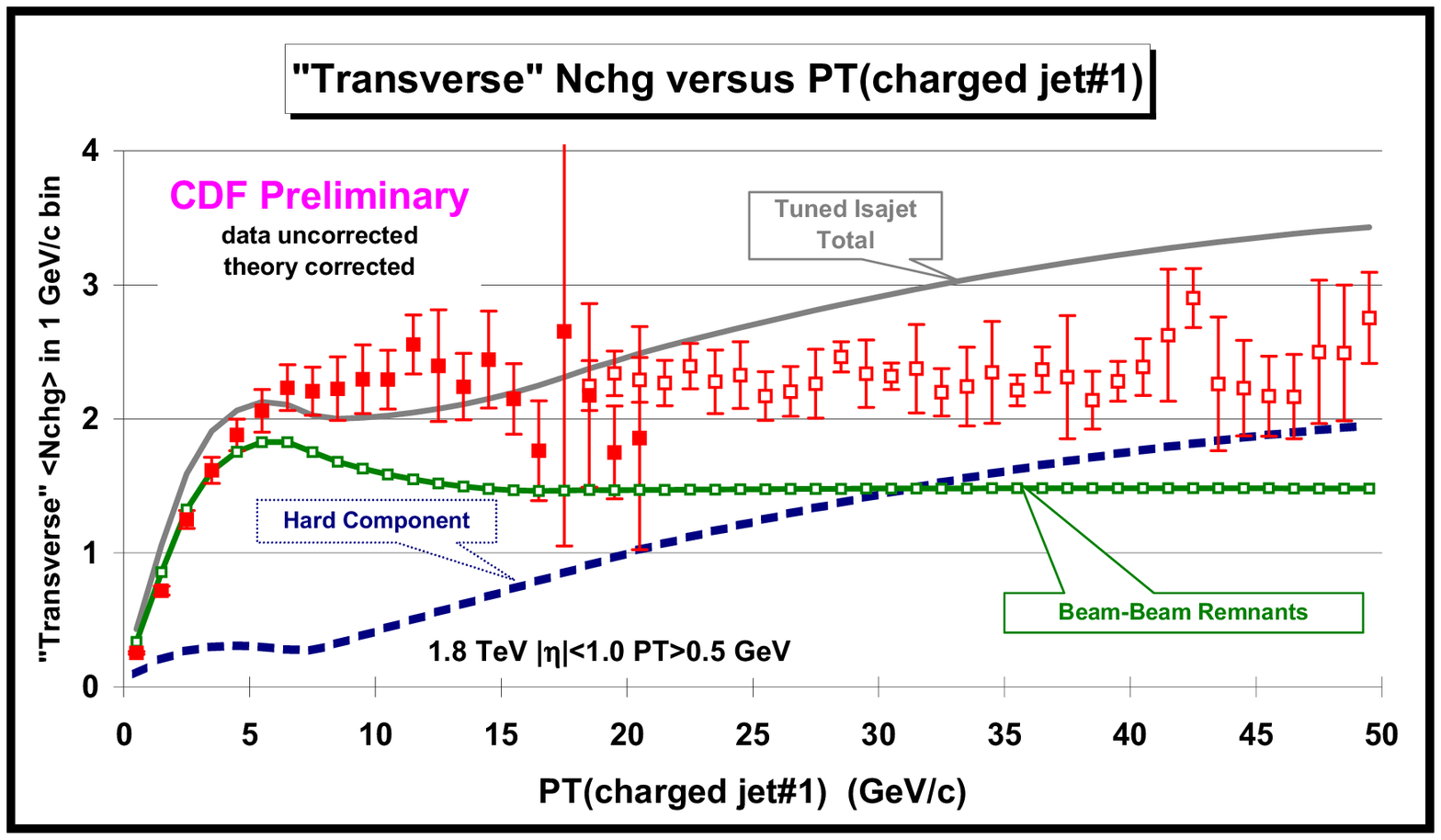}\end{center}
\caption{\em Data on the average number of charged particles (\ptcut, \etacut) in the ``transverse" region defined 
in Fig.~\ref{rdf_fig3} as a function of the transverse momentum of the leading charged jet compared with the 
QCD Monte-Carlo predictions of a ``tuned" version of {\tt ISAJET} 7.32 (CUTJET$=12\gevc$, \hardcut).  For the ``tuned'' version 
of {\tt ISAJET} the \pt\ distribution of the \BBR\ is generated according to $e^{-bp_T}$, where $b=2/(\gevc)$.
The predictions are divided into two categories: charged particles that arise from the break-up of the beam and 
target ({\it beam-beam remnants}), and charged particles that result from the outgoing jets plus initial and 
final-state radiation ({\it hard scattering component}).
\label{rdf_fig27}
}
\end{figure}

Fig.~\ref{rdf_fig27} compares the data on the average number of charged particles in the ``transverse" region 
with the ``tuned" version of {\tt ISAJET}, where the predictions for the ``transverseä region are divided into 
two categories: charged particles that arise from the break-up of the beam and target ({\it beam-beam remnants}), 
and charged particles that result from the outgoing jets plus initial and final-state radiation 
({\it hard scattering component}).  If one compares Fig.~\ref{rdf_fig27} with Fig.~\ref{rdf_fig7} one sees that 
the tuned version of {\tt ISAJET} has a larger ``beam-beam remnant" plateau and less particles from 
the ``hard-scattering" component (\ie initial-state radiation).  The ``tuned" version of {\tt ISAJET} does a much better
job fitting the \UE\ as can be seen by comparing Fig.~\ref{rdf_fig25} with Fig.~\ref{rdf_fig17}.  However,  
Fig.~\ref{rdf_fig27} shows that there are still too many charged particles in the ``transverse" region at large
\ptchj\ values.

\subsubsection{Tuning {\tt PYTHIA}}
Now that we have constructed collider observables that are sensitive to the \UE\ we would like to tune 
the multiple parton interaction parameters of {\tt PYTHIA} to fit the data.  There are many tunable parameters.  Here we 
consider only the parameters given in Table~1.  The default values of the parameters are given in 
Table~2.  Note that the {\tt PYTHIA} default values sometimes change as the version changes \footnote{ The latest versions of PYTHIA (6.120 and higher) include additional parameters that allow one to adjust the 
energy dependence of multiple parton interactions.}

\begin{table}
\begin{center}
\caption{\em {\tt PYTHIA} multiple parton scattering parameters.}
\label{rdf_table1}
\begin{tabular}{||c|c|l||}
\hline\hline
{\bf Parameter} & {\bf Value} & {\bf Description} \\
\hline\hline
MSTP(81) &      $0$  &  Multiple-Parton Scattering off \\
\hline
           &  $1$  &  Multiple-Parton Scattering on  \\
\hline
MSTP(82) &      $1$  &  Multiple interactions assuming the same probability, \\
& & with an abrupt cut-off $P_T{\rm min}$=PARP(81) \\
\hline
           &  $3$  &  Multiple interactions assuming a varying impact parameter \\
& & and a hadronic matter overlap consistent with a \\
& & single Gaussian matter distribution, with a smooth turn-off $P_{T0}$=PARP(82) \\
\hline
           &  $4$  &  Multiple interactions assuming a varying impact parameter \\
& & and a hadronic matter overlap consistent with a \\
& & double Gaussian matter distribution (governed by PARP(83) and PARP(84)) \\
& & with a smooth turn-off $P_{T0}$=PARP(82) \\
\hline\hline
\end{tabular}
\end{center}
\end{table}

\begin{table}
\begin{center}
\caption{\em Default values for some of the multiple parton scattering parameters of {\tt PYTHIA}.}
\label{rdf_table2}
\begin{tabular}{||c|c|c||}
\hline\hline
{\bf Parameter} & {\bf {\tt PYTHIA} 6.115} & {\bf {\tt PYTHIA} 6.125} \\
\hline\hline
MSTP(81) &      $1$  &  $1$ \\
\hline
MSTP(82) &  $1$  &  $1$  \\
\hline
PARP(81) &      $1.4\gevc$  & $1.9\gevc$ \\
\hline
PARP(82) &      $1.55\gevc$  & $2.1\gevc$ \\
\hline
PARP(83) &      $0.5$  & $0.5$ \\
\hline
PARP(84) &      $0.2$  & $0.2$ \\
\hline\hline
\end{tabular}
\end{center}
\end{table}

\begin{figure}
\begin{center}
\includegraphics[scale=0.5]{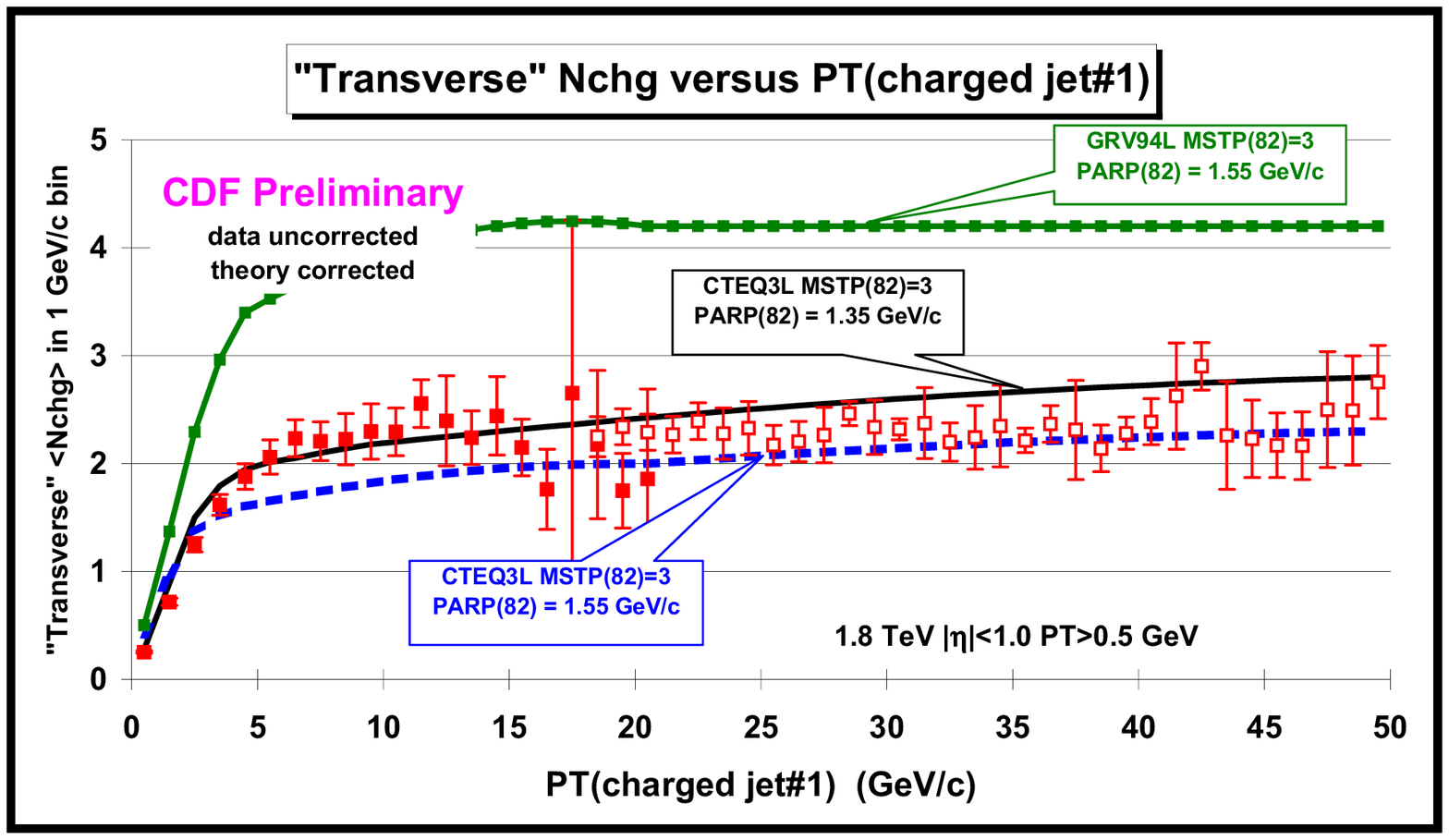}\\
\includegraphics[scale=0.5]{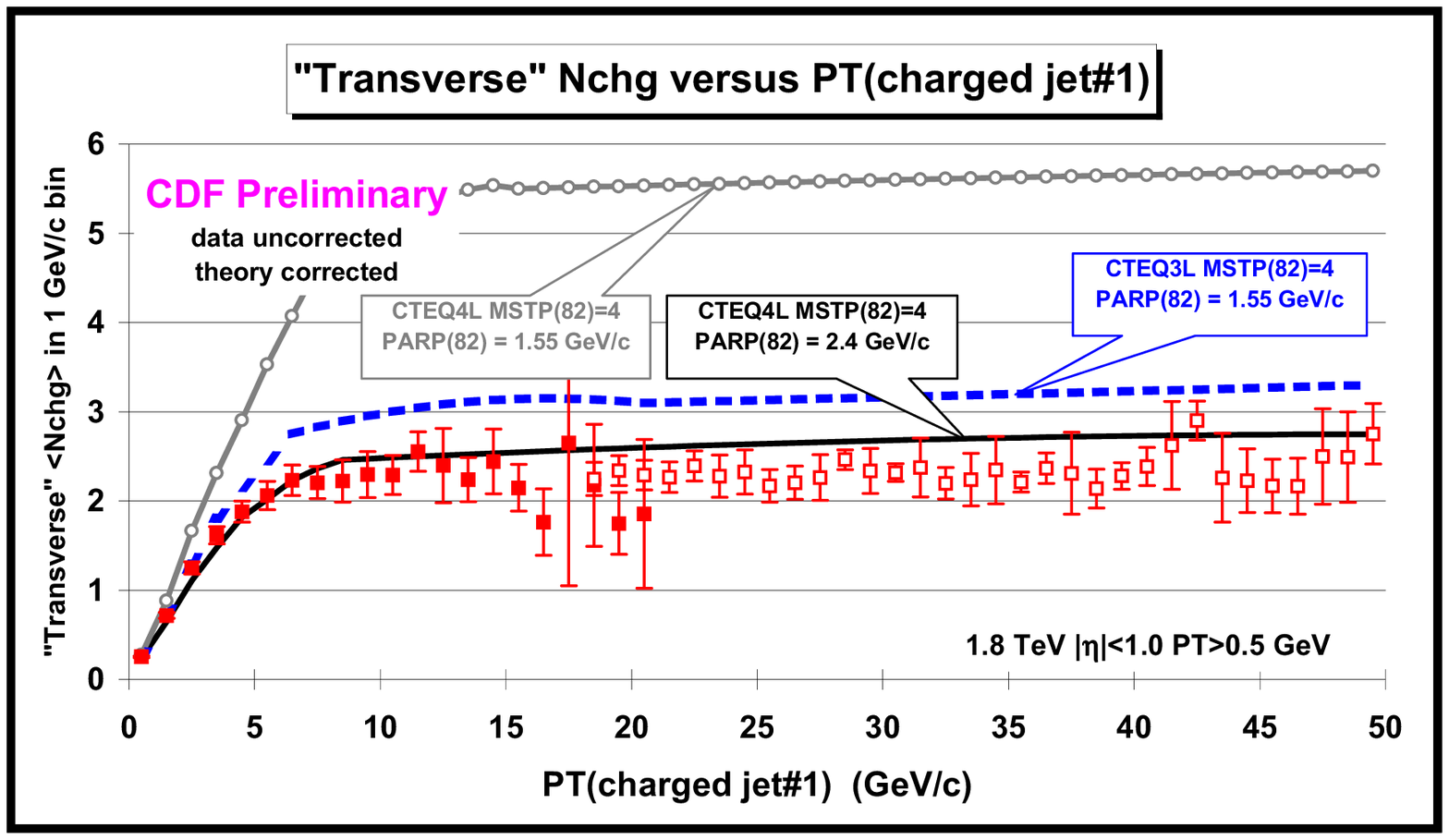}\end{center}
\caption{\em Data on the average number of charged particles (\ptcut, \etacut) in the ``transverse'' region 
as a function of the transverse momentum of the leading charged jet compared with the QCD Monte-Carlo predictions of 
{\tt PYTHIA} 6.115 with different structure functions and different multiple parton interaction parameters and 
with \hardzero. The theory curves are corrected for the track finding efficiency and have an 
error ({\it statistical plus systematic}) of around $5\%$.\label{rdf_fig28}
}
\end{figure}

Fig.~\ref{rdf_fig28}  shows data on the average number of charged 
particles in the ``transverse" region compared with 
the QCD Monte-Carlo predictions of {\tt PYTHIA} 6.115 with different structure functions and different multiple parton 
interaction parameters and with \hardzero.  For {\tt PYTHIA} the amount of multiple parton scattering depends 
on the parton distribution functions (\ie the structure functions) and hence the number of particles produced in the 
``transverse" region (\ie the \UE) changes if one changes the structure functions.  {\tt HERWIG} and 
{\tt ISAJET} do not include multiple parton scattering and for them the number of particles in the ``transverse" is 
essentially independent of the choice of structure functions.

\begin{figure}
\begin{center}
\includegraphics[scale=0.5]{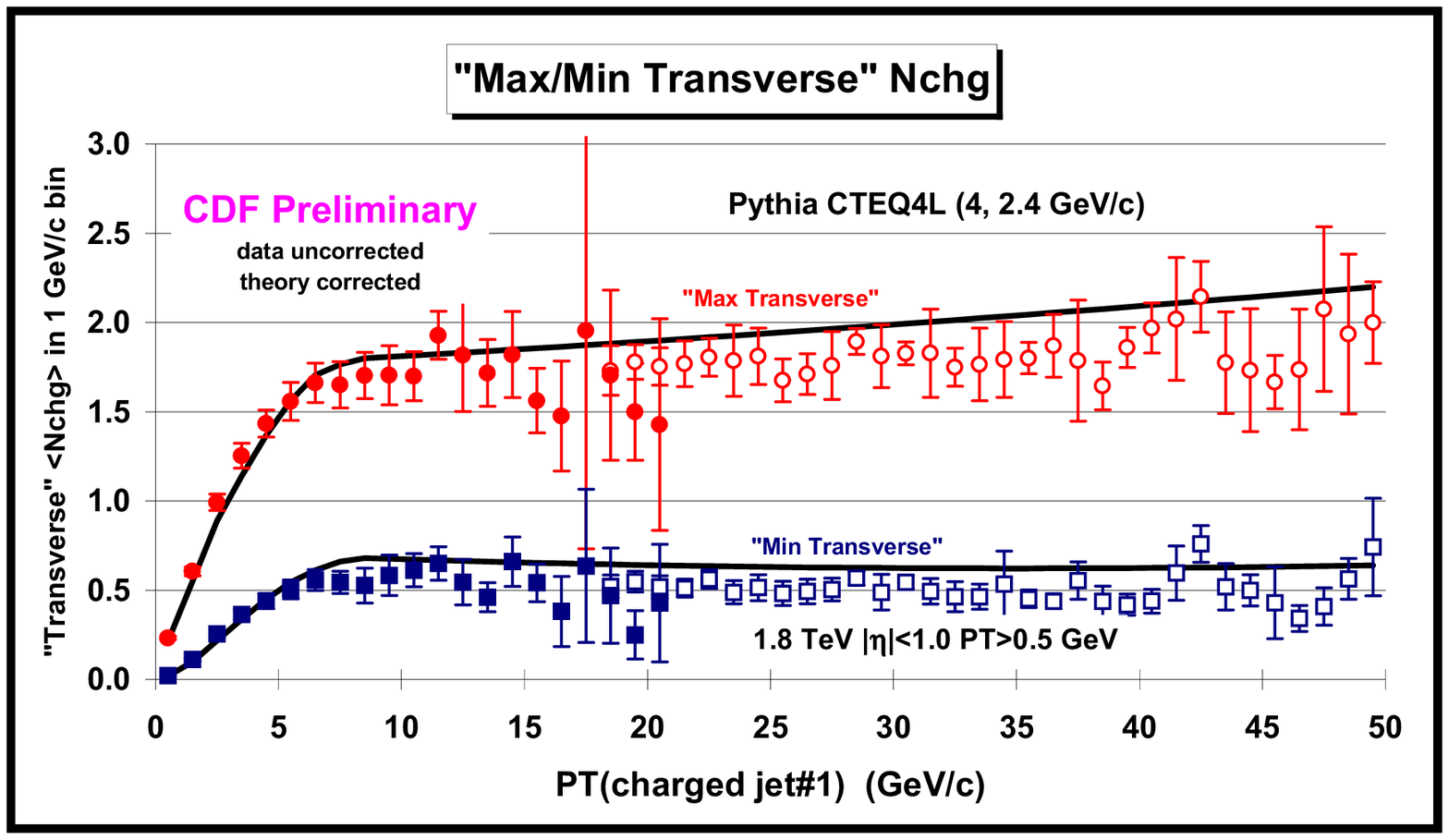}\end{center}
\caption{\em Data on the average number of ``transMAX" and ``transMIN" charged particles (\ptcut, \etacut) as a function 
of the transverse momentum of the leading charged jet defined compared with the QCD Monte-Carlo predictions of 
{\tt PYTHIA} 6.115 ({\it tuned version}, CTEQ4L, MSTP(82) = 4, PARP(82) = $1.4\gevc$, \hardzero). The theory curves are 
corrected for the track finding efficiency and have an error ({\it statistical plus systematic}) of around $5\%$. \label{rdf_fig30}
}
\end{figure}

\begin{figure}
\begin{center}
\includegraphics[scale=0.5]{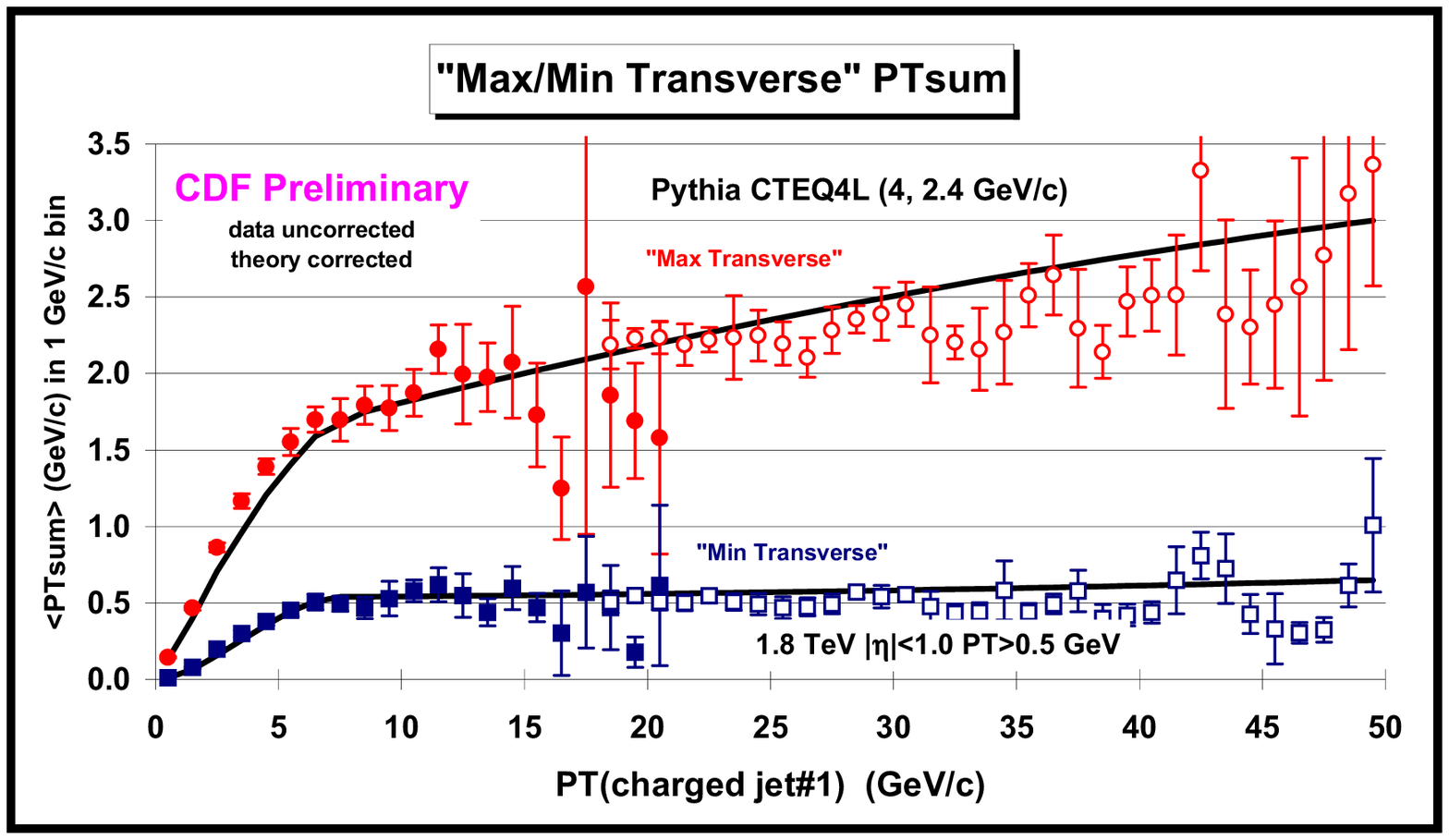}\end{center}
\caption{\em Data on the average {\it scalar} \pt\ sum of ``transMAX" and ``transMIN" charged particles (\ptcut, \etacut) 
as a function of the transverse momentum of the leading charged jet defined compared with the QCD Monte-Carlo 
predictions of {\tt PYTHIA} 6.115 ({\it tuned version}, CTEQ4L, MSTP(82) = 4, PARP(82) = $1.4\gevc$, \hardzero). The theory 
curves are corrected for the track finding efficiency and have an error ({\it statistical plus systematic}) of 
around $5\%$. \label{rdf_fig32}
}
\end{figure}

\begin{figure}
\begin{center}
\includegraphics[scale=0.4]{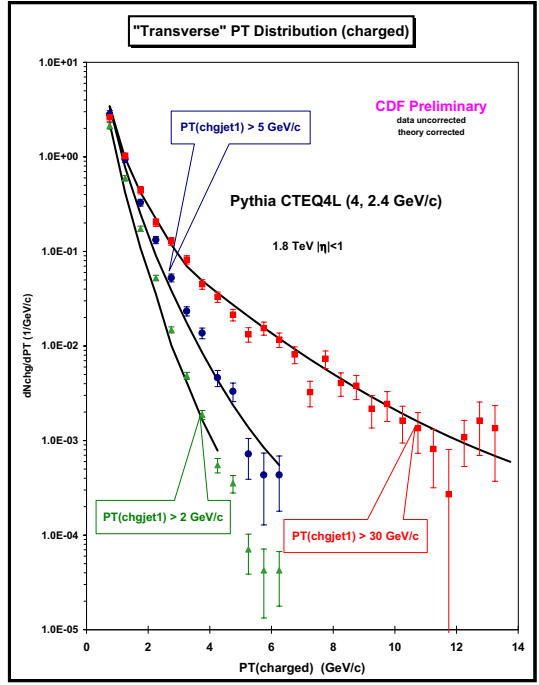}\end{center}
\caption{\em Data from Fig.~\ref{rdf_fig17} compared with the QCD 
Monte-Carlo model predictions of {\tt PYTHIA} 6.115 ({\it tuned version}, CTEQ4L, MSTP(82) = 4, PARP(82) = $1.4\gevc$, \hardzero).  
The theory curves are corrected for the track finding efficiency and have an error ({\it statistical plus systematic}) 
of around $5\%$.\label{rdf_fig34}
}
\end{figure}

Figs.~\ref{rdf_fig30}-\ref{rdf_fig34} show the results of a ``tuned" version of 
{\tt PYTHIA} 6.115 with MSTP(82) = $4$ and PARP(82) = $2.4\gevc$ 
using the CTEQ4L structure functions.  One must first choose a structure function and then tune the multiple parton 
scattering parameters for that structure function.  In generating the {\tt PYTHIA} curves in 
Figs.~30-35  we have taken 
\hardzero.  In general the perturbative $2$-to-$2$ parton scattering subprocesses diverge as \pthard\ goes to 
zero.  {\tt PYTHIA} regulates these divergences using the same cut-off parameters that are used to regulate the multiple 
parton scattering cross section (see Table~1).  This allows for the possibility of using {\tt PYTHIA} to simultaneously 
describe both ``soft" and ``hard" collisions.  Most of the CDF \MB\ events are ``soft", with less than $3\%$ of the 
events having \ptchj $>5\gevc$.  There is no clear separation between ``soft" and ``hard" collisions, but 
roughly speaking \ptchj $<2\gevc$ corresponds to ``soft" \MB\ collisions and demanding \ptchj $>5\gevc$ 
assures a ``hard" collision.  Figs.~\ref{rdf_fig30}-\ref{rdf_fig32} show that the ``tuned" version of 
{\tt PYTHIA} with \hardzero\ describes fairly well the transition between ``soft" and ``hard" collisions.  
The QCD Monte-Carlo models with \hardcut\ cannot describe the data for \ptchj $<3\gevc$ 
(see Fig.~\ref{rdf_fig5} and Fig.~\ref{rdf_fig6}), whereas {\tt PYTHIA} 
with \hardzero\ seems to do a good job on the ``transverse" observables as  \ptchj\ goes to zero.
Fig.~\ref{rdf_fig34} shows the data on the transverse momentum distribution of charged particles in the ``transverse" 
region compared with the ``tuned" version of {\tt PYTHIA} 6.115 (CTEQ4L, MSTP(82) = $4$, PARP(82) = $2.4\gevc$).  The fit is 
not perfect, but it is much better than the {\tt HERWIG} prediction shown in Fig.~\ref{rdf_fig17}.  Multiple parton 
scattering produces 
more large \pt\ particles in the ``transverse" region, which is what is needed to fit the data.  
The \pt\ distribution in the ``transverse" region, at low values of \ptchj, 
for the ``tuned" version of {\tt PYTHIA} is also 
dominated by the ``beam-beam remnant" contribution as is the case for {\tt HERWIG} (see Fig.~\ref{rdf_fig19}).  However, for 
{\tt PYTHIA} the ``beam-beam remnant" component includes contributions from multiple parton scattering, which results 
in a less steep \pt\ distribution. Also, unlike {\tt ISAJET} and {\tt HERWIG}  for {\tt PYTHIA} 
the ``beam-beam remnant" component increases as \ptchj\ increases due to multiple parton scattering.

\subsubsection{Tuning {\tt HERWIG}}
The latest version of {\tt HERWIG} includes a multiple-parton scattering option \cite{Butterworth:1996zw}, which Jon Butterworth 
has tuned to fit some of the data presented here \cite{Butterworth:2001ab}.

\subsubsection{Summary}

The \UE\ in a hard scattering process is a complicated and interesting object which involves aspects of 
both non-perturbative and perturbative QCD.  Studying the ``transMAX" and ``transMIN" pieces of the ``transverse" 
region provides additional information not contained in the sum.  In the QCD Monte-Carlo models the various 
components that make up the \UE\ are weighted differently in ``transMAX" and ``transMIN" terms.  
The ``transMAX" term preferentially selects the ``hard component" of the \UE\ ({\it outgoing jets plus 
initial and final-state radiation}) while the ``transMIN" term preferentially selects the ``beam-beam remnant" 
component.  Unfortunately one cannot cleanly isolate a single component of the \UE\ since all 
components contribute to both  ``transMAX", ``transMIN", and to the difference.  However, requiring the 
Monte-Carlo models to fit both ``transMAX" and ``transMIN" (or the sum and difference) puts additional constraints 
on the way the generators model the \UE.  

{\tt ISAJET} ({\it with independent fragmentation}) produces too many ({\it soft}) particles in the \UE\ with the 
wrong dependence on \ptchj. {\tt HERWIG} and {\tt PYTHIA} modify the leading-log picture to include ``color coherence 
effects" which leads to ``angle ordering" within the parton shower and they do a better job describing the ``underlying 
event".  Both {\tt ISAJET} and {\tt HERWIG} have the too steep of a \pt\ dependence of the ``beam-beam remnant" component 
of the \UE\ and hence do not have enough \BBR\ with \ptcut.  A modified version of {\tt ISAJET} in which the \BBR\ are
generated with an exponential distribution of the form $e^{-bp_T}$, where $b=2/(\gevc)$ improves the fit to the data.
{\tt PYTHIA} with multiple parton scattering does the best job at fitting the data.

The increased activity in the \UE\ of a hard scattering over that observed in  ``soft" collisions cannot be 
explained solely by initial-state radiation. Multiple parton interactions provide a natural way of explaining the 
increased activity in the \UE\ in a hard scattering.  A hard scattering is more likely to occur when 
the ``hard cores" of the beam hadrons overlap and this is also when the probability of a multiple parton interaction is 
greatest.  For a soft grazing collision the probability of a multiple parton interaction is small. However, multiple 
parton interactions are very sensitive to the parton structure functions (PDF).  You must first decide on a particular 
PDF and then tune the multiple parton interactions to fit the data.  

One should not take the ``tuned" version of {\tt PYTHIA} 6.115 (CTEQ4L, MSTP(82) = $4$, PARP(82) = $2.4\gevc$) 
presented here too seriously.  It is encouraging that it describes fairly well the "transverse" region over the 
range $0<$\ptchj$<50\gevc$ including the transition from ``soft" to ``hard" collisions.  However, it is still not 
quite right.  For example, it does not reproduce very well the multiplicity distribution of ``soft" collisions.  
More work needs to be done in tuning the Monte-Carlo models.  In addition, more work needs to be 
done before one can say for sure that the multiple parton interaction approach is correct.   {\tt HERWIG} without multiple 
parton scattering is not that far off the data.  Maybe we simply need to change and improve the way the 
Monte-Carlo models handle the ``beam-beam remnant" component.

\section*{Acknowledgments}

Work (ERW) supported in part by the European Community's Human Potential
Programme under contract HPRN-CT-2000-00149 Physics at Colliders. 
The work was supported in part by the Director, Office of Energy

Research, Office of High Energy and Nuclear Physics of the U.S. Department of Energy under Contracts
DE--AC03--76SF00098 (IH) and DE-AC02-98CH10886 (FP).  Accordingly, the U.S.
Government retains a nonexclusive, royalty-free license to publish or
reproduce the published form of this contribution, or allow others to
do so, for U.S. Government purposes.

\end{document}